\documentclass[aps,onecolumn,nofootinbib,amssymb,epsfig,eqsecnum,nobibnotes,floatfix]{revtex4-1}
\usepackage{amssymb,amsfonts,amsmath}
\usepackage{graphicx}
\usepackage{epstopdf}
\usepackage[all]{xy}
\usepackage{bm}
\usepackage{amsthm}
\usepackage[latin1]{inputenc}
\usepackage[english]{babel}
\usepackage{psfrag}

\usepackage{colordvi}
\usepackage{color}

\newcommand{\be}{\begin{equation}}
\newcommand{\ee}{\end{equation}}
\newcommand{\bea}{\begin{eqnarray}}

\newcommand{\eea}{\end{eqnarray}}

\newtheorem{Definition}{Definition}[section]
\newtheorem{Proposition}{Proposition}[section]

\def\be{\begin{equation}}
\def\ee{\end{equation}}
\def\bea{\begin{eqnarray}}
\def\eea{\end{eqnarray}}
\def\beq{\begin{eqnarray}}
\def\eeq{\end{eqnarray}}

\def\de#1/de#2{\frac{\partial {#1}}{\partial {#2}}}
\def\De#1/de#2{\dfrac{\partial {#1}}{\partial {#2}}}

\def\interior{\,\hbox{\vrule depth0pt height.6pt width4pt%
\vrule depth0pt height8pt}\;\,}

\begin{document}

\title{Extended Theories of Gravity}

\author{Salvatore Capozziello and Mariafelicia De Laurentis, }
\address{Dipartimento di Scienze Fisiche,  Università di Napoli "Federico II" and INFN Sezione di Napoli,\\ Complesso Universitario  di
Monte S. Angelo, Edificio G, Via Cinthia, I-80126, Napoli, Italy~}
\email{capozziello@na.infn.it}
\email{ felicia@na.infn.it}

\begin{abstract} 
Extended Theories of Gravity can be considered a new paradigm to cure shortcomings of General Relativity at infrared and ultraviolet scales. They are an approach that, by preserving the undoubtedly positive results of  Einstein's Theory,  is aimed to address conceptual and experimental problems recently emerged in Astrophysics,  Cosmology and High Energy Physics. In particular, the goal is to encompass, 
in a self-consistent scheme, problems like Inflation, Dark Energy, Dark Matter, Large Scale Structure and, first of all, to give at least an effective description of Quantum Gravity.
We review the basic principles that any gravitational theory  has to follow.  The geometrical interpretation is discussed in a broad perspective in order to highlight the basic assumptions of General Relativity and its possible extensions in the general framework of gauge theories. Principles of such modifications are presented, focusing on  specific classes of theories like $f(R)$-gravity and scalar-tensor gravity in the  metric and Palatini approaches.  The special role of torsion is also discussed. The conceptual features of these theories are fully explored and attention is payed to the issues of dynamical and conformal equivalence between them considering also the initial value problem.  A number of viability criteria are presented considering the post-Newtonian and the post-Minkowskian limits. In particular, we discuss the problems of neutrino oscillations and gravitational waves in Extended Gravity. Finally, future perspectives of Extended Gravity are considered with possibility to go beyond a trial and error approach.

\end{abstract}

\maketitle

\tableofcontents

\newpage

\section*{{\bf Part I : Generalities and open problems}}

\section{Introduction}
\label{uno}

Several issues and shortcomings emerged in the last thirty years leading to the conclusion that Einstein's General Relativity is not the final theory of gravitational interaction. The goal of this Report is to take into account some of these problems to show that more general approaches to gravity have to be pursued.

The Review article is organised in three parts. Each part has an introduction, a main body and is self-contained. 

Part I is devoted to a general discussion of gravitational interaction under the standard of gauge theories. Our aim is to show the path that led to General Relativity, its self-consistency and successes in addressing open problem at Einstein's time. A summary of shortcomings emerged later is given. In particular we discuss problems at infra-red and ultra-violet scales. After we review the Gauge Theory showing that General Relativity comes out from local gauge transformations, local Poincar\'e invariance and space-time symmetries. The role of space-time deformations and conformal transformations is discussed in view of Extended Theories of Gravity. We close this part by pointing out the physical meaning of General Relativity, in particular discussing the Equivalence Principle, the geodesic and metric structures, the post-Newtonian and post-Minkowskian limits.

Part II is the main part of this Report. After a general discussion of Extended Theories of Gravity, we provide a summary of their emergence in Quantum Field Theory formulated in curved space-time. We develop the variational principles and the field equations for some classes of Extended Theories in metric and Palatini formalism. Conformal transformations and their physical interpretation are widely discussed. The role of torsion in these theories is analysed. Besides, we develop their jet-bundle representation in order to put in evidence the role of symmetries and conserved quantities. After we present the Hamiltonian formulation showing how it can be connected to the problems of cosmological constant and renormalization at one-loop level. Finally the Initial Value Problem, in various formulations, is addressed showing that the well-formulation and well-position are still not available for any Extended Theory of Gravity.

Part III is devoted to the applications. We consider exact solutions in spherical and axial symmetry. After we develop post-Newtonian and post-Minkowskian limits. The main achievement is the fact that further features as corrections to Newton potential, new polarizations and new gravitational modes emerge with respect to General Relativity. In our opinion, these characteristics will be the test-bed capable of confirming or ruling out Extended Theories of Gravity.

We conclude the Report with some Appendices containing notations, zeta-function regularization and technicalities on Noether symmetries.

\subsection{A short history of  Theories of Gravity}
\label{uno.1}

It is remarkable that gravity is probably the fundamental interaction
which still remains the most enigmatic, even though it is related
to phenomena experienced in everyday life and is the one most easily conceived without any sophisticated knowledge.  As a matter of
fact, the gravitational interaction was the first one to have been put under
the microscope of experimental investigation, obviously due to
the simplicity of constructing a suitable experimental apparatus.

Galileo Galilei was the first to introduce pendula and inclined
planes to the study of terrestrial gravity at the end of the 16th
century \cite{galileo,galileo1,galileo2,galileo3,galileo4}. Gravity played an important role in the
development of Galileo's ideas about the necessity of experiment in
the study of Science, which had a great impact on modern scientific
thinking. However, it was not until 1665, when Isaac Newton
introduced the now renowned "inverse-square gravitational force
law", that terrestrial gravity was actually related to celestial
gravity in a single theory \cite{newtonG,newton1}. Newton's theory made correct predictions
for a variety of phenomena at different scales, including both
terrestrial experiments and planetary motion.

Obviously, Newton's contribution to gravity, quite apart from his
enormous contribution to physics overall, is not restricted to the
expression of the inverse square law. Much attention should be paid to
the conceptual basis of his gravitational theory, which incorporates
two key ideas:
\begin{enumerate}
\item The idea of absolute space, {\it i.e.} the view of
space as a fixed, unaffected structure; a rigid arena where physical phenomena take place.
\item The idea of what was later called
the Weak Equivalence Principle which, expressed in the language of
Newtonian theory, states that the inertial and the gravitational mass
coincide.
\end{enumerate}

Asking whether Newton's theory, or any other physical theory, is right or wrong, would be an  ill-posed question to begin with, since any
consistent theory is apparently "right". A more appropriate way to
pose the question would be to ask how suitable this theory is to
describe the physical world or, even better, how large a portion of
the physical world is sufficiently described by such a theory. Also, one
could ask how unique the specific theory is for the description of the
relevant phenomena. It was obvious, in the first 20 years after the
introduction of Newtonian gravity, that it did manage to explain all of
the aspects of gravity known at that time. However, all of the
questions above were posed sooner or later.

In 1855, Urbain Le Verrier observed a 35 arc-second excess precession
of Mercury's orbit and later on, in 1882, Simon Newcomb measured this
precession more accurately to be 43 arc-seconds \cite{lever,lever1,newc,mach,mach1}. This experimental
fact was not predicted by Newton's theory. It should be noted that Le
Verrier initially tried to explain the precession within the context
of Newtonian gravity, attributing it to the existence of another, yet
unobserved, planet whose orbit lies within that of Mercury. He was
apparently influenced by the fact that examining the distortion of the
planetary orbit of Uranus in 1846 had led him, and, independently,
John Couch Adams, to the discovery of Neptune and the accurate
prediction of its position and momenta. However, this innermost planet
was never found.

On the other hand, in 1893, Ernst Mach stated what was later called by
Albert Einstein "Mach's principle". This is the first constructive
attack to Newton's idea of absolute space after the 18th century debate between Gottfried Wilhelm von Leibniz and Samuel Clarke (Clarke was acting as Newton's spokesman) on the same subject, known as the Leibniz--Clarke Correspondence \cite{alexander}. Mach's idea can be
considered as rather vague in its initial formulation and it was
essentially brought into the mainstream of physics later on by Einstein along
the following lines:

\vspace{0.5cm}
\noindent {\it ``...inertia originates in a kind of interaction between bodies...".}
\vspace{0.5cm}

\noindent This is obviously in contradiction with Newton's
ideas, according to which inertia was always relative to the absolute
frame of space. There exists also a later, probably clearer
interpretation of Mach's Principle, which, however, also differs in substance. This was given by Dicke \cite{dicke}:
\vspace{0.5cm}

\noindent {\it "The gravitational constant should be a function of the mass distribution in the Universe".}

\vspace{0.5cm}

\noindent This is different from Newton's idea of the
gravitational constant as being universal and unchanging. Now Newton's
basic axioms have to be reconsidered.

But it was not until 1905, when Albert Einstein completed Special
Relativity, that Newtonian gravity would have to face a serious
challenge. Einstein's new theory, which managed to explain a series of
phenomena related to non-gravitational physics, appeared to be
incompatible with Newtonian gravity. Relative motion and all the
linked concepts had gone well beyond Galileo and Newton ideas
and it seemed that Special Relativity should somehow be generalised to
include non-inertial frames. In 1907, Einstein introduced the
equivalence between gravitation and inertia and successfully used it
to predict the gravitational redshift. Finally, in 1915, he completed
the theory of General Relativity (GR), a generalisation of Special
Relativity which included gravity and any accelerated frame.
 Remarkably, the theory matched
perfectly the experimental result for the precession of Mercury's
orbit, as well as other experimental findings like the Lense-Thirring \cite{LT,LT1,LT2,LT3}
gravitomagnetic precession (1918) and the gravitational deflection of
light by the Sun, as measured in 1919 during a Solar eclipse by Arthur
Eddington \cite{eddington}.

GR overthrew Newtonian gravity and continues to be up
to now an extremely successful and well-accepted theory for
gravitational phenomena. As mentioned before, and as often happens
with physical theories, Newtonian gravity did not lose its appeal to
scientists. It was realised, of course, that it is of limited validity
compared to GR, but it is still sufficient for most
applications related to gravity. What is more, in weak field limit of
gravitational field strength and velocities, GR
inevitably reduces to Newtonian gravity. Newton's equations for
gravity might have been generalised and some of the axioms of his
theory may have been abandoned, like the notion of an absolute frame,
but some of the cornerstones of his theory still exist in the
foundations of GR, the most prominent example being
the Equivalence Principle, in a more suitable formulation of course.

This brief chronological review, besides its historical interest, is
outlined here also for a practical reason. GR is bound
to face the same questions as were faced by Newtonian gravity and many people
would agree that it is actually facing them now. In the forthcoming
sections, experimental facts and theoretical problems will be
presented which justify that this is indeed the case. Remarkably,
there exists a striking similarity to the problems which Newtonian
gravity faced, {\it i.e.} difficulty in explaining several
observations, incompatibility with other well established theories and
lack of uniqueness. 

\subsection{What we mean for a "Good Theory of Gravity"}
\label{uno.2}

From a phenomenological point of view, there are some minimal
requirements that any relativistic theory of gravity has to match.
First of all, it has to explain the astrophysical observations
({\it e.g.} the orbits of planets, self-gravitating
structures).

This means that it has to reproduce the Newtonian dynamics in the
weak-energy limit. Besides, it has to pass the classical Solar
System tests which are all experimentally well founded
\cite{Will93}.
As second step, it should reproduce Galactic dynamics considering
the observed baryonic constituents ({\it e.g.} luminous components as
stars, sub-luminous components as planets, dust and gas),
radiation and Newtonian potential which is, by assumption,
extrapolated to Galactic scales.

Thirdly, it should address the problem of large scale structure
({\it e.g.} clustering of galaxies) and finally cosmological dynamics,
which means to reproduce, in a self-consistent way, the
cosmological parameters as the expansion rate, the Hubble
constant, the density parameter and so on. Observations and
experiments, essentially, probe the standard baryonic matter, the
radiation and an attractive overall interaction, acting at all
scales and depending on distance: the gravity.

The simplest theory which try to satisfies the above requirements
is the GR \cite{einstein}. It is firstly based, on the assumption that space and
time have to be entangled into a single space-time structure,
which, in the limit of no gravitational forces, has to reproduce
the Minkowski space-time structure. Einstein profitted also of
ideas earlier put forward by Riemann, who stated that the Universe
should be a curved manifold and that its curvature should be
established on the basis of astronomical observations
\cite{riemann}.

In other words, the distribution of matter has to influence point
by point the local curvature of the space-time structure. The
theory, eventually formulated by Einstein in 1915, was strongly
based on three assumptions that the Physics of Gravitation has to
satisfy.
\begin{quote}
The "{\it Principle of Relativity}", that requires all
frames to be good frames for Physics, so that no preferred
inertial frame should be chosen a priori (if any exist).
\end{quote}
\begin{quote}
The "{\it Principle of Equivalence}", that amounts to require
inertial effects to be locally indistinguishable from
gravitational effects (in a sense, the equivalence between the
inertial and the gravitational mass).
\end{quote}
\begin{quote}
The "{\it Principle of General Covariance}", that requires field
equations to be "generally covariant" (today, we would better say
to be invariant under the action of the group of all space-time
diffeomorphisms) \cite{schroedinger}.
\end{quote}
\begin{quote}
And - on the top of these three principles - the requirement that
causality has to be preserved (the "{\it Principle of Causality}",
{\it i.e.} that each point of space-time should admit a universally valid
notion of past, present and future).
\end{quote}
Let us also recall that the older Newtonian theory of space-time
and gravitation, that Einstein wanted to reproduce at least in
the limit of weak gravitational forces (what is called today the
"post-Newtonian approximation"), required space and time to be
absolute entities, particles moving in a preferred inertial frame
following curved trajectories, the curvature of which ({\it e.g.} the
acceleration) had to be determined as a function of the sources
({\it i.e.} the "forces").

On these bases, Einstein was  led to postulate that the
gravitational forces have to be expressed by the curvature of a
metric tensor field $ds^2 = g_{\mu\nu}dx^{\mu}dx^{\nu}$ on a
four-dimensional space-time manifold, having the same signature of
Minkowski metric, {\it e.g.}, the so-called "Lorentzian signature",
herewith assumed to be $(-,+,+,+)$. He also postulated that
space-time is curved in itself and that its curvature is locally
determined by the distribution of the sources, {\it e.g.}, being
space-time a continuum, by the four-dimensional generalization of
what in Continuum Mechanics is called the "matter stress-energy
tensor", {\it e.g.} a rank-two (symmetric) tensor  $T^{(m)}_{\mu\nu}$.

Hilbert and Einstein \cite{schroedinger} proved that the field equations for a metric tensor $g_{\mu\nu}$, related to a given distribution of matter-energy, can be achieved by starting from the Ricci curvature scalar $R$ which is an invariant. We will give details below.

The choice of Hilbert and Einstein was completely arbitrary  (as
it became clear a few years later), but it was certainly the
simplest one both from the mathematical and the physical point of view. As it was later clarified by Levi--Civita in 1919,
curvature is  not a "purely metric notion" but, rather, a notion
related to the   "linear connection"  to which "parallel
transport" and "covariant derivation" refer \cite{levicivita}.

In a sense, this is the precursor idea of what in the sequel would
be called a "gauge theoretical framework" \cite{gauge}, after the
pioneering work by Cartan in 1925 \cite{cartan}. But at the time
of Einstein, only metric concepts were at hands and his solution
was the only viable.

It was later clarified  that the three principles of relativity,
equivalence and covariance, together with causality, just require
that the space-time structure has to be determined by either one or
both of two fields, a Lorentzian metric $g$ and a linear
connection $\Gamma$, assumed at the beginning to be torsionless for the sake of
simplicity.

The metric  $g$  fixes the causal structure of space-time (the
light cones) as well as its metric relations (clocks and rods);
the connection  $\Gamma$  fixes the free-fall, {\it i.e.} the locally
inertial observers. They have, of course, to satisfy a number of
compatibility relations which amount to require that photons
follow null geodesics of $\Gamma$, so that $\Gamma$ and $g$ can be
independent, {\it a priori}, but constrained, {\it a posteriori},
by some physical restrictions. These, however, do not impose that
$\Gamma$ has necessarily to be the Levi-Civita connection of  $g$
\cite{palatiniorigin}.

This justifies - at least on a purely theoretical basis - the fact
that one can envisage the so-called "alternative theories of
gravitation", that we prefer  to call "{\it Extended Theories of
Gravitation}" (ETGs) since their starting points are exactly those
considered by Einstein and Hilbert: theories in which gravitation
is described by either a metric (the so-called "purely metric
theories"), or by a linear connection (the so-called "purely
affine theories") or by both fields (the so-called "metric-affine
theories", also known as "first order formalism theories"). In
these theories,  the Lagrangian is a scalar density  of the
curvature invariants constructed out of both $g$ and $\Gamma$.

The choice of Hilbert-Einstein Lagrangian is by no means unique and it turns out that
the Hilbert-Einstein Lagrangian is in fact the only choice that
produces an invariant that is linear in second derivatives of the
metric (or first derivatives of the connection). A Lagrangian
that, unfortunately, is rather singular from the Hamiltonian
point of view, in much than same way as Lagrangians, linear in
canonical momenta, are rather singular in Classical Mechanics (see
{\it e.g.} \cite{arnold}).

A number of attempts to generalize GR (and unify it to
Electromagnetism) along these lines were  followed by Einstein
himself and many others (Eddington, Weyl,
Schr\"{o}dinger, just to
quote the main contributors; see, {\it e.g.}, \cite{unification}) but
they were eventually given up in the fifties of XX Century, mainly
because of a number of difficulties related to the definitely more
complicated structure of a non-linear theory (where by
"non-linear" we mean here a theory that is based on non-linear
invariants of the curvature tensor), and also because of the new
understanding of physics that is currently based on four
fundamental forces and requires the more general "gauge framework"
to be adopted (see \cite{unification2}).

Still a number of sporadic investigations about "alternative
theories" continued even after 1960 (see \cite{Will93} and Refs.
quoted therein for a short history). The search for a coherent
quantum theory of gravitation or the belief that gravity has to be
considered as a sort of low-energy limit of string theories \cite{green}, something that we are not willing to enter
here in detail, has more or less recently revitalized the idea
that there is no reason to follow the simple prescription of
Einstein and Hilbert and to assume that gravity should be
classically governed by a Lagrangian linear in the curvature.

Further curvature invariants or non-linear functions of them
should be also considered, especially in view of the fact that
they have to be included in both the semi-classical expansion of a
quantum Lagrangian or in the low-energy limit of a string
Lagrangian.

Moreover, it is clear from the recent astrophysical observations
and from the current cosmological hypotheses that Einstein
equations are no longer a good test for gravitation  at Solar
System, Galactic, extra-galactic and cosmic scale, unless one does
not admit that the matter side of field equations contains some
kind of exotic matter-energy which is the "dark matter" and "dark
energy" side of the Universe.

The idea which we propose here is much simpler. Instead of
changing the matter side of Einstein field equations in
order to fit the "missing matter-energy" content of the currently
observed Universe (up to the $95\%$ of the total amount!), by
adding any sort of inexplicable and strangely behaving matter and
energy, we claim that it is simpler and more convenient to change
the gravitational side of the equations, admitting corrections
coming from non-linearities in the effective Lagrangian. However, this is
nothing else but a matter of taste and, since it is possible, such
an approach should be explored. Of course, provided that the
Lagrangian can be conveniently tuned up ({\it i.e.}, chosen in a huge
family of allowed Lagrangians) on the basis of its best fit with
all possible observational tests, at all scales (Solar, Galactic,
extragalactic and cosmic).

Something that, in spite of some commonly accepted but disguised
opinion, can and should be done before rejecting a priori a
non-linear theory of gravitation (based on a non-singular
Lagrangian) and insisting that the Universe has to be necessarily
described by a rather singular gravitational Lagrangian (one that
does not allow a coherent perturbation  theory from a good
Hamiltonian point of view) accompanied by matter that does not follow
the behaviour that standard baryonic matter, probed in our
laboratories, usually satisfies.

\subsection{General Relativity and its shortcomings}
\label{uno.3}

Considering the above discussion it is worth noticing that
in the last thirty years several shortcomings came out in the Einstein theory and people began
to investigate whether GR is the only fundamental theory capable of explaining
the gravitational interaction. Such issues come, essentially, from cosmology and quantum field
theory.
The  shortcomings are related both to many theoretical aspects
and to observational results. In this section we will try to summarize these problems. An
important issue has to be underlined: even if there are many problems, the reaction of scientific community is not uniform. In a very simple scheme we can summarize the guide lines.

Many people will agree that modern physics is based on two main pillars:
GR and Quantum Field Theory. Each of these two
theories has been very successful in its own arena of physical
phenomena: GR in describing gravitating systems and
non-inertial frames from a classical point of view on large enough
scales, and Quantum Field Theory at high
energy or small scale regimes where a classical description breaks
down.  However, Quantum Field Theory assumes that space-time is flat
and even its extensions, such as Quantum Field Theory in curved space
time, consider space-time as a rigid arena inhabited by quantum fields.
GR, on the other hand, does not take into account the
quantum nature of matter. Therefore, it comes naturally to ask what
happens if a strong gravitational field is present at quantum scales. How do quantum fields behave in the
presence of gravity? To what extent are these amazing theories
compatible?

Let us try to pose the problem more rigorously. Firstly, what needs to
be clarified is that there is no final proof that gravity should
have some quantum representation at high energies or small scales, or
even that it will retain its nature as an interaction. The
gravitational interaction is so weak compared with other interactions
that the characteristic scale under which one would expect to
experience non-classical effects relevant to gravity, the Planck
scale, is $10^{-33}$ cm. Such a scale is not of course accessible by
any current experiment and it is doubtful whether it will ever be
accessible to future experiments either
\footnote{This fact does not imply, of course, that imprints of Quantum Gravity phenomenology 
cannot be found in lower energy experiments.}. However, there are a number
of reasons for which one would prefer to fit together GR and Quantum Field Theory \cite{brill, isham}. Let us list
some of the most prominent ones here and leave the discussion about
how to address them for the next subsection.
Curiosity is probably the motivation leading scientific research. From
this perspective it would be at least unusual if the gravity research
community was so easily willing to abandon any attempt to describe the
regime where both quantum and gravitational effects are important. The
fact that the Planck scale seems currently experimentally inaccessible
does not, in any way, imply that it is physically irrelevant. On the
contrary, one can easily name some very important open issues of
contemporary physics that are related to the Planck scale.

A particular example is the Big Bang scenario in which the Universe
inevitably goes through an era in which its dimensions are smaller than
the Planck scale (Planck era). On the other hand, space-time in GR is a continuum and so in principle all scales are relevant.
From this perspective, in order to derive conclusions about the nature
of space-time one has to answer the question of what happens on very
small and very large scales.

\subsubsection{UV scales: the Quantum Gravity Problem}
\label{uno.3.1}

One of the main challenges of modern physics is to 
construct a theory able to describe the fundamental interactions 
of nature as different aspects of the same theoretical construct. 
This goal has led, 
in the past decades, to the formulation of several unification 
schemes which, {\it inter alia}, attempt  to describe gravity 
by putting it on the same footing as the other interactions. All 
these schemes try to describe the fundamental fields in terms
of the conceptual apparatus of Quantum Mechanics. This is based 
on
the fact that the states of a physical system are described by 
vectors in a Hilbert space ${\cal H}$ and the physical fields are 
represented by linear operators defined on domains of ${\cal H}$. 
Until now, any attempt to incorporate gravity in this scheme has 
either failed or been unsatisfactory.  The main 
conceptual  problem is that the gravitational field describes 
simultaneously the gravitational degrees of freedom  
and the background space-time in which these degrees of freedom 
live.

Owing to the difficulties of building a complete theory  
unifying interactions and particles, during the last decades the two fundamental theories of modern physics, GR and 
Quantum Mechanics, have been critically 
re-analyzed. On the one hand, one assumes that the matter fields 
(bosons and fermions) come out from superstructures 
({\em e.g.} Higgs bosons or superstrings) that, undergoing 
certain phase transitions, have generated the known particles. 
On 
the other hand, it is assumed that the geometry
({\em e.g.} the Ricci tensor or the Ricci scalar) interacts 
directly with quantum 
matter  fields which  back-react on it. This 
interaction necessarily  modifies the standard
gravitational theory, that is, the Lagrangian of 
gravity plus the effective 
fields is modified with respect to the Hilbert-Einstein one, and 
this fact can directly lead to the ETGs.

From the  point of view of cosmology, the modifications of 
standard gravity provide inflationary scenarios of 
interest. In any case, a condition that must be satisfied  
in order for such theories to be physically acceptable is that  
GR is recovered in the low-energy limit.

Although remarkable conceptual progress has been made following  
the introduction of generalized gravitational theories, at 
the same time the  mathematical difficulties have increased. 
The corrections introduced  into the Lagrangian augment the 
(intrinsic) non-linearity of the Einstein equations, making them 
more difficult to study because  differential 
equations of higher order than second are often obtained and 
because it is impossible to separate the geometric from the 
matter degrees of freedom. In order to overcome these
difficulties and simplify the equations of
motion, one often looks for symmetries of  the 
Lagrangian and identifies conserved quantities which allow  
exact solutions of dynamics to be discovered. The key step in 
the implementation of this  program consists of passing from 
the Lagrangian of the relevant  fields to a point-like 
Lagrangian or, in other words, in going from a system 
with an infinite number of degrees of freedom to one with a 
finite number of degrees
of freedom. Fortunately, this is feasible in cosmology because 
most models of physical interest are spatially homogeneous 
Bianchi models   
and the observed Universe is spatially homogeneous 
and isotropic to a high 
degree (Friedmann-Lemaitre-Robertson-Walker (FLRW) models).

The need for a quantum theory of gravity was recognized at  
the end of the 1950s, when physicist tried for the first time 
to treat all interactions at a fundamental level and to describe 
them in terms of Quantum Field Theory. Naturally, the first 
attempts to quantize gravity used the canonical
approach and the covariant 
approach, which had been applied 
with remarkable success to Electromagnetism. In the first 
approach applied to Electromagnetism, one considers the electric 
and magnetic fields  
satisfying the Heisenberg  Uncertainty Principle and the quantum 
states are gauge-invariant functionals generated by the 
vector potential defined on three-surfaces of constant time. In 
the  second approach, one quantizes 
the  two degrees of freedom of the Maxwell field without any  3+1 decomposition  
of the metric, while the quantum states are elements of a Fock 
space \cite{itzykson}. These procedures are equivalent in Electro-magnetism. The 
former allows for a well-defined measure, whereas  the latter is more 
convenient for perturbative calculations such as, for example, 
the computation of  the  $S$-matrix in Quantum Electro Dynamics (QED).

These methods have been applied also to GR, but 
many
difficulties arise in this case due to the fact that Einstein's 
theory cannot be formulated in terms of a quantum field 
theory on a fixed Minkowski background. To be more 
specific,  in 
GR the geometry of  the background space-time  
cannot be given {\em a priori}: space-time is the dynamical 
variable 
itself. In order to introduce the notions of 
causality, time, and 
evolution, one must first solve the equations of 
motion and then "build" the space-time. For
example, in order to know if particular initial conditions 
will give
rise to a black hole, it is necessary to fully 
evolve them by
solving the Einstein equations. Then, taking into account the
causal structure  of the obtained solution 
, one has to study the
asymptotic metric at future null infinity in order to assess 
whether it is related, in the causal past, with those initial 
conditions. 
This problem becomes intractable at the  quantum level. Due to 
the Uncertainty Principle, in non-relativistic Quantum Mechanics, 
particles do not move along well-defined trajectories and one  
can only calculate the probability amplitude $\psi (t, x)$ that 
a measurement at  time $t$ detects a particle around the 
spatial point $x$. Similarly, in Quantum Gravity, the evolution 
of an initial state does not provide a
specific space-time. In the absence of a space-time, how is it
possible to introduce basic concepts such as 
causality,  time,
elements of the scattering matrix, or black 
holes?

The canonical and  covariant 
approaches  provide different answers 
to 
these questions. The canonical approach is based on the
Hamiltonian formulation of GR and aims at using 
the canonical quantization procedure. The canonical commutation 
relations are the same that lead to the Uncertainty Principle; in 
fact, the commutation of certain operators on a 
spatial three-manifold of constant time is imposed, and this 
three-manifold is fixed in order to preserve the notion of  
causality. In the
limit of asymptotically flat space-time, the motion generated by
the Hamiltonian must be interpreted as temporal evolution (in
other words, when the background becomes the Minkowski space-time,
the Hamiltonian operator assumes again its role as the generator 
of  translations). The canonical approach
  preserves the 
geometric features of GR without the need to introduce   
perturbative methods \cite{Wheeler57, ADM62, DeWitt68, 
Misner70, Misner72}.

The covariant approach,  instead, employs  Quantum 
Field Theory concepts and methods. The basic idea is that the 
problems mentioned above can be  easily circumvented by splitting 
the metric $g_{\mu\nu}$ into a kinematical part
$\eta_{\mu\nu}$ (usually flat) and  a dynamical part 
$h_{\mu\nu}$, as in 
\be
 g_{\mu\nu}=\eta_{\mu\nu}+h_{\mu\nu}\, . 
\ee
The geometry of the background is given by the flat metric 
tensor and is the same as in Special Relativity and ordinary 
Quantum Field Theory, which allows one to define the concepts of 
causality, time, and  scattering. The quantization procedure is 
then  applied to the dynamical field, considered as a 
(small) deviation of the metric from the Minkowski background 
metric. Quanta are discovered to be particles with spin~two, 
called 
{\it gravitons}, which propagate in flat 
space-time and are 
defined by 
$h_{\mu\nu}$. Substituting the metric $g_{\mu\nu}$ into the 
Hilbert-Einstein action, it follows that the Lagrangian
of the gravitational sector contains a sum whose terms represent, 
at different 
orders of approximation, the interaction of  
gravitons and,   
eventually, terms describing   matter-graviton 
interaction  (if 
matter is present). Such terms are analyzed by using the 
standard techniques of  perturbative Quantum Field Theory.

These quantization programs  were both pursued during the 1960s 
and 1970s.  In the canonical approach, Arnowitt, Deser, and 
Misner \cite{ADM62} provided a Hamiltonian formulation of  
GR 
using methods proposed earlier by Dirac and Bergmann. In this  
Hamiltonian formalism, the canonical variables are the 
three-metric on the spatial submanifolds obtained by foliating  
the four-dimensional manifold (note that this foliation is 
arbitrary). The Einstein equations  
give constraints between the three-metrics and their conjugate 
momenta and the evolution equation for these fields, known as the 
{\em Wheeler-DeWitt} (WDW) 
{\em equation}. In this way, 
GR is interpreted as the dynamical theory of
the three-geometries ({\em 
geometrodynamics}).  The 
main difficulties arising from this approach are that the 
quantum equations involve products of operators defined at the 
same space-time point  and, in addition, they entail the  
construction of distributions whose physical meaning is unclear. 
In any case, the main problem is the absence of a Hilbert space 
of states and, as consequence, a probabilistic interpretation of 
the quantities calculated is missing.

The covariant quantization approach is closer to the known  
physics of particles and fields in the sense that it has been
possible to extend the perturbative methods of QED to 
gravitation. This has allowed the  analysis of the mutual  
interaction between  gravitons and of the 
matter-graviton
interactions. The formulation of Feynman rules for 
gravitons and 
the demonstration that the theory might be unitary at every order 
of the expansion was achieved by DeWitt \cite{DeWitt}.

Further progress was achieved with Yang-Mills theories,
which describe the strong, weak, and electromagnetic interactions
of quarks and leptons  by means of symmetries. Such theories
are renormalizable because it is possible to give the
fermions a mass through the mechanism of  Spontaneous Symmetry 
Breaking. Then, it is natural to attempt  to 
consider gravitation as a Yang-Mills theory in the covariant 
perturbation  approach and check whether it is
renormalizable. However, gravity does not fit into this scheme; 
it turns out to be non-renormalizable when one considers the 
graviton-graviton interactions  (two-loops diagrams) and 
graviton-matter interactions (one-loop  
diagrams).\footnote{Higher order
terms in the perturbative series imply an infinite number of
free parameters. At the one-loop level it is sufficient to
renormalize only the effective constants $G_{eff}$ and
$\Lambda_{eff}$ which, at low energy, reduce to Newton's 
constant $G_N$ and the cosmological constant $\Lambda$.} The 
covariant method  allows one to construct a theory of gravity 
which is  
renormalizable at one-loop in the perturbative series 
\cite{BirrellDavies}. Due to the non-renormalizability of gravity 
at different orders,
its validity is restricted only to the low-energy domain, 
{\em i.e.}, to large
scales, while it fails at high energy and small 
scales. This implies that the full unknown theory of gravity 
has  to be invoked near or at the Planck era   
and that, 
sufficiently far from 
the Planck scale,  GR and its first loop 
corrections describe the gravitational
interactions. In this context, it makes sense to add higher order
terms to the Hilbert-Einstein action as we will do in the second part of this Report. Besides, if the free 
parameters  are chosen appropriately, the theory has a better 
ultraviolet behavior and  is asymptotically free. Nevertheless, 
the Hamiltonian of these theories is not bounded  from below and 
they are unstable. In particular, unitarity is violated 
and probability is not conserved.

An alternative approach to the search for a theory of Quantum 
Gravity  comes from the study of the Electroweak 
interaction. In 
this  approach, gravity is treated  neglecting the other 
fundamental interactions. The unification of the  
Electromagnetic  and  the weak interactions suggests that it 
might be possible to obtain a consistent theory when 
gravitation is coupled  to some kind of matter. This is
the basic idea of {\it Supergravity} \cite{van}. In 
this class of theories, 
the divergences due to the bosons (gravitons in this case) are
cancelled exactly by those due to the fermions, leading to a
renormalized theory of gravity. Unfortunately, this scheme works
only at the two-loop level and  for matter-gravity couplings. 
The Hamiltonian is positive-definite and the  theory
turns out to be unitary. But, including higher order loops, the 
infinities re-appear, destroying the renormalizability of the 
theory.

Perturbative methods are also used in String Theories.  In 
this case, the approach is completely different from  
the previous one because the concept of particle is  replaced by 
that of an extended object, the fundamental string.  The usual 
physical particles, including the spin two graviton, correspond 
to 
excitations of the string. The theory has only one free parameter 
(the string tension)  and the  interaction couplings are  
determined uniquely. It follows that string theory  contains 
all fundamental physics and it is therefore considered as a 
candidate for the {\it Theory of Everything}.  String 
Theory  seems to be unitary and the
perturbative series converges implying finite terms. This property
follows from the fact that strings are intrinsically
extended objects, so that ultraviolet divergencies coming from 
small scales or from large transfer impulses, are naturally 
cured. In other words, the natural cutoff is given by the string 
length, which is of Planck size   $l_{Pl}$. 
At scales larger  than 
$l_{Pl}$, the effective string action can be rewritten in terms 
of non-massive vibrational modes, {\em i.e.}, in terms of scalar 
and tensor fields ({\it tree-level effective action}). This
eventually leads to an effective theory of gravity 
non-minimally coupled   with  scalar 
fields, the so-called {\it dilaton fields}.

To conclude, let us summarize the previous considerations:

\begin{itemize}

\item  a consistent ({\em i.e.}, unitary and renormalizable) 
theory of gravity  does not exist  yet. 

\item In the quantization program for gravity, two 
approaches have been used: the {\em covariant approach}  and the 
{\em perturbative approach}. They do not lead to a
definitive theory of Quantum Gravity.

\item   In the low-energy regime  (with respect to the Planck 
energy) at large scales, GR can be generalized by 
introducing into the Hilbert-Einstein action terms of higher 
order in the curvature invariants and non-minimal 
couplings  
between matter and gravity. These lead, at the one-loop level, to 
a consistent and renormalizable theory.

\end{itemize}

A part the lack of final theory, the Quantum Gravity Problem already contains some issues and shortcomings which could be already addresses by the today physics. We will summarize them in the forthcoming section. 

\subsubsection{Issues and shortcomings in Quantum Gravity }
\label{uno.3.2}

Considered  the status of art, are some predictions of Quantum Gravity already available?
Can remnants of Planck scale be detected at lower energy couplings and masses?
 As it is well known, only a fine-tuned
combination of the low-energy constants leads to an observable Universe like ours.
It would thus appear strange if a fundamental theory
possessed just the right constants to achieve this. Hogan \cite{hogan} has argued that
Grand Unified Theories constrain relations among parameters, but leave enough
freedom for a selection. In particular, he suggests that one coupling constant and
two light fermion masses are not fixed by the symmetries of the fundamental theory \footnote{String theory contains only one fundamental dimensionfull parameter, the string length.
The connection to low energies may nonetheless be non-unique due to the existence of many
different possible "vacua"..}.
 One could then determine this remaining free constants only by the (weak
form of the) Anthropic Principle: they have values such that a Universe like ours
is possible. The cosmological constant, for example, must not be much bigger
than the presently observed value, because otherwise the Universe would expand
 too fast to allow the formation of galaxies. The Universe is, however, too
special to be explainable on purely anthropic grounds. We know that the maximal entropy would be reached if all the matter in the
observable Universe were collected into a single giant black hole. This entropy
would be  about $10^{123}$, which is exceedingly more than the observed
entropy of about $10^{88}$. The "probability" for our Universe would then be
about $\frac{exp(10^{88})}{exp(10^{123})}$. From the Anthropic
Principle alone one would not need such a special Universe. As for the cosmological
constant, for example, one could imagine its calculation from a fundamental
theory. Taking the presently observed value for $\Lambda$, one can construct a mass
according to
\begin{equation}
\left(\frac{\hbar\Lambda^{\frac{1}{2}}}{G}\right)\simeq 15MeV\,,\nonumber\\
\end{equation}
which in elementary particle physics is not an unusally big or small value. The
observed value of $\Lambda$ could thus emerge together with medium-size particle mass
scales.
Since fundamental theories are expected to contain only one dimensionfull
parameter, low-energy constants emerge from fundamental quantum fields. An
important example in string theory, is the dilaton field from which
one can calculate the gravitational constant. In order that these fields mimic
physical constants, two conditions have to be satisfied. First, decoherence must
be effective in order to guarantee a classical behaviour of the field. Second, this
"classical" field must then be approximately constant in large-enough space-time
regions, within the limits given by experimental data. The field may still vary
over large times or large spatial regions and thus mimic a "time- or space-varying
constant". The last word on any physical theory has to be spoken by experiments (observations).
Apart from the possible determination of low-energy constants and
their dependence on space and time, what could be the main tests for Quantum
Gravity?
\begin{itemize}
\item Black-hole evaporation: A key test would be the final evaporation phase of
a black hole. To this end, it would be useful to observe signatures of primordial black holes.
These objects  are forming not at 
the end of stellar collapse, but they can result from strong density
perturbations in the early Universe. In the context of inflation, their initial
mass can be as small as 1 {\it g}. Primordial black holes with initial masses of
about $5\times10^{14}${\it g} would evaporate at the present age of the Universe. Unfortunately,
no such object has yet been observed. Especially promising may
be models of inflationary cosmology acting at different scales \cite{Bringmann}.
\item Cosmology: Quantum aspects of gravitational field may be observed
in the anisotropy spectrum of the cosmic microwave background. First,
future experiments may be able to observe the contribution of the gravitons
generated in the early Universe. This important effect was already emphasized
in \cite{staro}. The production of gravitons by the cosmological
evolution would be an effect of Quantum Gravity. Second, quantum-gravitational correction terms from the Wheeler-DeWitt equation
or its generalization in loop quantum cosmology may leave their impact
on the anisotropy spectrum. Third, a discreteness
in the inflationary perturbations could manifest itself in the spectrum \cite{hogan}.
\item Discreteness of of space and time: Both in String Theory and Quantum Gravity there are hints of a discrete structure of space-time. This quantum foam could
be seen through the observation of effects violating local Lorentz invariance
\cite{amelino}, for example, in the dispersion relation of
the electromagnetic waves coming from gamma-ray bursts. It has even be
suggested that space-time fluctuations could be seen in atomic interferometry
\cite{Percival}. However, there exist severe observational
constraints \cite{Peters}.
\item Signatures of higher dimensions: An important feature of String Theory is
the existence of additional space-time dimensions. They
could manifest themselves in scattering experiments at the Large Hadron
Collider (LHC) at CERN. It is
also imaginable that they cause observable deviations from the standard
cosmological scenario \cite{LHC}.
\end{itemize}

Some of these features are discussed in detail in \cite{Kimberly}. Of course, there may be other possibilities which are not yet known
and which could offer great surprises. It is, for example, imaginable that a fundamental
theory of Quantum Gravity is intrinsically non-linear \cite{Penrose,
Singh}. This is in contrast to most currently studied theories of Quantum
Gravity, which are linear.
Quantum Gravity has been studied since the end of the 1920s. No doubt,
much progress has been made since then. The final goal has not yet been reached. The belief expressed here is that a consistent and experimentally successful theory
of Quantum Gravity will be available in the future. However, it may still take a
while before this time is reached. In any case, ETGs could constitute a serious approach in this direction.

\subsubsection{IR scales: Dark Energy and Dark Matter}
\label{uno.3.3}

The  revision of standard early cosmological scenarios leading to 
inflation implies that a new approach is necessary also at
later epochs:   ETGs could play a fundamental role also in this
context. In fact,  the increasing bulk of data accumulated 
over the past few years has paved the way for a new 
cosmological model usually referred to as the {\it Concordance 
Model} or $\Lambda$ Cold Dark Matter ($\Lambda$CDM) model.

The Hubble diagram of type Ia 
supernovae (hereafter SNeIa) 
measured by both the Supernova Cosmology Project 
\cite{Knopetal03,  Perlmutteretal03} and
the High-$z$ Team \cite{Riessetal98, Tonryetal03} up to redshift 
$z \sim 1$, was 
the first piece of evidence that the Universe is 
currently undergoing a phase of accelerated expansion. Besides, 
balloon-born experiments such as {\em 
BOOMERANG}
\cite{deBernardisetal00} and {\em MAXIMA}
\cite{Maxima} determined the location of the first two 
Doppler peaks in 
the spectrum of Cosmic Microwave Background (CMB) anisotropies, strongly suggesting a Universe 
with  flat spatial sections. When  combined with the constraints 
on the matter density parameter $\Omega_M$, these data indicate that the Universe is 
dominated by an unclustered fluid with negative pressure 
commonly referred to as {\it dark energy}, 
which drives  
the accelerated expansion. This picture has been further 
strengthened by the recent precise measurements of the CMB  
spectrum by the {\em WMAP} satellite experiment 
\cite{WMAP,hinshaw, Spergel:2006hy}, and 
by the extension of the SNeIa Hubble diagram to redshifts larger 
than one \cite{Riess04}.

An overwhelming number  of papers appeared following these 
observational evidences,  which present a  large variety 
of models attempting to explain  the cosmic acceleration. The 
simplest explanation would be  the well known 
cosmological constant $\Lambda$ \cite{SahniStarobinsky00}. 
Although the 
latter  
provides the best-fit to most of the available astrophysical data 
\cite{WMAP}, the $\Lambda$CDM model fails egregiously in 
explaining why the inferred value of $\Lambda$ is so tiny
(120 orders of magnitude lower) in comparison with the typical
value of the vacuum energy density predicted by 
particle physics,  and why its present value  is comparable to 
the matter density, this is the so-called {\it coincidence 
problem}.

As a tentative solution, many authors have replaced the
cosmological constant with a scalar field $\phi$ rolling 
slowly down a 
flat section of a potential $V(\phi)$ and giving rise to the 
models known  as {\it quintessence}
\cite{QuintRev,tsu1}. Albeit 
successful in fitting the data with many models, the 
quintessence
approach to dark energy is still  plagued by 
the coincidence problem since the 
 dark energy and dark matter densities evolve 
differently and reach 
comparable values only during a very short time  of the 
history of the Universe,  coinciding in order of magnitude right at the present era. 
In other words, the quintessence dark energy
is tracking matter and evolves in the same way for a long time; 
at late times, somehow it  changes its behaviouor 
and no longer tracks the dark matter  but 
begins to dominate 
in the fashion of a (dynamical) cosmological constant. This is 
the coincidence problem of quintessence.

Furthermore, the origin of this quintessence  scalar 
field is mysterious, although various (usually rather exotic)   
models have been proposed, leaving a great deal of uncertainty 
on the choice of the scalar field potential $V(\phi)$ 
necessary to achieve 
the late-time acceleration of the Universe. The subtle and 
elusive nature of  dark energy has led
many authors to  look for a completely different explanation of  
the acceleration  behaviour of the cosmos without introducing  
exotic components. To this end, it is worth stressing that the
present-day cosmic acceleration  only requires a negative 
pressure component that comes to dominate the dynamics late in 
the matter era, but does 
not tell us anything about the  nature and the number of the 
cosmic 
fluids advocated to fill the Universe. This consideration  
suggests that it could be possible to explain the accelerated 
expansion with  a single cosmic fluid characterized by an 
equation 
of state causing it to act like dark matter  at
high densities, while behaving as dark energy at low 
densities. An attractive feature of these models, usually 
referred to as {\it Unified Dark
Energy} (UDE) or {\it Unified Dark 
Matter} (UDM) models, is that
the coincidence problem  is solved naturally,  at least at the 
phenomenological level, with such an 
approach \cite{odi2006,odi20062}. Examples are 
the generalized Chaplygin gas  
\cite{Chaplygin}, the tachyon field
\cite{tachyon}, and condensate cosmology \cite{Bassett}. A
different class of UDE models  with a 
single fluid has been 
proposed \cite{Hobbit1, Hobbit2}: its energy density scales with
the redshift $z$ in such a way that a radiation-dominated era, 
followed by a  matter  era and then by an accelerating phase can 
be naturally achieved. These models are extremely
versatile since they can be interpreted both in the framework of
UDE  or as two-fluid scenarios 
containing dark matter  and 
scalar field dark energy. A characteristic 
feature of this 
approach is  that a generalized equation of state can  always 
be obtained and the fit to the observational data can be 
attempted. However, such models explain the phenomenology but cannot be addressed to some fundamental physics.

There is another,  different, way to approach the problem of 
the cosmic  acceleration. As stressed in \cite{LSS03}, it is 
possible that the observed acceleration is not the manifestation 
of yet another ingredient of the cosmic pie, but rather the first 
signal of a breakdown, in the infrared limit, of the laws of 
gravitation as we understand them. From this point of view, it is 
tempting to modify the Einstein-Friedmann equations to see 
whether it is 
still  possible to fit the astrophysical data with models 
containing only standard matter without exotic fluids. 
Examples 
are the Cardassian expansion \cite{Cardassian} and 
Dvali-Gabadadze-Porrati (DGP) gravity 
\cite{DGP}. Within  the 
same conceptual framework, it is possible to  find 
alternative schemes in which  a quintessential behaviour is 
obtained by incorporating effective models coming from 
fundamental physics and giving rise to generalized or 
higher order gravity actions \cite{Capozziello02IJMPD} (see 
Ref. \cite{CF1,booksalv,defelice,bookfelix}. For instance, a 
cosmological constant may be recovered as a
consequence of a non-vanishing torsion field, 
leading to a
model consistent with both the SNeIa Hubble diagram and
observations of the Sunyaev--Zel'dovich 
effect in galaxy 
clusters 
\cite{torsion}. SNeIa data could also be efficiently fitted
by including in the gravitational sector  higher order 
curvature invariants  
\cite{CapozzielloCardoneCarloniTroisi03, 
Li:2007xn,CF1,Li:2007xw}. 
These alternative models provide naturally  a cosmological 
component with negative pressure originating in the 
geometry of the Universe, thus overcoming the problematic nature 
of quintessence scalar fields.
Cosmological models coming from ETGs are in the track of this philosophy.

The variety of cosmological models which are phenomenologically  
viable  candidates to explain the observed accelerated expansion 
is clear from this short review.  This  
overabundance signals  that only a  limited number 
of cosmological tests are available to discriminate 
between competing 
theories, and it is clear that there is a high degeneracy of 
models.
Let us remark that both the SNeIa Hubble diagram and the angular 
size-redshift relation of compact radio sources  
\cite{ AngTest2} are distance-based probes of the 
cosmological 
model and, therefore, systematic errors and biases could be 
iterated. With this point in mind, it is interesting to search 
for tests based on time-dependent observables.
 For example, one can take into account the {\it lookback 
time} 
to distant objects since this quantity can discriminate between 
different cosmological models. The lookback time is 
observationally estimated as the difference between the 
present-day  age of the Universe and the age of a given object at 
redshift $z$. This estimate is possible if the object is a 
galaxy observed in more than one photometric band since its colour 
is determined by its age as a consequence of stellar evolution. 
Hence, it is possible to obtain an estimate of the galaxy's age 
by  measuring its magnitude in different bands and then using 
stellar evolutionary codes to choose the model that 
best reproduces the observed colours \cite{andreon,ester}.

Coming to the weak-field limit, which essentially
means considering Solar System scales, any alternative relativistic theory of gravity is expected to
reproduce GR which, in any case, is firmly tested only in this
limit and at these scales \cite{Will93}. Even this limit  is a 
matter of debate since several relativistic theories  do not 
reproduce exactly the Einsteinian results in their Newtonian 
limit but, in some sense,
generalize them. As was first noticed by Stelle
\cite{Stelle:1976gc}, $R^2$-gravity gives rise to 
Yukawa-like
corrections  to the Newtonian potential with 
potentially interesting physical consequences. For example, it is 
claimed by some authors that  the flat rotation 
curves of 
galaxies can be explained by such
terms \cite{Sanders90}. Others \cite{mannheim} have shown that a
conformal theory of gravity is nothing else but a fourth order
theory containing such terms in the Newtonian limit. Reports of 
an apparent anomalous long-range acceleration
in the data  of the Pioneer 10/11, Galileo, and
Ulysses spacecrafts could be accommodated in a general 
theoretical
scheme incorporating Yukawa corrections
to the Newtonian 
potential  \cite{Anderson02, bertolami}.

In general, any relativistic theory of gravitation  yields
corrections to the weak-field gravitational potentials ({\em 
e.g.},  \cite{Qua91}) which, at the  post-Newtonian level and 
in the 
Parametrized Post-Newtonian (PPN) formalism,
could constitute a test of these theories \cite{Will93}. 
Furthermore, the newborn {\it gravitational lensing 
astronomy} 
\cite{ehlers} is providing additional tests of gravity over 
small, large, and very large scales which will soon provide 
direct measurements of the variation of the Newton coupling 
\cite{krauss}, the potential of galaxies, clusters of galaxies, 
and several other features of self-gravitating systems.
Very likely,  such data will be capable of confirming or ruling
out as GR or ETGs.

This short overview shows that several shortcomings point out that GR cannot to be the final theory of gravity notwithstanding its successes in addressing a large amount of theoretical and experimental issues. ETGs could be a viable approach to solve some of these problems at IR and UV scales without pretending to be the comprehensive and self-consistent fundamental theory of gravity but in the track outlined by GR and then in the range of gauge theories.
This review paper is mainly devoted to the theoretical foundation of ETGs trying to insert them in the framework of gauge theories and showing that they are nothing else but a straightforward extension of GR. The cosmological phenomenology and the genuinely astrophysical aspects of ETGs are not faced here. We refer the readers to the excellent reviews and books quoted in the bibliography \cite{mybook,booksalv,mauro,defelice,CF1,lucaamendola,tsu1,salzano,faraoni-noi,lobo,reportodi,cliftonrep}.

\section{A summary of gauge symmetries}
\label{due}

Modern gauge theory has emerged as one of the most significant developments of physics of XX century. It has allowed us to realize partially the issue of unifying the fundamental interactions of nature. We now believe that the electromagnetism, which has been long studied, has been successfully unified with the nuclear weak interaction, the force to which radioactive decay is due.
What is the most remarkable about this unification is that these two forces differ in strength by a factor of nearly $10^5$. This important accomplishment by the Weinberg-Salam gauge theory \cite{weinberg}, and insight gained from it, have encouraged the hope that also the other fundamental forces could be unified within a gauge theory framework. At the same time, it has been realized that the potential areas of application for gauge theory extended far beyond elementary particle physics. Although much of the impetus for a gauge theory came from new discoveries in particle physics, the basic ideas behind gauge symmetry have also appeared in other areas as seemingly unrelated, such as condensed matter physics, non-linear wave phenomena and even pure mathematics. This great interest in gauge theory indicates that it is in fact a very general area of study and not only limited to elementary particles.

Gauge invariance was recognized only recently as the physical principle governing the fundamental forces between the elementary particles. Yet the idea of gauge invariance was first proposed by  Hermann Weyl in 1919 when only the electron and proton \cite{weyl} were known as fundamental particles. It required nearly 50 years for gauge invariance to be "rediscovered" and its significance to be understood. The reason for this long hiatus was that Weyl's physical interpretation of gauge invariance was shown to be incorrect soon after he had proposed the theory.
Gauge invariance only managed to survive because it was known to be a symmetry of Maxwell's equations and thus became a useful mathematical help in order to simplify equations and thus became a useful mathematical device for simplifying many calculations in the electrodynamics. In view of present success of gauge theory, we can say that gauge invariance was the classical case of a good idea which was discovered long before its time.

In this section, we present a brief summary of gauge theory in view of the fact that any theory of gravity can be considered under the same standard. The early history of gauge theory can be divided  into old and new periods where the dividing can be set in the 1950's.  The most important question is what motivated Weyl to propose the idea of gauge invariance as a physical symmetry? How did he manage to express it in a mathematical form that has remained almost the same today although the physical interpretation has altered radically? And, how did the development of Quantum Mechanics lead Weyl himself to a rebirth of a gauge theory?
The new period of gauge theory begins with the pioneering attempt of Yang and Mills to extend gauge symmetry beyond the narrow limits of electromagnetism  \cite{Yang-Mills}. Here we will review the radically new interpretation of gauge invariance required by Yang-Mills theory and the reasons for the failure of original theory. By comparing the new theory with that of Weyl, we can see that many original ideas of Weyl have been rediscovered and incorporated into the modern theory  \cite{moriyasu}. 

In these next subsection, our purpose is to present an elementary introduction to a gauge theory in order to show that any relativistic theory of gravity is a gauge theory.

\subsection{What is a gauge symmetry?}
\label{due.1}
In physics, gauge invariance (also called gauge symmetry) is the property of a field theory where different configurations of the underlying fundamental but unobservable fields result in identical observable quantities. A theory with such a property is called a gauge theory. A transformation from a field configuration to another is called a gauge transformation.
Modern physical theories describe nature in terms of fields, {\it e.g.}, the electromagnetic field, the gravitational field, and fields for the electron and all other elementary particles. A general feature of these theories is that none of these fundamental fields, which are the fields that change under a gauge transformation, can be directly measured. On the other hand, the observable quantities, namely the ones that can be measured experimentally as charges, energies, velocities, etc.  do not change under a gauge transformation, even though they are derived from the fields that do change. This (and any) kind of invariance under a transformation is called a symmetry.
 
 For example, in classical electromagnetism the Electric field, $E$, and the magnetic field, $B$, are observable, while the underlying and more fundamental electromagnetic potentials $V$ and $A$ are not.
  Under a gauge transformation which jointly alters the two potentials, no change occurs either in $E$ or $B$ or in the motion of charged particles. In this example, the gauge transformation was just a mathematical feature without any physical relevance, except that gauge invariance is intrinsically connected to the fundamental law of charge conservation.
As shown above, with the advent of Quantum Mechanics in the 1920s, and with successive Quantum Field Theory, the importance of gauge transformations has steadily grown. Gauge theories constrain the laws of physics, because of the fact that all the changes induced by a gauge transformation have to cancel each other out when written in terms of observable quantities. Over the course of the 20th century, physicists gradually realized that all forces (fundamental interactions) arise from the constraints imposed by local gauge symmetries, in which case the transformations vary from point to point in space and time. Perturbative quantum field theory (usually employed for scattering theory) describes forces in terms of force mediating particles called gauge bosons. The nature of these particles is determined by the nature of the gauge transformations. The culmination of these efforts is the Standard Model, a quantum field theory explaining all of the fundamental interactions except gravity.

\subsection{The Einstein connection}
\label{due.2}

In 1919, people thought that only two fundamental forces of nature existed, Electromagnetism and Gravitation. In that same year, a group of scientists also made the first experimental observation of starlight bending in the gravitational field of the sun during a total eclipse \cite{kluber}.
The brilliant confirmation of Einstein's General Theory of Relativity inspired Weyl to propose his own revolutionary idea of gauge invariance in 1919. To see how this came about, let us first briefly recall some basic ideas on which Relativity was built.
The fundamental concept underlying both Special and GR is that are no absolute frames of reference in the Universe. The physical motion of any system must be described relatively to some arbitrary coordinate frame specified by an observer, and the laws of physics must be independent of the choice of frame.
In Special Relativity, one usually, defines convenient reference frames, which are called "inertial",  in motion with uniform velocity. For example, consider a particle which is moving with constant velocity $v$ with respect to an observer. Let $S$ be the rest frame of the observer and $S'$ be an inertial frame which is moving at the same velocity as the particle. The observer can either state that the particle is moving with velocity $v$ in $S$ or that it is at rest $S'$. The important point to be noted from this trivial example is that the inertial frame $S'$ can always be related by a simple Lorentz transformation to the observer frame $S$. The transformation depends only on the relative velocity between the and observers, not on their positions in space-time. The particle and observer can be infinitesimally close together or at opposite ends of the Universe; the Lorentz transformation is still the same. Thus the Lorentz transformation, or rather the Lorentz symmetry group of Special Relativity, is an example of "global" symmetry.
In GR, the description of relative motion is much more complicated because now one is dealing with the motion of a system embedded in a gravitational field. As an illustration, let us consider the following "gedanken" experiment for measuring the motion of a test particle which is moving through a gravitational field. The measurement is to be performed by a physicist in an elevator. The elevator cable as broken so that the elevator and the physicist are falling freely \cite{bergmann}. As the particle moves through the field, the physicist determines its motion with respect to the elevator. Since both particle and elevator are falling in the same field, the physicist can describe the particle motion as if there were no gravitational field. The acceleration of the elevator cancels out the acceleration of particle due to gravity. This example of the Principle of Equivalence, follows from the well-known fact that all bodies accelerate  at the same rate given the gravitational field ({\it e.g.} $9.8m/s^2$ on the surface of the Earth).
Let us now compare the physicist in the falling elevator with the observer in the inertial frame in Special Relativity. It could appear that the elevator corresponds to an accelerating or "non-inertial" frame that is analogous to the frame $S'$ in which the particle appeared to be at rest. However, it is not true that a real gravitational field does not produce the same acceleration at every point in space. As one moves infinitely far away from the source, the gravitational field will eventually vanish. Thus, the falling elevator can only be used to define a reference frame within an infinitesimally small region where the gravitational field can be considered uniform. Over a finite region, the variation of the field may be sufficiently large for the acceleration of the particle not to be completely cancelled. 
We see that an important difference between Special Relativity and GR is that a reference frame can only be defined "locally" or at a single point in a gravitational field. This creates a fundamental problem. To illustrate this difficult point, let us now suppose that there are many more physicists in nearby falling elevators. Each physicist makes an independent measurement so that the path of particle in the gravitational field can be determined. The measurements are made in separate elevators at different locations in the field. Clearly, one cannot perform an ordinary Lorentz transformation between the elevators. If the different elevators were related only by Lorentz transformation, the acceleration would have to be independent of position and the gravitational field could not decrease with distance from the source.
Einstein solved the problem of relating nearby falling frames by defining a new mathematical relation known as "connection". To understand the meaning of a connection, let us consider a $4$-vector $A_\mu$ which represents some physically measured quantity. Now suppose that the physicist in the elevator located at $x$ observes that $A_\mu$ changes by an amount $dA_\mu$ and a second physicist in a different elevator at $x'$ observes a change in $dA'_\mu$. In Special Relativity, the differential $dA_\mu$ is also a vector like $A_\mu$ itself. Thus, the differential $A_\nu$ in the elevator at $x'$ is given by the relation  \footnote{The components of the $4$-vector $A_\mu=(A^0, \bf{A})$ and $A^\mu=(A_0, \bf{A})$ with $A^0=-A_0$. Vector components with upper and lower indices are related by $x_\mu=g_{\mu\nu}x^\nu$, where $g_{\mu\nu}$ is the metric tensor which appears in the definition of the invariant space-time interval $ds^2=g_{\mu\nu}dx^\mu dx^\nu$. The components of $g_{\mu\nu}$ are $g_{11}=g_{22}=g_{33}=1$, $g_{00}=-1$ and all other components are zero.}
\begin{eqnarray}
dA'_\nu=\frac{\partial x^\mu}{\partial x'^{\nu}}dA_\mu\,,
\label{II-1}
\end{eqnarray}
where $\mu,\nu=0,1,2,3$. The relation (\ref{II-1}) follows from the fact that the Lorentz transformation between $x$ and $x'$ is a linear transformation. We can no longer assume that the transformation from $x$ to $x'$ is linear in GR. Thus, we must write for $dA'_\nu$ the general expression
\begin{eqnarray}
dA'_\nu=\frac{\partial x^\mu}{\partial x'^{\nu}}dA_\mu+A_\mu d\left(\frac{\partial x^\mu}{\partial x'^{\nu}}\right)=\frac{\partial x^\mu}{\partial x'^{\nu}}dA_\mu+A_\mu\frac{\partial^2 x^\mu}{\partial x'^{\nu}\partial x'^{\lambda}}dx'^{\lambda}\,.
\label{II-2}
\end{eqnarray}
Clearly, the second derivatives $\displaystyle{\frac{\partial^2 x^\mu}{\partial x'^{\nu}\partial x'^{\lambda}}}$ will vanish if the $x^\mu$ are linear functions of the $x'^\nu$.
 Such terms are actually quite familiar in physics. They occur in "curvilinear" coordinate systems.  These curvilinear coefficients are denoted by the symbol
\begin{eqnarray}
\Gamma^\mu_{\nu\lambda}=\frac{\partial^2 x^\mu}{\partial x'^{\nu}\partial x'^{\lambda}}\,,
\label{II-5}
\end{eqnarray}
and are called the components of a "connection". They are also called affine connections or Christoffel symbols \cite {gravitation}. 
 It is important to note that the gravitational connection is not simply the result of using a curvilinear coordinate system. The value of the connection at each point in space-time is dependent on the properties of the gravitational field. The field is important in the determination of the relative orientation of the different falling elevators in the same way that the "upward" direction on the surface of the earth varies from one position to another. The analogy with curvilinear coordinate systems merely indicates  that the mathematical descriptions of free-falling frames and curvilinear coordinates are similar. Einstein generalized this similarity and arrived at the revolutionary idea of replacing gravity by the curvature of space-time \cite{einstein-gauge}.
Let us briefly summarize the essential characteristics of GR that Weyl would have utilized for his new gauge theory. First of all, GR involves a specific force, gravitation, which is not studied in Special Relativity. However, by studying the properties of coordinates frames just as in Special Relativity, one learns that only local coordinates can be defined in a gravitational field. This local property is required by the physical behavior of the field and leads naturally to the idea of a connection between local coordinate frames. Thus the essential difference between Special Relativity and GR is that the former is a global theory while the latter is a local theory. This local property was the key to Weyl's gauge theory \cite{moriyasu}. In Sec. \ref{tre} we will develope extensively this idea.

\subsection{The Weyl Gauge}
\label{due.3}
Weyl went a step beyond GR and asked the following question: if the effects of a gravitational field can be described by a connection which gives the relative orientation between local frame in space-time, can other forces of nature such as Electromagnetism also be associated with similar connections? Generalizing the idea that all physical measurements are relative, Weyl proposed that absolute magnitude or norm of a physical vector also should not be an absolute quantity but should depend on its location in space-time. A new connection would then be necessary in order to relate the lengths of vectors at different positions. This is the scale or "gauge" invariance. It is important to note here that the true significance of Weyl's proposal lies on the local property of gauge symmetry and not in a special choice of the norm or "gauge" as a physical variable. The assumption of locality is a powerful condition that determines not only the general structure but many of the detailed features of gauge theory.
Weyl's gauge invariance can be easily expressed in mathematical form \cite{yang}.Let us consider a vector at position $x$ with norm given by $f(x)$. If we shift the vector or transform the coordinates so that the vector is now at $x+dx$, the norm becomes $f(x+dx)$. Expanding to first order in $dx$, we can write the norm as

\begin{eqnarray}
f(x+dx)=f(x)+\partial_\mu f dx^\mu\,.
\label{II-6}
\end{eqnarray}
 We now introduce a gauge change through a multiplicative scaling factor $S(x)$.  The factor $S(x)$ is defined for convenience to equal unity at the position $x$. The scale factor at $x+dx$ is then given by

\begin{eqnarray}
S(x+dx)=1+\partial_\mu S dx^\mu\,.
\label{II-7}
\end{eqnarray}
The norm of the vector at $x+dx$ is then equal to the product of Eqs. (\ref{II-6}) and (\ref{II-7}). Keeping only first order terms in $dx$, we obtain
\begin{eqnarray}
Sf=f+\left(\partial_\mu S\right)f dx^\mu+\partial_\mu f dx^\mu\,.
\label{II-8}
\end{eqnarray}
For a constant vector, we see that the norm has changed by an amount
\begin{eqnarray}
\left(\partial_\mu+\partial_\mu S\right) f dx^{\mu}\,.
\label{II-9}
\end{eqnarray}
Te derivative $\partial_\mu S$ is the new mathematical "connection" associated with the gauge change. 
 Weyl identified the gauge connection  $\partial_\mu S$ with the electromagnetic potential $A_\mu$. It is straightforward to show that a second gauge change with a scale factor $\Lambda$ will transform the connection as follows,
\begin{eqnarray}
\partial_\mu S \rightarrow \partial_\mu S+ \partial_\mu \Lambda\,.
\label{II-10}
\end{eqnarray}
From classical Electromagnetism, we known that the potential behaves under gauge transformation like
\begin{eqnarray}
A_\mu  \rightarrow A_\mu+ \partial_\mu \Lambda\,,
\label{II-11}
\end{eqnarray}
which leaves the electric and magnetic fields unchanged. Since the forms of (\ref{II-10}) and (\ref{II-11}) are identical, it appeared that Weyl's new interpretation of the potential as a gauge connection was perfectly compatible with Electromagnetism.
Unfortunately, it was soon pointed out that the basic idea of scale invariance itself would lead to conflict with known physical facts \cite{bergmann1}. Some years later, Bergmann noted, that Weyl's original interpretation of gauge invariance would also be in conflict with Quantum Theory. The wave description of matter defines a natural scale for a particle through its Compton wavelength $\displaystyle{\lambda=\frac{h}{Mc}}$. Since the wavelength is determined by the particle mass $M$, it cannot depend on position and thus contradicts Weyl's original assumption about scale invariance.
Despite the initial failure of Weyl's gauge theory, the idea of a local gauge symmetry survived. It was well known that Maxwell's equations were invariant under a gauge change.
However, without an acceptable interpretation, gauge invariance was regarded as only an "accidental" symmetry of Electromagnetism. The gauge transformation property in Eq. (\ref{II-11}) was interpreted as just a statement of the well known arbitrariness of the potential in classical physics. Only the electric and magnetic fields were considered to be real and observable. Gauge symmetry was retained largely because it was useful for calculations in both classical and quantum electrodynamics. In fact a lot of problems in electrodynamics can often be most easily solved by first choosing a suitable gauge, such as the Coulomb gauge or Lorentz gauge, in order to make the equations more tractable \cite{moriyasu}.

\subsection{Electromagnetism as a Gauge Theory }
\label{due.4}

 It is clear that the electromagnetic interaction of charged particles could be interpreted as a local gauge theory within the context of Quantum Mechanics. In analogy with Weyl's first theory, the phase of a particle wavefunction can be identified as a new physical degree of freedom which is dependent on the space-time position. The value can be changed arbitrarily by performing purely mathematical phase transitions on the wavefunction at each point. Therefore, there must be some connections between phase value nearby points. The role of this connection is payed by the electromagnetic potential. This strict relation between potential and the change in phase is clearly demonstrated by the Aharonov--Bohm effect \cite{aharonov}. Thus by using the phase of wavefunction as the local variable instead of the norm of a vector, Electromagnetism can be interpreted as a local gauge theory very much as Weyl envisioned.
 Gauge transformations can be viewed as merely phase changes so that they look more like a property of Quantum Mechanics than Electromagnetism. In addition, the symmetry defined by the gauge transformations does not appear to be "natural". The set of all gauge transformations forms a one-dimensional unitary group known as the $U(1)$ group. This group does not arise from any form of coordinate transformation like the more familiar spin-rotation group $SU(2)$ or Lorentz group. Thus, one has lost the original interpretation proposed by Weyl of a new space-time symmetry.
The status of gauge theory was also influenced by the historical fact that Maxwell had formulated Electromagnetism long before Weyl proposed the idea of gauge invariance. Therefore, unlike the GR, the gauge symmetry group did not play any essential role in defining the dynamical content of Electromagnetism. This sequence of events was to be completly reversed in the development of modern gauge theory  \cite{jackson,goldstein,moriyasu}.

\subsection{The Yang-Mills Gauge}
\label{due.5}
In 1954, C. N. Yang and R. Mills proposed that the strong nuclear interaction can be described by a field theory like Electromagnetism. They postulated that the local gauge group was the $SU(2)$ isotopic-spin group. This idea was revolutionary because it changed the very concept of the "identity" of an elementary particle. If the nuclear interaction is a local gauge theory like Electromagnetism, then there is a potential  conflict with the notion of how a particle state. For example, let us assume that we can "turn off" the electromagnetic interaction so that we cannot distinguish the proton and neutron by electric charge. We also ignore the small mass difference. We must then label the proton as the "up" state of isotopic spin $\displaystyle{\frac{1}{2}}$ and the neutron as the "down" state. But if isotopic spin invariance is an independent symmetry at each point in space-time, we cannot assume that the "up" state is the same at any other point. The local isotopic spin symmetry allows to choose arbitrarily which direction is "up" at each point without reference to any other point.
Given that the labelling of a proton or a neutron is arbitrary at each point, once the choice has been made at one location, it is clear that some rule is then needed in order to make a comparison with the choice at any other position. The required rule, as Weyl proposed originally, is supplied by a connection. A new isotopic spin potential field was therefore postulated by Yang and Mills in analogy with the electromagnetic potential. However, the greater complexity of the $SU(2)$ isotopic-spin group as compared to the $U(1)$ phase group means that Yang-Mills potential will be quite different from the electromagnetic field. In Electromagnetism, the potential provides a connection between the phase values of the wavefunction at different positions. In the Yang-Mills theory, the phase is replaced by a more complicated local variable that specifies the direction of the isotopic spin. In order to understand qualitatively how this leads to a connection, we need only to recall that the $SU(2)$ isotopic-spin group is also the group of rotations in a 3-dimensional space \footnote{Technically, the $SU(2)$ group is different from the group of 3-dimensional rotations, $O(3)$; the $SU(2)$ group is the "covering group" of  $O(3)$.}. As an example, let us visualize the "up" component of isotopic spin  $\displaystyle{\frac{1}{2}}$ as a vector in an abstract "isotopic spin space". An obvious way to relate the "up" states at different locations $x$ and $y$ is to ask how much the "up" state at $x$ needs to be rotated so that it is oriented in the same direction as the "up" state at $y$. This suggests that the connection between isotopic spin states at different points must act like isotopic spin rotation itself. In other words, if a test particle in the "up" state at $x$ is moved through the potential field to position $y$, its isotopic spin direction must be rotated by the field so that it is pointing in the "up" direction corresponding to the position $y$. We can immediately generalize this idea to states of arbitrary isotopic spin. Since the components of an isotopic spin state can be transformed into another one  by elements of the $SU(2)$ group, we can conclude that the connection must be capable of performing the same isotopic spin transformations as the $SU(2)$ group itself. This idea that the isotopic spin connection, and therefore the potential, acts like the $SU(2)$ symmetry group is the most important result of the Yang--Mills theory. This concept lies at the heart of the local gauge theory. It shows explicitly how the gauge symmetry group is built into the dynamics of the interaction between particles and fields.
How is ti possible for a potential to generate a rotation in an internal symmetry space? To answer this question, we must define the Yang--Mills potential more carefully in the language of the rotation group. A 3-dimentional rotation $R(\theta)$ of a wavefunction is written as
\begin{eqnarray}
R(\theta)\psi=e^{i\theta{\bf L}}\psi\,,
\label{II-28}
\end{eqnarray}
where $\theta$ is the angle of rotation and ${\bf L}$ is the angular momentum operator. Let us compare this rotation with the phase change of wavefunction after a gauge transformation. The rotation has the same mathematical form as the phase factor of the wavefunction. However, this does not mean that the potential itself is a rotation operator like  $R(\theta)$. We noted earlier that the amount of phase change must also be proportional to the potential in order to ensure that Schr\"{o}dinger equation remains gauge invariant. To satisfy this condition, the potential must be proportional to the angular momentum operator  ${\bf L}$ in (\ref{II-28}). Thus, the most general form of the Yang--Mills potential is a linear combination of the angular momentum operators 

\begin{eqnarray}
A_\mu=\sum_i A_{\mu}^i(x)L_i\,,
\label{II-29}
\end{eqnarray}
where the coefficients $A_{\mu}^i(x)$ depend on the space-time position and we explicitly write the sum over the components. This relation indicates that the Yang--Mills is not a rotation, but rather is a "generator" of a rotation. For the case of Electromagnetism, the angular momentum operator is replaced by a unit matrix and  $A_{\mu}^i(x)$ is just proportional to the phase change $\partial_\mu \lambda$. The relation (\ref{II-29}) explicitly displays the dual role of the Yang--Mills potential as both a field in space-time and an operator in the isotopic-spin space.

We can immediately deduce some interesting properties of the Yang--Mills potential. For example, the potential must have three charge components corresponding to the three independent angular momentum operators $L_+$,  $L_-$ and  $L_3$. The potential component which acts like a raising operator $L_+$ can transform a "down" state into a "up" state. We can associate this formal operation with a real process where a neutron absorbs a unit of isotopic spin from the gauge field and turns into a proton. This example indicates that the Yang--Mills gauge field must itself carry electric charge unlike electromagnetic potential. The Yang--Mills field also differs in other respects from the electromagnetic field but they both have one property in common, namely, they have zero mass. The zero mass of the photon is well known from Maxwell's equations, but local gauge invariance requires that the mass of the gauge potential field be identically zero for any gauge theory. The reason is that the mass of the potential must be introduced into a Lagrangian through a term of the form
\begin{eqnarray}
m^2 A_\mu A^\mu\,.
\label{II-30}
\end{eqnarray}
This guarantees that the correct equation of motion for a vector field will be obtained from the Euler-Lagrange equations. Unfortunately, the term given by (\ref{II-30}) is not invariant under a gauge transformation. The special transformation property of the potential will introduce extra terms in (\ref{II-30}) proportional to $A_\mu$, which are not cancelled by the transformation of the wavefunction. Thus, the standard mass term is not allowed in the  Yang--Mills gauge field must have exactly zero mass like the photon.  The Yang--Mills field will therefore exhibit long-range behaviour like Coulomb field and cannot reproduce the observed short range of the nuclear force. Since this conclusion appeared to be an inescapable consequence of a local gauge invariance, the  Yang--Mills theory was not considered to be an improvement on the already existing theories of the strong nuclear interaction.

Although the  Yang--Mills theory field in its original purpose, it established the foundation for modern gauge theory. The $SU(2)$ isotopic-spin gauge transformation could not be regarded as a mere phase change; it required an entirely new interpretation of a gauge invariance. Yang and Mills showed for the first time that local gauge symmetry was a powerful fundamental principle that could provide new insight into the newly discovered "internal" quantum numbers like isotopic spin. In the  Yang--Mills theory, isotopic spin was not just a label for the charge states of particles, but it was crucially involved in determining the fundamental form of the interaction \cite{moriyasu}.

\subsection{Geometry and Gauge}
\label{due.6}
 The  Yang--Mills theory revived the old ideas that elementary particles have degrees of freedom in some "internal" space. By showing how these internal degrees of freedom could be unified in a non-trivial way with the dynamical motion in space-time, Yang and Mills discovered a new type of geometry in physics.

 The geometrical structure of a gauge theory can be seen by comparing the  Yang--Mills theory with of GR. The essential role of the connection is evident in both gauge theory and relativity. There is an  analogy between non-inertial coordinate frame and gauge theory but the local frame has to be  is located in an abstract space associated with the gauge symmetry group. To see how the gauge group defines an internal space, let us examine the examples of the $U(1)$ phase group and the $SU(2)$ isotopic spin group. For the  $U(1)$ group, the internal space consists of all possible values of the phase of the wavefunction. These phase values can be interpreted as angular coordinates in a 2-dimensional space. The internal symmetry space of $U(1)$ thus looks like a ring, and the coordinate of each point in this space is just the phase value itself. The internal space defined by $SU(2)$ group is more complicated because it describes rotation in a 3-dimensional space. 
 
 We recall that the orientations of an isotopic spin state can be generated by starting from a fixed initial isotopic direction, which can be chosen as the isotopic spin "up" direction, and then rotating to the desired final direction. The values of the three angles which specify the rotation can be considered as the coordinates for a point inside an abstract 3-dimensional space. Each point corresponds to a distinct rotation so that the isotopic spin states themselves can be identified with the points in this angular space. Thus, the internal symmetry space of the $SU(2)$ group looks like the interior of a 3-dimensional sphere.

 The symmetry space of a gauge group provides the local non-inertial coordinate frame for the internal degrees of freedom. To an imaginary observer inside this internal space, the interaction between a particle and an external gauge field looks like a rotation of the local coordinates. The amount of the rotation is determined by the strength of the external potential, and the relative change in the internal coordinates between two space-time points is just given by the connection as stated before. Thus, we see that there is a similarity between the geometrical description of relativity and the internal space picture.

The internal space is called a "fiber" by mathematicians \cite{choque,dreschsler}.
The idea of using a gauge potential to "link" together space-time with internal symmetries space is a new concept in physics. The new space formed by the union of 4-dimensional space time with an internal space is called a "fiber bundle" space. The reason for this name is that the internal spaces or "fibers" at each space-time point can all be viewed as the same space because they can be transformed into each other by a gauge transformation. Hence, the total space is a collection or "bundle" of fibers.

Given that Yang and Mills developed their theory using the same terminology as electrodynamics, it is relevant to ask if there are good reasons to describe gauge theory in geometrical terms, other than to establish a historical link with relativity. The best reason for doing so is that the geometrical picture provides a valuable aid to the standard language of field theory. Most of the pedagogical aids in field theory are based on a long familiarity with electrodynamics. Modern gauge theory, on the other hand, requires a new approach in order to deal with all a of the fundamental forces between elementary particles. The geometrical picture can provide a common arena for discussing electromagnetism, the electromagnetism, the strong and weak nuclear forces, and even gravity, because it depends on only very general properties of gauge theory. The fiber bundle representation will be reconsidered in Sec. XII to discuss ETGs.

\subsection{Local gauge transformations}
\label{due.7}
 We have described qualitatively how the gauge group is associated with a connection. Any particle or system which is localized in a small volume and carries an internal quantum number like isotopic spin has a direction in the internal symmetry space. This internal direction can be arbitrarily chosen at each point in space-time. In order to compare these internal space directions at two different space-time points $x$ and $x+dx$, we need to define an appropriate connection which can tell us how much the internal direction at $x$ differs from the direction $x+dx$. This connection must be capable of relating all possible directions in the internal space to each other. The most obvious way to relate two directions is to find out how much one direction has to rotate so that it agrees with the other direction. The set of all such rotations forms a symmetry group; thus, the connection between inertial space directions at different points act like a symmetry group as well.
 Our problem now is to see how a symmetry group transformation can lead to a connection which we identify with a gauge potential field. Let us, begin by writing the general form of a local symmetry transformation for an arbitrary (non Abelian) group,
 \begin{eqnarray}
 U\psi=exp\left(-iq\sum_k \theta^k(x)F_k\right)\psi\,.
 \label{III-1}
 \end{eqnarray}

The "local" nature of the transformation is indicated by the parameters $\theta^k(x)$ which are continuous of $x$. The constant $q$ is the electric charge for the $U(1)$ gauge group or a general "coupling constant" for an arbitrary gauge group. This is the only way in which the charge enters directly into the calculation. The general transformation (\ref{III-1}) is identical to the usual form of an ordinary spatial rotation if we identify the position-dependent parameter $\theta^k(x)$ with rotation angles. The $F_k$ are the generators of the internal symmetry group and satisfy the usual commutation relations
 \begin{eqnarray}
 \left[F_i,F_j\right]=ic_{ijk}F_k\,,
 \label{III-2}
 \end{eqnarray}
 where the structure constant $c_{ijk}$ depend on the particular group. For the isotopic-spin rotation group $SU(2)$, the generators, $F_k$ are the angular momentum operators. To see how the transformation (\ref{III-1}) defines a connection, let us consider the following simple operation. We will take a test particle described by a wavefunction $\psi(x)$ and move it between two points $x$ and $x+dx$ in space-time, and analyze how its direction changes in the internal symmetry space. The internal direction at $x$ is initially chosen to have the angles  $\theta^k(x)$.  As the test particle moves away form $x$, the internal direction changes in some continuous way until it reaches $x+dx$ where it has new internal direction given angles $\theta^k (x+dx)$. For an infinitesimal distance $dx$, this change can be described by the transformation (\ref{III-1}) acting on $\psi(x)$ and producing a rotation of the internal direction equal to the difference $d\theta^k=\theta^k(x+dx)-\theta^k(x)$. This rotation gives us what we need, namely, a connection between internal space directions at different points in space-time. We also see that this connection involves the derivative of a quantity just like the connection defined by Weyl. In this case, the quantities are the internal rotation angles $\theta^k(x)$. This is a straightforward generalization of the phase of a wavefunction to a set of angles which specify the internal direction.

\subsection{Connections and potentials}
\label{due.8}
Let us see how to calculate the connection from the symmetry transformation (\ref{III-1}) by moving the charged test particle through an external potential field. We will explicitly separate  the particle wavefunction $\psi(x)$ into external and internal parts. Let us write
 \begin{eqnarray}
\psi(x)=\sum_\alpha \psi_\alpha(x)u_\alpha\,,
 \label{III-3}
 \end{eqnarray}
where $u_\alpha$ form a set o a "basis vectors" in the internal space. The index $\alpha$ is an internal label such as the components of isotopic spin. The basis $u_\alpha$ is analogous to the local non-inertial frame in relativity. The external part $ \psi_\alpha(x)$  is then a "component of $\psi(x)$ in the basis $u_\alpha$. Under an inertial symmetry transformation, they transform in the usual way
 \begin{eqnarray}
 \psi_\beta=U_{\alpha\beta}\psi_\alpha\,,
 \label{III-4}
 \end{eqnarray}
where $U_{\alpha\beta}$ is the matrix representation of the symmetry group. We assume, that the representations is irreducible so that the particle has a unique charge or isotopic spin. The decomposition in Eq. (\ref{III-3}) is particularly useful because it will allow us to interpret the effect of the external potential field on the particle as a precession of the internal basis.

Now, when the test particle moves from $x$ to $x+dx$ through the external potential field, $\psi(x)$ changes by an amount $d\psi$ given

\begin{eqnarray}
d\psi=\psi(x+dx)-\psi(x)\,.
 \label{III-5}
 \end{eqnarray}

In general, $d\psi$ must contain both the change in the external $x$-dependent part of $\psi_\alpha$ and change in the internal space basis $u_\alpha$. From Eq. (\ref{III-3}), we can expand $d\psi$ to first order in $dx $ as

\begin{eqnarray}
d\psi=\sum_\alpha\left[ \left(\partial_\mu\psi_\alpha\right)dx^\mu u_\alpha+\psi_\alpha du_\alpha\right] \,.
 \label{III-6}
 \end{eqnarray}

The second terms contains the change $du_\alpha$ in the internal space basis. This term is given by the connection which we discussed above; it describes the effect of the external potential field on the internal space direction of the particle.

We now need to calculated the change $du_\alpha$ in the internal space basis. The connection between the internal space direction at different space-time points is given by an internal rotation. In this case, the internal directions are specified by a set of basis vectors, so we must calculate the change $du_\alpha$ from an infinitesimal internal rotation $U(dx)$ which is associated with the external displacement $dx$.

From Eq. (\ref{III-1}), we calculate the infinitesimal internal rotation $U(dx)$,

\begin{eqnarray}
U(dx)=exp\left[ -iq \sum_k d\theta^k F_k\right]\,,
 \label{III-7a}
 \end{eqnarray}

\begin{eqnarray}
d\theta^k=(\partial_\mu \theta^k)dx_\mu\,
 \label{III-7b}
 \end{eqnarray}
 
 which rotates the internal basis $u$ by an amount $du$,
 
 \begin{eqnarray}
 U(dx)u=u+du\,.
  \label{III-8}
 \end{eqnarray}
 The generators $F_k$ act like matrix operators on the column basis vector $u_\alpha$ so we can write
 
 \begin{eqnarray}
U(dx)u_\alpha=exp\left[ -iq\sum_k\left(\partial_\mu \theta^k\right)dx^\mu \left(F_k\right)_{\alpha\beta}\right]u_\beta\,.
  \label{III-9}
 \end{eqnarray} 
 Expanding $U(x)$ to first order in $dx$, we obtain
 
  \begin{eqnarray}
u_\alpha+du_\alpha=\left[\delta_{\alpha\beta}-iq\sum_k\left(\partial_\mu \theta^k\right)dx^\mu \left(F_k\right)_{\alpha\beta}\right]u_\beta\,,
\label{III-10}
 \end{eqnarray}  
 which then gives for the change in the basis,
  \begin{eqnarray} 
 du_\alpha=-iq\sum_k\left(\partial_\mu \theta^k\right)dx^\mu \left(F_k\right)_{\alpha\beta}u_\beta\,.
 \label{III-11}
 \end{eqnarray}  
 The net change  $du_\alpha$ will give us the connection that we have been seeking. Let us therefore introduce the new "connection operator"
    \begin{eqnarray} 
 \left(A_\mu\right)_{\alpha\beta}=\sum_k   \left(\partial_\mu \theta^k\right) \left(F_k\right)_{\alpha\beta}\,.
   \label{III-12}
 \end{eqnarray}    
We thus finally obtain for the total change $d\psi$,
 \begin{eqnarray}    
 d\psi  =\sum_{\alpha\beta} \left[ \left(\partial_\mu \psi_\alpha\right)\delta_{\alpha\beta}-iq  \left(A_\mu\right)_{\alpha\beta}\psi_\alpha \right]d x^\mu u_\beta\,,
    \label{III-13}
 \end{eqnarray}      
 where we have put in $\delta_{\alpha\beta}$   in order to factor out the basis vector $u_\beta$. Now, we can factor the change $d\psi$ into its own external and internal parts
 
  \begin{eqnarray} 
 d\psi=\sum_\beta\left(d\psi\right)_\beta u_\beta \equiv \sum_\beta\left(D_\mu\psi_\beta\right)dx^\mu u_\beta\,.
  \label{III-14}
 \end{eqnarray}     
     The operator $D_\mu$  the gauge covariant derivative which describes the changes in both the external and internal parts of $\psi(x)$. Thus we get from Eq. (\ref{III-13})
 
   \begin{eqnarray}     
 D_\mu\psi_\beta=\sum_\alpha\left[\delta_{\alpha\beta}\partial_\mu-iq \left(A_\mu\right)_{\alpha\beta} \right]\psi_\alpha \,.
  \label{III-15}
 \end{eqnarray}  
     
For the case of the electromagnetic gauge group $U(1)$, the internal space is one dimensional so that Eq. (\ref{III-15}) reduces to
  \begin{eqnarray}  
 D_\mu\psi=\left(\partial_\mu-iq A_\mu\right)\psi\,.    
    \label{III-16}
 \end{eqnarray}     
This is the "canonical momentum" which is familiar from Electromagnetism. We can also deduce from the example of the $U(1)$ group, that the connection operator defined in Eq. (\ref{III-12}), should be identified as the generalized version of the vector potential field $A_\mu$. Thus, we conclude that the external potential field is indeed a connection in the internal symmetry space.

\subsection{Choosing a gauge}
\label{due.9}

 We know that the usual reason for selecting a particular gauge is to simplify a calculation or to explicitly display an interesting features of a problem. This freedom to choose a gauge is another example of the arbitrariness in the vector potential.

The choice of a gauge actually involves both gauge invariance and Lorentz invariance simultaneously. A particular gauge usually imposes a constraints on the vector potential, such as ${\bf \bigtriangledown}\cdot {\bf A}=0$ for the Coulomb gauge. In general, such equations are obviously not Lorentz invariant. Yet we know that two observers, each in a different inertial frame, can choose the Coulomb gauge for the same electromagnetism problem. The electric and magnetic fields observed in the two different frames can then  be related by the usual Lorentz transformations between two frames. This example points out the fact that the space-time location $x$ at which the internal coordinate $\theta(x)$ is evaluated is also not a fixed position. A Lorentz transformation which changes the spatial coordinate in the inertial frame also affects the value of the internal angles ad can be interpreted as an internal rotation. Thus, regardless of whether the coordinate change is associated with Lorentz transformation of observers or an actual movement of the particle in the external field, the effect on the internal space is the same: it is rotated by a gauge transformation.

A Lorentz transformation between two inertial frames is therefore always associated with a gauge transformation. Thus, the vector potential observed in the two frames are related by
  \begin{eqnarray} 
  A'_\mu=L_\mu\,^\nu A_\nu-\partial_\mu \lambda\,,
 \label{III-22}
 \end{eqnarray} 
where $L_\mu\,^\nu$ is the Lorentz transformation. This shows that the vector potential actually does not transform like an ordinary four-vector under a Lorentz transformation. It picks up an extra term $\partial_\mu \lambda$ due to the rotaion in the internal space. This interesting fact is well known in quantum field theory but it is rarely mentioned in ordinary electromagnetism \cite{bjorken}.

We can now see exactly what is involved for a particular choice of gauge. In the Lorentz gauge, $\partial^\mu A_\mu$ is required to be invariant,

 \begin{eqnarray} 
\partial'^\mu A'_\mu=\partial^\mu A_\mu\,,
 \label{III-23}
 \end{eqnarray} 
even though $A_\mu$ is not a true four-vector. From Eq. (\ref{III-22}), we see that this is possible only if

 \begin{eqnarray} 
\partial^\mu \partial_\mu \lambda=0\,,
 \label{III-24}
 \end{eqnarray}
or, equivalently, 

\begin{eqnarray}
 \bigtriangledown ^2 \lambda-\frac{\partial^2 \lambda}{\partial t^2}=0 \,,
 \label{III-25}
 \end{eqnarray}

 which is the familiar equation for $\lambda$ in the Lorentz gauge. Thus we see that the choice of Lorentz gauge is a requirement that the effect of the internal space precession be eliminated so that $\partial^\mu A_\mu$ can be treated as if it were a relativistic invariant. By the same reasoning, other gauges like the Coulomb gauge are not invariant because the gauge condition does not completely eliminate the extra internal precession term. Thus in the Coulomb gauge
 \begin{eqnarray}
 {\bf \bigtriangledown'}\cdot {\bf A'}={\bf \bigtriangledown}\cdot {\bf A}=0\,,
  \label{III-26}
 \end{eqnarray}
but ${\bf A}$ and ${\bf A'}$ are not related by a simple Lorentz transformation. An additional gauge rotation is required \cite{france}.

\section{Gravity from gauge invariants}
\label{tre}
The above considerations on gauge symmetry can be developed for gravity. In this section, before discussing the physical meaning of GR, we will derive in a very general way the field equations considering manifolds equipped with curvature and torsion. Specifically, we want to show the role of global and local Poincar\'{e} invariance and the relevance of conservation laws in any theory of gravity. The approach is completely general and suitable for spinor, vector, bivector and tetrad fields independently of their specific physical meaning. Our first issue is what can "generate" gravity.
 
\subsection{What can "generate" gravity?}
\label{tre.1}
Since the perturbative scheme is unsatisfactory because it fails over one loop level and cannot be renormalized \cite{book},  as we have seen above, we can ask what can to produce gravity or in other words if there exists invariance principles leading to the gravitation \cite{SF}.
Following the prescriptions of GR, the physical
space-time is assumed to be a four-dimensional differential
manifold. In Special Relativity, this
manifold is the Minkwoski flat-space-time $M_{4}$ while, in GR, the underlying space-time is assumed to be curved in
order to describe the effects of gravitation.

Utiyama \cite{Utiyama} was the first to propose that GR can be seen as a gauge theory based on the local
Lorentz group $SO(3, 1)$ in much the same way that the
Yang-Mills gauge theory
 \cite{Yang-Mills} was developed on the basis of the internal
isospin gauge group $SU(2)$. In this formulation the Riemannian
connection is the gravitational counterpart of the Yang-Mills
gauge fields. While $SU(2)$, in the Yang-Mills theory, is an
internal symmetry group, the Lorentz symmetry represents the local
nature of space-time rather than internal degrees of freedom. The
Einstein Equivalence Principle, asserted for GR,
requires that the local space-time structure can be identified with
the Minkowski space-time possessing Lorentz symmetry. In order to
relate local Lorentz symmetry to the external space-time, we need
to solder the local space to the external space. The soldering
tools are the tetrad fields. Utiyama regarded the tetrads as
objects given \textit{a priori}. Soon after, Sciama  \cite{Sciama}
recognized that space-time should necessarily be endowed with
torsion in order to accommodate spinor fields. In other words, the
gravitational interaction of spinning particles requires the
modification of the Riemann space-time of GR to be
a (non-Riemannian) curved space-time with torsion. Although Sciama
used the tetrad formalism for his gauge-like handling of
gravitation, his theory fell shortcomings in treating tetrad
fields as gauge fields. Kibble  \cite{Kibble} made a comprehensive
extension of the Utiyama gauge theory of gravitation by showing
that the local Poincar\'{e} symmetry $SO(3,1)\rtimes T\left(
3,1\right) $ ($\rtimes $ represents the semi-direct product) can
generate a space-time with torsion as well as curvature. The gauge
fields introduced by Kibble include the tetrads as well as the
local affine connection. There have been a variety of gauge
theories of gravitation based on different local symmetries which
gave rise to several interesting applications in theoretical
physics \cite{Grignani, hehlrev2, Inomata, Ivanov1, Ivanov2,
Mansouri1, Mansouri2, Ivanenko, Sardanashvily,
Chang}.

Following the Kibble approach, it can be demonstrated how
gravitation can be formulated starting from a pure gauge
point of view. In particular, the aim of this section is to show, in
details, how a theory of gravitation is a gauge theory which can
be obtained starting from the local Poincar\'{e} symmetry and this feature works not only for GR but also for ETGs.

 A gauge theory of gravity based on a
nonlinear realization  of the local conformal-affine  group of
symmetry transformations can be formulated in any case \cite{ali}. This means that the
coframe fields and gauge connections of the theory can be always
obtained. The tetrads and Lorentz group metric have been  used to
induce a space-time metric. The inhomogenously transforming (under
the Lorentz group) connection coefficients give rise to
gravitational gauge potentials used to define covariant
derivatives accommodating minimal coupling of matter and gauge
fields. On the other hand, the tensor valued connection forms can be used as auxiliary dynamical fields associated with the
dilation, special conformal and deformation (shear) degrees of
freedom inherent to the bundle manifold. This allowes to define
the bundle curvature of the theory. Then boundary topological
invariants have been constructed. They served as a prototype
(source free) gravitational Lagrangian. Finally the Bianchi
identities, covariant field equations and gauge currents are
obtained.

Here, starting from the Invariance Principle, we consider first
the Global Poincar\'{e} Invariance and then the Local Poincar\'{e}
Invariance. This approach lead to construct a given theory of
gravity as a gauge theory. This point of view, if considered in
detail, can avoid many shortcomings  and could be useful to
formulate self-consistent schemes for any theory of gravity.

\subsection{Invariance Principles and the Noether Theorem}
\label{tre.2}
As it is well-known, field equations and conservation laws can
be obtained from a least action principle. The same principle is
the basis of any gauge theory so we start from it to develop our
considerations. Let us start from a least action principle and the
Noether theorem.

Let $\chi (x)$ be a multiplet field defined at a space-time point $x$ and $
\mathcal{L}\{\chi (x)$, $\partial _{j}\chi (x)$; $x\}$ be the
Lagrangian density of the system. In order to make a distinction between the global transformations and the local transformations, for the moment we use the Latin indices ($i,k=0,1,2,3$) for the former and the greek indices ($\mu,\nu=0,1,2,3$) for the latter.
 The action integral of the
system over a given space-time volume $\Omega $ is defined by
\begin{equation}
I(\Omega )=\int_{_{\Omega }}\mathcal{L}\{\chi (x)\text{, }\partial _{j}\chi
(x)\text{; }x\}\,d^{4}x\text{.}
\end{equation}
Now let us consider the infinitesimal variations of the
coordinates
\begin{equation}
x^{i}\rightarrow x^{\prime }{}^{i}=x^{i}+\delta x^{i}\text{,}
\end{equation}
and the field variables
\begin{equation}
\chi (x)\rightarrow \chi ^{\prime }(x^{\prime })=\chi (x)+\delta \chi (x)
\text{.}
\end{equation}
Correspondingly, the variation of the action is given by
\begin{equation}
\delta I=\int_{_{\Omega ^{\prime }}}\mathcal{L}^{\prime }(x^{\prime
})\,d^{4}x^{\prime }-\int_{_{\Omega }}\mathcal{L}(x)\,d^{4}x=\int_{_{\Omega
}}\left[ \mathcal{L}^{\prime }(x^{\prime })||\partial _{j}x^{\prime
}{}^{j}||-\mathcal{L}(x)\right] \,d^{4}x\text{.}
\end{equation}
Since the Jacobian for the infinitesimal variation of coordinates becomes
\begin{equation}
||\partial _{j}x^{\prime }{}^{j}||=1+\partial _{j}(\delta x^{j})\text{,}
\end{equation}
 the variation of the action takes the form,
\begin{equation}
\delta I=\int_{_{\Omega }}\left[ \delta \mathcal{L}(x)+\mathcal{L}
(x)\,\partial _{j}(\delta x^{j})\right] \,d^{4}x  \label{Action1}
\end{equation}
where
\begin{equation}
\delta \mathcal{L}(x)=\mathcal{L}^{\prime }(x^{\prime })-\mathcal{L}(x)\text{
.}
\end{equation}

For any function $\Phi (x)$ of $x$, it is convenient to define the fixed
point variation $\delta _{0}$ by,
\begin{equation}
\delta _{0}\Phi (x):=\Phi ^{\prime }(x)-\Phi (x)=\Phi ^{\prime }(x^{\prime
})-\Phi (x^{\prime })\text{.}
\end{equation}
Expanding the function to first order in $\delta x^{j}$ as
\begin{equation}
\Phi (x^{\prime })=\Phi (x)+\delta x^{j}\,\partial _{j}\Phi (x)\text{,}
\end{equation}
we obtain
\begin{equation}
\delta \Phi (x)=\Phi ^{\prime }(x^{\prime })-\Phi (x)=\Phi ^{\prime
}(x^{\prime })-\Phi (x^{\prime })+\Phi (x^{\prime })-\Phi (x)=\delta
_{0}\Phi (x)+\delta x^{j}\,\partial _{j}\Phi (x)\text{,}
\end{equation}
or
\begin{equation}
\delta _{0}\Phi (x)=\delta \Phi (x)-\delta x^{j}\partial _{j}\Phi (x)\text{.}
\end{equation}
The advantage to have the fixed point variation is that $\delta
_{0}$ commutes with $\partial _{j}$:
\begin{equation}
\delta _{0}\partial _{j}\Phi (x)=\partial _{j}\delta _{0}\Phi (x)\text{.}
\end{equation}
For $\Phi (x)=\chi (x)$, we have
\begin{equation}
\delta \chi =\delta _{0}\chi +\delta x^{i}\partial _{i}\chi \text{,}
\end{equation}
and
\begin{equation}
\delta \partial _{i}\chi =\partial _{i}(\delta _{0}\chi )-\partial (\delta
x^{j})\partial _{i}\chi \text{.}
\end{equation}
Using the fixed point variation in the integrand of
(\ref{Action1}) gives
\begin{equation}
\delta I=\int_{_{\Omega }}\left[ \delta _{0}\mathcal{L}(x)+\partial
_{j}(\delta x^{j}\,\mathcal{L}(x))\right] \,d^{4}x\text{.}  \label{Action2}
\end{equation}
If we require the action integral defined over any arbitrary region $\Omega $
be invariant, that is, $\delta I=0$, then we must have
\begin{equation}
\delta \mathcal{L}+\mathcal{L}\partial _{j}(\delta x^{j})=\delta _{0}
\mathcal{L}+\partial _{j}(\mathcal{L}\delta x^{j})=0\text{.}
\end{equation}
If $\partial _{j}(\delta x^{j})=0$, then $\delta \mathcal{L}=0$, that is,
the Lagrangian density $\mathcal{L}$ is invariant. In general, however, $
\partial _{j}(\delta x^{j})\neq 0$, and $\mathcal{L}$ transforms like a
scalar density. In other words, $\mathcal{L}$ is a Lagrangian density unless
$\partial _{j}(\delta x^{j})=0$.

For convenience, let us introduce a function $h(x)$ that behaves like a
scalar density, namely
\begin{equation}
\delta h+h\partial _{j}(\delta x^{j})=0\text{.}
\end{equation}
We further assume $\mathcal{L}(\chi ,\partial _{j}\chi
;x)=h(x)L(\chi ,\partial _{j}\chi ;x)$. Then we see that
\begin{equation}
\delta \mathcal{L}+\mathcal{L}\partial _{j}(\delta x^{j})=h\delta L\text{.}
\end{equation}
Hence the action integral remains invariant if
\begin{equation}
\delta L=0\text{.}
\end{equation}
The newly introduced function $L(\chi ,\partial _{j}\chi ;x)$ is the scalar
Lagrangian of the system.

Let us calculate the integrand of (\ref{Action2}) explicitly. The fixed
point variation of $\mathcal{L}(x)$ is a consequence of a fixed point
variation of the field $\chi (x)$,
\begin{equation}
\delta _{0}\mathcal{L}=\frac{\partial \mathcal{L}}{\partial \chi }\delta
_{0}\chi +\frac{\partial \mathcal{L}}{\partial (\partial _{j}\chi )}\delta
_{0}(\partial _{j}\chi )
\end{equation}
which can be cast into the form,
\begin{equation}
\delta _{0}\mathcal{L}=[\mathcal{L}]_{\chi }\delta _{0}\chi +\partial
_{j}\left( \frac{\partial \mathcal{L}}{\partial (\partial _{j}\chi )}\delta
_{0}\chi \right)
\end{equation}
where
\begin{equation}
\lbrack \mathcal{L}]_{\chi }\equiv \frac{\partial \mathcal{L}}{\partial \chi
}-\partial _{j}\left( \frac{\partial \mathcal{L}}{\partial (\partial
_{j}\chi )}\right) \text{.}
\end{equation}
Consequently, we have the action integral in the form
\begin{equation}
\delta I=\int_{_{\Omega }}\left\{ [\mathcal{L}]_{\chi }\delta _{0}\chi
+\partial _{j}\left( \frac{\partial \mathcal{L}}{\partial (\partial _{j}\chi
)}\delta \chi -T_{k}^{j}\,\delta x^{k}\right) \right\} d^{4}x\text{,}
\end{equation}
where
\begin{equation}
T^{j}\,_{k}:=\frac{\partial \mathcal{L}}{\partial (\partial _{j}\chi )}
\partial _{k}\chi -\delta _{k}^{j}\,\mathcal{L}
\end{equation}
is the canonical energy-momentum tensor density. If the variations are
chosen in such a way that $\delta x^{j}=0$ over $\Omega $ and $\delta
_{0}\chi $ vanishes on the boundary of $\Omega $, then $\delta I=0$ gives us
the Euler-Lagrange equation,
\begin{equation}
\lbrack \mathcal{L}]_{\chi }=\frac{\partial \mathcal{L}}{\partial \chi }
-\partial _{j}\left( \frac{\partial \mathcal{L}}{\partial (\partial _{j}\chi
)}\right) =0\text{.}
\end{equation}
On the other hand, if the field variables obey the Euler-Lagrange equation, $
[\mathcal{L}]_{\chi }=0$, then we have
\begin{equation}
\partial _{j}\left( \frac{\partial \mathcal{L}}{\partial (\partial _{j}\chi )
}\delta \chi -T^{j}\,_{k}\,\delta x^{k}\right) =0\text{,}
\end{equation}
which gives rise, considering also the Noether theorem, to
conservation laws. These very straightforward considerations are
at the basis of our following discussion.

\subsection{The Global Poincar\'e invariance}
\label{tre.3}
As standard,  we assert that our space-time in the absence of
gravitation is a Minkowski space $M_{4}$. The isometry group of
$M_{4}$ is the group of Poincar\'{e} transformations (PT) which
consists of the Lorentz group $SO(3$, $1)$ and the translation
group $T(3$, $1)$. The Poincar\'{e} transformations of coordinates
are
\begin{equation}
x^{i}\overset{PT}{\rightarrow }x^{\prime }{}^{i}=a^{i}\,_{j}x^{j}+b^{i}\text{
,}  \label{PoincareTransf}
\end{equation}
where $a_{j}^{i}$ and $b^{i}$ are real constants, and $a_{j}^{i}$ satisfy
the orthogonality conditions $a_{k}^{i}a_{j}^{k}=\delta _{j}^{i}$. For
infinitesimal variations,
\begin{equation}
\delta x^{\prime }{}^{i}=\chi ^{\prime }(x^{\prime })-\chi (x)=\varepsilon
^{i}\,_{j}x^{j}+\varepsilon ^{i}  \label{PoincareTrasf1}
\end{equation}
where $\varepsilon _{ij}+\varepsilon _{ji}=0$. While the Lorentz
transformation forms a six parameter group, the Poincar\'{e} group has ten
parameters. The Lie algebra for the ten generators of the Poincar\'{e} group
is
\begin{eqnarray}
\lbrack \Xi _{ij}\text{, }\Xi _{kl}] &=&\eta _{ik}\,\Xi _{jl}+\eta
_{jl}\,\Xi _{ik}-\eta _{jk}\,\Xi _{il}-\eta _{il}\,\Xi _{jk}\text{,}  \notag
\\
&& \\
\lbrack \Xi _{ij}\text{, }T_{k}] &=&\eta _{jk}T_{i}-\eta _{ik}T_{j}\text{, \
}[T_{i}\text{, }T_{j}]=0\text{,}  \notag
\end{eqnarray}
where $\Xi _{ij}$ are the generators of Lorentz transformations,
and $T_{i}$ are the generators of four-dimensional translations.
Obviously, $\partial
_{i}(\delta x^{i})=0$ for the Poincar\'{e} transformations (\ref
{PoincareTransf}). Therefore, our Lagrangian density $\mathcal{L}$, which is
the same as $L$ with $h(x)=1$ in this case, is invariant; namely, $\delta
\mathcal{L}=\delta L=0$ for $\delta I=0$.

Suppose that the field $\chi (x)$ transforms under the infinitesimal Poincar
\'{e} transformations as
\begin{equation}
\delta \chi =\frac{1}{2}\varepsilon ^{ij}S_{ij}\chi \text{,}
\end{equation}
where the tensors $S_{ij}$ are the generators of the Lorentz group,
satisfying
\begin{equation}
S_{ij}=-S_{ji}\text{, \ }[S_{ij}\text{, }S_{kl}]=\eta _{ik}\,S_{jl}+\eta
_{jl}\,S_{ik}-\eta _{jk}\,S_{il}-\eta _{il}\,S_{jk}\text{.}
\label{LorentzGen}
\end{equation}
Correspondingly, the derivative of $\chi $ transforms as
\begin{equation}
\delta (\partial _{k}\chi )=\frac{1}{2}\varepsilon ^{ij}S_{ij}\partial
_{k}\chi -\varepsilon ^{i}\,_{k}\partial _{i}\chi \text{.}
\end{equation}
Since the choice of infinitesimal parameters $\varepsilon ^{i}$ and $
\varepsilon ^{ij}$ is arbitrary, the vanishing variation of the
Lagrangian density  $\delta \mathcal{L}=0$ leads to the
identities,
\begin{equation}
\frac{\partial \mathcal{L}}{\partial \chi }S_{ij}\chi +\frac{\partial
\mathcal{L}}{\partial (\partial _{k}\chi )}(S_{ij}\partial _{k}\chi +\eta
_{ki}\partial _{j}\chi -\eta _{kj}\partial _{i}\chi )=0\text{.}
\end{equation}
We also obtain the following conservation laws
\begin{equation}
\partial _{j}\,T_{k}^{j}=0,\ \partial _{k}\left(
S^{k}\,_{ij}-x_{i}T^{k}\,_{j}+x_{j}T^{k}\,_{i}\right) =0\text{,}
\end{equation}
where
\begin{equation}
S^{k}\,_{ij}:=-\frac{\partial \mathcal{L}}{\partial (\partial _{k}\chi )}
S_{ij}\chi \text{.}
\end{equation}
These conservation laws imply that the energy-momentum and angular momentum,
respectively
\begin{equation}
P_{l}=\int T_{l}^{0}\,d^{3}x,\ J_{ij}=\int \left[ S^{0}\,_{ij}\,-\left(
x_{i}T^{0}\,_{j}-x_{j}T^{0}\,_{i}\right) \right] d^{3}x\text{,}
\end{equation}
are conserved. This means that the system invariant under the ten
parameter symmetry group has ten conserved quantities. This is an
example of Noether symmetry. The first term of the angular
momentum integral corresponds to the spin angular momentum while
the second term gives the orbital angular momentum. The global
Poincar\'{e} invariance of a system means that, for the system,
the space-time is homogeneous (all space-time points are equivalent)
as dictated by the translational invariance and is isotropic (all
directions about a space-time point are equivalent) as indicated by
the Lorentz invariance. It is interesting to observe that the
fixed point variation of the field variables takes the form
\begin{equation}
\delta _{0}\chi =\frac{1}{2}\varepsilon ^{j}\,_{k}\Xi _{j}\,^{k}\,\chi
+\varepsilon ^{j}\,T_{j}\,\chi \text{,}
\end{equation}
where
\begin{equation}
\Xi _{j}\,^{k}=S_{j}\,^{k}+\left( x^{j}\partial _{k}-x^{k}\partial
_{j}\right) \text{, \ }T_{j}=-\partial _{j}\text{.}
\end{equation}
We remark that $\Xi _{j}\,^{k}$ are the generators of the Lorentz
transformation and $T_{j}$ are those of the translations.

\subsection{The Local Poincar\'e invariance}
\label{tre.4}

As next step,  let us consider a modification of the infinitesimal
Poincar\'{e}
transformations (\ref{PoincareTrasf1}) by assuming that the parameters $
\varepsilon _{k}^{j}$ and $\varepsilon ^{j}$ are functions of the
coordinates and by writing them altogether as
\begin{equation}
\delta x^{\mu }=\varepsilon ^{\mu }\,_{\nu }(x)\,x^{\nu }+\varepsilon ^{\mu
}(x)=\xi ^{\mu }\text{,}
\end{equation}
which we call the local Poincar\'{e} transformations (or the
general coordinate transformations). As before, in order to distinguish between global
transformations and local transformation, we use the Latin indices $(j$, $
k=0$, $1$, $2$, $3)$ for the former and the Greek indices $(\mu $,
$\nu =0$, $1$, $2$, $3)$ for the latter. The variation of the
field variables $\chi (x) $ defined at a point $x$ is still the
same as that of the global Poincar\'{e} transformations,
\begin{equation}
\delta \chi =\frac{1}{2}\varepsilon _{ij}S^{ij}\,\chi \text{.}
\end{equation}
The corresponding fixed point variation of $\chi $ takes the form,
\begin{equation}
\delta _{0}\chi =\frac{1}{2}\varepsilon _{ij}S^{ij}\chi -\xi ^{\nu }\partial
_{\nu }\chi .  \label{FixedPtVar}
\end{equation}
Differentiating both sides of (\ref{FixedPtVar}) with respect to $x^{\mu }$,
we have
\begin{equation}
\delta _{0}\partial _{\mu }\chi =\frac{1}{2}\varepsilon ^{ij}S_{ij}\partial
_{\mu }\chi +\frac{1}{2}(\partial _{\mu }\varepsilon ^{ij})\,S^{ij}\chi
-\partial _{\mu }(\xi ^{\nu }\partial _{\nu }\chi )\text{.}
\end{equation}
By using these variations, we obtain the variation of the
Lagrangian $L$,
\begin{equation}
\delta \mathcal{L}+\partial _{\mu }(\delta x^{\mu })\mathcal{L}=h\delta
L=\delta _{0}\mathcal{L}+\partial _{\nu }(\mathcal{L}\delta x^{\nu })=-\frac{
1}{2}(\partial _{\mu }\varepsilon ^{ij})\,S^{\mu }\,_{ij}-\partial _{\mu
}\xi ^{\nu }\,T_{\text{ }\nu }^{\mu \,}\text{,}
\end{equation}
which is no longer zero unless the parameters $\varepsilon ^{ij}$ and $\xi
^{\nu }$ become constants. Accordingly, the action integral for the given
Lagrangian density $\mathcal{L}$ is not invariant under the local Poincar\'{e} transformations. We notice that while $\partial _{j}(\delta
x^{j})=0$ for the local Poincar\'{e} transformations, $\partial
_{\mu }\xi ^{\mu }$
does not vanish under local Poincar\'{e} transformations. Hence, as expected $
\mathcal{L}$ is not a Lagrangian scalar but a Lagrangian density.
As mentioned earlier, in order to define the Lagrangian $L$, we
have to select an appropriate non-trivial scalar function $h(x)$
satisfying
\begin{equation}
\delta h+h\partial _{\mu }\xi ^{\mu }=0\text{.}
\end{equation}

Now we consider a minimal modification of the Lagrangian so as to
make the action integral invariant under the local Poincar\'{e}
transformations. It is rather obvious that if there is a covariant
derivative $\nabla _{k}\chi $ which transforms as
\begin{equation}
\delta (\nabla _{k}\chi )=\frac{1}{2}\varepsilon ^{ij}S_{ij}\nabla _{k}\chi
-\varepsilon ^{i}\,_{k}\nabla _{i}\chi \text{,}
\end{equation}
then a modified Lagrangian $L^{\prime }(\chi $, $\partial _{k}\chi $, $
x)=L(\chi $, $\nabla _{k}\chi $, $x)$, obtained by replacing
$\partial _{k}\chi $ of $L(\chi $, $\partial _{k}\chi $, $x)$ by
$\nabla _{k}\chi $, remains invariant under the local Poincar\'{e}
transformations, that is
\begin{equation}
\delta L^{\prime }=\frac{\partial L^{\prime }}{\partial \chi }\delta \chi +
\frac{\partial L^{\prime }}{\partial (\nabla _{k}\chi )}\delta (\nabla
_{k}\chi )=0\text{.}
\end{equation}
To find such a $k$-covariant derivative, we introduce the gauge fields $
A^{ij}\,_{\mu }=-A^{ji}\,_{\mu }$ and define the $\mu$-covariant derivative
\begin{equation}
\nabla _{\mu }\chi :=\partial _{\mu }\chi +\frac{1}{2}A^{ij}\,_{\mu
}S_{ij}\chi \text{,}  \label{mu-covariant}
\end{equation}
in such a way that the covariant derivative transforms as
\begin{equation}
\delta _{0}\nabla _{\mu }\chi =\frac{1}{2}S_{ij}\nabla _{\mu }\chi -\partial
_{\mu }(\xi ^{\nu }\nabla _{\nu }\chi )\text{.}
\end{equation}
The transformation properties of $A_{\;\;\;\mu }^{ab}$ are determined by $
\nabla _{\mu }\chi $ and $\delta \nabla _{\mu }\chi $. Making use of
\begin{equation}
\delta \nabla _{\mu }\chi =\frac{1}{2}\varepsilon ^{ij},_{\mu }S_{ij}\chi +
\frac{1}{2}\varepsilon ^{ij}S_{ij}\partial _{\mu }\chi -\left( \partial
_{\mu }\xi ^{\nu }\right) \partial _{\nu }\psi +\frac{1}{2}\delta
A_{\;\;\;\mu }^{ij}S_{ij}\chi +\frac{1}{4}A_{\;\;\;\mu
}^{ij}S_{ij}\varepsilon ^{kl}S_{kl}\chi
\end{equation}
and comparing with (\ref{mu-covariant}) we obtain,
\begin{equation}
\delta A_{\;\;\;\mu }^{ij}S_{ij}\chi +\varepsilon ^{ij},_{\mu }S_{ij}\chi +
\frac{1}{2}\left( A_{\;\;\;\mu }^{ij}\varepsilon ^{kl}-\varepsilon
^{ij}A_{\;\;\;\mu }^{kl}\right) S_{ij}S_{kl}\chi +\left( \partial _{\mu }\xi
^{\nu }\right) A_{\;\;\;\nu }^{ij}S_{ij}\chi =0\text{.}  \label{inter}
\end{equation}
Using the antisymmetry in $ij$ and $kl$ to rewrite the term in
parentheses on the {\it rhs} of (\ref{inter}) as $\left[
S_{ij},S_{kl}\right] A_{\;\;\;\mu }^{ij}\varepsilon ^{kl}\chi $,
we see the explicit appearance of the commutator $\left[
S_{ij},S_{kl}\right] $. Using the expression for the
commutator of Lie algebra generators
\begin{equation}
\left[ S_{ij}\text{, }S_{kl}\right] =\frac{1}{2}c_{\;\;\;\;\ \left[ ij\right]
\left[ kl\right] }^{\left[ ef\right] }S_{ef}\text{,}
\end{equation}
where $c_{\;\;\;\;\left[ ij\right] \left[ kl\right] }^{\left[ ef\right] }$
(the square brackets denote anti-symmetrization) is the structure constants
of the Lorentz group (deduced below), we have
\begin{equation}
\left[ S_{ij}\text{, }S_{kl}\right] A_{\;\;\;\mu }^{ij}\varepsilon ^{kl}=
\frac{1}{2}\left( A_{\mu }^{ic}\varepsilon _{c}^{j}-A_{\mu }^{cj}\varepsilon
_{c}^{i}\right) S_{ij}\text{.}
\end{equation}
The substitution of this equation and the consideration of the antisymmetry of $
\varepsilon _{c}^{\;b}=-\varepsilon _{\;c}^{b}$ enable us to write
\begin{equation}
\delta A^{ij}\,_{\mu }=\varepsilon ^{i}\,_{k}A^{kj}\,_{\mu }+\varepsilon
^{j}\,_{k}A^{ik}\,_{\mu }-(\partial _{\mu }\xi ^{\nu })A^{ij}\,_{\nu
}-\partial _{\mu }\varepsilon ^{ij}.
\end{equation}
We require the $k$-derivative and $\mu $-derivative of $\chi $ to be
linearly related as
\begin{equation}
\nabla _{k}\chi =V_{k}\,^{\mu }(x)\nabla _{\mu }\chi ,  \label{inter2}
\end{equation}
where the coefficients $V_{k}\,^{\mu }(x)$ are position-dependent and behave
like a new set of field variables. From (\ref{inter2}) it is evident that $
\nabla _{k}\chi $ varies as
\begin{equation}
\delta \nabla _{k}\chi =\delta V_{k}^{\mu }\nabla _{\mu }\chi +V_{k}^{\mu
}\delta \nabla _{\mu }\chi \text{.}
\end{equation}
Comparing with $\delta \nabla _{k}\chi =\frac{1}{2}\varepsilon
^{ab}S_{ab}\nabla _{k}\chi -\varepsilon _{\text{ }k}^{j}\nabla _{j}\chi $ we
obtain,
\begin{equation}
V_{\alpha }^{k}\delta V_{k}^{\mu }\nabla _{\mu }\chi -\xi ^{\nu },_{\alpha
}\nabla _{\nu }\chi +V_{\alpha }^{k}\varepsilon _{\text{ }k}^{j}\nabla
_{j}\chi =0\text{.}
\end{equation}
Exploiting $\delta \left( V_{\alpha }^{k}V_{k}^{\mu }\right) =0$ we find the
quantity $V_{k}\,^{\mu }$ transforms according to
\begin{equation}
\delta V_{k}\,^{\mu }=V_{k}\,^{\nu }\partial _{\nu }\xi ^{\mu }-V_{i}\,^{\mu
}\varepsilon ^{i}\,_{k}.
\end{equation}
It is also important to recognize that the inverse of $\det (V_{k}\,^{\mu })$
transforms like a scalar density as $h(x)$ does. For our minimal
modification of the Lagrangian density, we utilize this available quantity
for the scalar density $h$; namely, we let
\begin{equation}
h(x)=[\det (V_{k}\,^{\mu })]^{-1}\text{.}
\end{equation}
In the limiting case, when we consider Poincar\'{e}
transformations, that are not space-time dependent, $V_{k}\,^{\mu
}\rightarrow \delta _{k}^{\mu }$ so that $h(x)\rightarrow 1$. This
is a desirable property. Then we replace the Lagrangian density
$\mathcal{L}(\chi $, $\partial _{k}\chi $, $x)$, invariant under
the global Poincar\'{e} transformations, by a Lagrangian density
\begin{equation}
\mathcal{L}(\chi \text{, }\partial _{\mu }\chi \text{; }x)\rightarrow
h(x)L(\chi \text{, }\nabla _{k}\chi )\text{.}
\end{equation}
The action integral with this modified Lagrangian density remains
invariant under the local Poincar\'{e} transformations. Since the
local Poincar\'{e} transformations $\delta x^{\mu }=\xi ^{\mu
}(x)$ are nothing else but  generalized coordinate
transformations, the newly introduced gauge fields $V_{i}^{\lambda
}$ and $A^{ij}\,_{\mu }$ can be interpreted, respectively, as the
tetrad ({\it vierbein}) fields which set the local coordinate
frame and as a local affine connection with respect to the tetrad
frame (see also \cite{basini}).

\subsection{Spinors, vectors, bivectors and tetrads}
Let us consider first  the case where the multiplet field $\chi $
is the Dirac field $\psi (x)$ which behaves like a four-component
spinor under the Lorentz transformations and transforms as
\begin{equation}
\psi (x)\rightarrow \psi ^{\prime }(x^{\prime })=S(\Lambda )\psi (x)\text{,}
\end{equation}
where $S(\Lambda )$ is an irreducible unitary representation of the Lorentz
group. Since the bilinear form $v^{k}=i\overline{\psi }\gamma ^{k}\psi $ is
a vector, it transforms according to
\begin{equation}
v^{j}=\Lambda _{\text{ }k}^{j}v^{k}\text{,}
\end{equation}
where $\Lambda _{\text{ }i}^{j}$ is a Lorentz transformation matrix
satisfying
\begin{equation}
\Lambda _{ij}+\Lambda _{ji}=0\text{.}
\end{equation}
The invariance of $v^{i}$ (or the covariance of the Dirac
equation) under the
transformation $\psi (x)\rightarrow \psi ^{\prime }(x^{\prime })$ leads to%
\begin{equation}
S^{-1}(\Lambda )\gamma ^{\mu }S(\Lambda )=\Lambda _{\nu }^{\mu }\gamma ^{\nu
},  \label{Transf}
\end{equation}
where the $\gamma ^{\prime }s$ are the Dirac $\gamma $-matrices satisfying
the anticommutator,
\begin{equation}
\gamma _{i}\gamma _{j}+\gamma _{j}\gamma _{i}=\eta _{ij}\mathbf{1}\text{.}
\end{equation}
Furthermore, we notice that the $\gamma $-matrices have the following
properties:
\begin{equation}
\left\{
\begin{array}{c}
\left( \gamma _{0}\right) ^{\dag }=-\gamma _{0}\text{, }\left( \gamma
^{0}\right) ^{2}=\left( \gamma _{0}\right) ^{2}=-1\text{, }\gamma
_{0}=-\gamma ^{0}\text{ and }\gamma _{0}\gamma ^{0}=1 \\
\\
\left( \gamma _{k}\right) ^{\dag }=\gamma _{k}\text{ ,}\left( \gamma
^{k}\right) ^{2}=\left( \gamma _{k}\right) ^{2}=1\text{; }(k=1,2,3)\text{
and }\gamma _{k}=\gamma ^{k} \\
\\
\left( \gamma _{5}\right) ^{\dag }=-\gamma _{5}\text{, }\left( \gamma
_{5}\right) ^{2}=-1\text{ and }\gamma ^{5}=\gamma _{5}\text{.}%
\end{array}
\right.
\end{equation}
We assume the transformation $S(\Lambda )$ can be put into the form $
S(\Lambda )=e^{\Lambda _{\mu \nu }\gamma ^{\mu \nu }}$. Expanding
$S(\Lambda )$ about the identity and only retaining terms up to
the first order in the
infinitesimals and expanding $\Lambda _{\mu \nu }$ to the first order in $
\varepsilon _{\mu \nu }$
\begin{equation}
\Lambda _{\mu \nu }=\delta _{\mu \nu }+\varepsilon _{\mu \nu }\text{, }
\varepsilon _{ij}+\varepsilon _{ji}=0,  \label{LorentzTransf}
\end{equation}
we get
\begin{equation}
S(\Lambda )=1+\frac{1}{2}\varepsilon ^{ij}\gamma _{ij}\text{.}
\label{DiracTransf}
\end{equation}
In order to determine the form of $\gamma _{ij}$, we substitute
(\ref{LorentzTransf})
and (\ref{DiracTransf}) into (\ref{Transf}) to obtain
\begin{equation}
\frac{1}{2}\varepsilon _{ij}\left[ \gamma ^{ij}\text{, }\gamma ^{k}\right]
=\eta ^{ki}\varepsilon _{ji}\gamma ^{j}\text{.}  \label{inter3}
\end{equation}
Rewriting the {\it rhs} of (\ref{inter3}) using the antisymmetry of $\varepsilon
_{ij}$ as
\begin{equation}
\eta ^{ki}\varepsilon _{ji}\gamma ^{j}=\frac{1}{2}\varepsilon _{ij}\left(
\eta ^{ki}\gamma ^{j}-\eta ^{kj}\gamma ^{i}\right) \text{,}
\end{equation}
yields
\begin{equation}
\left[ \gamma ^{k}\text{, }\gamma ^{ij}\right] =\eta ^{ki}\gamma ^{j}-\eta
^{kj}\gamma ^{i}\text{.}
\end{equation}
Assuming the solution to have the form of an antisymmetric product of two
matrices, we obtain the solution
\begin{equation}
\gamma ^{ij}:=\frac{1}{2}\left[ \gamma ^{i}\text{, }\gamma ^{j}\right] \text{.}
\end{equation}
If $\chi =\psi $, the group generator $S_{ij}$ appearing in (\ref{LorentzGen}
) is identified with
\begin{equation}
S_{ij}\equiv \gamma _{ij}=\frac{1}{2}(\gamma _{i}\gamma _{j}-\gamma
_{j}\gamma _{i}).
\end{equation}
To be explicit, the Dirac field transforms under Lorentz
transformations  (LT) as
\begin{equation}
\delta \psi (x)=\frac{1}{2}\varepsilon ^{ij}\gamma _{ij}\psi (x).
\end{equation}
The Pauli conjugate of the Dirac field is denoted $\overline{\psi }$ and
defined by
\begin{equation}
\overline{\psi }(x):=i\psi ^{\dagger }(x)\,\gamma _{0}\text{, }i\in
\mathbb{C}
\text{.}
\end{equation}
The conjugate field $\overline{\psi }$ transforms under LTs as,
\begin{equation}
\delta \overline{\psi }=-\overline{\psi }\frac{1}{2}\varepsilon ^{ij}
\overline{\psi }\gamma _{ij}\text{.}
\end{equation}

Under local LTs, $\varepsilon _{ab}(x)$ becomes a function of
space-time. Now, unlike $\partial _{\mu }\psi (x)$, the derivative
of $\psi ^{\prime
}(x^{\prime })$ is no longer homogenous due to the occurrence of the term $
\gamma ^{ab}\left[ \partial _{\mu }\varepsilon _{ab}(x)\right] \psi (x)$ in $
\partial _{\mu }\psi ^{\prime }(x^{\prime })$, which is non-vanishing unless
$\varepsilon _{ab}$ is constant. When going from locally flat to curved
space-time, we must generalize $\partial _{\mu }$ to the covariant derivative $
\nabla _{\mu }$ to compensate for this extra term, allowing to
gauge the group of LTs. Thus, by using $\nabla _{\mu }$, we can
preserve the
invariance of the Lagrangian for arbitrary local LTs at each space-time point
\begin{equation}
\nabla _{\mu }\psi ^{\prime }(x^{\prime })=S(\Lambda (x))\nabla _{\mu }\psi
(x)\text{.}
\end{equation}
To determine the explicit form of the connection belonging to $\nabla _{\mu }
$, we study the derivative of $S(\Lambda (x))$. The transformation $
S(\Lambda (x))$ is given by
\begin{equation}
S(\Lambda (x))=1+\frac{1}{2}\varepsilon _{ab}(x)\gamma ^{ab}\text{.}
\end{equation}
Since $\varepsilon _{ab}(x)$ is only a function of space-time for local
Lorentz coordinates, we express this infinitesimal LT in terms of general
coordinates only by shifting all space-time dependence of the local
coordinates into tetrad fields as
\begin{equation}
\varepsilon _{ab}(x)=V_{a}^{\text{ \ }\lambda }(x)V_{\text{ \ }b}^{\nu
}(x)\varepsilon _{\lambda \nu }\text{.}
\end{equation}
Substituting this expression for $\varepsilon _{ab}(x)$, we obtain
\begin{equation}
\partial _{\mu }\varepsilon _{ab}(x)=\partial _{\mu }\left[ V_{a}^{\text{ \ }
\lambda }(x)V_{\text{ \ }b}^{\nu }(x)\varepsilon _{\lambda \nu }\right]
\text{.}
\end{equation}
However, since $\varepsilon _{\lambda \nu }$ has no space-time
dependence, this reduces to
\begin{equation}
\partial _{\mu }\varepsilon _{ab}(x)=V_{a}^{\text{ \ }\lambda }(x)\partial
_{\mu }V_{b\lambda }(x)-V_{b}^{\text{ \ }\nu }(x)\partial _{\mu }V_{a\nu }(x)
\text{.}  \label{inter4}
\end{equation}
Letting
\begin{equation}
\omega _{\mu ba}:=V_{b}^{\text{ \ }\nu }(x)\partial _{\mu }V_{a\nu }(x),
\label{inter9}
\end{equation}
the first and second terms in Eq.(\ref{inter4}) become
$V_{a}^{\lambda
}(x)\partial _{\mu }V_{b\lambda }(x)=\frac{1}{2}\omega _{\mu ab}$ and $%
V_{b}^{\nu }(x)\partial _{\mu }V_{a\nu }(x)=\frac{1}{2}\omega _{\mu ba}$
respectively. Using the identification
\begin{equation}
\partial _{\mu }\varepsilon _{ab}(x)=\omega _{\mu ab}\text{,}
\end{equation}
we write
\begin{equation}
\partial _{\mu }S(\Lambda (x))=-\frac{1}{2}\gamma ^{ab}\omega _{\mu ab}\text{%
.}
\end{equation}
According to (\ref{mu-covariant}), the covariant derivative of the Dirac
spinor is
\begin{equation}
\nabla _{\mu }\psi =\partial _{\mu }\psi +\frac{1}{2}A^{ij}\,_{\mu }\gamma
_{ij}\psi \text{.}  \label{del-mu-psi}
\end{equation}
Correspondingly, the covariant derivative of $\bar{\psi}$ is given by
\begin{equation}
\nabla _{\mu }\overline{\psi }=\partial _{\mu }\overline{\psi }-\frac{1}{2}%
A^{ij}\,_{\mu }\bar{\psi}\gamma _{ij}\text{.}
\end{equation}
Using the covariant derivatives of $\psi $ and $\bar{\psi}$, we can show
that
\begin{equation}
\nabla _{\mu }v_{j}=\partial _{\mu }v_{j}-A^{i}\,_{j\mu }v_{i}\text{.}
\end{equation}%
The same covariant derivative should be used for any covariant vector $v_{k}$
under the Lorentz transformation. Since $\nabla _{\mu }(v_{i}v^{i})=\partial
_{\mu }(v_{i}v^{i})$, the covariant derivative for a contravariant vector $%
v^{i}$ must be
\begin{equation}
\nabla _{\mu }v^{i}=\partial _{\mu }v^{i}+A^{i}\,_{j\mu }v^{j}\text{.}
\end{equation}%
Since the tetrad $V_{i}\,^{\mu }$ is a covariant vector under Lorentz
transformations, its covariant derivative must transform according to the
same rule. Using $\nabla _{a}=V_{a}^{\mu }(x)\nabla _{\mu }$, the covariant
derivatives of a tetrad in local Lorentz coordinates read%
\begin{equation}
\nabla _{\nu }V_{i}\,^{\mu }=\partial _{\nu }V_{i}\,^{\mu }-A^{k}\,_{i\nu
}V_{k}\,^{\mu }\text{, }\nabla _{\nu }V^{i}\,_{\mu }=\partial _{\nu
}V^{i}\,_{\mu }+A^{i}\,_{k\nu }V^{k}\,_{\mu }\text{.}
\end{equation}%
The inverse of $V_{i}\,^{\mu }$ is denoted by $V^{i}\,_{\mu }$ and satisfies
\begin{equation}
V^{i}\,_{\mu }V_{i}\,^{\nu }=\delta _{\mu }\,^{\nu },~~~V^{i}\,_{\mu
}V_{j}\,^{\mu }=\delta ^{i}\,_{j}\text{.}
\end{equation}%
To allow the transition to curved space-time, we take account of
the general coordinates of objects that are covariant under local
Poincar\'{e}
transformations. Here we define the covariant derivative of a quantity $%
v^{\lambda }$ which behaves like a controvariant vector under the local
Poincar\'{e} transformation. Namely
\begin{equation}
D_{\nu }v^{\lambda }\equiv V_{i}\,^{\lambda }\nabla _{\nu }v^{i}=\partial
_{\nu }v^{\lambda }+\Gamma ^{\lambda }\,_{\mu \nu }v^{\mu }\text{, \ }D_{\nu
}v_{\mu }\equiv V^{i}\,_{\mu }\nabla _{\nu }v_{i}=\partial _{\nu }v_{\mu
}-\Gamma ^{\lambda }\,_{\mu \nu }v_{\lambda }\text{,}
\end{equation}%
where%
\begin{equation}
\Gamma ^{\lambda }\,_{\mu \nu }:=V_{i}\,^{\lambda }\nabla _{\nu
}V^{i}\,_{\mu }\equiv -V^{i}\,_{\mu }\nabla _{\nu }V_{i}\,^{\lambda }.
\end{equation}%
The definition of $\Gamma ^{\lambda }\,_{\mu \nu }$ implies
\begin{eqnarray}
D_{\nu }V_{i}\,^{\lambda } &=&\nabla _{\nu }V_{i}\,^{\lambda }+\Gamma
^{\lambda }\,_{\mu \nu }V_{i}\,^{\mu }=\partial _{\nu }V_{i}\,^{\lambda
}-A^{k}\,_{i\nu }V_{k}\,^{\lambda }+\Gamma ^{\lambda }\,_{\mu \nu
}V_{i}\,^{\mu }=0\text{,}  \label{inter6} \\
&&  \notag \\
D_{\nu }V^{i}\,_{\mu } &=&\nabla _{\nu }V^{i}\,_{\mu }-\Gamma ^{\lambda
}\,_{\mu \nu }V^{i}\,_{\lambda }=\partial _{\nu }V^{i}\,_{\mu
}+A^{i}\,_{k\nu }V^{k}\,_{\mu }-\Gamma ^{\lambda }\,_{\mu \nu
}V^{i}\,_{\lambda }=0\text{.}  \notag
\end{eqnarray}%
From (\ref{inter6}) we find,
\begin{equation}
A^{i}\,_{k\nu }=V^{i}\,_{\lambda }\partial _{\nu }V_{k}\,^{\lambda }+\Gamma
^{\lambda }\,_{\mu \nu }V^{i}\,_{\lambda }V_{k}\,^{\mu }=-V_{k}\,^{\lambda
}\partial _{\nu }V^{i}\,_{\lambda }+\Gamma ^{\lambda }\,_{\mu \nu
}V^{i}\,_{\lambda }V_{k}\,^{\mu }\text{.}
\end{equation}%
or, equivalently, in terms of $\omega $ defined in (\ref{inter9}),%
\begin{equation}
A^{i}\,_{k\nu }=\omega _{\text{ }\nu k}^{i}+\Gamma ^{\lambda }\,_{\mu \nu
}V^{i}\,_{\lambda }V_{k}\,^{\mu }=-\omega _{k\nu }^{\text{ \ \ \ }i}+\Gamma
^{\lambda }\,_{\mu \nu }V^{i}\,_{\lambda }V_{k}\,^{\mu }.
\end{equation}%
Using this in (\ref{del-mu-psi}), we may write
\begin{equation}
\nabla _{\mu }\psi =(\partial _{\mu }-\Gamma _{\mu })\psi \text{,}
\end{equation}%
where%
\begin{equation}
\Gamma _{\mu }=\frac{1}{4}\left( \omega _{\text{ \ }j\mu }^{i}-\Gamma
^{\lambda }\,_{\mu \nu }V^{i}\,_{\lambda }V_{j}\,^{\nu }\right) \gamma
_{i}\,^{j},
\end{equation}%
which is known as the Fock-Ivanenko connection \cite{fock}.

We now study the transformation properties of $A_{\mu ab}$. Recall $\omega
_{\mu ab}=V_{a}^{\text{ \ }\lambda }(x)\partial _{\mu }V_{\beta \lambda }(x)$
and since $\partial _{\mu }\eta _{ab}=0$, we write%
\begin{equation}
\Lambda _{\text{ \ }\overline{a}}^{a}\eta _{ab}\partial _{\mu }\Lambda _{%
\overline{b}}^{\text{ \ }b}=\Lambda _{\text{ \ }\overline{a}}^{a}\partial
_{\mu }\Lambda _{a\overline{b}}\text{.}
\end{equation}%
Note that barred indices are equivalent to the primed indices used above.
Hence, the spin connection transforms as%
\begin{equation}
A_{\overline{a}\overline{b}\overline{c}}=\Lambda _{\overline{a}}^{\text{ \ }%
a}\Lambda _{\overline{b}}^{\text{ \ }b}\Lambda _{\overline{c}}^{\text{ \ }%
c}A_{abc}+\Lambda _{\overline{a}}^{\text{ \ }a}\Lambda _{\overline{c}}^{%
\text{ \ }c}V_{\text{ \ }a}^{\mu }(x)\partial _{\mu }\Lambda _{\overline{b}%
c}.
\end{equation}%
To determine the transformation properties of%
\begin{equation}
\Gamma _{abc}=A_{abc}-\left[ V_{\text{ \ }a}^{\mu }(x)\partial _{\mu }V_{%
\text{ \ }b}^{\nu }(x)\right] V_{\nu c}(x)\text{,}
\end{equation}%
we consider the local LT of $\left[ V_{a}^{\text{ \ }\mu }(x)\partial _{\mu
}V_{\text{ \ }b}^{\nu }(x)\right] V_{\nu c}(x)$ which is,%
\begin{equation}
\left[ V_{\text{ \ }\overline{a}}^{\mu }(x)\partial _{\mu }V_{\text{ \ }%
\overline{b}}^{\nu }\right] V_{\nu \overline{c}}(x)=\Lambda _{\overline{a}}^{%
\text{ \ }a}\Lambda _{\overline{b}}^{\text{ \ }b}\Lambda _{\overline{c}}^{%
\text{ \ }c}\left[ A_{\text{ }ab}^{\nu }V_{\nu c}(x)\right] +\Lambda _{%
\overline{a}}^{\text{ \ }a}\Lambda _{\overline{c}}^{\text{ \ }c}V_{\text{ \ }%
a}^{\mu }(x)\partial _{\mu }\Lambda _{c\overline{b}}\text{.}
\end{equation}%
From this result, we obtain the following transformation law,
\begin{equation}
\Gamma _{\overline{a}\overline{b}\overline{c}}=\Lambda _{\overline{a}}^{%
\text{ \ }a}\Lambda _{\overline{b}}^{\text{ \ }b}\Lambda _{\overline{c}}^{%
\text{ \ }c}\Gamma _{abc}.
\end{equation}

We now explore the consequence of the antisymmetry of $\omega _{abc}$ in $bc$%
. Recalling the equation for $\Gamma _{abc}$, exchanging $b$ and $c$ and
adding the two equations, we obtain%
\begin{equation}
\Gamma _{abc}+\Gamma _{acb}=-V_{\text{ \ }a}^{\mu }(x)\left[ \left( \partial
_{\mu }V_{\text{ \ }b}^{\nu }(x)\right) V_{\nu c}(x)+\left( \partial _{\mu
}V_{\text{ \ }c}^{\nu }(x)\right) V_{\nu b}(x)\right] \text{.}
\end{equation}%
We know however, that%
\begin{equation}
\partial _{\mu }\left[ V_{\text{ \ }b}^{\nu }(x)V_{\nu c}(x)\right] =V_{\nu
c}(x)\partial _{\mu }V_{\text{ \ }b}^{\nu }(x)+V_{\lambda b}(x)\partial
_{\mu }V_{\text{ \ }c}^{\lambda }(x)+V_{b}^{\text{ \ }\nu }(x)V_{\text{ \ }%
c}^{\lambda }(x)\partial _{\mu }g_{\lambda \nu }\text{.}
\end{equation}%
Letting $\lambda \rightarrow \nu $ and exchanging $b$ and $c$, we obtain%
\begin{equation}
\partial _{\mu }\left[ V_{\text{ \ }b}^{\nu }(x)V_{\nu c}(x)\right] =-V_{b}^{%
\text{ \ }\lambda }(x)V_{\text{ \ }c}^{\nu }(x)\partial _{\mu }g_{\nu
\lambda }\text{,}
\end{equation}%
so that, finally,%
\begin{equation}
\Gamma _{abc}+\Gamma _{acb}=V_{a}^{\text{ \ }\mu }(x)V_{b}^{\text{ \ }%
\lambda }(x)V_{c}^{\text{ \ }\nu }(x)\partial _{\mu }g_{\nu \lambda }\text{.}
\end{equation}%
This, however, is equivalent to%
\begin{equation}
\Gamma _{\overline{a}\overline{b}\overline{c}}+\Gamma _{\overline{a}\text{ }%
\overline{c}\overline{b}}=V_{\text{ \ }\overline{a}}^{\overline{\mu }}(x)V_{%
\text{ \ }\overline{b}}^{\overline{\lambda }}(x)V_{\text{ \ }\overline{c}}^{%
\overline{\nu }}(x)\partial _{\overline{\mu }}g_{\overline{\nu }\overline{%
\lambda }}\text{,}
\end{equation}%
and then%
\begin{equation}
\Gamma _{\mu \lambda \nu }+\Gamma _{\mu \nu \lambda }=\partial _{\mu }g_{\nu
\lambda }\text{,}  \label{inter5}
\end{equation}%
which we recognize as the general coordinate connection. It is known that
the covariant derivative for general coordinates is%
\begin{equation}
\nabla _{\mu }A_{\nu }^{\text{ \ }\lambda }=\partial _{\mu }A_{\nu }^{\text{
\ }\lambda }+\Gamma _{\text{ \ }\mu \sigma }^{\lambda }A_{\nu }^{\text{ \ }%
\sigma }-\Gamma _{\text{ \ }\mu \nu }^{\sigma }A_{\sigma }^{\text{ \ }%
\lambda }\text{.}
\end{equation}%
In a Riemannian manifold, the connection is symmetric under the
exchange of $\mu
\nu $, that is, $\Gamma _{\text{ \ }\mu \nu }^{\lambda }=\Gamma _{\text{ \ }%
\nu \mu }^{\lambda }$. Using the fact that the metric is a symmetric tensor
we can now determine the form of the Christoffel connection by cyclically
permuting the indices of the general coordinate connection equation (\ref%
{inter5}) yielding
\begin{equation}
\Gamma _{\mu \nu \lambda }=\frac{1}{2}\left( \partial _{\mu }g_{\nu \lambda
}+\partial _{\nu }g_{\lambda \mu }-\partial _{\lambda }g_{\mu \nu }\right) .
\end{equation}%
Since $\Gamma _{\mu \nu \lambda }=\Gamma _{\nu \mu \lambda }$ is valid for
general coordinate systems, it follows that a similar constraint must hold
for local Lorentz transforming coordinates as well, so we expect $\Gamma
_{abc}=\Gamma _{bac}$. Recalling the equation for $\Gamma _{abc}$ and
exchanging $a$ and $b$, we obtain%
\begin{equation}
\omega _{abc}-\omega _{bac}=V_{\nu c}(x)\left[ V_{\text{ \ }a}^{\mu
}(x)\partial _{\mu }V_{\text{ \ }b}^{\nu }(x)-V_{\text{ \ }b}^{\mu
}(x)\partial _{\mu }V_{\text{ \ }a}^{\nu }(x)\right] \text{.}
\end{equation}%
We now define the \textit{objects of anholonomicity} as%
\begin{equation}
\Omega _{cab}:=V_{\nu c}(x)\left[ V_{\text{ \ }a}^{\mu }(x)\partial _{\mu
}V_{\text{ \ }b}^{\nu }(x)-V_{\text{ \ }b}^{\mu }(x)\partial _{\mu }V_{\text{
\ }a}^{\nu }(x)\right] .
\end{equation}%
Using $\Omega _{cab}=-\Omega _{cba}$, we permute indices in a similar manner
as was done for the derivation of the Christoffel connection above yielding,
\begin{equation}
\omega _{ab\mu }=\frac{1}{2}\left[ \Omega _{cab}+\Omega _{bca}-\Omega _{abc}%
\right] V_{\text{ \ }\mu }^{c}\equiv \Delta _{ab\mu }.  \label{inter8}
\end{equation}%
For completeness, we determine the transformation law of the
Christoffel connection. Making use of $\Gamma _{\mu \nu }^{\lambda
}e_{\lambda
}=\partial _{\mu }e_{\nu }$ where%
\begin{equation}
\partial _{\overline{\mu }}e_{\overline{\nu }}=X_{\text{ \ }\overline{\mu }%
}^{\mu }X_{\text{ \ }\overline{\nu }}^{\nu }\partial _{\mu }e_{\nu }+X_{%
\text{ \ }\overline{\mu }}^{\mu }\left( \partial _{\mu }X_{\text{ \ }%
\overline{\nu }}^{\nu }\right) e_{\nu }\text{,}
\end{equation}%
we can show
\begin{equation}
\Gamma _{\text{ \ }\overline{\mu }\text{ }\overline{\nu }}^{\overline{%
\lambda }}=X_{\text{ \ }\overline{\mu }}^{\mu }X_{\text{ \ }\overline{\nu }%
}^{\nu }X_{\lambda }^{\text{ \ }\overline{\lambda }}\Gamma _{\text{ \ }\mu
\nu }^{\lambda }+X_{\text{ \ }\overline{\mu }}^{\mu }X_{\nu }^{\text{ \ }%
\overline{\lambda }}X_{\text{ \ }\mu \overline{\nu }}^{\nu },
\end{equation}%
where
\begin{equation}
X_{\text{ \ }\mu \overline{\nu }}^{\nu }\equiv \partial _{\mu }\partial _{%
\overline{\nu }}x^{\nu }\text{.}
\end{equation}

In the light of the above considerations, we may regard
infinitesimal local gauge transformations as local rotations of
basis vectors belonging to the tangent space \cite{Chang,
Mansouri3} of the manifold. For this reason, given a local frame
on a tangent plane to the point $x$ on the base manifold, we can
obtain all other frames on the same tangent plane by means of
local rotations of the original basis vectors. Reversing this
argument, we observe that by knowing all frames residing in the
horizontal tangent space to a point $x$ on the base manifold
enables us to deduce the corresponding gauge group of symmetry
transformations. These arguments are completely general and can be adopted for any theory of gravity coming from gauge symmetry.

\subsection{Curvature, torsion and metric}

From the definition of the Fock-Ivanenko covariant derivative, we
can find
the second order covariant derivative%
\begin{eqnarray}
D_{\nu }D_{\mu }\psi  &=&\partial _{\nu }\partial _{\mu }\psi +\frac{1}{2}%
S_{cd}\left( \psi \partial _{\nu }A_{\mu }^{\text{ \ }cd}+A_{\mu }^{\text{ \
}cd}\partial _{\nu }\psi \right) +\Gamma _{\text{ \ }\mu \nu }^{\rho
}D_{\rho }\psi +\frac{1}{2}S_{ef}A_{\nu }^{\text{ \ }ef}\partial _{\mu }\psi
\notag \\
&&+\frac{1}{4}S_{ef}S_{cd}A_{\nu }^{\text{ \ }ef}A_{\mu }^{\text{ \ }cd}\psi
\text{.}
\end{eqnarray}%
Recalling $D_{\nu }V^{c\mu }=0$, we can solve for the spin connection in
terms of the Christoffel connection%
\begin{equation}
A_{\mu }^{\text{ \ }cd}=-V_{\text{ \ }\lambda }^{d}\partial _{\mu
}V^{c\lambda }-\Gamma _{\mu }^{\text{ \ }cd}\text{.}
\end{equation}%
The derivative of the spin connection is then%
\begin{equation}
\partial _{\mu }A_{\text{ \ \ }\nu }^{cd}=-V_{\text{ \ }\lambda
}^{d}\partial _{\mu }\partial _{{\nu }}V^{c\lambda }-\left( \partial _{\nu
}V^{c\lambda }\right) \partial _{\mu }V_{\lambda }^{\text{ \ }d}-\partial
_{\mu }\Gamma _{\text{ \ \ }\nu }^{cd}\text{.}
\end{equation}%
Noting that the Christoffel connection is symmetric and partial derivatives
commute, we find%
\begin{equation}
\left[ D_{\mu }\text{, }D_{\nu }\right] \psi =\frac{1}{2}S_{cd}\left[ \left(
\partial _{\nu }A_{\text{ \ \ }\mu }^{cd}-\partial _{\mu }A_{\text{ \ \ }\nu
}^{cd}\right) \psi \right] +\frac{1}{4}S_{ef}S_{cd}\left[ \left( A_{\text{ \
\ }\nu }^{ef}A_{\text{ \ \ }\mu }^{cd}-A_{\text{ \ \ }\mu }^{ef}A_{\text{ \
\ }\nu }^{cd}\right) \psi \right] \text{,}
\end{equation}%
where%
\begin{equation}
\partial _{\nu }A_{\text{ \ \ }\mu }^{cd}-\partial _{\mu }A_{\text{ \ \ }\nu
}^{cd}=\partial _{\mu }\Gamma _{\text{ \ \ }\nu }^{cd}-\partial _{\nu
}\Gamma _{\text{ \ \ }\mu }^{cd}\text{.}
\end{equation}%
Relabeling running indices, we can write%
\begin{equation}
\frac{1}{4}S_{ef}S_{cd}\left( A_{\text{ \ \ }\nu }^{ef}A_{\text{ \ \ }\mu
}^{cd}-A_{\text{ \ \ }\mu }^{ef}A_{\text{ \ \ }\nu }^{cd}\right) \psi =\frac{%
1}{4}\left[ S_{cd}\text{, }S_{ef}\right] A_{\text{ \ }\mu }^{ef}A_{\text{ \
\ }\nu }^{cd}\psi \text{.}
\end{equation}%
Using $\left\{ \gamma _{a},\gamma _{b}\right\} =2\eta _{ab}$ to deduce%
\begin{equation}
\left\{ \gamma _{a}\text{, }\gamma _{b}\right\} \gamma _{c}\gamma _{d}=2\eta
_{ab}\gamma _{c}\gamma _{d}\text{,}
\end{equation}%
we find that the commutator of bi-spinors is given by%
\begin{equation}
\left[ S_{cd}\text{, }S_{ef}\right] =\frac{1}{2}\left[ \eta _{ce}\delta
_{d}^{a}\delta _{f}^{b}-\eta _{de}\delta _{c}^{a}\delta _{f}^{b}+\eta
_{cf}\delta _{e}^{a}\delta _{d}^{b}-\eta _{df}\delta _{e}^{a}\delta _{c}^{b}%
\right] S_{ab}\text{.}  \label{inter7}
\end{equation}%
Clearly the terms in brackets on the {\it rhs} of (\ref{inter7}) are
antisymmetric in $cd$ and $ef$ and  also antisymmetric under the
exchange of pairs of indices $cd$ and $ef$. Since the alternating
spinor is antisymmetric in $ab$, so it must be the terms in
brackets: this means that the commutator does not vanish. Hence,
the term in brackets is totally antisymmetric under interchange of
indices $ab$, $cd$ and $ef$ and exchange of these pairs of
indices. We identify this as the structure constant  of the
Lorentz group \cite{DeWitt}
\begin{equation}
\left[ \eta _{ce}\delta _{d}^{a}\delta _{f}^{b}-\eta _{de}\delta
_{c}^{a}\delta _{f}^{b}+\eta _{cf}\delta _{e}^{a}\delta _{d}^{b}-\eta
_{df}\delta _{e}^{a}\delta _{c}^{b}\right] ={c_{[cd][ef]}}^{[ab]}={c^{[ab]}}%
_{[cd][ef]},
\end{equation}%
with the aid of which we can write%
\begin{equation}
\frac{1}{4}\left[ S_{cd}\text{, }S_{ef}\right] A_{\text{ \ \ }\mu }^{ef}A_{%
\text{ \ \ }\nu }^{cd}\psi =\frac{1}{2}S_{ab}\left[ A_{\text{ \ }e\nu
}^{a}A_{\text{ \ \ }\mu }^{eb}-A_{\text{ \ }e\nu }^{b}A_{\text{ \ \ }\mu
}^{ae}\right] \psi \text{,}
\end{equation}%
where%
\begin{equation}
A_{\text{ \ }e\nu }^{a}A_{\text{ \ \ }\mu }^{eb}-A_{\text{ \ }e\nu }^{b}A_{%
\text{ \ \ }\mu }^{ae}=\Gamma _{\nu e}^{a}\Gamma _{\text{ \ \ }\mu
}^{eb}-\Gamma _{\nu e}^{b}\Gamma _{\text{ \ \ }\mu }^{ea}\text{.}
\end{equation}%
Combining these results, the commutator of two $\mu $-covariant
differentiations gives
\begin{equation}
\lbrack \nabla _{\mu },\nabla _{\nu }]\chi =-\frac{1}{2}R^{ij}\,_{\mu \nu
}S_{ij}\chi ,
\end{equation}%
where
\begin{equation}
R^{i}\,_{j\mu \nu }=\partial _{\nu }A^{i}\,_{j\mu }-\partial _{\mu
}A^{i}\,_{j\nu }+A^{i}\,_{k\nu }A^{k}\,_{j\mu }-A^{i}\,_{k\mu }A^{k}\,_{j\nu
}\text{.}
\end{equation}%
Using the Jacobi identities for the commutator of covariant
derivatives, it follows that the field strength  $R^{i}\,_{j\mu
\nu }$ satisfies the Bianchi identity
\begin{equation}
\nabla _{\lambda }R^{i}\,_{j\mu \nu }+\nabla _{\mu }R^{i}\,_{j\nu \lambda
}+\nabla _{\nu }R^{i}\,_{j\lambda \mu }=0\text{.}
\end{equation}%
Permuting indices, this can be put into the cyclic form
\begin{equation}
\varepsilon ^{\alpha \beta \rho \sigma }\nabla _{\beta }R_{\text{ \ }\rho
\sigma }^{ij}=0,
\end{equation}%
where $\varepsilon ^{\alpha \beta \rho \sigma }$ is the Levi-Civita
alternating symbol. Furthermore, $R^{ij}\,_{\mu \nu }=\eta
^{jk}R^{i}\,_{k\mu \nu }$ is antisymmetric with respect to both pairs of
indices,
\begin{equation}
R^{ij}\,_{\mu \nu }=-R^{ji}\,_{\mu \nu }=R^{ji}\,_{\nu \mu }=-R^{ij}\,_{\nu
\mu }.
\end{equation}%
This condition is known as the first curvature tensor identity.

To determine the analogue of $[\nabla _{\mu },\nabla _{\nu }]\chi
$ in local coordinates, we start from $\nabla _{k}\psi =V_{\text{
\ }k}^{\mu
}\nabla _{\mu }\psi $. From $\nabla _{k}\psi $ we obtain,%
\begin{equation}
\nabla _{l}\nabla _{k}\psi =V_{\text{ \ }l}^{\nu }\left( \nabla _{\nu }V_{%
\text{ \ }k}^{\mu }\right) \nabla _{\mu }\psi +V_{\text{ \ }l}^{\nu }V_{%
\text{ \ }k}^{\mu }\nabla _{\nu }\nabla _{\mu }\psi \text{.}
\end{equation}%
Permuting indices and recognizing%
\begin{equation}
V_{\mu }^{\text{ \ }a}\nabla _{\nu }V_{\text{ \ }k}^{\mu }=-V_{k}^{\text{ \ }%
\mu }\nabla _{\nu }V_{\text{ \ }\mu }^{a}\text{,}
\end{equation}%
(which follows from $\nabla _{\nu }\left( V_{\mu }^{a}V_{k}^{\mu }\right) =0$%
), we arrive at%
\begin{equation}
V_{\text{ \ }l}^{\nu }\left( \nabla _{\nu }V_{\text{ \ }k}^{\mu }\right)
\nabla _{\mu }\psi -V_{\text{ \ }k}^{\mu }\left( \nabla _{\mu }V_{\text{ \ }%
l}^{\nu }\right) \nabla _{\nu }\psi =\left( V_{\text{ \ }l}^{\mu }V_{\text{
\ }k}^{\nu }-V_{\text{ \ }k}^{\mu }V_{\text{ \ }l}^{\nu }\right) \left(
\nabla _{\nu }V_{\mu }^{\text{ \ }a}\right) \nabla _{a}\psi \text{.}
\end{equation}%
Defining%
\begin{equation}
C_{\;\;kl}^{a}:=\left( V_{\text{ \ }k}^{\mu }V_{\text{ \ }l}^{\nu }-V_{\text{
\ }l}^{\mu }V_{\text{ \ }k}^{\nu }\right) \nabla _{\nu }V_{\mu }^{\text{ \ }%
a},
\end{equation}%
the commutator of the $k$-covariant differentiations takes the final form
\cite{Kibble}%
\begin{equation}
\lbrack \nabla _{k},\nabla _{l}]\chi =-\frac{1}{2}R^{ij}\,_{kl}S_{ij}\chi
+C^{i}\,_{kl}\nabla _{i}\chi ,
\end{equation}%
where
\begin{equation}
R^{ij}\,_{kl}=V_{k}\,^{\mu }V_{l}\,^{\nu }R^{ij}\,_{\mu \nu }\text{.}
\end{equation}%
As  done for $R^{i}\,_{j\mu \nu }$ using the Jacobi identities for
the commutator of covariant derivatives, we find  the Bianchi
identity in Einstein-Cartan space-time  \cite{hehlrev, Blagojevic}
\begin{equation}
\varepsilon ^{\alpha \beta \rho \sigma }\nabla _{\beta }R_{\text{ }\rho
\sigma }^{ij}=\varepsilon ^{\alpha \beta \rho \sigma }C_{\beta \rho }^{\text{
\ \ \ }\lambda }R_{\text{ \ }\sigma \lambda }^{ij}.
\end{equation}%
The second curvature identity%
\begin{equation}
R_{\text{ \ }[\rho \sigma \lambda ]}^{k}=2\nabla _{\lbrack \rho }C_{\sigma
\lambda ]}^{\text{ \ \ \ \ }k}-4C_{[\rho \sigma }^{\text{ \ \ \ }%
b}C_{\lambda ]b}^{\text{ \ \ \ }k}
\end{equation}%
leads to,%
\begin{equation}
\varepsilon ^{\alpha \beta \rho \sigma }\nabla _{\beta }C_{\rho \sigma }^{%
\text{ \ \ \ }k}=\varepsilon ^{\alpha \beta \rho \sigma }R_{\text{ \ }j\rho
\sigma }^{k}V_{\text{ }\beta }^{j}.
\end{equation}%
Notice that if$\ $%
\begin{equation}
\Gamma _{\;\;\mu \nu }^{\lambda }=V_{i}^{\text{ \ }\lambda }\nabla _{\nu }V_{%
\text{ \ }\mu }^{i}=-V_{\mu }^{\text{ \ }i}\nabla _{\nu }V_{\text{ \ }%
i}^{\lambda }\text{,}
\end{equation}%
then%
\begin{equation}
\Gamma _{\;\;\mu \nu }^{\lambda }-\ \Gamma _{\;\;\nu \mu }^{\lambda
}=V_{i}^{\lambda }\left( \nabla _{\nu }V_{\text{ \ }\mu }^{i}-\nabla _{\mu
}V_{\text{ \ }\nu }^{i}\right) \text{.}
\end{equation}%
Contracting by $V_{k}^{\mu }V_{l}^{\nu }$, we obtain \cite{Kibble},%
\begin{equation}
C_{\;\text{\ }kl}^{a}=V_{k}^{\text{ \ }\mu }V_{l}^{\text{ \ }\nu }V_{\lambda
}^{\text{ \ }a}\left( \ \Gamma _{\;\;\mu \nu }^{\lambda }-\ \Gamma _{\;\;\nu
\mu }^{\lambda }\right) \text{.}
\end{equation}%
We therefore conclude that $C_{\;kl}^{a}$ is related to the antisymmetric
part of the affine connection
\begin{equation}
\Gamma _{\;\;\left[ \mu \nu \right] }^{\lambda }=V_{\mu }^{\text{ \ }%
k}V_{\nu }^{\text{ \ }l}V_{a}^{\text{ \ }\lambda }C_{\;\ kl}^{a}\equiv
T_{\;\;\mu \nu }^{\lambda },
\end{equation}%
which is usually interpreted as space-time torsion $T_{\;\;\mu \nu
}^{\lambda }$. Considering $\Delta _{ab\mu }$ defined in
(\ref{inter8}), we see that the most
general connection in the Poincar\'{e} gauge approach to gravitation is%
\begin{equation}
A_{ab\mu }=\Delta _{ab\mu }-K_{ab\mu }+\Gamma ^{\lambda }\,_{\nu \mu
}V_{a\lambda }V_{b}\,^{\nu },
\end{equation}%
where
\begin{equation}
K_{abc}=-\left( T^{\lambda }\,_{\nu \mu }-T_{\nu \mu }^{\text{ \ \ }\lambda
}+T_{\mu \text{ \ }\nu }^{\text{ \ }\lambda }\right) V_{a\lambda
}V_{b}\,^{\nu }V_{c}^{\text{ \ }\mu }\text{,}
\end{equation}%
is the contorsion tensor. Now, the quantity $R_{\sigma \mu \nu }^{\rho
}=V_{i}\,^{\rho }R^{i}\,_{\sigma \mu \nu }$ may be expressed as
\begin{equation}
R^{\rho }\,_{\sigma \mu \nu }=\partial _{\nu }\Gamma _{\text{ \ }\sigma \mu
}^{\rho }-\partial _{\mu }\Gamma _{\text{ \ }\sigma \nu }^{\rho }+\Gamma
^{\rho }\,_{\lambda \nu }\Gamma ^{\lambda }\,_{\sigma \mu }-\Gamma ^{\rho
}\,_{\lambda \mu }\Gamma ^{\lambda }\,_{\sigma \nu }\text{.}
\label{curvature}
\end{equation}%
Therefore, we can regard $R^{\rho }\,_{\sigma \mu \nu }$ as the
curvature tensor with respect the affine connection $\Gamma
^{\lambda }\,_{\mu \nu }$. By using the inverse of the tetrad, we
define the metric of the space-time manifold by
\begin{equation}
g_{\mu \nu }=V^{i}\,_{\mu }V^{j}\,_{\nu }\eta _{ij}.
\end{equation}%
From (\ref{inter6}) and the fact that the Minkowski metric is constant, it
is obvious that \ the metric so defined is covariantly constant, that is,
\begin{equation}
D_{\lambda }g_{\mu \nu }=0.
\end{equation}%
The space-time thus specified by the local Poincar\'{e} transformation is
said to be metric. It is not difficult to show that
\begin{equation}
\sqrt{-g}=[\det V^{i}\,_{\mu }]=[\det V_{i}\,^{\mu }]^{-1}\text{,}
\end{equation}%
where $g=\det g_{\mu \nu }$. Hence we may take $\sqrt{-g}$ for the density
function $h(x)$.

\subsection{The field equations of gravity}
\label{tre.6}

Finally, we are able to deduce the field equations for the
gravitational field.
From the curvature tensor $R^{\rho }\,_{\sigma \mu \nu }$, given in (\ref%
{curvature}), the Ricci tensor  follows
\begin{equation}
R_{\sigma \nu }=R^{\mu }\,_{\sigma \mu \nu }\text{.}
\end{equation}%
and the scalar curvature
\begin{equation}
R=R^{\nu }\,_{\nu }=\overset{\text{L}}{R}+\partial _{i}K_{a}^{\text{ \ }%
ia}-T_{a}^{\text{ \ }bc}K_{bc}^{\text{ \ \ }a}
\end{equation}%
where $\overset{\text{L}}{R}$ denotes the usual\ Ricci scalar of
GR. Using this scalar curvature $R$, we choose the
Lagrangian density for free Einstein-Cartan gravity
\begin{equation}
\mathcal{L}_{G}=\frac{1}{2\kappa }\sqrt{-g}\left( \overset{\text{L}}{R}%
+\partial _{i}K_{a}^{\text{ \ }ia}-T_{a}^{\text{ \ }bc}K_{bc}^{\text{ \ \ }%
a}-2\Lambda \right) ,
\end{equation}%
where $\kappa $ is a gravitational coupling constant, and $\Lambda
$ is the cosmological constant. These considerations can be easily
extended to any function of $\overset{\text{L}}{R}$ as in
\cite{cianci}. Observe that the second term is a divergence and
may be ignored. The field equation can be obtained from the total
action,
\begin{equation}
S=\int \left\{ \mathcal{L}_{\text{field}}(\chi ,\partial _{\mu }\chi
,V_{i}\,^{\mu },A^{ij}\,_{\mu })+\mathcal{L}_{G}\right\} d^{4}x\text{,}
\end{equation}%
where the matter Lagrangian density is taken to be%
\begin{equation}
\mathcal{L}_{\text{field}}=\frac{1}{2}\left[ \overline{\psi }\gamma
^{a}D_{a}\psi -\left( D_{a}\overline{\psi }\right) \gamma ^{a}\psi \right]
\text{.}
\end{equation}%
Modifying the connection to include Christoffel, spin connection and
contorsion contributions so as to operate on general, spinorial arguments, we
have%
\begin{equation}
\Gamma _{\mu }=\frac{1}{4}g_{\lambda \sigma }\left( \Delta _{\text{ \ }\mu
\rho }^{\sigma }-\overset{\text{L}}{\Gamma }\text{ }_{\text{ \ }\rho \mu
}^{\sigma }-K_{\text{ \ }\rho \mu }^{\sigma }\right) \gamma ^{\lambda \rho }.
\end{equation}%
It is important to keep in mind that $\Delta _{\text{ \ }\mu \rho }^{\sigma }
$ act only on multi-component spinor fields, while $\overset{\text{L}}{%
\Gamma }$ $_{\text{ \ }\rho \mu }^{\sigma }$ act on vectors and arbitrary
tensors. The gauge covariant derivative for a spinor and adjoint spinor is
then given by%
\begin{equation}
D_{\mu }\psi =\left( \partial _{\mu }-\Gamma _{\mu }\right) \psi \text{, \ }%
D_{\mu }\overline{\psi }=\partial _{\mu }\overline{\psi }-\overline{\psi }%
\Gamma _{\mu }\text{.}
\end{equation}%
The variation of the field Lagrangian is%
\begin{equation}
\delta \mathcal{L}_{\text{field}}=\overline{\psi }\left( \delta \gamma ^{\mu
}D_{\mu }+\gamma ^{\mu }\delta \Gamma _{\mu }\right) \psi \text{.}
\end{equation}%
We know that the Dirac gamma matrices are covariantly vanishing, so%
\begin{equation}
D_{\kappa }\gamma _{\iota }=\partial _{\kappa }\gamma _{\iota }-\Gamma
_{\iota \kappa }^{\mu }\gamma _{\mu }+\left[ \gamma _{\iota }\text{, }%
\widehat{\Gamma }_{\kappa }\right] =0\text{.}
\end{equation}%
The $4\times 4$ matrices $\widehat{\Gamma }_{\kappa }$\ are real matrices
used to induce similarity transformations on quantities with spinor
transformation \cite{Brill} properties according to%
\begin{equation}
\gamma _{i}^{\prime }=\widehat{\Gamma }^{-1}\gamma _{i}\widehat{\Gamma }%
\text{.}
\end{equation}%
Solving for $\widehat{\Gamma }_{\kappa }$ leads to,%
\begin{equation}
\widehat{\Gamma }_{\kappa }=\frac{1}{8}\left[ \left( \partial _{\kappa
}\gamma _{\iota }\right) \gamma ^{\iota }-\Gamma _{\text{ \ }\iota \kappa
}^{\mu }\gamma _{\mu }\gamma ^{\iota }\right] \text{.}
\end{equation}%
Taking the variation of $\widehat{\Gamma }_{\kappa }$,%
\begin{eqnarray}
\delta \widehat{\Gamma }_{\kappa } &=&\frac{1}{8}\left[
\begin{array}{c}
\left( \partial _{\kappa }\delta \gamma _{\iota }\right) \gamma ^{\iota
}+\left( \partial _{\kappa }\gamma _{\iota }\right) \delta \gamma ^{\iota
}-\left( \delta \Gamma _{\text{ \ }\iota \kappa }^{\mu }\right) \gamma _{\mu
}\gamma ^{\iota } \\
-\Gamma _{\text{ \ }\iota \kappa }^{\mu }\left( \left( \delta \gamma _{\mu
}\right) \gamma ^{\iota }+\gamma _{\mu }\delta \gamma ^{\iota }\right)
\end{array}%
\right]  \\
&=&\frac{1}{8}\left[ \left( \partial _{\kappa }\delta \gamma _{\iota
}\right) \gamma ^{\iota }-\left( \delta \Gamma _{\text{ \ }\iota \kappa
}^{\mu }\right) \gamma _{\mu }\gamma ^{\iota }\right] \text{.}  \notag
\end{eqnarray}%
Since we require the anticommutator condition on the gamma matrices $%
\{\gamma ^{\mu },\gamma ^{\nu }\}=2g^{\mu \nu }$ to hold, the variation of
the metric gives%
\begin{equation}
2\delta g^{\mu \nu }=\{\delta \gamma ^{\mu },\gamma ^{\nu }\}+\{\gamma ^{\mu
}\delta \gamma ^{\nu }\}\text{.}
\end{equation}%
One solution to this equation is,%
\begin{equation}
\delta \gamma ^{\nu }=\frac{1}{2}\gamma _{\sigma }\delta \gamma ^{\sigma \nu
}\text{.}
\end{equation}%
With the aid of this result, we can write%
\begin{equation}
\left( \partial _{\kappa }\delta \gamma _{\iota }\right) \gamma ^{\iota }=%
\frac{1}{2}\partial _{\kappa }\left( \gamma ^{\nu }\delta g_{\nu \iota
}\right) \gamma ^{\iota }\text{.}
\end{equation}%
Finally, exploiting the anti-symmetry in $\gamma _{\mu \nu }$ we obtain%
\begin{equation}
\delta \widehat{\Gamma }_{\kappa }=\frac{1}{8}\left[ g_{\nu \sigma }\delta
\Gamma _{\mu \kappa }^{\text{ \ \ }\sigma }-g_{\mu \sigma }\delta \Gamma
_{\nu \kappa }^{\text{ \ \ }\sigma }\right] \gamma ^{\mu \nu }\text{.}
\end{equation}%
The field Lagrangian defined in the Einstein-Cartan space-time can
be written \cite{Carroll, Hehl2, hehlrev, Shapiro, Blagojevic}
explicitly in terms of its Lorentzian and contorsion components as
\begin{equation}
\mathcal{L}_{\text{field}}=\frac{1}{2}\left[ \left( \overset{\text{L}}{D}%
_{\mu }\overline{\psi }\right) \gamma ^{\mu }\psi -\overline{\psi }\gamma
^{\mu }\overset{\text{L}}{D}_{\mu }\psi \right] -\frac{\hbar c}{8}K_{\mu
\alpha \beta }\overline{\psi }\left\{ \gamma ^{\mu }\text{, }\gamma ^{\alpha
\beta }\right\} \psi .
\end{equation}%
Using the following relations%
\begin{equation}
\left\{
\begin{array}{c}
-\frac{1}{4}K_{\mu \alpha \beta }\overline{\psi }\left\{ \gamma ^{\mu }\text{%
, }\gamma ^{\alpha \beta }\right\} \psi =\frac{1}{4}K_{\mu \alpha \beta }%
\overline{\psi }\gamma ^{\beta \alpha }\gamma ^{\mu }\psi -\frac{1}{4}K_{\mu
\alpha \beta }\overline{\psi }\gamma ^{\mu }\gamma ^{\alpha \beta }\psi
\text{,} \\
\gamma ^{\mu }\gamma ^{\nu }\gamma ^{\lambda }\varepsilon _{\mu \nu \lambda
\sigma }=\left\{ \gamma ^{\mu }\text{, }\gamma ^{\nu \lambda }\right\}
\varepsilon _{\mu \nu \lambda \sigma }=3!\gamma _{\sigma }\gamma _{5}\text{,
\ } \\
\left\{ \gamma ^{\mu }\text{, }\gamma ^{\nu \lambda }\right\} =\gamma
^{\lbrack \mu }\gamma ^{\nu }\gamma ^{\lambda ]}\text{, \ }%
\end{array}%
\right.
\end{equation}%
we obtain
\begin{equation}
K_{\mu \alpha \beta }\overline{\psi }\left\{ \gamma ^{\mu }\text{, }\gamma
^{\alpha \beta }\right\} \psi =\frac{1}{2i}K_{\mu \alpha \beta }\varepsilon
^{\alpha \beta \mu \nu }\left( \overline{\psi }\gamma _{5}\gamma _{\nu }\psi
\right) \text{.}
\end{equation}%
Here we define the contorsion axial vector%
\begin{equation}
K_{\nu }:=\frac{1}{3!}\varepsilon ^{\alpha \beta \mu \nu }K_{\alpha \beta
\mu }.
\end{equation}%
Multiplying through by the axial current $j_{\nu }^{5}=\overline{\psi }%
\gamma _{5}\gamma _{\nu }\psi $, we obtain%
\begin{equation}
\left( \overline{\psi }\gamma _{5}\gamma _{\nu }\psi \right) \varepsilon
^{\alpha \beta \mu \nu }K_{\mu \alpha \beta }=-6ij_{\nu }^{5}K^{\nu }\text{.}
\end{equation}%
Thus, the field Lagrangian density becomes%
\begin{equation}
\mathcal{L}_{\text{field}}=\frac{1}{2}\left[ \left( \overset{\text{L}}{D}%
_{\mu }\overline{\psi }\right) \gamma ^{\mu }\psi -\overline{\psi }\gamma
^{\mu }\overset{\text{L}}{D}_{\mu }\psi \right] +\frac{3i\hbar c}{8}K_{\mu
}j_{5}^{\mu }\text{.}
\end{equation}

The total action reads%
\begin{eqnarray}
\delta I &=&\delta \int \mathcal{L}_{G}\sqrt{-g}d^{4}x+\delta \int \mathcal{L%
}_{\text{field}}\sqrt{-g}d^{4}x \\
&=&\int \left( \delta \mathcal{L}_{G}+\delta \mathcal{L}_{\text{field}%
}\right) \sqrt{-g}d^{4}x\text{.}  \notag
\end{eqnarray}%

Writing the metric in terms of the tetrads $g^{\mu \nu }=V_{\;i}^{\mu
}V^{\nu i}$, we observe%
\begin{equation}
\delta \sqrt{-g}=-\frac{1}{2}\sqrt{-g}\left( \delta V_{\;i}^{\mu }V_{\mu
}^{\;i}+V_{\nu i}\delta V^{\nu i}\right) \text{.}
\end{equation}%
By using%
\begin{equation}
\delta V^{\nu i}=\delta \left( \eta ^{ij}V_{\;j}^{\nu }\right) =\eta
^{ij}\delta V_{\;j}^{\nu }\text{,}
\end{equation}%
we are able to deduce%
\begin{equation}
\delta \sqrt{-g}=-\sqrt{-g}V_{\mu }^{\;i}\delta V_{i}^{\;\mu }.
\end{equation}%
For the variation of the Ricci tensor $R_{i\nu }=V_{i}^{\;\mu }R_{\mu \nu }$
we have
\begin{equation}
\delta \overset{\text{L}}{R}_{i\nu }=\delta V_{i}^{\;\mu }\overset{\text{L}}{%
R}_{\mu \nu }+V_{i}^{\;\mu }\delta \overset{\text{L}}{R}_{\mu \nu }\text{.}
\end{equation}%
In an inertial frame, the Ricci tensor reduces to%
\begin{equation}
\overset{\text{L}}{R}_{\mu \nu }=\partial _{\nu }\overset{\text{L}}{\Gamma }%
\text{ }_{\beta \mu }^{\beta }-\partial _{\beta }\overset{\text{L}}{\Gamma }%
\text{ }_{\nu \mu }^{\beta }\text{,}
\end{equation}%
so that
\begin{equation}
\delta \overset{\text{L}}{R}_{i\nu }=\delta V_{i}^{\;\mu }\overset{\text{L}}{%
R}_{\mu \nu }+V_{i}^{\;\mu }\left( \partial _{\nu }\delta \overset{\text{L}}{%
\Gamma }\text{ }_{\beta \mu }^{\beta }-\partial _{\beta }\delta \overset{%
\text{L}}{\Gamma }\text{ }_{\nu \mu }^{\beta }\right) \text{.}
\end{equation}%
The second term can be converted into a surface term, so it may be ignored.
Collecting our results, we have%
\begin{equation}
\left\{
\begin{array}{c}
\delta g^{\mu \nu }=-g^{\mu \rho }g^{\nu \sigma }\delta g_{\rho \sigma }%
\text{,} \\
\delta \sqrt{-g}=-\frac{1}{2}\sqrt{-g}g_{\mu \nu }\delta g^{\mu \nu }=-\sqrt{%
-g}V_{\mu }^{\;i}\delta V_{i}^{\;\mu }\text{,} \\
\delta R_{\mu \nu }=g_{\rho \mu }\left( \nabla _{\lambda }\delta \Gamma _{%
\text{ \ \ }\nu }^{\lambda \rho }-\nabla _{\nu }\delta \Gamma _{\text{ \ \ }%
\lambda }^{\lambda \rho }\right) +T_{\lambda \mu }^{\text{ \ \ }\rho }\delta
\Gamma _{\text{ \ \ }\rho \nu }^{\lambda }\text{, \ }\delta \overset{\text{L}%
}{R}_{i\nu }=\delta V_{i}^{\;\mu }\overset{\text{L}}{R}_{\mu \nu } \\
\delta R=\overset{\text{L}}{R}\text{ }^{\mu \nu }\delta g_{\mu \nu }+g^{\mu
\nu }\left( \nabla _{\lambda }\delta \overset{\text{L}}{\Gamma }\text{ }_{%
\text{ \ \ }\mu \nu }^{\lambda }-\nabla _{\nu }\delta \overset{\text{L}}{%
\Gamma }\text{ }_{\text{ \ }\mu \lambda }^{\lambda }\right) -T_{a}^{\text{ \
}bc}\delta K_{bc}^{\text{ \ \ }a}\text{.}%
\end{array}%
\right.
\end{equation}%
From the above results, we obtain%
\begin{equation}
\delta I_{G}=\frac{1}{16\pi }\int \left[
\begin{array}{c}
\left( R_{i}^{\text{ \ }\mu }-\frac{1}{2}V_{i}^{\text{ \ }\mu }R-V_{i}^{%
\text{ \ }\mu }\Lambda \right) \delta V_{\text{ }\mu }^{i}+2g^{\rho \lambda
}T_{\mu \lambda }^{\text{ \ \ }\sigma }\delta \Gamma _{\text{ \ }\rho \sigma
}^{\mu } \\
+g^{\mu \nu }\left( \nabla _{\lambda }\delta \overset{\text{L}}{\Gamma }%
\text{ }_{\text{ \ \ }\mu \nu }^{\lambda }-\nabla _{\nu }\delta \overset{%
\text{L}}{\Gamma }\text{ }_{\text{ \ }\mu \lambda }^{\lambda }\right)
\end{array}%
\right] \sqrt{-g}d^{4}x\text{.}
\end{equation}%
The last term in the action can be converted into a surface term, so it may
be ignored. Using the four-current $v^{\mu }$ introduced earlier, the action
for the matter fields read \cite{Brill}%
\begin{eqnarray}
\delta I_{\text{field}} &=&\int \left[ \overline{\psi }\delta \gamma ^{\mu
}\nabla _{\mu }\psi +\overline{\psi }\gamma ^{\mu }\delta \widehat{\Gamma }%
_{\mu }\psi \right] \sqrt{-g}d^{4}x \\
&=&\int \left\{
\begin{array}{c}
\left[ \frac{1}{2}g^{\mu \nu }\overline{\psi }\gamma _{i}\left( \nabla _{\nu
}\psi \right) +T_{\text{ \ }\rho \sigma }^{\mu }T_{i}^{\text{ }\rho \sigma
}-\delta _{i}^{\mu }T_{\lambda \rho \sigma }T^{\lambda \rho \sigma }\right]
\delta V_{\text{ }\mu }^{i} \\
+\frac{1}{8}\left( g^{\rho \nu }v^{\mu }-g^{\rho \mu }v^{\nu }\right) \left(
g_{\mu \sigma }\delta \overset{\text{L}}{\Gamma }\text{ }_{\text{ \ }\nu
\rho }^{\sigma }-g_{\nu \sigma }\delta \overset{\text{L}}{\Gamma }\text{ }_{%
\text{ \ }\mu \rho }^{\sigma }\right)
\end{array}%
\right\} \sqrt{-g}d^{4}x\text{.}  \notag
\end{eqnarray}%
Removing the derivatives of variations of the metric appearing in $\delta
\Gamma _{\text{ \ }\nu \rho }^{\sigma }$ via partial integration, and
equating to zero the coefficients of $\delta g^{\mu \nu }$ and $\delta T_{%
\text{ \ }\nu \rho }^{\sigma }$\ in the variation of the action integral, we
obtain%
\begin{eqnarray}
0 &=&\frac{1}{16\pi }\left( R_{\mu \nu }-\frac{1}{2}g_{\mu \nu }R-g_{\mu \nu
}\Lambda \right) +\left( \frac{1}{2}\overline{\psi }\gamma _{\nu }\nabla
_{\mu }\psi -\frac{1}{4}\nabla _{\mu }v_{\nu }\right)   \label{Field-Eqn1} \\
&&+\nabla _{\sigma }T_{\mu \nu }^{\text{ \ \ }\sigma }+T_{\mu \rho \sigma
}T_{\nu }^{\text{ }\rho \sigma }-g_{\mu \nu }T_{\lambda \rho \sigma
}T^{\lambda \rho \sigma }  \notag
\end{eqnarray}%
and%
\begin{equation}
T_{\rho \sigma \lambda }=\kappa \tau _{\rho \sigma \lambda }.
\label{Field-Eqn2}
\end{equation}%
Eqs.(\ref{Field-Eqn1}) have the form of Einstein equations%
\begin{equation}
G_{\mu \nu }-g_{\mu \nu }\Lambda =\kappa \Sigma _{\mu \nu },
\end{equation}%
where $\displaystyle{\kappa=\frac{8\pi G_N}{c^4}}$ and the Einstein tensor and non-symmetric energy-momentum tensors are
\begin{equation}
G_{\mu \nu }=R_{\mu \nu }-\frac{1}{2}g_{\mu \nu }R\text{,}
\end{equation}%
\begin{equation}
\Sigma _{\mu \nu }=\Theta _{\mu \nu }+\mathfrak{T}_{\mu \nu }\text{,}
\end{equation}%
respectively \footnote{We will deduce below these some equations in the framework of GR.}. Here we identify $\Theta _{\mu \nu }$ as the canonical
energy-momentum%
\begin{equation}
\Theta _{\text{ \ }\nu }^{\mu }=\frac{\partial \mathcal{L}_{\text{field}}}{%
\partial (\nabla _{\mu }\chi )}\nabla _{\nu }\chi -\delta _{\text{ }\nu
}^{\mu }\mathcal{L}_{\text{field}}\text{,}
\end{equation}%
while $\mathfrak{T}_{\mu \nu }$ is the stress-tensor form of the
non-Riemannian manifold. For the case of spinor fields being
considered here the explicit form of the energy-momentum
components \cite{Schwinger} are (after symmetrization of
corresponding canonical source terms in the Einstein
equation),%
\begin{equation}
\Theta _{\mu \nu }=-\left[ \overline{\psi }\gamma _{\mu }\nabla _{\nu }\psi
-\left( \nabla _{\nu }\overline{\psi }\right) \gamma _{\mu }\psi +\overline{%
\psi }\gamma _{\nu }\nabla _{\mu }\psi -\left( \nabla _{\mu }\overline{\psi }%
\right) \gamma _{\nu }\psi \right]
\end{equation}%
and by using the second field equation (\ref{Field-Eqn2}), we determine%
\begin{equation}
\mathfrak{T}_{\mu \nu }=\nabla _{\sigma }T_{\mu \nu }^{\text{ \ \ }\sigma
}+T_{\mu \rho \sigma }\tau _{\nu }^{\text{ }\rho \sigma }-g_{\mu \nu
}T_{\lambda \rho \sigma }\tau ^{\lambda \rho \sigma },
\end{equation}%
where $\tau _{\mu \nu }^{\text{ \ \ }\sigma }$ is the so-called
spin - energy
potential \cite{Hehl2, hehlrev}%
\begin{equation}
\tau _{\mu \nu }^{\text{ \ \ }\sigma }:=\frac{\partial \mathcal{L}_{\text{%
field}}}{\partial (\nabla _{\sigma }\chi )}\gamma _{\mu \nu }\chi \text{.}
\end{equation}%
Explicitly, the spin energy potential reads $\tau ^{\mu \nu \sigma }=%
\overline{\psi }\gamma ^{\lbrack \mu }\gamma ^{\nu }\gamma ^{\sigma ]}\psi $%
. The equation of motion obtained  from the variation of the
action with respect to $\overline{\psi }$ reads \cite{Hehl2,
hehlrev}
\begin{equation}
\gamma ^{\mu }\nabla _{\mu }\psi +\frac{3}{8}T_{\mu \nu \sigma }\gamma
^{\lbrack \mu }\gamma ^{\nu }\gamma ^{\sigma ]}\psi =0\text{.}
\end{equation}%
It is interesting to observe that this generalized curved space-time Dirac
equation can be recast into the nonlinear equation of the Heisenberg-Pauli
type%
\begin{equation}
\gamma ^{\mu }\nabla _{\mu }\psi +\frac{3}{8}\varepsilon \left( \overline{%
\psi }\gamma ^{\mu }\gamma _{5}\psi \right) \gamma _{\mu }\gamma _{5}\psi =0.
\end{equation}

Although the gravitational field equation is similar in form to
the Einstein field equation, it differs from the original Einstein
equations because the  curvature tensor, containing space-time
torsion, is non-Riemannian. Assuming that the Euler-Lagrange
equations for the matter fields are satisfied, we obtain the
following conservation laws for the angular - momentum and
energy - momentum%
\begin{equation}
\left.
\begin{array}{c}
V_{\text{ \ }i}^{\mu }V_{\text{ \ }j}^{\nu }\Sigma _{\lbrack \mu \nu
]}=\nabla _{\nu }\mathcal{\tau }_{ij}^{\text{ \ }\nu }\text{,} \\
\\
V_{\mu }^{\text{ \ }k}\nabla _{\nu }\Sigma _{\text{ \ }\kappa }^{\nu
}=\Sigma _{\text{ \ }\kappa }^{\nu }T_{\text{ \ }\mu \nu }^{k}+\mathcal{\tau
}_{\text{ \ }ij}^{\nu }R_{\text{ \ }\mu \nu }^{ij}\text{.}%
\end{array}%
\right.
\end{equation}

\subsection{Concluding remarks}
\label{tre.sette}

In this section we have shown how all the necessary ingredients for a theory of gravitation can be obtained from a gauge theory of local Poincar\'{e} symmetry. In fact gauge fields can be obtained by requiring the invariance of the Lagrangian density under local Poincar\'{e} invariance. The resulting gravity theory describes a space endowed with non-vanishing curvature and torsion.
The simplest prototype of this approach is the Einstein-Cartan Theory which is the lowest order gravitational action linear in the curvature scalar and quadratic in torsion. This scheme, as will see in the second part of this Report, can be extended to more general gravitational theories as the ETGs. The Dirac spinors can be introduced as matter sources which couple with gravity via the torsion stress-tensor $\mathfrak{T}_{\mu \nu }$ which is a component of the total energy- momentum $\Sigma_{\mu\nu}$.

In view of the structure of the generalized energy-momentum-tensor, we remark that the gravitational field equations here obtained are of Yang-Mills type. It is worth noticing that the torsion tensor plays the role of the Yang-Mills field strength. Beside this gauge approach, we want to show now that gravitational field can be recovered also by space-time deformations. In some sense, this approach is more general and it is directly related to one of the main features that will be widely adopted in ETGs: the conformal transformations.

\section{Deformations and conformal transformations}
\label{quattro}

To complete the above considerations, it is worth noticing that also space-time deformations can be related to the generation of gravitational field. In this section, we develop the space-time deformation formalism considering the main quantities which can be related to gravity.
The issue to consider a general way to deform the space-time
metrics is not new. It has been posed in different ways and is
related to several physical problems ranging from the spontaneous
symmetry breaking of unification theories up to gravitational
waves, considered as space-time perturbations. In cosmology, for
example, one faces the problem to describe an observationally
lumpy Universe at small scales which becomes isotropic and
homogeneous at very large scales according to the Cosmological
Principle. In this context, it is crucial to find a way to connect
background and locally perturbed metrics \cite{Ellis84}. For
example, McVittie \cite{McVittie33} considered a metric which
behaves as a Schwarzschild one at short ranges and as a
Friedmann-Lemaitre-Robertson-Walker metric at very large scales.
Gautreau \cite{gautreau} calculated the metric generated by  a
Schwarzschild mass embedded in a Friedmann cosmological fluid
trying to address the same problem. On the other hand, the
post-Newtonian parameterization, as a standard, can be considered
as a deformation of a background, asymptotically flat Minkowski
metric.
In general,  the deformation problem has been explicitly  posed by
Coll and collaborators \cite{Coll1,Coll:2001wy,Llosa:2003di} who
conjectured the possibility to obtain any metric from the
deformation of a space-time with constant curvature. The problem
was solved only for  3-dimensional spaces but a straightforward
extension should be to achieve the same result for space-times of
any dimension.

In principle, new exact solutions of the Einstein field equations
can be obtained by studying perturbations. In particular, dealing
with perturbations as Lorentz matrices of scalar fields
$\Phi^{A}_{\phantom{A}C}$ reveals particularly useful. Firstly
they transform as scalars with respect the coordinate
transformations. Secondly, they are dimensionless and, in each
point, the matrix $\Phi^{A}_{\phantom{A}C}$ behaves as the element
of a group. As we shall see below, such an approach can be related
to the conformal transformations giving an ``extended"
interpretation  and a straightforward physical meaning of them
(see \cite{Faraoni,allemandi} and references therein for a
comprehensive review). Furthermore scalar fields related to
space-time deformations have a straightforward physical
interpretation which could contribute to explain several
fundamental issues  as the Higgs mechanism in unification
theories, the inflation in cosmology and  other  pictures where
scalar fields play a fundamental role in dynamics.
In this section, we are going to discuss  the properties of the
deforming matrices $\Phi^{A}_{\phantom{A}C}$ and we will derive, the field equations for them, showing
how them can parameterize the  deformed metrics, according to the
boundary and initial conditions and to the energy-momentum tensor.
The layout  is the following, we define the
space-time perturbations in the framework of the metric formalism
giving the notion of first and second deformation matrices.
In particular, we discuss how deformation matrices can be split in
their trace, traceless and skew parts. We derive the contributions
of deformation to the geodesic equation and, starting from the
Riemann curvature  tensor, the general equation of deformations. We discuss the notion of linear perturbations under the
standard of deformations. In particular, we recast the equation of propagating perturbations as the equation of propagating perturbations as the equation of
gravitational waves and the transverse traceless gauge under the
standard of deformations. After we discuss the action
of deformations on the Killing vectors. The result consists in
achieving a notion of approximate symmetry.

\subsection{Space-time deformations}
In order to start our considerations, let us take into account a
metric $\mathbf{g}$ on a space-time manifold $\mathcal{M}$. We can decompose it by a co-tetrad field $\omega^{A}(x)$. Here capital Latin indices are for tetrad fields and we are assuming the most general context to develope our considerations.
The metric is

\begin{equation}\label{tetrad}
\mathbf{g}=\eta_{AB}\omega^{A} \omega^{B}.
\end{equation}
Let us define now a new tetrad field
$\widetilde{\omega}=\Phi^{A}_{\phantom{A}C}(x)\,\omega^{C}$, with
$\Phi^{A}_{\phantom{A}C}(x)$ a matrix of scalar fields. Finally we
introduce a space-time  $\widetilde{\mathcal{M}}$ with the metric
$\widetilde{g}$  defined in the following way
\begin{equation}\label{deformed}
\mathbf{\widetilde{g}}=\eta_{AB}\Phi^{A}_{\phantom{A}C}\Phi^{B}_{\phantom{B}D}\,\omega^{C}
\omega^{D} = \gamma_{CD}(x)\omega^{C}
\omega^{D},
\end{equation}
where also $\gamma_{CD}(x)$ is a matrix of fields which are
scalars with respect to the coordinate transformations.

If $\Phi^{A}_{\phantom{A}C}(x)$ is  a Lorentz matrix in any point of  $\mathcal{M}$, then
\begin{equation}\label{i.3}
    \widetilde{g}\equiv g
\end{equation}
otherwise  we say that $\widetilde{g}$ is a  deformation of $g$
and $\widetilde{\mathcal{M}}$ is a deformed $\mathcal{M}$. If all
the functions of $\Phi^{A}_{\phantom{A}C}(x)$ are continuous, then
there is a {\it one - to - one} correspondence between the points
of $\mathcal{M}$ and the points of $\widetilde{\mathcal{M}}$.

In particular, if  $\xi$ is a Killing vector for $g$ on
$\mathcal{M}$, the  corresponding vector $\widetilde{\xi}$ on
$\widetilde{\mathcal{M}}$ could not  necessarily be a Killing
vector.

A particular subset of these deformation matrices is given by
\begin{equation}\label{conformal}
{\Phi}{^{A}_{C}}(x)=\Omega(x)\, \delta^{A}_{\phantom{A}C}.
\end{equation}
which define conformal transformations of the metric,
\begin{equation}\label{confmetric}
  \widetilde{g}=\Omega^{2}(x) g\,.
\end{equation}

In this sense, the  deformations defined by Eq. (\ref{deformed})
can be regarded as a generalization of the conformal
transformations which will be discussed in details in the next part of this Report.

We  call the matrices $\Phi^{A}_{\phantom{A}C}(x)$ {\it first
deformation matrices}, while we can refer to
\begin{equation}\label{seconddeformation}
 \gamma_{CD}(x)=\eta_{AB}\Phi^{A}_{\phantom{A}C}(x)\Phi^{B}_{\phantom{B}D}(x).
\end{equation}
as the {\it second deformation matrices}, which, as seen above,
are also matrices of scalar fields. They generalize  the Minkowski
matrix \(\eta_{AB}\) with constant elements in the definition of
the metric. A further restriction on the matrices
$\Phi^{A}_{\phantom{A}C}$ comes from the theorem proved by Riemann
by which an $n$-dimensional metric has $n(n-1)/2$ degrees of
freedom (see \cite{Coll:2001wy} for details). With this
definitions in mind, let us consider the main properties of
deforming matrices.

\subsection{Properties of deforming matrices}

Let us take into account a four dimensional space-time with
Lorentzian signature. A family of matrices
$\Phi^{A}_{\phantom{A}C}(x)$ such that
\begin{equation}\label{fi}
    \Phi^{A}_{\phantom{A}C}(x)\in GL(4)\, \forall x,
\end{equation}
are defined on such a space-time.

These functions are not necessarily continuous and can connect
space-times with different topologies. A singular scalar field
introduces a deformed manifold $\widetilde{\mathcal{M}}$ with a
space-time singularity.

As it is well known, the Lorentz matrices
$\Lambda^{A}_{\phantom{A}C}$ leave the Minkowski metric invariant
and then
\begin{equation}\label{deformlambda2}
\mathbf{g}=\eta_{EF}\Lambda^{E}_{\phantom{E}A}\Lambda^{F}_{\phantom{F}B}\Phi^{A}_{\phantom{A}C}\Phi^{B}_{\phantom{B}D}\,
\omega^{C}
\omega^{D} =\eta_{AB}\Phi^{A}_{\phantom{A}C}\Phi^{B}_{\phantom{B}D}\,\omega^{C}
\omega^{D}.
\end{equation}
It follows that $\Phi^{A}_{\phantom{A}C}$ give rise to right
cosets of the Lorentz group, {\it i.e.} they are the elements of  the
quotient group $GL(4,\mathbf{R})/SO(3,1)$. On the other hand,  a
right-multiplication of $\Phi^{A}_{\phantom{A}C}$ by a Lorentz
matrix induces a different deformation matrix.

The inverse deformed metric is
\begin{equation}\label{inversemetric}
     \widetilde{g}^{\alpha\beta}=\eta^{CD}{\Phi^{-1}}^{A}_{\phantom{A}C}{\Phi^{-1}}^{B}_{\phantom{B}D}e_{A}^{\alpha}e_{B}^{\beta}
\end{equation}
where ${\Phi^{-1}}^{A}_{\phantom{A}C}{\Phi}^{C}_{\phantom{C}B}=\Phi^{A}_{\phantom{A}C}{\Phi^{-1}}^{C}_{\phantom{C}B}=
\delta^{A}_{B}$.

Let us decompose now the matrix $\Phi_{AB}=\eta_{AC}\,
\Phi^{C}_{\phantom{C}B}$ in its symmetric and antisymmetric parts
\begin{equation}\label{decomposition}
     \Phi_{AB}= \Phi_{(AB)}+\Phi_{[AB]}= \Omega\,\eta_{AB} +  \Theta_{AB} + \varphi_{AB}
\end{equation}
where $ \Omega= \Phi^{A}_{\phantom{A}A}$,  $ \Theta_{AB} $ is the
traceless symmetric part   and $ \varphi_{AB}$ is the skew
symmetric part of  the first deformation matrix respectively. Then
standard conformal transformations are nothing else but
deformations with $\Theta_{AB}=\varphi_{AB}=0$ \cite{Wald84}.

Finding the inverse matrix ${\Phi^{-1}}^{A}_{\phantom{A}C}$ in
terms of $\Omega$, $\Theta_{AB}$  and $\varphi_{AB}$ is not
immediate, but as above, it can be split in the three terms
\begin{equation}\label{inversesplitting}
      {\Phi^{-1}}^{A}_{\phantom{A}C}=\alpha\delta^{A}_{\phantom{A}C}+\Psi^{A}_{\phantom{A}C}+\Sigma^{A}_{\phantom{A}C}
\end{equation}
where $\alpha$, $\Psi^{A}_{\phantom{A}C}$ and
$\Sigma^{A}_{\phantom{A}C}$ are  respectively the trace, the
traceless symmetric part and the antisymmetric part of the inverse
deformation matrix. The second deformation matrix, from the above
decomposition, takes the form
\begin{equation}\label{secondmatrix}
  \gamma_{AB}= \eta_{CD}(\Omega\, \delta_{A}^{C}+
  \Theta_{\phantom{C}A}^{C}+ \varphi_{\phantom{C}A}^{C})(\Omega\, \delta_{B}^{D}+
  \Theta_{\phantom{D}B}^{D}+ \varphi_{\phantom{D}B}^{D})
\end{equation}
and then
$$   \gamma_{AB}=  \Omega^{2}\,\eta_{AB} + 2\Omega\,\Theta_{AB}+  \eta_{CD}\, \Theta_{\phantom{C}A}^{C}\,\Theta_{\phantom{D}B}^{D} + \eta_{CD}\, (\Theta_{\phantom{C}A}^{C}\,\varphi_{\phantom{D}B}^{D}$$
\begin{equation}\label{secondmatrix1}
  +   \varphi_{\phantom{C}A}^{C}\,\Theta_{\phantom{D}B}^{D}) + \eta_{CD}\,\varphi_{\phantom{C}A}^{C}\,\varphi_{\phantom{D}B}^{D}.
\end{equation}
In general, the deformed metric can be split as
\begin{equation}\label{splits}
     {\tilde{g}}{_{\alpha\beta}}=\Omega^{2}{g}{_{\alpha\beta}}+{\gamma}{_{\alpha\beta}}
\end{equation}
where
\begin{equation}\label{acca}
{\gamma}{_{\alpha\beta}}=\left( 2\Omega\,\Theta_{AB}+  \eta_{CD}\, \Theta_{\phantom{C}A}^{C}\,\Theta_{\phantom{D}B}^{D} + \eta_{CD}\, (\Theta_{\phantom{C}A}^{C}\,\varphi_{\phantom{D}B}^{D}+   \varphi_{\phantom{C}A}^{C}\,\Theta_{\phantom{D}B}^{D})
 + \eta_{CD}\,\varphi_{\phantom{C}A}^{C}\,\varphi_{\phantom{D}B}^{D}\right){\omega}{^{A}_{\alpha}}{\omega}{^{B}_{\beta}}
  \end{equation}
In particular,  if $ \Theta_{AB}=0$, the deformed metric
simplifies to
\begin{equation}\label{split}
    \widetilde{g}_{\alpha\beta}=\Omega^{2}g_{\alpha\beta}+\eta_{CD}\,\varphi_{\phantom{C}A}^{\,C}\,\varphi_{\phantom{D}B}^{\,D}
    \omega^{A}_{\phantom{A}\alpha}\omega^{B}_{\phantom{B}\beta}
\end{equation}
and, if $\Omega=1$, the deformation of a metric consists in adding
to the background metric a tensor $\gamma_{\alpha\beta}$. We have to
remember that all these quantities are not independent as, by the
theorem mentioned in \cite{Coll:2001wy}, they have to form at most
six independent functions in a four dimensional space-time.

Similarly the controvariant deformed metric can be always
decomposed in the following way
\begin{equation}\label{controvariantdecomposition}
    \widetilde{g}^{\alpha\beta}= \alpha^{2}g^{\alpha\beta}+ \lambda^{\alpha\beta}
\end{equation}
Let us find the relation between ${\gamma}{_{\alpha\beta}}$ and $
\lambda^{\alpha\beta}$. By using
$\widetilde{g}_{\alpha\beta}\widetilde{g}^{\beta\gamma}=\delta_{\alpha}^{\gamma}$, we obtain
\begin{equation}\label{relationgammalambda}
     \alpha^{2}\Omega^{2}\delta_{\alpha}^{\gamma}+ \alpha^{2}\gamma_{\alpha}^{\gamma}+\Omega^{2}\lambda_{\alpha}^{\gamma}+
     {\gamma}{_{\alpha\beta}}\lambda^{\beta\gamma}=\delta_{\alpha}^{\gamma}
\end{equation}
if the deformations are conformal transformations, we  have
$\alpha=\Omega^{-1}$, so  assuming such a condition, one obtain
the following matrix equation
\begin{equation}\label{relationgammalambda1}
    \alpha^{2}\gamma_{\alpha}^{\gamma}+\Omega^{2}\lambda_{\alpha}^{\gamma}+
     {\gamma}{_{\alpha\beta}}\lambda^{\beta\gamma}=0\,,
\end{equation}
and
 \begin{equation}\label{lambdaaa}
 (\delta_{\alpha}^{\beta}+
     \Omega^{-2}{\gamma}_{\alpha}^{\beta})\lambda_{\beta}^{\gamma}=-\Omega^{-4}\gamma_{\alpha}^{\gamma}
 \end{equation}
 and finally
 \begin{equation}\label{lambdaaaa}
 \lambda_{\beta}^{\gamma}=-\Omega^{-4}{{(\delta+
     \Omega^{-2}{\gamma})^{-1}}}{^{\alpha}_{\beta}}\gamma_{\alpha}^{\gamma}
 \end{equation}
where   ${(\delta+ \Omega^{-2}{\gamma})^{-1}}$ is the inverse
tensor of $(\delta_{\alpha}^{\beta}+
     \Omega^{-2}{\gamma}_{\alpha}^{\beta})$.

To each matrix $\Phi^{A}_{\phantom{A}B}$,  we can associate a
(1,1)-tensor $\phi^{\alpha}_{\phantom{\alpha}\beta}$  defined by
\begin{equation}\label{3.1pp}
  \phi^{\alpha}_{\phantom{\alpha}\beta}= \Phi^{A}_{\phantom{A}B}\omega^{B}_{\beta}e_{A}^{\alpha}
\end{equation}
such that
\begin{equation}\label{3.2}
     \widetilde{g}_{\alpha\beta}=g_{\gamma\delta}\phi^{\gamma}_{\phantom{\gamma}\alpha}\phi^{\delta}_{\phantom{\delta}\beta}
\end{equation}
which can  be decomposed as in Eq.~(\ref{split}). Vice-versa from a
(1,1)-tensor $\phi^{\alpha}_{\phantom{\alpha}\beta}$, we can define a matrix of
scalar fields as
\begin{equation}\label{3.3gg}
     \phi^{A}_{\phantom{A}B} =  \phi^{\alpha}_{\phantom{\alpha}\beta} \omega_{\alpha}^{A}e_{B}^{\beta}.
\end{equation}

The  Levi Civita connection corresponding to the metric
(\ref{splits}) is related to the original connection by the
relation
\begin{equation}\label{1}
     {\widetilde{\Gamma}}{^{\gamma}_{\alpha\beta}}=  {\Gamma}{^{\gamma}_{\alpha\beta}} + {C}{^{\gamma}_{\alpha\beta}}
\end{equation}
(see \cite{Wald84}), where
\begin{equation}\label{2}
  {C}{^{\gamma}_{\alpha\beta}}=2\widetilde{g}^{\gamma\delta}{g}{_{d(\alpha}}{\nabla}{_{\beta)}} \Omega -g_{\alpha\beta}\widetilde{g}^{\gamma\delta} \nabla_{\delta}\Omega +\frac{1}{2} \widetilde{g}^{\gamma\delta}\left( \nabla_{\alpha}\gamma_{\delta\beta}+\nabla_{\beta}\gamma_{\alpha\delta}-\nabla_{\delta}\gamma_{\alpha\beta}\right).
\end{equation}
Therefore, in a deformed space-time, the connection deformation
acts like a force that deviates  the test particles from the
geodesic motion in the unperturbed space-time. As a matter of fact
the geodesic equation for the deformed space-time

\begin{equation}\label{geodesics1}
     \frac{d^{\,2}x^{\gamma}}{d\lambda^{2}}+ \tilde{\Gamma}^{\gamma}_{\phantom{\gamma}\alpha\beta}\frac{dx^{\alpha}}{d\lambda}
     \frac{dx^{\beta}}{d\lambda}=0
\end{equation}
becomes

\begin{equation}\label{geodesics2}
     \frac{d^{\,2}x^{\gamma}}{d\lambda^{2}}+ \Gamma^{\gamma}_{\phantom{\gamma}\alpha\beta}\frac{dx^{\alpha}}{d\lambda}
     \frac{dx^{\beta}}{d\lambda}=-C^{\gamma}_{\phantom{\gamma}\alpha\beta}\frac{dx^{\alpha}}{d\lambda}
     \frac{dx^{\beta}}{d\lambda}.
\end{equation}

The  deformed Riemann curvature tensor is then
\begin{equation}\label{deformedcurvature}
    \widetilde{R}_{\alpha\beta\gamma}^{\phantom{\alpha\beta\gamma}\delta}=  R_{\alpha\beta\gamma}^{\phantom{\alpha\beta\gamma}\delta}+
    \nabla_{\beta} C ^{\delta}_{\phantom{\delta}\alpha\gamma}-\nabla_{\alpha}C^{\delta}_{\phantom{\delta}\beta\gamma}+
     C^{\epsilon}_{\phantom{\epsilon}\alpha\gamma} C^{\delta}_{\phantom{\delta}\beta\epsilon}- C^{\epsilon}_{\phantom{\epsilon}\beta\gamma}C^{\delta}_{\phantom{\delta}\alpha\epsilon},
\end{equation}
while the deformed Ricci tensor obtained by contraction is
\begin{equation}\label{deformedricci}
  \widetilde{R}_{\alpha\beta}= R_{\alpha\beta} + \nabla_{\delta} C ^{\delta}_{\phantom{\delta}\alpha\beta}-\nabla_{\alpha}C^{\delta}_{\phantom{\delta}\delta\beta}+ C^{\epsilon}_{\phantom{\epsilon}\alpha\beta} C^{\delta}_{\phantom{\delta}\delta\epsilon}- C^{\epsilon}_{\phantom{\epsilon}\delta\beta}C^{\delta}_{\phantom{\delta}\alpha\epsilon}
\end{equation}
and the curvature scalar
\begin{equation}\label{deformedcurvaturescalar}
  \widetilde{R}= \widetilde{g}^{\alpha\beta} \widetilde{R}_{\alpha\beta}=
  \widetilde{g}^{\alpha\beta}{R}_{\alpha\beta}+\widetilde{g}^{\alpha\beta}
  \left[\nabla_{\delta} C ^{\delta}_{\phantom{\delta}\alpha\beta}-\nabla_{\alpha}C^{\delta}_{\phantom{\delta}\delta\beta}+
  C^{\epsilon}_{\phantom{\epsilon}\alpha\beta} C^{\delta}_{\phantom{\delta}\delta\epsilon}- C^{\epsilon}_{\phantom{\epsilon}\delta\beta}C^{\delta}_{\phantom{\delta}\alpha\epsilon}\right]
\end{equation}

From  the above curvature quantities, we obtain finally the
equations for the deformations. In the vacuum case,  we simply
have
\begin{equation}\label{eeq}
   \widetilde{R}_{\alpha\beta} = {R}_{\alpha\beta}+\nabla_{\delta} C ^{\delta}_{\phantom{\delta}\alpha\beta}-\nabla_{\alpha}C^{\delta}_{\phantom{\delta}\delta\beta}+ C^{e}_{\phantom{e}ab} C^{d}_{\phantom{d}de}- C^{e}_{\phantom{e}db}C^{d}_{\phantom{d}ae}=0
\end{equation}
where ${R}_{ab}$ must be regarded as a known function.  In
presence of matter, we consider the equations
\begin{equation}\label{eeqmatter}
 {R}_{\alpha\beta}+\nabla_{\delta} C ^{\delta}_{\phantom{\delta}\alpha\beta}-\nabla_{\alpha}C^{\delta}_{\phantom{\delta}\delta\beta}+ C^{\epsilon}_{\phantom{\epsilon}\alpha\beta} C^{\delta}_{\phantom{\delta}\delta\epsilon}- C^{\epsilon}_{\phantom{\epsilon}\delta\beta}C^{\delta}_{\phantom{\delta}\alpha\epsilon}=\kappa\left( \widetilde{T}_{\alpha\beta}-\frac{1}{2}\widetilde{g}_{\alpha\beta}\widetilde{T}\right)
\end{equation}
These
last equations can be improved by considering the field
equations\footnote{ We will discuss these equations in the next section deriving them
from  variational principle.}%
\begin{equation}\label{undeformedeeq}
 {R}_{\alpha\beta}=  \kappa\left(T_{\alpha\beta}-\frac{1}{2}g_{\alpha\beta}T\right)
\end{equation}
and then
\begin{equation}\label{definitivaeeqmatter}
     \nabla_{\delta} C ^{\delta}_{\phantom{\delta}\alpha\beta}-\nabla_{\alpha}C^{\delta}_{\phantom{\delta}\delta\beta}+ C^{\epsilon}_{\phantom{\epsilon}\alpha\beta} C^{\delta}_{\phantom{\delta}\delta\epsilon}- C^{\epsilon}_{\phantom{\epsilon}\delta\beta}C^{\delta}_{\phantom{\delta}\alpha\epsilon}= \kappa\left[\widetilde{T}_{\alpha\beta}-\frac{1}{2}\widetilde{g}_{\alpha\beta}\widetilde{T}-\left(T_{\alpha\beta}-\frac{1}{2}g_{\alpha\beta}T\right)\right]
\end{equation}
 are the most general equations for  deformations.
 
\subsection{The propagation of metric deformations}
Metric deformations can be used to describe perturbations.  To
this aim we can simply consider the deformations

\begin{equation}\label{2defor}
     \Phi^{A}_{\phantom{A}B}=\delta^{A}_{\phantom{A}B}+\varphi^{A}_{\phantom{A}B}
\end{equation}
with
\begin{equation}\label{2piccolezza}
    |\,\varphi^{A}_{\phantom{A}B}|\ll 1,
\end{equation}
together with their derivatives
\begin{equation}\label{2piccolezza1}
    |\,\partial\varphi^{A}_{\phantom{A}B}|\ll 1\,.
\end{equation}
With this approximation, immediately we find the inverse relation
\begin{equation}\label{3defor}
     (\Phi^{-1})^{A}_{\phantom{A}B}\simeq\delta^{A}_{\phantom{A}B}-\varphi^{A}_{\phantom{A}B}.
\end{equation}
As a remarkable example,  we have that gravitational waves are
generally described, in linear approximation, as perturbations of
the Minkowski metric
\begin{equation}\label{linearapproximation}
     g_{\alpha\beta}=\eta_{\alpha\beta}+\gamma_{\alpha\beta}.
\end{equation}
In our case, we can extend in a covariant way such an
approximation. If $ \varphi_{AB}$ is an antisymmetric matrix, we
have
\begin{equation}\label{covariantlinearapproximation}
\widetilde{g}_{\alpha\beta}=g_{\alpha\beta}+\gamma_{\alpha\beta}
\end{equation}
where  the first order terms in $\varphi^{A}_{\phantom{A}B}$
vanish and $\gamma_{\alpha\beta}$ is of second order
\begin{equation}\label{gammagw}
    \gamma_{\alpha\beta}=\eta_{AB}\varphi^{A}_{\phantom{A}C}\varphi^{B}_{\phantom{B}D}\omega^{C}_{\phantom{C}\alpha}
    \omega^{D}_{\phantom{D}\beta}.
\end{equation}
Consequently
\begin{equation}\label{controvariante}
    \widetilde{g}^{\alpha\beta}=g^{\alpha\beta}+\gamma^{\alpha\beta}
\end{equation}
where
\begin{equation}\label{gam}
    \gamma^{\alpha\beta}=\eta^{AB}(\varphi^{-1})_{\phantom{A}A}^{C}(\varphi^{-1})_{\phantom{B}B}^{D}e_{C}^{\phantom{A}\alpha}
    e_{D}^{\phantom{B}\beta}.
\end{equation}
Let us consider the background metric $g_{ab}$, solution of the
equations in the vacuum
\begin{equation}\label{vacein}
    R_{\alpha\beta}=0.
\end{equation}
We obtain the equation of perturbations considering  only the
linear terms in Eq.~(\ref{eeq}) and neglecting the contributions of
quadratic terms. We find
\begin{equation}\label{approx1}
   \widetilde{R}_{\alpha\beta} = \nabla_{\delta}  C ^{\delta}_{\phantom{\delta}\alpha\beta}
   -\nabla_{\alpha}C^{\delta}_{\phantom{\delta}\delta\beta}=0\,,
\end{equation}
and, by the explicit form of $C ^{\delta}_{\phantom{\delta}\alpha\beta}$, this
equation becomes
\begin{equation}\label{approx2}
    \left( \nabla_{\delta}\nabla_{\alpha} {\gamma}{^{\delta}_{\beta}}+ \nabla_{\delta}\nabla_{\beta} {\gamma}{^{\delta}_{\alpha}}-\nabla_{\delta}\nabla^{\delta} {\gamma}{_{\alpha\beta}}\right)-
    \left( \nabla_{\alpha}\nabla_{\delta} {\gamma}{^{\delta}_{\beta}}+ \nabla_{\alpha}\nabla_{\beta} {\gamma}{^{\delta}_{\delta}}-
    \nabla_{\alpha}\nabla^{\delta} {\gamma}{^{\delta}_{\beta}}\right)=0\,.
\end{equation}
Imposing the transverse traceless gauge on $\gamma_{\alpha\beta}$ , {\it i.e.}
the standard gauge conditions
\begin{equation}\label{gauge1}
    \nabla^{\alpha}\gamma_{\alpha}=0
\end{equation}
and
\begin{equation}\label{gauge2}
    \gamma={\gamma}{^{\alpha}_{\alpha}}=0
\end{equation}
Eq.~(\ref{approx2}) reduces to
\begin{equation}\label{eqwald}
    \nabla_{\beta}\nabla^{\beta}\gamma_{\alpha\gamma}-2{R}{{^{\,\beta}_{\alpha\gamma}}^{\delta}}\gamma_{\beta\delta}=0\,,
\end{equation}
see also \cite{Wald84}. In our context, this equation is a
linearized equation for  deformations and it is straightforward to
consider perturbations and, in particular, gravitational waves, as
small deformations of the metric. This result can be immediately
translated into  the above scalar field matrix equations. Note
that such an equation can be applied to the conformal part of the
deformation, when the general decomposition is considered.

As an example, let us take into account the deformation matrix
equations applied to the Minkowski metric, when the deformation
matrix assumes the form (\ref{2defor}). In this case, the Eq. (\ref{approx2}), become ordinary wave equations for
$\gamma_{ab}$. Considering the deformation matrices, these
equations become, for a tetrad field of constant vectors,
\begin{equation}\label{equationdeformation}
    \partial^{\delta} \partial_{\delta}\varphi^{C}_{\phantom{C}A}\varphi_{CB}+2\,
    \partial_{\delta}\varphi^{C}_{\phantom{A}A}  \partial^{\delta}\varphi_{CB}+ \varphi^{C}_{\phantom{C}A}\partial^{\delta}
    \partial_{\delta}\varphi_{CB}=0\,.
\end{equation}
The above gauge conditions are now
\begin{equation}\label{gauge1phi}
     \varphi_{AB}\varphi^{BA}=0
\end{equation}
and
\begin{equation}\label{gauge2phi}
     {e}{_{D}^{\delta}}\left[\partial_{\delta}\varphi_{CA}{\varphi}{^{C}_{B}}+
     \varphi_{CA}\partial_{\delta}{\varphi}{^{C}_{B}}\right]=0\,.
\end{equation}
This result shows that the gravitational waves can be fully
recovered starting from the scalar fields which describe the
deformations of the metric. In other words, such scalar fields can
assume the meaning of gravitational waves modes.
We will discuss in details gravitational waves in the next section and in the third part of this Report.

\subsection{Approximate Killing vectors}
Another important issue which can be addressed starting from
space-time deformations is related to the symmetries. In
particular, they assume a fundamental role in describing when a
symmetry is preserved or broken under the action of a given field.
In GR, the Killing vectors are always related to
the presence of given space-time symmetries \cite{Wald84}.

Let us take an exact solution of the Einstein equations, which
satisfies the Killing equation
\begin{equation}\label{killing}
    ( L_{\mathbf{\xi}}g)_{\alpha\beta}=0
\end{equation}
where $\mathbf{\xi}$, being the generator of an infinitesimal
coordinate transformation, is a Killing vector. If we take a
deformation of the metric with the scalar matrix
\begin{equation}\label{defor}
     \Phi^{A}_{\phantom{A}B}=\delta^{A}_{\phantom{A}B}+\varphi^{A}_{\phantom{A}B}
\end{equation}
with
\begin{equation}\label{piccolezza}
    |\,\varphi^{A}_{\phantom{A}B}|\ll 1\,,
\end{equation}
and
\begin{equation}\label{nokilling}
    ( L_{\mathbf{\xi}}\widetilde{g})_{\alpha\beta}\neq 0\,,
\end{equation}
being

\begin{equation}\label{killingtetrad}
    ( L_{\mathbf{\xi}}e^{A})_{\alpha}=0\,,
\end{equation}
we have
\begin{equation}\label{nokilling2}
    ( L_{\mathbf{\xi}} \varphi)^{A}_{\phantom{A}B} =
     \xi^{\alpha}\partial_{\alpha}\varphi^{A}_{\phantom{A}B}\neq 0\,.
\end{equation}
If  there is some region $\mathcal{D}$ of the deformed space-time
$\mathcal{M}_{deformed}$ where

\begin{equation}\label{nokilling3}
    |\,( L_{\mathbf{\xi}} \varphi)^{A}_{\phantom{A}B}|\ll 1
\end{equation}
we say that $\mathbf{\xi}$ is an {\it approximate Killing vector}
on $\mathcal{D}$. In other words, these approximate Killing
vectors allow to ``control" the space-time symmetries under the
action of a given deformation.
We can calculate the modified connection $\hat{\Gamma_{\alpha\beta}^{\gamma}}$
in many alternative ways. Let us introduce the tetrad $e_{A}$ and
cotetrad $\omega^{B} $ satisfying the orthogonality relation\begin{equation}\label{ieomegaappendix}
   i_{e_{A}}\omega^{B}=\delta_{A}^{B}
\end{equation}
and the non-integrability condition (anholonomy)
\begin{equation}\label{1appendix}
d\omega^{A}=\frac{1}{2}\Omega_{BC}^{A}\omega^{B}\wedge\omega^{C}.
\end{equation}
The corresponding connection is
\begin{equation}\label{gammaappendix}
    \Gamma_{BC}^{A}=\frac{1}{2}\left(\Omega_{BC}^{A}-\eta^{AA'}\eta_{BB'}\Omega_{A'C}^{B'}-
    \eta^{AA'}\eta_{CC'}\Omega_{A'B }^{C'} \right)
\end{equation}
If we deform the metric as in (\ref{deformed}), we have two
alternative ways to write this expression: either writing the
``deformation'' of the metric in the space of tetrads or
``deforming''   the tetrad field as in the following expression
\begin{equation}\label{deformedagainappendix}
 \hat{g} =\eta_{AB}\Phi^{A}_{\phantom{A}C}\Phi^{B}_{\phantom{B}D}\,\omega^{C}
\omega^{D} =
  \gamma_{AB} \,\omega^{A}
\omega^{B}=\eta_{AB} \,\hat{\omega}^{A}\hat{ \omega}^{B}.
\end{equation}
In the first case,  the contribution of the Christoffel symbols,
constructed by the metric $\gamma_{AB}$, appears
\begin{equation}\label{hatgammaappendix} \hat{\Gamma}_{BC}^{A}=\frac{1}{2}\left(\Omega_{BC}^{A}-\gamma^{AA'}\gamma_{BB'}\Omega_{A'C}^{B'}-
    \gamma^{AA'}\gamma_{CC'}\Omega_{A'B }^{C'} \right)+ \frac{1}{2}\gamma^{AA'}\left(i_{ e_{C}}  d\gamma_{BA'}-
i_{e_{B}}  d\gamma_{CA'}
     - i _{e_{A'}}  d\gamma_{BC}\right)
\end{equation}
In the second case,  using (\ref{1appendix}), we can define the
new anholonomy objects $\hat{C}_{BC}^{A}$.
\begin{equation}\label{2appendix}
d\hat{\omega}^{A}=\frac{1}{2}\hat{\Omega}_{BC}^{A}\hat{\omega}^{B}\wedge\hat{\omega}^{C}.
\end{equation}
After some  calculations, we have
\begin{equation}\label{newhatcappendix}
\hat{\Omega}_{BC}^{A}=\Phi^{A}_{\phantom{A}E}
{\Phi^{-1}}_{\phantom{D}B}^{D}
{\Phi^{-1}}_{\phantom{F}C}^{F}\,{\Omega}_{DF}^{E}+2\Phi^{A}_{\phantom{A}F}e_{G}^{a}
\left({\Phi^{-1}}_{\phantom{G}[B}^{G}\partial_{a}{\Phi^{-1}}_{\phantom{F}C]}^{F}\right)
\end{equation}
As we are assuming a constant metric in  tetradic space, the
deformed connection is
\begin{equation}\label{gammahat1appendix}
   \hat{ \Gamma}_{BC}^{A}=\frac{1}{2}\left(\hat{\Omega}_{BC}^{A}-\eta^{AA'}\eta_{BB'}\hat{\Omega}_{A'C}^{B'}-
    \eta^{AA'}\eta_{CC'}\hat{\Omega}_{A'B }^{C'} \right).
\end{equation}

Substituting (\ref{newhatcappendix}) in (\ref{gammahat1appendix}),
the final expression of $\hat{ \Gamma}_{BC}^{A}$, as a function of
$\Omega_{BC}^{A}$, $\Phi^{A}_{\phantom{A}B}$,
${\Phi^{-1}}_{\phantom{D}C}^{D}$ and $e_{G}^{\alpha}$ is
\begin{eqnarray}\label{gammahat2appendix}
\qquad  \hat{ \Gamma}_{ABC}=\Delta_{ABC}^{DEF}\bigg[\frac{1}{2}
 \eta_{DG}\,\Phi^{G}_{\phantom{G}G'}
{\Phi^{-1}}_{\phantom{E'}E}^{E'}
{\Phi^{-1}}_{\phantom{F'}F}^{F'}\,\Omega^{G'}_{E'F'} 
+
\eta_{DK}\Phi^{K}_{\phantom{D}H}e_{G}^{\alpha}
 {\Phi^{-1}}_{\phantom{G}
[E}^{G}\partial_{|\alpha|}{\Phi^{-1}}_{\phantom{H}F] }^{H}\bigg] \qquad
\end{eqnarray}
where
\begin{equation}\label{delta1}
\Delta_{ABC}^{DEF}=\delta_{A}^{D}\delta_{C}^{E}\delta_{B}^{F}-\delta_{B}^{D}\delta_{C}^{E}\delta_{A}^{F}+
\delta_{C}^{D}\delta_{A}^{E}\delta_{B}^{F}.
\end{equation}

\subsection{Discussion and conclusions}
In this section, we have proposed  a novel definition of space-time
metric deformations parameterizing them in terms of scalar field
matrices. The main result is that deformations can be described as
extended conformal transformations. This fact gives a
straightforward physical interpretation of conformal
transformations: conformally related metrics can be seen as the
"background" and the "perturbed" metrics. In other words, the
relations between the Jordan frame and the Einstein frame, which we will discuss for ETGs, can be
directly interpreted through the action of the deformation
matrices contributing to solve the issue of what the true physical
frame is \cite{Faraoni,allemandi}.

Besides, space-time metric deformations can be immediately recast
in terms of perturbation theory allowing a completely covariant
approach to the problem of gravitational waves.

Results related to those presented here has been proposed in
\cite{Coll1,Coll:2001wy}. There it is shown that any metric in a
three dimensional manifold can be decomposed in the form
\begin{equation}\label{coll1}
     \widetilde{g}_{\alpha\beta}= \sigma(x)h_{\alpha\beta}+\epsilon s_{\alpha}s_{\beta}
\end{equation}
where $h_{\alpha\beta}$ is a metric with constant curvature, $\sigma(x)$ is
a scalar function,  $s_{a}$ is a three-vector and $\epsilon=\pm
1$. A relation has to be imposed between $\sigma$ and $s_{a}$ and
then  the metric can be defined, at most, by three independent
functions.

In a subsequent paper \cite{Llosa:2004uf}, Llosa and Soler showed
that (\ref{coll1}) can be generalized to arbitrary dimensions by
the form
\begin{equation}\label{coll2}
     \widetilde{g}_{\alpha\beta}= \lambda(x)g_{\alpha\beta}+\mu(x)F_{\alpha\gamma}g^{\gamma\delta}F_{\alpha\beta}
\end{equation}
where $g_{ab}$ is a constant curvature metric, $F_{ab}$ is a
two-form, $\lambda(x)$ and $\mu(x)$ are two scalar functions.
These results are fully recovered and generalized from our
approach as soon as the deformation of a constant metric is
considered and suitable conditions on the tensor $\Theta_{AB}$ are
imposed.

In general, we have shown that, when we turn to the tensor
formalism, we can work with arbitrary metrics and arbitrary
deforming $\gamma_{\alpha\beta}$ tensors. In principle, by arbitrary
deformation matrices, not necessarily real, we can pass from a
given metric to any other metric.   As an example, a noteworthy
result has been achieved by Newman and Janis \cite{Newman:1965tw}:
They showed that, through a complex coordinate transformation, it
is always possible to achieve a Kerr metric from a Schwarzschild
one. In our language, this means that a space-time deformation
allows to pass from a spherical symmetry to a cylindrical one. In Part \ref{tre} of this Report, we will give an example of this approach.
Furthermore, it has been shown
\cite{Banados:1992wn,Banados:1992gq} that  three dimensional black
hole solutions can be found by identifying 3-dimensional anti-de
Sitter space on which acts a discrete subgroup of $SO(2,2)$.

In all these examples, the transformations which lead to the
results are considered as ``coordinate transformations''. We think
that this definition is a little bit misleading since one does not
covariantly perform the same transformations on {\it all} the
tensors defined on the manifold. On the other hand, our definition
of metric deformations and deformed manifolds can be
straightforwardly  related to the standard notion of perturbations
since, in principle, it works on a given region $\mathcal{D}$ of
the deformed space-time (see, for example,
\cite{Bardeen:1980kt,Mukhanov:1990me}).

Going back to the issue proposed in Sec. \ref{tre} "{\it What can generate gravity?}", we can say that space-time deformations, as gauge invariance are intimately related to the gravitational field.
In fact, deformation are fully connected to geometric nature of gravitational field but do not specify exactly a single theory. In other words GR is nothing else but a possible "realization" of gravity.
This means that such an interaction can be described in a wider context than the Einstein theory.
Before facing the ETGs, let us recall the main physical meanings related to GR.

\section{The physical meaning of General Relativity}
GR, as we discussed above, is a theory of gravitation that was developed by Albert Einstein between 1907 and 1915. According to GR, the observed gravitational attraction between masses results from their warping of space and time.
Up to the beginning of the 20th century, Newton's law of universal gravitation had been accepted for more than two hundred years as a valid description of the gravitational force between masses. In Newton's model, gravity is the result of an attractive force between massive objects. Although even Newton was bothered by the unknown nature of that force, the basic framework was extremely successful for describing motions.
Experiments and observations show that Einstein's description of gravitation accounts for several effects that are unexplained by Newton's law, such as  anomalies in the orbits of Mercury and other planets. GR also predicts novel effects of gravity, such as gravitational waves, gravitational lensing and an effect of gravity on time known as gravitational time dilation. Many of these predictions have been confirmed by experiments, while others are the subject of ongoing research. For example, although there is indirect evidence for gravitational waves, direct evidence of their existence is still being sought by several teams of scientists in experiments such as VIRGO, LIGO and GEO 600 \cite{LIGO2,acernese,LISA2,rates}.
GR has developed as an essential tool in modern astrophysics. It provides the foundation for the current understanding of black holes, regions of space where gravitational attraction is so strong that light can not escape. Their strong gravity is thought to be responsible for the intense radiation emitted by certain types of astronomical objects (such as active galactic nuclei or quasars). GR is also part of the framework of the standard Big Bang model of cosmology.
Although GR is not the only relativistic theory of gravity, it is the simplest theory that is consistent with the experimental data. Nevertheless, a number of open questions remain, the most fundamental of which is how GR can be reconciled with the laws of quantum physics to produce a complete and self-consistent theory of quantum gravity.
In this section we discuss 
the physical foundation of GR outlining some of the issues that must be recovered in any relativistic theory of gravity, in particular the ETGs.
\subsection{The Equivalence Principle}
Equivalence principle is the physical foundation  of any metric
theory of gravity \cite{Will93,salvatorenewton} and the starting point of our considerations.

The first formulation of  Equivalence principle
comes out from the theory of gravitation formulates by Galileo and
Newton; it is called the Weak Equivalence Principle and it
states that the ``inertial mass'' $m$ and the ``gravitational
mass'' $M$ of any object are equivalent. In Newtonian physics, the
``inertial mass'' $m$ is a coefficient which appears in the second
Newton law: $\vec{F}= m \, \vec{a}$ where $\vec{F}$ is the force
exerted on a mass $m$ with acceleration $\vec{a}$; in Special
Relativity  (without gravitation) the ``inertial mass'' of a
body appears to be proportional to the rest energy of the body:
$E=m \, c^2$. Considering the Newtonian gravitational attraction,
one introduces the ``gravitational mass'' $M$: the gravitational
attraction force between two bodies of ``gravitational mass'' $M$
and $M'$ is $F= G_N M M'/r^2$ where $G_N$ is the Newtonian
gravitational constant and $r$ the distance
 between the two bodies. Various experiments \cite{urbino} demonstrate
 that $m\equiv M$.
 The present accuracy of this relation in laboratory is of the order of $10^{-13}$;
 spatial projects
 are currently designed to achieve precision of $10^{-15}$
  \cite{microSCOPE} and $10^{-18}$ \cite{miniSTEP}.

 The Weak Equivalence Principle statement implies that it is impossible to distinguish
between the effects of a gravitational field from those
experienced in uniformly accelerated frames, using the simple
observation of the free-falling particles behavior. The Weak Equivalence Principle can be
formulated again in the following statement \cite{Will93}:

\begin{quote}
 {\it If an uncharged test body is placed at an initial event in space-time
 and given an initial velocity there, then its subsequent trajectory will be
 independent of its internal structure and composition}.
\end{quote}

\indent  A generalization of  Weak Equivalence Principle claims that the Special Relativity is only locally valid.
It has been achieved by Einstein  after the formulation of  Special Relativity
theory where the concept of mass looses some of its uniqueness:
the mass is reduced to a manifestation of energy and momentum.
According to Einstein, it is impossible to distinguish between
uniform acceleration and an external gravitational field, not only
for free-falling particles but whatever is the experiment. This
equivalence principle has been called Einstein Equivalence
Principle ; its main statements are the following
\cite{Will93}:

\begin{quote}
 \begin{itemize}
  \item   Weak Equivalence Principle is valid;
  \item the outcome of any local non-gravitational test
  experiment is independent of  velocity of  free-falling apparatus;
  \item  the outcome of any local non-gravitational
  test experiment is independent of where and when in
  the Universe it is performed.
 \end{itemize}
\end{quote}

\indent  One defines as ``local non-gravitational experiment" an
experiment performed in a small-size \footnote{In order to avoid
the inhomogeneities.} freely falling laboratory. From the Einstein Equivalence
Principle, one
gets that the gravitational interaction depends on the curvature
of space-time, {\it i.e.} the postulates of any metric theory of gravity
have to be satisfied \cite{Will93}:

 \begin{itemize}
  \item space-time is endowed with a metric $g_{\mu\nu}$;
  \item  the world lines of test bodies are geodesics of the metric;
  \item  in local freely falling frames, called local Lorentz frames,
  the non-gravitational laws of physics are those of Special Relativity.
 \end{itemize}

\indent  One of the predictions of this principle is the gravitational
red-shift, experimentally verified by Pound and Rebka in 1960
\cite{PoundRebka}. It is worth stressing that gravitational interactions are
specifically excluded from   Weak Equivalence Principle and Einstein Equivalence
Principle. In order to classify
alternative theories of gravity, the Gravitational Weak
Equivalence Principle and the Strong Equivalence Principle has to be introduced. The  Strong Equivalence Principle
 extends the  Einstein Equivalence
Principle by including
all the laws of physics in its terms \cite{Will93}:

\begin{quote}
 \begin{itemize}
  \item  Weak Equivalence Principle is valid for self-gravitating bodies as well as for test bodies (Gravitational Weak
Equivalence Principle);
  \item the outcome of any local test experiment is
  independent of the velocity of the free-falling apparatus;
  \item the outcome of any local test experiment is
  independent of where and when in the Universe it is performed.
 \end{itemize}
\end{quote}

\indent  Therefore, the Strong Equivalence Principle contains the  Einstein Equivalence
Principle, when gravitational forces
are ignored. Many authors claim that the only theory coherent with
the Strong Equivalence Principle is GR. An extremely important issue related to the consistency of Equivalence Principle with respect to the Quantum Mechanics. Some phenomena, like neutrino oscillations could violate it if induced by the gravitational field. In Part III of this Report, we discuss possible neutrino oscillation effects induced by corrections to GR.

\subsection{The geodesic structure}

In GR the space-time metric is related to geodesic motion because  
the Equivalence Principle requires that the motion of a 
point-like body  in free fall be described by the
geodesic equation. The latter can be derived  from
the variational principle
 \begin{eqnarray} \label{variationalprincipleforgeodesics}
 \delta S= \delta \int_{A}^{B}ds=0\,,
 \end{eqnarray}
where $ds$ is the line element and $A$ and $B$ are the initial 
and final points along the space-time trajectory, respectively.
It is instructive to deduce the geodesic equation in order to show how it is related to the metric structure ({\it i.e.} the casual structure) of GR.

The line element is written as
 \begin{eqnarray}
ds=\left| g_{\alpha\beta}dx^{\alpha}
dx^{\beta}\right|^{1/2}=\left| g_{\alpha\beta} 
\frac{dx^{\alpha}}{ds}
\frac{dx^{\beta}}{ds}\right|^{1/2}ds \,, 
 \end{eqnarray}
with $s$ playing the role of 
an affine parameter, and from which it follows that 
 \begin{eqnarray}
g_{\alpha\beta}u^{\alpha}u^{\beta}=-1 \,, 
 \end{eqnarray}
where ${\displaystyle u^{\alpha}=\frac{dx^{\alpha}}{ds}}$ is the 
four-velocity of the particle. Substitution into 
Eq.~(\ref{variationalprincipleforgeodesics}) yields 
  \begin{eqnarray}
\delta S= \delta
\int_{A}^{B}\left| g_{\alpha\beta} \frac{dx^{\alpha}}{ds}
\frac{dx^{\beta}}{ds} \right|^{1/2} ds=0 \,.
 \end{eqnarray}
By performing this variation, one obtains
 \begin{eqnarray}
  \delta
S=\int_{A}^{B}
\frac{1}{2\sqrt{\left| 
g_{\alpha\beta} u^{\alpha}u^{\beta} \right|}} 
\left[g_{\alpha\beta,\lambda}\delta
x^{\lambda}\frac{dx^{\alpha}}{ds} \frac{dx^{\beta}}{ds}+
2g_{\alpha\beta}\frac{d}{ds}\left(\delta
x^{\alpha}\right)\frac{dx^{\beta}}{ds} \right]ds=0 \,.
 \end{eqnarray}
The second term in square brackets is
$
g_{\alpha\beta} \delta \left(\frac{dx^{\alpha}}{ds}
\frac{dx^{\beta}}{ds}\right)$ as a consequence of  the fact that
$\delta \left(ds\right)=d\left(\delta s\right)$, hence 
\begin{equation}
 g_{\alpha\beta}\delta \left(\frac{dx^{\alpha}}{ds}
\frac{dx^{\beta}}{ds}\right)= g_{\alpha\beta}
\frac{dx^{\alpha}}{ds}\delta\left( \frac{dx^{\beta}}{ds}\right)+
g_{\alpha\beta}\frac{dx^{\beta}}{ds}\delta\left(
\frac{dx^{\alpha}}{ds}\right)=2g_{\alpha\beta}\frac{dx^{\beta}}{ds}
\frac{d}{ds}\left(\delta x^{\alpha}\right) \,. 
\end{equation}
Using  $g_{\alpha\beta}u^{\alpha}u^{\beta}=-1$, it is 
\begin{equation}
\delta S=\int_{A}^{B} 
\frac{1}{2}\left[g_{\alpha\beta,\lambda}\delta
x^{\lambda}\frac{dx^{\alpha}}{ds}
\frac{dx^{\beta}}{ds}+2g_{\alpha\beta}\frac{dx^{\beta}}{ds}
\frac{d}{ds}\left(\delta x^{\alpha}\right) \right]ds=0 
\end{equation}
and  integration by parts of the second term yields 
\begin{eqnarray}
\delta S & = &  \int_{A}^{B} 
\frac{1}{2}\left( g_{\alpha\beta,\lambda} \delta
x^{\lambda} \frac{dx^{\alpha}}{ds}\frac{dx^{\beta}}{ds}\right) ds +
\left[ g_{\alpha\beta}\frac{dx^{\beta}}{ds} \delta
x^{\alpha}\right]_{A}^{B}
- \int_{A}^{B}  
\frac{d}{ds}\left( g_{\alpha\beta}\frac{dx^{\beta}}{ds} \right)\delta
x^{\alpha} ds =  0 \,. 
\end{eqnarray} 
By imposing that, at the endpoints,  it is $
\delta x^{\alpha}\left(A\right)=\delta x^{\alpha}\left(B\right)=0 $,  
the second term vanishes  and  
\begin{eqnarray} 
\delta S & = & \int_{A}^{B}
\frac{1}{2}\left( g_{\alpha\beta,\lambda} 
\frac{dx^{\alpha}}{ds}\frac{dx^{\beta}}{ds}\delta
x^{\lambda}\right) ds- \int_{A}^{B}
\left( g_{\alpha\beta}\frac{d^{2}x^{\beta}}{ds^{2}} 
+g_{\alpha\beta,\lambda}\frac{dx^{\lambda}}{ds}\frac{dx^{\beta}}{ds} 
\right) \delta  x^{\alpha}ds =  0 \,. 
\end{eqnarray}
This  equation can be written as   
\begin{eqnarray} 
\delta  S=\int_{A}^{B} 
\left[\left(\frac{1}{2}g_{\alpha\beta,\lambda} 
-g_{\lambda\beta,\alpha}\right)\frac{dx^{\alpha}}{ds}\frac{dx^{\beta}}{ds}
-g_{\lambda\beta}\frac{d^{2}x^{\beta}}{ds^{2}}\right]\delta 
x^{\lambda}ds=0 
\,. 
\end{eqnarray}
This integral vanishes for all variations $\delta x^{\lambda}$ 
with fixed endpoints if  
\begin{eqnarray} 
g_{\lambda\beta}\frac{d^{2}x^{\beta}}{ds^{2}} 
=\left(\frac{1}{2}g_{\alpha\beta,\lambda} 
-g_{\lambda\beta,\alpha}\right)u^{\alpha}u^{\beta} \,.
\end{eqnarray}
Since
\begin{equation}
g_{\lambda\beta,\alpha}u^{\alpha}u^{\beta}=
g_{\lambda\alpha,\beta} 
u^{\beta}u^{\alpha}= 
\frac{1}{2}\left(g_{\lambda\beta,\alpha} 
+g_{\lambda\alpha,\beta}\right)u^{\alpha}u^{\beta} \,,
\end{equation} 
whereas
\begin{eqnarray} 
g_{\lambda\beta} \, \frac{d^{2}x^{\beta}}{ds^{2}} 
=\frac{1}{2}\left(g_{\alpha\beta,\lambda}-g_{\lambda\beta,\alpha}
-g_{\lambda\alpha,\beta}\right)u^{\alpha}u^{\beta}  
\end{eqnarray}
we have 
\begin{equation}
\left\{\lambda,\alpha\beta\right\}=\frac{1}{2}\left(g_{\lambda\alpha,\beta} 
+g_{\lambda\beta,\alpha}-g_{\alpha\beta,\lambda}\right) 
\end{equation}
and 
\begin{eqnarray} 
g_{\lambda\beta}\frac{d^{2}x^{\beta}}{ds^{2}} 
+\left\{\lambda,\alpha\beta\right\}u^{\alpha}u^{\beta}=0 \,.
\end{eqnarray}
Multiplying by $g^{\lambda\tau}$ and remembering that
\begin{eqnarray} 
g^{\lambda\tau}g_{\lambda\beta}= \delta^{\tau}_{\beta} \,, 
\;\;\;\;\;\;\;  g^{\lambda\tau} \left\{\lambda,\alpha\beta\right\}= 
\Gamma_{\alpha\beta}^{\tau} 
\,,\end{eqnarray} 
one has  
\begin{eqnarray} 
\frac{d^{2}x^{\tau}}{ds^{2}}+ 
\Gamma_{\alpha\beta}^{\tau}u^{\alpha}u^{\beta} = 0 \,,
 \end{eqnarray}
which is the geodesic equation describing  the free fall motion 
of a point-like body  in the gravitational  field represented by the Christoffel symbols
$\Gamma_{\alpha\beta}^{\tau}$.

This result means that the connection is Levi-Civita and then $g$ and $\Gamma$ are intimately related in GR. This feature, as stressed by Palatini \cite{palatiniorigin} and Einstein \cite{palaeinstein}, shows that geodesic structure ($\Gamma$) and casual structure ($g$) are not independent. This is not true, in general, in ETGs as we will discuss in the next part of this Report.

\subsection{The Einstein field equations in metric and Palatini formalisms}
\label{FEQ}

The Einstein field equations are a particular case of those deduced in previous section for the Einstein-Cartan theory.
Here we report their derivation in details in view of the variational principles that we will adopt for ETGs. In other words, the field equations in the cases of non-minimal couplings and non-linear functions of curvature invariants. We illustrate the derivation of the Einstein field equations in vacuo as the starting point.

Let us consider 
\begin{equation}
\delta \int d\Omega \sqrt{-g} \, {\cal L} =0\,,
\end{equation}
where   $ \sqrt{-g} \, d\Omega $ 
is the invariant volume element and ${\cal L}$ is the 
Lagrangian density. In fact, under the coordinate 
transformation 
$ \overline{x}\hspace{0.01pt}^{\alpha}\rightarrow   
x^{\alpha}=x^{\alpha} 
\left(\overline{x}\hspace{0.01pt}^{\mu}\right)$, 
where $\overline{x}\hspace{0.01pt}^{\mu}$ are the "initial'' 
local coordinates, we have
\begin{equation}
d\Omega = J \, d\overline{\Omega}\,, \;\;\;\;\;\;  
J=\mbox{det} \left(\frac{\partial x^{\alpha}}{\partial
\overline{x}^{\mu}}\right) \,,
\end{equation} 
with $J$ the  Jacobian determinant
of the transformation. Moreover, we have
\begin{eqnarray}
\overline{g}_{\alpha\beta} & = & \mbox{diag } \left( -1, 1, 1, 1 
\right) \, ,\\
&&\nonumber\\
\overline{g}_{\alpha\beta} & = & \frac{\partial x^{\mu}}{\partial
\overline{x}^{\alpha}} \, \frac{\partial x^{\nu}}{\partial
\overline{x}^{\beta}} \, g_{\mu\nu} \,,
\end{eqnarray}
$ \overline{g}=-1=J^{2}g $ and, therefore,  
\begin{eqnarray}
d\overline{\Omega}=\frac{d\Omega  }{J}=\sqrt{-g} \, d\Omega \,.
\end{eqnarray} 
Since we want the Euler-Lagrange equations deriving from  
the variational principle to be of second order, the Lagrangian 
must be quadratic in the first order derivatives of
$g_{\mu\nu}$.  These first order derivatives contain the 
Christoffel symbols, which are not coordinate-invariant. Then we 
have to choose for the Lagrangian density  ${\cal L}$  
expressions containing higher order derivatives and, {\em 
a priori}, this brings the danger  that the field equations 
could become  of order higher than second (we will discuss 
in detail this point for  ETGs). The obvious choice of Hilbert 
and Einstein  
for the Lagrangian density ${\cal L}$ was the Ricci scalar
curvature $R$. The variational principle is then
\begin{eqnarray} \label{urcailprincipiodiHilbert-Einstein!}
\delta \int d\Omega \, \sqrt{-g} \, R=0 \,. 
\end{eqnarray}
The relations  
\begin{eqnarray}
\delta g =g \, g^{\mu\nu} \delta g_{\mu\nu} = - g \, 
g_{\mu\nu} \delta g^{\mu\nu}
\end{eqnarray}
yield 
\begin{eqnarray}
\delta \left( \sqrt{-g} \, \right)=-\frac{\delta g }{2\sqrt{-g}}
=-\frac{1}{2}\sqrt{-g} \, g_{\alpha\beta}\delta 
g^{\alpha\beta} \,, 
\end{eqnarray}
from which it follows that 
\begin{eqnarray}
\lefteqn { \int\left[\left(\delta \sqrt{-g}\right)R + \sqrt{-g} 
\, R_{\mu\nu} \delta g^{\mu\nu} + \sqrt{-g} \, g^{\mu\nu} \delta 
R_{\mu\nu}\right]d\Omega={}} \nonumber\\
&& \nonumber\\
 & & {}=\int\sqrt{-g} \, \delta g^{\mu\nu}{\textstyle 
\left[R_{\mu\nu}-\frac{1}{2}Rg_{\mu\nu}\right]}d\Omega+
 \int\sqrt{-g} \, g^{\mu\nu}\delta R_{\mu\nu}d\Omega=0 \,. {} 
\label{urca253} 
\end{eqnarray}
The second integral can be evaluated in the local inertial 
frame, obtaining
\begin{eqnarray}
&& R_{\mu\nu}(0) =  G_{\mu\nu}^{\alpha}\hspace{0.01pt}_{,\alpha}-
  G_{\mu\alpha}^{\alpha}\hspace{0.01pt}_{,\nu} \,,\\
&&\nonumber\\
&& \delta  R_{\mu\nu}(0)=\frac{\partial}{\partial x^{\alpha}} \left( 
\delta   G_{\mu\nu}^{\alpha} \right) -\frac{\partial}{\partial x^{\nu}} \left( 
\delta
G_{\mu\alpha}^{\alpha} \right) \, ,\\
&&\nonumber\\
&& g^{\mu\nu}(0) \delta 
 R_{\mu\nu}(0)=g^{\mu\nu}(0) \, \frac{\partial}{\partial x^{\alpha}} \left( 
\delta  G_{\mu\nu}^{\alpha} \right) -g^{\mu\nu}(0) 
\, \frac{\partial}{\partial x^{\nu}} \left( \delta  G_{\mu\alpha}^{\alpha} 
\right) \nonumber\\
&&\nonumber\\
&& =g^{\mu\nu}(0) \, \frac{\partial}{\partial x^{\rho}} \left( \delta 
  G_{\mu\nu}^{\rho} \right) -g^{\mu\rho}(0) \, \frac{\partial}{\partial 
x^{\rho}} \left( \delta  G_{\mu\alpha}^{\alpha} \right) \nonumber\\
&&\nonumber\\
&& = \frac{\partial}{\partial x^{\rho}}\left[ g^{\mu\nu}(0) \delta  
G_{\mu\nu}^{\rho}-g^{\mu\rho}(0) \delta
    G_{\mu\alpha}^{\alpha}\right] \, .
\end{eqnarray}
Then, we can write  
\begin{eqnarray}
g^{\mu\nu}(0)
\delta R_{\mu\nu}(0)=\frac{\partial W^{\rho}}{\partial x^{\rho}} \,, 
\;\;\;\;\;\;
W^{\rho}=g^{\mu\nu}(0) \delta  G_{\mu\nu}^{\rho}
-g^{\mu\rho}(0)\delta G_{\mu\alpha}^{\alpha} \,. 
\end{eqnarray}
The second integral in Eq.~(\ref{urca253}) can be discarded since  
its
argument is a pure divergence; in 
fact, in general coordinates it is
\begin{eqnarray}
 \int\sqrt{-g} \, g^{\mu\nu} \delta R_{\mu\nu}  
d\Omega = \int\sqrt{-g} \, \, \frac{\partial
W^{\rho}}{\partial x^{\rho}}\, d\Omega 
=\int\sqrt{-g} \,\, W^{\rho}\hspace{0.01pt}_{;\rho}d\Omega 
=\int \frac{\partial }{\partial  x^{\rho}} 
\left( \sqrt{-g} \, W^{\rho}\right) d\Omega=0\,,
\end{eqnarray}
and then
\begin{eqnarray}
\int\sqrt{-g} \, 
\delta g^{\mu\nu} \left( 
R_{\mu\nu}-\frac{1}{2}Rg_{\mu\nu}\right) d\Omega=0 \,,
\end{eqnarray}
from which we obtain the vacuum field equations  of GR 
\begin{eqnarray} \label{urcaleequazionidicamponelvuoto!}
G_{\mu\nu} \equiv R_{\mu\nu}-\frac{1}{2} \, g_{\mu\nu}R=0\,,  
\end{eqnarray}
as Euler-Lagrange equations of the Hilbert-Einstein action. 
{\em Vice-versa},  
starting 
from Eq.~(\ref{urcaleequazionidicamponelvuoto!})  and  retracing 
the previous steps in inverse order ({\em i.e.}, integrating the 
Einstein equations), one 
can re-obtain the Hilbert-Einstein 
action~(\ref{urcailprincipiodiHilbert-Einstein!}), thus  
demonstrating  the equivalence between this action and the field 
equations~(\ref{urcaleequazionidicamponelvuoto!}). Introducing 
matter fields as sources is straightforward, once we introduce suitable forms of stress-energy tensor. A suitable choice is the stress-energy tensor of a perfect fluid, that is 

\begin{equation}
T^{(m)}_{\mu\nu} = \left( P + \rho \right) u_{\mu}u_{\nu} +  P\,  
g_{\mu\nu} \,,
\end{equation}
where $u^{\mu} $ is the four-velocity of the fluid particles and  
$P$ and $\rho$ are the pressure and  energy density of the 
fluid, respectively,  the continuity equation requires 
$T^{(m)}_{\mu\nu}$ to be 
covariantly constant, {\em i.e.}, to
satisfy the conservation law 
\begin{equation}\label{Introconservation}
\nabla^{\mu} T^{(m)}_{\mu\nu} = 0\,,
\end{equation}
where  $\nabla_{\alpha}$ denotes the covariant derivative 
operator 
of the metric $g_{\mu\nu}$. In fact,  $\nabla^{\mu} R_{\mu\nu}$ 
does not vanish, except in the special case  $R \equiv 
0$. Einstein and Hilbert  independently concluded  
that the field equations had 
to be 
\begin{equation}\label{Introfield}
G_{\mu\nu} = \kappa \, T^{(m)}_{\mu\nu}\,,
\end{equation}
where
$G_{\mu\nu}$ is now called the {\em Einstein tensor} of $g_{\mu\nu}$.
These equations 
satisfy the conservation law (\ref{Introconservation}) since the  
relation 
\beq
\nabla^{\mu} G_{\mu\nu} = 0\,,
\eeq 
holds as a contraction  of the  Bianchi identities that the 
curvature tensor of $g_{\mu\nu}$ has to satisfy 
\cite{Weinberg72}.

The Lagrangian that, when varied, produces the field equations
(\ref{Introfield}) is the sum of a "matter'' Lagrangian  
density ${\cal L}^{(m)}$,  the variational derivative of which is 
\beq 
T^{(m)}_{\mu\nu} = -\, \frac{2}{\sqrt{-g}} \, \frac{\delta  
\left( \sqrt{-g} \, {\cal  L}_{(m)} \right)}{\delta g^{\mu\nu}} ,
\eeq 
and of the above  gravitational  ({\em Hilbert-Einstein}) {\em 
Lagrangian} 
\beq \label{IntroHE} 
L_{HE}\equiv \sqrt{-g} \, {\cal L}_{HE} = \sqrt{-g} \, R \,,
\eeq 
where
$g$ is the determinant of the metric $g_{\mu\nu}$. Solving the Einstein field equations means to determine the form of the metric $g_{\mu\nu}$ starting from the distribution of matter-energy in a given region of space. In other words, this achievement fixes the (causal) metric  structure of the space-time.
 This is the derivation of the Einstein field equations in the so called metric  formalism.

 However it is possible to obtain the field equations without resorting to a local inertial frame \cite{gravitation}. Following the Weyl attempt to 
 unify Gravity with Electromagnetism  \cite{weyl}, as discussed in Sec.\ref{due},  
 an affine connection defined over the space-time,  can be assumed  as a dynamical field
non-trivially depending on a metric.  As we have seen, Weyl's idea failed because of a wrong choice
of the Lagrangian and few more issues, but it generated however a keypoint: connections may have a
physically interesting dynamics.

Einstein  soon showed a great interest in Weyl's idea.  He too
began to play with connections, in order to seek for a
``geometrically'' Unified Theory.  But he never arrived to
``dethronize'' $g$ in the description of the gravitational field.
He was not so happy with the fact that
the gravitational field is not the fundamental object, but just a
by-product of the metric; however, he never really changed his
mind about the physical and mathematical role of $g$.

In 1925 Einstein constructed a theory depending on a metric $g$
and a symmetric affine connection $\Gamma$ to be varied
independently (the so-called "Palatini method", because of a
misunderstanding with W.\ Pauli,  see \cite{frafe,palatini}); he defined a
Lagrangian theory in which the  Palatini-Einstein
gravitational Lagrangian is

\begin{equation}
 \sqrt{-g} \mathcal{L}_{\rm PE}(g,\Gamma,\partial\Gamma):\equiv \stackrel{\Gamma}R \,\sqrt{-g}\,.
 \label{5p}
\end{equation}
 The Lagrangian density $\mathcal{L}_{\rm PE}$
  contains (first order) derivatives of $\Gamma$ but no derivatives of $g$. In other words,  it is
  of order zero in the metric while it is first order in the connection.
  By a well-known
  feature of the Lagrangian formalism, this implies that {\it a
  priori} $g$ has no independent dynamics, being a sort of "Lagrange multiplier"; if we would assume that
  $\Gamma$ is frozen from the beginning, $g$ would be frozen too.
  On the other hand since $g$ has dynamics,  this happens because $g$ is related to $\Gamma$, which is the true dynamical
  variable of this Lagrangian theory. This time is the metric
$g$ to gain a dynamical meaning from $\Gamma$, that plays the role of a fundamental field.

 In $4$ dimension, there are $10 + 40$ independent variables and the field
equations, in vacuum, are:

\begin{eqnarray}
\left\{
\begin{array}{rl}
   \stackrel{\Gamma}R_{(\mu\nu)}-\frac{1}{2}\stackrel{\Gamma}R g_{\mu\nu}=0\\
    \\
   \stackrel{\Gamma} \nabla_\alpha(\sqrt{g}\,g^{\mu\nu})=0
   \end{array}
   \right.
\label{6p}%
\end{eqnarray}
where $  \stackrel{\Gamma}R_{(\mu\nu)}$ is the symmetric part of $  \stackrel{\Gamma}R_{\mu\nu}$ and $\nabla{\Gamma}$
denotes the covariant derivative with respect to $\Gamma$.
 The second field equation (\ref{6p})
constrains the connection $\Gamma$, which is {\it a priori} arbitrary, to coincide {\it a posteriori}
with the Levi-Civita connection of the metric $g$ (Levi-Civita Theorem).  By substituting this
information into the first field equation (\ref{6p}), the vacuum Einstein equation for $g$ is
 obtained.  In the  "Palatini formalism",  the metric $g$ determines rods, clocks and
protractors, while the connection $\Gamma$ determines the free-fall, but since {\it a posteriori} the
same result of GR is found, Einstein soon ceased to show a real interest in this formalism.

The situation does not change if matter is present through a matter
Lagrangian $\mathcal{L}_{(m)}$ (independent of $\Gamma$ but just depending on $g$ and other external matter
fields), that generates the above energy-momentum tensor $T_{\mu\nu}^{(m)}$.  If
the total Lagrangian is then assumed to be $\mathcal{L}_{\rm tot}:\equiv \mathcal{L}_{\rm PE}+\mathcal{L}_{(m)}$, field equations
(\ref{6p}) are replaced by
\begin{equation}
  \stackrel{\Gamma}R_{(\mu\nu)}-\frac{1}{2}\stackrel{\Gamma}Rg_{\mu\nu}=\kappa\,T_{\mu\nu}
  \label{7p}
  \end{equation}
and again (\ref{6p}) implies,  {\it a posteriori}, that (\ref{7p})
reduces  to the Einstein equations.

  Let us remark that the dynamical coincidence between $\Gamma$ and the Levi-Civita connection of $g$
  is entirely due to the particular Lagrangian considered by Einstein, which is the {\it simplest},
  but not the only possible one. Furthermore, the Palatini method privileges the affine structure with respect to the metric structure.
 Notice that, in this case ({\it i.e.} in "Palatini formalism''), the
relations
\begin{equation}
\Gamma^\lambda_{\mu\nu}=\left\{^\lambda_{\mu\nu}\right\}_g
\label{8p}
\end{equation}
are field equations: the fact that $\Gamma$ is the Levi-Civita
connection of $g$ is no longer an assumption {\it a priori} but
it is the outcome of the field equations  \cite{gravitation}.
As  soon as the Lagrangian is not the Palatini-Einstein  but a more general one, the affine connection does not coincide, in general,  with the Levi-Civita one.
This fact gives rise to extremely interesting results. Below we will discuss this point considering the curvature quantities derived in the Palatini formalism as different from those derived in the metric approach.

\subsection{The  Newtonian limit of  General Relativity}
GR is not the only theory of gravitation and, several alternative
 theories of gravity have been investigated from the 60's, considering the
 space-time to be ``special relativistic''
 at a background level and treating gravitation as a Lorentz-invariant
 field on the background.
 
 Two different classes of experiments have been studied: the
first ones testing the foundations of gravitation theory -- among
them the Equivalence Principle -- the second one testing the metric theories of
gravity where space-time is endowed with a metric tensor and where
the Einstein Equivalence Principle is valid. However as discussed in the Introduction, for several fundamental  reasons extra fields might be necessary to
describe the gravitation, {\it e.g.} scalar fields or higher-order
corrections in curvature invariants.

  Two sets of
equations can be distinguished \cite{urbino}.
 The first ones couple the gravitational fields to the
 non--gravitational contents of the Universe,
{\it i.e.} the matter distribution, the electromagnetic fields, etc...
 The second set of equations gives the evolution
 of non--gravitational fields. Within the framework of metric
  theories, these laws
 depend only on the metric: this
 is a consequence of the EEP and the so-called  "minimal
 coupling".
  In most theories,
 including GR, the second set of equations is derived from the first
 one,
  which is the only
 fundamental one; however, in many situations, the two sets are decoupled.
 The gravitational field studied in these approaches (without
cosmological considerations) is mainly due to the Sun and the
Eddington-Robertson expansion gives the corresponding metric.
Assuming spherical symmetry and a static gravitational field, one
can prove that there exists a coordinate system such as

\begin{eqnarray}
 d s^2 \;=\; -A(r) \, dt^2 +B(r) \, dr^2 \,+ r^2( \sin^2 \theta \, d\phi^2+ d\theta^2) \;,
\label{eq:isotropic} \end{eqnarray}
 $d t$ being the proper time between two neighbouring
events.  The Newtonian gravitational field does not exceed
$G_N M_{\odot}/R_{\odot} c^2 \sim 2 \times 10^{-6}$, where $c$ is
the speed of light, $M_{\odot}$ is the mass of the Sun and
$R_{\odot}$ its radius. The metric is quasi-Minkowskian, $A(r)$
and $B(r)$ are dimensionless functions which depend only on $G_N$,
$M$, $c$ and $r$. Indeed, the only pure number that can be built
with these four quantities is $G_N M /r c^2$. The
Eddington-Robertson metric is a Taylor expansion of $A$ and $B$
which gives

\begin{eqnarray}
 d s^2 &\simeq& -\left( 1- 2 \, \alpha \, \frac{G_N \, M}{r \, c^2} + 2 \, \beta \,
 \left( \frac{G_N \, M}{r c^2} \right)^2 + ...\right) dt^2 + \left(1+ 2 \, \gamma \, \frac{G_N \, M}{r c^2} + ...\right)\,dr^2 +\,r^2(
     \sin^2 \theta \, d\phi^2+d\theta^2)\,.\nonumber\\
\label{eq:expansion}\end{eqnarray}
which in isotropic coordinates is
\begin{eqnarray}
 d s^2 &\simeq& -\left( 1- 2 \, \alpha \, \frac{G_N \, M}{r \, c^2} + 2 \, \beta \,
 \left( \frac{G_N \, M}{r c^2} \right)^2 + ...\right) dt^2 + \left(1+ 2 \, \gamma \, \frac{G_N \, M}{r c^2} + ...\right)\,\left(dx^2+dy^2+dz^2\right)\,.\nonumber\\
\label{eq:expansion1}\end{eqnarray}
 where $r,\,\theta,\,\phi$ are related to $x\,,y\,,z$ in the usual manner:
\begin{eqnarray}
r=\sqrt{x^2+y^2+z^2}\,,\qquad \theta=tan^{-1}\left(\frac{z}{\sqrt{x^2+y^2+z^2}}\right)\,, \qquad \phi=tan^{-1}\left(\frac{y}{x}\right)
\end{eqnarray}
where $r$ is the new "isotropic" radial coordinate, not to be confused with the Schwarzschild $r$ (see \cite{gravitation} for a detailed discussion of this important issue).
 The coefficients $\alpha$, $\beta$, $\gamma$ are called the post-
Newtonian parameters and their values depend on the considered
theory of gravity: in GR, one has $\alpha=\beta=\gamma=1$. The post
Newtonian parameters can be measured not only in the Solar System
but also in relativistic binary neutron stars such as
$PSR~1913+16$.

A generalization of the previous formalism is
known as {\it Parametrized Post-Newtonian} (PPN) formalism.
Comparing metric theories of gravity among them and with
experimental results becomes particularly easy if the PPN
approximation is used. The following requirements are needed:

\begin{itemize}
 \item particles are moving slowly with respect to the speed of light;
 \item gravitational field is weak and considered as a perturbation of the flat space-time;
 \item gravitational field is also static, {\it i.e.} it does not change with time.
\end{itemize}

 The PPN limit of metric theories of gravity is characterized
by a set of 10 real-valued parameters; each metric theory of
gravity corresponds to particular values of PPN parameters. The
PPN framework has been used first for the analysis of Solar System
gravitational experiments, then for the definition and the
analysis of new tests of gravitation and finally for the
analysis
and the classification of alternative metric theories of gravity \cite{Will93}.

 By the middle 1970's, the Solar System was no more considered
as the unique testing ground of gravitation theories. Many
alternative theories of gravity agree with GR in the
Post-Newtonian limit and thus with Solar System experiments;
nevertheless, they do not agree for other predictions (such as
cosmology, neutron stars, black holes and gravitational radiation)
for which the post-Newtonian limit is not adequate.  In addition,
the possibility that experimental probes, such as gravitational
radiation detectors,
 would be available in the future to perform extra-Solar System tests led to the abandon of the
 Solar System as the only arena to test gravitation theories.

 The study of the binary pulsar $PSR~1913+16$, discovered by R.
Hulse and J. Taylor \cite{HT}, showed that this system combines
large post-Newtonian gravitational effects, highly relativistic
gravitational fields (associated with the pulsar) with the
evidence of an emission of gravitational radiation (by the binary
system itself). Relativistic gravitational effects allowed one to
do accurate measurements of astrophysical parameters of the
system, such as the mass of the two neutron stars. The measurement
of the orbital period rate of change agreed with the prediction of
the gravitational waves (GW) emission in the GR framework, in
contradiction with the predictions of most alternative theories,
even those with PPN limits identical to GR. However, the evidence
was not conclusive to rule out other theories since several
shortcomings remain, up to now, unexplained and , as we see below, other forms of gravitational radiation ({\it e.g.} polarizations and massive states) with respect to those predicted by GR.

\subsection{The Minkowskian limit of General Relativity}
Replacing the Newtonian limit by a less restrictive hypothesis
leads to the weak field approximation: practically, the field is
still weak, but it is allowed to change in time and there is no
more restriction on the test particles motion. New phenomena are
associated with this hypothesis like the emission of gravitational
radiation  and the deflection of light. This framework allows one
to split the metric $g_{\mu\nu}$ into two parts: the flat Minkowski
metric $\eta_{\mu\nu} = \text{diag}(-1,1,1,1)$ plus a perturbative
term $h_{\mu\nu}$, assumed to be small. This linearized version of GR
describes the propagation of a symmetric tensor $h_{\mu\nu}$ on a flat
background space-time. So, the metric reads

\begin{eqnarray}
 g_{\mu\nu} \;=\; \eta_{\mu\nu} + h_{\mu\nu} \qquad \text{with} \qquad \left |h_{\mu\nu}\right| \ll 1 \, .
\label{eq:metr_decomp} \end{eqnarray}

 As $h_{\mu\nu}$ is small, one can neglect terms higher than the first
order in the perturbation $h_{\mu\nu}$; in particular, one can
raise/lower indexes with $\eta_{\mu\nu}$ and $\eta^{\mu\nu}$ as the
corrections are of higher order in the perturbation

\begin{eqnarray}
 g^{\mu\nu} \;=\; \eta^{\mu\nu} - h_{\mu\nu} \qquad \text{with} \qquad h^{\mu\nu} \;=\; \eta^{\mu\rho}\eta^{\nu\sigma} h_{\rho\sigma} \; .
 \end{eqnarray}

 The aim is to find the equations of motion to which the
perturbations $h_{\mu\nu}$ obey by investigating the Einstein equations
to the first order. Inserting the new metric
(\ref{eq:metr_decomp}) in the Einstein tensor, we obtain \begin{eqnarray}
 G_{\mu\nu} \;=\; \frac{1}{2} \left( \partial_{\sigma} \partial_{\nu} h^{\sigma}_{\mu} + \partial_{\sigma} \partial_{\mu} h^{\sigma}_{\nu} -
 \partial_{\mu} \partial_{\nu} h - \
 h_{\mu\nu}- \eta_{\mu\nu} \, \partial_{\rho} \partial_{\sigma} h^{\rho\sigma} + \eta_{\mu\nu} h \right)\,,
 \label{eq:new_tensor_Eins}
 \end{eqnarray}

 where $h = \eta^{\mu\nu} h_{\mu\nu} = h^\mu_\mu$ is the trace of the
perturbation and $ =  -\frac{1}{2} \partial_{tt} + \partial_{xx} + \partial_{yy} +
\partial_{zz}$ is the d'Alembertian of the
flat space-time, using from now on (unless  otherwise stated) geometrical units for which $c=1$.\\
 The stress-energy tensor is computed at the 0-order in $h_{\mu\nu}$:
the energy and the momentum have to be small too, according to the
weak field approximation and the lowest non-vanishing order in
$T_{\mu\nu}$ is of the same order of magnitude as the perturbation.
Therefore, the conservation law becomes $\partial^\mu T_{\mu\nu} = 0$.

\subsubsection{Gravitational Waves} 

 GW are weak ripples in the curvature of space-time, produced by the
 motions of matter.
 They propagate at the speed of light. The linearized
 Einstein equations allow wave solutions, in a way similar to Electromagnetism.
 These GW are transverse to the propagation direction and show two independent polarization states.

The new metric (\ref{eq:metr_decomp}) does not fix the
space-time frame completely; two possible gauges can be applied in
addition to simplify the Einstein equations. Using the Lorentz
gauge $\partial_\mu h^\mu_\lambda -\frac{1}{2} \partial_\lambda h=0$, the Einstein
 are linearized and can be written as

\begin{eqnarray}
 \Box \, h_{\mu\nu} - \frac{1}{2} \, \eta_{\mu\nu} \, \Box \, h \;=\; \frac{16\pi G_N}{c^4} \, T_{\mu\nu}^{(m)} \; .
 \label{eq:general_GW}
\end{eqnarray}
where standard units have been restored.
The trace-reversed perturbation is defined as

\begin{eqnarray}
 \bar{h}_{\mu\nu} \;=\; h_{\mu\nu} - \frac{1}{2} \, \eta_{\mu\nu} \, h \; .
 \label{eq:trace_renv}
\end{eqnarray}

One can choose a frame in which the harmonic gauge condition,
$\partial_\mu \bar{h}^\mu_\nu = 0,$ is verified. Then, the Einstein
field equations become

\begin{eqnarray}
 \Box \bar{h}_{\mu\nu} \;=\;  \frac{16\,\pi \,G_N}{c^4} \, T_{\mu\nu}^{(m)} \; .
 \label{eq:GW_eq}
\end{eqnarray}

 and, in vacuum, one has simply

\begin{eqnarray}
  \Box \bar{h}_{\mu\nu} \;=\; 0 \; .
 \label{eq:vacuum}
\end{eqnarray}

These equations are similar to of the Electromagnetism field
equations and one can use the same method to solve them. Indeed,
looking at the Einstein equations in vacuum,
Eq.~(\ref{eq:vacuum}), one can note that they are in the form
of a wave equation for $\bar{h}_{\mu\nu}$, the d'Alembertian
reduces to the form $\Box = - \frac{1}{2} \partial_{tt} + \nabla^2$.
Therefore, in the absence of matter, one looks for plane waves
solutions
\begin{eqnarray}
 \bar{h}_{\mu\nu} \;=\; C_{\mu\nu} e^{ik_\sigma x^{\sigma}}\,,
 \label{eq:vacuum_sol}
\end{eqnarray}

where $C_{\mu\nu}$ is a constant and symmetric tensor of rank 2 and $k$
is a time-like vector, the wave vector. The plane wave in Eq.~(\ref{eq:vacuum_sol})
is a solution of the linearized equations in vacuum, Eq.~(\ref{eq:vacuum}), if the
wave vector $k$ is null, {\it i.e.} satisfies the condition $k^\sigma k_\sigma = 0$ and shows that
GW propagate to the speed of light.

 The four conditions of the harmonic gauge $k_\mu C^{\mu\nu} = 0$
lead to six independent components for the symmetric tensor
$C^{\mu\nu}$. As there are still some unused degrees of freedom, one
can make another gauge choice on the tensor $C_{\mu\nu}$:

\begin{eqnarray}
 C^\mu_\mu &=& 0 \quad \text{Traceless}  \, ;\\
 C^{0\mu}  &=& 0 \quad \text{Transverse} \, .
\end{eqnarray}

 One has, in this way, the so called Transverse-Traceless (TT)
gauge. These five relations give four new constraints on
$C^{\mu\nu}$ in addition to the harmonic gauge condition; therefore,
only two independent components remain in $C^{\mu\nu}$. As the wave
is traceless, one can check
from Eq.~(\ref{eq:trace_renv}) that $\bar{h}_{\mu\nu}^{TT} = h_{\mu\nu}^{TT}$.
 Therefore, the general form of the symmetric tensor $C_{\mu\nu}$
is finally

\begin{eqnarray}
 C_{\mu\nu} =\left(
 \begin{array}{cccc}
  0 & 0 & 0 & 0\\
  0 & C_{11} & C_{12} & 0\\
  0 & C_{12} & - \, C_{11} & 0\\
  0 & 0 & 0 & 0\\
 \end{array}
 \right)
\end{eqnarray}

 Let us define $C_{11} = h_+$ and $C_{12} = h_\times$; GW
appears to have two polarized states, "$+$'' and "$\times$'', 
which modify the space-time curvature in  different ways. In
tensorial form, one can write

\begin{eqnarray}
 h \;=\; \left[ h_+ \left( \vec{e_1} \otimes \vec{e_1} - \vec{e_2} \otimes \vec{e_2} \right)
 + 2 \, h_\times \left( \vec{e_1} \otimes \vec{e_2} \right) \right] \, e^{i\omega( t - x / c )} \; .
\end{eqnarray}

  Being $\vec{\xi}=(\xi^1,\xi^2,\xi^3)$  the separation between
two free particles and taking into account the geodesic deviation
\cite{Wald84} which describes the evolution of two free-falling
particles, if the GW propagates in the direction $x^3$,  only
$\xi^1$ and $\xi^2$ are involved in its passage.
 Assuming a  polarized GW, the integration of
the geodesic deviation equation gives:

\begin{itemize}
 \item Polarization "+'' ($h_\times = 0$)~:
  \beq\left(
 \begin{array}{c}
  \xi^1 ( t )\\
  \xi^2 ( t )
 \end{array}
 \right)
  =\left(
 \begin{array}{cc}
  1 + \frac{1}{2} \, h_+e^{i k^\sigma x_\sigma} & 0 \\
  0 & 1 - \frac{1}{2} \, h_+e^{i k^\sigma x_\sigma}
 \end{array}
 \right)\left(
 \begin{array}{c}
  \xi^1 ( 0 )\\
  \xi^2 ( 0 )
 \end{array}
 \right)
  \eeq
 \item Polarization "$\times$'' ($h_+ = 0$)~:
  \beq
  \left(
 \begin{array}{c}
  \xi^1 ( t )\\
  \xi^2 ( t )
 \end{array}
 \right)
  =\left(
\begin{array}{cc}
 1 & \frac{1}{2} \, h_\times e^{i k^\sigma x_\sigma} \\
\frac{1}{2} \, h_\times e^{i k^\sigma x_\sigma} & 1
\end{array}\right)
\left(\begin{array}{cc}
\xi^1 ( 0 )\\
\xi^2 ( 0 )
\end{array}\right)
 \eeq
\end{itemize}

 Let us consider now a test-mass ring (a massless and
free--falling set of particles) interacting with GW, lying in an
plane orthogonal to the direction of the wave propagation. Its
oscillations depend on the GW polarization.

 After having found a solution to Einstein equations in vacuum,
let us solve now Eq.~(\ref{eq:GW_eq}) with a non-zero source term.
The solution is computed using the retarded Green function

\begin{eqnarray}
 \bar{h}_{\mu\nu} \left( t, \vec{x} \right) \; = \; \frac{4 \, G_N}{c^4} \int_{\vec{y} \,
 \epsilon \, \text{Source}} \frac{1}{\left|\, \vec{x} \, - \, \vec{y} \, \right|} \, T_{\mu\nu}^{(m)} \,
 \left( t_r, \vec{y} \right) d^3\vec{y}\,,
\label{eq:pot_retard} \end{eqnarray}

with $\left| \, \vec{x}-\vec{y} \, \right| = \sqrt{\delta_{ij}(x^i-y^i)(x^j-y^j)}$
(Euclidean distance) and $t_r = t - \left| \, \vec{x}-\vec{y} \, \right| / c$ (retarded time).
 Let us consider an isolated source with a density $\rho$ and a
characteristic dimension $\delta R$, located at a distance $R$ from
 the observation point $\vec{x}$. One assumes $\delta R \ll R$ so, in particular,
 one gets $\left| \, \vec{x}-\vec{y} \, \right| \approx R$ and one can move this constant term outside
 the integral in Eq.~(\ref{eq:pot_retard}). As the stress-energy tensor
 verifies the conservation of energy $\partial^\mu T_{\mu\nu}^{(m)} = 0$, the harmonic
 gauge condition $\partial_\mu \bar{h}^\mu_\nu = 0$ is also verified. Moreover,
 the radiation is mostly emitted at frequencies $\omega / 2\pi$, so that
 ${\displaystyle \frac{\delta R}{ c} \ll \frac{1}{\omega}}$. Then, it is
 possible to demonstrate that only the spatial coordinates
 of  tensor $\bar{h}_{\mu\nu}$
 are different from zero.
The quadrupole momentum tensor $q_{ij}$ of the source energy
density is defined as

\begin{eqnarray}
 q_{ij}(t) \;=\; \int_{\vec{y} \, \epsilon \,
 \text{source}} \, y_i \, y_j \, T_{00}^{(m)}( t, \vec{y} ) \, d^3\vec{y} \qquad
 \text{with } \qquad T_{00}^{(m)} \;\approx\; \rho c^2 \; .
\end{eqnarray}

The metric perturbation is given by

\begin{eqnarray}
 \bar{h}_{ij}(t,{\bf x}) = \frac{2 \, G_N}{R \, c^4} \, \ddot{q}_{ij} (t_r) \; .
\label{eq:h_bar} \end{eqnarray}

So, GW, generated by an isolated non-relativistic object, are
proportional to
 the second derivative of the quadrupolar momenta of the energy density. Eq.~(\ref{eq:h_bar})
 shows that the metric perturbation amplitude $h$ varies as
 the inverse of  distance to the source $R$; a faster decreasing with the distance, {\it e.g.} $1/R^2$,
 would make vain the hope of any GW detection. Fortunately,
 GW detectors are sensitive to the amplitude of the signal.

The energy emitted by gravitational radiation is difficult to
define. A way to overcome this difficulty is to define the
stress-energy tensor by developing the metric $g_{\mu\nu}$ and the
Einstein tensor $G_{\mu\nu}$ at the second order:

\begin{eqnarray}
 g_{\mu\nu} &=& \eta_{\mu\nu} + h_{\mu\nu} + h^{(2)}_{\mu\nu}         \; ,             \\
 \nonumber\\
 G_{\mu\nu} &=& G^{(1)}_{\mu\nu} \; [\eta + h^{(2)}] + G^{(2)}_{\mu\nu} \; [\eta + h] \; .
\end{eqnarray}

 Einstein equations in the vacuum $G_{\mu\nu} = 0$ can be written in
the form

\begin{eqnarray}
 G^{(1)}_{\mu\nu} \; [\eta + h^{(2)}] \; = \; \frac{8 \, \pi \, G_N}{c^4} \; t_{\mu\nu}\,,
\end{eqnarray}

with the definition

\begin{eqnarray}
 t_{\mu\nu} \; = \;  \frac{c^4}{8 \, \pi \, G_N} \; G^{(2)}_{\mu\nu} \; [\eta + h] \; .
\end{eqnarray}

 The Bianchi identity says that $\partial_\mu t^{\mu\nu} = 0$,
therefore $t_{\mu\nu}$ can be considered as the stress-energy tensor
of a gravitational field, yet, it is only a pseudo-tensor
\cite{LandauLifschitz}. One can compute the energy density $t_{00}$ by
averaging over many cycles (because the energy is not localized)
the GW energy:

\begin{eqnarray}
 t_{00} \; = \; \frac{c^2}{16 \, \pi \, G} \; \langle \; \dot{h_+}^2 + \dot{h_\times}^2 \; \rangle \; .
\end{eqnarray}

 Then, the source luminosity is ${\displaystyle L = r^2
\int_{\text{S}} c \, t_{00} \, d\Omega}$, where the integration is performed on sphere $\text{S}$ of radius $r$.
Introducing the reduced quadrupolar momenta

\begin{eqnarray}
 Q_{ij} \; = \; q_{ij} - \frac{1}{3} \delta_{ij} \delta^{kl} q_{kl}\,,
\end{eqnarray}

one obtains the  Einstein quadrupole formula

\begin{eqnarray}
 L \; = \; \frac{G}{c^5} \; \langle \; \frac{d^3 Q_{ij}}{dt^3} \frac{d^3 Q_{ij}}{dt^3} \; \rangle \; .
\label{eq:quadrupole} \end{eqnarray}

 It is worth noticing that a GW emission requires a variation
of the quadrupolar momentum, as shown already in
Eq.~(\ref{eq:h_bar}). This typical feature of  gravitational
radiation, added to the weakness of the coupling constant between
gravitation and matter, here
$G_N/c^5~\approx~10^{-53}~\text{W}^{-1}$, explains why GW
amplitudes are so small compared to those produced by
electromagnetic radiation. In conclusion, the gravitational
radiation is quadrupolar and a symmetric spherical body does not
emit GWs because its reduced quadrupolar momenta are zero.

The corresponding quantum field is the graviton with zero mass and
spin $2$.
 Within the framework of more general theories, the gravitation can be described
 as a combination of states with a spin $2$ and spin zero or as a
 particle with spin maximum $2$. One can also imagine that the mass
 of graviton is not zero or that the state mass spectrum is complex.
 Presently, there is no observational reason to doubt that
 the present observational bounds on the mass of the graviton are
 severe.

 Discovering GWs would open {\it "a new window onto the
Universe''}  \cite{thorne}. It is clear that being sensitive to an
additional radiation would lead to major discoveries like when the
Universe became observed through radio, X or gamma waves. Then, it
would allow physicists to test GR in the strong field regime, to
check the gravity velocity (assumed to be the speed of light in
the Einstein theory) or to verify that GWs only change distances
perpendicular to their direction of propagation.

 Alternative
relativistic theories of gravity also predict the existence of GWs.
However, many essential features of the radiation are different:
the number of polarization states, the propagation speed, the
efficiency of wave generation, the possible existence of massive states \cite{thorne,maggiorebook}.

In this section, we have outlined, with no claim to completeness, some physical issues which constitute the test-bed of GR and any alternative relativistic theory of gravity. As we shall see below, ETGs enlarge the possibilities of viable theories in the track of GR with the aim to address, partially, some of the open questions at infrared and ultra-violet scales.

\newpage

\section*{\bf{Part II: Extended Theories of Gravity}}

\section{Introduction}

As widely discussed above, due to the problems of Standard Cosmological Model, and, first of all, to the lack of a definitive Quantum Gravity Theory, alternative
theories have been considered in order to attempt, at least, a semi-classical scheme where GR and its positive results could be recovered.
 One of the most
fruitful approaches is that of {\it Extended Theories of
Gravity} (ETGs) which have become a sort of paradigm in the study
of gravitational interaction. They are based on corrections and
enlargements of the  Einstein theory. The paradigm consists,
essentially,  in adding higher-order curvature invariants and
minimally or non-minimally coupled scalar fields into dynamics
which come out from  the effective action of Quantum Gravity
\cite{CF1,mauro}.  This approach is coherent to the fact that these generalized theories emerge, like Einstein's gravity, from the Gauge Theory, as we have seen above, and can be framed in a bundle structure as we will show below.

Other motivations to modify GR, as discussed, come from the issue of a full recovering of the Mach Principle which leads to assume a varying
gravitational coupling. This principle states that the local
inertial frame is determined by some average of the motion of
distant astronomical objects \cite{bondi}. This fact implies that
the gravitational coupling can be scale-dependent and related to
some scalar field. As a consequence,  the concept of "inertia''
and the Equivalence Principle have to be revised.

Besides, every unification scheme as Superstrings, Supergravity or
Grand Unified Theories, takes into account effective actions where
non-minimal couplings to the geometry or higher-order terms in the
curvature invariants are present. Such contributions are due to
one-loop or higher-loop corrections in the high-curvature regimes
near the full (not yet available) Quantum Gravity regime
\cite{CF1}. Specifically, this scheme was adopted in order to
deal with the quantization on curved space-times and the result was
that the interactions among quantum scalar fields and background
geometry or the gravitational self-interactions yield corrective
terms in the Hilbert-Einstein Lagrangian \cite{BirrellDavies}. Moreover,
it has been realized that such corrective terms are inescapable in
order to obtain the effective action of Quantum Gravity at scales
closed to the Planck one \cite{quant3}. All these approaches
are not the "{\it full quantum gravity}" but are needed as
working schemes toward it. In the next subsection, we will discuss the Quantum Field Theory approach in curved space-time and the emergence of curvature corrections.

In summary, higher-order terms in curvature invariants (such as
$R^{2}$, $R_{\mu\nu} R^{\mu\nu}$, 
$R^{\mu\nu\alpha\beta}R_{\mu\nu\alpha\beta}$, $R \,\Box R$, or $R
\,\Box^{k}R$) or non-minimally coupled terms between scalar fields
and geometry (such as $\phi^{2}R$) have to be added to the effective
Lagrangian of gravitational field when quantum corrections are
considered. For instance, one can notice that such terms occur in
the effective Lagrangian of strings or in Kaluza-Klein theories,
when the mechanism of dimensional reduction is used
\cite{gaspven}.

On the other hand, from a conceptual point of view, there are no {\it
a priori} reason to restrict the gravitational Lagrangian to a
linear function of the Ricci scalar $R$, minimally coupled with
matter \cite{mauro}. Furthermore, the idea that there are
no "exact'' laws of physics could be taken into serious account:
in such a case, the effective Lagrangians of physical interactions
are "stochastic'' functions. This feature means that the local
gauge invariances ({\it i.e.} conservation laws) are well approximated
only in the low energy limit and the fundamental physical
constants can vary \cite{BarrowOttewill83}.

Beside fundamental physics motivations, all these theories have
acquired a huge interest in cosmology due to the fact that they
"naturally" exhibit inflationary behaviours able to overcome the
shortcomings of  Cosmological Standard Model (based on GR). The
related cosmological models seem  realistic and capable of
matching with the Cosmic Microwave Background Radiation observations \cite{starobinsky,kerner,la}.
Furthermore, it is possible to show that, via conformal
transformations, the higher-order and non-minimally coupled terms
always correspond to the Einstein gravity plus one or more than
one minimally coupled scalar fields
\cite{TeyssandierTourrenc83,Maeda89,Wands94,wands,Gottetal90,adams}.

More precisely, higher-order terms appear always  as contributions
of order two in the field equations in metric formalism. For example, a term like
$R^{2}$ gives fourth order equations \cite{ruzmaikin}, $R \ \Box
R$ gives sixth order equations \cite{Gottetal90,sixth,adams,Buchdahl,berkin}, $R
\,\Box^{2}R$ gives eighth order equations \cite{eight} and so on.
By a conformal transformation, any 2nd-order  derivative term
corresponds to a scalar field\footnote{The dynamics of such scalar
fields is usually given by the corresponding Klein-Gordon
Equation, which is  second order.}: for example, fourth-order
gravity gives Einstein plus one scalar field, sixth-order gravity
gives Einstein plus two scalar fields and so on
\cite{Gottetal90,schmidtNL}.

Considering a mathematical point of view, the problem of reducing more
general theories to Einstein standard form has been extensively
treated; one can see that, through a "Legendre'' transformation, higher-order theories, under suitable regularity
conditions on the Lagrangian, take the form of the Einstein one in
which a scalar field (or more than one) is the source of the
gravitational field (see for example
\cite{mauro,sokolowski,ordsup,MagnanoSokolowski94}); on the other
side, as discussed above, it has been studied the mathematical
equivalence between models with variable gravitational coupling
with the Einstein gravity through suitable conformal
transformations (see \cite{dicke,Dicke62,DamourFarese92,damour1}).

In any case, the debate on the physical meaning of conformal
transformations is far to be solved [see \cite{Faraoni,faraoni-noi} and
references therein for a comprehensive review]. Several authors
claim for a true physical difference between Jordan frame
(higher-order theories and/or variable gravitational coupling)
since there are experimental and observational evidences which
point out that the Jordan frame could be suitable to better match
solutions with data. Others state that the true physical frame is
the Einstein one according to the energy theorems
\cite{MagnanoSokolowski94}. However, the discussion is open and no
definitive statement has been formulated up to now.

 The problem should be faced from a more general point of view and the
Palatini approach to gravity, introduced in Sec.\ref{FEQ}, could be useful to this goal \cite{palaeinstein, buchdahl,frafe}. 
In \cite{olmoreview}, this approach is widely discussed for ETGs and several important applications are reported.

The fundamental idea  of the Palatini formalism, as we have seen in Sec. V, is to consider the
 connection $\Gamma$, entering the
definition of the Ricci tensor, to  be independent of the metric
$g$ defined on the space-time ${\cal M}$. The Palatini formulation
for the standard Hilbert-Einstein  theory results to be equivalent
to the purely metric theory: this follows from the fact that the
field equations for the connection  $\Gamma$, firstly considered
to be independent of the metric,  give the Levi-Civita connection
of the metric $g$. As a consequence, there is  no reason to impose
the Palatini variational principle in the standard
Hilbert-Einstein theory instead of the metric  variational
principle.

However, the situation completely changes if we consider the ETGs,
depending on functions of  curvature invariants, as $f(R)$, or
non-minimally coupled  to some scalar field. In these cases, the
Palatini and the metric variational principle provide different
field equations and the theories thus derived differ
\cite{MagnanoSokolowski94,Ferraris:1994af}. The relevance of  Palatini approach, in
this framework, has been recently proven in relation to
cosmological applications
\cite{Capozziello02IJMPD,NojiriOdintsov03,CF1,Vollick:2003,Li:2006vi,Li:2006ag,olmoreview,silvio}.

It has also been studied the crucial problem of the Newtonian
potential in alternative theories of gravity and its relations
with the conformal factor \cite{meng_rev}. 
 In \cite{olmoreview} interesting revision of the literature on the Newtonian limit of Palatini theories is provided. From a physical
point of view, considering the metric $g$ and the connection $\Gamma$
as independent fields means to decouple the metric structure of
space-time and its geodesic structure (being, in general, the
connection $\Gamma$ not the Levi-Civita connection of $g$). The
chronological structure of space-time is governed by $g$ while the
trajectories of particles, moving in the space-time, are governed
by $\Gamma$.

This decoupling enriches the geometric structure of space-time and
generalizes the purely metric formalism. This metric-affine
structure of space-time  is naturally translated, by means of the
same (Palatini) field equations, into a bi-metric structure of
space-time. Beside the \textit{physical} metric $g$, another metric
$h$ is involved. This new metric is related, in the case of
$f(R)$-gravity, to the connection. As a matter of fact, the
connection $\Gamma$ results to be the Levi-Civita connection of
$h$ and thus provides the geodesic structure of space-time \cite{allemandi}.

If we consider the case of non-minimally coupled interaction in
the gravitational Lagrangian (scalar-tensor theories), the new
metric $h$ is  related to the non-minimal coupling. The new metric
$h$ can be thus related to a different geometric and physical
aspect of the gravitational theory. Thanks to the Palatini
formalism, the non-minimal coupling and the scalar field, entering
the evolution of the gravitational fields, are separated from the
metric structure of space-time. The situation mixes when we
consider the case of higher-order-scalar-tensor theories. Due to
these features, the Palatini approach could greatly contribute to
clarify the physical meaning of conformal transformation
\cite{allemandi}. A part the issue of the physical frame, as we have said before, higher-order corrections in curvature invariants and non-minimal couplings emerge from the formulation of Quantum Field Theory in a curved space-time. In the next subsection, we will sketch this approach giving some fundamental motivations to extended GR.

\section{Quantum field theory in curved space-time}

At small scales and high energies, an hydrodynamic  
description of matter as a  perfect fluid  is inadequate: a more 
accurate description requires quantum field theory formulated on 
a  curved space,  in the framework of either GR or another 
relativistic  theory of gravity. Since, at  scales 
comparable to  the Compton  wavelength of the relevant 
particles, matter must be quantized, one
can employ a semiclassical description of gravitation in which 
the Einstein equations assume  the form

 \begin{eqnarray} 
G_{\mu\nu}  \equiv R_{\mu\nu} -\frac{1}{2} \, g_{\mu\nu} R =<T_{\mu\nu}> \,,
 \label{Intro1.2.1}
\end{eqnarray}
where the usual Einstein tensor $G_{\mu\nu}$ appears on the 
left hand side whereas the right hand side contains the 
expectation value of a quantum stress-energy tensor sourcing  
the gravitational field. Here the coupling constant has been incorporated in the average process so we have not to distinguish between effective and bare couplings. More precisely, if $|\psi>$ is a  
quantum state  describing the early Universe, then

 \begin{eqnarray} 
<T_{\mu\nu}> \equiv  <\psi|\hat{T} _{\mu\nu}|\psi>\,, 
\label{Intro1.2.2}
\end{eqnarray}

where $\hat{T} _{\mu\nu}$ is the quantum operator  associated with 
the classical energy-momentum tensor of the 
matter field and the right hand side is an appropriately 
regularized expectation value. In
general, a quantized matter field $\hat{\phi}$ is subject to 
self-interactions and it interacts also with other fields and 
with the  gravitational background. Such interaction terms may be  
included in the definition of an effective 
potential\footnote{Hereafter, scalar fields and potentials
are understood as their effective values, obtained averaging over 
quantum states. In this sense, classical fields and  potentials 
are the expectation values of quantum fields and potentials.} 

\begin{eqnarray} 
V_{eff}(\phi)= <a|\hat{\cal H}|a>  \label{Intro1.2.3}
\end{eqnarray}
with
 \begin{eqnarray}  \phi = <a|\hat{\phi}|a>\,, \label{Intro1.2.4}
\end{eqnarray}
where $|a>$  represents a normalized state of the theory under 
consideration ({\em i.e.}, $<a|a>=1$) and  $\hat{\cal H}$ is the 
Hamiltonian 
operator satisfying  $\delta <a|\hat{\cal 
H}|a>=0$, where $\delta$ is the variation on the average of 
${\cal H}$-eigenstates. This condition corresponds to energy 
conservation.

In a curved space-time, even in the absence of classical matter 
and
radiation, quantum fluctuations of matter fields give  
non-vanishing contributions to $ <T_{\mu\nu}> $, an effect  
similar to the vacuum of QED  
\cite{BirrellDavies,parkerbook}. 
When matter fields are free, massless and conformally 
invariant, these corrections assume the form 
 \begin{eqnarray} 
<T_{\mu\nu}> = k_{1}\,  ^{(1)}H_{\mu\nu} + k_{3}\,^{(3)}H_{\mu\nu} \,.
 \label{Intro1.2.5}
\end{eqnarray}
Here $k_1$ and $k_3$ are numerical coefficients, while
\begin{eqnarray}
^{(1)}H_{\mu\nu} & = &   2R_{;\mu\nu}-2g_{\mu\nu} \Box R +2RR\_{\mu\nu}
-\frac{1}{2} \, g_{\mu\nu} R^{2}\,, \label{Intro1.2.6} \\
&&\nonumber\\
 ^{(3)}H_{\mu\nu} & = &  {R^{\sigma}}_{\mu} R_{\nu\sigma} 
-\frac{2}{3} \, R R_{\mu\nu} -\frac{1}{2} \, g_{\mu\nu}
 R^{\sigma\tau}R_{\sigma\tau} +\frac{1}{4} \, g_{\mu\nu} R^{2}\,.
 \label{Intro1.2.7}
\end{eqnarray}
 The divergence of the tensor $^{(1)}H_{\mu\nu}$ vanishes,
\begin{eqnarray}
 ^{(1)}H_{\mu;\nu}^{\nu}=0\,.  \label{Intro1.2.8}
\end{eqnarray}
This tensor can be obtained by varying a quadratic 
contribution to the local action,
\begin{eqnarray}
^{(1)}H_{\mu\nu}
=\frac{2}{\sqrt{-g}} \frac{\delta}{\delta g^{\mu\nu}} 
\left( \sqrt{-g}\,  
R^{2} \right) \,.    \label{Intro1.2.9}
\end{eqnarray}
In order to remove the infinities coming from  $ <T_{\mu\nu}>$ and 
obtain a renormalizable theory, one has to
introduce infinitely many counterterms in the Lagrangian 
density of 
gravity. One of these terms is $C R^{2}\sqrt{-g}$, where $C$ 
is a parameter that diverges logarithmically. Eq. 
(\ref{Intro1.2.1}) cannot be generated by a finite action because then 
the  gravitational field would be completely renormalizable, 
{\em i.e.}, it would suffice to eliminate a finite number of 
divergences to make gravity similar to QED. Instead, one can 
only  construct a truncated quantum theory of gravity. The 
expansion in  loops is done in terms of $\hbar$, so the truncated 
theory at the one-loop level contains all terms of order 
$\hbar$. In this sense, this is the first quantum correction to 
GR. It assumes that matter fields are {\it free} 
and, due to the Equivalence Principle, all forms of matter couple 
in the same way to gravity. It also implies an {\it intrinsic} 
non-linearity of gravity, so that a number of loops are needed 
in order to take into account self-interactions or 
mutual interactions between matter and gravitational 
fields. At the one-loop level, divergences can be removed by 
renormalizing the cosmological constant $\Lambda_{eff}$ and the 
gravitational constant $G_{eff}$.
The one-loop contributions to $<T_{\mu\nu}>$ are the  
quantities $ ^{(1)}H_{\mu\nu}$ and $^{(3)}H_{\mu\nu}$ above. In 
addition, one has to consider
\begin{eqnarray}
 ^{(2)}H_{\mu\nu} = 2{R^{\sigma}}_{\mu ; \nu \sigma}-\Box
R_{\mu\nu}- \frac{1}{2} \, g_{\mu\nu}\Box R +{R^{\sigma}}_{\mu} R_{\sigma
\nu}-\frac{1}{2} \, R^{\sigma\tau} R_{\sigma\tau}g_{\mu\nu}\,.
\end{eqnarray}
It is shown in Refs.~\cite{BirrellDavies,parkerbook} that the relation
\begin{eqnarray}
 ^{(2)}H_{\mu\nu} = \frac{1}{3}^{(1)} H_{\mu\nu}
 \end{eqnarray}
holds in conformally flat space-times. In this case, only the 
first and third  terms of $H_{\mu\nu} $ are present in Eq. 
(\ref{Intro1.2.5}).
Since  one can add to the parameter $C$ in the Lagrangian term 
$C \sqrt{-q} \, R^2  $ an arbitrary constant, the 
coefficient $k_{1}$ can assume any value --- the latter should be 
determined experimentally \cite{BirrellDavies,parkerbook}.

The tensor $ ^{(3)}H_{\mu\nu} $ is conserved only in conformally 
flat space-times (for example,  FLRW spaces) and it 
cannot be obtained by varying a local action. Finally, one has
\begin{eqnarray}
k_{3}=\frac{1}{1440\pi^{2}} \left( N_{0}+\frac{11}{2} \, N_{1/2} 
+31 N_{1}\right)\,,   \label{Intro1.2.10}
 \end{eqnarray}
where the  $N_i$'s ($i=0, 1/2, 1$) are determined by the 
number of  quantum  fields with spin
$0,1/2$, and $1$. Vector fields contribute more to $k_3$ due to 
the larger coefficient $ 31 $ of $N_1$. These massless fields,  
as well as the spinorial ones, are described by conformally
invariant equations and appear in $ <T_{\mu\nu}>$ in the  form 
(\ref{Intro1.2.5}).
The trace of the energy-momentum tensor vanishes for 
conformally invariant classical fields while, owing to 
the term weighted by $k_{3}$, one finds that the expectation 
value of the  tensor (\ref{Intro1.2.5}) has non-vanishing trace. 
This 
fact is at the origin of the so-called {\it trace 
anomaly}.

Let us discuss briefly how the conformal anomalies are generated
when the origin of the tensor $T_{\mu\nu}$ is not classical, 
{\em i.e.},  when quantum field theories are formulated in curved 
space-time. As we will see in more detail later, if a theory is 
conformally invariant, under the conformal 
transformation
\begin{eqnarray}
 g_{\mu\nu}(x)\rightarrow \tilde{g}_{\mu\nu}(x) \equiv \Omega^{2}(x)g_{\mu\nu}(x) 
\,.
 \label{Intro1.2.11}
 \end{eqnarray}
the  action in $(n+1)$ space-time dimensions satisfies the  
functional equation

\begin{eqnarray}
S [\tilde{g}_{\mu\nu}]= S [g_{\mu\nu}]+
 \int d^{n+1}x \,\, \frac{\delta 
S [\tilde{g}_{\mu\nu}]}{\delta\tilde{g}^{\rho\sigma}}
 \delta\tilde{g}^{\rho\sigma}  \,,
 \label{Intro1.2.12}
 \end{eqnarray}
where the use of 
\begin{eqnarray}
\delta\tilde{g}^{\mu\nu}(x) 
= -2 \Omega^{-1}(x) \, \tilde{g}^{\mu\nu}(x) \delta\Omega(x)\, ,
 \label{Intro1.2.13}
 \end{eqnarray}
and of the classical variational principle
\begin{eqnarray}
T_{\mu\nu} ^{(m)} = -\,  \frac{2}{\sqrt{-g}} \, \frac{\delta 
S^{(m)} }{\delta g} 
\,,  \label{Intro1.2.14}
 \end{eqnarray}
yields 
\begin{eqnarray}
S [\tilde{g}^{\mu\nu}]=S [g_{\mu\nu}]-\int d^{n+1}x 
\, \sqrt{-\tilde{g}}\, \, 
 {T^{\rho}}_{\rho}(\tilde{g}_{\mu\nu})\Omega^{-1}\delta
 \Omega \,{.}
 \label{Intro1.2.15}
  \end{eqnarray}
From this, it follows that 
\begin{eqnarray}
{T^{\rho}}_{\rho}[g_{\mu\nu}(x)]=-\left.\frac{\Omega(x)}{\sqrt{-g}}
 \, \frac{\delta S [\tilde{g}^{\mu\nu}]}{\delta 
\Omega(x)}\right
 |_{\Omega =1}  \,.  \label{Intro1.2.16}
 \end{eqnarray}
Hence, if the classical action is invariant under conformal
transformations, the trace of the energy-momentum tensor 
vanishes. At
the quantum level this situation could not occur for the 
following reason. A conformal transformation is, essentially, 
a rescaling of lengths with a different rescaling factor at each 
space-time point $x$; the presence of a mass, and 
hence of a length scale, in the theory breaks  conformal 
invariance and generates the trace anomaly.  To preserve 
conformal 
invariance  one has to consider massless fields,  as done in 
(\ref{Intro1.2.5}). In this case one obtains the condition
\begin{eqnarray}
 <T_{\rho}^{\rho}> = 0\,,
 \label{Intro1.2.17}
 \end{eqnarray}
which allows one to consider a conformally
invariant theory.  Note that gravity is not 
renormalizable in the usual way; because of this,  divergences 
appear as soon as quantum effects are considered. A  
loop  expansion yields
\begin{eqnarray}
<T_{\rho}^{\rho}>=   <T_{\rho}^{\rho}>_{div} 
+<T_{\rho}^{\rho}>_{ren}=0\,,
 \label{Intro1.2.18}
 \end{eqnarray}
confirming the validity of Eq. (\ref{Intro1.2.17}). In this case  
conformal invariance is preserved only if the 
divergent part  is equal (up to the sign) to the 
renormalized tensor. An  anomalous trace term will appear on 
the right hand side of  the 
field  equations (\ref{Intro1.2.1}) which, at one-loop  and 
in the zero mass limit of the fields, is given by 
\begin{eqnarray}
<T_{\rho}^{\rho}>_{div} 
=\left[\tilde{k}_{1}\left(M-\frac{2}{3}\Box R\right) 
+\tilde{k}_{3} {\cal G} \right]
=-<T_{\rho}^{\rho}>_{ren}\,,
 \label{Intro1.2.19}
 \end{eqnarray}
for a  four-dimensional theory. Here $\tilde{k}_{1}$ and 
$\tilde{k}_{3}$ are proportional to
$k_{1}$ and $k_{3}$, while $M$ and ${\cal G}$ are obtained from
$^{(1)}H_{\mu\nu}$ and $^{(3)}H_{\mu\nu}$ as
 \begin{eqnarray}
 M & = & R^{\alpha\beta\gamma\delta}R_{\alpha\beta\gamma\delta}- 
2R^{\alpha\beta}
 R_{\alpha\beta}+\frac{1}{3} \, R^{2}\,,
 \label{Intro1.2.20}\\
&&\nonumber\\
{\cal G} & = & R^{2} - 4R ^{\alpha\beta}
R_{\alpha\beta} + R^{\alpha\beta\gamma \delta} 
R_{\alpha\beta\gamma\delta} \,.
 \label{Intro1.2.21}
\end{eqnarray}
${\cal G}$ is the {\it Gauss-Bonnet} term. In four dimensions, 
the integral
 \begin{eqnarray}
\int d^4x \sqrt{-g} \,\, {\cal G}
 \label{Intro1.2.22}
 \end{eqnarray}
is an invariant ({\it Euler characteristic}) which provides 
information about  the topology of the space-time manifold on 
which
the theory is formulated ({\it Gauss-Bonnet 
theorem}). In a FLRW
background $M$ vanishes identically but $ {\cal G}$ gives 
non-vanishing contributions to (\ref{Intro1.2.5}) even if the 
variation of (\ref{Intro1.2.22}) is zero (in four dimensions).

In general, by summing all the geometric terms deduced from  
the Riemann tensor and of the same order in  
$ <T_{\rho}^{\rho}>_{ren}$, one derives the right hand side of 
(\ref{Intro1.2.5}). If the background metric is conformally flat,  
this  can be expressed by means of eqs.  (\ref{Intro1.2.6}) and 
(\ref{Intro1.2.7}). Then, one can conclude that the trace anomaly 
due 
to the geometric terms arises because the one-loop approach 
is an attempt to formulate quantum field theories on curved 
space-time.\footnote{Eqs. (\ref{Intro1.2.6}) and  
(\ref{Intro1.2.7}) can  
include  terms containing derivatives of the metric of order 
higher than fourth (fourth order corresponding to the $R^{2}$ 
term) 
if all 
possible Feynman diagrams are included. For example, 
corrections such as $R\Box R$ or $R^{2}\Box
R$ can be present in $^{(3)}H_{\mu\nu}$ implying equations of motion
that contain  sixth order derivatives of the metric. 
Also these terms can be treated by making use of conformal 
transformations \cite{sixth}.} Cosmological models arising from 
(\ref{Intro1.2.5}) are studied in \cite{Bilicetal07}.
The masses of the matter fields and their mutual interactions  
can be neglected in the high curvature limit because $R>>
m^{2}$. The matter-graviton interactions generate   
non-minimal 
coupling terms in the effective 
Lagrangian. The 
one-loop contributions of such terms are comparable to the ones 
due to the trace anomaly and generate, from the conformal point 
of view, the same effects on gravity. The simplest effective 
Lagrangian that takes into account these corrections is
\begin{eqnarray}
{\cal L}_{NMC} =- \frac{1}{2} \nabla^{\alpha} \phi 
\nabla_{\alpha} \phi -V(\phi)  - \frac{\xi}{2} \,  
R\phi^{2} \,, \label{Intro1.2.23}
 \end{eqnarray}
where $\xi$ is a dimensionless coupling constant between 
the scalar and the gravitational fields. The scalar field 
stress-energy  tensor  will be modified accordingly but  a conformal 
transformation can be found 
such that the modifications due to curvature terms can, 
at least formally, be cast in the form of a matter-curvature 
interaction. The same argument holds for the trace anomaly.
Certain Grand-Unified theories 
lead to a polynomial coupling 
of the form $ 1+\xi \phi^{2}+\zeta \phi^{4}$ generalizing the one
of (\ref{Intro1.2.23}), while an exponential coupling 
$\mbox{e}^{-\alpha \varphi} R$ between a 
scalar field (dilaton) $\varphi$ and the Ricci scalar appears  
instead in the effective Lagrangian of string theories. 
The field equations obtained by varying  the Lagrangian 
density $ \sqrt{-g} \, {\cal 
L}_{NMC}$ are
\begin{eqnarray}
\left( 1-\kappa \xi  \phi^2 \right) G_{\mu\nu} & =& \kappa 
\left\{
\nabla_{\mu}\phi
\nabla_{\nu} \phi -\frac{1}{2} \, g_{\mu\nu} \, 
\nabla^{\alpha}\phi \, \nabla_{\alpha} \phi -V\,
g_{\mu\nu} \xi \, \left[ g_{\mu\nu} \Box \left( \phi^2 \right)
-\nabla_{\mu} \nabla_{\nu}
\left( \phi^2 \right) \right] \right\} \,,  \label{pippa2} \\
&& \nonumber\\
& \Box \phi & -\frac{dV}{d\phi} -\xi R \phi =0 \, .   
\label{nmKG}
\end{eqnarray}
The non-minimal coupling of the 
scalar field is reminiscent of 
that exhibited by the four-vector potential of curved space  
Maxwell  theory.
below.
Motivation  for the 
non-minimal  coupling in the 
Lagrangian ${\cal L}_{NMC}$ comes from many directions. A  
nonzero $\xi$ is  generated 
by first loop corrections even if it 
is absent in the classical action 
\cite{BirrellDavies, 
BirrellDavies80, NelsonPanangaden82, FordToms82,
ParkerToms85,Ford87,parkerbook}. Renormalization of a 
classical theory with 
$\xi=0$ shifts  this coupling 
constant to a value which is typically small 
\cite{Allen83, Ishikawa83} but can, however, 
affect drastically an inflationary cosmological 
scenario and determine its success or failure 
\cite{Abbott81, FutamaseMaeda89,Futamaseetal89, 
FaraoniPRD96, Calgary, mybook}. 
A non-minimal  coupling term is 
expected at
high curvatures \cite{FordToms82,Ford87}, and it has 
been argued that classicalization of the Universe in quantum 
cosmology indeed requires $\xi \neq 0$ \cite{Okamura98}. 
Moreover, non-minimal coupling   can 
solve potential problems of 
primordial nucleosynthesis 
\cite{Chen01, Chenetal01} and  
the absence  of pathologies in the  propagation of 
$\phi$-waves seems to 
require  conformal coupling for all non-gravitational 
scalar fields  
(\cite{SonegoFaraoni93, 
FaraoniSonego94, GribPoberii95, GribRodrigues95}, see also 
\cite{DeserNepomechie84, FaraoniGunzigIJMPD}).\footnote{Note,  
however, that the  distinction between gravitational and 
non-gravitational fields  becomes representation-dependent in 
ETGs, together with  the various formulations of the EP.} 
The conformal value $\xi=1/6$ is also an infrared  fixed point of 
the renormalization group in finite GUTs \cite{Buchbinder86, 
BuchbinderOdintsov83, BuchbinderOdintsov85,
Buchbinderetal86, 
Odintsov91, MutaOdintsov91, ElizaldeOdintsov94, book,
Buchbinderetal89, Reuter94, Bonanno95, BonannoZappala97}. 
Non-minimally coupled scalar fields 
have been widely used 
in relation with specific inflationary scenarios 
\cite{Barrosoetal92,
FakirHabib93,
GarciaLinde95, 
KomatsuFutamase98,
 FutamaseTanaka99, 
KomatsuFutamase99, 
Leeetal99,Salopeketal89, FakirUnruh90a,FakirUnruh90b,
FakirHabibUnruh92,HwangNoh98}.  The approach adopted was largely 
one in which  $\xi$ is regarded as a free parameter  to be used 
at will in order to fix possible problems of specific 
inflationary scenarios; see \cite{think, mybook} for more 
general treatments. Geometric 
reheating of the Universe with  strong 
coupling $\xi \gg 1$ has also been studied 
\cite{Tsujikawaetal00,TsujikawaBassett02} and non-minimally 
coupled  
scalar fields have been considered in relation with 
wormholes
\cite{HalliwellLaflamme89,
CouleMaeda90,Coule92}, black holes  
\cite{Hiscock90, VanderBijRadu00}, and boson stars
\cite{Jetzer92,VanderBijGleiser87, LiddleMadsen92}. The 
value of  the coupling $\xi$ is not, in general, a free 
parameter but it depends on the physical nature of the 
particular scalar field 
$\phi$ \cite{VoloshinDolgov82, HillSalopek92,  
Hosotani85,ParkerToms85b, FordToms82, Ford87} (see 
\cite{FaraoniPRD96, think,  mybook}  for reviews of the 
available theoretical prescriptions for the value of $\xi$).

To conclude, any attempt to formulate quantum field theory
on a curved space-time necessarily leads to modifying the 
Hilbert-Einstein action. This means adding terms containing 
non-linear invariants of the curvature tensor or non-minimal 
couplings between matter and the 
curvature originating  
in the perturbative expansion. In cosmology, all these 
modifications may affect deeply  inflationary 
scenarios originally proposed using minimally coupled scalars 
\cite{think, FaraoniPRD96}. 
Although rare and very speculative alternatives have been 
proposed 
to the inflationary paradigm, the latter is currently accepted by 
most authors as the ``canonical'' cure to the 
shortcomings of the Standard Big Bang Model, with the added 
bonus of  providing a viable mechanism for the generation of 
density  perturbations  to seed the structures observed today in 
the Universe. However, the effects of non-minimal 
coupling on the 
inflationary paradigm need to be assessed carefully.

On the other hand,   the vacuum energy of  free quantized fields of very low masses can significantly alter also the recent expansion of the Universe as shown in \cite{parker1,parker2}. In fact, the effective action  can be obtained from  non-perturbative sums of scalar curvature terms in the propagator. As a result of non-perturbative quantum effects, the scalar curvature of the matter-dominated Universe stops decreasing and approaches a constant value. The Universe evolves from an open matter-dominated epoch to a mildly inflating de Sitter expansion. The Hubble constant, during the present de Sitter epoch, as well as the time at which the transition occurs from matter-dominated to de Sitter expansion, are determined by the mass of the field and by the present matter density. These models provide a theoretical explanation of the observed recent acceleration of the Universe, and gives a good fit to data from high-redshift Type Ia Supernovae, with  masses of about $10^{-33}$ eV, and a current ratio of matter density to critical density, $\Omega_0 <0.4$ . The age of the Universe then follows with no further free parameters in the theory, and turns out to be greater than $13$ Gyr. The Universe is spatially open and consistent with the possibility of inflation in the very early Universe. Furthermore, such models arise from  standard renormalizable theories of  free quantum fields in curved space-time, and do not require a cosmological constant or the associated fine-tuning. In this perspective, ETGs represent also a valid alternative to Dark Energy models.

\section{Variational principles and field equations in metric formalism}
Having discussed the more general aspects of gravitation theory and
briefly reviewed or mentioned some alternatives to
GR, we will concentrate now on a number of specific
gravitation theories that have received attention lately. We begin by
devoting this section to the exploration of their theoretical aspects.
In the following, their phenomenological aspects will be
studied as well.
The theories considered can come from an action as can many of the
interesting theories of gravity. We concentrate on theories which
include a scalar field as an extra field mediating the gravitational
interaction (such as scalar-tensor theories), theories whose action
includes higher order curvature invariants. We
also extensively consider theories with a connection which is
independent of the metric.
The actions of these theories are presented and in many cases their
resemblance with effective low-energy actions coming from more
fundamental theories is briefly discussed. We also present the
derivation of the field equations through the application of a
suitable variational principle and analyse the basic characteristics
of the theory, as expressed through the field equations.

\subsection{The Brans-Dicke gravity as the first extension}
\label{sec:chap3sec1}

The  Brans-Dicke  theory of gravity 
\cite{Jordan38, Jordan52, 
Fierz56, BransDicke61}  is the prototype of gravitational 
theories alternative to GR. The action in the Jordan 
frame (the 
set of  variables $\left( g_{\mu\nu}, \phi \right)$) is
\begin{eqnarray}\label{BDvar1}
S_{(BD)}=\frac{1}{16\pi} \, \int d^4 x \, \sqrt{-g} \left[
\phi R -\frac{\omega}{\phi} \, g^{\mu\nu} \nabla_{\mu}\phi \, 
\nabla_{\nu}\phi -V( \phi) \right] +S^{(m)} \,,
\end{eqnarray}
where
 \begin{eqnarray}
S^{(m)}=\int d^4 x \, \sqrt{-g} \, {\cal L}^{(m)}
\end{eqnarray}

is the action of  ordinary matter and $\omega $ is the  
dimensionless  Brans-Dicke parameter. The factor $\phi$ in
the denominator of the kinetic term of $\phi$ in the  action 
(\ref{BDvar1}) is purely conventional and has the only 
purpose of making  $\omega$ dimensionless. 
Matter does not couple directly to $\phi$, {\em i.e.}, the 
Lagrangian density ${\cal L}^{(m)}$ is independent of  $\phi$ 
(``minimal coupling'' of matter). However,  $\phi$ couples  
directly to the Ricci scalar. The gravitational field is
described by both the metric tensor $g_{\mu\nu}$ and the 
Brans-Dicke scalar $\phi$ which, together with the matter 
variables, constitute the degrees of freedom of the theory. As 
usual for scalar fields, the potential $V( \phi)$   
generalizes  the cosmological constant 
and may reduce to a constant, or to a mass term.\footnote{Due 
to the particular equation (\ref{BDvar3}) satisfied by the 
Brans-Dicke field $\phi$, its   mass  is not the coefficient of 
the  quadratic term in the expansion of $V(\phi)$, as for 
minimally coupled scalar fields, but rather the quantity $m$ 
defined by $ {\displaystyle  m^2=\frac{1}{2\omega+3} \left( \phi 
\, \frac{  d^2V}{d\phi^2}-\frac{dV}{d\phi} \right) } $ 
\cite{VFphimassCQG09}.}

The original motivation 
for introducing  Brans-Dicke  theory  
was  the implementation of  
Mach's Principle. This is achieved in 
Brans-Dicke 
theory by 
making the  effective gravitational 
coupling strength $G_{eff} 
\sim  \phi^{-1}$ depend on the space-time position  and being 
governed by distant matter sources, as in Eq.~(\ref{BDvar3}) 
below. As already remarked, modern  interest in 
Brans-Dicke  and  scalar-tensor theories is motivated by the 
fact that they are obtained as low-energy limits of string 
theories.
The  variation of  the action (\ref{BDvar1}) with respect to 
$g^{\mu\nu} $ and  the well 
known properties \cite{LandauLifschitz}
\begin{eqnarray} 
\delta \left( \sqrt{-g} \, \right) & = &  -\frac{1}{2} \sqrt{-g} 
\,  
g_{\mu\nu} \, \delta g^{\mu\nu} \,,\label{BDvar1bis} \\
&&\nonumber\\
\delta \left( \sqrt{-g} \, R \right) & =& \sqrt{-g} \left( 
R_{\mu\nu}-\frac{1}{2}
\, g_{\mu\nu} R \right) \delta g^{\mu\nu} 
\equiv \sqrt{-g} \, G_{\mu\nu} \, \delta
g^{\mu\nu} \,,  \label{BDvar1ter}
\end{eqnarray}
yield the field equation
\begin{eqnarray}  
 G_{\mu\nu} &=& \frac{8\pi}{\phi} \, T_{\mu\nu}^{(m)}
+ \frac{\omega}{\phi^2} \left( \nabla_{\mu} \phi \, \nabla_{\nu} 
\phi  -\, \frac{1}{2} g_{\mu\nu} \nabla^{\alpha} \phi 
\nabla_{\alpha}  \phi \right) 
+\frac{1}{\phi} \left( \nabla_{\mu} \nabla_{\nu} \phi
 - g_{\mu\nu} \, \Box \phi \right) 
- \, \frac{V}{2\phi} \, g_{\mu\nu}\, , \label{BDvar2}
\end{eqnarray}
where 
\begin{eqnarray}   \label{BDvar3bis}
T_{\mu\nu}^{(m)} \equiv \frac{-2}{\sqrt{-g}}\, 
\frac{\delta }{\delta g^{\mu\nu} }
\left( \sqrt{-g} \, {\cal L}^{(m)} \right)
\end{eqnarray}
is the energy-momentum tensor of ordinary matter. 
By varying the action with respect to $\phi$, one obtains
\begin{eqnarray}   \label{BDvar3bisbis}
\frac{2\omega}{\phi} \, \Box \phi +R -\frac{\omega}{\phi^2} \, 
\nabla^{\alpha} \phi \nabla_{\alpha} \phi -\frac{dV}{d\phi}=0 \,.
\end{eqnarray}
Taking now the trace of Eq.~(\ref{BDvar2}),
\begin{eqnarray}   \label{BDvar3bisbisbis}
R=\frac{-8 \pi \, T^{(m)} }{\phi } \, 
+\frac{\omega}{\phi^2} \, \nabla^{\alpha} \phi 
\nabla_{\alpha} \phi + 
\frac{3\Box \phi}{\phi}  + \frac{2V}{\phi} \, ,
\end{eqnarray}
and using the resulting Eq.~(\ref{BDvar3bisbisbis}) to eliminate 
$R$ from Eq.~(\ref{BDvar3bisbis}) leads to
\begin{eqnarray}     \label{BDvar3}
\Box \phi  = \frac{1}{ 2\omega +3 }
  \left( 8\pi \, T^{(m)}  +   \phi \, \frac{dV}{d\phi} -2V 
\right) \, .
\end{eqnarray}
According to this equation, the scalar $\phi$ is sourced 
by non-conformal matter ({\em i.e.}, by matter with trace 
$T^{(m)} \neq 0 $),   however the scalar does  not couple 
directly to ${\cal  L}^{(m)}$: the Brans-Dicke scalar  
$\phi$ reacts on ordinary matter only indirectly through the 
metric tensor $g_{\mu\nu}$, as dictated by Eq.~(\ref{BDvar2}). 
The term proportional  to $\phi \,  dV/d\phi -2V$ on 
the right hand side of  Eq.~(\ref{BDvar3}) vanishes if the 
potential has the form $V(\phi)=m^2\phi^2/2$ familiar 
from the Klein-Gordon equation and from particle physics.
The action (\ref{BDvar1}) and the field equation 
(\ref{BDvar2}) suggest that the field $\phi$  be identified 
with the inverse of the effective gravitational 
coupling
\be  \label{BDvar4}
G_{eff} \left( \phi \right)= \frac{1}{\phi} \,,
\ee
a function of the space-time location. In order to guarantee a 
positive gravitational coupling, only the range of values $\phi 
> 
0$ corresponding to attractive gravity is considered. The 
dimensionless Brans-Dicke parameter $\omega$  is a 
free  parameter of the theory: a 
value of $\omega $  of order unity  would be natural in 
principle (and it does appear in the low-energy limit of the 
bosonic  string theory). However,  
values of $\omega$ of this 
order of magnitude are excluded by Solar System experiments, 
for a  massless or light field $\phi$ ({\em i.e.}, one that has a 
range larger than the size of the Solar System).

The larger the value of $\omega$, the closer Brans-Dicke
gravity  is to GR \cite{Weinberg72}; 
there are, however, 
exceptions such as  vacuum Brans-Dicke solutions,\footnote{One 
should keep in mind, however, that the limit of particular 
space-time solutions of the field equations of a gravitational 
theory  should be taken in a  coordinate-independent way 
\cite{Geroch69, PaivaRomero93,Paivaetal93}.} and  solutions 
sourced by conformal 
matter \cite{Matsuda72,
RomeroBarros92,
RomeroBarros93a,
RomeroBarros93b,
ScheelShapiroTeukolsky1,
BanerjeeSen97,
Faraoni98PLA,
Anchordoquietal98,
Faraoni99PRD,
mybook}. The most stringent  experimental limit, $\omega > 40,000 
$,  was  set by the Cassini probe in 2003    
\cite{BertottiIessTortora}. 

Brans-Dicke theory with a free or 
light scalar field  is 
viable in the limit of large  $\omega$, but the
large value of this parameter required to satisfy the 
experimental 
bounds is certainly fine-tuned and  makes Brans-Dicke 
theory 
unappealing. However, this fine-tuning becomes unnecessary if the 
scalar field is given a sufficiently large mass and, 
therefore, a short range. This means that a self-interaction 
potential $V(\phi)$ has to be considered in discussing the limits 
on $\omega$ and this fact is an adjustment of the original 
Brans-Dicke theory \cite{BransDicke61}.
%
\subsection{$f(R)$-gravity in metric formalism}
\label{sec:chap3sec2}

We now examine the variational principle and the field
equations of another class of ETGs,   $f(R)$-gravity 
in the metric formalism. The salient feature of these ETGs  
is that the field equations are of
fourth order and, therefore,  more complicated than those 
of  GR (which is recovered as the special case  $f(R)=R$). 
Due to their higher order, these field equations admit  a much 
richer variety of solutions than the Einstein equations. For 
simplicity, we begin by  discussing quadratic corrections to the 
Hilbert-Einstein theory,   which provide interesting cosmology.

\subsubsection{The case of $ f(R)=R +\alpha  R^2 $} 
\label{sec:chap3sec2subsec1}

Quadratic corrections  in the Ricci scalar motivated by 
attempts to renormalize GR, 
constitute a  straightforward extension of GR and have been 
particularly  relevant in cosmology  since 
they allow  a self-consistent inflationary model to be 
constructed \cite{starobinsky}. We will use this model as 
an example before discussing general metric  $f(R)$-gravity.

Let us begin by deriving the field equations for the Lagrangian
density 
 \begin{eqnarray}  \label{VAR12.1} 
{\cal L}=R+\alpha R^{2}+2\kappa {\cal L}^{(m)}
\end{eqnarray} 

from the variational principle  $ \delta\int d^{4}x \sqrt{-g} \, 
{\cal  L}=0 $.  We consider vacuum first. The  
variation gives
 \begin{eqnarray}   \label{VAR12.3}
\int d^{4}x  \sqrt{-g} \, G_{\alpha\beta}\delta
g^{\alpha\beta} + \alpha \, \delta\int d^{4}x  \sqrt{-g} 
\, R^{2} =0\,, 
\end{eqnarray} 
in which the variation of $R\sqrt{-g} $ produces the Einstein 
tensor.  We now  compute the second term on the right hand 
side of Eq. (\ref{VAR12.3}). We have 
 \begin{eqnarray} \label{VAR12.4}
\delta \int d^{4}x \sqrt{-g} \, R^{2}  = -\frac{1}{2}\int d^{4}x  
\sqrt{-g} \,\,  g_{\alpha\beta}\delta g^{\alpha\beta}R^{2} + 
2\int d^{4}x   \sqrt{-g} \, R\delta R 
\end{eqnarray} 
and 
 \begin{eqnarray}  \label{VAR12.5} 
\int d^{4}x  \sqrt{-g} \, R\delta  R = \int d^{4}x  \sqrt{-g} \, 
R\left(\delta g^{\alpha\beta} R_{\alpha\beta} + 
g^{\alpha\beta}\delta R_{\alpha\beta} \right) \,.
\end{eqnarray} 
By using the fact that 
 \begin{eqnarray}\label{VAR12.6}
g^{\alpha\beta}\delta
R_{\alpha\beta} =  
\nabla_{\alpha}\nabla_{\beta} 
h^{\alpha\beta}  -\Box h \,,
\end{eqnarray} 
where 
 \begin{eqnarray} \label{VAR12.9} 
h^{\alpha\beta} \equiv -\delta g^{\alpha\beta},\
\;\;\;\;\; h \equiv -g_{\alpha\beta}\delta g^{\alpha\beta} \,, 
\end{eqnarray} 
one has 
 \begin{eqnarray}\label{VAR12.7}
\int d^{4}x  \sqrt{-g} \, R \, g^{\alpha\beta}\delta
R_{\alpha\beta} = 
\int d^{4}x \sqrt{-g} \, R 
\left( \nabla_{\alpha}\nabla_{\beta} h^{\alpha\beta} -\Box
h\right) \,. 
\end{eqnarray} 
Integrating by parts twice, the operators 
$\nabla_{\alpha}\nabla_{\beta}$ and $\Box$ acting  on   
$ h^{\alpha\beta}$ and $h$, respectively, transfer their action 
onto $R$ and  
 \begin{eqnarray}\label{VAR12.8}
\int d^{4}x \sqrt{-g} \, R \, g^{\alpha\beta}\delta
R_{\alpha\beta} = \int d^{4}x  \sqrt{-g} \, \left(h
^{\alpha\beta}  \nabla_{\alpha}\nabla_{\beta}  R  
-h\Box R\right) \,. 
\end{eqnarray} 
Using Eq.~(\ref{VAR12.9}), Eq. (\ref{VAR12.8}) becomes 
 \begin{eqnarray}\label{VAR12.10} 
\int d^{4}x \sqrt{-g} \, R \, g^{\alpha\beta}\delta
 R_{\alpha\beta}  
=\int d^{4}x \sqrt{-g} \left(-\delta g^{\alpha\beta} 
\nabla_{\alpha}\nabla_{\beta} R 
+g_{\alpha\beta}\Box R\delta g^{\alpha\beta}\right) \,. 
\end{eqnarray} 
Upon substitution of Eq.~(\ref{VAR12.10}) into  
Eq.~(\ref{VAR12.5}), one obtains
 \begin{eqnarray} \label{VAR2.11} 
\int d^{4}x \sqrt{-g} \, R\delta 
R = \int d^4x \, \sqrt{-g} \left(R\delta 
g^{\alpha\beta}R_{\alpha\beta}-\delta 
g^{\alpha\beta} \nabla_{\alpha}\nabla_{\beta}  R 
+g_{\alpha\beta}\Box R\delta
g^{\alpha\beta}\right)  
\end{eqnarray} 
and Eq.~(\ref{VAR12.4}) takes the form
\begin{eqnarray}
\delta \int d^{4}x \sqrt{-g} \, R^{2} 
& = & -\frac{1}{2}\int d^{4}x \sqrt{-g} \, g_{\alpha\beta}\delta
g^{\alpha\beta} R^{2}+ 2\int d^{4}x \sqrt{-g} \, 
\left(R\delta
g^{\alpha\beta}R_{\alpha\beta}-\delta  
g^{\alpha\beta} \nabla_{\alpha}\nabla_{\beta}  R
+g_{\alpha\beta} \Box R\delta
g^{\alpha\beta}\right) \nonumber\\
&&\nonumber\\
& =&  \int d^{4}x \sqrt{-g} \, {\textstyle\left(2RR_{\alpha\beta} 
-\frac{1}{2} \, g_{\alpha\beta}R^{2}\right)}\delta
g ^{\alpha\beta} +  2   \int d^{4}x \sqrt{-g} 
\, \left(g_{\alpha\beta}\Box R- \nabla_{\alpha}\nabla_{\beta}  
R   \right)\delta  
g^{\alpha\beta}  \, . \label{VAR12.12}
\end{eqnarray}
Substituting  this equation into Eq.~(\ref{VAR12.3})  and 
including the matter part of the Lagrangian ${\cal 
L}^{(m)}$ which produces the energy-momentum tensor 
$T_{\mu\nu}^{(m)}$,  the field  equations
\begin{eqnarray}\label{VAR12.13}
G_{\alpha\beta} + \alpha\left[  
2R\left( R_{\alpha\beta} 
-\frac{1}{4} g_{\alpha\beta}R\right)  + 2\left( g_{\alpha\beta}\Box
R-\nabla_{\alpha}\nabla_{\beta}  R\right)\right] = \kappa \, 
T_{\alpha\beta}^{(m)}  
\end{eqnarray}

are obtained; they  are fourth-order equations for the metric 
components.
The trace of Eq. (\ref{VAR12.13}) is
\begin{eqnarray}\label{VAR12.14} 
\Box R -\frac{1}{6\alpha}\left(R+\kappa \,  T^{(m)} \right)=0\,,
\end{eqnarray}
which shows that  $\alpha$ must  be  
positive.  One can also define an angular frequency $\omega$ 
(equivalent to a mass $m$) so that   
\begin{eqnarray}\label{VAR12.15}
\frac{1}{6\alpha}=\omega^{2}=m^{2} \,. 
\end{eqnarray}
Following this definition, 
Eq. (\ref{VAR12.14}) becomes 
\begin{eqnarray}\label{VAR12.16} 
\Box R - m^{2}\left(R+\kappa \,  T^{(m)} \right)=0 \,. 
\end{eqnarray} 
 Eq. (\ref{VAR12.16}) can be seen 
as  an effective Klein-Gordon equation for the effective scalar 
field degree of freedom $R$ (sometimes called {\it
scalaron}).

\subsubsection{$f(R)$-gravity: the general case}
\label{sec:chap3sec2subsec2}

Let us discuss now a generic analytical\footnote{This 
assumption is  not, strictly speaking, necessary and is 
sometimes relaxed in the literature.} function  $ f(R)$ in 
the metric  formalism, beginning with the vacuum case, as 
described by 
the Lagrangian density $ \sqrt{-g} \, {\cal L}=\sqrt{-g} \, f(R) 
$ obeying the variational principle $  
\delta \int d^{4}x  \sqrt{-g} \, f(R) =0$. We have
\begin{eqnarray}
&& \delta  \int d^{4}x \sqrt{-g} \,  f(R)  = \int 
d^{4}x \left[  \delta \left( \sqrt{-g} \, 
f(R) \right) + \sqrt{-g} \, \delta\left(f(R) \right)\right]   
\nonumber\\
&&\nonumber\\
&&   = \int d^{4}x \sqrt{-g} \, {\textstyle\left[f'(R)R_{\mu\nu} 
-\frac{1}{2}g_{\mu\nu}f(R)\right]}\delta
g^{\mu\nu} + \int d^{4}x \sqrt{-g} \, f'(R)g^{\mu\nu}\delta 
R_{\mu\nu} \, ,\nonumber\\
&& \label{VAR12.19}
\end{eqnarray}
where the prime denotes  differentiation with respect to $R$.
We now compute these integrals in the local inertial frame. By 
using   
\begin{eqnarray}\label{VAR12.20} 
g^{\mu\nu}\delta  R_{\mu\nu} 
=g^{\mu\nu}\partial_{\sigma} 
\left(\delta G_{\mu\nu}^{\sigma}\right) 
-g^{\mu\sigma}\partial_{\sigma}\left(\delta G_{\mu\nu}^{\nu}\right)
\equiv\partial_{\sigma}W^{\sigma}  
\end{eqnarray}
where  
\begin{eqnarray}\label{VAR12.21} 
W^{\sigma}\equiv g^{\mu\nu} \delta G_{\mu\nu}^{\sigma} 
-g^{\mu\sigma}\delta G_{\mu\nu}^{\nu} \,,
\end{eqnarray}  
the second integral in Eq. (\ref{VAR12.19}) can be written  as
\begin{eqnarray}
\int d^{4}x \sqrt{-g} \, f'(R)g^{\mu\nu}\delta
R_{\mu\nu} =\int d^{4}x \sqrt{-g} 
\, f'(R) \partial_{\sigma}W^{\sigma} \, . 
\end{eqnarray}
Integration  by parts yields
\begin{eqnarray}
\int d^{4}x \sqrt{-g} \, f'(R)g^{\mu\nu}\delta
R_{\mu\nu} &= &  
\int d^{4}x \frac{\partial}{\partial x^{\sigma} }  \left[ \sqrt{-g} \, 
f'(R)W^{\sigma}\right] 
-   \int d^{4}x \partial_{\sigma}\left[\sqrt{-g}f'(R)\right]W^{\sigma} 
\, . 
\end{eqnarray}
The first integrand is a total  divergence and can be discarded 
by assuming that the fields vanish at infinity, obtaining   
\begin{eqnarray}\label{VAR12.22} 
\int d^{4}x \sqrt{-g} \, f'(R) g^{\mu\nu}\delta
R_{\mu\nu} = -\int d^{4}x 
\partial_{\sigma} \left[\sqrt{-g} \, f'(R)\right]W^{\sigma} \,. 
\end{eqnarray}

Let us calculate now the term $W^{\sigma}$ appearing in
Eq. (\ref{VAR12.22}). We have
 \begin{eqnarray}\label{VAR12.23} 
\delta G_{\mu\nu}^{\sigma} 
 =  \delta\left[\frac{1}{2}g^{\sigma\alpha}\left(\partial_{\mu}g_{\alpha\nu}
+\partial_{\nu}g_{\mu\alpha}-\partial_{\alpha}g_{\mu\nu}\right)\right] 
=  \frac{1}{2}g^{\sigma\alpha}\left[\partial_{\mu}\left(\delta
g_{\alpha\nu}\right)+\partial_{\nu}\left(\delta
g_{\mu\alpha}\right)-\partial_{\alpha}\left(\delta g_{\mu\nu}\right)\right] 
\, ,
\end{eqnarray}

since in the locally inertial frame considered here it is 
\begin{eqnarray}\label{VAR12.24}
\partial_{\alpha}g_{\mu\nu}=\nabla_{\alpha}g_{\mu\nu}= 0\,.
\end{eqnarray}
Similarly, it is 
\begin{eqnarray}\label{VAR12.25}
\delta G_{\mu\nu}^{\nu}=\frac{1}{2} \, g^{\nu\alpha}\partial_{\mu}\left(\delta
g_{\nu\alpha}\right) \,.
\end{eqnarray}
By combining Eqs. (\ref{VAR12.24}) and 
(\ref{VAR12.25}), one obtains 
\begin{eqnarray}
 g^{\mu\nu}\delta G_{\mu\nu}^{\sigma} 
& = & \frac{1}{2} \, g^{\mu\nu}\left[ 
-\partial_{\mu}\left(g_{\alpha\nu}\delta
g^{\alpha\sigma}\right)-\partial_{\nu}\left(g_{\mu\alpha}\delta
g^{\sigma\alpha}\right)-g^{\sigma\alpha}\partial_{\alpha}\left(\delta
g_{\mu\nu}\right)\right] = \frac{1}{2} \, \partial^{\sigma}\left(g_{\mu\nu}\delta g^{\mu\nu}
\right)-\partial^{\mu}\left(g_{\alpha\mu}\delta g^{\nu\alpha}\right) \,,
\label{VAR12.26} \\
&&\nonumber\\
g^{\mu\sigma}\delta G_{\mu\nu}^{\nu} & = & 
-\frac{1}{2} \, \partial^{\sigma}\left(g_{\nu\alpha}\delta g^{\nu\alpha}\right) 
\, ,  \label{VAR12.27}
\end{eqnarray}
from which it follows immediately  that   
\begin{eqnarray}\label{VAR12.28}
W^{\sigma}=\partial^{\sigma}\left(g_{\mu\nu}\delta
g^{\mu\nu}\right)-\partial^{\mu}\left(g_{\mu\nu}\delta 
g^{\sigma\nu}\right) \,.
\end{eqnarray}
Using this equation one can write 
\begin{eqnarray} 
 \int d^{4}x \sqrt{-g} \, f'(R)g^{\mu\nu}\delta
R_{\mu\nu} 
 =     \int d^{4}x 
\partial_{\sigma}\left[\sqrt{-g} \, 
f'(R)\right] \left[\partial^{\mu}\left(g_{\mu\nu}\delta
g^{\sigma\nu}\right)-\partial^{\sigma}\left(g_{\mu\nu}\delta
g^{\mu\nu}\right)\right]  \,. \label{VAR12.29}
\end{eqnarray}
Integrating by parts and discarding total divergences, one 
obtains 
\begin{eqnarray}
 \int d^{4}x \, \sqrt{-g} \, f'(R)g^{\mu\nu}\delta 
R_{\mu\nu} =    \int d^{4}x \, 
g_{\mu\nu} \partial^{\sigma}\partial_{\sigma}\left[\sqrt{-g}f'(R)\right]\delta
g^{\mu\nu} 
 - \int d^{4}x \,
g_{\mu\nu}\partial^{\mu}\partial_{\sigma}\left[\sqrt{-g} \, f'(R)\right] 
\delta g^{\sigma\nu}  \,. \label{VAR12.30}
\end{eqnarray}
The variation of the action is then
\begin{eqnarray}\label{VAR12.31} 
\lefteqn{  \delta\int d^{4}x  \sqrt{-g} \, f(R) 
=\int d^{4}x  \sqrt{-g} \, {\textstyle\left[ 
f'(R)R_{\mu\nu}-\frac{1}{2}
f(R) g_{\mu\nu}\right]}\delta g^{\mu\nu} } \nonumber\\ 
& & {}+\int d^{4}x \left[
g_{\mu\nu}\partial^{\sigma}\partial_{\sigma} 
\left(\sqrt{-g} \, f'(R)\right)-g_{\sigma\nu}\partial^{\mu} 
\partial_{\sigma}\left(\sqrt{-g}f'(R)\right)
\right]\delta g^{\mu\nu} \, . {} 
\end{eqnarray}
The vanishing of the variation implies the fourth order 
vacuum field 
equations
\begin{eqnarray}\label{VAR12.32} 
f'(R)R_{\mu\nu}-\frac{f(R)}{2} \, g_{\mu\nu} =\nabla_{\mu} 
\nabla_{\nu}f'(R)-g_{\mu\nu}\Box 
f'(R) \, .
\end{eqnarray}
These equations  can be re-arranged in the Einstein-like form
\begin{eqnarray}\label{VAR12.33}
f'(R) R_{\mu\nu}-\frac{ f'(R)}{2} \, g_{\mu\nu} R 
+\frac{f'(R)}{2} \, g_{\mu\nu}R-\frac{f(R)}{2} \, g_{\mu\nu}= 
\nabla_{\mu} \nabla_{\nu} f'(R)-
g_{\mu\nu}\Box f'(R) \, , 
\end{eqnarray}
and then 
\begin{eqnarray}\label{VAR12.34}
G_{\mu\nu}=\frac{1}{f'(R)} \left\{
\nabla_{\mu}\nabla_{\nu} f'(R) - g_{\mu\nu}\Box f'(R)
+ g_{\mu\nu} \frac{ \left[ f(R)-f'(R) R \right]}{2}  \right\}  
\end{eqnarray}
The right hand side of Eq. (\ref{VAR12.34}) is then regarded  as 
an  
effective stress-energy tensor, which we call {\it 
curvature fluid}  energy-momentum tensor 
$T_{\mu\nu}^{(curv)}$ 
sourcing the effective Einstein equations. Although this 
interpretation is questionable in principle because the field 
equations describe a theory different from GR, and one is 
forcing upon them the interpretation as effective Einstein 
equations, this approach is quite useful in practice.
%

\subsection{A more general class of Extended Theories of Gravity}
\label{sec:chap3sec3}

ETGs exhibit two main features: first, the geometry can 
couple non-minimally to some scalar 
field; second,  
derivatives of the metric components of order higher than second  
may appear. In the first case, we say that we have  
scalar-tensor theories of gravity, and in the second case we have  
higher order theories. Combinations of non-minimally 
coupled   and higher order terms can 
also emerge in effective 
Lagrangians, producing mixed   higher order/scalar-tensor 
gravity.

 A general
class of higher-order-scalar-tensor theories in four dimensions is
 \footnote{For the aims of this review, we do
not need more complicated invariants like $R_{\mu\nu}R^{\mu\nu}$,
$R_{\mu\nu\alpha\beta}R^{\mu\nu\alpha\beta}$,
$C_{\mu\nu\alpha\beta}C^{\mu\nu\alpha\beta}$ which are also
possible. We will consider such invariants for the discussion of new polarizations and gravitational modes in Part III. } given by the action
 \begin{eqnarray} \label{V3.1} {\cal S}=\int
d^{4}x\sqrt{-g}\left[F(R,\Box R,\Box^{2}R,..\Box^kR,\phi)
 -\frac{\epsilon}{2}
g^{\mu\nu} \phi_{; \mu} \phi_{; \nu}+ 2\kappa{\cal L}^{(m)}\right], \end{eqnarray} where $F$ is
an unspecified function of curvature invariants and of a scalar
field $\phi$. The term ${\cal L}^{(m)}$, as above, is the minimally
coupled ordinary matter contribution; $\epsilon$ is a constant which specifies the theory. Actually its
 values can be $\epsilon =\pm 1,0$ fixing the nature and the
 dynamics of the scalar field which can be a standard scalar
 field, a phantom field or a field without dynamics (see
 \cite{valerio,CP,odi2005} for details).
In the metric approach, the field equations are obtained by
varying (\ref{V3.1}) with respect to  $g_{\mu\nu}$.  We get
  \begin{eqnarray} 
\label{3.2cc} G^{\mu\nu}&=&\frac{1}{{\cal
G}}\left[\kappa T^{\mu\nu}+\frac{1}{2}g^{\mu\nu} (F-{\cal G}R)+
(g^{\mu\lambda}g^{\nu\sigma}-g^{\mu\nu} g^{\lambda\sigma})
{\cal G}_{;\lambda\sigma}\right.\nonumber\\
& & +\frac{1}{2}\sum_{i=1}^{k}\sum_{j=1}^{i}(g^{\mu\nu}
g^{\lambda\sigma}+
  g^{\mu\lambda} g^{\nu\sigma})(\Box^{j-i})_{;\sigma}
\left(\Box^{i-j}\frac{\partial F}{\partial \Box^{i}R}\right)_{;\lambda}\nonumber\\
& &\left.-g^{\mu\nu} g^{\lambda\sigma}\left((\Box^{j-1}R)_{;\sigma}
\Box^{i-j}\frac{\partial F}{\partial \Box^{i}R}\right)_{;\lambda}\right]\,,
\end{eqnarray} where $G^{\mu\nu}$ is the above Einstein tensor and 

 \begin{eqnarray} 
\label{3.4gg}
  {\cal G}\equiv\sum_{j=0}^{n}\Box^{j}\left(\frac{\partial F}{\partial \Box^{j} R}
\right)\;. \end{eqnarray}

 The differential Eqs.(\ref{3.2cc}) are of order
$(2k+4)$. The stress-energy tensor is due to the kinetic part of
the scalar field and to the ordinary matter:  \begin{eqnarray}  \label{3.51}
T_{\mu\nu}=T^{(m)}_{\mu\nu}+\frac{\epsilon}{2}[\phi_{;\mu}\phi_{;\nu}-
\frac{1}{2}\phi_{;}^{\alpha}\phi_{; \alpha}]\;.  \end{eqnarray} The (eventual) contribution of
a potential $V(\phi)$ is contained in the definition of $F$. From now
on, we shall indicate by a capital $F$ a Lagrangian density
containing also the contribution of a potential $V(\phi)$ and by
$F(\phi)$, $f(R)$, or $f(R,\Box R)$ a function of such fields
 without potential.
By varying with respect to the scalar field $\phi$, we obtain the
Klein-Gordon equation
  \begin{eqnarray} \label{3.62} \epsilon\Box\phi=-\frac{\partial
F}{\partial\phi}\,.  \end{eqnarray} 
 Several approaches can be used to deal with such
equations. For example, as we said, by a conformal transformation,
it is possible to reduce an ETG to a (multi) scalar-tensor theory
of gravity \cite{Qua91,Wands94,wands,Gottetal90,DamourFarese92}.
The simplest extension of GR is achieved assuming, as discussed,   \begin{eqnarray}
\label{fr}
F=f(R)\,,\qquad \epsilon=0\,, \end{eqnarray}  in the action (\ref{V3.1}).
  The standard
Hilbert-Einstein action is, of course, recovered for $f(R)=R$.
Varying with respect to $g_{\alpha\beta}$, we get Eq. (\ref{VAR12.32}) and, after some manipulations, Eq. (\ref{VAR12.34})
where the gravitational contribution due to higher-order terms can
be simply reinterpreted as a stress-energy tensor contribution.
This means that additional and higher-order terms in the
gravitational action act, in principle, as a stress-energy tensor,
related to the  form of $f(R)$.  Considering also the standard
perfect-fluid matter contribution, we have
 \begin{equation}\label{h4}
G_{\alpha\beta}=\frac{1}{f'(R)}\left\{\frac{1}{2}g_{\alpha\beta}\left[f(R)-Rf'(R)\right]
+f'(R)_{;\alpha\beta} -g_{\alpha\beta}\Box f'(R)\right\}+ \frac{\kappa T^{(m)}_{\alpha
\beta}}{f'(R)}=T^{(curv)}_{\alpha\beta}+\frac{\kappa T^{(m)}_{\alpha
\beta}}{f'(R)}\,,
\end{equation}
where $T^{(curv)}_{\alpha\beta}$ is an effective stress-energy
tensor constructed by the extra curvature terms.  In the case of
GR,   $T^{(curv)}_{\alpha\beta}$ identically vanishes while the
standard, minimal coupling is recovered for the matter
contribution. The peculiar behaviour of $f(R)=R$ is  due to the
particular form of the Lagrangian itself which, even though it is
a second order Lagrangian, can be non-covariantly rewritten as the
sum of a first order  Lagrangian plus a pure divergence term. The
Hilbert-Einstein Lagrangian can be in fact recast as follows:
\begin{equation}
L_{HE}= {\cal L}_{HE} \sqrt{-g}  =\Big[ p^{\alpha \beta}
(\Gamma^{\rho}_{\alpha \sigma} \Gamma^{\sigma}_{\rho
\beta}-\Gamma^{\rho}_{\rho \sigma} \Gamma^{\sigma}_{\alpha
\beta})+ \nabla_\sigma (p^{\alpha \beta} {u^{\sigma}}_{\alpha
\beta}) \Big]
\end{equation}
\noindent where:
\begin{equation}
 p^{\alpha \beta} =\sqrt{-g}  g^{\alpha \beta} = \frac{\partial {\cal{L}}}{\partial R_{\alpha \beta}}
\end{equation}
$\Gamma$ is the Levi-Civita connection of $g$ and
$u^{\sigma}_{\alpha \beta}$ is a quantity constructed out with the
variation of $\Gamma$ \cite{weinberg}. Since $u^{\sigma}_{\alpha
\beta}$ is not a tensor, the above expression is not covariant;
however a standard procedure has been studied to recast covariance
in the first order theories. This clearly shows that
the field equations should consequently be second order  and the
Hilbert-Einstein Lagrangian is thus degenerate.
From the action (\ref{V3.1}), it is possible to obtain another
interesting case by choosing \begin{eqnarray} F=F(\phi)R-V(\phi)\,,\qquad \epsilon
=-1\,.\end{eqnarray} In this case, we get
\begin{equation} \label{s1} {\cal S}= \int V(\phi) \left[F(\phi) R+ \frac{1}{2} g^{\mu\nu}
\phi_{;\mu}\phi_{;\nu}- V(\phi) \right] \end{equation} $V(\phi)$ and
$F(\phi)$ are generic functions describing respectively the
potential and the coupling of a scalar field $\phi$.  The
Brans-Dicke theory of gravity is a particular case of the action
(\ref{s1}) for $V(\phi)$=0. The variation with
respect to $g_{\mu\nu}$ gives the second-order field equations
\begin{equation} \label{s2} 
F(\phi) G_{\mu\nu}= F(\phi)\left[R_{\mu\nu}-
\frac{1}{2} R _{\mu\nu} \right]= -\frac{1}{2} T^{\phi}_{\mu\nu}- g_{\mu\nu} \Box_{g} F(\phi)+F(\phi)_{;\mu\nu}\,, \end{equation} 
 here $\Box_{g}$ is the d'Alembert
operator with respect to the metric $g$.  The energy-momentum
tensor relative to the scalar field is
\begin{equation} \label{s4} T^{\phi}_{\mu\nu}= \phi_{;\mu} \phi_{;\nu}- \frac{1}{2} g_{\mu\nu} \phi_{;\alpha} \phi_{;}^{\alpha}+g_{\mu\nu} V(\phi) \end{equation} The variation with respect
to $\phi$ provides the Klein - Gordon equation, {\it i.e.} the field
equation for the scalar field:  \begin{equation} \label{s5}
\Box_{g} \phi- R F_{\phi}(\phi)+ V_{\phi}(\phi)= 0
\end{equation}
where $\displaystyle{F_{\phi}(\phi)= \frac{dF(\phi)}{d\phi}}$, $\displaystyle{V_{\phi}(\phi)= \frac{dV(\phi)}{d\phi}}$. This last equation
is equivalent to the Bianchi contracted identity \cite{cqg}.
Standard fluid matter can be treated as above.


\section{The Palatini formalism}
\label{sec:chap3sec5}

 The fundamental idea  of the Palatini 
formalism is to regard  the
(usually torsion-free) connection $\Gamma^{\mu}_{\nu\alpha}$  
entering the definition of the Ricci tensor as a 
variable independent 
of the space-time metric $ g_{\mu\nu}$. The Palatini 
formulationof GR is  equivalent
to the metric version of this theory as a consequence of the 
fact that the field equations for the connection  
$\Gamma^{\alpha}_{\mu\nu}$  give the Levi-Civita connection
of the metric $g_{\mu\nu}$ \cite{Wald84}. As a consequence, there 
is  no 
particular reason 
to impose the Palatini variational principle in GR instead of the 
metric  variational principle.

As we said, the situation is different in ETGs depending on functions of  
curvature invariants or for gravity 
non-minimally coupled   to a scalar 
field. In these cases, the
Palatini  and the metric variational  
principle yield different
field equations and different physics 
\cite{MagnanoSokolowski94, Ferraris:1994af}. 
The  Palatini approach  in the context of ETGs  has been the 
subject of much interest in 
cosmological applications \cite{Capozziello02IJMPD,CF1, 
Vollick:2003, Li:2006vi, Li:2006ag}.
As discussed above, considering the metric $g_{\mu\nu}$ and 
the connection  $ \Gamma^{\alpha}_{\mu\nu}$ as independent fields 
amounts to decoupling the metric structure of space-time and its 
geodesic structure with the connection 
$\Gamma^{\alpha}_{\mu\nu}$ being distinct from  the Levi-Civita 
connection of $g_{\alpha\beta}$. In principle, this decoupling 
enriches the geometric structure of space-time and generalizes 
the purely  metric formalism. By means of the Palatini field 
equations, this dual structure of space-time  is naturally 
translated into a bimetric structure of the theory: instead of a 
metric and an independent connection, the Palatini 
formalism  can 
be seen as containing two independent metrics  $g_{\mu\nu}$ and 
$h_{\mu\nu}=f'(R) \, g_{\mu\nu}  $. In Palatini $f(R)$-gravity 
the new metric 
$h_{\mu\nu}$ determining the geodesics is related to the 
connection  $\Gamma^{\alpha}_{\mu\nu} $ by the fact that 
the latter turns out to be the Levi-Civita connection of
$h_{\mu\nu}$.
In scalar-tensor gravity, the second metric $h_{\mu\nu}$ is  
related to the non-minimal coupling   
of the Brans-Dicke-like 
scalar. In the Palatini formalism  the 
non-minimal 
coupling   and 
the scalar field are separated from the
metric structure of space-time. Physical consequences of this fact are discussed in \cite{silvio,amendola10}.
 However, also other geometrical invariants, besides $R$, can be considered in the Palatini formalism. In \cite{limota}, microscopic and macroscopic behaviors of Palatini modified gravity theories are discussed, in particular a detailed study of $f(R,R^{\mu\nu}R_{\mu\nu})$-models is reported. In \cite{olmoalepuz} dynamical aspects of Palatini $f(R)$, $f(R^{\mu\nu}R_{\mu\nu})$, and   $f(R,R^{\mu\nu}R_{\mu\nu})$-theories are studied. Isotropic and anisotropic bouncing cosmologies in Palatini $f(R)$ and $f(R^{\mu\nu}R_{\mu\nu})$-theories are discussed in \cite{barraolmo}.  A Lagrangian of the type $\mathcal{L}=R+\alpha R_{\mu\nu}R^{\mu\nu} $ is also studied in Palatini formalism in the classical paper \cite{buch1}.
Here, for the sake of simplicity, we will discuss only the cases of $f(R)$-gravity and non-minimally coupled theories in Palatini formalism.
 
%
%
\subsection{The Palatini approach and the conformal structure}

Let us work out examples showing the  role of  conformal 
transformations in the Palatini approach  to 
ETGs  \cite{allemandi}, beginning with fourth order gravity in which the 
difference between metric and Palatini variational principles is  
evident.  The Ricci scalar in $f({\cal 
R})$ is 
$ {\cal R} \equiv {\cal R}( g,\Gamma) \equiv 
g^{\alpha\beta} {\cal R}_{\alpha 
\beta}(\Gamma )$  and is a generalized  Ricci scalar, whereas 
$ {\cal R}_{\mu \nu  }(\Gamma 
)$ is the Ricci tensor of a torsion-free connection 
$\Gamma^{\alpha}_{\mu\nu}$ which,
{\it a priori}, has no relations with the space-time metric 
$g_{\mu\nu}$. The gravitational sector of the theory is 
described by the analytical function $f({\cal R})$, while
$\sqrt{-g}$ denotes the usual  scalar density of weight $1$. 
The field equations derived with the Palatini variational 
principle are
\begin{eqnarray}
&& f^{\prime }({\cal 
R}) {\cal R}_{(\mu\nu)}(\Gamma)-\frac{ f({\cal R})}{2}\,  g_{\mu 
\nu }= T^{(m)}_{\mu\nu} \,, \label{PALffv1} \\
&&\nonumber\\
&& \nabla _{\alpha }^{\Gamma } \left[ \sqrt{-g} \, f' ({\cal  
R} )g^{\mu \nu  } \right] =0 \,, \label{PALffv2}
\end{eqnarray}
where  $\nabla^{\Gamma}_{\mu} $ is the covariant derivative of 
the non-metric connection $\Gamma^{\alpha}_{\mu\nu}$, and we use 
units in which  $8\pi G=1$.  
It is important to stress that Eq. (\ref{PALffv2}) is obtained 
under the assumption that  the matter sector described by 
${\cal L}^{(m)}$ is functionally independent of the 
(non-metric) connection $\Gamma^{\alpha}_{\mu\nu}$; 
however it may contain metric covariant derivatives 
$\stackrel{g}{\nabla}$ of the matter fields. This means that the 
matter stress-energy tensor 
$ T^{(m)}_{\mu\nu} \left[ g,\Psi \right]$ 
depends on the metric $g_{\mu\nu}$ and on the matter fields 
collectively denoted by  $\Psi$, together with 
their covariant derivatives with respect to the 
Levi-Civita connection of $g_{\mu\nu}$. 
It is easy to see from Eq.~(\ref{PALffv2})  that 
$\sqrt{-g} \, f' ({\cal R} ) g^{\mu \nu }$ is a 
symmetric  tensor density of weight $1$, which  naturally 
leads to the introduction of a new metric $h_{\mu \nu}$ 
conformally related to $g_{\mu\nu}$ by  
\cite{Ferraris:1994af, allemandi}
\begin{equation}\label{PALh_met}
\sqrt{-g} \, f' ({\cal R}) \, g^{\mu \nu} = \sqrt{-h} \,\, h^{\mu 
\nu }\,.
\end{equation}
With this definition $\Gamma^{\alpha}_{\mu\nu} $ is  the 
Levi-Civita connection of the metric $h_{\mu\nu}$, with the only 
restriction that the conformal factor
$\sqrt{-g} \, f' ({\cal 
R}) g^{\mu \nu}$ relating $g_{\mu\nu}$ and $h_{\mu\nu}$ be  
non-degenerate. In the case of the Hilbert-Einstein Lagrangian  
it is $f' ({\cal R})=1$ and the statement is trivial.

The conformal transformation 
\begin{equation}\label{PALh_met1}
g_{\mu\nu} \longrightarrow h_{\mu \nu }=f' ({\cal R}) \, g_{\mu 
\nu }
\end{equation}
implies that  ${\cal R}_{(\mu\nu)}(\Gamma)={\cal R}_{\mu \nu 
}(h) $. 
It is useful to consider the trace of the field 
equations (\ref{PALffv1}) 
\begin{equation} \label{PALstructR}
f' ({\cal R}) {\cal R} -2f( {\cal R})= 
g^{\alpha\beta} T^{(m)}_{\alpha\beta}\equiv
 T^{(m)} \,,
\end{equation}
which controls the solutions of Eq. (\ref{PALffv2}). We  
refer to this scalar equation as the \textit{structural equation} 
of space-time. {\em In vacuo}  and in the presence of conformally 
invariant matter with  $T^{(m)}=0$, this scalar equation
admits  constant solutions. In these cases, Palatini $f({\cal 
R})$-gravity reduces to GR with a cosmological constant  
\cite{Ferraris:1994af, SahniStarobinsky00}.
In the case of interaction with matter fields, the structural
equation (\ref{PALh_met1}), if explicitly solvable, provides in 
principle an expression  ${\cal R}=F( T^{(m)} )$  and, 
as a result, both $f({\cal R})$ and $f' ({\cal 
R})$ can be expressed in
terms of $T^{(m)}$. This fact allows one to express, at least 
formally, ${\cal R}$ in terms of $T^{(m)} $, which has 
deep consequences for the description of physical systems, 
as we will see later.  Matter rules  the bimetric structure of 
space-time and, consequently,  both the geodesic and metric 
structures which are intrinsically different. This behaviour 
generalizes the vacuum case. 

Let us now  extend the Palatini formalism to 
non-minimally coupled  scalar-tensor 
theories, with the goal of 
understanding the bimetric structure of space-time  in these 
theories  and  its possible geometric and physical
interpretation.  We denote by $S_1$ the action functional of  
Palatini scalar-tensor theories, while non-minimal 
interaction between scalar-tensor and $f(R)$ gravities will be 
considered later, calling $S_2$ the respective 
action. Then, we will finally consider the case of
 scalar fields $\phi$ non-minimally coupled to   
the gravitational fields $\left( g_{\mu\nu}, 
\Gamma^{\alpha}_{\mu\nu} \right)$, denoting by  $S_3$ the
corresponding action. In this case,  the  low curvature limit  
${\cal R} \rightarrow  0$, which is relevant for the present 
epoch of the history of the  Universe, is particularly 
significant.

The  scalar-tensor action  can be generalized, in
order to better develop the Palatini approach, as
\begin{equation} \label{PALlagrfR1}
S_1=\int d^{4}x \sqrt{-g} \left[  F(\phi) {\cal R} - \frac{\epsilon}{2} \stackrel{g}{\nabla}_\mu  \phi  \stackrel{g}{\nabla}^{ 
\mu} \phi -V(\phi) + {\cal L}^{(m)} \left( \Psi, 
\stackrel{g}{\nabla} \Psi 
\right) \right] \,,
\end{equation}
with  $\epsilon=\pm 1$ corresponding to an ordinary   
scalar or a  phantom  field, respectively.
The field equations for the metric $g_{\mu\nu}$ 
and the connection $\Gamma^{\alpha}_{\mu\nu}$ are 
\begin{eqnarray}\label{PAL40}
&& F(\phi) \left( {\cal R}_{(\mu\nu)}-\frac{1}{2} \, g_{\mu \nu} 
{\cal R}  \right) = T^{(\phi )}_{\mu \nu }+T^{(m)}_{\mu \nu }  
\,, \\
&&\nonumber\\
&& \nabla_{ \alpha }^{\Gamma }\left[ \sqrt{-g} \, F (\phi)g^{\mu 
\nu } \right] =0  \,,
\end{eqnarray}
where ${\cal R}_{(\mu\nu)}$ is defined by Eq.~(\ref{PALffv1}). 
The 
equation  of motion of the matter fields is 
 \begin{eqnarray}\label{PAL42}
&& \epsilon \Box \phi=  V_{\phi} (\phi) + F_{\phi} (\phi) {\cal R} 
\,, \\
&&\nonumber\\
&& \frac{\delta {\cal L}^{(m)}}{\delta \Psi}=0 \,.
\end{eqnarray}
In this case, the structural equation of space-time implies that
 \begin{equation} \label{PALstru1a}
{\cal R}=-\frac{\left( T^{(\phi) } + T^{(m)} \right) }{F(\phi)} 
\,,
\end{equation}
where we must  require that $F(\phi) > 0$. The bimetric
structure of space-time is thus defined by the {\em ansatz}
 \begin{equation}\label{PAL41}
\sqrt{- g} \,\, F (\phi) g^{\mu \nu }=\sqrt{- h} \, \, h^{\mu \nu 
}
 \end{equation}
so that $h_{\mu\nu}$ is conformal to  $g_{\mu\nu}$,
\begin{equation} \label{PALbimetri1}
h_{\mu \nu }=  F (\phi) \, g_{\mu \nu }\,.
 \end{equation}
It follows  from Eq. (\ref{PALstru1a})  that {\em in vacuo} 
$T^{( \phi )}=0$ and $T^{(m)}=0$ this theory is equivalent 
to vacuum GR. If $F(\phi)=F_0=$~const. 
we recover GR with  a minimally coupled scalar
field, which means that the Palatini approach intrinsically gives
rise to the conformal structure (\ref{PALbimetri1}) of the theory
which is trivial in the Einsteinean, minimally coupled, case.
As a further step, let us generalize the previous results to the 
case of non-minimal coupling in the 
framework of
$f(R)$ theories. The action functional can be written as
\begin{equation} \label{PALlagrfR2}
S_2=\int d^{4}x \sqrt{ -g}  \left[ F(\phi) f({\cal R}) -\frac{\epsilon}{2} \, \stackrel{g}{\nabla}_\mu \phi  
\stackrel{g}{\nabla}^{ \mu} 
\phi -V(\phi)+ {\cal L}^{(m)}(\Psi, \stackrel{g}{\nabla} \Psi) 
\right] 
\end{equation}
where $f({\cal R})$ is, as usual, an analytical function of 
${\cal R}$.  
The Palatini field equations  for the gravitational sector are
\begin{eqnarray}
&& F(\phi) \left[ f' ({\cal R} ) {\cal R}_{(\mu\nu)}- 
\frac{f({\cal R})}{2} g_{\mu \nu} \right] = T^{( \phi)}_{\mu \nu 
} + 
T^{(m)}_{\mu 
\nu } \,, \\
&&\nonumber\\
&& \nabla_{\alpha }^{\Gamma }\left[ \sqrt{ -g} F(\phi) 
f' ({\cal R})  g^{\mu \nu } \right]=0 \,.
\end{eqnarray}
The equations of motion for the scalar and matter fields are
\begin{eqnarray}
&& \epsilon \Box \phi=  V_{\phi} (\phi) +  
F_{\phi} (\phi) f({\cal R}) \,, \\
&&\nonumber\\
&&  \frac{\delta {\cal L}^{(m)}}{\delta  \Psi}=0 \,, 
\end{eqnarray}
in which  the non-minimal interaction term enters the 
modified
Klein-Gordon equations. In this case, the structural equation of
space-time implies that
 \begin{equation} \label{PALstru1}
f' ({\cal R}) {\cal R}-2 f({\cal R})=\frac{T^{(\phi)} 
+T^{(m)}}{F(\phi)}\,.
\end{equation}
 The bimetric structure of space-time is given by 
 \begin{equation}\label{PAL47}
\sqrt{- g} \, F (\phi) f' ({\cal R}) g^{\mu \nu }=\sqrt{- h} \,  
h^{\mu \nu}
 \end{equation}
with $g_{\mu\nu}$ and $h_{\mu\nu}$ again conformally 
related, 
 \begin{equation}
h_{\mu \nu }=  F (\phi) f' ({\cal R}) \, g_{\mu \nu }\,.
\end{equation}
Once the structural equation is solved, the conformal 
factor
depends  on the values of the matter fields ($\phi, \Psi$) or,
more precisely, on the traces of their stress-energy tensors and 
the value of $\phi$. {\em In vacuo}, Eq.~(\ref{PALstru1}) implies  
that the theory reduces again to Einstein gravity as for  
minimally interacting $f(R)$ theories \cite{Ferraris:1994af}. The
validity of this property is related to the decoupling of the
scalar field from the metric.

Finally, let us  discuss  the situation in which  the 
gravitational Lagrangian is a general function of $\phi$ and 
${\cal R}$, 
as in 
\begin{equation} \label{PALlagrfR3}
S_3=\int d^{4}x \sqrt{ -g} \left[  K\left( \phi, {\cal R} \right) 
- \frac{\epsilon}{2} \stackrel{g}{\nabla}_\mu \phi  
\stackrel{g}{\nabla}^{ \mu} \phi
-V(\phi)+  {\cal L}^{(m)} \left( \Psi, \stackrel{g}{\nabla} \Psi 
\right) \right] \,,
\end{equation}
which yields the gravitational field equations 
\begin{eqnarray}
&&   \frac{\partial \; K \left( \phi, {\cal R} \right)}{ \partial 
{\cal R}}
  \, 
{\cal R}_{(\mu\nu)}- \frac{ K(\phi, {\cal R})}{2} \,   g_{\mu 
\nu}  =  T^{(\phi)}_{\mu \nu }+T^{(m)}_{\mu \nu } \,,\\
&&\nonumber\\
&& \nabla _{\alpha }^{\Gamma } \left[
\sqrt{ -g} \, \frac{\partial \; K(\phi, {\cal R})}{\partial  
{\cal R}} \,
g^{\mu \nu } \right] =0\,, 
\end{eqnarray}
while the scalar and matter fields obey
\begin{eqnarray}
&& \epsilon \Box \phi=  V_{\phi} (\phi) + 
\frac{\partial \; K(\phi, {\cal R})}{\partial \phi }  \
\,,\\ 
&&\nonumber\\
&&\frac{\delta {\cal L}^{(m)} }{\delta \Psi}=0\,. 
\end{eqnarray}
The structural equation of space-time can be expressed as
 \begin{equation} \label{PALstru2}
\frac{\partial K(\phi, {\cal R})}{\partial {\cal R}} \, {\cal 
R}-2 K(\phi, {\cal R}) =
T^{(\phi)} + T^{(m)} \,.
\end{equation}
When  solved, Eq. (\ref{PALstru2}) provides again the form of 
the Ricci scalar  in terms of the traces of the 
stress-energy tensors of matter and of the scalar field (with   
$K(\phi,{\cal R}) >  0$). 
The 
bimetric structure of space-time is defined by 
 \begin{equation}
\sqrt{- g} \,\, \frac{\partial K(\phi, {\cal R})}{\partial {\cal 
R}}  \, g^{\mu \nu }=\sqrt{- h} \,\, h^{\mu \nu }
 \end{equation}
with 
 \begin{equation} \label{PALconf3}
h_{\mu \nu }=  \frac{\partial K(\phi, {\cal R})}{\partial {\cal 
R}} \, g_{\mu 
\nu }\,.
 \end{equation}
The conformal factor   depends  on  the 
matter 
fields only through the traces of their stress-energy tensors. 
The 
conformal factor   and the bimetric 
structure are ruled by 
these traces and by the value of the scalar field $\phi$.
In this case, in general, one does not recover GR, as is 
evident from Eq. (\ref{PALstru2}) in which the strong 
coupling between ${\cal R}$ and $\phi$ prevents,  even  {\em in 
vacuo}, the possibility of obtaining  constant
solutions.

Let us discuss  the  $R \rightarrow 0 $ regime,  a good 
approximation to  the present epoch of the observed Universe.  
The linear expansion of the analytical function $K \left( \phi, 
{\cal R} \right)$ 
\begin{equation}
K(\phi, {\cal R})=K_0 (\phi)+K_1 (\phi) {\cal R}+ {\mbox O}( 
{\cal R}^2 )
 \end{equation}
with 
 \begin{equation}
 K_0 (\phi) =  K(\phi, {\cal R}) \left.\right|_{{\cal R}=0} \,, 
\;\;\;\;\;  K_1 
(\phi)=\left(
\frac{\partial K(\phi,{\cal  R})}{ \partial {\cal R} } 
\right) \left.\right|_{{\cal R}=0} \,,
 \end{equation}
can be substituted into Eqs. (\ref{PALstru2}) and 
(\ref{PALconf3}) obtaining, to first order,  the structural
equation and the bimetric structure. The structural equation 
yields 
\begin{equation} \label{PALRgenapr}
{\cal  R}=\frac{-1}{K_1 (\phi) } \left[ T^{(\phi)} + T^{(m)}) + 2
K_0 (\phi) \right]
 \end{equation}
and the value of the Ricci scalar is always determined, in
the linear  approximation, in terms of 
$T^{(\phi)}$, $ T^{(m)}$, and $ \phi$. The bimetric 
structure is, otherwise, simply defined by  the first 
term of the Taylor expansion, which is
\begin{equation}
h_{\mu \nu }=  K_1 (\phi) \, g_{\mu \nu } 
\end{equation}
reproducing, as expected, the scalar-tensor case
(\ref{PALbimetri1}).  Scalar-tensor theories  can then be
recovered as the linear  approximation of a general theory in 
which  gravity and the non-minimal couplings  are arbitrary 
(cf.  Eqs. (\ref{PALRgenapr}) and (\ref{PALstru1})). This fact 
agrees with the above considerations when the Lagrangians of 
physical interactions can be considered as locally 
gauge-invariant stochastic functions \cite{BarrowOttewill83}.
Finally, there exist also bimetric theories which cannot be 
conformally related \cite{Will93} and torsion
will also appear  in the most general framework  \cite{hehlrev2, 
Capozziello:2001mq}. These  more  general theories will not be discussed 
here.

\subsection{Equivalence between $f(R)$ and scalar-tensor 
gravity}
\label{sec:chap3sec6}

Metric and Palatini $f(R)$ gravities are equivalent to 
scalar-tensor theories  with the 
derivative of the  function $f(R)$ 
playing the role of the Brans-Dicke scalar, as has been 
re-discovered several times  \cite{Higgs59, 
TeyssandierTourrenc83,  Whitt84, Wands94, Chiba03PLB}. We 
illustrate this equivalence beginning with the metric 
formalism.  

\subsubsection{Equivalence between scalar-tensor and metric $f(R)$-gravity} 

In metric $f(R)$-gravity, we introduce the scalar $
\phi \equiv R $; then the action
\begin{equation}
\label{EQUIV6}
S = \frac{1}{2\kappa}\int d^{4}x\sqrt{-g} \, f(R) + S^{(m)}
\end{equation}
is rewritten in the form 
\cite{Higgs59, TeyssandierTourrenc83,  Whitt84, Wands94, 
Chiba03PLB}
\begin{equation}
\label{EQUIV7}
S = \frac{1}{2\kappa}\int d^{4}x\sqrt{-g}\left[ \psi(\phi)R - 
V(\phi)  \right] +  S^{(m)}
\end{equation}
when $f^{\prime\prime}(R) \neq 0$, where
\begin{eqnarray}
\label{EQUIV8}
\psi = f^\prime(\phi) \,, &
\,\,\,\,\,\,\,\,\,\,\,\,\,\,\;\;\;  V(\phi) = \phi 
f^\prime(\phi) - f(\phi)  \,.
\end{eqnarray}
It is trivial to see that the action~(\ref{EQUIV7}) coincides 
with (\ref{EQUIV6}) if $\phi = R$. {\em Vice-versa},   let us 
vary the  action~(\ref{EQUIV7}) with respect to $\phi$, which 
leads to 
\begin{equation}\label{EQUIVquesta}  
R \, \frac{d\psi}{d\phi}-\frac{dV}{d\phi} = 
\left(R-\phi\right)f^{\prime\prime}(R) = 0 \,.
\end{equation}
Eq.~(\ref{EQUIVquesta}) implies  that  $\phi = R$  when 
$f^{\prime\prime}(R)  
\neq 0$. The action~(\ref{EQUIV7}) has the Brans-Dicke  form 
\begin{equation}
S = \frac{1}{2\kappa}\int d^{4}x\sqrt{-g}\left[ \psi R - 
\frac{\omega}{2} \, 
\nabla^{\mu}\psi\nabla_{\mu}\psi - U(\psi)\right] + S^{(m)}
\end{equation}
with Brans-Dicke field $\psi$, Brans-Dicke parameter $\omega=0$,  
and potential  $U(\psi) =  V\left[  \phi(\psi) \right]$.  An 
$\omega=0$ Brans-Dicke theory   was 
originally studied for the purpose of obtaining  a Yukawa 
correction to the 
Newtonian potential in the weak-field limit \cite{OHanlon72} and 
called ``O'Hanlon theory'' or ``massive 
dilaton gravity''. 
The variation of the action~(\ref{EQUIV7})  yields the field 
equations  
\begin{eqnarray}
&& G_{\mu\nu} = \frac{\kappa}{\psi} \, T_{\mu\nu}^{(m)} 
-\frac{1}{2\psi}\,  U(\psi)g_{\mu\nu}+\frac{ 1}{\psi} 
\left(\nabla_{\mu}\nabla_{\nu}\psi-g_{\mu\nu}\Box \psi\right) 
\,,\\
&&\nonumber\\
&& 3\Box \psi+2  U(\psi)-\psi\,\frac{dU}{d\psi}=\kappa \, 
T^{(m)} \,.
\end{eqnarray}
%

\subsubsection{Equivalence between scalar-tensor and Palatini $f(R)$- gravity} 

Palatini $f(R)$-gravity   is also 
equivalent to a special 
Brans-Dicke theory  with a scalar field 
potential. The Palatini 
action
\begin{equation}
S = \frac{1}{2\kappa}\int d^{4}x\sqrt{-g}\, f({\cal R}) + 
S^{(m)} 
\end{equation}
is equivalent to 
\begin{equation}
\label{EQUIV14}
S = \frac{1}{2\kappa} \int d^{4}x \sqrt{-g}\left[ 
f(\chi)+f^\prime(\chi)\left 
(\cal{R}-\chi\right)\right] + S^{(m)} \,.
\end{equation}
It is straightforward to see that the variation of this action  
with respect to $\chi$ yields  $ \chi = 
\cal{R}$.  We can now  use the field $\phi \equiv  
f^\prime(\chi)$ and 
the  fact that the  curvature $\cal{R}$ is the (metric) Ricci 
curvature of the new metric $ h_{\mu\nu} = 
f^\prime({\cal R}) \, g_{\mu\nu} $ conformally related   to 
$ g_{\mu\nu} $, as already  explained.  Using now the  
well known transformation  property of  the Ricci scalar under 
conformal rescalings  \cite{Synge55,Wald84}
\begin{equation}
{\cal R}= R+ \frac{3}{2\phi}\nabla^{\alpha} 
\phi\nabla_{\alpha}\phi-\frac{3}{2}\Box \phi  
\end{equation}
and discarding a  boundary term, the action~(\ref{EQUIV14}) can 
be presented in the form
\begin{equation}
\label{EQUIV17}
S = \frac{1}{2\kappa} \int d^{4}x\sqrt{-g} \left[ \phi R + 
\frac{3}{2\phi} \, \nabla^{\alpha}\phi\nabla_{\alpha}\phi - 
V(\phi)\right] + S^{(m)} \,,
\end{equation}
where 
\begin{equation}
V(\phi)=\phi\chi(\phi)-f\left[ \chi(\phi)\right] \,.
\end{equation}
This action is clearly that of  a Brans-Dicke 
theory 
with  Brans-Dicke parameter $\omega= 
-3/2$ and a potential. This theory has been studied occasionally 
in the literature \cite{Anderson71, 
OHanlonTupper72, Deser70, OHanlon72, 
Davidson05, DabrowskiDenkiewiczBlaschke07}, but it turns out to 
be a pathological case \cite{Flanagan:2003rb,
Olmo:2008ye, Barausse:2007pn, Barausse:2007ys, Barausse:2008nm, 
FaraoniPLB08}.

A study of anisotropic singularities in non-minimally coupled modified gravity models is reported in \cite{ferraz}.

\section{Conformal transformations and Extended Theories of
Gravity}
\label{sec:chap3sec7}

We have already mentioned  the Jordan  and 
the Einstein frame 
on several occasions: it is now time to look in detail at the 
conformal 
transformations providing different representations of ETGs and 
a solution-generating technique. 
In this subsection  we present of conformal transformations
application of conformal transformations to 
Brans--Dicke gravity first, and  then to more general 
scalar-tensor and $f(R)$ theories.

\subsection{The case of Brans-Dicke gravity}
\label{sec:chap3sec7subsec1}

In Brans-Dicke theory the choice of 
conformal factor  
\cite{Dicke62}
\begin{eqnarray} \label{cft45} 
\Omega=\sqrt{ G\phi} 
\end{eqnarray} 
in  the conformal transformation $g_{\mu\nu}\rightarrow 
\tilde{g}_{\mu\nu}=\Omega^2 g_{\mu\nu}$  brings the gravitational 
sector of the  
Brans-Dicke action
\begin{eqnarray}
S_{BD}=\int d^4x \, \sqrt{-g} \left[ \phi R -\frac{\omega}{\phi} 
\, g^{\alpha\beta} \nabla_{\alpha}\phi \nabla_{\beta}\phi 
-V(\phi)  \right] +S^{(m)}
\end{eqnarray} 
into the Einstein frame form. Then the  
scalar field redefinition
\begin{eqnarray} \label{cft46} 
\tilde{\phi}( \phi)= \sqrt{ \frac{2\omega+3}{16\pi G} } \,
\ln \left( \frac{\phi}{\phi_0} \right) \,,
\end{eqnarray} 
with  $\phi > 0$ and  $\omega > -3/2$ transforms  the 
scalar field kinetic energy density into canonical form. In 
terms of the variables $\left( \tilde{g}_{\mu\nu}, 
\tilde{\phi} \right)$, the Brans-Dicke action assumes its {\em 
Einstein frame}  form
\begin{eqnarray} 
S_{BD}& =& \int d^4 x \, \left\{ \sqrt{ -\tilde{g}} \left[
\frac{ \tilde{R}}{16\pi G} -\frac{1}{2} \, \tilde{g}^{\alpha\beta}
\tilde{\nabla}_{\alpha}\tilde{\phi} 
\tilde{\nabla}_{\beta}\tilde{\phi} -U\left(
\tilde{\phi} \right) \right] + \exp \left( -8\sqrt{ \frac{\pi G}{2\omega +3} } \,\,
\tilde{\phi} \right) {\cal L}^{(m)} 
\left[ \tilde{g} \right] \right\} \,,
\label{cft47} 
\end{eqnarray} 
where $\tilde{\nabla}_{\alpha}$ is the covariant derivative
operator of the rescaled metric $\tilde{g}_{\alpha\beta}$ and 
\begin{eqnarray}  \label{cft47bis}
U\left( \tilde{\phi} \right) = V\left[ \phi \left( \tilde{\phi} 
\right) \right] \exp \left( -8 \sqrt{\frac{\pi G}{2\omega+3} } \, 
\tilde{\phi}
\right) =\frac{V(\phi)}{\left( G\phi \right)^2} 
\end{eqnarray}
is the Einstein frame potential.   The restriction of the 
parameter range to $\omega > -3/2$ is sometimes attributed to the 
need of guaranteeing that it is possible to perform  
the conformal transformation. However,  one 
could take the absolute value $\left| 2\omega +3 \right| $ 
there, but in actual fact $\omega$ cannot cross the barrier 
$-3/2$: the $\omega=-3/2$ Brans-Dicke theory is pathological.  With a special potential, 
$\omega=-3/2$ Brans-Dicke theory  is 
equivalent to Palatini $f(R)$ 
gravity.
The Jordan frame  scalar has the dimensions 
of $G^{-1}$, while the 
Einstein frame   scalar $\tilde{\phi}$ has the dimensions of 
$G^{-1/2}$ and is usually   measured in Planck  masses.
In the GR limit $\phi \rightarrow$ const., the
Jordan  and the Einstein frames  coincide.
 However an important remark is necessary at this point. Though the original Brans-Dicke theory, for which the potential $V(\phi)$ is set to zero, is pathological for $\omega=-3/2$, it is not necessarily true in general. In fact, as soon as the potential is different from zero, the pathology can be removed. In this general sense, some models within the  $\omega=-3/2$ are pathological but not all of them. It is worth noticing that some successful  applications to the early Universe of this type of theories exist in literature showing that not all $\omega=-3/2$ are pathological, see {\em e.g.}  \cite{barra1,Kainulainen:2007bt,kai1,olmo3}. 

It is also important to stress that the Einstein frame representation of Brans-Dicke theory with $\omega=-3/2$ or $\omega<-3/2$ can be perfectly formulate as long as one provides the suitable generalization of Eq. (\ref{cft46}). In this sense the case $\omega=-3/2$ sets a frontier between standard scalar fields and phantom fields, because, for $\omega<-3/2$, the kinetic term of the scalar field has the "wrong" sign in front of it. A detailed discussion of this point is reported in \cite{DamourFarese92}.

The inspection of  the action (\ref{cft47}) often leads people to 
state 
that, in the Einstein frame,  gravity is described by GR, but 
there are  two important differences between Einstein 
frame  
Brans-Dicke gravity  and Einstein's 
theory. First, the free scalar  $\tilde{\phi}$ acting as a source 
of gravity on the right hand  side of the field equations is  
always present, {\em i.e.}, in the Einstein frame   solutions of 
the vacuum  field equations $\tilde{R}_{\mu\nu}=0$ cannot be 
obtained as in vacuum GR because the scalar  $\tilde{\phi}$ 
pervades the  space-time manifold and  cannot be removed. This 
persistence is a reminder of the 
cosmological origin of $\phi \sim G_{eff}^{-1}$  in the original 
(Jordan frame)  Brans-Dicke 
theory 
\cite{BransDicke61}. The 
scalar  $\tilde{\phi}$ is always present even if,  
formally, the gravitational field is only  described by the 
metric 
tensor   $\tilde{g}_{\alpha\beta}$ in the
Einstein frame.  The conformal transformation shifts 
the Jordan frame  gravitational variable 
$\phi$ into Einstein
frame   matter\footnote{This property makes 
it 
clear that the distinction between gravitational and 
non-gravitational degrees of freedom depends on the conformal 
representation of a gravitational theory.} $\tilde{\phi}$.

The second difference between GR and Einstein 
frame Brans-Dicke 
theory  consists of the fact that  the 
matter  Lagrangian ${\cal 
L}^{(m)}$ is now multiplied by the exponential factor in 
Eq.~(\ref{cft47}). This factor is  described as an 
anomalous coupling of  matter to the scalar 
$\tilde{\phi}$ which has no counterpart in  GR. It is because 
of this coupling 
that the  matter energy-momentum tensor
$\tilde{T}_{\alpha\beta}^{(m)}$  in the 
Einstein frame  obeys $\tilde{\nabla}^{\beta} \, \tilde{T}_{\alpha\beta}^{(m)} = - \tilde{T}^{(m)} 
  \,\tilde{\nabla}_{\alpha} \left(  \ln \Omega \right)$
 instead of the  GR 
conservation equation  $\nabla^{\beta}
\,  T_{\alpha\beta}^{(m)} =0$. The modified conservation equation  
implies changes to the geodesic
equation and to the equation of geodesic deviation, and the
violation of
the  Equivalence Principle   in the 
Einstein frame.
Under the conformal transformation, the 
matter energy-momentum tensor $T_{\mu\nu}^{(m)}$ scales as 
\begin{eqnarray}
\tilde{T}^{\alpha\beta}_{(m)}=\Omega^s \, \,  
T^{\alpha\beta}_{(m)} \,, \;\;\;\;\;\;\;\;\;
\tilde{T}_{\alpha\beta}^{(m)}=\Omega^{s+4} \,\,  
T_{\alpha\beta}^{(m)} \,, 
\end{eqnarray}
where $s$ is an appropriate conformal weight. The conservation
equation $\nabla^{\beta} \,  T_{\alpha\beta}^{(m)}= 0$ transforms 
(in four space-time dimensions) as \cite{Wald84}
\begin{eqnarray} \label{cftx1}
\tilde{\nabla}_{\alpha} \left( \Omega^s \,  T^{\alpha\beta}_{(m)} 
\right) =\Omega^s
\, \nabla_{\alpha} T^{\alpha\beta}_{(m)}
+\left( s+6 \right) \Omega^{s-1} \,  
T^{\alpha\beta}_{(m)}\nabla_a \Omega -
\Omega^{s-1} g^{\alpha\beta} \, T^{(m)} \nabla_{\alpha} \Omega 
\,.
\end{eqnarray}
It is
convenient to choose the conformal weight $s=-6$ which yields,
consistently with $\tilde{T}^{(m)}= \Omega^{-4} \, T^{(m)}$,
\begin{eqnarray} \label{cftx3}
\tilde{T}^{(m)} \equiv    \tilde{g}^{\alpha\beta} \, 
\tilde{T}_{\alpha\beta}^{(m)}= \Omega^{-4}  \,\, T^{(m)} \,,
\end{eqnarray}
and  $\tilde{T}^{(m)}$ vanishes if and only if $T^{(m)}=0$. 
Eq.~(\ref{cftx1}) assumes the form
\begin{eqnarray} \label{cftx3bis}
\tilde{\nabla}_{\alpha}  \tilde{T}^{\alpha\beta}_{(m)}  =- 
\tilde{T}^{(m)}\,
\tilde{g}^{\alpha\beta} \, \tilde{\nabla}_{\alpha}
\left(  \ln \Omega \right)  \,  . 
\end{eqnarray}
Since $\Omega=\sqrt{G\phi}$ it is  
\begin{eqnarray} \label{cftx4}
\tilde{\nabla}_{\alpha}  \tilde{T}^{\alpha\beta}_{(m)}   =- 
\frac{1}{2\phi} \,\, \tilde{T}^{(m)}
 \,  \tilde{\nabla}^{\beta} \phi
\end{eqnarray}
or, in terms of the Einstein frame   scalar 
\cite{Wagoner70},
\begin{eqnarray} \label{cftx5}
\tilde{\nabla}_{\alpha}  \tilde{T}^{\alpha\beta}_{(m)} = - \sqrt{ 
\frac{4\pi G}{2\omega
+3} } \, \, \tilde{T}^{(m)} \,\, \tilde{\nabla}^{\beta} 
\tilde{\phi} \,.
\end{eqnarray}
The geodesic equation receives corrections as a consequence 
of  Eq.~(\ref{cftx5}). Consider a dust fluid with  
energy-momentum tensor
\begin{eqnarray} \label{cftx6}
\tilde{T}_{\alpha\beta}^{(m)}=\tilde{\rho}^{(m)} \, 
\tilde{u}_{\alpha} \, \tilde{u}_{\beta} \,;
\end{eqnarray}
Eq.~(\ref{cftx5}) yields
\begin{eqnarray} \label{cftx7}
\tilde{u}_{\alpha} \, \tilde{u}_{\beta} \, \tilde{\nabla}^{\beta}  
\tilde{\rho}^{(m)} 
+\tilde{\rho}^{(m)} 
\, \tilde{u}_{\alpha} \, \tilde{\nabla}^{\beta}  
\tilde{u}_{\beta}
+\tilde{\rho}^{(m)} \,
\tilde{u}_{\gamma} \,\tilde{\nabla}^{\gamma}\, \tilde{u}_{\alpha}
= \sqrt{ \frac{4\pi G}{2\omega +3} } \, \,
\tilde{\rho}^{(m)} \, \tilde{\nabla}_{\alpha} \tilde{\phi}  \,.
\end{eqnarray}
Using an affine parameter $\lambda$ along the fluid 
worldlines with tangent $\tilde{u}^{\mu}$, Eq.~(\ref{cftx7}) 
becomes 
\begin{eqnarray} \label{cftx8}
\tilde{u}_{\alpha} \left(
\frac{d\tilde{\rho}^{(m)} }{d\lambda}+\tilde{\rho}^{(m)} \,  
\tilde{\nabla}^{\gamma}
\tilde{u}_{\gamma} \right) +\tilde{\rho}^{(m)} 
\left(  \frac{ d\tilde{u}_a}{d\lambda}
-\, \sqrt{ \frac{4\pi G}{2\omega+3}} \,  
\tilde{\nabla}_{\alpha}\phi \right)=0 \,.
\end{eqnarray}
This equations splits into the two equations
\begin{eqnarray} \label{cftx9}
\frac{d\tilde{\rho}^{(m)} }{d\lambda} + \tilde{\rho}^{(m)} \, 
\tilde{\nabla}^{\gamma}
\tilde{u}_{\gamma} =0 
\end{eqnarray}
and
\begin{eqnarray} \label{cftx10}
\frac{d\tilde{u}^{\alpha} }{d\lambda} =\sqrt{ \frac{4\pi 
G}{2\omega+3}} \, \,   
\tilde{\nabla}^{\alpha} \tilde{\phi} \,.
\end{eqnarray}
The geodesic equation is then modified in the Einstein 
frame as 
\cite{Wagoner70,Cho92,Cho97}
\begin{eqnarray} \label{cftx11}
\frac{d^2 x^{\mu}}{d\lambda^2} +\tilde{\Gamma}_{\nu\sigma}^{\mu} 
\,  
\frac{d x^{\nu}}{d\lambda} \, 
\frac{d x^{\sigma}}{d\lambda} = 
\sqrt{ \frac{4\pi G}{2\omega + 3}} \, \, \tilde{\nabla}^{\mu} 
\tilde{\phi} \,.
\end{eqnarray}
The correction on the right hand side  is often described as a 
fifth force proportional to the gradient $\tilde{\nabla}^{\mu} 
\tilde{\phi}$ that couples 
universally  to all massive test particles. The Weak Equivalence 
Principle (universality of 
free 
fall)  is violated by this fifth force because of the 
space-time dependence of $\tilde{\nabla}^{\mu} 
\tilde{\phi}$.  Due to this coupling, scalar-tensor  
theories in the Einstein
frame   appear to be  non-metric 
theories.
On the other hand,  it 
is well known  that all metric theories of gravity satisfy the 
Weak Equivalence Principle 
\cite{Will93}, and the (non-)metricity 
becomes a statement on whether a theory satisfies or not it. Therefore, the metric  character 
of ETGs, and whether they satisfy 
or not the Equivalence Principle, 
become  properties dependent 
on the conformal frame representation. This fact leaves the 
foundation of relativistic gravity on a rather shaky ground, 
which is a problem especially when trying to isolate the 
fundamental properties of classical gravity which should be 
preserved in approaches to quantum or emergent 
gravity. 

As expected, the equation of null geodesics is left unaffected 
by the conformal transformation: null geodesics receive no fifth 
force correction  in the
Einstein frame.   This invariance is 
consistent with the fact  that  
the equation of null geodesics can be derived from the 
Maxwell
equations in the  high frequency limit of the geometric optics 
approximation, in conjunction with the fact that Maxwell's 
equations are 
conformally invariant in a four-dimensional manifold.   
A more direct way of looking at conformal invariance for null 
geodesics   is by noting  that the  electromagnetic field 
stress-energy tensor has vanishing trace $T=0$ and the 
corresponding  conservation equation  $\nabla^{\beta} 
\, T_{\alpha\beta}=0$ is unaffected by the conformal 
rescaling $g_{\mu\nu}\rightarrow \tilde{g}_{\mu\nu}=\Omega^2 \, 
g_{\mu\nu}$, together 
with the geodesic equation for a null dust described by 
Eq.~(\ref{cftx6}) when $\tilde{u}_{\mu} \tilde{u}^{\mu} =0$.

A correction to the timelike geodesic equation similar to the one 
discovered in  Brans-Dicke theory 
appears in the low-energy limit 
of  string theory
\cite{TaylorVeneziano88,Gasperini99}, in which 
the dilaton replaces the Brans-Dicke field and  a similar 
coupling violates the Equivalence Principle
\cite{TaylorVeneziano88, Cho92, 
DamourPolyakov94, Cho97}. The violation is  kept small in order 
not to violate the Solar System  bounds \cite{Will93}. In 
the low-energy limit of string theory   the  
dilaton  couples  with 
different strengths to bodies of different nuclear composition  
which carry a dilatonic charge  $q$, contrary to the Brans-Dicke 
field which couples universally to all forms of 
non-conformal matter. The formal substitution of the dilatonic  
charge $q$ with the factor $2\sqrt{ \pi G/  \left( 2\omega +3 
\right)} $ allows  a parallel between the two 
theories, but  in string theory   it may be 
possible 
to  eliminate the coupling  by setting the dilatonic charge $q$ 
to  zero in certain cases, whereas the coupling of the Einstein 
frame  Brans-Dicke scalar  cannot be 
eliminated.
\subsection{Scalar-tensor theories}  

More general scalar-tensor theories   are  described by the 
Jordan frame action
\begin{eqnarray} 
S_{ST} &  = &\int d^4 x \, \sqrt{-g} \, \left\{ \frac{1}{16 \pi 
G}  \left[  \phi
R -\frac{ \omega (\phi)}{\phi} \, g^{\alpha\beta}\nabla_{\alpha} 
\phi \, \nabla_{\beta} \phi
\right] - V( \phi) +  \alpha_m \, {\cal L}^{(m)} \right\} \,, 
\label{cft71}
\end{eqnarray}
where  $\alpha_m$ is the coupling constant of ordinary 
matter.  The conformal factor is still 
given by Eq.~(\ref{cft45}) 
while the Einstein frame scalar field is 
defined by the 
differential relation
\begin{eqnarray}  \label{cft72}
d \tilde{\phi}  =\sqrt{  \frac{  2\omega ( \phi )+3}{16\pi G} } 
\,\, \frac{d\phi}{\phi} \,.
\end{eqnarray}
The Einstein frame   scalar-tensor action 
is 
\begin{eqnarray}
S_{ST}&=& \int d^4 x \, \sqrt{-\tilde{g}} \, \left[ 
\frac{\tilde{R}}{16 \pi G}
 -\frac{1}{2} \, \tilde{g}^{\alpha\beta} \, 
\tilde{\nabla}_{\alpha} \tilde{\phi}
\, \tilde{\nabla}_{\beta} \tilde{\phi} -U \left( \tilde{\phi} 
\right)
+\tilde{\alpha}_m \left( \phi \right)  \, {\cal L}^{(m)} \right]
\nonumber \\
&&         \label{cft72bis}
\end{eqnarray}
with scalar field potential 
\begin{eqnarray} \label{cftciuccio}
U \left( \tilde{\phi} \right)= \frac{ V\left[ \phi ( \tilde{\phi} )
\right]}{\left( G\phi \right)^2} 
\end{eqnarray}
and coupling 
\begin{eqnarray}
\tilde{\alpha}_m \left(  \tilde{\phi} \right)= \frac{ \alpha_m }{ 
\left( G\phi \right)^2 } \,.
\end{eqnarray}
Again, Eq.~(\ref{cft72bis}) can be seen as  the action for 
GR  with a  canonical scalar field which has 
positive-definite  kinetic energy  density, but with the
important difference that the matter Lagrangian density  
is multiplied by the factor $\Omega^{-4}=\left( G\phi 
\right)^{-2} $, which can be interpreted as  a variation of  
the  coupling constant $\alpha_m$ with space and/or time.   
Again, this matter-$\tilde{\phi}$ coupling is  responsible for 
the non-conservation of  $\tilde{T}_{\mu\nu}^{(m)}$ 
as in Eq.~(\ref{cftx5}), and for violating   the
Equivalence Principle.
The conformal transformation technique has been used as a tool  
for  generating exact solutions of a scalar-tensor   
theory beginning from known solutions of GR \cite{Harrison72, 
BelinskiiKhalatnikov73,  Lorentz84, 
VandenBergh80, VandenBergh82, VandenBergh83a, VandenBergh83b, 
VandenBergh83c,  VandenBergh83d, BarrowMaeda90},  and for  
deriving approximate solutions  of the linearized theory 
\cite{BarrosRomero98}. This solution-generating technique is most  
convenient for solutions  with vanishing potential: in fact, when 
a potential $V( \phi)$ is present,  solutions 
that correspond to a  physically well  motivated potential in 
one frame generate, via  
the conformal mapping, solutions in the other frame which  rarely 
correspond to a physical potential. Consider,  
for example, Brans-Dicke theory   with a mass 
term  $V( \phi)=m^2 \phi^2/2$ in the Jordan 
frame. The Einstein frame  potential is 
\begin{eqnarray} \label{cft74}
U=\frac{1}{2} \, \left( \frac{m}{G} \right)^2 \,,
\end{eqnarray}
{\em i.e.}, a cosmological constant. A given functional form of
 $V( \phi)$ in the Jordan frame
corresponds to a very different form of $U \left( \tilde{\phi} 
\right) $ in the Einstein frame.  
Reversing the problem, which  
Jordan  frame potential $V(  \phi)$  produces a 
mass term $U\left( \tilde{\phi} \right) =m^2 \tilde{\phi}^2 /2$ 
in the Einstein frame?   
Eqs.~(\ref{cft46}) and 
(\ref{cftciuccio})  yield the answer  
\begin{eqnarray} \label{cft77}
V \left( \phi \right)=U \left( \tilde{\phi} \right) \phi^2=
\frac{m^2 G}{32\pi }  \, \left( 2\omega+3 \right) \left[ \phi \ln 
\left(
\frac{\phi}{\phi_1}  \right) \right]^2 \,,
\end{eqnarray}
with $\phi_1$ is a constant. It would be   difficult 
to motivate this potential from a known theory of particle 
physics. As a conclusion, it is  legitimate to  use exact  
solutions in the Einstein frame   to 
generate solutions in the  
Jordan frame, but this procedure usually produces solutions of 
limited physical interest.

In $D> 2 $ space-time dimensions, the 
scalar-tensor  theory described by the action
\begin{eqnarray}   
S_{ST}^{(D)} =\int d^Dx\,\, \sqrt{-g} \left[ f( \phi ) R -\omega( 
\phi ) 
\nabla^{\alpha}\phi
\nabla_{\alpha} \phi -V(\phi) + \alpha_m \, {\cal L}^{(m)} 
\right] \,,
\end{eqnarray}

can be conformally transformed according to 
\begin{eqnarray}  \label{1B2}
g_{\alpha\beta} \longrightarrow \tilde{g}_{\alpha\beta} =f( 
\phi)^{ \frac{2}{D-2}} 
\, g_{\alpha\beta} \,,
\end{eqnarray}

and 
\begin{eqnarray}  \label{1B3}
d\tilde{\phi}= \frac{d\phi}{f( \phi)} \, \sqrt{ f\left( \phi 
\right)  +\frac{D-1}{D-2}
\left( \frac{df}{d\phi} \right)^2} 
\end{eqnarray}

producing the new scalar field potential in the Einstein 
frame 
\begin{eqnarray}
U \left( \tilde{\phi} \right)=\frac{  V\left[ \phi \left( 
\tilde{\phi}
\right) \right]}{f(\phi)^{\frac{D}{D-2}} } \,.
\end{eqnarray}

%
%

\subsection{$f(R,\phi)$-gravity}

The action of generalized scalar-tensor  gravity  
\begin{eqnarray} \label{cft79}
S=\frac{1}{16\pi} \int d^4 x \, \sqrt{-g} \, \left[ f\left( 
\phi ,  R \right) -\, \frac{ \epsilon}{2} \, 
g^{\alpha\beta}\nabla_{\alpha} \phi \nabla_{\beta} 
\phi \right] 
\end{eqnarray}

is mapped  into its Einstein frame form  
by a conformal 
transformation   which 
was rediscovered many times  in particular realizations 
\cite{Whitt84, TeyssandierTourrenc83, Schmidt87, Starobinsky87, 
BarrowCotsakis88, Maeda89, Gottetal90, CotsakisSaich94, Wands94}.  
The conformal factor is 
\begin{eqnarray} \label{cft82}
\Omega=\left[  16 \pi G \left| \frac{\partial f}{\partial R} \right|
+\mbox{constant} \right]^{1/2}
\end{eqnarray}

and, together with  the scalar field redefinition
\begin{eqnarray} \label{cft83}
\tilde{\phi}= \frac{1}{\sqrt{ 8\pi G} } \,\sqrt{ \frac{3}{2}} \, \ln
\left[ \sqrt{32\pi} \, G \left| 
\frac{\partial f}{\partial R} \right| \, \right] \,,
\end{eqnarray}

allows the action (\ref{cft79}) to be rewritten in the 
Einstein frame form
\begin{eqnarray} \label{cft84}
S & = & \alpha \int d^4 x \, \sqrt{-\tilde{g}} \, \left\{
\frac{\tilde{R}}{16\pi
G} -\frac{1}{2} \, \tilde{g}^{\alpha\beta} 
\tilde{\nabla}_{\alpha} 
\tilde{\phi}
\tilde{\nabla}_{\beta} \tilde{\phi}- \frac{\epsilon \, \alpha}{2} \exp \left[
-\sqrt{\frac{16\pi G}{3} }\,\,\tilde{\phi} \right] -U \left( \phi,
\tilde{\phi} \right) \right\} \,,
\end{eqnarray}
where \cite{Maeda89,Hwang90CQG}
\begin{eqnarray} \label{cft85}
\alpha = \mbox{sign} \left(  
\frac{\partial f}{\partial R} \right) \,,
\end{eqnarray}

and 
\begin{eqnarray} 
U \left( \phi, \tilde{\phi} \right) & = & \alpha 
 \, \mbox{e}^{ 
-8\sqrt{ \frac{\pi G}{3} } \,\,\tilde{\phi} }  
\left[ \frac{\alpha R\left( \phi , \tilde{\phi} \right)}{16\pi G } 
\exp  \left( \sqrt{  \frac{ 16\pi G}{ 3} } \, \tilde{\phi}  \right) 
- F\left( \phi ,
\tilde{\phi} \right) \right] \,,  \label{cft86}\\
&&\nonumber\\
F\left( \phi , \tilde{\phi} \right) & = & f \left[ \phi ,
R \left( \phi , \tilde{\phi} \right) \right] \,.\label{cft87}
\end{eqnarray}
This action describes a non-linear $\sigma$-model with 
canonical gravity and two scalar fields $\phi$ and 
$\tilde{\phi}$ which reduce to a single one if  $f\left( \phi , 
\tilde{\phi} \right)$ is 
linear in $R$. In this   case the Einstein frame action is
\begin{eqnarray} \label{cft89}
S=\frac{ \left| f \right|}{f} \, \int d^4 x \, \sqrt{-\tilde{g}} \,
\left[ \frac{\tilde{R} }{ 16\pi G} -\frac{1}{2} \, 
\tilde{g}^{\alpha\beta}
 \, \tilde{\nabla}_{\alpha} \tilde{\phi} \, 
\tilde{\nabla}_{\beta} \tilde{\phi} 
 -U \left( \tilde{\phi} \right) \right] \,,
\end{eqnarray}

where
\begin{eqnarray}
\tilde{\phi} & = & \frac{1}{\sqrt{ 8\pi G} } \int d\phi \, \left[
\frac{ 2 \, \epsilon f( \phi )+ 6 \left( df/d\phi \right)^2 }{4 
f^2( \phi) }  \right]^{1/2} \,, \label{cft90}\\
&&\nonumber\\
U \left( \tilde{\phi}  \right) & = &  
\frac{  \mbox{sign} (f) \,  V \left(  \phi  \right) }{ \left( 16 
\pi G  \,f \, \right)^2 }      \,, \label{cft91}
\end{eqnarray}
in which $\phi =\phi\left(  \tilde{\phi} \right)$. 
%
\subsection{The interpretation of conformal frames}
\label{sec:chap3sec7subsec4}

Some considerations are in order at this point.  The conformal 
transformation from the Jordan to the Einstein 
frame
is a mathematical map which allows one to study several  aspects
of scalar-tensor gravity, $f(R)$-gravity and, in general, any
ETG. However, having now available both the Jordan and the
Einstein conformal frames (and infinitely many other conformal
frames could be defined by choosing the conformal factor $\Omega$
arbitrarily), one wonders whether the two frames are also {\em
physically} equivalent or only mathematically related. In other
words,  the problem  is whether the physical meaning of the
theory is ``preserved'' or not by the use of  conformal
transformations. One has now the metric $g_{\mu\nu} $ and its
conformal cousin $\tilde{g}_{\mu\nu}$ and the question has been
posed of which one is the ``physical metric'', {\em i.e.}, the
metric from which curvature, geometry, and physical effects
should be calculated and compared with experiment
\cite{silvio}. The issue of "which  frame is the physical
one'' has been debated for a long time and it regularly
resurfaces in the literature, with authors arguing in
favor of one frame against the other, and others supporting the
view that the two frames are physically equivalent and that the
issue is a pseudo-problem. Many errors in the literature over the
years, from advocates of both points of view, have contributed to
confusion.

The first to approach this issue seems to have been Fierz (see
\cite{GrumillerKummerVassilevich03}) but the first popular
argument is due to Dicke, who presented it  in the paper
introducing the conformal transformation for Brans-Dicke 
theory
\cite{Dicke62}. Dicke's argument is that physics must be 
invariant
under a rescaling of units and the conformal 
transformation is
merely a local rescaling: units are not changed rigidly over the
entire space-time manifold, but by amounts which are different at
different space-time points. In Dicke's view, the two  frames are
equivalent provided that the units of mass, length, and time, and
quantities derived from them  scale with appropriate powers of 
the conformal factor in the Einstein frame 
\cite{Dicke62}.

With this view in mind, it is not difficult to see why many
authors consider the issue of which  conformal frame is physical
a pseudo-problem. In  principle, it  is difficult to object to
this argument, but there are two difficulties:

\begin{enumerate}

\item even though Dicke's argument is clear in principle, its
application to practical situations is a  different matter. The
view that the  two conformal frames are merely different {\em
representations} of the same theory, similar to different gauges
of a gauge  theory, should be checked explicitly using
the equations
describing the physics. "Physical equivalence'' is a vague
concept  because one can consider many different matter (or test)
fields in curved space-time and different types of physics, or
different  physical aspects of a problem. When checking 
explicitly
the physical equivalence between the two frames, one has to
specify which physical field, or physical process is considered 
and
the equations describing it. The equivalence could then be shown
explicitly, but there is no proof that holds for all of physics,
for example for  Klein-Gordon fields, spinors, for cosmology,
black holes, {\em etc.} While physical 
equivalence has been  proved
for various physical aspects, no proof comprehensive of  all
physical fields and different physical applications exists.

\item Dicke's argument is purely classical. In  cosmology, black
hole physics, and ETGs quantum fields in curved 
space play a
significant role and the equivalence of conformal frames is not
clear at all at the quantum level. Of course, not much is known
about this equivalence in quantum gravity due to the lack of a
definitive quantum gravity theory, but when the metric
$g_{\mu\nu}$  is quantized in full quantum gravity approaches,
inequivalent quantum theories are found \cite{Flanagan04b,
AshekarCorichi03,  ElizaldeNaftulinOdintsov94,
GrumillerKummerVassilevich03}. One  can consider the 
semiclassical
regime in which gravity is  classical and the matter fields are
quantized: again, one would  expect the conformal frames to be
inequivalent  because the conformal 
transformation can be seen as
a Legendre transformation \cite{MagnanoSokolowski94},  similar to
the Legendre  transformation of the classical mechanics of point
particles which  switches from the canonical Lagrangian
coordinates $q$  to the variables $\left\{q, p\right\}$  of the
Hamiltonian  formalism.  Now, it is well known that Hamiltonians
that are classically equivalent become inequivalent when
quantized, producing different energy spectra and scattering
amplitudes \cite{DegasperisRuijsenaars01, CalogeroDegasperis04,
GlassScanio77}.  However, the conformal equivalence between 
Jordan and Einstein frame seems to hold to some extent at the
semiclassical level \cite{Flanagan04b}.\footnote{A common 
argument among particle physicists relies on the equivalence 
theorem of Lagrangian field theory stating that the $S$-matrix is 
invariant under local (nonlinear) field redefinitions 
\cite{Dyson48, Chisolm61, Kamefuchietal61, ColemanWessZumino69, 
SalamStrathdee70, Lam73, Blasietal99}. Since the conformal 
transformation is,
essentially, a field redefinition, it  would seem that quantum
physics is invariant under change of the conformal frame. 
However,
the field theory  in which the equivalence theorem is derived
applies to gravity only in the perturbative regime in which the
fields  deviate slightly from Minkowski space. In this regime,
tree level quantities can be calculated in any conformal frame
with the same result, but in the non-perturbative regime field
theory  and the equivalence theorem do not apply.} Again, only
a particular kind of physics has been considered and one
cannot make statements about all possible physical situations.

\end{enumerate}

Unfortunately, the scaling of units in the Einstein frame is 
often forgotten, producing results that either do  not make sense 
or are partially or totally incorrect, or sometimes the error is
inconsequential,\footnote{Dicke himself applied the conformal
transformation and the scaling of units incorrectly
\cite{DickePeebles64} in GR  cosmology (see
Ref.~\cite{FaraoniNadeauconfo, CarreraGiuliniRevModPhys}).} 
reinforcing the opposite view  that the two frames are completely
equivalent. While Dicke's  explanation is very appealing and
several claims supporting the view that the two frames are 
inequivalent
turned out to be incorrect because they simply neglected the
scaling of units in the Einstein frame, one should not forget 
that Dicke's argument is not inclusive of all areas of physics 
and it is better to  check explicitly that the physics of  a  
certain field does not  depend on the conformal representation  
and not make sweeping statements. Certain points have  been 
raised in the literature which either constitute a problem for 
Dicke's  view, or, at least, indicate that this viewpoint cannot 
be   applied blindly, including the following.

\begin{itemize}

\item Massive test particles follow timelike
geodesics in the Jordan frame, while they 
deviate from geodesic motion in the Einstein frame due to a force proportional  to the
gradient of the scalar field (equivalently, of the conformal
factor or of the varying mass unit \cite{FaraoniNadeauconfo}).
Hence, the Weak Equivalence Principle is satisfied in the Jordan
frame but not in the Einstein frame due to the coupling of the
scalar field to ordinary matter, or to the variation of  units. 
Since the Equivalence  
Principle is the
foundation of relativistic gravity, this aspect is important and
there are two ways to look at it. One can cherish the view that
the two conformal frames are equivalent also with respect to the
Equivalence Principle, which implies 
that the latter is
formulated in a way that depends on the conformal frame
representation. Or, one could view the
violation of  the Weak Equivalence Principle in the  Einstein
frame more  pragmatically  by saying that 
"physical  equivalence''
of the  two frames is a vague term which must be defined 
precisely and this concept cannot be used  blindly, in fact the 
Equivalence Principle of 
standard  textbooks holds only in one frame  but
not in the other. This fact could be used as an argument against
the physical equivalence of the frames.

\item The  Brans-Dicke-like scalar  field easily violates all
of the energy conditions in the Jordan frame, but satisfies them
in the Einstein frame. While this fact does not eliminate
singularities in one frame leaving them in the other
\cite{FaraoniNadeauconfo} ({\em i.e.}, the two frames are
equivalent with respect to the presence of singularities), one
cannot say that the two frames are "equivalent'' with respect to
the energy conditions. This difficulty arises because part of the
matter sector of the theory, in the Einstein frame, comes from 
the conformal factor; in  other words, the 
conformal transformation
mixes matter and  geometric degrees of freedom, which is the
source of many  interpretational problems
\cite{Capozzielloetal97,bazeia}. Thus, even if 
the theory turns out to be independent of the conformal 
representation, its interpretation is not.

\item There are studies of  FLRW cosmology in which  the Universe
accelerates in one frame  but not in the other. From  the
pragmatic point of view of an astronomer  attempting to fit
observational data (for example, type Ia  
supernovae data to a model of the present acceleration of 
the Universe),  the two frames certainly do not appear to be 
"physically  equivalent'' \cite{nodicap2, CapozzielloPrado}.

\end{itemize}

To approach correctly the problem  of physical equivalence under
conformal transformations, one 
can  compare physics in different conformal frames  at the level 
of the Lagrangian, of the field equations, and of their 
solutions. This comparison may not always
be easy but, in certain cases, it is extremely useful
to discriminate between frames. It has been adopted, for
example, in  Ref.~\cite{cno}, to compare
cosmological models in the Einstein and  the Jordan frame.
Specifically, it has been shown that solutions of $f(R)$  and
scalar-tensor gravity cannot be assumed to be physically
equivalent to those in the Einstein frame 
when matter fields are given by generalized Equations of State 
(EoS). The situation is summarized in Table~\ref{framesTable}.

\begin{table}
\begin{center}
\begin{tabular}{|ccccc|} \hline
  &  &  &   & \\
  &  &  &   & \\
  $EoS$ & $\longleftrightarrow$ & ${\cal L}_{ST}$ &  
$\longleftrightarrow$ &
     ${\cal L}_{f(R)}$ \\
  &  &  &   & \\
  $\updownarrow$ &  & $\updownarrow$ &  & $\updownarrow$ \\
  &  &  &   & \\
  ~~~~Einstein eqs. & $\longleftrightarrow$ & ST eqs. & 
$\longleftrightarrow$ & $f(R)$ eqs. \\
  &  &  &   & \\
  $\updownarrow$ &  & $\updownarrow$ &  & $\updownarrow$ \\
  &  &  &   & \\
  E frame sol. & $\longleftrightarrow$ & E frame sol.+$\phi$ & 
$\longleftrightarrow$ & J frame sol.~~~~ \\ 
  &  &  &   & \\
  &  &  &   & \\
\hline
\end{tabular}
\end{center}
\caption{\label{framesTable}Three approaches (EoS, 
scalar-tensor and $f(R)$) compared at the level of Lagrangians, 
field equations, and their solutions.  Mathematical 
equivalence of the three levels does not
automatically imply physical equivalence of the solutions.}
\end{table}

In these, and in other situations, one must specify precisely 
what ``physical equivalence'' means. In certain situations 
physical equivalence is demonstrated simply by taking into 
account the coupling of the Brans-Dicke-like scalar field to 
matter and the varying units in the Einstein 
frame, but in other cases the
physical equivalence is not obvious and it does not seem to hold.
At the very least, this equivalence, if it is valid at all,  must
be defined in precise terms and discussed in ways that are far
from obvious. For this reason, it would be too simplistic to
dismiss the issue of the conformal frame entirely as a
pseudo-problem that has been solved  for all physical situations
of interest. It is fair to say that there have been surprises and
non-trivial difficulties have been uncovered.

\section{ Extended Theories with torsion}
In this section, we want to face the problem to study $f(R)$-gravity
considering also torsion. Torsion theories have been taken into
account firstly by Cartan and then where introduced by Sciama and
Kibble in order to deal with spin in GR (see
\cite{hehlrev2} for a review). Being the spin as fundamental as the
mass of the particles, torsion was introduced in order to complete
the following scheme: the mass (energy) as the source of curvature
and the spin as the source of torsion.

Up to some time ago, torsion  did not seem to produce models with
observable effects since  phenomena implying  spin and gravity
were considered to be significant only in the very early Universe.
After, it has been proven  that spin  is not the only source of
torsion. As a matter of fact, torsion field can be decomposed in
three irreducible tensors, with different properties. In
\cite{Capozziello:2001mq}, a systematic classification of these
different types of torsion and their possible sources is
discussed. This means that a wide class of torsion models could be
investigate  independently of spin as their source.

In principle, torsion could be constrained at every astrophysical
scale and, as recently discussed, data coming from Gravity Probe B
could contribute to this goal also  at Solar System level
\cite{tegmark}.

In above section, a systematic
discussion of metric-affine $f(R)$-gravity has been pursued. 
Here, following the same philosophy, we want to show
that, starting from a generic $f(R)$ theory, the curvature and the
torsion can give rise to an effective curvature-torsion
stress-energy tensor capable, in principle, to address the problem
of the Dark Side of the Universe in a very general geometric
scheme \cite{CCSV3}. We do not consider the possible microscopic distribution
of spin but a general  torsion vector field in $f(R)$-gravity.

\subsection{The $f(R)$-field equations with torsion}
Let us discuss the main features of a $f(R)$-gravity considering
the most general case in which torsion is present in a ${\cal
U}_4$ manifold \footnote{We indicate with ${\cal V}_4$ a $4D$ pseudo-Riemaniann manifold without torsion and with ${\cal U}_4$ a $4D$  manifold with torsion. Some of the notions already introduced in Sec. III will be adopted here.} \cite{CCSV1}. In a  metric-affine formulation, the metric ${g}$ and the connection ${ \Gamma}$ can be, in general, considered
independent fields. More precisely, the dynamical fields are pairs
$(g,\Gamma)\/$ consisting of a pseudo-Riemannian metric $g\/$ and
a metric compatible linear connection $\Gamma\/$ on the
space--time manifold $M\/$. The corresponding field equations are
derived by varying separately with respect to the metric and the
connection the action functional
\begin{equation}\label{00.1}
{\cal S}\/(g,\Gamma)=\int{\sqrt{-g}f\/({\cal R})\,ds}\,,
\end{equation}
where $f\/$ is a real function, ${\cal R}\/(g,\Gamma) = g^{\mu\nu}{\cal R}_{\mu\nu}\/$
(with ${\cal R}_{\mu\nu}:= {\cal R}^\sigma_{\;\;\mu\sigma\nu}\/$) is the scalar curvature
associated with the connection $\Gamma\/$ and $ds :=
dx^1\wedge\dots\wedge dx^4\/$. We use the
index notation
\begin{equation}\label{00.1bis}
{\cal R}^\sigma_{\;\;\gamma\mu\nu}=\frac{\partial{\Gamma_{\nu\gamma}^{\;\;\;\sigma}}}{\partial{x^\mu}} -
\frac{\partial{\Gamma_{\mu\gamma}^{\;\;\;\sigma}}}{\partial{x^\nu}} +
\Gamma_{\mu\phi}^{\;\;\;\sigma}\Gamma_{\nu\gamma}^{\;\;\;\phi} -
\Gamma_{\nu\phi}^{\;\;\;\sigma}\Gamma_{\mu\gamma}^{\;\;\;\phi}\,,
\end{equation}
for the curvature tensor and
\begin{equation}\label{00.1tris}
\nabla_{\frac{\partial} {\partial{x^\mu}}}\frac{\partial}{\partial{x^\nu}} = \Gamma_{\mu\nu}^{\;\;\;\sigma}\,\frac{\partial
}{\partial{x^\sigma}}\,,
\end{equation}
for the connection coefficients.
In order to evaluate the variation $\delta{\cal S}\/$ under
arbitrary deformations of the connection, we recall that, given a
metric tensor $g_{\mu\nu}\/$, every metric connection $\Gamma\/$ may
be expressed as
\begin{equation}\label{00.2}
\Gamma_{\mu\nu}^{\;\;\;\sigma} =\tilde{\Gamma}_{\mu\nu}^{\;\;\;\sigma} -
K_{\mu\nu}^{\;\;\;\sigma}\,,
\end{equation}
where (in the holonomic basis ${\displaystyle \left\{
\frac{\partial}{\partial x^\mu}, dx^\mu \right \}}$)
$\tilde{\Gamma}_{\mu\nu}^{\;\;\;\sigma}$ denote the coefficients  of the
Levi-Civita connection associated with the metric $g_{\mu\nu}\/$,
while $K_{\mu\nu}^{\;\;\;\sigma}\/$ indicate the components of a tensor
satisfying the antisymmetry property $K_{\mu}^{\;\;\nu\sigma} = -
K_{\mu}^{\;\;\sigma\nu}\/$. This last condition ensures the metric
compatibility of the connection $\Gamma\/$.
In view of this, we can identify the actual degrees of freedom of
the theory with the (independent) components of the metric $g\/$
and the tensor $K\/$. Moreover, it is easily seen that the
curvature and the contracted curvature tensors associated with
every connection \eqref{00.2} can be expressed respectively as
\begin{subequations}
\begin{equation}\label{00.3}
{\cal R}^\sigma_{\;\;\mu\eta\nu}= \tilde{ R}^\sigma_{\;\;\mu\eta\nu} +
\tilde{\nabla}_\nu K_{\eta\mu}^{\;\;\;\sigma} -
\tilde{\nabla}_\eta K_{\nu\mu}^{\;\;\;\sigma} +
K_{\nu\mu}^{\;\;\;\phi}K_{\eta\phi}^{\;\;\;\sigma} -
K_{\eta\mu}^{\;\;\;\phi}K_{\nu\phi}^{\;\;\;\sigma}\,,
\end{equation}
and
\begin{equation}\label{00.4}
{\cal R}_{\mu\nu}={\cal R}^\sigma_{\;\;\mu\sigma\nu}=\tilde{R}_{\mu\nu} +
\tilde{\nabla}_\nu K_{\sigma\mu}^{\;\;\;\sigma} -
\tilde{\nabla}_\sigma K_{\nu\mu}^{\;\;\;\sigma} +
K_{\nu\mu}^{\;\;\;\phi}K_{\sigma\phi}^{\;\;\;\sigma} -
K_{\sigma\mu}^{\;\;\;\phi}K_{\nu\phi}^{\;\;\;\sigma}\,,
\end{equation}
\end{subequations}
where $\tilde{R}^\sigma_{\;\;\mu\eta\nu}\/$ and
$\tilde{ R}_{\mu\nu}=\tilde{R}^\sigma_{\;\;\mu\sigma\nu}\/$ are respectively the
Riemann and the Ricci tensors of the Levi--Civita connection
$\tilde\Gamma\/$ associated with the given metric $g\/$, and
$\tilde\nabla\/$ indicates the Levi-Civita covariant derivative.

Making use of the identities \eqref{00.4}, the action functional
\eqref{00.1} can be written in the equivalent form
\begin{equation}\label{00.5}
{\cal S}\/(g,\Gamma)=\int\sqrt{-g}f\/(g^{\mu\nu}(\tilde{R}_{\mu\nu} +
\tilde{\nabla}_\nu K_{\sigma\mu}^{\;\;\;\sigma} -
\tilde{\nabla}_\sigma K_{\mu\nu}^{\;\;\;\sigma} +
K_{\nu\mu}^{\;\;\;\phi}K_{\sigma\phi}^{\;\;\;\sigma} -
K_{\sigma\mu}^{\;\;\;\phi}K_{\nu\phi}^{\;\;\;\sigma}))\,ds\,,
\end{equation}
more suitable for variations in the connection. Taking the metric
$g$ fixed, we have the identifications
$\delta\Gamma_{\mu\nu}^{\;\;\;\sigma}=\delta K_{\mu\nu}^{\;\;\;\sigma}\/$ and then
the variation
\begin{equation}\label{00.6}
\delta{\cal S}= \int \sqrt{-g}f'\/({\cal R})g^{\mu\nu}(
\tilde{\nabla}_\nu\delta{K}_{\sigma\mu}^{\;\;\;\sigma} -
\tilde{\nabla}_\sigma\delta{K}_{\nu\mu}^{\;\;\;\sigma} +
\delta{K}_{\nu\mu}^{\;\;\;\phi}K_{\sigma\phi}^{\;\;\;\sigma} +
K_{\nu\mu}^{\;\;\;\phi}\delta{K}_{\sigma\phi}^{\;\;\;\sigma} -
\delta{K}_{\sigma\mu}^{\;\;\;\phi}K_{\nu\phi}^{\;\;\;\sigma} -
K_{\sigma\mu}^{\;\;\;\phi}\delta{K}_{\nu\phi}^{\;\;\;\sigma})ds\,.
\end{equation}
Using the divergence theorem, taking the antisymmetry properties
of $K\/$ into account and renaming finally some indexes, we get
the expression
\begin{equation}\label{00.7}
\delta{\cal S}= \int \sqrt{-g}\left[ -\frac{\partial{f'}}{\partial{x^\mu}}\delta^\sigma_\nu +
\frac{\partial{f'}}{\partial{x^\nu}}\delta^\sigma_\mu + f'K_{\phi\nu}^{\;\;\;\phi}\delta^\sigma_\mu -
f'K_{\phi\mu}^{\;\;\;\phi}\delta^\sigma_\nu - f'K_{\mu\nu}^{\;\;\;\sigma} 
 + f'K_{\nu\mu}^{\;\;\;\sigma}\right]\delta{K}_\sigma^{\;\;\mu\nu}\,ds\,.
\end{equation}
The requirement $\delta{\cal S}=0\/$ yields therefore a first set
of field equations given by
\begin{equation}\label{00.8}
K_{\phi\nu}^{\;\;\;\phi}\delta^\sigma_\mu - K_{\phi\mu}^{\;\;\;\phi}\delta^\sigma_\nu -
K_{\mu\nu}^{\;\;\;\sigma} + K_{\nu\mu}^{\;\;\;\sigma} =
\frac{1}{f'}\frac{\partial{f'}}{\partial{x^\phi}}\/\left(\delta^\phi_\mu\delta^\sigma_\nu -
\delta^\phi_\nu\delta^\sigma_\mu\right)\,.
\end{equation}
Considering that the torsion coefficients of the connection
$\Gamma\/$ are $T_{\mu\nu}^{\;\;\;\sigma}:= \Gamma_{\mu\nu}^{\;\;\;\sigma} -
\Gamma_{\nu\mu}^{\;\;\;\sigma} = -K_{\mu\nu}^{\;\;\;\sigma} + K_{\nu\mu}^{\;\;\;\sigma}\/$
and thus (due to antisymmetry)
$T_{\phi\mu}^{\;\;\;\phi}=-K_{\phi\mu}^{\;\;\;\phi}\/$, eqs.~\eqref{00.8} can be
rewritten as
\begin{equation}\label{00.9a}
T_{\mu\nu}^{\;\;\;\sigma} + T_{\nu\phi}^{\;\;\;\phi}\delta^\sigma_\mu -
T_{\mu\phi}^{\;\;\;\phi}\delta^\sigma_\nu =
\frac{1}{f'}\frac{\partial{f'}}{\partial{x^\phi}}\/\left(\delta^\phi_\mu\delta^\sigma_\nu -
\delta^\phi_\nu\delta^\sigma_\mu\right)\,,
\end{equation}
or, equivalently, as
\begin{equation}\label{00.9b}
T_{\mu\nu}^{\;\;\;\sigma} = -
\frac{1}{2f'}\frac{\partial{f'}}{\partial{x^\phi}}\/\left(\delta^\phi_\mu\delta^\sigma_\nu -
\delta^\phi_\nu\delta^\sigma_\mu\right)\,.
\end{equation}
In order to study the variation $\delta{\cal S}\/$ under arbitrary
deformations of the metric, it is convenient to resort to the
representation \eqref{00.1}. Indeed, from the latter, we have
directly
\begin{equation}\label{00.10}
\delta{\cal S}=\int{\sqrt{-g}\/\left[ f'\/({\cal R}){\cal R}_{\mu\nu} -
\frac{1}{2}f\/({\cal R})g_{\mu\nu} \right]\delta{g^{\mu\nu}}\,ds}\,,
\end{equation}
thus getting the second set of field equations
\begin{equation}\label{00.11}
f'\/({\cal R}){\cal R}_{(\mu\nu)} - \frac{1}{2}f\/({\cal R})g_{\mu\nu}=0\,.
\end{equation}
Of course, one can obtain the same equations \eqref{00.11}
starting from the representation \eqref{00.5} instead of
\eqref{00.1}. In that case, the calculations are just longer.
As a remark concerning Eqs. \eqref{00.11}, it is worth noticing
that any connection satisfying Eqs. \eqref{00.2} and \eqref{00.9b}
gives rise to a contracted curvature tensor ${\cal R}_{\mu\nu}\/$
automatically symmetric. Indeed, since the tensor $K\/$ coincides
necessarily with the contorsion tensor, namely
\begin{equation}\label{00.12}
K_{\mu\nu}^{\;\;\;\sigma} = \frac{1}{2}\/\left( - T_{\mu\nu}^{\;\;\;\sigma} +
T_{\nu\;\;\;\mu}^{\;\;\sigma} - T^\sigma_{\;\;\mu\nu}\right)\,,
\end{equation}
from Eqs. \eqref{00.9b} we have
\begin{equation}\label{00.13}
K_{\mu\nu}^{\;\;\;\sigma} = \frac{1}{3}\/\left(T_\nu\delta^\sigma_\mu -
T_\phi g^{\phi\sigma}g_{\mu\nu}\right)\,,
\end{equation}
being
\begin{equation}\label{00.14}
T_\mu := T_{\mu\sigma}^{\;\;\;\sigma} = -\frac{3}{2f'}\frac{\partial{f'}}{\partial{x^\mu}}\,.
\end{equation}
Inserting Eq. \eqref{00.13} in Eq \eqref{00.4},  the contracted
curvature tensor can be represented as
\begin{equation}\label{00.15}
{\cal R}_{\mu\nu} = \tilde{R}_{\mu\nu} + \frac{2}{3}\tilde{\nabla}_{\nu}T_\mu+
\frac{1}{3}\tilde{\nabla}_\sigma T^\sigma g_{\mu\nu} + \frac{2}{9}T_\mu T_\nu -
\frac{2}{9}T_\sigma T^\sigma g_{\mu\nu}\,.
\end{equation}
The last expression, together with Eqs. \eqref{00.14}, entails the
symmetry of the indexes $\mu$ and $\nu$. Therefore, in Eq.
\eqref{00.11} we can omit the symmetrization symbol and write
\begin{equation}\label{00.16}
f'\/({\cal R}){\cal R}_{\mu\nu} - \frac{1}{2}f\/({\cal R})g_{\mu\nu}=0\,.
\end{equation}
Now, considering the trace of the equation \eqref{00.16}, we get
\begin{equation}\label{00.17}
f'\/({\cal {\cal R}}){\cal {\cal R}}  - 2f\/({\cal {\cal R}})=0\,.
\end{equation}
The latter is an identity automatically satisfied by all possible
values of ${\cal R}\/$ only in the special case $f\/({\cal R})=\alpha {\cal R}^2\/$. In
all other cases, Eq.\eqref{00.17} represents a constraint on the
scalar curvature ${\cal R}\/$.
As a conclusion follows that, if $f\/({\cal R})\not = \alpha {\cal R}^2\/$, the
scalar curvature ${\cal R}\/$ has to be constant (at least on connected
domains) and coincides with a given solution value of
\eqref{00.17}. In such a circumstance, Eqs.\eqref{00.9b} imply
that the torsion $T^{\;\;\;\sigma}_{\mu\nu}\/$ has to be zero and the
theory reduces to a $f\/({\cal R})$-theory without torsion. In
particular, we notice that in the case $f\/({\cal R})={\cal R}\/$, eq.
\eqref{00.17} yields ${\cal R}=0\/$ and therefore Eqs. \eqref{00.16} are
equivalent to Einstein's equations in empty space ${\cal R}_{\mu\nu}=0\/$. On
the other hand, if we assume $f\/({\cal {\cal R}})=\alpha {\cal R}^2\/$, we can have
non--vanishing torsion. In this case, by replacing Eq.
\eqref{00.17} in Eqs. \eqref{00.9b} and \eqref{00.16}, we obtain
field equations of the form
\begin{subequations}\label{00.18}
\begin{equation}\label{00.18a}
{\cal R}_{\mu\nu} - \frac{1}{4}{\cal R}g_{\mu\nu}=0\,,
\end{equation}
\begin{equation}\label{00.18b}
T^{\;\;\;\sigma}_{\mu\nu}= -\frac{1}{2{\cal R}}\frac{\partial{{\cal R}}}{\partial{x^\mu}}\delta^\sigma_\nu +
\frac{1}{2{\cal R}}\frac{\partial{{\cal R}}}{\partial{x^\nu}}\delta^\sigma_\mu\,.
\end{equation}
\end{subequations}
Finally, making use of Eq. \eqref{00.15} and the consequent
relation
\begin{equation}\label{00.19}
{\cal R}= \tilde{R} + 2\tilde{\nabla}_\sigma T^\sigma - \frac{2}{3}T_\sigma T^\sigma\,,
\end{equation}
in Eqs. \eqref{00.18}, we can separately point out the
contribution due to the metric and that due to the torsion. In
fact, directly from eqs. \eqref{00.18a} we have
\begin{equation}\label{00.20}
\tilde{R}_{\mu\nu} - \frac{1}{4}\tilde{R}g_{\mu\nu} =
-\frac{2}{3}\tilde{\nabla}_\nu T_\mu +
\frac{1}{6}\tilde{\nabla}_\sigma T^\sigma g_{\mu\nu} -\frac{2}{9}T_\mu T_\nu +
\frac{1}{18}T_\sigma T^\sigma g_{\mu\nu}\,,
\end{equation}
while from the "trace'' $\displaystyle{T_\mu := T_{\nu\sigma}^{\;\;\;\sigma} =
-\frac{3}{2{\cal R}}\frac{\partial{{\cal R}}}{\partial{x^\mu}}\/}$ of Eqs. \eqref{00.18b}, we derive
\begin{equation}\label{00.21}
\frac{\partial}{\partial{x^\mu}}\/\left( \tilde {R} + 2\tilde{\nabla}_\sigma T^\sigma -
\frac{2}{3}T_\sigma T^\sigma \right) = -\frac{2}{3}\left( \tilde{R} +
2\tilde{\nabla}_\sigma T^\sigma - \frac{2}{3}T_\sigma T^\sigma \right)\/T_\mu\,,
\end{equation}
Eqs. \eqref{00.20} and \eqref{00.21} are the  coupled field
equations in vacuum for  metric and  torsion in the $f\/({\cal R})=\alpha
{\cal R}^2\/$ gravitational theory.

The presence of matter is embodied in the action functional
\eqref{00.1} by adding to the gravitational Lagrangian a suitable
material Lagrangian density ${\cal L}_m\/$, namely
\begin{equation}\label{000.1}
{\cal S}\/(g,\Gamma)=\int{\left(\sqrt{-g}f\/({\cal R}) + {\cal
L}_m\right)\,ds}\,.
\end{equation}
Throughout this Review Paper we shall consider material Lagrangian density
${\cal L}_m\/$ not containing terms depending on torsion degrees
of freedom. The physical meaning
of this assumption will be discussed later. In this case, the
field equations take the form
\begin{subequations}
\begin{equation}\label{000.2a}
f'\/({\cal R}){\cal R}_{\mu\nu} - \frac{1}{2}f\/({\cal R})g_{\mu\nu}=\Sigma_{\mu\nu}\,,
\end{equation}
\begin{equation}\label{000.2b}
T_{\mu\nu}^{\;\;\;\sigma} = -
\frac{1}{2f'\/({\cal R})}\frac{\partial{f'\/({\cal R})}}{\partial{x^p}}\/\left(\delta^\phi_i\delta^\sigma_\nu
- \delta^\phi_\nu\delta^\sigma_\mu\right)\,,
\end{equation}
\end{subequations}
where ${\displaystyle \Sigma_{\mu\nu}:= -
\frac{1}{\sqrt{-g}}\frac{\delta{\cal L}_m}{\delta g^{\mu\nu}}\/}$
plays the role of the energy--momentum tensor. As in Sec. III, the notation for the stress-energy tensor has been changed in order to avoid confusion. Also here we are adopting physical units. From the trace of
Eq. \eqref{000.2a}, we obtain a fundamental relation between the
curvature scalar ${\cal R}\/$ and the trace
$\Sigma:=g^{\mu\nu}\Sigma_{\mu\nu}\/$, which is
\begin{equation}\label{000.3}
f'\/({\cal R}){\cal R} -2f\/({\cal R}) = \Sigma\,,
\end{equation}
(see also \cite{mauro} and references therein). In what follows,
we shall systematically suppose that the relation \eqref{000.3} is
invertible and that $\Sigma \not= const\/$, thus allowing to
express the curvature scalar $R\/$ as a suitable function of
$\Sigma\/$, namely
\begin{equation}\label{000.4}
{\cal R}=F\/(\Sigma)\,.
\end{equation}
With this assumption in mind, using Eqs. \eqref{000.3} and
\eqref{000.4} we can rewrite equations \eqref{000.2a} and
\eqref{000.2b} in the form
\begin{subequations}
\begin{equation}\label{000.5a}
{\cal R}_{\mu\nu} -\frac{1}{2}{\cal R}g_{\mu\nu}= \frac{1}{f'\/(F\/(\Sigma))}\left(
\Sigma_{\mu\nu} - \frac{1}{4}\Sigma g_{\mu\nu} \right) -
\frac{1}{4}F\/(\Sigma)g_{\mu\nu}\,,
\end{equation}
\begin{equation}\label{000.5b}
T_{\mu\nu}^{\;\;\;\sigma} = -
\frac{1}{2f'\/(F\/(\Sigma))}\frac{\partial{f'\/(F\/(\Sigma))}}{\partial{x^\phi}}\/\left(\delta^\phi_\mu\delta^\sigma_\nu
- \delta^\phi_\nu\delta^\sigma_\mu\right)\,.
\end{equation}
\end{subequations}
Moreover, making use of Eqs. \eqref{00.15} and \eqref{00.19}, in
Eq. \eqref{000.5a} we can decompose the contracted curvature
tensor and the curvature scalar in their Christoffel and torsion
dependent terms, thus getting an Einstein-like equation of the
form
\begin{equation}\label{000.6}
\tilde{{R}}_{\mu\nu} -\frac{1}{2}\tilde{{R}}g_{\mu\nu}= \frac{1}{f'\/(F\/(\Sigma))}\left( \Sigma_{\mu\nu} - \frac{1}{4}\Sigma g_{\mu\nu} \right) - \frac{1}{4}F\/(\Sigma)g_{\mu\nu} - \frac{2}{3}\tilde{\nabla}_{\nu}T_\nu \\
+ \frac{2}{3}\tilde{\nabla}_\sigma T^\sigma g_{\mu\nu} - \frac{2}{9}T_\mu T_\nu -
\frac{1}{9}T_\sigma T^hg_{\mu\nu}\,.
\end{equation}
Now, setting
\begin{equation}\label{000.7}
\varphi := f'\/(F\/(\Sigma))\,,
\end{equation}
from the trace of Eqs. \eqref{000.5b}, we obtain
\begin{equation}\label{000.8}
T_\mu := T_{\mu\sigma}^{\;\;\;\sigma} = - \frac{3}{2\varphi}\frac{\partial\varphi}{\partial{x^\mu}}\,.
\end{equation}
Therefore, substituting in Eqs. \eqref{000.6}, we end up with the
final equations
\begin{equation}\label{000.9}
\tilde{{R}}_{\mu\nu} -\frac{1}{2}\tilde{{R}}g_{\mu\nu}= \frac{1}{\varphi}\Sigma_{\mu\nu} + \frac{1}{\varphi^2}\left( - \frac{3}{2}\frac{\partial\varphi}{\partial {x^\mu}}\frac{\partial \varphi}{\partial{x^\nu}} + \varphi\tilde{\nabla}_{\mu}\frac{\partial\varphi}{\partial{x^\mu}} + \frac{3}{4}\frac{\partial\varphi}{\partial{x^\sigma}}\frac{\partial\varphi}{\partial{x^\gamma}}g^{\sigma\gamma}g_{\mu\nu} 
- \varphi\tilde{\nabla}^\sigma\frac{\partial\varphi}{\partial{x^\sigma}}g_{\mu\nu} -
V\/(\varphi)g_{\mu\nu} \right)\,,
\end{equation}
where we defined the effective potential
\begin{equation}\label{000.10}
V\/(\varphi):= \frac{1}{4}\left[ \varphi
F^{-1}\/((f')^{-1}\/(\varphi)) +
\varphi^2\/(f')^{-1}\/(\varphi)\right]\,,
\end{equation}
Eqs. \eqref{000.9} may be difficult to solve, neverthless we can
simplify this task finding solutions for a conformally related
metric. Indeed, performing a conformal transformation of the kind
$\bar{g}_{\mu\nu} = \varphi g_{\mu\nu}\/$, eqs. \eqref{000.9} may be
rewritten in the easier form (see, for example,
\cite{mauro,Olmo:2005zr,German})
\begin{equation}\label{000.10.1}
\bar{{ R}}_{\mu\nu} - \frac{1}{2}\bar{{R}}\bar{g}_{\mu\nu} =
\frac{1}{\varphi}\Sigma_{\mu\nu} -
\frac{1}{\varphi^3}V\/(\varphi)\bar{g}_{\mu\nu}\,,
\end{equation}
where $\bar{{R}}_{\mu\nu}\/$ and $\bar{{R}}\/$ are respectively the Ricci
tensor and the Ricci scalar curvature associated with the
conformal metric $\bar{g}_{\mu\nu}\/$.
Concerning the connection $\Gamma\/$, solution of the variational
problem $\delta{\cal S}=0\/$, from eqs.~\eqref{00.2},
\eqref{00.13} and \eqref{000.8}, one gets the explicit expression
\begin{equation}\label{000.11}
\Gamma_{\mu\nu}^{\;\;\;\sigma} =\tilde{\Gamma}_{\mu\nu}^{\;\;\;\sigma} +
\frac{1}{2\varphi}\frac{\partial\varphi}{\partial{x^\nu}}\delta^\sigma_\mu -
\frac{1}{2\varphi}\frac{\partial\varphi}{\partial{x^\phi}}g^{\phi\sigma}g_{\mu\nu}\,.
\end{equation}
We can now compare these results with those obtained for $f(R)\/$
theories in the Palatini formalism
\cite{allemandi,Vollick:2003,allemandi,Sotiriou06,Olmo:2005zr}.
If both the theories (with torsion and Palatini--like) are
considered as "metric'', in the sense that the dynamical
connection $\Gamma\/$ is not coupled with matter ${\displaystyle
\left(\frac{\delta{\cal L}_m}{\delta\Gamma}=0\right)\/}$ and it
does not define parallel transport and covariant derivative in
space--time, then the two approaches are completely equivalent.
Indeed, in the "metric'' framework, the true connection of
space--time is the Levi-Civita one associated with the metric
$g\/$ and the role played by the dynamical connection $\Gamma\/$
is just to generate the right Einstein-like equations of the
theory. Now, surprisingly enough, our field equations
\eqref{000.9} are identical to the Einstein-like equations
derived within the Palatini formalism  \cite{Olmo}.
On the other hand, if the theories are genuinely metric--affine,
then they are different even though the condition ${\displaystyle
\frac{\delta{\cal L}_m}{\delta\Gamma}=0\/}$ holds. In order to
stress this point, we recall that in a metric--affine theory the
role of  dynamical connection is not only that of generating
Einstein--like field equations but also defining parallel
transport and covariant derivative in space-time. Therefore,
different connections imply different space-time properties. This
means that the geodesic structure and the causal structures could
not obviously coincide. For a discussion on this point see
\cite{allemandi}. Furthermore, it can be  easily shown that the
dynamical connection \eqref{000.11} differs from that derived
within the Palatini formalism. Indeed the latter results to be the
Levi-Civita connection $\bar{\Gamma}\/$ associated with the
conformal metric $\bar{g}= \varphi g\/$ \cite{Sotiriou06,Olmo:2005zr},
while clearly \eqref{000.11} is not. More precisely, Eq.
\eqref{000.11} is related to $\bar{\Gamma}\/$ by the projective
transformation
\begin{equation}\label{000.12}
\bar{\Gamma}_{\mu\nu}^{\;\;\;\sigma} =\Gamma_{\mu\nu}^{\;\;\;\sigma} +
\frac{1}{2\varphi}\frac{\partial\varphi}{\partial{x^\mu}}\delta^h_\nu\,,
\end{equation}
which is not allowed in the present theory because, for a fixed
metric $g$, the connection \eqref{000.12} is no longer metric
compatible.

To conclude, we notice that Eqs. \eqref{000.10.1} are deducible
from an Einstein-Hilbert like action functional only under
restrictive conditions. More precisely, let us suppose that the
material Lagrangian depends only on the components of the metric
and not on its derivatives as well as that the trace
$\Sigma=\Sigma_{\mu\nu}g^{\mu\nu}\/$ is independent of the metric and its
derivatives. Then, from the identities
\begin{equation}\label{000.10.2}
\sqrt{-\bar{g}}=\varphi^2\sqrt{-g}, \quad \frac{ \partial
}{\partial{g^{\mu\nu}}}=\frac{1}{\varphi}\frac{\partial}{\partial{\bar{g}^{\mu\nu}}} \quad {\rm and}
\quad  \Sigma_{\mu\nu}=-\frac{1}{\sqrt{-g}}\frac{\delta{\cal
L}_m}{\delta g^{\mu\nu}}=-\frac{1}{\sqrt{-g}}\frac{\partial{{\cal
L}_m}}{\partial{g^{\mu\nu}}}\,,
\end{equation}
we have the following relation
\begin{equation}\label{000.10.3}
\Sigma_{\mu\nu} = - \varphi\frac{1}{\sqrt{-\bar{g}}}\frac{\partial{{\cal
L}_m}}{\partial{\bar{g}^{\mu\nu}}}:= \varphi\bar{\Sigma}_{\mu\nu}\,.
\end{equation}
In view of this, and being $\varphi=\varphi\/(\Sigma)\/$, it is
easily seen that Eqs. \eqref{000.10.1} may be derived by varying
with respect to $\bar{g}^{\mu\nu}\/$ the action functional
\begin{equation}\label{000.10.4}
{\bar{\cal S}}\/(\bar{g})=\int{\left[\sqrt{-\bar{g}}\left(\bar{{R}}
-\frac{2}{\varphi^3}V\/(\varphi) \right)+ {\cal L}_m\right]\,ds}\,.
\end{equation}
Therefore, under the stated assumptions, $f({\cal R})\/$-gravity with
torsion in the metric framework is conformally equivalent to an
Einstein-Hilbert like theory.

%
\subsection{Equivalence of $f(R)$-gravity with torsion and scalar-tensor  theories}
The above considerations directly lead to study the relations
between $f(R)\/$-gravity with torsion and scalar-tensor theories
with the aim to  investigate their possible equivalence. To this
end, we recall that the action functional of a (purely metric)
scalar-tensor theory is
\begin{equation}\label{00051}
{\cal S}\/(g,\varphi)=\int{\left[\sqrt{-g}\left(\varphi\tilde{R}
-\frac{\omega_0}{\varphi}\varphi_\mu\varphi^\mu - U\/(\varphi)
\right)+ {\cal L}_m\right]\,ds}\,,
\end{equation}
where $\varphi\/$ is the scalar field which, depending on the sign
of the kinetic term, could assume also the role of a phantom field
\cite{valerio},  ${\displaystyle \varphi_\mu :=
\frac{\partial\varphi}{\partial{x^\mu}\/}}$ and $U\/(\varphi)\/$ is the potential of
$\varphi\/$. For $U\/(\varphi)=0\/$ such a theory reduces to the
standard Brans--Dicke theory \cite{BransDicke61}. The matter Lagrangian
${\cal L}_m\/(g_{\mu\nu},\psi)\/$ is a function of the metric and some
matter fields $\psi\/$; $\omega_0\/$ is the so called Brans--Dicke
parameter. The field equations derived by varying with respect to
the metric and the scalar field are
\begin{equation}\label{0000.2}
\tilde{R}_{\mu\nu} -\frac{1}{2}\tilde{ R}g_{\mu\nu}=
\frac{1}{\varphi}\Sigma_{\mu\nu} + \frac{\omega_0}{\varphi^2}\left(
\varphi_\mu\varphi_\nu  - \frac{1}{2}\varphi_\sigma\varphi^\sigma\/g_{\mu\nu}
\right) + \frac{1}{\varphi}\left( \tilde{\nabla}_{j}\varphi_\mu -
\tilde{\nabla}_\sigma\varphi^\sigma\/g_{\mu\nu} \right) -
\frac{U}{2\varphi}g_{\mu\nu}\,,
\end{equation}
and
\begin{equation}\label{0000.3}
\frac{2\omega_0}{\varphi}\tilde{\nabla}_\sigma\varphi^\sigma + \tilde{ R} -
\frac{\omega_0}{\varphi^2}\varphi_\sigma\varphi^\sigma - U' =0\,,
\end{equation}
where ${\displaystyle \Sigma_{\mu\nu}:= -
\frac{1}{\sqrt{-g}}\frac{\delta{\cal L}^{(m)}}{\delta g^{\mu\nu}}\/}$ and
${\displaystyle U' :=\frac{dU}{d\varphi}\/}$.

Taking the trace of Eq. \eqref{0000.2} and using it to replace
$\tilde R\/$ in Eq. \eqref{0000.3}, one obtains the equation
\begin{equation}\label{0000.4}
\left( 2\omega_0 + 3 \right)\/\tilde{\nabla}_\sigma\varphi^\sigma = \Sigma +
\varphi U' -2U\,.
\end{equation}
By a direct comparison, it is immediately seen that for $\omega_0
=-\frac{3}{2}\/$ and ${\displaystyle U\/(\varphi)
=\frac{2}{\varphi}V\/(\varphi)\/}$ (where $V\/(\varphi)\/$ is
defined as in Eq. \eqref{000.10}), Eqs. \eqref{0000.2} become
formally identical to the Einstein-like equations \eqref{000.9}
for a $f(R)\/$ theory with torsion. Moreover, in such a
circumstance, Eq. \eqref{0000.4} reduces to the algebraic equation
\begin{equation}\label{0000.5}
\Sigma + \varphi U' -2U =0\,,
\end{equation}
relating the matter trace $\Sigma\/$ to the scalar field
$\varphi\/$, exactly as it happens for $f(R)\/$-gravity. In
particular, it is a straightforward matter to verify that (under
the condition $f''\not= 0\/$) Eq. \eqref{0000.5} expresses {\it
exactly} the inverse relation of \eqref{000.7}, namely
\begin{equation}\label{0000.6}
\Sigma=F^{-1}\/((f')^{-1}\/(\varphi))\qquad \Leftrightarrow \qquad
\varphi=f'\/(F(\Sigma))
\end{equation}
being $F^{-1}\/(X) = f'\/(X)X - 2f\/(X)\/$. In fact we have
\begin{equation}\label{0000.6.2}
U\/(\varphi) =\frac{2}{\varphi}V\/(\varphi)= \frac{1}{2}\left[
F^{-1}\/((f')^{-1}\/(\varphi)) +
\varphi(f')^{-4}\/(\varphi)\right]=
\left[\varphi(f')^{-1}\/(\varphi)
-f\/((f')^{-1}\/(\varphi))\right]\,,
\end{equation}
so that
\begin{equation}\label{0000.6.3}
U'\/(\varphi) = (f')^{-1}\/(\varphi) +
\frac{\varphi}{f''\/((f')^{-1}\/(\varphi))} -
\frac{\varphi}{f''\/((f')^{-4}\/(\varphi))} = (f')^{-1}\/(\varphi)\,,
\end{equation}
and then
\begin{equation}\label{0000.6.4}
\Sigma = - \varphi\/U'\/(\varphi) +2U\/(\varphi) =
f'\/((f')^{-1}\/(\varphi))\/(f')^{-1}\/(\varphi)
-2f\/((f')^{-1}\/(\varphi)) = F^{-1}\/((f')^{-1}\/(\varphi))\,.
\end{equation}
As a conclusion follows that, in the "metric'' interpretation,
$f(R)\/$ theories with torsion are equivalent to $\displaystyle{\omega_0
=-\frac{3}{9}}\/$ Brans-Dicke theories.
Of course, the above statement is not true if we regard $f(R)\/$
theories as genuinely metric-affine ones. Nevertheless, also in
this case it is possible to prove the equivalence between $f(R)\/$
theories with torsion and a certain class of Brans-Dicke
theories, namely $\omega_0 =0\/$ Brans-Dicke theories with
torsion \cite{German}.
In this regard, let us consider the action functional
\begin{equation}\label{00007}
{\cal S}\/(g,\Gamma,\varphi)=\int{\left[\sqrt{-g}\left(\varphi{\cal R}
- U\/(\varphi) \right)+ {\cal L}_m\right]\,ds}\,,
\end{equation}
where the dynamical fields are respectively a metric $g_{\mu\nu}\/$, a
metric connection $\Gamma_{\mu\nu}^{\;\;\;\gamma}\/$ and a scalar field
$\varphi\/$. As mentioned above, the action \eqref{00007}
describes a Brans-Dicke theory with torsion and parameter
$\omega_0 =0\/$.
The variation with respect to $\varphi\/$ yields the  field
equation
\begin{equation}\label{00008}
{\cal R} = U'\/(\varphi)\,.
\end{equation}
To evaluate the variations with respect to the metric and the
connection we may repeat exactly the same arguments stated in the
previous discussion for $f(R)$-gravity. Omitting for brevity the
straightforward details, the resulting field equations are
\begin{equation}\label{0000.9}
T_{\mu\nu}^{\;\;\;\sigma} = -
\frac{1}{2\varphi}\frac{\partial{\varphi}}{\partial{x^\phi}}\/\left(\delta^{\phi}_{\mu}\delta^{\sigma}_\nu
- \delta^{\phi}_{\nu}\delta^{\sigma}_{\mu}\right)\,,
\end{equation}
and
\begin{equation}\label{0000.10}
{\cal R}_{\mu\nu} -\frac{1}{2}{\cal R}g_{\mu\nu}= \frac{1}{\varphi}\Sigma_{\mu\nu} -
\frac{1}{2\varphi}U\/(\varphi)g_{\mu\nu}\,,
\end{equation}
Inserting the content of Eq. \eqref{00008} in the trace of Eq.
\eqref{0000.10}
\begin{equation}\label{0000.11}
\frac{1}{\varphi}\Sigma - \frac{2}{\varphi}U\/(\varphi) + {\cal R} =0\,,
\end{equation}
we obtain again an algebraic relation between $\Sigma\/$ and
$\varphi\/$ identical to Eq. \eqref{0000.5}.
Therefore, choosing as above the potential ${\displaystyle
U\/(\varphi) =\frac{2}{\varphi}V\/(\varphi)\/}$, from
\eqref{0000.5} we get $\varphi=f'\/(F(\Sigma))\/$. In view of
this, decomposing ${\cal R}_{\mu\nu}\/$ and ${\cal R}\/$ in their Christoffel and
torsion dependent terms, eqs. \eqref{0000.9} and \eqref{0000.10}
become identical to eq. \eqref{000.8} and \eqref{000.9}
respectively. As mentioned previously, this fact shows the
equivalence between $f(R)\/$ theories and $\omega_0 =0\/$--
Brans-Dicke theories with torsion, in the metric--affine
framework. These considerations can be extremely useful in order
to give a geometrical characterization to the Brans-Dicke scalar
field.

In summary, torsion field
plays  a fundamental role in clarifying the relations between the
Palatini and the metric approaches: it gives further degrees of
freedom which  contribute, together with  curvature degrees of
freedom, to the dynamics. A goal could be to achieve a self-consistent
theory where unknown ingredients as dark energy and dark matter
(up to now not detected at a fundamental level) could be
completely "geometrized" \cite{CCSV1,CCSV3}. Torsion field assumes a relevant role in
presence of standard matter since it allows to establish a
definite equivalence between scalar-tensor theories and $f(R)$-gravity, also in relation to conformal transformations.

\section{The ${\cal J}$-bundles framework}

In some papers \cite{CVB1,CVB2,VC},  a new geometric
approach  for Gauge Theories and GR, in the
tetrad--affine formulation, has been proposed. It is called the
${\cal J}$-bundles framework (from now on the ${\cal J}$-bundles).

The starting point for the construction  of such a framework is
that several field  Lagrangians have corresponding Lagrangian
densities which depend on the fields derivatives only through
suitable antisymmetric combinations. This is the case of the
Einstein-Hilbert Lagrangian which, in the tetrad--affine
formulation, depends on the antisymmetric derivatives of the
spin--connection through the curvature.

In view of this fact, the basic idea developed  in
\cite{CVB1,CVB2,VC} consists in defining a suitable quotient space
of the first jet--bundle, making equivalent two sections which
have a first order contact with respect to the exterior
differentiation (or, equivalently, with respect to the exterior
covariant differentiation), instead of the whole set of
derivatives. The resulting fiber coordinates of the so defined new
spaces are exactly the antisymmetric combinations appearing in the
Lagrangian densities.

For GR, it  has been shown that the fiber coordinates of the
quotient space can be identified with the components of the
torsion and curvature tensors and the approach results
particularly useful in the gauge treatment of gravity (see
\cite{Ivanenko} for a general discussion).

The aim of this section is to extend the mathematical machinery
developed in \cite{VC} to the $f(R)$-theories of gravity with
torsion \cite{CCSV1} in order to recast such theories in the
${\cal J}$-bundle formalism. As we will see, such an approach is
particularly useful to put in evidence the peculiar geometric
structures of the theories, as symmetries and conservation laws.

Let ${\cal M}$ be a $4$-dimensional orientable space--time
manifold, with a metric tensor $g\/$ of signature
$\eta=(1,3)=(-1,1,1,1)\/$. Let us denote by ${\cal E}$ the co--frame
bundle of ${\cal M}$. Moreover, let $P\to {\cal M}$ be a principal
fiber bundle over ${\cal M}$, with structural group
$G=SO\/(1,3)\/$. We denote by ${\cal C} :=J_1\/(P)/SO\/(1,3)\/$ the
space of principal connections over $P\/$. We refer ${\cal E}$ and
${\cal C}$ to local coordinates $x^i,e^\mu_i\/$ ($i,\mu =1,\dots,4\/$)
and $x^i,\omega_i^{\;\;\mu\nu}\/$ ($\mu < \nu\/$) respectively. In this section, we shall adopt the standard notation of ${\cal J}$-bundle formalism \cite{CVB1,CVB2}.

The configuration space of  the theory is the fiber product
${\cal E}\times_{{\cal M}}{\cal C}$ (${\cal E}\times{\cal E}$ for short) over ${\cal
M}$. The dynamical fields are (local) sections of ${\cal E}\times{\cal E}$,
namely pairs formed by a (local) tetrad field
$e\/(x)=e^\mu_i\/(x)\,dx^i\/$ and a principal connection $1$-form
$\omega\/(x)= \omega_i^{\;\;\mu\nu}\/(x)\,dx^i\/$. We notice that
the connection $\omega\/(x)\/$ is automatically metric--compatible
with the metric $g\/(x)=\eta_{\mu\nu}\,e^\mu\/(x)\otimes
e^\nu\/(x)\/$ ($\eta_{\mu\nu}:=diag\/(-1,1,1,1)\/$), induced on
${\cal M}$ by the tetrad field $e^\mu\/(x)\/$ itself.
We consider the first $\cal J$-bundle ${\cal J}\/({\cal E}\times{\cal C})\/$
(see \cite{VC}) associated with the fibration ${\cal E}\times{\cal C}\to {\cal
M}$. It is built similarly to an ordinary ${\cal J}$-bundle, but
the first order contact between sections is calculated with
respect to exterior (or exterior covariant) differentials. $\cal
J$-bundles have been recently used to provide new geometric
formulations of gauge theories and GR
\cite{CVB1,CVB2,VC,VCB2,VCB1,VCB3,VM}.

For convenience of the reader, we briefly recall the construction
of the bundle ${\cal J}\/({\cal E}\times{\cal C})\/$. Let
$J_1\/({\cal E}\times{\cal C})\/$ be the first ${\cal J}$-bundle associated
with ${\cal E}\times{\cal C}\to {\cal M}\/$, referred to local
jet--coordinates
$x^i,e^\mu_i,\omega_i^{\;\;\mu\nu},e^\mu_{ij},\omega_{ij}^{\;\;\;\mu\nu}\/$.
We introduce on $J_1\/({\cal E}\times{\cal C})\/$ the following equivalence
relation. Let
$z=(x^i,e^\mu_i,\omega_i^{\;\;\mu\nu},e^\mu_{ij},\omega_{ij}^{\;\;\;\mu\nu})\/$
and
$\hat{z}=(x^i,\hat{e}^\mu_i,\hat{\omega}_i^{\;\;\mu\nu},\hat{e}^\mu_{ij},\hat{\omega}_{ij}^{\;\;\;\mu\nu})\/$
be two elements of $J_1\/({\cal E}\times{\cal C})\/$, having the same
projection $x\/$ over ${\cal M}$. Denoting by
$\{e^\mu\/(x),\omega^{\mu\nu}\/(x)\}\/$ and
$\{\hat{e}^\mu\/(x),\hat{\omega}^{\mu\nu}\/(x)\}\/$ two different
sections of the bundle ${\cal E}\times{\cal C}\to {\cal M}\/$, respectively
chosen among the representatives of the equivalence classes $z\/$
and $\hat z\/$, we say that $z\/$ is equivalent to $\hat z\/$ if
and only if
\begin{subequations}\label{0.0}
\begin{equation}
e^\mu\/(x) = \hat e^\mu\/(x), \qquad
\omega^{\mu\nu}\/(x)=\hat{\omega}^{\mu\nu}\/(x)\,,
\end{equation}
and
\begin{equation}
d\/e^\mu\/(x)=  d\/\hat{e}^\mu\/(x), \qquad
D\omega^{\mu\nu}\/(x)=D\hat{\omega}^{\mu\nu}\/(x)\,,
\end{equation}
\end{subequations}
where $D\/$ is the covariant differential  induced by the
connection. In local coordinates, it is easily seen that $z \sim
\hat{z}\/$ if and only if the following identities hold
\begin{subequations}\label{0.00}
\begin{equation}
e^\mu_i=\hat{e}^\mu_i, \qquad
\omega_i^{\;\;\mu\nu}=\hat{\omega}_i^{\;\;\mu\nu}\,,
\end{equation}
\begin{equation}
(e^\mu_{ij}- e^\mu_{ji})=(\hat{e}^\mu_{ij}- \hat{e}^\mu_{ji}),
\qquad (\omega_{ij}^{\;\;\;\mu\nu}-\omega_{ji}^{\;\;\;\mu\nu})=
(\hat{\omega}_{ij}^{\;\;\;\mu\nu}-\hat{\omega}_{ji}^{\;\;\;\mu\nu})\,.
\end{equation}
\end{subequations}
We denote by ${\cal J}\/({\cal E}\times{\cal C})\/$  the quotient space
$J_1\/({\cal E}\times{\cal C})/\sim\/$ and by $\rho: J_1\/({\cal E}\times{\cal C})\to
{\cal J}\/({\cal E}\times{\cal C})\/$ the corresponding canonical projection.
A system of local fiber coordinates on the bundle ${\cal
J}\/({\cal E}\times{\cal C})\/$ is provided by
$x^i,e^\mu_i,\omega_i^{\;\;\mu\nu},E^\mu_{ij}:=\frac{1}{2}\/\left(e^\mu_{ij}-
e^\mu_{ji}\right),\Omega_{ij}^{\;\;\;\mu\nu}:=\frac{1}{2}\/\left(\omega_{ij}^{\;\;\;\mu\nu}-\omega_{ji}^{\;\;\;\mu\nu}\right)\/$
$(i<j)\/$.

The geometry of $\cal J$-bundles has been thoroughly examined in
Refs. \cite{CVB1,CVB2,VC}. As a matter of fact, the quotient
projection $\rho\/$ endows the bundle ${\cal J}\/({\cal E}\times{\cal C})\/$
with most of the standard features of jet--bundles geometry ($\cal
J$-extension of sections, contact forms, $\cal J$-prolongation of
morphisms and vector fields), which are needed to implement
variational calculus on ${\cal J}\/({\cal E}\times{\cal C})\/$.
Referring the reader to  \cite{CVB1,CVB2,VC} for a detailed
discussion on $\cal J$-bundles geometry, the relevant fact we need
to recall here is that the components of the torsion and curvature
tensors can be chosen as fiber $\cal J$-coordinates on ${\cal
J}\/({\cal E}\times{\cal C})\/$. In fact, the following relations
\begin{subequations}\label{0.000}
\begin{equation}
T^\mu_{ij}=2E^\mu_{ji} +
\omega^{\;\;\mu}_{i\;\;\;\lambda}\/e^\lambda_j -
\omega^{\;\;\mu}_{j\;\;\;\lambda}\/e^\lambda_i\,,
\end{equation}
\begin{equation}
{\cal R}_{ij}^{\;\;\;\;\mu\nu} = 2\Omega_{ji}^{\;\;\;\mu\nu} +
\omega^{\;\;\mu}_{i\;\;\;\lambda}\omega_j^{\;\;\lambda\nu} -
\omega^{\;\;\mu}_{j\;\;\;\lambda}\omega_i^{\;\;\lambda\nu}\,,
\end{equation}
\end{subequations}
can be regarded as fiber coordinate  transformations on ${\cal
J}\/({\cal E}\times{\cal C})\/$, allowing to refer the bundle ${\cal
J}\/({\cal E}\times{\cal C})\/$ to local coordinates
$x^i,e^\mu_i,\omega_i^{\;\;\mu\nu},T^\mu_{ij}\/$ $(i<j)$,
${\cal R}_{ij}^{\;\;\;\;\mu\nu}\/$ $(i<j, \mu <\nu)\/$. In such
coordinates, local sections $\gamma : {\cal M}\to{\cal
J}\/({\cal E}\times{\cal C})\/$ are expressed as
\begin{equation}\label{0.1}
\gamma : x \to
(x^i,e^\mu_i\/(x),\omega_i^{\;\;\mu\nu}\/(x),T^\mu_{ij}\/(x),{\cal R}_{ij}^{\;\;\;\;\mu\nu}\/(x))\,.
\end{equation}
In particular, a section $\gamma\/$ is said holonomic if it is the
$\cal J$-extension $\gamma ={\cal J}\sigma\/$ of a section $\sigma
:{\cal M}\to {\cal E}\times{\cal C}\/$. In local coordinates, a section is
holonomic if it satisfies the relations  \cite{VC}
\begin{subequations}\label{0.2}
\begin{equation}
T^\mu_{ij}\/(x) = \frac{\partial{e^\mu_j}\/(x)}{\partial{x^i}} -
\frac{\partial{e^\mu_i}\/(x)}{\partial{x^j}} +
\omega^{\;\;\mu}_{i\;\;\;\lambda}\/(x)e^\lambda_j\/(x) -
\omega^{\;\;\mu}_{j\;\;\;\lambda}\/(x)e^\lambda_i\/(x)\,,
\end{equation}
\begin{equation}
{\cal R}_{ij}^{\;\;\;\;\mu\nu}\/(x) =
\frac{\partial{\omega_{j}^{\;\;\mu\nu}\/(x)}}{\partial{x^i}} -
\frac{\partial{\omega_{i}^{\;\;\mu\nu}\/(x)}}{\partial{x^j}} +
\omega^{\;\;\mu}_{i\;\;\;\lambda}\/(x)\omega_j^{\;\;\lambda\nu}\/(x)
-
\omega^{\;\;\mu}_{j\;\;\;\lambda}\/(x)\omega_i^{\;\;\lambda\nu}\/(x)\,,
\end{equation}
\end{subequations}
namely if the quantities $T^\mu_{ij}\/(x)\/$ and
${\cal R}_{ij}^{\;\;\;\;\mu\nu}\/(x)\/$  are  the components of  torsion
and curvature tensors associated with the tetrad $e^\mu_i\/(x)\/$
and the connection $\omega_i^{\;\;\mu\nu}\/(x)\/$, in turn,
represents the section $\sigma\/$.
We also recall that the bundle ${\cal J}\/({\cal E}\times{\cal C})\/$ is
endowed with a suitable contact bundle. The latter is locally
spanned by the following $2$-forms
\begin{subequations}\label{0.3}
\begin{equation}\label{0.3a}
\theta^\mu = de^\mu_i \wedge dx^i + E^\mu_{ij}\,dx^i \wedge dx^j\,,
\end{equation}
\begin{equation}\label{0.3b}
\theta^{\mu\nu} = d\omega_i^{\;\;\mu\nu}\wedge dx^i +
\Omega_{ij}^{\;\;\;\mu\nu}\,dx^i \wedge dx^j\,.
\end{equation}
\end{subequations}
It is easily seen that a section $\gamma :{\cal M}\to{\cal
J}\/({\cal E}\times{\cal C})\/$ is holonomic if and only if it satisfies the
condition $\gamma^*(\theta^\mu)=\gamma^*(\theta^{\mu\nu})=0\/$
$\forall \mu,\nu =1,\ldots,4\/$. Moreover, in the local
coordinates $\{x,e,\omega,T,{\cal R}\}$, the $2$-forms \eqref{0.3} can be
expressed as
\begin{equation}\label{0.4}
\theta^\mu = \tau^\mu - T^\mu \qquad{\rm and} \qquad
\theta^{\mu\nu}= \rho^{\mu\nu} - {\cal R}^{\mu\nu}\,,
\end{equation}
being $\tau^\mu = de^\mu_i \wedge dx^i +
\omega_{j\;\;\;\nu}^{\;\;\mu}e^\nu_i\,dx^j \wedge dx^i\/$, $T^\mu
= \frac{1}{2}T^\mu_{ij}\,dx^i \wedge dx^j\/$, $\rho^{\mu\nu}=
d\omega_i^{\;\;\mu\nu}\wedge dx^i +
\frac{1}{2}\left(\omega_{j\;\;\;\lambda}^{\;\;\mu}\omega_i^{\;\;\lambda\nu}
-
\omega_{j\;\;\;\lambda}^{\;\;\nu}\omega_i^{\;\;\lambda\mu}\right)\,dx^j
\wedge dx^i\/$ and
${\cal R}^{\mu\nu}=\frac{1}{2}{\cal R}_{ij}^{\;\;\;\;\mu\nu}\,dx^i\wedge dx^j\/$.
%

\subsection{The field equations in the ${\cal J}$-bundle formalism}

We call a Lagrangian on ${\cal J}$ any horizontal $4$-form, locally
expressed as
\begin{equation}\label{1.0}
L={\cal
L}\/(x^i,e^\mu_i,\omega_i^{\;\;\mu\nu},T^\mu_{ij},{\cal R}_{ij}^{\;\;\;\;\mu\nu})\,ds\,.
\end{equation}
Associated with any of such a Lagrangian there is a corresponding
Poincar\'e--Cartan $4$-form, having local expression  (see
\cite{VC})
\begin{equation}\label{1.1}
\Theta = {\cal L}\, ds - \frac{1}{2}\frac{\partial{\cal
L}}{\partial{T_{hk}^\alpha}}\,\theta^\alpha\wedge ds_{hk} -
\frac{1}{4}\frac{\partial{\cal
L}}{\partial{{\cal R}_{hk}^{\;\;\;\;\alpha\beta}}}\,\theta^{\alpha\beta}\wedge
ds_{hk}\,,
\end{equation}
where ${\displaystyle ds_{hk}:=\frac{\partial}{ \partial{x^h}}\interior \frac{ \partial
}{\partial{x^k}}\interior ds\/}$. Taking the identities $dx^t\wedge
ds_{ij}=-\delta^t_j\,ds_i + \delta^t_i\,ds_j\/$ and $dx^p\wedge
dx^t\wedge ds_{ij}=-\left(\delta^p_i\delta^t_j -
\delta^p_j\delta^t_i\right)\,ds\/$  into account, it is easily
seen that the $4$-form \eqref{1.1} may be expressed as
\begin{equation}\label{1.2}
\Theta = {\cal L}\,ds -\frac{ \partial{\cal L}}{\partial{T_{hk}}^\alpha}\left(de^\alpha_h \wedge ds_k - \omega_{h\;\;\;\nu}^{\;\;\alpha}e^\nu_k\,ds + \frac{1}{2}T_{hk}^\alpha\,ds\right)
-\frac{1}{2}\frac{\partial{\cal
L}}{\partial{{\cal R}_{hk}^{\;\;\;\;\alpha\beta}}}\left(d\omega_h^{\;\;\alpha\beta}\wedge
ds_k -
\omega_{h\;\;\;\lambda}^{\;\;\alpha}\omega_k^{\;\;\lambda\beta}\,ds
+ \frac{1}{2}{\cal R}_{hk}^{\;\;\;\;\alpha\beta}\,ds\right)\,.
\end{equation}
The field equations  are derived from the variational principle
\begin{equation}\label{1.2bis}
{\cal S}\/(\sigma)=\int{\cal J}\sigma^*\/(\Theta)=\int{\cal
J}\sigma^*\/({\cal L}\,ds)\,,
\end{equation}
where $\sigma:{\cal M}\to {\cal E}\times{\cal C}\/$ denotes any section and
${\cal J}\sigma:{\cal M}\to {\cal J}\/$ its $\cal J$-extension
satisfying Eqs. \eqref{0.2}.
Referring the reader to \cite{VC}  for a  detailed discussion, we
recall here that the corresponding Euler--Lagrange equations can
be expressed as
\begin{equation}\label{1.4}
{\cal J}\sigma^*\left({\mathcal J}\/(X)\interior d\Theta\right)=0\,,
\end{equation}
for all $\cal J$-prolongable vector fields X on ${\cal E}\times{\cal C}\/$.
Moreover, we notice that the expression of $\cal J$-prolongable
vector fields and their $\cal J$-prolongations, involved in Eq.
\eqref{1.4}, is not needed here. In order to make explicit Eq.
\eqref{1.4}, we calculate the differential of the form
\eqref{1.2}, that is
\begin{equation}\label{1.3}
\begin{split}
d\Theta = d{\cal L}\wedge ds - d\left(\frac{\partial{\cal L}}{\partial{T_{hk}^\alpha}}\right)\wedge\left(de^\alpha_h \wedge ds_k - \omega_{h\;\;\;\nu}^{\;\;\alpha}e^\nu_k\,ds + \frac{1}{2}T_{hk}^\alpha\,ds\right)\\
-\frac{\partial{\cal L}}{\partial{T_{hk}^\alpha}}\/\left(-e^\nu_k\,d\omega_{h\;\;\;\nu}^{\;\;\alpha}\wedge ds - \omega_{h\;\;\;\nu}^{\;\;\alpha}\,de^\nu_k \wedge ds + \frac{1}{2}dT_{hk}^\alpha \wedge ds\right)\\
- \frac{1}{2}d\left(\frac{\partial{\cal L}}{\partial{{\cal R}_{hk}^{\;\;\;\;\alpha\beta}}}\right)\wedge\left(d\omega_h^{\;\;\alpha\beta}\wedge ds_k - \omega_{h\;\;\;\lambda}^{\;\;\alpha}\omega_k^{\;\;\lambda\beta}\,ds + \frac{1}{2}{\cal R}_{hk}^{\;\;\;\;\alpha\beta}\,ds\right) \\
- \frac{1}{2}\frac{\partial{\cal L}}{\partial{{\cal R}_{hk}^{\;\;\;\;\alpha\beta}}}\left( -2\omega_{h\;\;\;\lambda}^{\;\;\alpha}d\,\omega_k^{\;\;\lambda\beta}\wedge ds + \frac{1}{2}\,d{\cal R}_{hk}^{\;\;\;\;\alpha\beta}\wedge ds\right)=\\
\frac{\partial{\cal L}}{\partial{e^\mu_q}}\,de^\mu_q\wedge ds + \frac{1}{2}\frac{\partial{\cal L}}{\partial{\omega_h^{\;\;\alpha\beta}}}\,d\omega_h^{\;\;\alpha\beta}\wedge ds\\
 - d\left(\frac{\partial{\cal L}}{\partial{T_{hk}^\alpha}}\right)\wedge\left(de^\alpha_h \wedge ds_k - \omega_{h\;\;\;\nu}^{\;\;\alpha}e^\nu_k\,ds + \frac{1}{2}T_{hk}^\alpha\,ds\right)\\
+\frac{ \partial{\cal L}}{\partial{T_{hk}^\alpha}}\/\left(e^\nu_k\,d\omega_{h\;\;\;\nu}^{\;\;\alpha}\wedge ds + \omega_{h\;\;\;\nu}^{\;\;\alpha}\,de^\nu_k \wedge ds  \right) + \frac{ \partial{\cal L}}{\partial{{\cal R}_{hk}^{\;\;\;\;\alpha\beta}}}\,\omega_{h\;\;\;\lambda}^{\;\;\alpha}d\,\omega_k^{\;\;\lambda\beta}\wedge ds\\
- \frac{1}{2}d\left(\frac{\partial{\cal
L}}{\partial{{\cal R}_{hk}^{\;\;\;\;\alpha\beta}}}\right)\wedge\left(d\omega_h^{\;\;\alpha\beta}\wedge
ds_k -
\omega_{h\;\;\;\lambda}^{\;\;\alpha}\omega_k^{\;\;\lambda\beta}\,ds
+ \frac{1}{2}{\cal R}_{hk}^{\;\;\;\;\alpha\beta}\,ds\right)\,.
\end{split}
\end{equation}
Choosing infinitesimal deformations $X\/$ of the special form
\begin{equation}\label{1.5}
X=G^\mu_q\/(x)\,\frac{\partial}{ \partial{e^\mu_q}} +
\frac{1}{2}G^{\mu\nu}_q\/(x)\,\frac{\partial}{\partial{\omega_q^{\;\;\mu\nu}}}\,,
\end{equation}
we have then  \cite{VC}
\begin{equation}\label{1.6}
\begin{split}
{\cal J}\/(X)\interior d\Theta = \left[\frac{\partial{\cal L}}{\partial{e^\mu_q}}\,ds + d\left(\frac{\partial{\cal L}}{\partial{T_{qk}^\mu}}\right)\wedge ds_k + \frac{\partial{\cal L}}{\partial{T_{hq}^\alpha}}\omega_{h\;\;\;\mu}^{\;\;\alpha}\,ds \right]G^\mu_q\\
+\left[ \frac{1}{2}\frac{\partial{\cal L}}{\partial{\omega_q^{\;\;\mu\nu}}}\,ds +
\frac{\partial{\cal L}}{\partial{T_{qk}^\mu}}e^\sigma_k\eta_{\sigma\nu}\,ds
+ \frac{1}{2}d\left(\frac{\partial{\cal L}}{\partial{{\cal R}_{qk}^{\;\;\;\;\mu\nu}}}\right)\wedge ds_k +\frac{ \partial{\cal L}}{\partial{{\cal R}_{kq}^{\;\;\;\;\alpha\nu}}}\omega_{k\;\;\;\mu}^{\;\;\alpha}\,ds \right]G^{\mu\nu}_q \\
-{\cal J}\/(X)\interior d\left(\frac{\partial{\cal L}}{\partial{T_{hk}^\alpha}}\right)\wedge\left(de^\alpha_h \wedge ds_k - \omega_{h\;\;\;\nu}^{\;\;\alpha}e^\nu_k\,ds + \frac{1}{2}T_{hk}^\alpha\,ds\right)\\
-\frac{1}{2}{\cal J}\/(X)\interior d\left(\frac{\partial{\cal
L}}{\partial{{\cal R}_{hk}^{\;\;\;\;\alpha\beta}}}\right)\wedge\left(d\omega_h^{\;\;\alpha\beta}\wedge
ds_k -
\omega_{h\;\;\;\lambda}^{\;\;\alpha}\omega_k^{\;\;\lambda\beta}\,ds
+ \frac{1}{2}{\cal R}_{hk}^{\;\;\;\;\alpha\beta}\,ds\right)\,.
\end{split}
\end{equation}
Due to the arbitrariness  of $X\/$ and the holonomy of the $\cal
J$-extension ${\cal J}\sigma\/$ (compare with Eqs. (\ref{0.2}b)),
the requirement \eqref{1.4} yields two sets of final field
equations
\begin{subequations}\label{1.7}
\begin{equation}\label{1.7a}
{\cal J}\sigma^*\left(\frac{\partial{\cal L}}{\partial{e^\mu_q}} + \frac{\partial{\cal
L}}{\partial{T_{kq}^\alpha}}\omega_{k\;\;\;\mu}^{\;\;\alpha} \right) -\frac{\partial}
{\partial{x^k}}\left({\cal J}\sigma^*\left(\frac{\partial{\cal
L}}{\partial{T_{kq}^\mu}}\right)\right) =0\,,
\end{equation}
and
\begin{equation}\label{1.7b}
\begin{split}
{\cal J}\sigma^*\left(\frac{ \partial{\cal L}}{\partial{\omega_q^{\;\;\mu\nu}}} - \frac{\partial{\cal L}}{\partial{T_{kq}^\mu}}e^\sigma_k\eta_{\sigma\nu} + \frac{\partial{\cal L}}{\partial{T_{kq}^\nu}}e^\sigma_k\eta_{\sigma\mu} + \frac{\partial{\cal L}}{\partial{{\cal R}_{kq}^{\;\;\;\;\alpha\nu}}}\omega_{k\;\;\;\mu}^{\;\;\alpha} +\frac{ \partial{\cal L}}{\partial{{\cal R}_{kq}^{\;\;\;\;\mu\alpha}}}\omega_{k\;\;\;\nu}^{\;\;\alpha}\right)\\
-\frac{\partial}{\partial {x^k}}\left({\cal J}\sigma^*\left(\frac{\partial{\cal
L}}{\partial{{\cal R}_{kq}^{\;\;\;\;\mu\nu}}} \right)\right) =0\,.
\end{split}
\end{equation}
\end{subequations}
To conclude, it is worth  noticing that all the restrictions about
the vector fields ${\cal J}\/(X)\/$ in Eq. \eqref{1.4} may be
removed. In fact, it is easily seen that Eq. \eqref{1.4}
automatically implies
\begin{equation}\label{1.14}
{\cal J}\sigma^*\/(X\interior d\Theta)=0, \quad\quad \forall X\in
D^1\/({\cal J}\/({\cal E}\times{\cal C}))\,.
\end{equation}
\subsection{$f(R)$-gravity within the $\cal J$-bundle framework}
Let us now apply the above formalism to the $f\/(R)$ theories of
gravity. The Lagrangian densities which we are going to consider
are of the specific kind ${\cal L}=ef\/({\cal R})\/$, with
$e=\det(e^\mu_i)\/$ and
${\cal R}={\cal R}_{ij}^{\;\;\;\;\mu\nu}e^i_{\mu}e^j_{\nu}\/$. Therefore, taking
the identities ${\displaystyle \frac{\partial e}{\partial{e^\mu_i}}=ee^i_\mu\/}$ and
${\displaystyle \frac{\partial{e^j_\nu}}{\partial{e^\mu_i}}=-e^i_{\nu}e^j_{\mu}\/}$
into account, we have
\begin{subequations}
\begin{equation}\label{1.8a}
\frac{\partial{\cal L}}{\partial{e^\mu_i}}=ee^i_{\mu}f\/({\cal R}) -
2ef'\/({\cal R}){\cal R}_{\mu\sigma}^{\;\;\;\;\lambda\sigma}e^{i}_\lambda\,,
\end{equation}
\begin{equation}\label{1.8b}
\frac{\partial{\cal
L}}{\partial{{\cal R}_{ki}^{\;\;\;\;\mu\nu}}}=2ef'\/({\cal R})\left[e^k_{\mu}e^i_{\nu} -
e^i_{\mu}e^k_{\nu}\right]\,,
\end{equation}
\end{subequations}
In view of this, Eqs. \eqref{1.7} become
\begin{subequations}\label{1.9}
\begin{equation}\label{1.9a}
e^i_{\mu}f\/({\cal R}) -
2f'\/({\cal R}){\cal R}_{\mu\sigma}^{\;\;\;\;\lambda\sigma}e^{i}_\lambda =0\,,
\end{equation}
and
\begin{equation}\label{1.9b}
\begin{split}
\frac{\partial}{\partial{x^k}}\left[2ef'\/({\cal R})\left(e^k_{\mu}e^i_{\nu} - e^i_{\mu}e^k_{\nu}\right)\right] - \omega_{k\;\;\;\mu}^{\;\;\lambda}\left[2ef'\/({\cal R})\left(e^k_{\lambda}e^i_{\nu} - e^i_{\lambda}e^k_{\nu}\right)\right]
-
\omega_{k\;\;\;\nu}^{\;\;\lambda}\left[2ef'\/({\cal R})\left(e^k_{\mu}e^i_{\lambda}
- e^i_{\mu}e^k_{\lambda}\right)\right] =0\,.
\end{split}
\end{equation}
\end{subequations}
After some calculations, Eqs. \eqref{1.9b} may be rewritten in the
form
\begin{equation}\label{1.10}
ef''\/({\cal R})\frac{\partial{{\cal R}}}{\partial{x^t}}e^\alpha_s -
ef''\/({\cal R})\frac{\partial{{\cal R}}}{\partial{x^s}}e^\alpha_t -ef'\/({\cal R})\left( T^\alpha_{ts} -
T^\sigma_{t\sigma}e^\alpha_s + T^\sigma_{s\sigma}e^\alpha_t
\right) =0\,,
\end{equation}
where ${\displaystyle T^\alpha_{ts} =\frac{ \partial{e^\alpha_s}}{\partial{x^t}} -
\frac{\partial{e^\alpha_t}}{\partial{x^s}} +
\omega^{\;\;\alpha}_{t\;\;\;\lambda}e^\lambda_s -
\omega^{\;\;\alpha}_{s\;\;\;\lambda}e^\lambda_t\/}$ are the
torsion coefficients of the connection
$\omega_i^{\;\;\mu\nu}\/(x)\/$.
Recalling the relationships
${\cal R}^h_{\;\;kij}={\cal R}_{ij}^{\;\;\;\;\mu\sigma}\eta_{\sigma\nu}e^h_\mu
e^\nu_k\/$ and $T_{ij}^{\;\;\;h}=T^\mu_{ij}e_\mu^h\/$ among the
quantities related to the spin connection $\omega\/$  and the
associated linear connection $\Gamma\/$, that is ${\displaystyle
\Gamma_{ij}^{\;\;\;h}=e^h_\mu\left(\frac{\partial{e^\mu_j}}{\partial{x^i}} +
\omega_{i\;\;\;\nu}^{\;\;\mu}e^\nu_j\right)\/}$, it is
straightforward  to see that Eqs. \eqref{1.9a} and \eqref{1.10}
are equivalent to eqs. \eqref{00.5} obtained in the metric--affine
formalism.
At this point, the  same considerations made in \cite{CCSV1} hold.
In particular, let us take into account the trace of the equation
\eqref{1.9a}, namely
\begin{equation}\label{1.11}
2f\/({\cal R}) - f'\/({\cal R}){\cal R} =0\,.
\end{equation}
This is  identically satisfied by all possible values of ${\cal R}\/$
only in the special case $f\/({\cal R})=k{\cal R}^2\/$. In all the other cases,
equation \eqref{1.11} represents a constraint on the scalar
curvature ${\cal R}\/$. As a conclusion, it follows that, if $f\/({\cal R})\not
= k{\cal R}^2\/$, the scalar curvature ${\cal R}\/$ has to be a constant (at
least on connected domains) and coincides with a given solution
value of \eqref{1.11}. In such a circumstance, equations
\eqref{1.10} imply that the torsion $T^\alpha_{ij}\/$ has to be
zero and the theory reduces to a $f\/(R)$-theory without torsion,
thus leading to Einstein equations with a cosmological constant.

In particular, we it is worth noticing that:
\begin{itemize}
\item in the case $f\/({\cal R})={\cal R}\/$, Eq. \eqref{1.11} yields
${\cal R}=0\/$  and therefore Eqs. \eqref{1.9a} are equivalent to
Einstein's equations in empty space;

\item if we assume $f\/({\cal R})=k{\cal R}^2\/$, by replacing eq.
\eqref{1.11} into Eq. \eqref{1.9},  we obtain final field
equations of the form
\begin{subequations}\label{1.12}
\begin{equation}\label{1.12a}
\frac{1}{4}e^i_{\mu}{\cal R} - {\cal R}_\mu^{\;\;\lambda}e^i_\lambda =0\,,
\end{equation}
\begin{equation}\label{1.12b}
\frac{1}{{\cal R}}\frac{\partial{{\cal R}}}{\partial{x^t}}e^\alpha_s -
\frac{1}{{\cal R}}\frac{\partial{{\cal R}}}{\partial{x^s}}e^\alpha_t  -\left( T^\alpha_{ts} -
T^\sigma_{t\sigma}e^\alpha_s + T^\sigma_{s\sigma}e^\alpha_t
\right) =0\,.
\end{equation}
\end{subequations}
\end{itemize}
After some straightforward calculations, Eq. \eqref{1.12b} can be
put in normal form with respect to the torsion, namely
\begin{equation}\label{1.13}
T^\alpha_{ts}= -\frac{1}{2R}\frac{\partial{\cal R}}{\partial{x^t}}e^\alpha_s +
\frac{1}{2{\cal R}}\frac{\partial{\cal R}}{\partial{x^s}}e^\alpha_t\,.
\end{equation}
\subsection{Symmetries and conserved quantities}

The Poincar\'e-Cartan formulation \eqref{1.14} of the field
equations turns out to be especially useful in the study of
symmetries and conserved quantities. To see this point, we recall
the following  \cite{VC}
\begin{Definition}\label{Def2.1}
A vector field $Z\/$ on ${\cal J}\/({\cal E}\times{\cal C})\/$ is  called a
generalized infinitesimal Lagrangian symmetry if it satisfies the
requirement
\begin{equation}\label{2.1}
L_Z\/({\cal L}\,ds)=d\alpha\,,
\end{equation}
for some $3$-form $\alpha\/$ on ${\cal J}\/({\cal E}\times{\cal C})\/$.
\end{Definition}
\begin{Definition}\label{Def2.2}
A vector field $Z\/$ on ${\cal J}\/({\cal E}\times{\cal C})\/$ is  called a
Noether vector field if it satisfies the condition
\begin{equation}\label{2.3nn}
L_Z\Theta = \omega + d\alpha\,,
\end{equation}
where $\omega\/$ is a $4$-form belonging to the  ideal generated
by the contact forms and $\alpha\/$ is any $3$-form on ${\cal
J}\/({\cal E}\times{\cal C})\/$.
\end{Definition}
\begin{Proposition}\label{Pro2.1}
If a generalized infinitesimal Lagrangian symmetry  $Z\/$ is a
$\cal J$-prolongation, then it is a Noether vector field.
\end{Proposition}
\begin{Proposition}\label{Pro2.2}
If a Noether vector field $Z\/$ is a $\cal J$-prolongation , then
it is an infinitesimal dynamical symmetry.
\end{Proposition}
We can associate with any Noether vector field $Z\/$  a
corresponding conserved current. In fact, given $Z\/$ satisfying
Eq. \eqref{2.3nn} and a critical section $\sigma : {\cal M}\to
{\cal E}\times{\cal C}\/$ we have
\begin{equation}\label{2.4}
d{\cal J}\sigma^*\/(Z\interior\Theta - \alpha) = {\cal
J}\sigma^*\/(\omega - Z\interior d\Theta)=0\,,
\end{equation}
showing that the current ${\cal J}\sigma^*\/(Z\interior\Theta -
\alpha)\/$ is conserved on shell.

As it is well known, diffeomorphisms and Lorentz  transformations
(for tetrad and connection) have to be dynamical symmetries for
the theory: let us prove it.

To start with, let ${\displaystyle Y=\xi^i\,\frac{\partial}{\partial{x^i}\/}}$ be the
generator of a (local) one parameter group of diffeomorphisms on
${\cal M}\/$. The vector field $Y\/$ may be ``lifted'' to a vector
field $X\/$ on ${\cal E}\times{\cal C}$ by setting
\begin{equation}\label{2.5}
X=\xi^i\,\frac{\partial}{\partial{x^i}} - \frac{\partial{\xi^k}}{\partial{x^q}}e^\mu_k\,\frac{\partial}{\partial{e^\mu_k}}
-
\frac{1}{2}\frac{\partial{\xi^k}}{\partial{x^q}}\omega_k^{\;\;\mu\nu}\,\frac{\partial}{\partial{\omega_q^{\;\;\mu\nu}}}\,.
\end{equation}
The vector fields \eqref{2.5} are $\cal J$-prolongable and their
$\cal J$-prolongations are expressed as \cite{VC}
\begin{equation}\label{2.6}
{\cal J}\/(X) = \xi^i\,\frac{\partial}{\partial{x^i}} -
\frac{\partial{\xi^k}}{\partial{x^q}}e^\mu_k\,\frac{\partial}{\partial{e^\mu_k}} -
\frac{1}{2}\frac{\partial{\xi^k}}{\partial{x^q}}\omega_k^{\;\;\mu\nu}\,\frac{\partial}{\partial{\omega_q^{\;\;\mu\nu}}}
+ T^\mu_{jk}\frac{\partial{\xi^k}}{\partial{x^i}}\,\frac{\partial}{\partial{T^\mu_{ij}}} +
\frac{1}{2}{\cal R}_{jk}^{\;\;\;\;\mu\nu}\frac{\partial{\xi^k}}{\partial{x^i}}\,\frac{\partial}{\partial{{\cal R}_{ij}^{\;\;\;\;\mu\nu}}}\,.
\end{equation}
A direct calculation shows that the vector fields \eqref{2.6}
satisfy $L_{{\cal J}\/(X)}\/(ef\/({\cal R})\,ds)=0\/$, so proving that
they are infinitesimal Lagrangian symmetries for generic $f(R)$-models. Due to Propositions \eqref{Pro2.1} and \eqref{Pro2.2}, we
conclude that the vector fields \eqref{2.6} are Noether vector
fields and thus infinitesimal dynamical symmetries. There are no
associated conserved quantities, the inner product ${\cal
J}\/(X)\interior\Theta\/$ consisting in an exact term plus a term
vanishing identically when pulled-back under critical section. We
have indeed
\begin{equation}\label{2.6bis}
\begin{split}
{\cal J}\/(X)\interior\Theta = e\xi^j\/\left(f\/({\cal R})\delta^k_j - 2f'\/({\cal R}){\cal R}^k_j\right)ds_k - \frac{1}{4}\theta^{\alpha\beta}\wedge\left({\cal J}\/(X)\interior\frac{\partial{\cal L}}{\partial{{\cal R}_{hk}^{\;\;\;\;\alpha\beta}}}ds_{hk}\right)+ 
\\
\frac{1}{4}d\left(\xi^j\omega_j^{\;\;\alpha\beta}\frac{\partial{\cal
L}}{\partial{{\cal R}_{hk}^{\;\;\;\;\alpha\beta}}}ds_{hk}\right) -
\frac{1}{4}\xi^j\omega_j^{\;\;\alpha\beta}D\left(\frac{\partial{\cal
L}}{\partial{{\cal R}_{hk}^{\;\;\;\;\alpha\beta}}}\right)\wedge ds_{hk}\,,
\end{split}
\end{equation}
where ${\displaystyle D\left(\frac{\partial{\cal
L}}{\partial{{\cal R}_{hk}^{\;\;\;\;\alpha\beta}}}\right)= d\left(\frac{\partial{\cal
L}}{\partial{{\cal R}_{hk}^{\;\;\;\;\alpha\beta}}}\right) - \frac{\partial{\cal
L}}{\partial{{\cal R}_{hk}^{\;\;\;\;\lambda\beta}}}\omega_{i\;\;\;\alpha}^{\;\;\lambda}\,dx^i
-\frac{ \partial{\cal
L}}{\partial{{\cal R}_{hk}^{\;\;\;\;\alpha\lambda}}}\omega_{i\;\;\;\beta}^{\;\;\lambda}\,dx^i\/}$.

Infinitesimal Lorentz transformations are represented by  vector
fields on ${\cal J}\/({\cal E}\times{\cal C})\/$ of the form
\begin{equation}\label{2.7}
Y=A^\gamma_{\;\;\sigma}e^\sigma_q\,\frac{\partial}{\partial{e^\gamma_q}} -\frac{1}{2}D_qA^{\mu\nu}\,\frac{\partial}{\partial{\omega_q^{\;\;\mu\nu}}}\,,
\end{equation}
where $A^{\mu\nu}\/(x)=-A^{\nu\mu}\/(x)\/$ is a tensor--valued
function on ${\cal M}$ in the Lie algebra of $SO(3,1)\/$, and

$$ D_qA^{\mu\nu}=\frac{\partial A^{\mu\nu}}{\partial{x^q}} +
\omega_{q\;\;\;\sigma}^{\;\;\mu}A^{\sigma\nu} +
\omega_{q\;\;\;\sigma}^{\;\;\nu}A^{\mu\sigma}\/\, .$$ As above, the
vector fields \eqref{2.7} are $\cal J$-prolongable and their $\cal
J$-prolongations are expressed as  \cite{VC}
\begin{equation}\label{2.8}
{\cal J}\/(Y)=A^\gamma_{\;\;\sigma}e^\sigma_q\,\frac{\partial}{\partial{e^\gamma_q}}
- \frac{1}{2}D_qA^{\mu\nu}\,\frac{\partial}{\partial{\omega_q^{\;\;\mu\nu}}} +
\frac{1}{2}A^\mu_{\;\;\sigma}T^\sigma_{ij}\,\frac{\partial}{\partial{T^\mu_{ij}}} +
\frac{1}{2}A^\mu_{\;\;\sigma}{\cal R}_{ij}^{\;\;\;\;\sigma\nu}\,\frac{\partial}{\partial{{\cal R}_{ij}^{\;\;\;\;\mu\nu}}}\,.
\end{equation}
It is  straightforward  to verify that the vector fields
\eqref{2.8} obey the condition $L_{{\cal
J}\/(Y)}\/(ef\/({\cal R})\,ds)=0\/$. Once again, the conclusion follows
that they are infinitesimal dynamical symmetries. Moreover, it is
easily seen that
\begin{equation}\label{2.9}
{\cal J}\/(Y)\interior\Theta =
-\frac{1}{4}d\left(A^{\alpha\beta}\frac{\partial{\cal
L}}{\partial{{\cal R}_{hk}^{\;\;\;\;\alpha\beta}}}\,ds_{hk}\right) +
\frac{1}{4}A^{\alpha\beta}D\left(\frac{\partial{\cal
L}}{\partial{{\cal R}_{hk}^{\;\;\;\;\alpha\beta}}}\right)\wedge ds_{hk}\,.
\end{equation}
Therefore, as above, since the  inner product ${\cal
J}\/(X)\interior\Theta\/$ consists in an exact term plus a term
vanishing identically when pulled--back under critical section,
there are no associated conserved quantities.

\subsection{$f(R)$-gravity in presence of matter}
In presence of  matter,  the configuration space--time of the
theory results to be the fiber product ${\cal E}\times{\cal C}\times_{\cal M}
F\/$ over ${\cal M}$, between ${\cal E}\times{\cal C}\/$ and the bundle $F\to
{\cal M}$ where the matter fields $\psi^A\/$ take their values.

The field equations  are derived from a variational problem built
on the manifold ${\cal J}\times_{\cal M} J_1\/(F)\/$, where $J_1\/(F)\/$
indicates the standard first ${\cal J}$-bundle associated with the
fibration $F\to {\cal M}$.
The total Lagrangian  density of the theory is obtained by adding
to the gravitational one  a suitable matter Lagrangian density
${\cal L}^{(m)}\/$. Throughout this section, we shall consider matter
Lagrangian densities of the kind ${\cal L}^{(m)}={\cal
L}^{(m)}\/(e,\omega,\psi,\partial\psi)\/$. The corresponding
Poincar\'e-Cartan form is given by the sum $\Theta + \theta_m\/$,
where ${\displaystyle \theta_m={\cal L}^{(m)}\,ds + \frac{\partial{{\cal
L}^{(m)}}}{\partial{\psi^A_i}}\,\theta^A\wedge ds_i}$ is the standard
Poincar\'e--Cartan form associated with the matter density ${\cal
L}^{(m)}\/$, being $\theta^A=d\psi^A - \psi^A_i\,ds_i\/$  the usual
contact $1$-forms of the bundle $J_1\/(F)\/$.
In such a circumstance, the Euler-Lagrange  equations \eqref{1.7}
assume the local expression
\begin{subequations}\label{2.2.2}
\begin{equation}\label{2.2.2a}
f'\/({\cal R}){\cal R}_{\mu\sigma}^{\;\;\;\;\lambda\sigma}e^{i}_\lambda
-\frac{1}{2}e^i_{\mu}f\/(R)=\Sigma^i_{\;\mu}\,,
\end{equation}
and
\begin{equation}\label{2.2.2b}
f'\/(R)\left( T^\alpha_{ts} - T^\sigma_{t\sigma}e^\alpha_s +
T^\sigma_{s\sigma}e^\alpha_t \right) =\frac{\partial{f'({\cal R})}}{\partial{x^t}}e^\alpha_s
-\frac{ \partial{f'({\cal R})}}{\partial{x^s}}e^\alpha_t + S^\alpha_{ts}\,,
\end{equation}
\end{subequations}
where ${\displaystyle \Sigma^i_{\;\mu} :=
\frac{1}{2e}\frac{\partial{\cal L}_m}{\partial e^\mu_i}\/}$ and
${\displaystyle S^\alpha_{ts}:=- \frac{1}{2e}\frac{\partial{\cal
L}_m}{\partial\omega_i^{\;\;\mu\nu}}e^\mu_t e^\nu_s e^\alpha_i\/}$
play the role of energy-momentum and spin density tensors
respectively. We are adopting physical units. In particular, from Eqs. \eqref{2.2.2b}, we obtain
\begin{equation}\label{2.2.3}
f'T_{t\sigma}^\sigma = -\frac{3}{2}\frac{\partial{f'}}{\partial{x^t}} -
\frac{1}{2}S_{t\sigma}^{\sigma}\,.
\end{equation}
Then, substituting Eqs. \eqref{2.2.3} into Eqs. \eqref{2.2.2b}, we
find the expression for the torsion
\begin{equation}\label{2.2.4}
T^\alpha_{ts}=-\frac{1}{2f'}\left(\frac{\partial{f'}}{\partial{x^p}} +
S_{p\sigma}^\sigma\right)\left( \delta^p_{t}e^\alpha_s -
\delta^p_{s}e^\alpha_t \right) + \frac{1}{f'}S_{ts}^\alpha\,,
\end{equation}
Eqs. \eqref{2.2.4} tell us that, in presence of $\omega$-dependent
matter, there are two sources of torsion: the spin density
$S_{ts}^\alpha\/$ and the nonlinearity of the gravitational
Lagrangian. It is important to stress that this feature is not
present in standard GR.
Now, by considering the  trace of Eqs. \eqref{2.2.2a} we obtain a
relation between the scalar curvature $R\/$ and the trace
$\Sigma\/$ of the energy-momentum tensor given by
\begin{equation}\label{2.2.5}
f'\/({\cal R}){\cal R} -2f\/({\cal R}) = \Sigma\,.
\end{equation}
When the trace $\Sigma\/$ is allowed to assume only a constant
value,  the present theory amounts to an Einstein--like (if
$S^\alpha_{ts}=0$,{\it i.e.} $\omega$-independent matter) or an
Einstein-Cartan-like theory (if $S^\alpha_{ts}\not= 0$, {\it i.e.}
$\omega$-dependent matter) with cosmological constant. In fact, in
such a circumstance, Eq. \eqref{2.2.5} implies that the scalar
curvature ${\cal R}\/$ also is constant. As a consequence, Eqs.
\eqref{2.2.2a} and \eqref{2.2.4} can be expressed as
\begin{subequations}\label{2.2.6}
\begin{equation}\label{2.2.6a}
{\cal R}_{\;\mu}^{i} - \frac{1}{2}\/\left({\cal R} + \Lambda \right)\/e^i_\mu =
k\Sigma_{\;\mu}^{i}\,,
\end{equation}
\begin{equation}\label{2.2.6b}
T^\alpha_{ts}= \frac{k}{2}\/\left( 2S_{ts}^\alpha -
S_{t\sigma}^{\sigma}e^\alpha_s + S_{s\sigma}^{\sigma}e^\alpha_t
\right)\,,
\end{equation}
\end{subequations}
where $\Lambda= kf\/({\cal R}) - {\cal R}\/$ and ${\displaystyle
k=\frac{1}{f'\/({\cal R})}\/}$, ${\cal R}\/$ being the constant value determined
by Eq.\eqref{2.2.5}, provided that $f'({\cal R})\not= 0\/$. As in the previous section, this result
holds with the exception of the particular case
$\Sigma=0\/$ and $f({\cal R})=\alpha\/{\cal R}^2\/$. Indeed, under these
conditions, Eq. \eqref{2.2.5} is a trivial identity which imposes
no restriction on the scalar curvature ${\cal R}\/$.
From now on,  we shall  suppose that $\Sigma\/$ is not forced to
be a constant  when the matter field equations are satisfied.
Besides, we shall suppose that the relation \eqref{2.2.5} is
invertible so that the scalar curvature can be thought as a
suitable function of $\Sigma\/$, namely
\begin{equation}\label{2.2.7}
{\cal R}=F\/(\Sigma)
\end{equation}
With this assumption in mind, defining the  tensors
${\cal R}^i_{\;j}:={\cal R}_{\mu\sigma}^{\;\;\;\;\lambda\sigma}e^{i}_{\lambda}e^\mu_j$,
$\Sigma^i_{\;j}:=\Sigma^i_{\;\mu}e^\mu_j$,
$T_{ij}^{\;\;\;h}:=T^\alpha_{ij}e_\alpha^h\/$ and
$S^{\;\;\;h}_{ij}:=S^\alpha_{ij}e_\alpha^h\/$, we rewrite Eqs.
\eqref{2.2.2a} and \eqref{2.2.4} in the equivalent form, in strict analogy to the results of the previous section\footnote{The Latin indices have here the same meaning of Greek indices of previous section.}
\begin{subequations}\label{2.2.8}
\begin{equation}\label{2.2.8a}
{\cal R}_{ij} -\frac{1}{2}{\cal R}g_{ij}= \frac{1}{f'\/(F\/(\Sigma))}\left(
\Sigma_{ij} - \frac{1}{4}\Sigma g_{ij} \right) -
\frac{1}{4}F\/(\Sigma)g_{ij}\,,
\end{equation}
\begin{equation}\label{2.2.8b}
T_{ij}^{\;\;\;h} = -
\frac{1}{2f'\/(F\/(\Sigma))}\left(\frac{\partial{f'\/(F\/(\Sigma))}}{\partial{x^p}} +
S_{p\sigma}^\sigma\right)\left(\delta^p_i\delta^h_j -
\delta^p_j\delta^h_i\right) +
\frac{1}{f'\/(F\/(\Sigma))}S^{\;\;\;h}_{ij}\,,
\end{equation}
\end{subequations}
In Eqs. \eqref{2.2.8a} one has to distinguish the order of the
indexes since, in general, the tensors ${\cal R}_{ij}\/$ and
$\Sigma_{ij}\/$ are not symmetric.

Moreover, following \cite{CCSV1}, in the {\it lhs} of Eqs.
\eqref{2.2.8a}, we can distinguish the contribution due to the
Christoffel terms from that due to the torsion dependent terms. To
see this point, from Eqs. \eqref{00.2} and \eqref{00.4}, we first
get the following representation for the contracted curvature
tensor
\begin{equation}\label{2.2.9}
{\cal R}_{ij}=\tilde{R}_{ij} + \tilde{\nabla}_jK_{hi}^{\;\;\;h} -
\tilde{\nabla}_hK_{ji}^{\;\;\;h} +
K_{ji}^{\;\;\;p}K_{hp}^{\;\;\;h} -
K_{hi}^{\;\;\;p}K_{jp}^{\;\;\;h}\,,
\end{equation}
where $\tilde{R}_{ij}\/$ is the Ricci tensor of the Levi--Civita
connection $\tilde\Gamma\/$ associated with the metric
$g_{ij}=\eta_{\mu\nu}e^\mu_{i}e^\nu_{j}\/$, and $\tilde{\nabla}\/$
denotes the Levi-Civita covariant derivative. Then, recalling the
expression of the contortion tensor \cite{hehl}
\begin{equation}\label{2.2.10}
K_{ij}^{\;\;\;h} = \frac{1}{2}\/\left( - T_{ij}^{\;\;\;h} +
T_{j\;\;\;i}^{\;\;h} - T^h_{\;\;ij}\right)\,,
\end{equation}
and using the second set of field equations \eqref{2.2.8b}, we
obtain the following representations
\begin{subequations}\label{2.2.11}
\begin{equation}\label{2.2.11a}
K_{ij}^{\;\;\;h}= \hat{K}_{ij}^{\;\;\;h} + \hat{S}_{ij}^{\;\;\;h}\,,
\end{equation}
\begin{equation}\label{2.2.11b}
\hat{S}_{ij}^{\;\;\;h}:=\frac{1}{2f'}\/\left( - S_{ij}^{\;\;\;h} +
S_{j\;\;\;i}^{\;\;h} - S^h_{\;\;ij}\right)\,,
\end{equation}
\begin{equation}\label{2.2.11c}
\hat{K}_{ij}^{\;\;\;h} := -\hat{T}_j\delta^h_i +
\hat{T}_pg^{ph}g_{ij}\,,
\end{equation}
\begin{equation}\label{2.2.11d}
\hat{T}_j:=\frac{1}{2f'}\/\left( \frac{\partial{f'}}{\partial{x^j}} +
S^\sigma_{j\sigma} \right)\,.
\end{equation}
\end{subequations}
Inserting eqs. \eqref{2.2.11} in eq. \eqref{2.2.9} we end up with
the final expression for ${\cal R}_{ij}\/$
\begin{equation}\label{2.2.12}
\begin{split}
{\cal R}_{ij}=\tilde{R}_{ij} + \tilde{\nabla}_j\hat{K}_{hi}^{\;\;\;h} +
\tilde{\nabla}_j\hat{S}_{hi}^{\;\;\;h} -
\tilde{\nabla}_h\hat{K}_{ji}^{\;\;\;h} - \tilde{\nabla}_h\hat{S}_{ji}^{\;\;\;h} + \hat{K}_{ji}^{\;\;\;p}\hat{K}_{hp}^{\;\;\;h} + \hat{K}_{ji}^{\;\;\;p}\hat{S}_{hp}^{\;\;\;h} \\
+ \hat{S}_{ji}^{\;\;\;p}\hat{K}_{hp}^{\;\;\;h} +
\hat{S}_{ji}^{\;\;\;p}\hat{S}_{hp}^{\;\;\;h} -
\hat{K}_{hi}^{\;\;\;p}\hat{K}_{jp}^{\;\;\;h} -
\hat{K}_{hi}^{\;\;\;p}\hat{S}_{jp}^{\;\;\;h} -
\hat{S}_{hi}^{\;\;\;p}\hat{K}_{jp}^{\;\;\;h} -
\hat{S}_{hi}^{\;\;\;p}\hat{S}_{jp}^{\;\;\;h}\,.
\end{split}
\end{equation}
The last step is the substitution of Eqs. \eqref{2.2.12} into Eqs.
\eqref{2.2.8a}. Explicit examples of the  described procedure are
given in the next subsections. We will discuss specific cases of
fields  coupled with gravity acting as matter sources.
%
\subsubsection{The case of Dirac fields}
As a first example of matter, we consider  the case of Dirac
fields $\psi\/$. The matter Lagrangian density is given by
\begin{equation}\label{2.4.1}
{\cal L}_D = e\/\left[ \frac{i}{2}\/\left( \bar\psi\gamma^iD_i\psi
- D_i\bar\psi\gamma^i\psi \right) - m\bar\psi\psi \right]
\end{equation}
where ${\displaystyle D_i\psi = \frac{\partial\psi}{\partial{x^i}} +
\omega_i^{\;\;\mu\nu}S_{\mu\nu}\psi\/}$ and ${\displaystyle
D_i\bar\psi = \frac{\partial{\bar\psi}}{\partial{x^i}} -
\bar\psi\omega_i^{\;\;\mu\nu}S_{\mu\nu}\/}$ are the covariant
derivatives of the Dirac fields,\\
 ${\displaystyle
S_{\mu\nu}=\frac{1}{8}\left[\gamma_\mu,\gamma_\nu\right]\/}$,
$\gamma^i =\gamma^{\mu}e^i_\mu\/$; $\gamma^\mu\/$ denotes the
Dirac matrices.
The field equations for the Dirac fields are
\begin{equation}\label{2.4.1bis}
i\gamma^hD_h\psi - m\psi =0, \qquad iD_h\bar{\psi}\gamma^h +
m\bar\psi =0
\end{equation}
Since the Lagrangian \eqref{2.4.1} vanishes  for $\psi\/$ and
$\bar\psi\/$ satisfying the equation \eqref{2.4.1bis}, the
corresponding energy-momentum and spin density tensors are
expressed, respectively, as
\begin{equation}\label{2.4.2}
\Sigma_{ij} = \frac{i}{4}\/\left[ \bar\psi\gamma_iD_j\psi -
\left(D_j\bar\psi\right)\gamma_i\psi \right]
\end{equation}
and
\begin{equation}\label{2.4.3}
S_{ij}^{\;\;\;h}=
-\frac{i}{2}\bar\psi\left\{\gamma^h,S_{ij}\right\}\psi
\end{equation}
with ${\displaystyle
S_{ij}=\frac{1}{8}\left[\gamma_i,\gamma_j\right]\/}$. Now, using
the properties of the Dirac matrices, it is easily seen that
${\displaystyle
\left\{\gamma^h,S^{ij}\right\}=\frac{1}{2}\gamma^{[i}\gamma^j\gamma^{h]}\/}$.
This fact implies the total antisymmetry of the spin density
tensor $S_{ij}^{\;\;\;h}\/$. As a consequence, the contracted
curvature and  scalar curvature assume the simplified expressions
(compare with Eq. \eqref{2.2.12})
\begin{subequations}\label{2.4.4}
\begin{equation}\label{2.4.4a}
{\cal R}_{ij} = \tilde{R}_{ij} - 2\tilde{\nabla}_{j}\hat{T}_i -
\tilde{\nabla}_h\hat{T}^hg_{ij} + 2\hat{T}_i\hat{T}_j -
2\hat{T}_h\hat{T}^hg_{ij} - \tilde{\nabla}_h\hat{S}_{ji}^{\;\;\;h}
- \hat{S}_{hi}^{\;\;\;p}\hat{S}_{jp}^{\;\;\;h}
\end{equation}
and
\begin{equation}\label{2.4.4b}
{\cal R} = \tilde{R} - 6\tilde{\nabla}_{i}\hat{T}^i - 6\hat{T}_i\hat{T}^i
- \hat{S}_{hi}^{\;\;\;p}\hat{S}_{\;\;p}^{i\;\;\;h}
\end{equation}
\end{subequations}
where now ${\displaystyle \hat{T}_i
=\frac{1}{2f'}\de{f'}/de{x^i}\/}$ and ${\displaystyle
\hat{S}_{ij}^{\;\;\;h}:=-\frac{1}{2f'}S_{ij}^{\;\;\;h}\/}$.
Inserting eqs. \eqref{2.4.4} into eqs. \eqref{2.2.8a} and  using
of the above expression for $\hat{T}_i\/$, we obtain the final
Einstein-like equations
\begin{equation}\label{2.4.5}
\begin{split}
\tilde{R}_{ij} -\frac{1}{2}\tilde{R}g_{ij}= \frac{1}{\varphi}\Sigma_{ij} + \frac{1}{\varphi^2}\left( - \frac{3}{2}\de\varphi/de{x^i}\de\varphi/de{x^j} + \varphi\tilde{\nabla}_{j}\de\varphi/de{x^i} + \frac{3}{4}\de\varphi/de{x^h}\de\varphi/de{x^k}g^{hk}g_{ij} \right. \\
\left. - \varphi\tilde{\nabla}^h\de\varphi/de{x^h}g_{ij} -
V\/(\varphi)g_{ij} \right) +
\tilde{\nabla}_h\hat{S}_{ji}^{\;\;\;h} +
\hat{S}_{hi}^{\;\;\;p}\hat{S}_{jp}^{\;\;\;h} -
\frac{1}{2}\hat{S}_{hq}^{\;\;\;p}\hat{S}_{\;\;p}^{q\;\;\;h}g_{ij}
\end{split}
\end{equation}
where we have defined the scalar field
\begin{equation}\label{2.4.6}
\varphi := f'\/(F\/(\Sigma))
\end{equation}
and the effective potential
\begin{equation}\label{2.4.7}
V\/(\varphi):= \frac{1}{4}\left[ \varphi
F^{-1}\/((f')^{-1}\/(\varphi)) +
\varphi^2\/(f')^{-1}\/(\varphi)\right]
\end{equation}
To conclude, we notice that Eqs. \eqref{2.4.5} can be simplified
by performing a conformal transformation. Indeed, setting
$\bar{g}_{ij}:=\varphi\eta_{\mu\nu}e^\mu_ie^\nu_j\/$, Eqs.
\eqref{2.4.5} can be rewritten in the easier form
\begin{equation}\label{2.4.8}
\bar{{R}}_{ij} - \frac{1}{2}\bar{{R}}\bar{g}_{ij} =
\frac{1}{\varphi}\Sigma_{ij} -
\frac{1}{\varphi^3}V\/(\varphi)\bar{g}_{ij} +
\tilde{\nabla}_h\hat{S}_{ji}^{\;\;\;h} +
\hat{S}_{hi}^{\;\;\;p}\hat{S}_{jp}^{\;\;\;h} -
\frac{1}{2\varphi}\hat{S}_{hq}^{\;\;\;p}\hat{S}_{\;\;p}^{q\;\;\;h}g_{ij}
\end{equation}
where $\bar{R}_{ij}\/$  and $\bar{R}\/$ are respectively the Ricci
tensor and the Ricci scalar curvature associated with the
conformal metric $\bar{g}_{ij}\/$.

\subsubsection{The case of Yang--Mills fields}
As it has been shown in some previous works
\cite{CVB1,CVB2,VC,VCB2}, also gauge theories can be formulated
within the framework of ${\cal J}$-bundles. Therefore, we can
describe $f(R)$-gravity coupled with Yang--Mills fields  in the
new geometric setting.

To see this point,  let $Q\to {\cal M}$ be a principal fiber
bundle over space-time, with structural group a semi-simple Lie
group $G\/$. We consider the affine bundle $J_1\/Q/G\to {\cal
M}\/$ (the space of principal connections of $Q\to {\cal M}\/$)
and refer it to local coordinates $x^i,a^A_i\/$,
$A=1,\ldots,r=\dim G\/$.

In a combined theory of $f\/(R)$-gravity and Yang--Mills fields,
additional dynamical fields are principal connections of $Q\/$
represented by sections of the bundle $J_1\/Q/G\to {\cal M}\/$.
The extended configuration space of the theory is then the fiber
product ${\cal E}\times{\cal C}\times J_1\/Q/G\to {\cal M}\/$ over ${\cal M}$.

Following the approach illustrated above, we may construct
the quotient space ${\cal J}\/({\cal E}\times{\cal C}\times J_1\/Q/G)\/$  (see
\cite{VC} for details). As additional $\cal J$-coordinates the
latter admits the components $F^A_{ij}\/$ of the curvature tensors
of the principal connections of $Q\to {\cal M}\/$. Holonomic
sections of the bundle ${\cal J}\/({\cal E}\times{\cal C}\times J_1\/Q/G)\/$
are of the form \eqref{0.2} together with
\begin{equation}\label{3.1oo}
F^A_{ij}\/(x)=\de{a^A_j\/(x)}/de{x^i} - \de{a^A_i\/(x)}/de{x^j} +
a^B_j\/(x)a^C_i\/(x)C^A_{CB}
\end{equation}
where $C^A_{CB}\/$ are the structure coefficients of the Lie
algebra of $G\/$.

The Poincar\'e-Cartan $4$-form associated with  a Lagrangian on
${\cal J}\/({\cal E}\times{\cal C}\times J_1\/Q/G)\/$ of the form $L={\cal
L}\/(x^i,e^\mu_i,a^A_i,{\cal R}_{ij}^{\;\;\;\;\mu\nu},F^A_{ij})\,ds\/$ is
\begin{equation}\label{3.2uu}
\Theta = {\cal L}\, ds - \frac{1}{4}\de{\cal
L}/de{{\cal R}_{hk}^{\;\;\;\;\alpha\beta}}\,\theta^{\alpha\beta}\wedge
ds_{hk} - \frac{1}{2}\de{\cal L}/de{F_{hk}^{A}}\,\theta^A\wedge
ds_{hk}
\end{equation}
where $\theta^A = \Phi^A - F^A\/$,  being $F^A :=
\frac{1}{2}F^A_{ij}\,dx^i\wedge dx^j\/$ and $\Phi^A :=
da^A_i\wedge dx^i + \frac{1}{2}a^B_ia^C_jC^A_{CB}\,dx^j\wedge
dx^i\/$.

Variational field equations are still of the form
\begin{equation}\label{3.3vv}
{\cal J}\sigma^*\left({\cal J}\/(X)\interior d\Theta\right)=0
\end{equation}
for all $\cal J$-prolongable vector  fields X on ${\cal E}\times{\cal C}\times
J_1\/Q/G\/$. Due to the arbitrariness of the infinitesimal
deformations $X\/$, Eq. \eqref{3.3vv} splits into three sets of
final equations, respectively given by Eqs. \eqref{1.7} together
with
\begin{equation}\label{3.4dd}
{\cal J}\sigma^*\left(\de{\cal L}/de{a^A_i} - D_k\de{\cal
L}/de{F^A_{kj}}\right)=0
\end{equation}
In particular, if the Lagrangian density  is
\begin{equation}
{\cal L}=e\/\left(f\/(R)
-\frac{1}{4}F^A_{ij}F^B_{pq}\gamma_{AB}\eta^{\mu\nu}e^p_{\mu}e^i_{\nu}\eta^{\lambda\sigma}e^q_{\lambda}e^j_{\sigma}
\right)\/\end{equation} expressing $f\/(R)\/$-gravity coupled with
a free Yang-Mills field, Eqs. \eqref{1.7} and \eqref{3.4dd} assume
the explicit form
\begin{subequations}\label{3.52}
\begin{equation}\label{3.5a}
e^i_{\mu}f\/({\cal R}) -
2f'\/({\cal R}){\cal R}_{\mu\sigma}^{\;\;\;\;\lambda\sigma}e^{i}_\lambda
=\frac{1}{4}F^A_{jk}F_A^{jk}e^i_\mu - F^{Ai}_jF^j_{Ak}e^k_\mu
\end{equation}
\begin{equation}\label{3.5b}
\begin{split}
\de/de{x^k}\left[2ef'\/({\cal R})\left(e^k_{\mu}e^i_{\nu} - e^i_{\mu}e^k_{\nu}\right)\right] - \omega_{k\;\;\;\mu}^{\;\;\lambda}\left[2ef'\/({\cal R})\left(e^k_{\lambda}e^i_{\nu} - e^i_{\lambda}e^k_{\nu}\right)\right]
-
\omega_{k\;\;\;\nu}^{\;\;\lambda}\left[2ef'\/({\cal R})\left(e^k_{\mu}e^i_{\lambda}
- e^i_{\mu}e^k_{\lambda}\right)\right] =0
\end{split}
\end{equation}
\begin{equation}\label{3.5c}
D_k\/(eF^{ik}_A)=0
\end{equation}
\end{subequations}
where ${\displaystyle D_k\/(eF^{ik}_A)=\de{(eF^{ik}_A)}/de{x^k}
-a^B_k\/(eF^{ik}_C)C^C_{BA}\/}$.

Since  the trace of the energy-impulse tensor ${\displaystyle
T^i_\mu := \frac{1}{4}F^A_{jk}F_A^{jk}e^i_\mu -
F^{Ai}_jF^j_{Ak}e^k_\mu\/}$ vanishes identically, we have again:

\begin{itemize}
\item if $f\/({\cal R}) =k{\cal R}\/$ we recover the Einstein-Yang-Mills
theory;

\item if $f\/({\cal R}) \not = k{\cal R}^2\/$ the torsion is necessarily
zero and we recover a $f\/({\cal R})$-theory without torsion coupled with
a Yang-Mills field;

\item We can have non--vanishing torsion only in the case
$f\/({\cal R})=k{\cal R}^2\/$.
\end{itemize}
\subsubsection{The case of spin fluid matter}
As a last example of matter source, we consider the case of a
semiclassical spin fluid. This is characterized by an
energy--momentum tensor of the form
\begin{subequations}\label{2.5.1}
\begin{equation}\label{2.5.1a}
\Sigma^{ij}= (\rho + p)u^iu^j + pg^{ij}
\end{equation}
and a spin density tensor given by
\begin{equation}\label{2.5.1b}
S_{ij}^{\;\;\;h}=S_{ij}u^h
\end{equation}
\end{subequations}
where $u^i\/$  and $S_{ij}\/$ denote, respectively, the
$4$-velocity and the spin density of the fluid  (see, for example,
\cite{Hehl-Heyde-Kerlick} and references therein). However, the
constraint $u^iu_i=-1\/$ must hold. Other models of spin fluids
are possible, where, due to the treatment of spin as a
thermodynamical variable, different expressions for the
energy--momentum tensor may be taken into account
\cite{Ray-Smalley,Gasperini,deRitis}.

The $4$-velocity and the spin density satisfy by the so called
{\it convective condition}
\begin{equation}\label{2.5.2}
S_{ij}u^j =0
\end{equation}
It is easily seen that  the relations \eqref{2.5.2} imply the
identities
\begin{equation}\label{2.5.2bis}
\hat{S}_i^{\;\;ih}=-\hat{S}_i^{\;\;hi}
\end{equation}
obtained inserting Eq. \eqref{2.5.1b} in Eq. \eqref{2.2.11b} and
using \eqref{2.5.2}. Making use of \eqref{2.5.2bis} as well as of
Eqs. \eqref{2.2.11b}, \eqref{2.2.12}, \eqref{2.5.1b} and
\eqref{2.5.2}, we can express the  Ricci curvature tensor and
scalar respectively as
\begin{subequations}\label{2.5.3}
\begin{equation}\label{2.5.3a}
\begin{split}
{\cal R}_{ij} = \tilde{R}_{ij} - 2\tilde{\nabla}_{j}\hat{T}_i - \tilde{\nabla}_h\hat{T}^hg_{ij} + 2\hat{T}_i\hat{T}_j - 2\hat{T}_h\hat{T}^hg_{ij} - \frac{1}{f'}\hat{T}_hS^h_{\;\;j}u_i \\
- \frac{1}{2f'}\tilde{\nabla}_h\/\left(S_{ji}u^h + S_i^{\;\;h}u_j
- S^h_{\;\;j}u_i\right) + \frac{1}{4(f')^2}S^{pq}S_{pq}u_iu_j
\end{split}
\end{equation}
and
\begin{equation}\label{2.5.3b}
{\cal R} = \tilde{R} - 6\tilde{\nabla}_{i}\hat{T}^i - 6\hat{T}_i\hat{T}^i
- \frac{1}{4(f')^2}S^{pq}S_{pq}
\end{equation}
\end{subequations}
where ${\displaystyle \hat{T}_i =\frac{1}{2f'}\de{f'}/de{x^i}\/}$.
In view of this, substituting Eqs. \eqref{2.5.3} in Eqs.
\eqref{2.2.8a},  and using the definitions \eqref{2.4.6} and
\eqref{2.4.7}, we obtain Einstein-like equations of the form
\begin{equation}\label{2.5.4}
\begin{split}
\tilde{R}_{ij} -\frac{1}{2}\tilde{R}g_{ij}= \frac{1}{\varphi}\Sigma_{ij} + \frac{1}{\varphi^2}\left( - \frac{3}{2}\de\varphi/de{x^i}\de\varphi/de{x^j} + \varphi\tilde{\nabla}_{j}\de\varphi/de{x^i} + \frac{3}{4}\de\varphi/de{x^h}\de\varphi/de{x^k}g^{hk}g_{ij} \right. \\
\left. - \varphi\tilde{\nabla}^h\de\varphi/de{x^h}g_{ij} - V\/(\varphi)g_{ij} \right) + \frac{1}{\varphi}\hat{T}_hS^h_{\;\;j}u_i + \frac{1}{2\varphi}\tilde{\nabla}_h\/\left(S_{ji}u^h + S_i^{\;\;h}u_j - S^h_{\;\;j}u_i\right) \\
- \frac{1}{4\varphi^2}S^{pq}S_{pq}u_iu_j -
\frac{1}{8\varphi^2}S^{pq}S_{pq}g_{ij}
\end{split}
\end{equation}
Eqs. \eqref{2.5.4} are the "microscopic" field equations for
$f(R)$-gravity with torsion, coupled  with a semiclassical spin
fluid. In this form, all the source contributions are put  in
evidence and their role is clearly defined into dynamics.

\subsection{Equivalence with scalar-tensor theories}

The results of the last subsection lead to take into account the
analogies between $f(R)$-gravity with torsion and scalar-tensor
theories with torsion, as discussed, for example, in
\cite{German,Sung}. To this end, let us consider a Lagrangian
density of the form
\begin{equation}\label{2.6.1}
{\cal L}= \varphi\/e{\cal R} -eU\/(\varphi) + {\cal L}_m
\end{equation}
where  $\varphi\/$ is a scalar field, $U\/(\varphi)$ is a suitable
potential and ${\cal L}_m\/$ is a  matter Lagrangian density.

The Euler-Lagrange equations \eqref{1.7} applied to the
Lagrangian density  \eqref{2.6.1}, yield the corresponding field
equations
\begin{subequations}\label{2.6.2}
\begin{equation}\label{2.6.2a}
{\cal R}_{\mu\sigma}^{\;\;\;\;\lambda\sigma}e^{i}_\lambda
-\frac{1}{2}{\cal R}e^i_{\mu}=\frac{1}{\varphi}\Sigma^i_\mu -
\frac{1}{2\varphi}U\/(\varphi)e^i_{\mu}
\end{equation}
\begin{equation}\label{2.6.2b}
\varphi\left( T^\alpha_{ts} - T^\sigma_{t\sigma}e^\alpha_s +
T^\sigma_{s\sigma}e^\alpha_t \right)
=\de{\varphi}/de{x^t}e^\alpha_s - \de{\varphi}/de{x^s}e^\alpha_t +
S^\alpha_{ts}
\end{equation}
while the Euler-Lagrange equation for the scalar field is given
by
\begin{equation}\label{2.6.2c}
{\cal R}=U'\/(\varphi)
\end{equation}
\end{subequations}
Inserting Eq. \eqref{2.6.2c} in the trace of  Eqs. \eqref{2.6.2a},
we obtain an algebraic relation between the matter trace
$\Sigma\/$ and the scalar field $\varphi\/$ expressed as
\begin{equation}\label{2.6.3}
\Sigma -2U\/(\varphi) + \varphi\/U'\/(\varphi)=0
\end{equation}
Now, under the conditions ${\displaystyle
U\/(\varphi)=\frac{2}{\varphi}V\/(\varphi)\/}$, where
$V\/(\varphi)\/$ is defined as in Eq. \eqref{2.4.7} and $f''\not
=0\/$. It is easily seen that the relation \eqref{2.6.3}
represents exactly the inverse of \eqref{2.4.6}. In fact, from the
definition of the effective potential \eqref{2.4.7} and the
expression $F^{-1}\/(X)=f'\/(X)X - 2f\/(X)\/$ we have
\begin{equation}\label{2.6.4}
U\/(\varphi) =\frac{2}{\varphi}V\/(\varphi)= \frac{1}{2}\left[
F^{-1}\/((f')^{-1}\/(\varphi)) +
\varphi(f')^{-1}\/(\varphi)\right]=
\left[\varphi(f')^{-1}\/(\varphi)
-f\/((f')^{-1}\/(\varphi))\right]
\end{equation}
so that
\begin{equation}\label{2.6.5}
U'\/(\varphi) = (f')^{-1}\/(\varphi) +
\frac{\varphi}{f''\/((f')^{-1}\/(\varphi))} -
\frac{\varphi}{f''\/((f')^{-1}\/(\varphi))} = (f')^{-1}\/(\varphi)
\end{equation}
and then
\begin{equation}\label{2.6.6}
\Sigma = - \varphi\/U'\/(\varphi) +2U\/(\varphi) =
f'\/((f')^{-1}\/(\varphi))\/(f')^{-1}\/(\varphi)
-2f\/((f')^{-1}\/(\varphi)) = F^{-1}\/((f')^{-1}\/(\varphi))
\end{equation}
In view of the latter relation,  Eqs. \eqref{2.6.2} result to be
equivalent to Eqs. \eqref{2.2.7} and \eqref{2.2.8}. This fact
proves the  equivalence between $f(R)\/$-gravity and scalar-tensor
theories with torsion  obtained also in the ${\cal J}$-bundles
framework.
\subsection{Discussion}
In this Section, we have discussed the $f(R)$-theories of gravity
with torsion in the ${\cal J}$-bundles framework.

This formalism gives rise to a new geometric picture which allows
to put in evidence several features of the theories, in particular
their symmetries and conservation laws. In particular, due to the
fact that the components of the torsion and curvature tensors can
be chosen as fiber $\cal J$-coordinates on ${\cal
J}\/({\cal E}\times{\cal C})\/$, the field equations can be easily obtained in
suitable forms where the role of the geometry and the sources is
clearly defined.

Furthermore, such a representation allows to classify the
couplings with respect to the various matter fields whose global
effect is that to enlarge and characterize the $\cal J$-bundle.

We have given specific examples of couplings taking into account
Dirac fields, Yang-Mills fields and spin fluids. In any case,
the $\cal J$-vector fields allow to write the $f(R)$-field
equations in such a way that curvature, torsion and matter
components have a clear and distinct role into dynamics. This distinction emerges since we have adopted here a tetrad-affine formallism.

\section{The Hamiltonian formulation}
%
Here we want to face the problem to study $f(R)$-gravity
at a fundamental level. In particular, in the framework of the
Arnowitt-Deser-Misner ($\mathcal{ADM}$) formalism \cite{ADM62}, we
want to investigate the possibility to find out cosmological terms
as eigenvalues of generalized $f(R)$-Hamiltonians in a
Sturm-Liouville-like problem\footnote{See Ref.\cite{Remo1}, for
the application of the Sturm-Liouville problem in the case
of $f\left(  R\right)  =R$, even in presence of a massive
graviton.}. This issue is particularly relevant from several
viewpoints. First of all, our aim is to show that vacuum energy of
gravitational field is not a particular feature of GR where the
cosmological constant has to be added \textit{by hand} into
dynamics. At a classical level, it is well known that $f(R)$-gravity, for the Ricci scalar $R$ equal to a constant, exhibit
several de Sitter solutions \cite{BarrowOttewill83} but a definite
discussion, at a fundamental level, considering cosmological terms
as eigenvalues of such theories is lacking. Besides, the
computation of the Casimir energy, the seeking for zero point
energy in different backgrounds\footnote{For different $f(R)$, we
expect different zero point energies and, obviously, different
vacuum states.} give a track to achieve one-loop energy
regularization and renormalization for this kind of theories
\cite{zerbini,zerbini2}. On the other hand, these issues can be
considered in a multigravity approach to space-time foam if the
$\mathcal{N}$ space-times constituting the foam are supposed to
evolve, in general, with different curvature laws and ground
states (cosmological constants) \cite{Remo}.

\subsection{The Hamiltonian constraint of General Relativity}

Let us briefly report how to compute the Hamiltonian constraint for GR
considering the standard Hilbert-Einstein theory $f\left(  R\right)  =R$ and
the Arnowitt-Deser-Misner (ADM) $3+1$ decomposition \cite{ADM62}. In
terms of these variables, the line element is\footnote{Here we adopt again the standard notations and the signature used throughout this Report.}
\[
ds^{2}=g_{\mu\nu}\left(  x\right)  dx^{\mu}dx^{\nu}=\left(  -N^{2}+N_{i}%
N^{i}\right)  dt^{2}+2N_{j}dtdx^{j}+g_{ij}dx^{i}dx^{j}.
\]
$N$ is the \textit{lapse function}, while $N_{i}$ the \textit{shift function}.
In terms of these variables, the gravitational Lagrangian, with the boundary
terms neglected, can be written as
\begin{equation}
\mathcal{L}\left[  N,N_{i},g_{ij}\right]  =\sqrt{-g}R=\frac{N{}\,\sqrt{^{3}g}%
}{2\kappa}\text{ }\left[  K_{ij}K^{ij}-K^{2}+\,\left(  ^{3}R-2\Lambda
_{c}\right)  \right]  ,
\end{equation}
where $K_{ij}$ is the second fundamental form, $K=$ $g^{ij}K_{ij}$ is the
trace, $^{3}R$ is the three dimensional scalar curvature and $\sqrt{^{3}g}$ is
the three dimensional determinant of the metric. The conjugate momentum is
simply%
\begin{equation}
\pi^{ij}=\frac{\delta\mathcal{L}}{\delta\left(  \partial_{t}g_{ij}\right)
}=\left(  ^{3}g^{ij}K-K^{ij}\text{ }\right)  \frac{\sqrt{^{3}g}}{2\kappa
}.\label{mom}%
\end{equation}
By a Legendre transformation, we calculate the Hamiltonian
\begin{equation}
H=%
{\displaystyle\int}
d^{3}x\left[  N\mathcal{H+}N_{i}\mathcal{H}^{i}\right]  ,
\end{equation}
where%
\begin{equation}
\mathcal{H}=\left(  2\kappa\right)  G_{ijkl}\pi^{ij}\pi^{kl}-\frac{\sqrt
{^{3}g}}{2\kappa}\left(  ^{3}R-2\Lambda_{c}\right)  \label{cla1}%
\end{equation}
and%
\begin{equation}
\mathcal{H}^{i}=-2\nabla_{j}\pi^{ji}.\label{cla2}%
\end{equation}
where $\Lambda_{c}$ is the bare cosmological constant. The equations of motion
lead to two classical constraints%
\begin{equation}
\left\{
\begin{array}
[c]{c}%
\mathcal{H}=0\\
\mathcal{H}^{i}=0
\end{array}
\right.  ,
\end{equation}
representing invariance under \textit{time} re-parameterization and invariance
under diffeomorphism, respectively. $G_{ijkl}$ is the \textit{supermetric}
defined as%
\begin{equation}
G_{ijkl}=\frac{1}{2\sqrt{g}}(g_{ik}g_{jl}+g_{il}g_{jk}-g_{ij}g_{kl}).
\end{equation}
When $\mathcal{H}$ and $\mathcal{H}^{i}$ are considered as operators acting on
some wave function, we have%
\begin{equation}
\mathcal{H}\Psi\left[  g_{ij}\right]  =0\label{WDW}%
\end{equation}
and%
\begin{equation}
\mathcal{H}_{i}\Psi\left[  g_{ij}\right]  =0.\label{diff}%
\end{equation}
Eq.$\left(  \ref{WDW}\right)  $ is the Wheeler-De Witt equation (WDW)\cite{De
Witt}. Eqs.$\left(  \ref{WDW}\right)  $ and $\left(  \ref{diff}\right)  $
describe the \textit{wave function of the Universe }$\Psi\left[
g_{ij}\right]  $. The WDW equation represents invariance under \textit{time}
re-parameterization in an operatorial form. This standard lore can be applied
to a generic $f(R)$ theory of gravity with the aim to achieve a cosmological
term as an eigenvalue of the WDW equation.

\subsection{The Hamiltonian constraint for $f(R)$-gravity}

Let us consider now the Lagrangian density describing a generic $f(R)$ theory
of gravity, namely
\begin{equation}
\mathcal{L}=\sqrt{-g}\left(  f\left(  R\right)  -2\Lambda_{c}\right)
,\qquad\text{with}\;f^{\prime\prime}\neq0,\label{lag}%
\end{equation}
where $f\left(  R\right)  $ is an arbitrary smooth function of the scalar
curvature and primes denote differentiation with respect to the scalar
curvature. A cosmological term is added also in this case for the sake of
generality. Obviously $f^{\prime\prime}=0$ corresponds to GR. The generalized
Hamiltonian density for the $f\left(  R\right)  $ theory assumes the
form\footnote{See also Ref.\cite{Querella} for technical details.}%
\begin{equation}
\mathcal{H}=\frac{1}{2\kappa}\left[  \frac{\mathcal{P}}{6}\left(  {}^{\left(
3\right)  }R-2\Lambda_{c}-3K_{ij}K^{ij}+K^{2}\right)  +V(\mathcal{P})-\frac
{1}{3}g^{ij}\mathcal{P}_{\mid ij}-2p^{ij}K_{ij}\right]  ,
\end{equation}
where%
\begin{equation}
V(\mathcal{P})=\sqrt{g}\left[  Rf^{\prime}\left(  R\right)  -f\left(
R\right)  \right]  .\label{V(P)}%
\end{equation}
Henceforth, the superscript $3$ indicating the spatial part of the metric will
be omitted on the metric itself. When $f\left(  R\right)  =R$, $V(\mathcal{P}%
)=0$ as it should be. Since%
\begin{equation}
\mathcal{P}^{ij}=-2\sqrt{g}g^{ij}f^{\prime}\left(  R\right)  \qquad
\Longrightarrow\qquad\mathcal{P=}-6\sqrt{g}f^{\prime}\left(  R\right)  ,
\end{equation}
we have%
\begin{equation}
\mathcal{H}=\frac{1}{2\kappa}\left[  -\sqrt{g}f^{\prime}\left(  R\right)
\left(  {}^{\left(  3\right)  }R-2\Lambda_{c}-3K_{ij}K^{ij}+K^{2}\right)
+V(\mathcal{P})+2g^{ij}\left(  \sqrt{g}f^{\prime}\left(  R\right)  \right)
_{\mid ij}-2p^{ij}K_{ij}\right]  .\label{Hamf(R)}%
\end{equation}
With the help of Eq.$\left(  \ref{mom}\right)  $, Eq.$\left(  \ref{Hamf(R)}%
\right)  $ becomes%
\begin{equation}
\mathcal{H}=f^{\prime}\left(  R\right)  \left[  \left(  2\kappa\right)
G_{ijkl}\pi^{ij}\pi^{kl}{}-\frac{\sqrt{g}}{2\kappa}{}\left(  ^{\left(
3\right)  }R-2\Lambda_{c}\right)  \right]  +\frac{1}{2\kappa}\left[  \sqrt
{g}f^{\prime}\left(  R\right)  \left(  2K_{ij}K^{ij}\right)  +V(\mathcal{P}%
)+2g^{ij}\left(  \sqrt{g}f^{\prime}\left(  R\right)  \right)  _{\mid
ij}-2p^{ij}K_{ij}\right]  .
\end{equation}
However%
\begin{equation}
p^{ij}=\sqrt{g}K^{ij},
\end{equation}
then we obtain%
\begin{equation}
\mathcal{H}=f^{\prime}\left(  R\right)  \left[  \left(  2\kappa\right)
G_{ijkl}\pi^{ij}\pi^{kl}{}-\frac{\sqrt{g}}{2\kappa}{}\left(  ^{\left(
3\right)  }R-2\Lambda_{c}\right)  \right]  +\frac{1}{2\kappa}\left[  2\sqrt
{g}K_{ij}K^{ij}\left(  f^{\prime}\left(  R\right)  -1\right)  +V(\mathcal{P}%
)+2g^{ij}\left(  \sqrt{g}f^{\prime}\left(  R\right)  \right)  _{\mid
ij}\right]
\end{equation}
and transforming into canonical momenta, one gets
\begin{eqnarray}
\mathcal{H}=f^{\prime}\left(  R\right)  \left[  \left(  2\kappa\right)
G_{ijkl}\pi^{ij}\pi^{kl}{}-\frac{\sqrt{g}}{2\kappa}{}\left(  ^{\left(
3\right)  }R-2\Lambda_{c}\right)  \right]  &+& 2\left(  2\kappa\right)  \left[
G_{ijkl}\pi^{ij}\pi^{kl}+\frac{\pi}{4}^{2}\right]  \left(  f^{\prime}\left(
R\right)  -1\right)+  \nonumber\\ && 
+\frac{1}{2\kappa}\left[  V(\mathcal{P})+2g^{ij}\left(
\sqrt{g}f^{\prime}\left(  R\right)  \right)  _{\mid ij}\right]
.\label{Hamf(R)_1}%
\end{eqnarray}
By imposing the Hamiltonian constraint, we obtain
\begin{equation}
f^{\prime}\left(  R\right)  \left[  \left(  2\kappa\right)  G_{ijkl}\pi
^{ij}\pi^{kl}{}-\frac{\sqrt{g}}{2\kappa}{}^{\left(  3\right)  }R\right]  +{}%
\end{equation}%
\begin{equation}
+2\left(  2\kappa\right)  \left[  G_{ijkl}\pi^{ij}\pi^{kl}+\frac{\pi}{4}%
^{2}\right]  \left(  f^{\prime}\left(  R\right)  -1\right)  +\frac{1}{2\kappa
}\left[  V(\mathcal{P})+2g^{ij}\left(  \sqrt{g}f^{\prime}\left(  R\right)
\right)  _{\mid ij}\right]  =-f^{\prime}\left(  R\right)  \sqrt{g}%
\frac{\Lambda_{c}}{\kappa}%
\end{equation}
If we assume that $f^{\prime}\left(  R\right)  \neq0$ the previous expression
becomes%
\begin{eqnarray}
\left[  \left(  2\kappa\right)  G_{ijkl}\pi^{ij}\pi^{kl}{}-\frac{\sqrt{g}%
}{2\kappa}{}^{\left(  3\right)  }R\right]  +\left(  2\kappa\right)  \left[
G_{ijkl}\pi^{ij}\pi^{kl}+\frac{\pi}{4}^{2}\right]  \frac{2\left(  f^{\prime
}\left(  R\right)  -1\right)  }{f^{\prime}\left(  R\right)  }&+&\frac{1}{2\kappa
f^{\prime}\left(  R\right)  }\left[  V(\mathcal{P}) + 2g^{ij}\left(  \sqrt
{g}f^{\prime}\left(  R\right)  \right)  _{\mid ij}\right] = \nonumber\\ && =-\sqrt{g}%
\frac{\Lambda_{c}}{\kappa}{}.
\end{eqnarray}
Now, we integrate over the hypersurface $\Sigma$ to obtain%
\begin{eqnarray}
\int_{\Sigma}d^{3}x\left\{  \left[  \left(  2\kappa\right)  G_{ijkl}\pi
^{ij}\pi^{kl}{}-\frac{\sqrt{g}}{2\kappa}{}^{\left(  3\right)  }R\right]
+\left(  2\kappa\right)  \left[  G_{ijkl}\pi^{ij}\pi^{kl}+\frac{\pi}{4}%
^{2}\right]  \frac{2\left(  f^{\prime}\left(  R\right)  -1\right)  }%
{f^{\prime}\left(  R\right)  }\right\}
\end{eqnarray}%
\begin{equation}
+\int_{\Sigma}d^{3}x\frac{1}{2\kappa f^{\prime}\left(  R\right)  }\left[
V(\mathcal{P})+2g^{ij}\left(  \sqrt{g}f^{\prime}\left(  R\right)  \right)
_{\mid ij}\right]  =-\frac{\Lambda_{c}}{\kappa}\int_{\Sigma}d^{3}x\sqrt{g}.
\end{equation}
The term%
\begin{equation}
\frac{1}{\kappa}\int_{\Sigma}d^{3}x\frac{1}{f^{\prime}\left(  R\right)
}g^{ij}\left(  \sqrt{g}f^{\prime}\left(  R\right)  \right)  _{\mid ij}%
\end{equation}
appears to be a three-divergence and therefore will not contribute to the
computation. The remaining equation simplifies into%
\begin{equation}
\int_{\Sigma}d^{3}x\left\{  \left[  \left(  2\kappa\right)  G_{ijkl}\pi
^{ij}\pi^{kl}{}-\frac{\sqrt{g}}{2\kappa}{}^{\left(  3\right)  }R\right]
+\left(  2\kappa\right)  \left[  G_{ijkl}\pi^{ij}\pi^{kl}+\frac{\pi}{4}%
^{2}\right]  \frac{2\left(  f^{\prime}\left(  R\right)  -1\right)  }%
{f^{\prime}\left(  R\right)  }+\frac{V(\mathcal{P})}{2\kappa f^{\prime}\left(
R\right)  }\right\}  =-\frac{\Lambda_{c}}{\kappa}\int_{\Sigma}d^{3}x\sqrt
{g}.\label{GWDW}%
\end{equation}
By a canonical procedure of quantization, we want to obtain the vacuum state
of a generic $f(R)$ theory.

\subsection{The cosmological constant as an eigenvalue}

The standard WDW equation $\left(  \ref{WDW}\right)  $ can be cast into the
form of an eigenvalue equation
\begin{equation}
\hat{\Lambda}_{\Sigma}\Psi\left[  g_{ij}\right]  =\Lambda\left(  \vec
{x}\right)  \Psi\left[  g_{ij}\right]  , \label{WDW1}%
\end{equation}
where
\begin{equation}
\hat{\Lambda}_{\Sigma}=\left(  2\kappa\right)  G_{ijkl}\pi^{ij}\pi^{kl}%
-\frac{\sqrt{g}}{2\kappa}\!^{3}R.
\end{equation}
If we multiply Eq.$\left(  \ref{WDW1}\right)  $ by $\Psi^{\ast}\left[
g_{ij}\right]  $ and we functionally integrate over the three spatial metric
$g_{ij}$, we get%
\begin{equation}
\int\mathcal{D}\left[  g_{ij}\right]  \Psi^{\ast}\left[  g_{ij}\right]
\hat{\Lambda}_{\Sigma}\Psi\left[  g_{ij}\right]  =\int\mathcal{D}\left[
g_{ij}\right]  \Lambda\left(  \vec{x}\right)  \Psi^{\ast}\left[
g_{ij}\right]  \Psi\left[  g_{ij}\right]
\end{equation}
and after integrating over the hypersurface $\Sigma$, one can formally
re-write the modified WDW equation as%
\begin{equation}
\frac{1}{V}\frac{\int\mathcal{D}\left[  g_{ij}\right]  \Psi^{\ast}\left[
g_{ij}\right]  \int_{\Sigma}d^{3}x\hat{\Lambda}_{\Sigma}\Psi\left[
g_{ij}\right]  }{\int\mathcal{D}\left[  g_{ij}\right]  \Psi^{\ast}\left[
g_{ij}\right]  \Psi\left[  g_{ij}\right]  }=\frac{1}{V}\frac{\left\langle
\Psi\left\vert \int_{\Sigma}d^{3}x\hat{\Lambda}_{\Sigma}\right\vert
\Psi\right\rangle }{\left\langle \Psi|\Psi\right\rangle }=-\frac{\Lambda_{c}%
}{\kappa}, \label{WDW2}%
\end{equation}
where the explicit expression of $\Lambda\left(  \vec{x}\right)  $ has been
used and we have defined the volume of the hypersurface $\Sigma$ as%
\begin{equation}
V=\int_{\Sigma}d^{3}x\sqrt{g}.
\end{equation}
The formal eigenvalue equation $\left(  \ref{WDW2}\right)  $ is a simple
manipulation of Eq.$\left(  \ref{WDW}\right)  $. We can gain more information
considering a separation of the spatial part of the metric into a background
term, $\bar{g}_{ij}$, and a quantum fluctuation, $h_{ij}$,%
\begin{equation}
g_{ij}=\bar{g}_{ij}+h_{ij}.
\end{equation}
Thus Eq.$\left(  \ref{WDW2}\right)  $ becomes%
\begin{equation}
\frac{\left\langle \Psi\left\vert \int_{\Sigma}d^{3}x\left[  \hat{\Lambda
}_{\Sigma}^{\left(  0\right)  }+\hat{\Lambda}_{\Sigma}^{\left(  1\right)
}+\hat{\Lambda}_{\Sigma}^{\left(  2\right)  }+\ldots\right]  \right\vert
\Psi\right\rangle }{\left\langle \Psi|\Psi\right\rangle }=-\frac{\Lambda_{c}%
}{\kappa}\Psi\left[  g_{ij}\right]  , \label{WDW3}%
\end{equation}
where $\hat{\Lambda}_{\Sigma}^{\left(  i\right)  }$ represents the $i^{th}$
order of perturbation in $h_{ij}$. By observing that the kinetic part of
$\hat{\Lambda}_{\Sigma}$ is quadratic in the momenta, we only need to expand
the three-scalar curvature $\int d^{3}x\sqrt{g}R^{\left(  3\right)  }$ up to
the quadratic order and we get%
\[
\int_{\Sigma}d^{3}x\sqrt{\bar{g}}\left[  -\frac{1}{4}h\triangle h+\frac{1}%
{4}h^{li}\triangle h_{li}-\frac{1}{2}h^{ij}\nabla_{l}\nabla_{i}h_{j}%
^{l}+\right.
\]%
\begin{equation}
\left.  +\frac{1}{2}h\nabla_{l}\nabla_{i}h^{li}-\frac{1}{2}h^{ij}R_{ia}%
h_{j}^{a}+\frac{1}{2}hR_{ij}h^{ij}+\frac{1}{4}h\left(  R^{\left(  0\right)
}\right)  h\right]  \label{rexp}%
\end{equation}
where $h$ is the trace of $h_{ij}$ and $R^{\left(  0\right)  }$ is the three
dimensional scalar curvature. By repeating the same procedure for the
generalized WDW equation Eq.$\left(  \ref{GWDW}\right)  $, we obtain%
\begin{equation}
\frac{1}{V}\frac{\left\langle \Psi\left\vert \int_{\Sigma}d^{3}x\left[
\hat{\Lambda}_{\Sigma}^{\left(  2\right)  }\right]  \right\vert \Psi
\right\rangle }{\left\langle \Psi|\Psi\right\rangle }+\frac{2\kappa}{V}%
\frac{2\left(  f^{\prime}\left(  R\right)  -1\right)  }{f^{\prime}\left(
R\right)  }\frac{\left\langle \Psi\left\vert \int_{\Sigma}d^{3}x\left[
G_{ijkl}\pi^{ij}\pi^{kl}+\frac{\pi}{4}^{2}\right]  \right\vert \Psi
\right\rangle }{\left\langle \Psi|\Psi\right\rangle }+\frac{1}{V}%
\frac{\left\langle \Psi\left\vert \int_{\Sigma}d^{3}x\frac{V(\mathcal{P}%
)}{2\kappa f^{\prime}\left(  R\right)  }\right\vert \Psi\right\rangle
}{\left\langle \Psi|\Psi\right\rangle }=-\frac{\Lambda_{c}}{\kappa}.
\label{GWDW1}%
\end{equation}
From Eq.$\left(  \ref{GWDW1}\right)  $, we can define a \textquotedblleft%
\textit{modified}\textquotedblright\ $\hat{\Lambda}_{\Sigma}^{\left(
2\right)  }$ operator which includes $f^{\prime}\left(  R\right)  $. Thus, we
obtain%
\begin{equation}
\frac{\left\langle \Psi\left\vert \int_{\Sigma}d^{3}x\left[  \hat{\Lambda
}_{\Sigma,f\left(  R\right)  }^{\left(  2\right)  }\right]  \right\vert
\Psi\right\rangle }{\left\langle \Psi|\Psi\right\rangle }+\frac{2\kappa}%
{V}\frac{2\left(  f^{\prime}\left(  R\right)  -1\right)  }{f^{\prime}\left(
R\right)  }\frac{\left\langle \Psi\left\vert \int_{\Sigma}d^{3}x\left[
\frac{\pi}{4}^{2}\right]  \right\vert \Psi\right\rangle }{\left\langle
\Psi|\Psi\right\rangle }+\frac{1}{V}\frac{\left\langle \Psi\left\vert
\int_{\Sigma}d^{3}x\frac{V(\mathcal{P})}{2\kappa f^{\prime}\left(  R\right)
}\right\vert \Psi\right\rangle }{\left\langle \Psi|\Psi\right\rangle }%
=-\frac{\Lambda_{c}}{\kappa}, \label{GWDW2}%
\end{equation}
where%
\begin{equation}
\hat{\Lambda}_{\Sigma,f\left(  R\right)  }^{\left(  2\right)  }=\left(
2\kappa\right)  h\left(  R\right)  G_{ijkl}\pi^{ij}\pi^{kl}-\frac{\sqrt{g}%
}{2\kappa}\text{ }^{3}R^{lin},
\end{equation}
with%
\begin{equation}
h\left(  R\right)  =1+\frac{2\left[  f^{\prime}\left(  R\right)  -1\right]
}{f^{\prime}\left(  R\right)  } \label{h(R)}%
\end{equation}
and where $^{3}R^{lin}$ is the linearized scalar curvature whose expression is
shown in square brackets of Eq.$\left(  \ref{rexp}\right)  $. Note that when
$f\left(  R\right)  =R$, consistently it is $h\left(  R\right)  =1$. From
Eq.$\left(  \ref{GWDW2}\right)  $, we redefine $\Lambda_{c}$%
\begin{equation}
\Lambda_{c}^{\prime}=\Lambda_{c}+\frac{1}{2V}\frac{\left\langle \Psi\left\vert
\int_{\Sigma}d^{3}x\frac{V(\mathcal{P})}{f^{\prime}\left(  R\right)
}\right\vert \Psi\right\rangle }{\left\langle \Psi|\Psi\right\rangle }%
=\Lambda_{c}+\frac{1}{2V}\int_{\Sigma}d^{3}x\sqrt{g}\frac{Rf^{\prime}\left(
R\right)  -f\left(  R\right)  }{f^{\prime}\left(  R\right)  },
\label{NewLambda}%
\end{equation}
where we have explicitly used the definition of $V(\mathcal{P})$. In order to
make explicit calculations, we need an orthogonal decomposition for both
$\pi_{ij\text{ }}$ and $h_{ij}$ to disentangle gauge modes from physical
deformations. We define the inner product%

\begin{equation}
\left\langle h,k\right\rangle :=\int_{\Sigma}\sqrt{g}G^{ijkl}h_{ij}\left(
x\right)  k_{kl}\left(  x\right)  d^{3}x,
\end{equation}
by means of the inverse WDW metric $G_{ijkl}$, to have a metric on the space
of deformations, {\it i.e.} a quadratic form on the tangent space at $h_{ij}$, with%

\begin{equation}%
\begin{array}
[c]{c}%
G^{ijkl}=\frac{1}{2}(g^{ik}g^{jl}+g^{il}g^{jk}-2g^{ij}g^{kl})\text{.}%
\end{array}
\end{equation}
The inverse metric is defined on cotangent space and it assumes the form%

\begin{equation}
\left\langle p,q\right\rangle :=\int_{\Sigma}\sqrt{g}G_{ijkl}p^{ij}\left(
x\right)  q^{kl}\left(  x\right)  d^{3}x,
\end{equation}
so that%

\begin{equation}
G^{ijnm}G_{nmkl}=\frac{1}{2}\left(  \delta_{k}^{i}\delta_{l}^{j}+\delta
_{l}^{i}\delta_{k}^{j}\right)  .
\end{equation}
Note that in this scheme the \textquotedblleft inverse
metric\textquotedblright\ is actually the WDW metric defined on phase space.
The desired decomposition on the tangent space of 3-metric
deformations\cite{BergerEbin,York,MazurMottola,Vassilevich} is:%

\begin{equation}
h_{ij}=\frac{1}{3}hg_{ij}+\left(  L\xi\right)  _{ij}+h_{ij}^{\bot}
\label{p21a}%
\end{equation}
where the operator $L$ maps $\xi_{i}$ into symmetric tracefree tensors%

\begin{equation}
\left(  L\xi\right)  _{ij}=\nabla_{i}\xi_{j}+\nabla_{j}\xi_{i}-\frac{2}%
{3}g_{ij}\left(  \nabla\cdot\xi\right)  .
\end{equation}
Thus the inner product between three-geometries becomes
\[
\left\langle h,h\right\rangle :=\int_{\Sigma}\sqrt{g}G^{ijkl}h_{ij}\left(
x\right)  h_{kl}\left(  x\right)  d^{3}x=
\]%
\begin{equation}
\int_{\Sigma}\sqrt{g}\left[  -\frac{2}{3}h^{2}+\left(  L\xi\right)
^{ij}\left(  L\xi\right)  _{ij}+h^{ij\bot}h_{ij}^{\bot}\right]  . \label{p21b}%
\end{equation}
With the orthogonal decomposition in hand we can define the trial wave
functional as%
\begin{equation}
\Psi\left[  h_{ij}\left(  \overrightarrow{x}\right)  \right]  =\mathcal{N}%
\Psi\left[  h_{ij}^{\bot}\left(  \overrightarrow{x}\right)  \right]
\Psi\left[  h_{ij}^{\Vert}\left(  \overrightarrow{x}\right)  \right]
\Psi\left[  h_{ij}^{trace}\left(  \overrightarrow{x}\right)  \right]  ,
\label{twf}%
\end{equation}
where
\begin{equation}%
\begin{array}
[c]{c}%
\Psi\left[  h_{ij}^{\bot}\left(  \overrightarrow{x}\right)  \right]
=\exp\left\{  -\frac{1}{4}\left\langle hK^{-1}h\right\rangle _{x,y}^{\bot
}\right\} \\
\\
\Psi\left[  h_{ij}^{\Vert}\left(  \overrightarrow{x}\right)  \right]
=\exp\left\{  -\frac{1}{4}\left\langle \left(  L\xi\right)  K^{-1}\left(
L\xi\right)  \right\rangle _{x,y}^{\Vert}\right\} \\
\\
\Psi\left[  h_{ij}^{trace}\left(  \overrightarrow{x}\right)  \right]
=\exp\left\{  -\frac{1}{4}\left\langle hK^{-1}h\right\rangle _{x,y}%
^{Trace}\right\}
\end{array}
.
\end{equation}
The symbol \textquotedblleft$\perp$\textquotedblright\ denotes the
transverse-traceless tensor (TT) (spin 2) of the perturbation, while the
symbol \textquotedblleft$\Vert$\textquotedblright\ denotes the longitudinal
part (spin 1) of the perturbation. Finally, the symbol \textquotedblleft%
$trace$\textquotedblright\ denotes the scalar part of the perturbation.
$\mathcal{N}$ is a normalization factor, $\left\langle \cdot,\cdot
\right\rangle _{x,y}$ denotes space integration and $K^{-1}$ is the inverse
\textquotedblleft\textit{propagator}\textquotedblright. We will fix our
attention to the TT tensor sector of the perturbation representing the
graviton. Therefore, representation $\left(  \ref{twf}\right)  $ reduces to%
\begin{equation}
\Psi\left[  h_{ij}\left(  \overrightarrow{x}\right)  \right]  =\mathcal{N}%
\exp\left\{  -\frac{1}{4}\left\langle hK^{-1}h\right\rangle _{x,y}^{\bot
}\right\}  . \label{tt}%
\end{equation}
Actually there is no reason to neglect longitudinal and trace perturbations.
However, following the analysis of Refs.\cite{MazurMottola,GPY,VolkovWipf} on
the perturbation decomposition, we can discover that the relevant components
can be restricted to the TT modes and to the trace modes. Moreover, for
certain backgrounds, TT tensors can be a source of instability as shown in
Refs.\cite{GPY,VolkovWipf,Instability}. Even the trace part can be regarded as
a source of instability. Indeed this is usually termed \textit{conformal
}instability. The appearance of an instability on the TT modes is known as
non-conformal instability. This means that does not exist a gauge choice that
can eliminate negative modes. To proceed with Eq.$\left(  \ref{WDW3}\right)
$, we need to know the action of some basic operators on $\Psi\left[
h_{ij}\right]  $. The action of the operator $h_{ij}$ on $|\Psi\rangle
=\Psi\left[  h_{ij}\right]  $ is realized by \cite{Variational}
\begin{equation}
h_{ij}\left(  x\right)  |\Psi\rangle=h_{ij}\left(  \overrightarrow{x}\right)
\Psi\left[  h_{ij}\right]  .
\end{equation}
The action of the operator $\pi_{ij}$ on $|\Psi\rangle$, in general, is%

\begin{equation}
\pi_{ij}\left(  x\right)  |\Psi\rangle=-i\frac{\delta}{\delta h_{ij}\left(
\overrightarrow{x}\right)  }\Psi\left[  h_{ij}\right]  ,
\end{equation}
while the inner product is defined by the functional integration:
\begin{equation}
\left\langle \Psi_{1}\mid\Psi_{2}\right\rangle =\int\left[  \mathcal{D}%
h_{ij}\right]  \Psi_{1}^{\ast}\left[  h_{ij}\right]  \Psi_{2}\left[
h_{kl}\right]  .
\end{equation}
We demand that
\begin{equation}
\frac{1}{V}\frac{\left\langle \Psi\left\vert \int_{\Sigma}d^{3}x\hat{\Lambda
}_{\Sigma,f\left(  R\right)  }^{\left(  2\right)  }\right\vert \Psi
\right\rangle }{\left\langle \Psi|\Psi\right\rangle }=\frac{1}{V}\frac
{\int\mathcal{D}\left[  g_{ij}\right]  \Psi^{\ast}\left[  h_{ij}\right]
\int_{\Sigma}d^{3}x\hat{\Lambda}_{\Sigma,f\left(  R\right)  }^{\left(
2\right)  }\Psi\left[  h_{ij}\right]  }{\int\mathcal{D}\left[  g_{ij}\right]
\Psi^{\ast}\left[  h_{ij}\right]  \Psi\left[  h_{ij}\right]  } \label{vareq}%
\end{equation}
be stationary against arbitrary variations of $\Psi\left[  h_{ij}\right]  $.
Note that Eq.$\left(  \ref{vareq}\right)  $ can be considered as the
variational analog of a Sturm-Liouville problem with the cosmological constant
regarded as the associated eigenvalue. Therefore the solution of Eq.$\left(
\ref{WDW2}\right)  $ corresponds to the minimum of Eq.$\left(  \ref{vareq}%
\right)  $. The form of $\left\langle \Psi\left\vert \hat{\Lambda}_{\Sigma
}\right\vert \Psi\right\rangle $ can be computed with the help of the wave
functional $\left(  \ref{tt}\right)  $ and with the help of%
\begin{equation}
\frac{\left\langle \Psi\left\vert h_{ij}\left(  \overrightarrow{x}\right)
\right\vert \Psi\right\rangle }{\left\langle \Psi|\Psi\right\rangle }=0
\end{equation}
and%
\begin{equation}
\frac{\left\langle \Psi\left\vert h_{ij}\left(  \overrightarrow{x}\right)
h_{kl}\left(  \overrightarrow{y}\right)  \right\vert \Psi\right\rangle
}{\left\langle \Psi|\Psi\right\rangle }=K_{ijkl}\left(  \overrightarrow
{x},\overrightarrow{y}\right)  .
\end{equation}
Extracting the TT tensor contribution, we get%
\begin{equation}
\hat{\Lambda}_{\Sigma,f\left(  R\right)  }^{\left(  2\right)  ,\bot}=\frac
{1}{4V}\int_{\Sigma}d^{3}x\sqrt{\bar{g}}G^{ijkl}\left[  \left(  2\kappa
\right)  h\left(  R\right)  K^{-1\bot}\left(  x,x\right)  _{ijkl} +\frac
{1}{\left(  2\kappa\right)  }\left(  \triangle_{2}\right)  _{j}^{a}K^{\bot
}\left(  x,x\right)  _{iakl}\right]  . \label{p22}%
\end{equation}
The propagator $K^{\bot}\left(  x,x\right)  _{iakl}$ can be represented as
\begin{equation}
K^{\bot}\left(  \overrightarrow{x},\overrightarrow{y}\right)  _{iakl}:=%
{\displaystyle\sum_{\tau}}
\frac{h_{ia}^{\left(  \tau\right)  \bot}\left(  \overrightarrow{x}\right)
h_{kl}^{\left(  \tau\right)  \bot}\left(  \overrightarrow{y}\right)
}{2\lambda\left(  \tau\right)  }, \label{proptt}%
\end{equation}
where $h_{ia}^{\left(  \tau\right)  \bot}\left(  \overrightarrow{x}\right)  $
are the eigenfunctions of $\triangle_{2}$. $\tau$ denotes a complete set of
indices and $\lambda\left(  \tau\right)  $ are a set of variational parameters
to be determined by the minimization of Eq.$\left(  \ref{p22}\right)  $. The
expectation value of $\hat{\Lambda}_{\Sigma}^{\bot}$ is easily obtained by
inserting the form of the propagator into Eq.$\left(  \ref{p22}\right)  $%
\begin{equation}
-\frac{\Lambda_{c}^{\prime}\left(  \lambda_{i}\right)  }{\kappa}=\frac{1}{4}%
{\displaystyle\sum_{\tau}}
{\displaystyle\sum_{i=1}^{2}}
\left[  \left(  2\kappa\right)  h\left(  R\right)  \lambda_{i}\left(
\tau\right)  +\frac{\omega_{i}^{2}\left(  \tau\right)  }{\left(
2\kappa\right)  \lambda_{i}\left(  \tau\right)  }\right]  .
\end{equation}
By minimizing with respect to the variational function $\lambda_{i}\left(
\tau\right)  $, we obtain the total one loop energy density for TT tensors%
\begin{equation}
\Lambda_{c}^{\prime}\left(  \lambda_{i}\right)  =-\kappa\sqrt{h\left(
R\right)  }\frac{1}{4}%
{\displaystyle\sum_{\tau}}
\left[  \sqrt{\omega_{1}^{2}\left(  \tau\right)  }+\sqrt{\omega_{2}^{2}\left(
\tau\right)  }\right]  , \label{lambda1loop}%
\end{equation}
where $\Lambda_{c}^{\prime}$ is expressed by the Eq.$\left(  \ref{NewLambda}%
\right)  $. The above expression makes sense only for $\omega_{i}^{2}\left(
\tau\right)  >0$. It is the main formal result. It is true for
generic $f(R)$ functions since $h(R)$ explicitly appears in it.

\subsection{The transverse traceless (TT) spin 2 operator for the Schwarzschild metric and the WKB approximation}

The above considerations can be specified choosing a given metric.
For example, the quantity $\Lambda_{c}^{\prime}$ can be calculated
for a Schwarzschild metric in the WKB approximation. Apparently,
there is no a strong motivation to consider a Schwarzschild metric
as a probe for a cosmological problem. Nevertheless, every quantum
field induces a \textquotedblleft cosmological
term\textquotedblright\ by means of vacuum expectation values and
the variational approach we have considered is particularly easy
to use for a spherically symmetric metric. The Schwarzschild
metric is the simplest sourceless solution of the Einstein field
equations which can be used to compute a cosmological constant
spectrum. Of course, also Minkowski space can be put in the form
of a spherically symmetric metric, but in that case there is no
gravity at all. The other solution need a source which is not
considered in the present Review Paper. In this sense, the computation is
a real vacuum contribution to the cosmological term. The spin-two
operator
for the Schwarzschild metric is defined by%
\begin{equation}
\left(  \triangle_{2}h^{TT}\right)  _{i}^{j}:=-\left(  \triangle_{T}%
h^{TT}\right)  _{i}^{j}+2\left(  Rh^{TT}\right)  _{i}^{j},\label{spin2}%
\end{equation}
where the transverse-traceless (TT) tensor for the quantum fluctuation is
obtained by the following decomposition%
\begin{equation}
h_{i}^{j}=h_{i}^{j}-\frac{1}{3}\delta_{i}^{j}h+\frac{1}{3}\delta_{i}%
^{j}h=\left(  h^{T}\right)  _{i}^{j}+\frac{1}{3}\delta_{i}^{j}h.
\end{equation}
This implies that $\left(  h^{T}\right)  _{i}^{j}\delta_{j}^{i}=0$. The
transversality condition is applied on $\left(  h^{T}\right)  _{i}^{j}$ and
becomes $\nabla_{j}\left(  h^{T}\right)  _{i}^{j}=0$. Thus%
\begin{equation}
-\left(  \triangle_{T}h^{TT}\right)  _{i}^{j}=-\triangle_{S}\left(
h^{TT}\right)  _{i}^{j}+\frac{6}{r^{2}}\left(  1-\frac{2MG}{r}\right)
,\label{tlap}%
\end{equation}
where $\triangle_{S}$ is the scalar curved Laplacian, whose form is%
\begin{equation}
\triangle_{S}=\left(  1-\frac{2MG}{r}\right)  \frac{d^{2}}{dr^{2}}+\left(
\frac{2r-3MG}{r^{2}}\right)  \frac{d}{dr}-\frac{L^{2}}{r^{2}}\label{slap}%
\end{equation}
and $R_{j\text{ }}^{a}$ is the mixed Ricci tensor whose components are:
\begin{equation}
R_{i}^{a}=\left\{  -\frac{2MG}{r^{3}},\frac{MG}{r^{3}},\frac{MG}{r^{3}%
}\right\}  ,
\end{equation}
This implies that the scalar curvature is traceless. We are therefore led to
study the following eigenvalue equation
\begin{equation}
\left(  \triangle_{2}h^{TT}\right)  _{i}^{j}=\omega^{2}h_{j}^{i}\label{p31}%
\end{equation}
where $\omega^{2}$ is the eigenvalue of the corresponding equation. In doing
so, we follow Regge and Wheeler in analyzing the equation as modes of definite
frequency, angular momentum and parity\cite{Regge Wheeler}. In particular, our
choice for the three-dimensional gravitational perturbation is represented by
its even-parity form%
\begin{equation}
\left(  h^{even}\right)  _{j}^{i}\left(  r,\vartheta,\phi\right)  =diag\left[
H\left(  r\right)  ,K\left(  r\right)  ,L\left(  r\right)  \right]
Y_{lm}\left(  \vartheta,\phi\right)  ,\label{pert}%
\end{equation}
with%
\begin{equation}
\left\{
\begin{array}
[c]{c}%
H\left(  r\right)  =h_{1}^{1}\left(  r\right)  -\frac{1}{3}h\left(  r\right)
\\
K\left(  r\right)  =h_{2}^{2}\left(  r\right)  -\frac{1}{3}h\left(  r\right)
\\
L\left(  r\right)  =h_{3}^{3}\left(  r\right)  -\frac{1}{3}h\left(  r\right)
\end{array}
\right.  .
\end{equation}
From the transversality condition, we obtain $h_{2}^{2}\left(  r\right)
=h_{3}^{3}\left(  r\right)  $. Then $K\left(  r\right)  =L\left(  r\right)  $.
For a generic value of the angular momentum $L$, representation $\left(
\ref{pert}\right)  $ joined to Eq.$\left(  \ref{tlap}\right)  $ lead to the
following system of PDE's%

\begin{equation}
\left\{
\begin{array}
[c]{c}%
\left(  -\triangle_{S}+\frac{6}{r^{2}}\left(  1-\frac{2MG}{r}\right)
-\frac{4MG}{r^{3}}\right)  H\left(  r\right)  =\omega_{1,l}^{2}H\left(
r\right) \\
\\
\left(  -\triangle_{S}+\frac{6}{r^{2}}\left(  1-\frac{2MG}{r}\right)
+\frac{2MG}{r^{3}}\right)  K\left(  r\right)  =\omega_{2,l}^{2}K\left(
r\right)
\end{array}
\right.  . \label{p33}%
\end{equation}
Defining the "reduced" fields%

\begin{equation}
H\left(  r\right)  =\frac{f_{1}\left(  r\right)  }{r};\qquad K\left(
r\right)  =\frac{f_{2}\left(  r\right)  }{r},
\end{equation}
and passing to the proper geodesic distance from the \textit{throat} of the
bridge%
\begin{equation}
dx=\pm\frac{dr}{\sqrt{1-\frac{2MG}{r}}}, \label{throat}%
\end{equation}
the system $\left(  \ref{p33}\right)  $ becomes%

\begin{equation}
\left\{
\begin{array}
[c]{c}%
\left[  -\frac{d^{2}}{dx^{2}}+V_{1}\left(  r\right)  \right]  f_{1}\left(
x\right)  =\omega_{1,l}^{2}f_{1}\left(  x\right) \\
\\
\left[  -\frac{d^{2}}{dx^{2}}+V_{2}\left(  r\right)  \right]  f_{2}\left(
x\right)  =\omega_{2,l}^{2}f_{2}\left(  x\right)
\end{array}
\right.  \label{p34}%
\end{equation}
with
\begin{equation}
\left\{
\begin{array}
[c]{c}%
V_{1}\left(  r\right)  =\frac{l\left(  l+1\right)  }{r^{2}}+U_{1}\left(
r\right)  +m_{g}^{2}\\
\\
V_{2}\left(  r\right)  =\frac{l\left(  l+1\right)  }{r^{2}}+U_{2}\left(
r\right)  +m_{g}^{2}%
\end{array}
\right.  ,
\end{equation}
where we have defined $r\equiv r\left(  x\right)  $ and%
\begin{equation}
\left\{
\begin{array}
[c]{c}%
U_{1}\left(  r\right)  =\left[  \frac{6}{r^{2}}\left(  1-\frac{2MG}{r}\right)
-\frac{3MG}{r^{3}}\right] \\
\\
U_{2}\left(  r\right)  =\left[  \frac{6}{r^{2}}\left(  1-\frac{2MG}{r}\right)
+\frac{3MG}{r^{3}}\right]
\end{array}
\right.  .
\end{equation}
Note that%
\begin{equation}
\left\{
\begin{array}
[c]{c}%
U_{1}\left(  r\right)  \geq0\qquad\text{when }r\geq\frac{5MG}{2}\\
U_{1}\left(  r\right)  <0\qquad\text{when }2MG\leq r<\frac{5MG}{2}\\
\\
U_{2}\left(  r\right)  >0\text{ }\forall r\in\left[  2MG,+\infty\right)
\end{array}
\right.  . \label{negU}%
\end{equation}
In order to use the WKB approximation, we define two r-dependent radial wave
numbers $k_{1}\left(  r,l,\omega_{1,nl}\right)  $ and $k_{2}\left(
r,l,\omega_{2,nl}\right)  $%
\begin{equation}
\left\{
\begin{array}
[c]{c}%
k_{1}^{2}\left(  r,l,\omega_{1,nl}\right)  =\omega_{1,nl}^{2}-\frac{l\left(
l+1\right)  }{r^{2}}-m_{1}^{2}\left(  r\right) \\
\\
k_{2}^{2}\left(  r,l,\omega_{2,nl}\right)  =\omega_{2,nl}^{2}-\frac{l\left(
l+1\right)  }{r^{2}}-m_{2}^{2}\left(  r\right)
\end{array}
\right.  , \label{rwn}%
\end{equation}
where we have defined two r-dependent effective masses $m_{1}^{2}\left(
r\right)  $ and $m_{2}^{2}\left(  r\right)  $. The WKB approximation we will
use to evaluate Eq.$\left(  \ref{lambda1loop}\right)  $ is equivalent to the
scattering phase shift method and to the entropy computation in the brick wall
model. We begin by counting the number of modes with frequency less than
$\omega_{i}$, $i=1,2$. This is given approximately by%
\begin{equation}
\tilde{g}\left(  \omega_{i}\right)  =\int\nu_{i}\left(  l,\omega_{i}\right)
\left(  2l+1\right)  , \label{p41}%
\end{equation}
where $\nu_{i}\left(  l,\omega_{i}\right)  $, $i=1,2$ is the number of nodes
in the mode with $\left(  l,\omega_{i}\right)  $, such that $\left(  r\equiv
r\left(  x\right)  \right)  $
\begin{equation}
\nu_{i}\left(  l,\omega_{i}\right)  =\frac{1}{2\pi}\int_{-\infty}^{+\infty
}dx\sqrt{k_{i}^{2}\left(  r,l,\omega_{i}\right)  }. \label{p42}%
\end{equation}
Here it is understood that the integration with respect to $x$ and $l$ is
taken over those values which satisfy $k_{i}^{2}\left(  r,l,\omega_{i}\right)
\geq0,$ $i=1,2$. With the help of Eqs.$\left(  \ref{p41},\ref{p42}\right)  $,
we obtain the one loop total energy for TT tensors which is%
\begin{equation}
\frac{1}{8\pi}\sum_{i=1}^{2}\int_{-\infty}^{+\infty}dx\left[  \int
_{0}^{+\infty}\omega_{i}\frac{d\tilde{g}\left(  \omega_{i}\right)  }%
{d\omega_{i}}d\omega_{i}\right]  .
\end{equation}
By extracting the energy density contributing to the cosmological constant, we
get%
\begin{equation}
\Lambda_{c}^{\prime}=\Lambda_{c,1}^{\prime}+\Lambda_{c,2}^{\prime}=\rho
_{1}+\rho_{2} =-\sqrt{h\left(  R\right)  }\frac{\kappa}{16\pi^{2}}\left\{
\int_{0}^{+\infty}\omega_{1}^{2}\sqrt{\omega_{1}^{2}-m_{1}^{2}\left(
r\right)  }d\omega_{1}+\int_{0}^{+\infty}\omega_{2}^{2}\sqrt{\omega_{2}%
^{2}-m_{2}^{2}\left(  r\right)  }d\omega_{2}\right\}  , \label{tote1loop}%
\end{equation}
where we have included an additional $4\pi$ coming from the angular integration.

\subsection{One-loop energy Regularization and Renormalization}

In this subsection, we will use the zeta function regularization method to
compute the energy densities $\rho_{1}$ and $\rho_{2}$ (see Appendix B). Note that this
procedure is completely equivalent to the subtraction procedure of the Casimir
energy computation where the zero point energy (ZPE) in different backgrounds
with the same asymptotic properties is involved. To this purpose, we introduce
the additional mass parameter $\mu$ in order to restore the correct dimension
for the regularized quantities. Such an arbitrary mass scale emerges
unavoidably in any regularization scheme. Then we have%
\begin{equation}
\rho_{i}\left(  \varepsilon\right)  =-\sqrt{h\left(  R\right)  }\frac{\kappa
}{16\pi^{2}}\mu^{2\varepsilon}\int_{0}^{+\infty}d\omega_{i}\frac{\omega
_{i}^{2}}{\left(  \omega_{i}^{2}-m_{i}^{2}\left(  r\right)  \right)
^{\varepsilon-\frac{1}{2}}},\label{zeta}%
\end{equation}
where%
\begin{equation}
\left\{
\begin{array}
[c]{c}%
\rho_{1}\left(  \varepsilon\right)  =-\sqrt{h\left(  R\right)  }\frac{\kappa
}{16\pi^{2}}\int_{0}^{+\infty}\omega_{1}^{2}\sqrt{\omega_{1}^{2}-m_{1}%
^{2}\left(  r\right)  }d\omega_{1}\\
\\
\rho_{2}\left(  \varepsilon\right)  =-\sqrt{h\left(  R\right)  }\frac{\kappa
}{16\pi^{2}}\int_{0}^{+\infty}\omega_{2}^{2}\sqrt{\omega_{2}^{2}-m_{2}%
^{2}\left(  r\right)  }d\omega_{2}%
\end{array}
\right.  .\label{edens}%
\end{equation}
The integration has to be meant in the range where $\omega_{i}^{2}-m_{i}%
^{2}\left(  r\right)  \geq0$\footnote{Details of the calculation can be found
in the Appendix B.}. One gets%
\begin{equation}
\rho_{i}\left(  \varepsilon\right)  =\sqrt{h\left(  R\right)  }\kappa
\frac{m_{i}^{4}\left(  r\right)  }{256\pi^{2}}\left[  \frac{1}{\varepsilon
}+\ln\left(  \frac{\mu^{2}}{m_{i}^{2}\left(  r\right)  }\right)  +2\ln
2-\frac{1}{2}\right]  ,\label{zeta1}%
\end{equation}
$i=1,2$. In order to renormalize the divergent ZPE, we write%
\begin{equation}
\Lambda_{c}^{\prime}=8\pi G\left[  \rho_{1}\left(  \varepsilon\right)
+\rho_{2}\left(  \varepsilon\right)  +\rho_{1}\left(  \mu\right)  +\rho
_{2}\left(  \mu\right)  \right]  ,
\end{equation}
where we have separated the divergent part from the finite part. For practical
purposes, it is useful to divide $\Lambda_{c}^{\prime}$ with the factor
$\sqrt{h\left(  R\right)  }$. To this aim, we define%
\begin{equation}
\frac{\Lambda_{c}^{\prime}}{\sqrt{h\left(  R\right)  }}=\left[  \Lambda
_{c}+\frac{1}{2V}\int_{\Sigma}d^{3}x\sqrt{g}\frac{Rf^{\prime}\left(  R\right)
-f\left(  R\right)  }{f^{\prime}\left(  R\right)  }\right]  \frac{1}%
{\sqrt{h\left(  R\right)  }}%
\end{equation}
and we extract the divergent part of $\Lambda$, in the limit $\varepsilon
\rightarrow0$, by setting%
\begin{equation}
\Lambda^{div}=8\pi G\left[  \rho_{1}\left(  \varepsilon\right)  +\rho
_{2}\left(  \varepsilon\right)  \right]  =\frac{G}{32\pi\varepsilon}\left[
m_{1}^{4}\left(  r\right)  +m_{2}^{4}\left(  r\right)  \right]  .
\end{equation}
Thus, the renormalization is performed via the absorption of the divergent
part into the re-definition of the bare classical cosmological constant
$\Lambda_{c}$, that is
\begin{equation}
\Lambda_{c}\rightarrow\Lambda_{0}+\sqrt{h\left(  R\right)  }\Lambda^{div}.
\end{equation}
The remaining finite value for the cosmological constant reads\footnote{Since
$m_{1}^{2}\left(  r\right)  $ can change in sign, when we integrate over
$\omega_{1}$ we can use either $I_{+}$ or $I_{-}$. This leads to the
appearance of the absolute value.}%
\[
\frac{\Lambda_{0}^{\prime}\left(  \mu\right)  }{8\pi G}=\rho_{1}\left(
\mu\right)  +\rho_{2}\left(  \mu\right)  =\frac{1}{256\pi^{2}}\left\{
m_{1}^{4}\left(  r\right)  \left[  \ln\left(  \frac{\mu^{2}}{\left\vert
m_{1}^{2}\left(  r\right)  \right\vert }\right)  +2\ln2-\frac{1}{2}\right]
+\right.
\]%
\begin{equation}
\left.  +m_{2}^{4}\left(  r\right)  \left[  \ln\left(  \frac{\mu^{2}}%
{m_{2}^{2}\left(  r\right)  }\right)  +2\ln2-\frac{1}{2}\right]  \right\}
=\rho_{eff}^{TT}\left(  \mu,r\right)  ,\label{lambda0}%
\end{equation}
where%
\begin{equation}
\Lambda_{0}^{\prime}\left(  \mu\right)  =\frac{1}{\sqrt{h\left(  R\right)  }%
}\left[  \Lambda_{0}\left(  \mu\right)  +\frac{1}{2V}\int_{\Sigma}d^{3}%
x\sqrt{g}\frac{Rf^{\prime}\left(  R\right)  -f\left(  R\right)  }{f^{\prime
}\left(  R\right)  }\right]
\end{equation}
is the modified cosmological constant. The quantity in Eq.$\left(
\ref{lambda0}\right)  $ depends on the arbitrary mass scale $\mu.$ It is
appropriate to use the renormalization group equation to eliminate such a
dependence. To this aim, we impose that \cite{RGeq}%
\begin{equation}
\frac{1}{8\pi G}\mu\frac{\partial\Lambda_{0}^{\prime}\left(  \mu\right)
}{\partial\mu}=\mu\frac{d}{d\mu}\rho_{eff}^{TT}\left(  \mu,r\right)
.\label{rg}%
\end{equation}
Solving it, we find that the renormalized constant $\Lambda_{0}$ should be
treated as a running one in the sense that it varies, provided that the scale
$\mu$ is changing
\begin{equation}
\Lambda_{0}^{\prime}\left(  \mu,r\right)  =\Lambda_{0}^{\prime}\left(  \mu
_{0},r\right)  +\frac{G}{16\pi}\left[  m_{1}^{4}\left(  r\right)  +m_{2}%
^{4}\left(  r\right)  \right]  \ln\frac{\mu}{\mu_{0}}.\label{lambdamu}%
\end{equation}
Substituting Eq.$\left(  \ref{lambdamu}\right)  $ into Eq.$\left(
\ref{lambda0}\right)  $ we find%
\begin{equation}
\frac{\Lambda_{0}^{\prime}\left(  \mu_{0},r\right)  }{8\pi G}=-\frac{1}%
{256\pi^{2}}\left\{  m_{1}^{4}\left(  r\right)  \left[  \ln\left(
\frac{\left\vert m_{1}^{2}\left(  r\right)  \right\vert }{\mu_{0}^{2}}\right)
-2\ln2+\frac{1}{2}\right]  +m_{2}^{4}\left(  r\right)  \left[  \ln\left(
\frac{m_{2}^{2}\left(  r\right)  }{\mu_{0}^{2}}\right)  -2\ln2+\frac{1}%
{2}\right]  \right\}  .\label{lambdamu0}%
\end{equation}
It is worth remarking that while $m_{2}^{2}\left(  r\right)  $ is constant in
sign, $m_{1}^{2}\left(  r\right)  $ is not. Indeed, for the critical value
$\bar{r}=5MG/2$, $m_{1}^{2}\left(  \bar{r}\right)  =m_{g}^{2}$ and in the
range $\left(  2MG,5MG/2\right)  $ for some values of $m_{g}^{2}$, $m_{1}%
^{2}\left(  \bar{r}\right)  $ can be negative. It is interesting therefore
concentrate in this range. To further proceed, we observe that $m_{1}%
^{2}\left(  r\right)  $ and $m_{2}^{2}\left(  r\right)  $ can be recast into a
more suggestive and useful form, namely%
\begin{equation}
\left\{
\begin{array}
[c]{c}%
m_{1}^{2}\left(  r\right)  =U_{1}\left(  r\right)  =m_{1}^{2}\left(
r,M\right)  -m_{2}^{2}\left(  r,M\right)  \\
\\
m_{2}^{2}\left(  r\right)  =U_{2}\left(  r\right)  =m_{1}^{2}\left(
r,M\right)  +m_{2}^{2}\left(  r,M\right)
\end{array}
\right.  ,
\end{equation}
where $m_{1}^{2}\left(  r,M\right)  \rightarrow0$ when $r\rightarrow\infty$ or
$r\rightarrow2MG$ and $m_{2}^{2}\left(  r,M\right)  =3MG/r^{3}$. Nevertheless,
in the above mentioned range $m_{1}^{2}\left(  r,M\right)  $ is negligible
when compared with $m_{2}^{2}\left(  r,M\right)  $. So, in a first
approximation we can write%
\begin{equation}
\left\{
\begin{array}
[c]{c}%
m_{1}^{2}\left(  r\right)  \simeq-m_{2}^{2}\left(  r_{0},M\right)  =-m_{0}%
^{2}\left(  M\right)  \\
\\
m_{2}^{2}\left(  r\right)  \simeq m_{2}^{2}\left(  r_{0},M\right)  =m_{0}%
^{2}\left(  M\right)
\end{array}
\right.  ,
\end{equation}
where we have defined a parameter $r_{0}>2MG$ and $m_{0}^{2}\left(  M\right)
=3MG/r_{0}^{3}$. The main reason for introducing a new parameter resides in
the fluctuation of the horizon that forbids any kind of approach. Of course
the quantum fluctuation must obey the uncertainty relations. Thus Eq.$\left(
\ref{lambdamu0}\right)  $ becomes%
\begin{equation}
\frac{\Lambda_{0}^{\prime}\left(  \mu_{0},r\right)  }{8\pi G}=-\frac{m_{0}%
^{4}\left(  M\right)  }{128\pi^{2}}\left[  \ln\left(  \frac{m_{0}^{2}\left(
M\right)  }{4\mu_{0}^{2}}\right)  +\frac{1}{2}\right]  .\label{lambdamu0a}%
\end{equation}
Now, we compute the maximum of $\Lambda_{0}^{\prime}$, by setting%
\begin{equation}
x=\frac{m_{0}^{2}\left(  M\right)  }{4\mu_{0}^{2}}.
\end{equation}
Thus $\Lambda_{0}^{\prime}$ becomes%
\begin{equation}
\Lambda_{0}^{\prime}\left(  \mu_{0},x\right)  =-\frac{G\mu_{0}^{4}}{\pi}%
x^{2}\left[  \ln\left(  x\right)  +\frac{1}{2}\right]  .\label{LambdansM}%
\end{equation}
As a function of $x$, $\Lambda_{0}\left(  \mu_{0},x\right)  $ vanishes for
$x=0$ and $x=\exp\left(  -\frac{1}{2}\right)  $ and when $x\in\left[
0,\exp\left(  -\frac{1}{2}\right)  \right]  $, $\Lambda_{0}^{\prime}\left(
\mu_{0},x\right)  \geq0$. It has a maximum for
\begin{equation}
\bar{x}=\frac{1}{e}\qquad\Longleftrightarrow\qquad m_{0}^{2}\left(  M\right)
=\frac{4\mu_{0}^{2}}{e}%
\end{equation}
and its value is%
\begin{equation}
\Lambda_{0}^{\prime}\left(  \mu_{0},\bar{x}\right)  =\frac{G\mu_{0}^{4}}{2\pi
e^{2}}\qquad\text{or\qquad}\frac{1}{\sqrt{h\left(  R\right)  }}\left[
\Lambda_{0}\left(  \mu_{0},\bar{x}\right)  +\frac{1}{2V}\int_{\Sigma}%
d^{3}x\sqrt{g}\frac{Rf^{\prime}\left(  R\right)  -f\left(  R\right)
}{f^{\prime}\left(  R\right)  }\right]  =\frac{G\mu_{0}^{4}}{2\pi e^{2}}.
\end{equation}
Isolating $\Lambda_{0}\left(  \mu_{0},\bar{x}\right)  $, we get%
\begin{equation}
\Lambda_{0}\left(  \mu_{0},\bar{x}\right)  =\sqrt{h\left(  R\right)  }%
\frac{G\mu_{0}^{4}}{2\pi e^{2}}-\frac{1}{2V}\int_{\Sigma}d^{3}x\sqrt{g}%
\frac{Rf^{\prime}\left(  R\right)  -f\left(  R\right)  }{f^{\prime}\left(
R\right)  }.
\end{equation}
Note that $\Lambda_{0}\left(  \mu_{0},\bar{x}\right)  $ can be set to zero
when%
\begin{equation}
\sqrt{h\left(  R\right)  }\frac{G\mu_{0}^{4}}{2\pi e^{2}}=\frac{1}{2V}%
\int_{\Sigma}d^{3}x\sqrt{g}\frac{Rf^{\prime}\left(  R\right)  -f\left(
R\right)  }{f^{\prime}\left(  R\right)  }.\label{lambda0_fin}%
\end{equation}
Let us see what happens when $f\left(  R\right)  =\exp\left(
-\alpha R\right)  $. This choice is simply suggested by the
regularity of the function at every scale and by the fact
that any power of $R$, considered as a correction to GR, is
included. In this case, Eq.$\left( \ref{lambda0_fin}\right)
$ becomes%
\begin{equation}
\sqrt{\frac{3\alpha\exp\left(  -\alpha R\right)  +2}{\alpha\exp\left(  -\alpha
R\right)  }}\frac{G\mu_{0}^{4}}{\pi e^{2}}=\frac{1}{\alpha V}\int_{\Sigma
}d^{3}x\sqrt{g}\left(  1+\alpha R\right)  .
\end{equation}
For Schwarzschild, it is $R=0$, then%
\begin{equation}
\frac{G\mu_{0}^{4}}{\pi e^{2}}=\sqrt{\frac{1}{\left(  3\alpha+2\right)
\alpha}}.
\end{equation}
By setting $\alpha=G$, we have the relation%
\begin{equation}
\mu_{0}^{4}=\frac{\pi e^{2}}{G}\sqrt{\frac{1}{\left(  3G+2\right)  G}}.
\end{equation}
Note that in any case, the maximum of $\Lambda$ corresponds to
the minimum of the energy density.

\subsection{Concluding remarks}

Despite of its successes, GR can only be considered as a step toward a much more complete and
comprehensive structure due to a large number of weaknesses as discussed in this Report. Among
them, the issue to find out the fundamental gravitational vacuum
state is one of the main problem to achieve a definite Quantum
Gravity theory which, till now, is lacking. However, several
semiclassical approaches have been proposed and, from several
points of view, it is clear that the former Hilbert-Einstein
scheme has to be enlarged. The $f(R)$ theories of gravity are a
minimal but well founded extension of GR where the form of the
function $f(R)$ is not supposed {\it "a priori"} but is reconstructed by
the observed dynamics at galactic and cosmological scales
\cite{mimicking,prl}. Also if they seem a viable scheme from
cosmology and astrophysics viewpoints, their theoretical
foundation has to be sought  at a fundamental level. In
particular, one has to face the possibility to encompass the
$f(R)$-gravity in the general framework of Quantum Field Theory on
curved space-time. Here, we have dealt with the problem to
find out vacuum states for $f(R)$-gravity via the $(3+1)$ ADM
formalism. Analogously to GR, we have constructed the Hamiltonian
constraint of a generic $f(R)$ theory and then achieved a
canonical quantization giving the $f(R)$-WDW equation. In this
context, the cosmological constant (vacuum state) emerges as a WDW
eigenvalue. The related wave functional can be split by an
orthogonal decomposition and then, constructing the
transverse-traceless propagator, it is possible to obtain, after a
variational minimization, the total one-loop energy density for
the TT tensors. Such a quantity explicitly depends on the form of
$f(R)$. As an application, we derived the energy density
contributions to the cosmological constant for a TT spin 2
operator in the Schwarzschild metric and in the WKB approximation.
The one-loop energy regularization and renormalization are
achieved by the zeta function regularization method. The resulting
renormalized $\Lambda_{0}$ is a running constant which can be set
to zero depending on the value of an arbitrary mass scale
parameter $\mu$. As explicit calculation, we find out the value of
such a parameter for a theory of the form $f(R)=\exp(-\alpha R)$
in the Schwarzschild metric. This case can be used for
several applications at cosmological and astrophysical scales. In
particular, truncated versions of such an exponential function,
power law $f(R)$, have been used for galactic dynamics
\cite{noipla,prl,CapCardTro07}. In those cases, a corrected Newtonian
potential, derived from the $f(R)$ Schwarzschild solution, has
been used to fit, with great accuracy,
 data from low surface brightness galaxies without using dark
matter haloes. This approach has allowed to fix a suitable mass
scale comparable with the core size of galactic systems. Such a
mass can be directly related to the above parameter $\alpha$
depending on the core radius $r_c$ (see also \cite{CapCardTro07}).

In summary, the application of Quantum Field Theory methods to $f(R)$-gravity
seems a viable scheme and gives positive results toward the issue to select
vacuum states (eigenvalues) which can be interpreted as the cosmological
constant. However, further studies are needed in order to generalize such
results to other metrics and other ETGs.

%
\section{The initial value problem}

Another important problem is the initial value formulation.  
There are several criteria for the viability of a theory of 
gravity, and one of them is certainly the property of having a 
well-posed initial value problem  in order 
to guarantee that the 
theory has predictive power \cite{Wald84}.  The Cauchy 
problem of theories described by quadratic actions in curvature invariants 
was studied in \cite{Noakes83, TeyssandierTourrenc83} and found 
to be  well-posed (in four  space-time  dimensions, the Gauss-Bonnet 
identity  allows one to drop  the Kretschmann scalar 
$R_{\mu\nu\rho\sigma}R^{\mu\nu\rho\sigma}$ from the 
action). Ref.~\cite{TremblayFaraoni07} discusses  the initial 
value   problem of metric  and Palatini 
$f(R)$-gravity with  a 
general  ({\em i.e.}, not  restricted to the quadratic form 
considered in  \cite{Noakes83,TeyssandierTourrenc83}) 
function $f(R)$ but dropping 
$R_{\mu\nu} R^{\mu\nu}$ and  
$R_{\mu\nu\rho\sigma}  R^{\mu\nu\rho\sigma}$. 
  It is worth noticing that in this paper, the Cauchy problem for the Palatini approach of $f(R)$-gravity is  ill-formulated and ill-posed leading to the wrong result that the ADM formulation is inadequate for the Initial Value Problem. This shortcoming has recently been removed in  \cite{olmoalepuz2}, where the Hamiltonian formulation of $f(R)$-gravity, both in metric and in Palatini formalism, is fully consistent with the well formulation of the Cauchy problem and there is no reason to believe that it is not well-posed in general.

It is useful to 
begin with the Cauchy 
problem of scalar-tensor gravity, which 
was studied in 
\cite{Salgado06}, and then to use the results and the equivalence 
of metric and Palatini $f(R)$-gravity  with an $\omega=0, -3/2$ 
Brans-Dicke theory.

Let us provide some  terminology 
\cite{Solin06}:    the 
system of $3+1$ equations of motion of GR or ETGs is  
{\em well-formulated} if it can be recast as  a  system of  only 
first order equations in time and space  in the 
scalar field variables. The goal is to write this system in  the 
full  first order form 
\begin{equation}
\partial_t \vec{u} + M^i \nabla_i \vec{u}=\vec{S}\left( 
\vec{u}\right) \,,
\end{equation}
where $\vec{u}$  denotes the fundamental variables 
$h_{ij}, K_{ij}$, {\em etc.} of the usual $3+1$ 
 ADM splitting, 
$M^i$ is  called the {\em characteristic matrix} of the system, 
and  $\vec{S}\left(  \vec{u} \right)$ describes source terms and 
contains only the fundamental variables but not their 
derivatives. The initial  value formulation 
is then said to 
be {\em well-posed} if the system of partial 
differential equations is {\em symmetric hyperbolic} ({\em 
i.e.}, $M^i$ is symmetric)  and {\em strongly hyperbolic} ({\em 
i.e.}, $ s_iM^i$ has a real set of eigenvalues and a complete 
set of eigenvectors for  any one-form $s_i$, and obeys some 
boundedness conditions). 

For a physical theory to be viable, it must admit an 
appropriate initial value formulation to guarantee its 
predictability \cite{courant}. This means that, starting from  
suitably prescribed initial data, the subsequent dynamical 
evolution of the physical system is completely and uniquely 
determined. In this case, the problem is said to be {\it
well-formulated}. For example, in classical mechanics, given the
initial positions and velocities of the particles (or of the
constituents) composing  a physical system with a finite 
number of degrees of freedom and knowing the 
interactions between them, if the system   
evolves without external interferences the dynamical evolution 
is determined. This is true also for  field theories, for 
example, for the Maxwell field. However, 
even if the initial  value  problem is well-formulated, the 
theory must possess additional  
properties in order to be viable. 
First, small changes of the initial data must produce only small
perturbations in the subsequent dynamics over reasonably short  
time scales, in other words  the evolution equations   
should exhibit  a continuous dependence on the initial data 
in order to be predictive.  Second, for hyperbolic equations, 
changes in the initial data 
must  preserve the  causal structure of the theory. If both 
these requirements are  satisfied, the initial value problem
of the theory is also {\it well-posed}.
GR has been shown to admit a well-formulated and
well-posed initial value problem {\em in vacuo} and in the 
presence of "reasonable'' forms of matter (perfect fluids, 
minimally coupled scalar fields, {\em etc.})  but, for other 
relativistic field theories, the initial value formulation must 
be studied 
carefully. One  needs to satisfy constraints between the initial 
data  and perform wise gauge choices in order to cast the field 
equations in a form  suitable to correctly formulate the Cauchy 
problem. The  consequence  of 
well-posedness is that GR  is a 
"stable'' theory with a robust causal structure in which  
singularities can be classified (for a detailed discussion  
see Refs.~ \cite{Synge55, Wald84}).
Here we focus on  whether the initial value problem 
of ETGs (including  scalar-tensor and  $f(R)$ theories in both 
the metric and metric-affine formulation) is well-formulated. It 
is not {\em a priori} obvious that  standard GR methods are 
suitable for the discussion of the Cauchy  
problem in every ETG and it is doubtful that the full Cauchy 
problem can be properly addressed using only  the results 
available in the literature for the  fourth order theories 
described by a  quadratic  Lagrangian  
\cite{TeyssandierTourrenc83, kerner}. $f(R)$-gravity, like GR,
is a gauge theory with constrained dynamics and  establishing 
results on the initial value formulation relies on solving  the 
constraints on the initial data and on finding suitable gauges,   
coordinate choices in which the Cauchy problem
 can be 
demonstrated to well-formulated and, possibly, well-posed. In 
\cite{TeyssandierTourrenc83, Noakes83} the initial value 
problem is studied for quadratic Lagrangians in
the metric approach with the conclusion that it is well-posed. 
The Cauchy problem for generic $f(R)$ models is studied below in 
the metric and Palatini 
approaches  with  the result  that the problem is
well-formulated for the metric theory  in the presence of 
"reasonable'' matter and well-posed {\em in vacuo}.
It is shown below that the Cauchy problem of metric-affine
$f(R)$-gravity is well-formulated and well-posed {\em in vacuo}, 
while it can be at least well-formulated for various forms of 
matter including  perfect fluids, Klein-Gordon, and Yang-Mills 
fields. We use the $3+1$ ADM formulation and  the Gaussian normal coordinates approach, both of which  prove useful   
in the discussion of whether the Cauchy problem is well-formulated.  Of course, 
in order to prove the complete viability of a  theory, also  
well-posedness has to be demonstrated.

\subsection{The Cauchy problem of scalar-tensor gravity}

Early work on the initial value  problem 
of scalar-tensor gravity  
includes Refs.~\cite{Noakes83, TeyssandierTourrenc83,  
CockeCohen68}. 
Noakes \cite{Noakes83} proved well-posedness of the  Cauchy 
problem for a  non-minimally coupled  
scalar field $\phi$ with 
vacuum action 
\begin{equation}\label{CauchyNMC}
S_{NMC} = \int d^{4}x\sqrt{-g}\left[ \left(\frac{1}{2\kappa} - 
\xi\phi^{2}\right)  R - 
\frac{1}{2}\nabla^{\mu}\phi\nabla_{\mu}\phi - V(\phi)\right] \,.
\end{equation}
Cocke and Cohen \cite{CockeCohen68}  used  Gaussian normal  
coordinates to study the Cauchy problem of 
Brans-Dicke theory without potential  
$V(\phi)$. A systematic  approach to the  
Cauchy problem of  scalar-tensor theories of 
the form
\begin{equation}
S = \int d^{4}x\sqrt{-g}\left[ \frac{f(\phi) R}{2\kappa}  - 
\frac{1}{2}\nabla^{\alpha}\phi\nabla_{\alpha}\phi  - 
V(\phi)\right] + S^{(m)}
\end{equation}
independent of particular coordinate choices was proposed 
by Salgado \cite{Salgado06}, obtaining the result that   
the  Cauchy  problem  is well-posed {\em in vacuo} and  
well-formulated  otherwise. Slightly more  general  
scalar-tensor theories of the form
\begin{equation}
S = \int d^{4}x\sqrt{-g}\left[ \frac{f(\phi) R}{2\kappa} - 
\frac{\omega (\phi)}{2}\nabla^{\mu}\phi\nabla_{\mu} \phi - 
V(\phi)\right] + S^{(m)}\,,
\end{equation}
containing the additional coupling function $\omega(\phi)$ were 
studied in \cite{TremblayFaraoni07}.
In the notation of Ref.~\cite{Salgado06}, and setting $\kappa=1 
$ in this section, the field equations are 
\begin{eqnarray} 
G_{\mu\nu} & = &  \frac{1}{f}\left[ f^{\prime\prime} \left( 
\nabla_{\mu}\phi\nabla_{\nu}\phi  - 
g_{\mu\nu}\nabla^{\alpha}\phi\nabla_{\alpha}\phi \right) + 
f^\prime \left( \nabla_{\mu}\nabla_{\nu} \phi - 
g_{\mu\nu}\Box\phi \right)\right] \nonumber \\
&&\nonumber \\
 & +& \frac{1}{f}\left[ \omega \left( 
\nabla_{\mu}\phi\nabla_{\nu}\phi - \frac{1}{
2} g_{\mu\nu}\nabla^{\alpha}\phi\nabla_{\alpha}\phi \right) - 
V(\phi)g_{\mu\nu} + 
T_{\mu\nu}^{(m)}
\right] \,,\label{Cauchy24}\\
&&\nonumber \\
 \omega\Box\phi &+& \frac{f^\prime}{2}  R -  
V^\prime (\phi) +
\frac{\omega^\prime}{2}\nabla^{\alpha} 
\phi\nabla_{\alpha}\phi = 0 \,,
\end{eqnarray}
where a prime denotes differentiation with respect to $\phi$.  
Eq.~(\ref{Cauchy24}) is in the form of an effective Einstein 
equation \cite{Salgado06}
\be
G_{\mu\nu} =  T_{\mu\nu}^{(eff)} = \frac{1}{f(\phi)}\left( 
T_{\mu\nu}^{\left(   f \right)} + T_{\mu\nu}^{\left(  \phi  
\right)} + T_{\mu\nu}^{(m)}\right) \,,  \label{Cauchy26} 
\ee 
where 
\begin{equation}
T_{\mu\nu}^{\left( f \right)} = f^{\prime\prime}(\phi) \left( 
\nabla_{\mu} \phi \nabla_{\nu} \phi - 
g_{\mu\nu}\nabla^{\alpha}\phi\nabla_{\alpha}\phi \right) + 
f^\prime (\phi)\left( 
\nabla_{\mu}\nabla_{\nu} \phi - 
g_{\mu\nu}\Box\phi \right) \,,
\end{equation}
and 
\begin{equation}
T_{\mu\nu}^{(\phi)} = \omega(\phi) \left( 
\nabla_{\mu} \, \phi\nabla_{\nu}\phi - 
\frac{1}{2}g_{\mu\nu}\nabla^{\alpha}\phi \, \nabla_{\alpha}\phi 
\right) - V(\phi)g_{\mu\nu}
\end{equation}
has canonical structure.  The trace of Eq.~(\ref{Cauchy26}) yields   
\begin{equation}
\Box\phi = \frac{ \frac{f^\prime T^{(m)}}{2} -2f^\prime 
V(\phi) + f V^\prime (\phi) 
+ \left[  -\frac{\omega^\prime f}{2} -  \frac{f^\prime}{2} 
\left( \omega + 
3f^{\prime\prime} \right) \right] 
\nabla^{\alpha}\phi\nabla_{\alpha }\phi }{ f\left[ 
\omega+\frac{3\left(f^\prime\right)^{2}}{2f}\right] }\,.
\end{equation}
One then proceeds in the usual $3+1$  ADM formulation of the 
theory  in terms of  lapse, shift,  
extrinsic curvature, and  gradient of $\phi$ 
\cite{Wald84, Reula98,  Salgado06}. It is assumed that a 
time 
function $t$ is defined so  that the space-time 
$ \left( M, g_{\mu\nu}, \phi \right)$ is foliated by a family of 
hypersurfaces $\Sigma_{t}$  of constant  $t$  with unit timelike 
normal $n^{\mu}$.  The  three-dimensional metric  is $h_{\mu\nu} 
\equiv g_{\mu\nu} + n_{\mu} n_{\nu}$, ${h^{\mu}}_{\nu}$
 is the projection operator on $\Sigma_{t}$, and $n^{\mu} $ and 
${ h_{\mu}}^{\nu}$ satisfy 
\begin{eqnarray}
n^{\mu} n_{\mu} = -1 \,, &\,\,\,\,\,\,\,\,\,\,\,\,\,\,\,\, 
h_{\mu\nu} n^{\mu} = h_{\mu\nu} n^{\nu} = 0 \,, 
&\,\,\,\,\,\,\,\,\,\,\,\,\,\,\,\, {h_{\mu}}^{\nu} 
h_{\nu\rho} = h_{\mu\rho} \,.
\end{eqnarray}
The metric decomposition in terms of lapse function $N$ 
and shift  vector $ N^{\mu }$ is
\begin{equation}
ds^{2} = -\left( N^{2} - N^{i}N_{i} \right) dt^{2} -  
2N_{i}dtdx^{i} + h_{ij}dx^{i}dx^{j} \,,
\end{equation}
where $ i,j , k $ are spatial indices assuming the  values $ 
1,2$, and~3,  $N > 0$, $n_{\mu} = - N \nabla_{\mu} t
$, and 
\begin{equation}
N^{\mu} = -{h^{\mu}}_{\nu}t^{\nu} \,,
\end{equation}
and where the time flow vector $t^{\mu }$ obeys  
\begin{eqnarray}
&& t^{\mu}\nabla_{\mu} t = 1 \,,\\
&&\nonumber\\
&& t^{\mu} = -N^{\mu}+N n^{\mu} \,.
\end{eqnarray}
As a consequence, $N = -n_{\mu} t^{\mu}$ and $N^{\mu} n_{\mu} = 
0$. The extrinsic  curvature  of the hypersurfaces $\Sigma_{t}$ 
is
\begin{equation}
K_{\mu\nu} = -{h_{\mu}}^{\rho} 
{h_{\nu}}^{\sigma}\nabla_{\rho}n_{\sigma} 
\end{equation}
and the three-dimensional covariant derivative of $h_{\mu\nu}$ on 
$\Sigma_{t}$ is given by 
\begin{equation}
D_{i} ^{\left(3\right)}{T^{\mu_{1}\ldots}}_{\nu_{1}\ldots} = { 
h^{\mu_{1}}  }_{\rho_{1}}  \ldots 
{h^{\sigma_{1}}}_{\nu_{1}}\ldots {h^{f}}_{i}\nabla_{ f} ^{ 
\left(3\right)} {T^{\rho_{1}\ldots}}_{\sigma_{1}\ldots}
\end{equation}
for any three-tensor $^{(3)} {T^{\mu_{1} \ldots} 
}_{\nu_{1}\ldots}$ , 
with $D_i h_{\mu\nu}= 0$.  The 
 spatial gradient of the scalar field is
\begin{equation}
Q_{\mu} \equiv  D_{\mu} \phi \,,
\end{equation}
its momentum is
\begin{equation}
\Pi = {\cal L}_n \phi = n^{\mu}\nabla_{\mu}\phi 
\end{equation}
and
\begin{equation}
K_{ij} = -\nabla_i n_j =  -\frac{1}{2N}\left(\frac{\partial 
h_{ij}}{\partial t}  +  D_i N_j 
+ D_j N_i \right) \,,
\end{equation}
\begin{equation} 
\Pi = \frac{1}{N}\left(\partial_t\phi+N^{\alpha} 
Q_{\alpha}\right) \,,
\end{equation}
while 
\begin{equation}
\partial_t Q_i +N^l\partial_lQ_i+Q_l\partial_iN^l 
= D_i\left(N\Pi\right)   \,.
\end{equation}
The ADM  decomposition of the 
effective energy-momentum  tensor 
$T_{\mu\nu}^{(eff)}$ is then  
\begin{equation}
T_{\mu\nu}^{(eff)} = \frac{1}{f}\left( 
S_{\mu\nu}+J_{\mu} n_{\nu} +J_{\nu} n_{\mu}+ E n_{\mu} n_{\nu} 
\right) \,,
\end{equation}
where
\begin{equation} \label{Cauchy41}
S_{\mu\nu} \equiv {h_{\mu}}^{\rho} {h_{\nu}}^{\sigma} 
T_{\rho\sigma}^{(eff)} = 
\frac{1}{f}\left( S_{\mu\nu}^{(f)}+S_{\mu\nu}^{(\phi)} 
+S_{\mu\nu}^{(m)}\right) \,,
\end{equation}
\begin{equation}
\label{Cauchy42}
J_{\mu} \equiv -{h_{\mu}}^{\rho} T_{\rho\sigma}^{(eff)} 
n^{\sigma}= \frac{1}{f}\left(J_{\mu}^{(f)}+J_{\mu}^{(
\phi)}+J_{\mu}^{(m)}\right) \,,
\end{equation}
\begin{equation}\label{Cauchy43}
E \equiv n^{\mu} n^{\nu} T_{\mu\nu}^{(eff)} = 
\frac{1}{f}\left(E^{(f)}+E^{(\phi)}+E^{(m)}\right) \,,
\end{equation}
and $T^{(eff)} = S-E$, where $T^{(eff)} 
\equiv {{T^{(eff)}}^{\mu}}_{\mu}$ and $ S \equiv 
{S^{\mu}}_{\mu}$.
Using  the Gauss-Codazzi equations \cite{Wald84},   the effective 
Einstein equations  projected tangentially and orthogonally to 
$\Sigma_t$ yield the Hamiltonian 
constraint  \cite{Wald84, 
Salgado06}
\begin{equation}
^{(3)}R + K^2 - K_{ij}K^{ij} = 2E \,,
\end{equation}
the momentum constraint 
\begin{equation}
\label{Cauchy45}
D_l {K^l}_i - D_i K = J_i \,,
\end{equation}
and the dynamical equations
\begin{eqnarray}
\label{Cauchy46}
&&  \partial_t {K^i}_j + N^l \partial_l {K^i}_j + {K^i}_l 
\partial_j N^l - 
{K^l}_j 
\partial_l N^i + D^i D_j N \nonumber \\
&&\nonumber \\
&&- ^{(3)}{R^i}_j N - NK{K^i}_j = \frac{N}{2}\left[ \left(S-E\right) 
\delta^i_j -2S^i_j 
\right] \,,  \label{Cauchyquesta}
\end{eqnarray}
where $K \equiv {K^i}_i$. The trace of eq.~(\ref{Cauchyquesta})  
leads to 
\begin{equation}
\partial_t  K + N^l \partial_l K + ^{(3)}\Delta N - NK_{ij}K^{ij} 
= \frac{N}{2} \left( S + 
E \right) \,,
\end{equation}
where $^{(3)}\Delta \equiv D^i D_i$. The second order derivatives  
of $\phi$ are in principle troublesome because they  could make 
the  initial value  problem   
ill-formulated, but they  can be 
eliminated in most cases \cite{Salgado06}. The $f$- and 
$\phi$-quantities  of 
eqs.~(\ref{Cauchy41})--(\ref{Cauchy43}) turn out to be 
\begin{equation}
E^{(f)} = f^\prime\left( D^{\alpha} Q_{\alpha} + K\Pi \right) + 
f^{\prime\prime}Q^2\,,
\end{equation}
\begin{equation}
J_{\mu}^{(f)} = -f^\prime\left( K_{\mu}^{\rho} Q_{\rho} 
+D_{\mu} \Pi \right) - f^{\prime\prime
}\Pi Q_{\mu} \,,
\end{equation}
\begin{equation}
S_{\mu\nu}^{(f)} = f^\prime\left( D_{\mu} Q_{\nu} + \Pi 
K_{\mu\nu} - h_{\mu\nu}\Box\phi  \right) - f^{\prime\prime} 
\left[ h_{\mu\nu}\left( Q^2 - \Pi^2 \right) - Q_{\mu} Q_{\nu} 
\right] \,,
\end{equation}
where $Q^2 \equiv Q^{\alpha} Q_{\alpha}$. The quantities 
\begin{equation}
S^{(f)} = f^\prime \left( D_{\alpha} Q^{\alpha}  + K\Pi - 
3\Box\phi \right) 
+  f^{\prime\prime}\left(  3\Pi^2 - 2Q^2 \right) \,,
\end{equation}
\begin{equation}
S^{(f)}-E^{(f)} = -3f^\prime\Box\phi - 
3f^{\prime\prime}\left( Q^2 -   \Pi^2 \right) \,,
\end{equation}
are also useful \cite{Salgado06}, and the introduction of  
$\omega$ 
and $\omega^\prime$ yields the further quantities
\begin{eqnarray}
E^{(\phi)} & = &  \frac{\omega}{2}\left( \Pi^2 + Q^2 \right) + 
V(\phi)  \,,\\
&&\nonumber\\
J_{\mu} ^{(\phi)} & = &  -\omega\Pi Q_{\mu} \,,
\end{eqnarray}
\begin{equation}
S_{\mu\nu}^{(\phi)} = \omega Q_{\mu} Q_{\nu} - h_{\mu\nu}\left[ 
\frac{\omega}{2}\left(  Q^2 - \Pi^2 \right) + V(\phi) \right] \,,
\end{equation}
while
\begin{equation}
S^{(\phi)} = \frac{\omega}{2}\left( 3\Pi^2 - Q^2 \right) - 
3V(\phi) 
\end{equation}
and
\begin{equation}
S^{(\phi)}-E^{(\phi)} = \omega\left( \Pi^2 - Q^2 \right) - 
4V(\phi) \,.
\end{equation}
Finally, the quantities appearing  on the right hand sides  of 
the $3+1$ field  equations are 
\begin{eqnarray}
E & = &  \frac{1}{f} \left[ f^\prime\left( D^{\alpha} Q_{\alpha} 
+  K\Pi \right) +  \left( 
f^{\prime\prime}  +  \frac{\omega}{2}\right)Q^2 + 
\frac{\omega}{2}\Pi^2 + V(\phi) + 
 E^{(m)} \right] \,, \\
&&\nonumber\\
J_{\mu}  &=& \frac{1}{f}\left[ -f^\prime\left( {K_{\mu}}^{\alpha} 
Q_{\alpha} + D_{\mu} \Pi \right) -  
\left(  f^{\prime\prime} + \omega \right) \Pi Q_{\mu} + 
J_{\mu}^{(m)} \right] \,,\\
&&\nonumber\\
S_{\mu\nu} & = &  \frac{1}{f}\left\{ f^\prime \left( D_{\mu} 
Q_{\nu} + \Pi K_{\mu\nu} \right) 
 -h_{\mu\nu}\left[ \left( f^{\prime\prime}+\frac{\omega}{2} 
\right) \left( Q^2 - 
\Pi^2 \right) + V(\phi)  + f^\prime\Box\phi \right]\right\} 
\nonumber \\
&&\nonumber \\
& +& \frac{1}{f}\left[ \left( \omega + f^{\prime\prime} \right)  
Q_{\mu} Q_{\nu} + S_{\mu\nu}^{(m)} \right] \,,
\end{eqnarray}
while
\begin{eqnarray} 
 S & = & \frac{1}{f}\left[ f^{\prime} \left( D^{\alpha} 
Q_{\alpha} 
+ \Pi K  \right) - 3V(\phi) - 
3f^\prime\Box\phi \right] \nonumber \\
&&\nonumber \\
&- & \frac{1}{f}\left[ \left( 2f^{\prime\prime}  + 
\frac{\omega}{2}\right)Q^2
 - 3\left( f^{\prime\prime} + \frac{\omega}{2} \right)\Pi^2 + 
S^{(m)}\right] 
\,,\\
&&\nonumber\\
 S - E &=&  \frac{1}{f}\left[ \left(3f^{\prime\prime}+\omega\right) 
\left(\Pi^2-Q^2\right) - 4V(\phi) - 3f^\prime\Box\phi + 
S^{(m)} - E^{(m)}\right] \,,\nonumber\\
&& \\
 S + E &=& \frac{1}{f}\left[ 2f^\prime\left( D^{\alpha} 
Q_{\alpha} + K\Pi \right) 
-f^{\prime \prime}Q^2 
+ \left( 3f^{\prime\prime} + 2\omega \right)\Pi^2 \right] 
\nonumber \\
&&\nonumber \\
& +& \frac{1}{f}\left( - 2V(\phi)  - 3f^\prime\Box\phi + 
S^{(m)}+ E^{(m)}\right) 
\,.
\end{eqnarray}
The Hamiltonian and the momentum constraints  assume the form
\begin{eqnarray}
 ^{(3)}R &+& K^2 - K_{ij}K^{ij}  - \frac{2}{f}\left[ 
f^\prime\left( D_{\alpha} Q^{\alpha} +   K\Pi \right) + 
\frac{\omega}{2}\Pi^2 + \frac{Q^2}{2}\left( \omega + 2f^{
\prime\prime} \right) \right] \nonumber \\
&&\nonumber \\
& =& \frac{2}{f}\left( E^{(m)}+V(\phi) \right) 
\,,\label{Cauchy64}\\
&&\nonumber\\ 
D_l {K^l}_i &-& D_i K + \frac{1}{f}\left[ f^\prime \left( 
{K_i}^{\alpha} Q_{\alpha} + D_i\Pi 
\right) + \left( \omega + f^{\prime\prime} \right) \Pi Q_i 
\right] =  \frac{J_i^{(m)}}{f} \,,\label{Cauchy65} 
\end{eqnarray}
respectively, and  the dynamical equation~(\ref{Cauchy46}) is 
written as
\begin{eqnarray}
&& \partial_t {K^i}_j + N^l \partial_l {K^i}_j + {K^i}_l 
\partial_j N^l - {K_j}^l  \partial_l N^i 
+ D^i D_j N - ^{(3)}{R^i}_j N - NK{K^i}_j \nonumber \\
&&\nonumber \\
&& + \frac{N}{2f}\left[ f^{\prime\prime}\left( Q^2 - \Pi^2 
\right) + 2V(\phi) +  f^\prime\Box \phi  \right] 
\delta^i_j + \frac{Nf^\prime}{f}\left( D^i Q_j 
+ \Pi {K^i}_j \right) \nonumber \\
&&\nonumber \\
&&+ \frac{N}{f}\left( \omega + f^{\prime\prime} \right) Q^i Q_j 
= \frac{N}{2f} \left[\left( S^{(m)} - E^{(m)} \right) \delta^i_j 
- 2{S^{(m) \,\, i}}_j \right] \,.
\end{eqnarray}
The  trace of this equation is 
\begin{eqnarray}
\label{Cauchy67} 
&& \partial_t K + N^l \partial_l K + 
^{(3)}\Delta N - NK_{ij}K^{ij} - \frac{
Nf^\prime}{f}\left( D^{\alpha} Q_{\alpha} + \Pi K \right) 
\nonumber \\
&&\nonumber \\
&& + \frac{N}{2f}\left[ f^{\prime\prime}Q^2 - \left( 2\omega + 
3f^{\prime\prime}  
\right)\Pi^2 \right] = \frac{N}{2f}\left( -2V(\phi)  
-3f^\prime\Box \phi + 
S^{(m)} +  E^{(m)} \right)  \nonumber\\
&& 
\end{eqnarray}
where \cite{Salgado06}
\begin{eqnarray}
\label{Cauchy68}
&& {\cal L}_n\Pi - \Pi K - Q^{\alpha} D_{\alpha} \left( \ln N 
\right) - D_{\alpha} Q^{\alpha}  = -\Box \phi
 \nonumber \\
&& =  -\frac{1}{f\left[ \omega + \frac{3{(f^\prime)}^2}{2f} 
\right]}\left(  \frac{f^\prime T^{(m)}}{2} -2f^\prime 
V(\phi) + fV^\prime (\phi) \right) \nonumber \\
&&\nonumber \\
&& -\frac{1}{f \left[ \omega + \frac{3{(f^\prime)}^2}{2f} 
\right]}\left\{ \left[
\frac{-\omega^\prime f}{2} - \left( \omega + 3f^{\prime\prime} 
\right)
\frac{f^\prime}{2} \right] \nabla^{\alpha} 
\phi\nabla_{\alpha} \phi \right\} \,.
\end{eqnarray}
The  initial data {\em in vacuo} $(h_{ij}, K_{ij}, \phi, Q_i, 
\Pi)$ on an initial  hypersurface $\Sigma_0$ must satisfy the  
constraints~(\ref{Cauchy64}) and (\ref{Cauchy65}), in addition to  
\begin{equation}
Q_i - D_i\phi = 0 \,,
\end{equation}
\begin{equation}
D_i Q_j = D_j Q_i \,. 
\end{equation}
When  matter is present, the additional variables $E^{(m)}$, 
$J_{\mu}^{(m)}$, $S_{\mu\nu}^{(m)}$ must be assigned  on the  
initial hypersurface.  Prescribing  lapse  $N$ and shift 
$N^{\mu}$ is equivalent to fixing a 
gauge\footnote{Various gauge 
conditions employed in the literature are surveyed 
in~\cite{Salgado06,dyer}.}.  The differential system 
(\ref{Cauchy64})--(\ref{Cauchy67}) contains only  first-order  
derivatives  in both space and time once the d'Alembertian 
$\Box \phi$ is written in terms of 
$\phi, \nabla^{\mu} \phi\nabla_{\mu} \phi$, $f$,  and its 
derivatives by 
means of  Eq.~(\ref{Cauchy68}) \cite{Salgado06, 
TremblayFaraoni07, olmoalepuz2}.  
Following \cite{Salgado06}, the reduction to  a first-order 
system  
indicates that the Cauchy problem is  well-posed {\em in vacuo} 
and  well-formulated  in the presence of matter. 

\subsection{The initial value problem of $f(R)$-gravity in the ADM formulation}

 In order to deal with the Initial Value Problem for $f(R)$-gravity, let us consider the analogy of such a theory with scalar tensor gravity. Remembering the results of the previous subsection, metric $f(R)$-gravity models  
 are equivalent to Brans-Dicke gravity with $\omega = 0$,  while Palatini  $f(R)$-gravity models  are equivalent to Brans-Dicke 
gravity with $\omega = -3/2$. Following the discussion in  \cite{olmoalepuz2}, the Hamiltonian formulation of Brans-Dicke theories with a non-trivial potential can be carried out paying special attention to the case $\omega=-3/2$.  The scalar field, in this case,  presents a degenerate momentum which is proportional to a linear combination of the momenta of the induced metric $h_{\mu\nu}$. This degeneracy requires, following Dirac's algorithm for constrained systems, the introduction of a new constraint in the Hamiltonian. Consistency of the evolution of that constraint leads to a secondary constraint which establishes an algebraic relation between the scalar field and the trace of the energy-momentum tensor of the matter. In \cite{olmoalepuz2}, the constraint and evolution equations are written in a way that allows a comparison between the general case $\omega\neq -3/2$ and $\omega=-3/2$. The resulting constraint and evolution equations of the case $\omega=-3/2$ do not contain any higher-order time derivative of the matter fields, and only spatial derivatives of the scalar field appear up to the second-order. This implies that the spatial profiles of the matter sources must satisfy stronger differentiability requirements than in GR.  This property can be interpreted as a requirement due to the existence of a conformal geometry related with the matter fields.  By comparing the constraint and evolution equations of the theory with those of GR, the initial value Cauchy problem is well-formulated because of the intrinsic geometry of space-time is determined uniquely by an initial choice of $h_{\mu\nu}$ and  $\Pi^{\mu\nu}$, plus the corresponding positions and momenta of  matter on the initial Cauchy surface. This result indicates that also the well-posedness of the problem is possible in contrast with the results in \cite{TremblayFaraoni07}.

\subsection{The Gaussian normal coordinates approach in General Relativity}

A different approach to the Initial Value Problem uses 
Gaussian normal coordinates
(also called {\em synchronous coordinates}) instead of the ADM 
decomposition.  Before discussing ETGs, we recall the initial 
value formulation of GR in these coordinates, which  is  
well-formulated and   well-posed,  as shown in \cite{Wald84}. We 
adopt the formalism  developed  in Ref.~\cite{Synge55}.

Let us consider a system of 
Gaussian normal coordinates
\cite{Wald84}, in which the metric tensor has components  
$g_{00}=-1$ and $g_{0i}=0$. These coordinates serve the purpose 
of  splitting the space-time manifold ${\cal M}$ into a spatial 
hypersurface $ \Sigma_3 $ of constant time from the orthogonal 
time direction.

Given a second rank symmetric tensor   $W_{\mu\nu}$ 
on the globally hyperbolic space-time  $ \left( {\cal M}, g_{\mu\nu} 
\right) $, we define its  (symmetric) conjugate 
tensor  
\be
 W^*_{\mu\nu}=W_{\mu\nu} -\frac{W}{2} \, g_{\mu\nu} \,,
\ee
where $|$ denotes the covariant derivative with respect to the
Levi-Civita connection induced by $g_{\mu\nu}$ and 
$W \equiv W^{\mu\nu} g_{\mu\nu}$ is
the trace of $W_{\mu\nu}$. If $V_0$ is a space-time
domain in $ {\cal M} $ in which  $g_{00}\not =0 $ and  $ \Sigma_3$ is 
the three-surface of  equation $x^0=0$, then the following  
statements are equivalent:
\begin{enumerate}
    \item $W_{\mu\nu}=0$ in $ V_0 $;
    \item $W^*_{ij}=0$ and $W_{0\alpha}=0$ in $ V_0$;
    \item $W^*_{ij}=0$ and $W^{\mu}_{\nu | \mu}=0$ in $ V_0 $  
with $W_{o\mu}=0$ in $ \Sigma_3 $.
\end{enumerate}
Let us consider the Einstein equations ${\displaystyle 
G_{\mu\nu}= \kappa T_{\mu\nu}^{(m)} }$ 
and  the contracted Bianchi  identities $ 
\nabla^{\nu} T_{\mu\nu}^{(m)} =0 $; introducing  the tensor
\begin{equation}\label{A.2}
W_{\mu\nu} \equiv G_{\mu\nu} - \kappa \, T_{\mu\nu}^{(m)} \,,
\end{equation}
the conjugate tensor is
\begin{equation}\label{A.3}
W^*_{\mu\nu }=R_{\mu\nu} - \kappa \, T^*_{\mu\nu}\,,
\end{equation}
and the Einstein equations are 
\begin{equation}\label{A.4}
W_{\mu\nu}=0\,.
\end{equation}
These are 10 equations  for the  20 unknown functions 
$g_{\mu\nu}$ and $ T_{\mu\nu}^{(m)}$. We assign  the 10 functions 
$g_{0\mu}$ and $T_{ij}^{(m)}$; the remaining 10 functions 
$g_{ij}$ and $ T_{0\mu}^{(m)}$ are determined by Eq.~(\ref{A.4}). 
These functions can be expressed in the  equivalent form
\begin{equation}\label{A.5}
R_{ij} - \kappa \, T^*_{ij}= 0\,, \qquad\qquad
W^{\mu} _{\nu | \mu} = T^{\mu} _{\nu | \mu}=0\,,
\end{equation}
with the  condition
\begin{equation}\label{A.6}
G_{0\mu }- \kappa \, T_{0\mu}^{(m)}= 0 
\end{equation}
on the hypersurface $ x^0 =0 $. Eqs.~(\ref{A.5}) can be 
rewritten 
as 
\begin{eqnarray}
g_{ij ,00}=  2  \, \bar{R}_{ij} -
\frac{A}{2} \, g_{ij, 0} +
g^{lm} g_{ i l, 0}g_{j m, 0} + 2\kappa \,  
T^*_{ij}\,, \label{A.7a}\\
&&\nonumber\\
T_{0\nu ,0}^{(m)}= - {T^{(m)}}^0_{\nu, 0} = {T^{(m)}}^{i}_{\nu, 
i} + \Gamma_{i \mu }^{\;\;\; i} {T^{(m)}}^{\mu}_{\nu} -
\Gamma_{i \nu}^{\;\;\;\mu} {T^{(m)}}^{i}_{\nu} \,,\label{A.7b}
\end{eqnarray}
where $\bar{R}_{ij}$ is the intrinsic  Ricci tensor
of the hypersurface $x^0 = 0 $, $\Gamma_{\mu\nu}^{\;\;\; \rho}$ 
is the
Levi-Civita connection of the metric $g_{\mu\nu}$, and
\begin{equation}\label{A.8}
A \equiv g^{ij} g_{ij, 0}\,.
\end{equation}
In the same way, the constraint equation~(\ref{A.6}) becomes
\begin{eqnarray}
A_{, i} - D^{j}g_{ij, 0} + 2 \kappa \, T_{0i}^{(m)} = 0\,,
\label{A.9a}\\
&&\nonumber\\
\tilde{R} - \frac{A^2}{4} + \frac{B }{4} + 
2 \kappa \, T_{00}=0\,,\label{A.9b}
\end{eqnarray}
where $\bar{R}$ is the intrinsic Ricci scalar  of  the 
hypersurface $x^0 = 0 $, $D_i$ denotes the covariant derivative 
operator  on  this hypersurface associated with the Levi-Civita 
connection of the intrinsic metric $g_{ij} 
\left.\right|_{\Sigma_0}$ and
\begin{equation}\label{A.10}
B = g^{ij} g^{lm} g_{i l, 0}g_{i m, 0}\,.
\end{equation}
Let us assign now the  Cauchy data 
\begin{equation}\label{A.11}
g_{ij}, \quad g_{ij, 0 }, \quad T_{\mu 0}^{(m)} 
\end{equation}
on the hypersurface $x^0 =0$; they must satisfy the 
constraint equations~(\ref{A.9a}), (\ref{A.9b}),  (\ref{A.7a}), 
and eq.~(\ref{A.7b}) gives  the quantities
\begin{equation}\label{A.12}
g_{ij , 00}, \qquad T^{(m)}_{0\mu , 0}\,,
\end{equation}
as functions of the Cauchy data. By  differentiating 
Eqs.~(\ref{A.7a}) and (\ref{A.7b}),  it is straightforward to 
obtain 
time derivatives of higher order as functions of the initial  
data. This procedure allows one to locally reconstruct the 
solution of the field equations as a power series of  $x^0 $. The 
initial three-surface $ \Sigma_3$  is  then a Cauchy  
hypersurface for the globally hyperbolic space-time 
$ \left( {\cal M}, g_{\mu\nu} \right) $ and  the initial value  problem  
is 
well-formulated in GR. Our goal is now to  extend these results 
to $f(R)$-gravity in the metric-affine formalism.

\subsection{The Cauchy problem of $f(R)$-gravity in Gaussian normal coordinates}
\subsubsection{The vacuum case}

In the metric-affine formulation of $f( R)$-gravity the 
independent variables are  $ \left( 
g_{\mu\nu}, \Gamma_{\mu\nu}^{\alpha} \right) $, where
$g_{\mu\nu} $ is the metric and  $\Gamma_{\mu\nu}^{\alpha} $ is 
the linear connection.  {\em In vacuo}, the field equations are 
obtained by varying  the  action
\begin{equation}\label{2.0}
S \left[ g, \Gamma \right] = \int d^4x \, \sqrt{-g} \, 
f( {\cal R}) 
\end{equation}
with respect to the metric and  the connection, 
where  $ {\cal R} \left( g, \Gamma \right) = g^{\mu\nu} 
{\cal R}_{\mu\nu} $ is the scalar curvature of the  connection 
$\Gamma_{\mu\nu}^{\alpha}$ and $ {\cal R}_{\mu\nu}  $ is the 
Ricci tensor constructed with this connection. The  metric 
connection $\Gamma_{\mu\nu}^{\alpha}$  can have a non-vanishing 
torsion while, in the Palatini approach,   
$\Gamma_{\mu\nu}^{\alpha}$ is a non-metric but torsion-free 
connection  \cite{CCSV1}.

{\em In vacuo}, the field equations of $f(R)$-gravity with 
torsion are \cite{CCSV1,CCSV2,CCSV3}
\begin{eqnarray}
&& f'( {\cal R}) \, {\cal R}_{\mu\nu} - \frac{f({\cal R})}{2}  \, 
g_{\mu\nu}  = 0\,,\label{2.1a}\\
&&\nonumber\\
&& T_{\mu\nu}^{\;\;\;\alpha} = -
\frac{1}{2f'} \, \frac{\partial f'}{\partial x^{\rho}} \left( 
\delta^{\rho}_{\mu} \delta^{\sigma}_{\nu} -
\delta^{\rho} _{\nu} \delta^{\sigma}_{\mu}\right)\,, 
\label{2.1b}
\end{eqnarray}
while the field equations of $f(R)$-gravity  {\it \`{a} 
la} Palatini are  \cite{palatini,  francaviglia2,Olmo} 
\begin{eqnarray}
&& f'( {\cal R}) \, {\cal R}_{\mu\nu} - \frac{f({\cal R})}{2}  \, 
g_{\mu\nu}  = 0\,,\label{2.2a} \\
&&\nonumber\\
&& \nabla_{\mu}  \left[ f'( {\cal R}) g_{\mu\nu} \right] = 0 
\,.\label{2.2b}
\end{eqnarray}
In both cases, the trace of the field equations~(\ref{2.1a}) 
and~(\ref{2.2a}) yields
\begin{equation}\label{2.3mm}
f'( {\cal R}) \, {\cal R}  - 2 f( {\cal R})=0\,.
\end{equation}
When this equation admits solutions, the  scalar curvature ${\cal 
R} $ is  a  constant; then eqs.~(\ref{2.1b}) and~(\ref{2.2b}) 
imply that 
both  connections coincide with the Levi-Civita connection  
of the metric $g_{\mu\nu}$ which solves the 
field equations and both theories reduce to GR with a 
cosmological constant, for which the  Cauchy problem  is 
well-formulated and well-posed  \cite{Wald84}.

\subsubsection{The case with matter} 

Let us allow now a perfect fluid and study the 
Cauchy problem of $f(R)$-gravity. We discuss simultaneously 
the Palatini approach and a non-vanishing torsion,  
but we assume 
that the matter Lagrangian does not couple explicitly to  the  
connection. Then the field equations  are
\begin{eqnarray}
&& f'( {\cal R}) \, {\cal R}_{\mu\nu} - \frac{f({\cal R})}{2}  \, 
g_{\mu\nu  }=T^{(m)}_{\mu\nu}\,,  \label{3.1a}\\
&&\nonumber\\
&& T_{\mu\nu}^{\;\;\;\alpha} = - 
\frac{1}{2f'({\cal R})} 
\,  \frac{\partial f'( {\cal R})}{ \partial x^{\rho} } 
\left(\delta^{\rho}_{\mu} 
\delta^{\sigma}_{\nu} -  
\delta^{\rho}_{\nu} \delta^{\alpha}_{\mu}\right) \label{3.1b}
\end{eqnarray}
in the case of $f(R)$-gravity with 
torsion, and
\begin{eqnarray}
&& f'( {\cal R}) \, {\cal R}_{\mu\nu} - \frac{f({\cal R})}{2} \, 
g_{\mu\nu}=T^{(m)}_{\mu\nu}\,, \label{3.2a} \\
&&\nonumber\\
&& \nabla_{\alpha} \left[ f'({\cal R}) g_{\mu\nu} \right] 
=0\,,\label{3.2b} 
\end{eqnarray}
for Palatini  $f(R)$-gravity, where $ {\displaystyle 
T^{(m)}_{\mu\nu} \equiv  -
\frac{2}{\sqrt{-g } } \, \frac{\delta\left( \sqrt{-g} \,  {\cal 
L}^{(m)} \right) }{ \delta  g^{\mu\nu} } }$ is the matter 
energy-momentum tensor. The trace of 
eqs.~(\ref{3.1a}) and~(\ref{3.2a}) yields the relation between   
$ {\cal R} $ and  $ T^{(m)} \equiv g^{\mu\nu} T^{(m)}_{\mu\nu}$ 
\begin{equation}\label{3.3dd}
f'({\cal R}) \, {\cal R} -2f({\cal R}) = T^{(m)}\,.
\end{equation}
When  $ T^{(m)} =$~const. the  theory reduces to GR with a 
cosmological constant and the initial value problem is identical 
to the vacuum case. Assuming that the relation~(\ref{3.3dd}) is 
invertible and $ T^{(m)} \neq $~const., the Ricci scalar   can 
be expressed as a function of  $ T^{(m)} $ 
\begin{equation}\label{3.4ee}
{\cal R} =F \left( T^{(m)} \right) \,.
\end{equation}
It is then  easy to show that the field equations of both 
the Palatini and the metric-affine theory with 
torsion can be 
expressed  in the form \cite{CCSV1,CCSV2, Olmo} 
\begin{eqnarray}
{\cal R}_{\mu\nu} -\frac{1}{2} \, g_{\mu\nu} \, {\cal R} &  =&  
\frac{1}{\varphi} \, T^{(m)}_{\mu\nu}
+ \frac{1}{\varphi^2} \left[ -  
\frac{3}{2} \,  \frac{ \partial \varphi}{\partial x^{\mu} }  \, 
\frac{\partial \varphi}{\partial x^{\nu} }
+ \varphi\tilde{\nabla}_{\nu} \frac{ 
\partial \varphi}{ \partial x^{\mu} } \right. \nonumber\\
&&\nonumber\\
&+&  \left.  \frac{3}{4} \, 
\frac{\partial \varphi}{\partial x^{\alpha}} 
\, \frac{ \partial \varphi}{ \partial  x^{\beta}} 
\, g^{\alpha\beta} g_{\mu\nu} -
\varphi\tilde{\nabla}^{\alpha} 
\frac{\partial \varphi}{ \partial x^{\alpha}} \, g_{\mu\nu} -
V \left( \varphi \right) g_{\mu\nu} \right]\,, \label{3.53}
\end{eqnarray}
where 
\begin{equation}\label{3.61}
V \left( \varphi \right ) \equiv  \frac{1}{4} \left[ \varphi
F^{-1} \left( \left( f' \right )^{-1} \left( \varphi \right) 
\right) + \varphi^2 \left( f' \right)^{-1} 
\left( \varphi \right) \right]\,,
\end{equation}
is an effective potential for the scalar field
\begin{equation}\label{3.7}
\varphi \equiv f' \left( F\/\left( T^{(m)} \right) \right)\,.
\end{equation} 
By performing  the conformal transformation
$ g_{\mu\nu} 
\longrightarrow \tilde{g}_{\mu\nu}=\varphi \, g_{\mu\nu}$, 
Eq.~(\ref{3.53}) assumes 
the  simpler form \cite{CCSV1, Olmo,German}
\begin{equation}\label{3.8}
\tilde{R}_{\mu\nu} - \frac{1}{2} \,  \, \tilde{g}_{\mu\nu} \, 
\tilde{R}  =
\frac{1}{\varphi} \, T^{(m)}_{\mu\nu} -
\frac{1}{\varphi^3} \, V \left( \varphi \right)  
\tilde{g}_{\mu\nu}\,,
\end{equation}
where $\tilde{R}_{\mu\nu} $ and $\tilde{R} $ are  the 
Ricci tensor and the Ricci scalar of  the conformal 
metric $\tilde{g}_{\mu\nu}$, respectively. 

The connection ${\Gamma_{\mu\nu}}^{\alpha}$, solution of 
the field equations with 
torsion, is 
\begin{equation}\label{3.9}
\Gamma_{\mu\nu}^{\;\;\;\alpha} 
=\tilde{\Gamma}_{\mu\nu}^{\;\;\;\alpha} +
\frac{1}{2\varphi} \, 
\frac{\partial \varphi}{ \partial x^{\nu} } 
\, \delta^{\alpha}_{\mu} - \frac{1}{2\varphi} \, 
\frac{\partial 
\varphi}{\partial x^{\rho}} \, g^{\rho\alpha}g_{\mu\nu}\,,
\end{equation}
where $\tilde{\Gamma}_{\mu\nu}^{\;\;\;\alpha}$ is the 
Levi-Civita connection  of the metric $g_{\mu\nu}$ while  
${ \tilde{\Gamma}_{\mu\nu}}^{\alpha}$,   solution of the 
Palatini field  equations, coincides with the Levi-Civita 
connection of the 
conformal metric  $\tilde{g}_{\mu\nu}$.   $ 
{\Gamma_{\mu\nu}}^{\alpha} $ 
and $ { \tilde{\Gamma}_{\mu\nu}}^{\alpha} $ satisfy the relation
\begin{equation}\label{3.10}
\tilde{\Gamma}_{\mu\nu}^{\;\;\;\alpha} 
=\Gamma_{\mu\nu}^{\;\;\;\alpha} +
\frac{1}{2\varphi} \, \frac{\partial \varphi}{ \partial x^{\mu} 
} \, \delta^{\alpha}_{\nu} 
\end{equation}
and the Levi-Civita connections induced by the metrics 
$g_{\mu\nu}$ and  $\tilde{g}_{\mu\nu}$ are related by the  
identity
\begin{equation}\label{3.11}
\tilde{\Gamma}_{\mu\nu}^{\;\;\;\alpha}= 
\Gamma_{\mu\nu}^{\;\;\;\alpha} +
\frac{1}{2\varphi} \, 
\frac{ \partial \varphi}{ \partial x^{\nu}} 
\, \delta^{\alpha}_{\mu} - \frac{1}{2\varphi} \, 
\frac{\partial \varphi}{ 
\partial x^{\rho}} \, g^{\rho\alpha}g_{\mu\nu} +
\frac{1}{2\varphi} \, \frac{ \partial \varphi}{\partial x^{\mu} } 
\, \delta^{\alpha}_{\nu}\,.
\end{equation}
The field equations~(\ref{3.53}) have to be
considered together with the matter field equations and it must 
be kept  in mind that the conservation equations
for both the metric-affine theories (with torsion and {\it \`{a}
la} Palatini) coincide with the standard conservation laws of
GR \cite{CapozzielloVignolo3136}
\begin{equation}
\label{3.12} 
\tilde{\nabla}_{\nu} T^{\mu\nu}=0\,.
\end{equation}
It is straightforward to show that eq.~(\ref{3.12}) is
equivalent to the conservation law 
\begin{equation}\label{3.12a}
\tilde{\nabla}_{\nu} T^{\mu\nu}=0 
\end{equation}
 where 
\begin{equation}
T_{\mu\nu}=\frac{1}{\varphi} \, T^{(m)}_{\mu\nu} -
\frac{1}{\varphi^3} \, V \left( \varphi \right) 
\tilde{g}_{\mu\nu} 
\,,
\end{equation}
for the conformally transformed theories~(\ref{3.8}). In fact, 
by an explicit calculation of the divergence  
$\tilde{\nabla}_{\nu} T^{\mu\nu}$ where the 
relations~(\ref{3.11}) have been used, we obtain 
\begin{equation}\label{3.12b}
\tilde{\nabla}^{\nu} T_{\mu\nu}= \frac{1}{\varphi^2} 
\, \tilde\nabla^{\nu} T^{(m)}_{\mu\nu}
+ \frac{1}{\varphi^3} \, \frac{\partial 
\varphi }{ \partial x^{\mu} } \left[ -\frac{ 
T^{(m)} }{2} + \frac{3 V(\varphi)}{\varphi} - 
V'(\varphi)\right] \,.
\end{equation}
The constraint equations~(\ref{3.12}) and~(\ref{3.12a}) are then
mathematically equivalent in view  of the  relation 
\begin{equation}\label{3.12c}
T^{(m)} -\frac{6 V(\varphi) }{\varphi}   + 2V'(\varphi)=0\,,
\end{equation}
which is equivalent to the definition $\varphi=f'(F( T^{(m)} ))$
\cite{CCSV1}.

With these results in mind, the Cauchy problem  for  
Eq.~(\ref{3.53}) and the related equations of motion for matter 
can be approached  by discussing the equivalent initial value 
problem of the conformally transformed theories. Using, as in 
GR,  Gaussian normal coordinates  and beginning with  
Eqs.~(\ref{3.8}) and~(\ref{3.12a}), it is easy to conclude  that 
the Cauchy problem is well-formulated also 
in this case.

In general,  the equations of motion for matter imply the  
Levi-Civita  connection of the metric $g_{\mu\nu}$ and not 
the  connection induced from the conformal metric 
$\tilde{g}_{\mu\nu}$. Thanks to Eq.~(\ref{3.11}), this 
is not a problem since the  connection ${ 
\tilde{\Gamma}_{\mu\nu}}^{\alpha} $ can be 
expressed in terms of $ {\Gamma_{\mu\nu}}^{\alpha} $  and  the  
scalar field  $\varphi$  which, on the other hand, is a function  
of the 
matter fields. As a result, we could obtain slightly more 
complicated  equations implying further constraints on the 
initial data but,  in any case, the same equations can always 
be rewritten in "normal  form'' with respect to the maximal 
order time derivatives  of  the matter fields, determining a 
well-formulated Cauchy problem 
\cite{Wald84}.

As an example, let us examine in detail the perfect fluid case 
with barotropic equation of state $ P= P(\rho)$. The  
corresponding energy-momentum  tensor is
\begin{equation}\label{3.13}
T^{(m)}_{\mu\nu} = \left( P+ \rho \right) u_{\mu} u_{\nu} +  
P g_{\mu\nu} \,,
\end{equation}
and satisfies Eq.~(\ref{3.12}) with the normalization 
\begin{equation}\label{3.14}
g_{\mu\nu} u^{\mu} u^{\nu}  =-1\,,
\end{equation} 
of the fluid four-velocity. Eq.~(\ref{3.12})  gives
\begin{equation}\label{3.15.0}
\left(\rho + P \, u^{\nu} \right)_{|\nu}u_{\mu} + \left(\rho + 
P \right) u_{\mu| \nu} u^{\nu}
+ \frac{\partial  P}{\partial x^{\mu}} =0\,.
\end{equation}
 Contraction with $ u^{\alpha}$ yields
\begin{equation}\label{3.15a}
\left(\rho  u^{\nu} \right)_{|\nu} = -P \, u^{\nu}_{|\nu}
\end{equation}
while, substituting Eq.~(\ref{3.15a}) into Eq.~(\ref{3.15.0})
for $\alpha=1,2$, and~3, we obtain
\begin{equation}\label{3.15b}
\left(\rho + P \right) u^{\nu} u^i_{|\nu}= -\frac{\partial 
P}{ \partial x^{\nu} }\left( u^{i} u^{\nu} + g^{i \nu}\right)\,.
\end{equation}
Eqs.~(\ref{3.14}), (\ref{3.15a}), and 
(\ref{3.15b})  involve the metric $g_{\mu\nu}$ and its first 
derivatives;  using  the relation~(\ref{3.11}) we can rewrite 
them  in terms of  the conformal metric 
$\tilde{g}_{\mu\nu}$,  the scalar $\varphi=\varphi (\rho)$, and 
their  first derivatives, obtaining  
\begin{eqnarray}
&& \frac{1}{\varphi} \, \tilde{g}_{u} u^{\mu} u^{\nu} 
=-1\,,\label{3.16a}\\
&&\nonumber\\
&& \frac{\partial }{ \partial x^{\nu}} \left(\rho\/V^{\nu} 
\right) +  \tilde{\Gamma}_{\nu \sigma}^{\;\;\;\nu}\rho \, 
u^{\sigma} - \frac{2}{\varphi} \, \frac{\partial 
\varphi}{\partial x^{\sigma} } \, \rho u ^{\sigma} \nonumber\\
&&\nonumber\\
& = &   -P \left( \frac{\partial  u^{\nu}}{ \partial x^{\nu} } + 
\tilde{\Gamma}_{\nu \sigma}^{\;\;\;\nu} u^{\sigma} -
\frac{2}{\varphi}  \, \frac{\partial \varphi}{ \partial 
x^{\sigma} } u^{\sigma}  \right)\,,\label{3.16b}\\ 
&&\nonumber\\
&& \left(\rho + P \right) u^{\nu} \left[\frac{\partial u^i 
}{ \partial x^{\nu}} +
\tilde{\Gamma}_{\nu\sigma}^{\;\;\;i} u^{\sigma} + 
\frac{1}{2\varphi} 
\left(- \frac{\partial \varphi}{ 
\partial x^{\sigma} } \, \delta^i_{\nu} +
\frac{\partial \varphi}{\partial x^{\nu}} \, \delta^i_{\sigma}  - 
\frac{\partial \varphi}{ \partial x^{\rho}} \, g^{\rho 
i}g_{\nu\sigma } \right) u^{\sigma}\right] \nonumber\\
&&\nonumber\\
&  = & - \frac{\partial  P}{ \partial x^{\nu}} \left( u^{i} 
u^{\nu} 
+ g^{i  \nu}\right)\,.\label{3.16c} 
\end{eqnarray}
In Gaussian normal coordinates in which  $\tilde{g}_{00}=- 1$
(assuming $\varphi>0$) and $\tilde{g}_{0i }=0$,
 Eq.~(\ref{3.16a}) yields the expression of $ u^0$ in
terms of the remaining components $ u^i$. Eqs.~(\ref{3.16a}) and
\eqref{3.16b} can be regarded as linear equations for the
functions $ {\displaystyle  \partial u^i/ \partial  x^0 }$ and 
${\displaystyle \partial \rho/ \partial x^0 }$. The
explicit solution of these equations, in terms of the unknown
functions, could originate further constraints on the initial
data and on the form of the  function $f( {\cal R} )$. In 
Gaussian normal
coordinates, 
Eqs.~(\ref{3.16a}) and~(\ref{3.16b}) allow one  
to obtain  $\partial u^i/ \partial x^0 $ and 
$\partial \rho / \partial x^0 $ as 
functions of the initial data $\tilde{g}_{ij}$, 
$\partial {\tilde{g}_{ij}}/\partial x^0 $, $u_i $, and $\rho$ 
allowing the equations of motion of matter to be cast in normal 
form, hence   the Cauchy problem  is 
well-formulated.

Consider, as another  example, the initial value
formulation of $f(R)$-gravity coupled with Yang-Mills fields,  in
particular with the electromagnetic field. Also in this case, the
problem is well-formulated. In fact, the stress-energy tensor of a
Yang-Mills field has vanishing trace. Using Eq.~(\ref{3.3dd}), it 
is 
easy to prove that the Ricci scalar is  constant and 
then, using Eqs.~(\ref{3.1b}) and~(\ref{3.2b}) one concludes that 
the connection coincides with the Levi-Civita connection of  
$g_{\mu\nu}$. In this situation, both theories ({\it \`{a} la} 
Palatini and with torsion)  reduce to GR with a 
cosmological 
constant and  the  Cauchy problem  is 
well-formulated (this 
conclusions was already reached for the Maxwell 
field).  Moreover, 
the initial value problem is well-posed for  any theory in which 
the trace of the matter energy-momentum  tensor is constant, 
which 
is reduced to Einstein gravity with a  cosmological
constant. The Cauchy problem for  perfect 
fluid and scalar field 
sources is discussed in  \cite{CapozzielloVignolo3136, 
CapozzielloVignolo09}.

To conclude,  we have shown that the initial value 
problem  for ETGs can be at least well-formulated, passing 
another test for the viability of these theories.  Well-posedness 
is also  necessary in order to  achieve a complete control of the
dynamics but it depends on  the specific matter  fields adopted  
and the discussion becomes specific to them.

Since ETGs, like GR, are gauge theories, the choice of suitable 
coordinates may be crucial to show that the Cauchy 
problem  is formulated correctly. We have 
discussed the  two approaches using  the $3+1$  ADM 
decomposition and  
Gaussian normal coordinates, 
which can be defined when  
the covariant derivative operator $\nabla_{\mu}$ arises from a 
metric. These coordinates  are useful for calculations on a 
given non-null surface $ \Sigma_3 $, {\em i.e.},  a 
three-dimensional  embedded submanifold of the four-dimensional 
manifold ${\cal M}$. Gaussian normal 
coordinates  allow one to define 
uniquely timelike geodesics orthogonal  to $ \Sigma_3$ and to    
formulate correctly the conditions for the validity of the 
Cauchy-Kowalewski  theorem \cite{Wald84}.

In the metric-affine formalism  a given $f(R)$ theory  {\em in 
vacuo} is equivalent to GR  plus a cosmological constant, hence  
the initial value problem is well-formulated and well-posed. 
The same conclusion holds with matter sources whenever   the 
trace of the energy-momentum tensor is constant. As shown in 
\cite{CCSV1,CCSV2,CCSV3}, by introducing matter fields
in the Palatini and in the metric-affine approach with 
torsion,
one can define  $ {\cal R} =F \left(  T^{(m)} \right )$ and then the 
scalar field $ \varphi \equiv f' \left(  F \left(  T^{(m)} 
\right) \right) $, which allows one to reduce the theory to 
scalar-tensor gravity and to relate the
form of $f(R)$ to the trace of the matter energy-momentum 
tensor. In this case, it is always possible to show that the 
initial value problem is well-formulated  according to the results in \cite{olmoalepuz2}. Moreover, matter fields could induce 
further 
constraints on the Cauchy hypersurface  $ x^0=0 $ which, if 
suitably 
defined, lead to the normal form of the equations of motion for 
the matter sources. This is one of the main requirements 
for a well-formulated initial value problem. However different  
sources 
of the gravitational field, such as  perfect fluids,
Yang-Mills, and Klein-Gordon  fields, could generate different
constraints on the initial hypersurface $ \Sigma_3$. These 
constraints could also  imply restrictions on the possible form 
of $f(R)$. In conclusion,  as in GR, the choice of gauge is 
essential for a correct formulation of the initial value 
problem,  while the source  fields 
have to be  discussed carefully.

As final consideration, we can say that through the Lagrangian formulation, we have obtained the field 
equations of various theories of gravity. We have seen how the 
metric and Palatini variations 
produce different field 
equations in ETGs, contrary to what happens in GR. Conformal 
transformations have been applied to ETGs, and an overview of 
the 
Cauchy problem has been given. It is 
beginning to be clear that 
several aspects of a gravitational theory need to be taken into 
account  before the latter can be claimed to be viable. 

\newpage
\section* {{\bf Part III: Applications}}
\label{III}
In Part III of this Report we will discuss some remarkable applications which, beside cosmology and astrophysics, could constitute test-bed probes for ETGs. In particular, we will take into account the problems of spherical and axial symmetry in $f(R)$-gravity which give rise to solutions more general than the standard ones in GR. For example the Jebsen-Birkhoff theorem does not always hold and this fact could have remarcable physical consequences on self-gravitating systems \cite{newton,noi-prd}.

 Furthermore we consider the post-Newtonian limit showing that the recovering of Newtonian potential is nothing else but a particular case of GR while, in general, Yukawa-like terms come out from the weak field approximation. Such a new feature, beside addressing the problem of dark matter in galaxies and clusters of galaxies \cite{mnras2}, could give rise to new interesting phenomenology like neutrino oscillations induced by gravity.

Moreover the post-Minkowskian limit will be studied. In theories alternative to GR, it is straightforward to show that gravitational radiation presents further modes and polarizations which cannot be simply ignored if the problem of gravitational waves has to be seriously addressed.

\section{Spherical and axial symmetry}

In all areas of physics and mathematics it is common to 
search for insight into a theory by finding exact solutions of 
its fundamental equations and by studying these 
solutions in detail. This goal is particularly 
difficult in non-linear theories and the usual approach 
consists of assuming particular symmetries and searching for 
solutions with these symmetries.  Stripped of inessential 
features and simplified in this way, the search for exact 
solutions becomes easier. In a sense, this approach betrays a 
reductionist point of view but, pragmatically, it is often 
crucial to gain an understanding of 
the theory that cannot be obtained otherwise and that no 
physicist or mathematician would want to renounce to. In this 
section we discuss exact solutions of ETGs with spherical 
symmetry. Finally, the 
section ends with a discussion  from spherical to axially symmetric 
solutions. An example is given.

\subsection{Spherically symmetric solutions in $f(R)$-gravity}
The physically relevant spherically symmetric solutions of GR 
include the asymptotically flat Schwarzschild 
solution describing 
an  isolated body, which was derived in the very early days of 
GR. Other relevant solutions include the Schwarzschild-de 
Sitter (or Kottler) metric  
representing  a black hole embedded in a   de 
Sitter Universe, 
and the Lemaitre-Tolman-Bondi class of 
solutions 
describing spherical objects embedded in a dust-dominated  
cosmological  background
\cite{Lemaitre33, Tolman34, Bondi47, Stephanietal03}, to which 
one should add the McVittie metric and 
its generalizations 
\cite{McVittie33} (see \cite{Krasinskibook} for a survey of 
inhomogeneous solutions including spherically symmetric ones, 
and \cite{Stephanietal03} for exact solutions of GR). 
Spherical solutions 
representing black holes  have been 
instrumental in the 
development of black hole  mechanics and 
thermodynamics 
\cite{Wald84, WaldQFTbook}.

\subsubsection{Generalities of spherical symmetry}

The classical tests of GR pertain to the realm  
of  spherically symmetric solutions  and the weak-field limit 
\cite{Will93}. One of the fundamental properties of a 
gravitational  theory is the possibility of asymptotic flatness, 
{\em i.e.},  
space-time being Minkowskian far away 
from a localized distribution of mass-energy.  Alternative  
gravitational theories may or may not exhibit this  
physical property which allows for a  consistent comparison with  
GR. 
This point  is sometimes forgotten in the study  of 
the weak-field limit of alternative theories of gravity and  
can  be discussed in general  by considering 
the meaning of spherical solutions in ETGs 
when the standard results of GR are recovered in the limits
$r\rightarrow\infty$ and $f(R)\rightarrow R$.  Spherical 
solutions can be classified using the Ricci curvature  
$R$ as 
\begin{itemize} 

\item  solutions with $R=0$,
\item solutions with constant Ricci scalar $R=R_0\neq 0$, 
\item solutions with Ricci scalar $R(r)$ depending only on the 
radial coordinate $r$, and 
\item solutions with time-dependent $R(t,r)$.
\end{itemize}

In the first three cases the Jebsen-Birkhoff 
theorem is valid \cite{HawkingEllis}, 
meaning that  
stationary spherically symmetric solutions are necessarily  
static. However, as shown in the following, 
this theorem does not  hold for every situation in $f(R)$-gravity 
because temporal  evolution can emerge already in perturbation 
theory at some order of approximation.

A crucial role for the existence of exact spherical solutions 
is played by the relation between the metric 
potentials and by the relations  between  the latter  and 
the  Ricci scalar. The relation between the metric potentials and 
$R$ can be regarded as a constraint which assumes the form of a   
Bernoulli equation \cite{Capozziello:2007id}.  In principle,  
spherically 
symmetric solutions can be obtained
for any analytic  function $f(R)$ by solving this Bernoulli 
equation, for both the case of 
constant Ricci scalar and  
$R=R(r)$. These  spherically symmetric solutions can be used as 
backgrounds   to test how generic $f(R)$-gravity 
may deviate from GR. Theories that  imply $f(R)\rightarrow
R$ in the weak-field limit are particularly interesting. In such 
cases, the comparison with GR is straightforward 
and the experimental results evading the GR constraints can be 
framed in a  self-consistent picture
\cite{bertolami}. Finally, a perturbation approach can be 
developed to obtain  
spherical solutions at zero order, after which first order 
solutions are searched for. This scheme is iterative and
can, in principle, be extended to any order in the 
perturbations. It is crucial to consider $f(R)$ theories which
can be Taylor-expanded about a constant  value $R_0$ of the 
curvature scalar $R$.

\subsubsection{The Ricci scalar in spherical symmetry}

By imposing that the space-time metric is spherically 
symmetric, 
\be 
ds^2= - A(t,r) dt^2 + B(t,r)dr^2 + r^2d\Omega_2^2 \,,
\label{sphericalwhat?}
\ee
the Ricci  scalar  can be expressed  as 
\begin{eqnarray}
R \left( t, r \right) =  \left\{ B \left( \dot{A} 
\dot{B}-A'^2 \right) r^2 
+A\biggl[r 
\left(\dot{B}^2-A'B'\right)+2B\left(2A'+rA''-r\ddot{B}\right)\biggr] 
\right. 
\nonumber\\
\nonumber\\
\left.   -  4A^2\biggl[B^2-B+r B'\biggr] \right\}
\left( 2r^2A^2B^2 \right)^{-1}  \,,\label{riccispher}
\end{eqnarray}
where a  prime and an overdot denote differentiation  
with respect to $r$ and $t$, respectively. If the 
 metric~(\ref{sphericalwhat?}) is time-independent, {\em i.e.}, 
$ A(t,r)=a(r)$ and $B(t,r)=b(r)$, then Eq.~(\ref{riccispher}) 
assumes the simple form
\begin{eqnarray}
R(r) & = &   \left\{ a(r)\biggl[2b(r) 
\biggl(2a'(r)+ra''(r)\biggr)-ra'(r)b'(r)\biggr]-b(r)a'(r)^2r^2 
\right. \nonumber\\
&& \nonumber\\
& - & \left.  4a^2(r)\biggl(b(r)^2-b(r)+rb'(r)\biggr) 
\right\} \left( 2r^2a^2(r) b^2(r) \right)^{-1} \,.
\label{ricscalin}
\end{eqnarray}
One can see  Eq.~(\ref{ricscalin}) as a constraint on the 
functions $a(r)$ and $b(r)$ once a specific form of the Ricci
scalar is given. Eq.~(\ref{ricscalin}) reduces to the Bernoulli 
equation of index two 
\cite{Capozziello:2007id} 
\begin{equation}
b'(r)+h(r)b(r)+l(r)b^2(r)=0 
\end{equation}
for the metric component $b(r)$, {\em i.e.}, 
\begin{eqnarray}
 b'(r) & + &  \biggl\{\frac{r^2a'(r)^2 
-4a(r)^2-2ra(r)[2a(r)'+ra(r)'']}{ra(r)[4a(r)
+ra'(r)]}\biggr\}b(r) \nonumber\\
&& \nonumber\\ 
& +& \biggl\{\frac{2a(r)}{r} 
\biggl[\frac{2+r^2R(r)}{4a(r)+ra'(r)} 
\biggr]\biggr\}b(r)^2 =0 \,. \label{eqric}
\end{eqnarray}
The general solution of Eq.~(\ref{eqric}) is
\begin{equation}\label{gensol}
b(r)\,=\,\frac{\exp\left[-\int dr\,h(r) \right]}{K+\int 
dr\,l(r)\,\exp\left[-\int
dr\,h(r) \right]}\,,
\end{equation}
where $K$ is an integration constant and $h(r)$ and $l(r)$ 
are the coefficients  of the linear and quadratic  
terms in  $b(r)$, respectively. Inspection of 
this  Bernoulli equation
reveals that solutions corresponding to  $l(r)=0$ exist, which 
have a Ricci curvature scaling as ${\displaystyle
R\sim -2/r^2 } $ as spatial infinity is approached. No 
real solutions exist if  $h(r)$ vanishes identically.
The limit $r\rightarrow + \infty$ deserves special care: in order 
for the  gravitational potential $b(r)$ to have  the correct 
Minkowskian limit, both functions $h(r)$ and $l(r)$ must go  
to zero provided that the quantity $r^2R(r)$ is  
constant. This fact  implies that $b'(r)=0$, and, finally, also 
the metric potential $b(r)$ has the correct Minkowskian limit.

If asymptotic flatness of the metric is imposed,  the Ricci 
curvature must scale as $r^{-n}$ when $r\rightarrow +\infty$,    
where $n \geqslant  2$ is an integer,
\begin{equation}\label{condricc}
r^2R(r) \simeq \,r^{-n} \;\;\;\; \mbox{as} \;\;r\rightarrow 
+ \infty\,.
\end{equation}
Any other behavior of the Ricci scalar
would compromise  asymptotic flatness, as can be seen from 
Eq.~(\ref{eqric}). In fact,
let us consider  the simplest spherically symmetric case in 
which 
\begin{equation}\label
{me3}ds^2 = - a(r)dt^2 + \frac{dr^2}{a(r)} + r^2d\Omega_2^2\,.
\end{equation}
The Bernoulli equation~(\ref{eqric})
is easily  integrated 
and the most general metric potential $a(r)$ compatible with 
the constraint~(\ref{ricscalin}) is
\begin{equation}
a(r) = 1+\frac{k_1}{r}+\frac{k_2}{r^2}+\frac{1}{r^2}\int 
dr \biggl[\int r^2 R(r)dr\biggr] \,,
\end{equation}
where $k_1$ and $k_2$ are integration constants. The 
Minkowskian limit   $a(r) \rightarrow 1$ as $r\rightarrow 
\infty$ is 
obtained  only if the condition~(\ref{condricc}) is satisfied,  
otherwise the gravitational potential diverges.

\subsubsection{Spherical symmetry in $f(R)$-gravity}

Let us specialize now to metric $f(R)$ theories by considering 
an analytic function $f(R)$,  the fourth 
order field equations
\begin{equation}\label{HOEQ}
f'(R)R_{\mu\nu}-\frac{1}{2}f(R)g_{\mu\nu}-f'(R)_{;\mu\nu} 
+g_{\mu\nu}\Box
f'(R)\,=\kappa \, T_{\mu\nu}\,,
\end{equation}
and the corresponding 
trace equation
\begin{equation}\label{TrHOEQ}
3\Box f'(R)+f'(R)R-2f(R)\,=\kappa \, T\,.
\end{equation}
By rewriting Eq.~(\ref{HOEQ}) as  
\begin{eqnarray} 
G_{\mu \nu} & = & 
T^{(curv)}_{\mu \nu} + T^{(m)}_{\mu \nu}\,,
\label{eq: field}\\
&&\nonumber\\
T^{(curv)}_{\mu\nu} & = &  \frac{1}{f'(R)} \left\{ g_{\mu\nu}
\left[ f(R) - R f'(R)\right] + f'(R)^{; \rho \sigma}
\left( g_{\mu\rho} g_{\nu\sigma} - g_{\rho\sigma} g_{\mu
\nu}\right)\right\} \label{eq: curvstress}
\end{eqnarray}
matter enters Eq.~(\ref{eq: field}) through the modified
stress-energy tensor
\begin{equation}\label{eq:
mattstress} T^{(m)}_{\mu\nu} =
\frac{ \kappa \, T_{\mu\nu}}{f'(R)}\,.
\end{equation}
The most general spherically symmetric metric can be written as
\begin{equation}\label{me0}
ds^2 
= - m_1(t',r')dt'^2+m_2(t',r')dr'^2+m_3(t',r')dt'dr'+m_4(t',r')
d\Omega_2^2 \,,
\end{equation}
where  $m_i$ are functions of the radius $r'$ and of the time $t'$. 
A coordinate transformation
$t\,=\,U_1(t',r')\,, \ r\,=\,U_2(t',r')$  diagonalizes 
the metric~(\ref{me0}) and introduces the areal radius $r$ such 
that  $m_4(t',r') =r^2$, giving
\begin{equation}\label{sphericalme}
ds^2= - A(t,r) dt^2 + B(t,r)dr^2 + r^2d\Omega_2^2 \,,
\end{equation}
hence Eq.~(\ref{sphericalme}) can be taken  as the most general 
torsion-free  Lorentzian spherically 
symmetric  metric without 
loss of generality. The field equations (\ref{HOEQ}) and 
(\ref{TrHOEQ}) for this metric reduce to
\begin{equation}\label{fe4}
f'(R)R_{\mu\nu}-\frac{1}{2}f(R)g_{\mu\nu}
+\mathcal{H}_{\mu\nu}= \kappa \, T_{\mu\nu}\,,
\end{equation}
\begin{equation}\label{trfe4}
g^{\sigma\tau}H_{\sigma\tau} 
=f'(R)R-2f(R)+\mathcal{H}= \kappa \,T\,,
\end{equation}
where 
\begin{eqnarray}\label{highterms1}
 \mathcal{H}_{\mu\nu} & = & -f''(R)\biggl\{R_{,\mu\nu} 
-\Gamma^t_{\mu\nu}R_{,t}-\Gamma^r_{\mu\nu}R_{,r}-
g_{\mu\nu}\biggl[\biggl({g^{tt}}_{,t}+g^{tt}
\ln\sqrt{-g}_{,t}\biggr)R_{,t} \nonumber\\
&&\nonumber\\
& +& \biggl({g^{rr}}_{,r}+g^{rr} 
\ln\sqrt{-g}_{,r}\biggr)R_{,r}  
+g^{tt}R_{,tt}
+g^{rr}R_{,rr}\biggr]\biggr\} \nonumber\\
&&\nonumber\\
& -& f'''(R)\biggl[R_{,\mu}R_{,\nu}
-g_{\mu\nu}\biggl(g^{tt}{R_{,t}}^2+g^{rr}
{R_{,r}}^2\biggr)\biggr] \,, \label{highterms2}\\
&&\nonumber\\
 \mathcal{H} & = &  
g^{\sigma\tau}\mathcal{H}_{\sigma\tau}= 
3f''(R)\biggl[\biggl({g^{tt}}_{,t}+g^{tt}
\ln\sqrt{-g}_{,t}\biggr)R_{,t}
+\biggl({g^{rr}}_{,r}+g^{rr}\ln\sqrt{-g}_{,r}\biggr)
R_{,r} \nonumber\\
&& \nonumber\\
& + & g^{tt}R_{,tt}
+g^{rr}R_{,rr}\biggr]
+ 3f'''(R)\biggl[g^{tt}{R_{,t}}^2+g^{rr}{R_{,r}}^2\biggr]\,.
\end{eqnarray}
In these equations the derivatives of $f(R)$ with respect to $R$ 
are distinct from the time and spatial derivatives of $R$, 
a feature which will allow us to better
understand the dynamical behavior of the solutions.

\subsubsection{Solutions with constant Ricci scalar}

Let us assume that the Ricci scalar is constant, $R=R_0$. 
The field equations~(\ref{fe4}) and (\ref{trfe4}) 
with $\mathcal{H}_{\mu\nu}=0$ are 
\begin{equation}\label{fe2}
f'_0R_{\mu\nu}-\frac{1}{2} 
f_0g_{\mu\nu}= \kappa \,T_{\mu\nu}^{(m)}\,,
\end{equation}
\begin{equation}\label{fe2bis}
f'_0R_0-2f_0= \kappa \, T ^{(m)} \,,
\end{equation}
where $f_0 \equiv f(R_0) $ and $f'_0 \equiv f'(R_0)$, and they 
can be rewritten as
\begin{equation}\label{fe3}
R_{\mu\nu}+\lambda g_{\mu\nu}=q \kappa \, 
 T_{\mu\nu}^{(m)} \,,
\end{equation}
\begin{equation}\label{fe3tr}
R_0=q \kappa \, T-4\lambda 
\,,
\end{equation}
where ${\displaystyle \lambda=-\frac{f_0}{2f'_0}}$ ~and
$q^{-1}=f'_0$. We restrict to  Lagrangians
which reduce to the Hilbert-Einstein one as $R\rightarrow{0}$ 
and do not contain a  cosmological constant $\Lambda$, 
\begin{equation}
f(R)\simeq R \;\;\;\; \mbox{as} \;\; R\rightarrow 0 \,.
\end{equation}
Then, the trace equation~(\ref{fe2bis}) indicates that
{\em in vacuo} ($T_{\mu\nu}^{(m)}=0$) one  obtains a class of
solutions with constant Ricci curvature $R =R_0$. In particular, 
there exist solutions with $R_0=0$.

Let us suppose now that the above Lagrangian density  reduces to 
a constant for small  curvature values,  $  \lim_{R\rightarrow  
0} f=\Lambda  $. Interesting 
features emerge again from the trace
equation: using Eq.~(\ref{fe2bis}) and the definition 
of $f(R)$, it is seen that  zero curvature 
solutions do not exist in this case because 
\begin{equation}
\Psi'R-2\Psi-\Psi_0 R-2\Lambda =\kappa \, T^{(m)} \,.
\end{equation}
Contrary to GR, even in absence of matter there are no
Ricci-flat  solution of the field equations since the higher 
order  derivatives  give  constant  curvature solutions 
corresponding
to a sort of effective cosmological constant.
In fact, in GR, solutions with non-vanishing constant curvature  
occur only in the presence of matter because of the 
proportionality of the Ricci
scalar to  the trace of the matter energy-momentum tensor.
A similar situation can be obtained in the presence of a
cosmological constant $\Lambda$. The difference between GR and 
higher order gravity
is that the Schwarzschild-de Sitter 
solution is not
necessarily generated by a $\Lambda$-term, while the effect of 
an "effective" cosmological constant can be achieved by the 
higher order derivative contributions, as discussed extensively
in \cite{BarrowOttewill83, Multamaki:2006zb, Multamaki:2006ym, 
noether,QuintRev}.

Let us  consider now the problem of finding the general 
solution of Eqs.~(\ref{fe2}) and (\ref{fe2bis}) for the 
spherically  symmetric metric~(\ref{sphericalme}).  
The substitution of this 
metric 
into  the $\left( t,r\right) $  component  of~(\ref{fe2}) yields
$\displaystyle\frac{\dot{B}(t,r)}{rB(t,r)}=0$, which  means that 
$B(t,r)$ must be time-independent,   $B(t,\,r)=b(r)$. On the 
other 
hand, the
$\left( \theta,\theta\right) $ component of Eq.~(\ref{fe2}) 
yields  $\displaystyle\frac{A'(t,r)}{A(t,r)}=\zeta(r)$, where 
$\zeta(r)$ is a time-independent function and 
\begin{equation}
A(t,r)\,=\,\tilde{a}(t) \, \exp\biggl[\int\zeta(r)dr\biggr]
=\tilde{a}(t) \, \frac{b}{r^2} \, 
\exp\biggl[\int dr \, \frac{[2- r^2(2 
\lambda+ 2 q\kappa \, p)]b(r)}{r}\biggr] 
\end{equation}
where $P$ is  the
pressure of a  perfect fluid with stress-energy tensor
\begin{equation}\label{enmomten}
T_{\mu\nu}^{(m)} =\left( P+ \rho \right) u_\mu u_\nu + P 
g_{\mu\nu}\,.
\end{equation}
The function $A(t,r)$ is separable,   
$A(t,r)=\tilde{a}(t)a(r)$, and the line 
element~(\ref{sphericalme}) becomes
\begin{equation}\label{me2}
ds^2 = - \tilde{a}(t)a(r)dt^2 + b(r)dr^2 + r^2d\Omega_2^2 
\end{equation}
and is rewritten as
\begin{equation}\label{me4}
ds^2= - a(r) d\tilde{t}^2 + b(r)dr^2 + r^2d\Omega_2^2 
\end{equation}
by redefining the  time coordinate $t\rightarrow \tilde{t}$ 
as $ d\tilde{t}=\sqrt{\tilde{a}(t)} \, dt$. From now on, the 
tilde  will be dropped from this time coordinate.

To summarize, in a space-time with constant scalar curvature, any 
spherically symmetric background is necessarily static or,  
the Jebsen-Birkhoff theorem holds 
for $f(R)$-gravity with 
constant curvature (cf. Ref.~\cite{HawkingEllis}).

A remark is in order at this point. We have assumed 
a space-time with constant Ricci scalar and deduced  
conditions on the form of the gravitational potentials. 
The inverse problem can also be considered: whenever the 
gravitational potential $a(t,r)$ is a separable function and 
$b(t,r)$ is  time-independent, using the
definition of the Ricci scalar, it is $R=R_0=$~const. and at
the same time the solutions of the field equations will be
static if spherical symmetry is invoked. For a complete 
analysis of this problem, one should take
into account  the remaining field equations contained 
in~(\ref{fe3}) and (\ref{fe3tr}) which have to be satisfied by 
taking into account the expression of the Ricci scalar
(\ref{ricscalin}). One must then solve the system
\begin{eqnarray}
&& R_{tt}+\lambda\, a(r)-q \kappa  \left[ 
\rho+P \left( 1-a(r)\right)\right] = 0 \,,\\
&&\nonumber\\
&& R_{rr}-\lambda b(r)-q\kappa \, P 
b(r)=0\,,\\
&&\nonumber\\
&& R_0 -q \kappa \left( \rho-3P 
\right)+4\lambda = 
0 \,,\\
&&\nonumber\\
&& R\left( a(r),b(r) \right) = R_0 \,,
\end{eqnarray}
which takes the form
\begin{eqnarray}
&& \mbox{e}^{\int\frac{2-r^2 \left(2\lambda+ 2q \kappa \, P 
\right) b(r)}{r} \,dr}
\biggl\{ \left[ r^2\left( 2\lambda+ 2 q\kappa \,P -2 \right) 
\right]^2 
b(r)^4-4b(r)^3 \nonumber\\
&&\nonumber\\
&& -3r \left[ r^2 \left( 2 \lambda + 2 q \kappa \, P \right)-2 
\right] 
b'(r)b(r)^2\nonumber\\
&&\nonumber\\
&& +2r \left[ b'(r)+rb''(r) \right] b(r)-2r^2b'(r)^2\biggr\} 
-4r^4 q \kappa  \left( P+\rho \right) b(r)^2=0 \,,\\
&&\nonumber\\
&& \biggl\{3r \left[ r^2 \left( 2\lambda+ 2q \kappa \,P 
\right)b'(r)-2  \right] 
-8\biggr\}b(r)^2-4 \left[ r^2 \left( 2\lambda+ 2q\kappa \, P 
\right)-3 \right]b(r)^3 \nonumber\\
&&\nonumber\\
&& -\left[ r^2 \left( 2\lambda+2q \kappa \, P \right)
-2 \right] b(r)^4  +2r^2b'(r)^2-2rb(r) \left[ rb(r)''-3b'(r) 
\right]=0 \,,\\
&& \nonumber\\
&& \biggl\{r^2 \left[4\lambda+2q \kappa \, \left( P-\rho 
\right)\right] 
-8\biggr\}b(r)^3-\biggl\{3r 
\left[ r^2 \left( 2\lambda+ 2 q\kappa \, P \right)-2 \right]b'(r) 
-4\biggr\}b(r)^2 \nonumber\\
&&\nonumber\\
&& + \left[ r^2 \left(2
\lambda+ 2 q\kappa \, P \right)-2 \right] b(r)^4
 -2r^2b'(r)^2+2r \left[ rb''(r)-b'(r) \right] b(r)=0  
\end{eqnarray}
where, using Eq.~(\ref{ricscalin}),  the only unknown 
potential is now $ b(r)$.  A general solution  is found for  the 
particular equation of state 
$P=-\rho$:
\begin{equation}
ds^2= -\biggl(1+\frac{k_1}{r}+\frac{ 2 q\kappa \, 
\rho-2\lambda}{6} \, r^2\biggr)dt^2 + 
\frac{dr^2}{1+\frac{k_1}{r}+
\frac{ q \kappa \,\rho- \lambda}{3}r^2} + r^2d\Omega_2^2 \,.
\end{equation}
In the case of constant Ricci scalar $R=R_0$, all $f(R)$ 
theories admit  solutions with de 
Sitter-like behavior even in the weak-field
limit. This is one of the reasons why dark energy 
can  be replaced by $f(R)$-gravity  
\cite{BarrowOttewill83, 
Capozziello02IJMPD,
CapozzielloCarloniTroisi03,
CapozzielloCardoneCarloniTroisi03,
CDTT, NojiriOdintsov03PLB, NojiriOdintsov03, mimicking, 
Carloni:2004kp}.

Let us consider now $f(R)$-gravity with an analytic Lagrangian 
function $f(R)$, which we write as
\begin{equation}\label{f}
f(R)=\Lambda+\Psi_0 \, R+\Psi(R) \,,
\end{equation}
where $\Psi_0$ is a constant, $\Lambda$ plays the role 
of the cosmological constant, and $\Psi(R)$ is an analytic
function of $R$ satisfying the condition
\begin{equation}\label{psi}
\lim_{R\rightarrow 0} \, \, \frac{ \Psi(R)}{R^2} =\Psi_1
\end{equation}
with $\Psi_1$ another constant. By neglecting the cosmological
constant $\Lambda$ and setting $\Psi_0$ to zero,  
a new class of theories is obtained which, in the limit 
$R\rightarrow{0}$, does not  reproduce GR (Eq.~(\ref{psi}) 
implies that  $ f(R) \sim R^2$ as $R\rightarrow 0$). In this  
case, analyzing the complete set of
equations~(\ref{fe2}) and (\ref{fe2bis}), one can observe that 
both zero and constant (but non-vanishing) curvature 
solutions are possible. In particular,  if $ R=R_0=0$ the  
field equations are solved for all forms of the gravitational 
potentials appearing in  the spherically
symmetric background~(\ref{me4}),  provided that the Bernoulli
equation~(\ref{eqric}) relating these 
functions is satisfied for
$R(r)=0$. The  solutions are thus defined by the relation
\begin{equation}\label{gensol0}
b(r) = \frac{\exp \left [-\int
dr\,h(r) \right]}{ K+4\int\frac{dr\,a(r)\,\exp\left[ -\int
dr\,h(r) \right]}{r \left[ a(r)+ra'(r) \right]}}\,.
\end{equation}
Table~\ref{tableR0} provides examples of $f(R)$ theories
admitting solutions with constant but non-zero values of $R$ or  
null $R$. Each model admits Schwarzschild and Schwarzschild-de
Sitter solutions, in addition to the class of solutions given 
by~(\ref{gensol0}).

\begin{table}
\begin{center}
\begin{tabular}{|ccccc|}
  \hline
  & & & & \\
  & $ \mathbf{ f(R)}$ {\bf theory} & & {\bf Field equations} & \\
  & & & & \\
  \hline
  & & & & \\
  & $R$ &  & $R_{\mu\nu}=0$  & \\
  & & & & \\
  \hline
  & & & & \\
  & $\xi_1 R+\xi_2 R^n$ & & $\begin{cases}
                R_{\mu\nu}=0 & \text{with}\,\,R=0,\,\,\,\xi_1\neq 0\\
                R_{\mu\nu}+ \lambda g_{\mu\nu}=0 &
                \text{with}\,\,R=\biggl[\frac{\xi_1}{(n-2)\xi_2}\biggr]^{\frac{1}{n-1}}
                ,\,\,\,\xi_1\neq 0,\,\,\,n\neq2\\
                0=0 & \text{with}\,\,R=0,\,\,\,\xi_1=0\\
                R_{\mu\nu}+\lambda g_{\mu\nu}=0 & \text{with}\ R=R_0,\ \xi_1=0,\ n=2
                \end{cases}$& \\
  & & & & \\
  \hline
  & & & & \\
  & $\xi_1R+\xi_2R^{-m}$ &  & $R_{\mu\nu}+\lambda g_{\mu\nu}=0$ with $R=\biggl[-\frac{(m+2)\xi_2}{\xi_1}\biggr]^{\frac{1}{m+1}}$ & \\
  & & & & \\
  \hline
  & & & & \\
  & $\xi_1 R+\xi_2 R^n+\xi_3 R^{-m}$ &  & $R_{\mu\nu}+\lambda 
        g_{\mu\nu}=0$, \text{with}\ $R=R_0$\ so that & \\ 
   &&&   $\xi_1 R_0^{m+1}+(2-n)\xi_2R_0^{n+m}
  +(m+2)\xi_3=0$ & \\
  & & & & \\
  \hline
  & & & & \\
  & $\frac{R}{\xi_1+R}$ &  & $\begin{cases}
                                         R_{\mu\nu}=0 & \text{with}\ R=0\\
                                         R_{\mu\nu}+\lambda g_{\mu\nu}=0 & \text{with}\ R=-\frac{\xi_1}{2}
                                         \end{cases}$ & \\
  & & & & \\
  \hline
  & & & & \\
  & $\frac{1}{\xi_1+R}$ &  & $R_{\mu\nu}+\lambda g_{\mu\nu}=0$ \text{with}\ $R=-\frac{2\xi_1}{3}$ & \\
  & & & & \\
  \hline
\end{tabular}
\end{center}
\caption{\label{tableR0} Examples of $f(R)$ models admitting
constant or zero scalar curvature solutions. The powers
$n$ and $m$ are integers while the $\xi_i$ are real
constants.}
\end{table}

\subsubsection{Solutions with Ricci scalar depending on the radial coordinate}

Thus far, we have discussed the behavior of $f(R)$-gravity 
searching for spherically symmetric solutions with  constant 
Ricci curvature. In GR this situation is well known to give  
rise to the  Schwarzschild  
($R=0$) and the 
Schwarzschild-de Sitter   
($R=R_0\neq 0$) solutions. The search for 
spherically symmetric solutions  can be generalized to $f(R)$-gravity by allowing the Ricci scalar to depend on the radial 
coordinate. This approach is  interesting because, in general, 
higher order  theories of gravity admit naturally this kind 
of solution, with several examples reported in  the literature
\cite{Stelle:1976gc,prl,CapCardTro07,Multamaki:2006zb, 
Multamaki:2006ym,noether}. 
In the following we approach  the problem from a general point of 
view.

If we choose the Ricci scalar as a generic function $R(r)$ of the
radial coordinate, it is possible to show that also in
this case the  solution of the field equations~(\ref{fe4}) and
(\ref{trfe4}) is time-independent (if $T_{\mu\nu}^{(m)}=0$). In 
other
words, the Jebsen-Birkhoff theorem
holds. As in GR, it is 
crucial to study  the off-diagonal $\left( t,r\right) $ 
component  
of~(\ref{fe4}) which, for a generic $f(R)$,  reads
\begin{equation}
\frac{d}{dr}\biggl[ r^2f'(R)\biggr] \dot{B}(t,r) = 0\,,
\end{equation}
and two possibilities can occur. First, one can choose
$\dot{B}(t,r) \neq 0$,  implying that ${\displaystyle
f'(R)\sim 1/ r^2 }$. In this case the remaining field
equation is not satisfied and there is  incompatibility. The only 
possible solution is then given by  $\dot{B} \left(t, r \right) = 
0$ and  $B \left( t,r \right) = 
b(r)$. The $\left( \theta,\theta\right) $ equation is then  
used to determine 
that the potential $A(t,r)$ can be
factorized with respect to time, the  solutions are of
the type~(\ref{me2}), and the metric can be recast in the 
stationary spherically symmetric form~(\ref{me4}) by a 
suitable coordinate transformation.

Even the more general radial-dependent case
admits  time-independent solutions. From the trace equation 
and the $\left( \theta,\theta\right) $ equation, the relation 
\begin{equation}\label{birk}
 a(r) =b(r)\, \frac{ \mbox{e}^{\frac{2}{3} 
\int\frac{(Rf'-2f)b(r)}{R'f''}dr}}{r^4R'^2f''^2}
\end{equation}
(with $f'' >  0$) linking $a(r)$ and $b(r)$ can  be obtained,   
in addition to  \cite{Multamaki:2006zb, Multamaki:2006ym} 
\begin{equation}\label{rebf}
b(r) =\frac{6 \left[ f'(rR'f'')' 
-rR'^2f''^2 \right]}{rf \left( rR'f''-4f' \right)+ 
2f' \left[ rR \left( f'-rR'f'' \right)-3R'f'' \right] }\,.
\end{equation}
Again,  three more equations have to be satisfied in order 
to completely solve the system (respectively the $\left( 
t,t\right)$ and
$\left( r,r\right)$ components of the field equations plus the 
Ricci scalar
constraint), while  the only unknown functions are $f(R)$ and the
Ricci scalar $R(r)$.

 If we now consider a fourth order theory described by 
$f(R) = R+\Phi(R)$ with $\Phi(R)\ll R$ we are
able to satisfy the complete  set of equations up to third
order in $\Phi$. In particular,  we can solve the full set of
equations; the relations~(\ref{birk}) and~(\ref{rebf}) will 
provide the general solution  depending only on the forms of 
the functions $\Phi(R)$ and $R(r)$, {\em i.e.},
\begin{eqnarray}\label{solRr}
{}a(r) &=& b(r) \, 
\displaystyle\frac{ \mbox{e}^{\displaystyle-\frac{2}{3} 
\int\frac{\left[  R+(2\Phi-R\Phi') \right] 
b(r)}{R'\Phi''}dr}}{r^4R'^2\Phi''^2} \,, \\
\nonumber\\
b(r) &=& -\displaystyle\frac{3 (rR'\Phi'')_{,r}}{rR}\,.
\end{eqnarray}
Once the radial dependence of the scalar curvature is obtained,
Eq.~(\ref{solRr}) allows one to obtain the solution of the 
field equations  and  the  gravitational potential related to 
the function $a(r)$. The  physical relevance of this potential 
can be assessed by comparison with astrophysical
data ({\em e.g.},  \cite{Kainulainen:2007bt}).

\subsection{The Noether approach and spherical symmetry}
\label{sec:chap5sec5}

Exact spherically symmetric solutions with constant Ricci 
scalar in  $f(R)$-gravity can be found using the 
 Noether symmetry \cite{noether}. To begin, one 
needs to derive a point-like Lagrangian from the action of 
modified gravity  by imposing spherical symmetry, while enforcing 
the constancy of the Ricci scalar by means of a suitable  
Lagrange multiplier.  With the previous considerations in mind, a 
spherically symmetric space-time is described by the line element
\begin{equation}\label{NSme2}
{ds}^2= - A(r){dt}^2 + B(r){dr}^2 + M(r)d\Omega_2^2\,.
\end{equation}
The Schwarzschild solution of GR 
is obtained if $M(r)=r^2$ and
$ A(r)=B^{-1}(r)=1-2M/r$, with $r$ an areal radius. 
In the presence of spherical symmetry the action 
\begin{equation}
S = \int dr \, L \left( A, A', B, B', M, M', R, R' \right)
\end{equation}
contains only   a finite number of degrees of freedom, the 
Ricci scalar $R$ and the potentials   $A$, $B$, and $M$ 
defining the configuration space. The point-like 
Lagrangian is obtained by writing the action as
\begin{equation}\label{NSlm}
S =\int d^4x\sqrt{-g}\biggl[f(R)-\lambda(R-\bar{R})\biggr]\,,
\end{equation}
where $\lambda$ is a Lagrangian multiplier and $\bar{R}$ is the
Ricci scalar of the metric~(\ref{NSme2})
\begin{eqnarray}
\bar{R} & = &  \frac{A''}{AB}  +  \frac{2 M''}{BM} 
+\frac{A'M'}{ABM}-  \frac{A'^2}{2A^2B}-\frac{M'^2}{2BM^2} 
-\frac{A'B'}{2AB^2}-\frac{B'M'} {B^2M}-\frac{2}{M} \nonumber\\
&&\nonumber\\
& \equiv &  
R^*+\frac{A''}{AB}+ \, \frac{2M''}{BM}\,, \label{NSrs}
\end{eqnarray}
where $R^*$ collects the terms containing first order 
derivatives.  The Lagrange multiplier $\lambda$ is obtained by 
varying  the  action~(\ref{NSlm}) with respect to $R$, which 
yields  $\lambda= f_R(R)$. By expressing the
metric determinant $g$ and $\bar{R}$ as functions of  $A$, $B$,  
and $M$,  Eq.~(\ref{NSlm}) gives
\begin{eqnarray}
S & = & \int dr \, \sqrt{AB} \, M \biggl[ 
f-f_{R}\biggl(R-R^*-\frac{A''}{AB}-\frac{2M''}{BM}\biggr)\biggr] 
\nonumber\\
&&\nonumber\\
&= & \int dr\biggl\{ \sqrt{AB} \, 
M\biggl[ f-f_{R} \left( R-R^* \right)\biggr] 
-\biggl( \frac{f_R M}{ \sqrt{AB} } \biggr)'A'-
\biggl( \frac{2 \sqrt{A}}{\sqrt{B}} \, f_R \biggr)'M' \biggr\} 
\,. \nonumber\\
&& \label{NSac1}
\end{eqnarray} 
The last two integrals  differ by  a total divergence which 
can be  discarded, and the point-like Lagrangian becomes
\begin{eqnarray}\label{NSlag}
L &= & -\frac{\sqrt{A} f_{R}}{2M \sqrt{B} }{(M')}^2 
-\frac{f_{R}}{ \sqrt{AB} } \, A'M'-\frac{Mf_{RR}}{ 
\sqrt{ AB}} \, A'R' \nonumber\\
&& \nonumber\\
& - & \frac{2 
\sqrt{A} f_{RR}}{ 
\sqrt{ B} } \, R'M'- \sqrt{AB} \left[ \left( 2+MR 
\right)f_{R}-Mf \right]\,.
\end{eqnarray}
The canonical Lagrangian~(\ref{NSlag}) is written in  
compact form using matrix notation as 
\begin{equation}\label{NSla}
L ={\underline{q}'}^t\hat{T}\underline{q}'+V \,,
\end{equation}
where $\underline{q}=\left( A,B,M,R \right)$ and 
$\underline{q}'= \left( A',B',M',R' \right)$
are the generalized Lagrangian coordinates  and velocities.  The 
index "{\it t}" denotes the 
transposed 
vector.  The kinetic tensor is  
\begin{equation} 
\hat{T}_{ij} = \frac{\partial^2 L}{\partial 
q'_i\partial q'_j} 
\end{equation}
and $V(q)$ is the potential energy depending only on the
generalized coordinates. The Euler-Lagrange equations are 
\begin{eqnarray}
&& \frac{d}{dr} \left( \nabla_{q'} L \right) -\nabla_{q} 
L =2 \,\, \frac{d}{dr} \biggl(\hat{T} 
\underline{q}'\biggr)-\nabla_{q}V-{\underline
{q}'}^t\biggl(\nabla_{q}\hat{T}\biggr)\underline{q}' \nonumber\\
&& \nonumber\\
&&  =  \,2\hat{T}\underline{q}''+2\biggl(\underline{q}' 
\cdot\nabla_{q}\hat{T}\biggr)
\underline{q}'-\nabla_{q}V 
-\underline{q}'^t\biggl(\nabla_{q}\hat{T}\biggr)\underline{q}'=0 
\,. \label{NSfe2}
\end{eqnarray}
$ R$ obeys a  constraint relating the Lagrangian coordinates. The 
Hessian determinant of (\ref{NSlag}),
${\displaystyle \left|\left|\frac{\partial^2 L}{\partial
q'_i\partial q'_j}\right|\right|}$, vanishes because 
the point-like Lagrangian does not depend on the generalized 
velocity $B'$. The metric component $B$ does not
contribute to the dynamics, but its  equation of motion  has to 
be taken into account as a further constraint.  The definition of 
the  energy 
\begin{equation}\label{NSene}
E_L={\underline {q}'}\cdot\nabla_{q'} L- L
\end{equation}
coincides with the  Euler-Lagrangian equation for the component 
$B$ of the  generalized coordinate $\underline{q}$. Then,  the 
Lagrangian~(\ref{NSlag}) contains only three  degrees of 
freedom and not four, as expected {\em a
priori}. Now, since the equation of motion for $B$ does not 
contain the  derivative $B'$, it can be  solved explicitly in 
term of $B$ as a function of the other Lagrangian coordinates:
\begin{equation}\label{NSeqb}
B=\frac{2M^2f_{RR}A'R'+ 2Mf_{R}A'M' 
+4 AMf_{RR}M'R'+Af_{R}M'^2}{2AM \left[ \left( 2+MR 
\right)f_{R}-Mf \right]}\,.
\end{equation}
By inserting Eq.~(\ref{NSeqb}) into the Lagrangian~(\ref{NSlag}),  
we obtain a non-vanishing Hessian matrix removing the singular
dynamics. The new Lagrangian reads\footnote{Lowering  the dimension
of the configuration space through the substitution~(\ref{NSeqb}) 
leaves the dynamics unaffected because $B$ is non-dynamical. In
fact, if  Eq.~(\ref{NSeqb}) is introduced  into the 
set of dynamical equations~(\ref{NSlag}), these   coincide with 
the equation   derived from (\ref{NSlag2}).}
\begin{equation}
L^*= \sqrt{ {\bf L} }
\end{equation}
with
\begin{eqnarray}\label{NSlag2}
{\bf L} & = &  \underline{q'}^t\hat{{\bf
L}} \underline{q'}= 
\frac{[(2+MR)f_{R}-fM]}{M} \nonumber\\
&& \nonumber\\\
& \cdot&  \left[ 2M^2f_{RR}A'R' +2MM'(f_{R}A'+2Af_{RR}R') 
+Af_{R}M'^2 
\right]\,.
\end{eqnarray}
Since ${\displaystyle \frac{\partial{\bf L}}{\partial r}=0}$,
${\bf L}$ is canonical (${\bf L}$ is a quadratic form 
in the generalized velocities $A'$, $M'$ and $R'$ and  
coincides with the Hamiltonian), hence ${\bf 
L}$ can be regarded as the
new Lagrangian with  three  degrees of freedom. It is crucial 
that the Hessian determinant  
\begin{equation}
\left|\left|\frac{\partial^2 {\bf L}}{\partial q'_i\partial
q'_j}\right|\right| =3AM \left[ 
(2+MR)f_{R}-Mf \right]^3f_{R}{f_{RR}}^2\,.
\end{equation}
now does not vanish. It is assumed that $(2+MR)f_{R}-Mf\neq0$, 
otherwise the
above definitions of $B$, and ${\bf L}$ (Eqs.~(\ref{NSeqb})  
and~(\ref{NSlag2})) are meaningless. Moreover, it is assumed that  
$f_{RR}\neq 0$ allows for a wide class of fourth order gravity 
models. The GR case $f(R)=R$  is special: the GR point-like 
Lagrangian requires  a further reduction of the number of  
degrees of freedom  and the  previous results cannot be applied 
directly.  Eq.~(\ref{NSlag}) yields
\begin{equation}\label{NSlagr}
L_{GR}=-\frac{ \sqrt{A} }{2M \sqrt{B}} \,{(M')}^2 
-\frac{1}{ \sqrt{AB} } \, A'M'-2 \sqrt{AB} \,,
\end{equation}
which, through the Euler-Lagrange equations,  provides the 
standard GR equations  for the Schwarzschild 
metric. The absence 
of the generalized velocity $B'$ in Eq.~(\ref{NSlagr}) is 
evident. Again, the  Hessian determinant vanishes. Nevertheless, 
considering again the constraint~(\ref{NSeqb}) for $B$, it is 
possible to obtain a Lagrangian  with non-vanishing Hessian. In 
particular, it is 
\begin{eqnarray}
B_{GR} & = & \frac{(M')^2}{4M}+\frac{A'M'}{2A}\,, \label{NSeqbgr} 
\\
&&\nonumber\\
L_{GR}^* &= &  \sqrt{ {\bf 
L}_{GR}} =\sqrt{\frac{M'(2MA'+AM')}{M}}\,,
\label{NSlag2gr}
\end{eqnarray}
and the Hessian determinant is
\begin{equation}
\left|\left|\frac{\partial^2 {\bf L}_{GR}}{\partial q'_i\partial 
q'_j}\right|\right|=-1\,,
\end{equation}
a non-vanishing sub-matrix of the $f(R)$ Hessian matrix.

The Euler-Lagrange equations derived from  Eqs.~(\ref{NSeqbgr}) 
and~(\ref{NSlag2gr}) yield the vacuum solutions of  GR
\begin{equation}\label{NSschsol}
A= k_{4}-\frac{k_{3}}{r+k_{1}}\,,\ \ \ \ \ \ \
B=\frac{k_{2}k_{4}}{A}\,,\ \ \ \ \ \ M=k_{2}(r+k_{1})^2\,.
\end{equation}
In particular, the standard form of the Schwarzschild 
solution is 
recovered for  $k_{1}=0$, $k_{2}=1$, ${\displaystyle 
k_{3}= 2GM /c^2 }$, and
$k_{4}=1$.

Table~\ref{tableNoetherspherical}  summarizes  the field 
equations associated with 
the point-like Lagrangians and their relation with respect to 
the ones of the standard approach.

\begin{table}[h]
\begin{center}
\begin{tabular}{|ccc|} \hline
  & &\\
  {\bf Field equations approach} &  & {\bf Point-like Lagrangian 
approach} \\
  & &\\
   &  &   \\
  & &\\
  $\delta\int d^4x\sqrt{-g} \, f=0$  & $\leftrightarrows$ &  
$\delta\int dr \, L=0$  \\
  $\downarrow$ &  &  $\downarrow$ \\
  $H_{\mu\nu} 
=\partial_\rho \left[ \frac{\partial(\sqrt{-g} \, 
f)}{\partial_\rho 
g^{\mu\nu}} \right]-\frac{\partial (\sqrt{-g} \, f)}{\partial 
g^{\mu\nu}}=0$ &
  & $\frac{d}{dr} \left( \nabla_{q'} L \right) -\nabla_{q} L=0$ 
\\  & 
$\leftrightarrows$ &   \\
  & &\\
  $H=g^{\mu\nu}H_{\mu\nu}=0$ &
  & $E_L={\underline
  {q}'}\cdot\nabla_{q'} L- L$ \\
  $\downarrow$ &  & $\downarrow$  \\
  & &\\
  $H_{00}=0$ & $\leftrightarrows$ & $\frac{d}{dr} 
\left( \frac{\partial 
L}{\partial A'} \right) -\frac{\partial L}{\partial A}=0$ \\
  & &\\
  $H_{rr}=0$ & $\leftrightarrows$ & $\frac{d}{dr} 
\left( \frac{\partial 
L}{\partial B'} \right) -\frac{\partial L}{\partial B}\propto 
E_L=0$ \\
  & &\\
  $H_{\theta\theta}={\csc^{2}\theta}H_{\phi\phi}=0$ &  
$\leftrightarrows$ & $\frac{d}{dr} \left( \frac{\partial 
L}{\partial M'} \right) -\frac{\partial L}{\partial M}=0$ \\
  & &\\  
$H=A^{-1}H_{00}-B^{-1}H_{rr} 
-2M^{-1}{\csc^{2}\theta}H_{\phi\phi}=0$ & $ \leftrightarrows$ & a 
combination of the above equations \\
  & &\\
\hline
\end{tabular}
\end{center}
\caption{\label{tableNoetherspherical} The field equations 
approach and the point-like
Lagrangian approach differ because spherical symmetry 
can be imposed either in the field equations after standard 
variation with respect to the metric, or directly
into the Lagrangian, which then becomes point-like. The energy
$E_L$ corresponds to the $\left( 0, 0 \right)$ component of
$H_{\mu\nu}$.  The absence of $B'$ in the Lagrangian implies the
proportionality between the constraint equation for $B$ and the
energy function $E_L$. As a consequence, there are only three 
independent equations and three unknown
functions. The $ \left( \theta, \theta \right) $ component 
corresponds to the field  equation for $M$. $H_{\mu\nu}$ is given 
in Appendix C. 
}\end{table}

In  spherical symmetry, the  areal radius $r$ plays the role of 
an affine parameter.  Then,  the 
configuration space is $\mathcal{Q}=\left( A, M, R 
\right)$ and  the tangent space is  $\mathcal{TQ}=\left( A, A', 
M, M', R, R'\right)$. According to the Noether 
theorem, the 
existence of a symmetry for the dynamics described by the 
Lagrangian~(\ref{NSlag2})  implies the existence of a conserved 
quantity. The Lie differentiation  of Eq.~(\ref{NSlag2})  
yields\footnote{From now on,  $\underline{q}$ denotes the vector 
$(A,M,R)$.} 
\begin{equation}
{\cal L}_{\mathbf{X}}{\bf L} = 
\underline{\alpha}\cdot\nabla_{q}{\bf L
}+\underline{\alpha}'\cdot\nabla_{q'}{\bf L}
=\underline{q}'^t\biggl[\underline{\alpha}\cdot\nabla_{q}\hat{{\bf
L }}+ 2\biggl(\nabla_{q}\alpha\biggr)^t\hat{{\bf L
}}\biggr]\underline{q}'\,.
\end{equation}
This Lie derivative vanishes if the functions 
${\underline{\alpha}}$ satisfy the system 
\begin{equation}\label{NSsys}
 \alpha_{i}\frac{\partial
\hat{{\bf L}}_{km}}{\partial 
q_{i}}+2 \, \frac{\partial\alpha_{i}}{\partial q_{k}}\hat{{\bf L
}}_{im}=0\,.
\end{equation}
Solving the system~(\ref{NSsys})  means finding the functions
$\alpha_{i}$ which identify the Noether vector. However the 
system~(\ref{NSsys}) depends implicitly  on the form of the 
function $f(R)$  and, by solving it, one obtains the  forms 
of the function $f(R)$ which are compatible with spherical
symmetry. Alternatively, by choosing  the form of $f(R)$, 
(\ref{NSsys}) can be solved explicitly. As an example,  
the system~(\ref{NSsys}) is satisfied if we choose
\begin{equation}\label{NSsolsy}
f(R) = f_0 R^s\,, \ \ \ \  
\underline{\alpha}=(\alpha_1,\alpha_2,\alpha_3)=
\biggl((3-2s)kA,\ -kM,\ kR\biggr)
\end{equation}
with $s$ a real number, $k$ an integration constant, and $f_0$ a
dimensional coupling constant.\footnote{The dimensions are given
by $R^{1-s}$ in term of the Ricci scalar. For simplicity,  
$f_0$ is set to unity in the following.}
This means that for  $f(R)=R^s$ there exist at least one  
Noether symmetry 
 and  a related conserved quantity 
\begin{eqnarray}\label{NScm}
\Sigma_{0} &=& \underline{\alpha}\cdot\nabla_{q'}{\bf L
} \nonumber\\
&&\nonumber\\
&= & 2sk MR^{2s-3} \left[ 2s+(s-1)MR  \right] \left[ 
(s-2)RA'-(2s^2-3s+1)AR' \right]\,.
\end{eqnarray}
A physical interpretation of $\Sigma_{0}$  is possible  in GR. In 
this case, obtained 
for $s=1$, the above procedure must be applied to the
Lagrangian~(\ref{NSlag2gr}), obtaining the solution
\begin{equation}\label{NSsolsygr}
\underline{\alpha}_{GR}=(-kA,\ kM)\,.
\end{equation}
The functions $A$ and $M$ provide the Schwarzschild 
solution~(\ref{NSschsol}), and 
then the constant of motion takes 
the form
\begin{equation}\label{NScmgr}
\Sigma_{0}= \frac{2GM}{c^2} 
\end{equation}
in standard units;  the conserved quantity  is the Schwarzschild 
radius (or the mass of the gravitating system).

Another solution can be found for constant Ricci scalar $R=R_0$ 
\cite{Multamaki:2006zb}, for which the field
equations reduce to 
\beq \label{NSfe1} 
R_{\mu\nu}+k_0
g_{\mu\nu}=0\,,\eeq where ${\displaystyle
k_0=-\frac{1}{2}f(R_0)/f_R(R_0)}$. The general solution is
\beq\label{NSscwdes}
A(r)=\frac{1}{B(r)}=1+\frac{k_0}{r}+\frac{R_0}{12} \, 
r^2\,,\qquad
M=r^2
\eeq 
which includes the special case 
\beq
A(r)=\frac{1}{B(r)}=1+\frac{k_0}{r}\,,\qquad M=r^2\,,\qquad
R=0\,.
\eeq 
The solution~(\ref{NSscwdes}) is the well known
Schwarzschild-de Sitter metric.

In the general case $f(R)=R^s$, the Lagrangian (\ref{NSlag2})
becomes
\begin{eqnarray}
{\bf L} & = & \frac{sR^{2s-3}[2s+(s-1)MR]}{M}\nonumber\\
&& \nonumber\\ 
& \cdot & \left[ 2(s-1)M^2A'R'+2MRM'A'+4(s-1)AMM'R'+ARM'^2 
\right] 
\end{eqnarray} 
and the expression~(\ref{NSeqb}) of $B$ is 
\begin{equation}
B=\frac{s\left[ 2(s-1) M^2A'R' 
+2MRM'A'+4(s-1)AMM'R'+ARM'^2 
\right]}{2AMR \left[ 2s+(s-1)MR \right]} \,,
\end{equation}
and GR is recovered for $s=1$.

Using the constant of motion~(\ref{NScm}),  one can solve for 
$A$,  obtaining
\begin{equation}
A=R^{\frac{2s^2-3s+1}{s-2}} 
\biggl\{k_1+\Sigma_{0}\int\frac{R^{\frac{4s^2-9s+5}{2-s}} 
dr}{2ks(s-2)M \left[ 2s+(s-1)MR \right]}\biggr\}
\end{equation}
for $s\neq2$, where $k_1$ is an integration constant. For 
$s=2$ one finds
\begin{equation}
A=- \frac{\Sigma_{0}}{12kr^2(4+r^2R)RR'}\,.
\end{equation}
These relations allow one to find general solutions of the field
equations regulating the function $R(r)$. For example, the 
solution  corresponding to 
\begin{equation}
s=5/4\,,\;\;\; \ \ \ M=r^2\,,\ \;\;\ \ \ R= 5 r^{-2}\,,
\end{equation}
is the spherically symmetric metric given by 
\begin{equation}
ds^2=- \frac{1}{\sqrt{5}} \left( k_2+k_1 r \right)dt^2 + 
\frac{1}{2}\biggl(\frac{1}{1+\frac{k_2}{k_1r}}\biggr)dr^2 
+ r^2d\Omega_2^2  \label{textbf4dot89}
\end{equation}
with ${\displaystyle k_2=\frac{32\Sigma_0}{225 k}}$. The value of 
$s$ for this solution is  ruled out by Solar System experiments 
\cite{BarrowClifton2, BarrowClifton3, BarrowClifton1}.

To summarize, the Noether symmetry 
approach provides a general 
method to find  spherically symmetric exact solutions of   ETGs, 
and of metric $f(R)$-gravity in particular.  The procedure 
consists of~$i)$ obtaining  the point-like $f(R)$ Lagrangian 
with  spherical symmetry; $ii)$ writing the Euler-Lagrange 
equations; $iii)$ searching for a Noether vector field; and $iv)$ 
reducing the dynamics and then integrating the equations of 
motion using the constants of motion. {\em Vice-versa}, this 
approach  also allows one to select families
of $f(R)$ models with spherical symmetry.  The method can 
be generalized. If a symmetry exists,
the Noether approach allows  
transformations of variables to  
cyclic ones, reducing the dynamics to obtain  exact solutions. 
For example, since we know that $f(R)=R^s$-gravity admits a 
constant of motion,   the Noether symmetry suggests 
the coordinate transformation
\begin{equation}
{\bf L}\left( A, M, R, A', M', R' 
\right) \rightarrow \widetilde{{\bf L
}} \left( \widetilde{M}, \widetilde{R}, \widetilde{A}', 
\widetilde{M}', \bar{R}' \right)\,, 
\end{equation} for the 
Lagrangian~(\ref{NSlag2}), 
where the conserved quantity  
corresponds to the cyclic variable $\widetilde{A}$. In the 
presence of multiple   symmetries  one can find multiple cyclic 
variables.   If three Noether symmetries  
exist, 
the Lagrangian
${\bf L}$ can be mapped  into a Lagrangian with three cyclic 
coordinates  
$\widetilde{A}=\widetilde{A}({\underline{q}})$,
$\widetilde{M}=\widetilde{M}({\underline{q}})$ and
$\widetilde{R}=\widetilde{R}({\underline{q}})$ which are 
functions of the old generalized coordinates. These new functions 
must satisfy the  system
\begin{eqnarray}
\left( 3-2s \right) A \, \frac{\partial\widetilde{A}}{\partial
A}-M \, \frac{\partial\widetilde{A}}{\partial
M}+R \, \frac{\partial\widetilde{A}}{\partial R}=1\,, \\
\nonumber\\
\left( 3-2s \right) A \, \frac{\partial\widetilde{q}_i}{\partial
A}-M \, \frac{\partial\widetilde{q}_i}{\partial
M}+R \, \frac{\partial\widetilde{q}_i}{\partial R} = 0\,,
\label{NSsys2}
\end{eqnarray}
with $i=2,3$ (we have set  $k=1$). A solution of~(\ref{NSsys2}) 
for $s\neq 3/2$ is 
\begin{eqnarray}
\widetilde{A} & = & \frac{\ln 
A}{ (3-2s)}+F_{A} \left( A^{\frac{\eta_A}{3-2s}}
M^{\eta_A} \, A^{\frac{\xi_A}{2s-3}}M^{\xi_A} \right)  \,, \\
&&\nonumber\\
\widetilde{q}_i & = & F_{i} \left( A^{\frac{\eta_i}{3-2s}} 
M^{\eta_i},A^{\frac{\xi_i}{2s-3}}M^{\xi_i} \right) \label{NSso1}
\end{eqnarray}
while, if $s=3/2$,
\begin{eqnarray}
\widetilde{A} & = & -\ln M+F_{A}(A)G_A(MR) \,, \\
&&\nonumber\\
\widetilde{q}_i & = & F_{i}(A)G_i(MR) \,, \label{NSso2}
\end{eqnarray}
where $F_A$, $F_i$, $G_A$ and $G_i$ are arbitrary functions  and
$\eta_A$, $\eta_i$, $\xi_A$, and $\xi_i$ are integration 
constants.

The considerations of this section make it clear once again 
that the Jebsen-Birkhoff theorem  
does not hold, in general, for  metric $f(R)$-gravity.

\subsection{From spherical to axial symmetry in
$f(R)$-gravity}

Here, we want to seek for a general metod to find out
axially symmetric solutions  by performing a complex coordinate transformation
on the spherical metrics. Since the discovery of the Kerr solution \cite{pap:kerr},  many attempts have been made to find a
physically reasonable interior matter distribution that may be
considered as its source. For a review on these approaches
see \cite{pap:dmm, pap:ak}. Though much progress has been made, results have
been generally disappointing. As far as we know,  nobody  has
obtained a physically satisfactory interior solution. This seems
surprising given the success of matching internal spherically symmetric solutions to the Schwarzschild metric. The problem is
 that the loss of a degree of symmetry makes the
derivation of analytic results  much more difficult.  Severe
restrictions are placed on  the interior metric by maintaining that it
must be joined smoothly to the external axially symmetric metric.
Further restrictions are placed on the  interior solutions to ensure that they correspond to
physical objects.

Furthermore
since the axially symmetric metric has no radiation field associated with it,
its source should  be also non-radiating. This places even
further constraints on the structure of the interior
solution  \cite{Stephanietal03}. Given the strenuous nature of these limiting
conditions, it  is not surprising to learn that  no
satisfactory solution to the problem of finding  sources
for the Kerr metric has been obtained. In general,  the failure is  due to  internal
structures whose physical properties are unknown. This shortcoming makes hard to find consistent
 boundary conditions.

Newman and Janis
showed that it  is possible to  obtain an  axially symmetric  solution (like the
Kerr metric) by making an elementary complex transformation on the
Schwarzschild solution~\cite{pap:nj1}. This same method has been
used to obtain a new stationary and axially symmetric
solution known as the
Kerr-Newman metric~\cite{pap:nj2}. The Kerr-Newman space-time is
associated to the exterior geometry of a rotating massive and
charged black-hole. For a  review on the Newman-Janis method
 to obtain both the Kerr and Kerr-Newman metrics
see~\cite{bk:ier}.

By means of very elegant
mathematical arguments, Schiffer et al.~\cite{pap:mms} have  given a rigorous
proof to show how the Kerr metric can be derived starting from a complex
transformation on the Schwarzschild solution. We will not go
into the details of this demonstration, but point out  that
 the proof  relies on two main assumptions. The first  is that the metric
belongs to the same algebraic class of the Kerr-Newman solution,
namely the Kerr-Schild class \cite{pap:gcd}. The second assumption is
that the metric corresponds to an empty solution of the Einstein
field equations. In the case we are going to study, these assumptions are not
considered and hence the proof  in \cite{pap:mms} is not applicable. It is clear, by the
generation of the Kerr-Newman metric, that all the components of
the stress-energy tensor need to be non-zero for the Newman-Janis method to be
successful. In fact, G\"{u}rses and G\"{u}rsey, in
1975~\cite{pap:mg}, showed that if a metric can be written in the Kerr-Schild
form, then a complex transformation "is allowed in General
Relativity.'' Here,  we will show that  such a transformation
can be  extended to $f(R)$-gravity.

We now show how it is possible to obtain an axially symmetric 
solution starting from a spherically symmetric one, using a   
method developed by Newman and Janis 
in GR 
\cite{pap:nj1,pap:nj2}. This method can be applied to a static 
spherically symmetric metric adopted as a ``seed'' metric. In 
principle, the procedure could be applied whenever Noether 
symmetries are present. We apply this procedure to solutions of 
metric $f(R)$-gravity \cite{cqgaxially}.

In general, the approach is not  straightforward since, if  
$f(R)\neq R$, the field equations are of fourth order  and the 
relevant existence theorems and boundary conditions are 
different from those of GR. However, the existence of a Noether 
symmetry guarantees the consistency of the chosen $f(R)$ model 
with the field equations.

Let us consider a  spherically symmetric metric of the suitable form
\begin{equation}
ds^2 = - \mbox{e}^{2\phi(r)}dt^2 + \mbox{e} ^{2\lambda(r)}dr^2 +  
r^2 d\Omega_2^2 \,. \label{eqn:ssm}
\end{equation}
Following Newman and Janis, the line 
element~(\ref{eqn:ssm}) can 
be written in Eddington-Finkelstein coordinates 
$ \left( u,r,\theta,\varphi \right)$, {\em i.e.},  the $ g_{rr}$ 
component is eliminated by the coordinate change and a cross term 
is introduced \cite{thorne}. We set  $ dt = du +  F(r) dr $ with  
$F(r)= \pm \mbox{e} ^{\lambda(r)-\phi(r)}$, turning the 
line element~(\ref{eqn:ssm}) into
\begin{equation}\label{nullelemn}
ds^2 = -\mbox{e}^{2\phi(r)} du^2 \mp 2 \,  
\mbox{e}^{\lambda(r) - \phi(r)}dudr +
r^2 d\Omega_2^2 \,.
\end{equation}
The surface $u\,=\,$costant is a  light cone with vertex in  
$r\,=\,0$. The inverse metric tensor in null coordinates is
\begin{equation}\label{metrictensorcontro}
g^{\mu\nu}  = \left(
\begin{array}{cccc}
0 & \mp \mbox{e}^{-\lambda(r)-\phi(r)} & 0 & 0     \\
&&&\\
\mp \mbox{e}^{-\lambda(r)-\phi(r)} & \mbox{e}^{-2\lambda(r)} & 0 
& 0 \\
&&&\\
0 & 0 & \frac{1}{r^2} & 0 \\
&&&\\
0 & 0 & 0 & \frac{1}{ r^2\sin^2{\theta} } 
\end{array}    \right) \,.
\end{equation}
The matrix~(\ref{metrictensorcontro}) can be written in terms 
of a null tetrad as
\begin{equation}\label{eq:gmet}
g^{\mu\nu} = - l^\mu n^\nu -  l^\nu n^\mu + m^\mu\bar{m}^\nu + 
m^\nu\bar{m}^\mu \,,
\end{equation}
where $l^\mu$, $n^\mu$, $m^\mu$, and $\bar{m}^\mu$ satisfy the 
conditions
\begin{equation}
l_\mu l^\mu\,=\,m_\mu m^\mu\,=\,n_\mu n^\mu\,=\,0 \,, \,\,\,\,\,
l_\mu n^\mu\,=\,-m_\mu\bar{m}^\mu\,= -1 \, , \,\,\,\,\, l_\mu
m^\mu\,=n_\mu m^\mu\,=\,0 \,, 
\end{equation} 
and where an overbar denotes complex conjugation. At any 
spatial point, the tetrad can be chosen
in the following manner: $l^\mu$ is the outward null vector
tangent to the light cone, $n^\mu$ is the inward null vector 
pointing toward the origin, and $m^\mu$ and $\bar{m}^\mu$ are  
vectors tangent to the two-dimensional sphere defined by 
constant 
$r$ and $u$. For the space-time~(\ref{metrictensorcontro}), the 
null tetrad  can be chosen as 
\begin{eqnarray}
l^\mu\, &= & \,\delta^\mu_1 \,, \\
&&\nonumber\\
n^\mu\, &= & \,-\frac{1}{2} \, 
\mbox{e}^{-2\lambda(r)} \, \delta^\mu_1+
\mbox{e}^{-\lambda(r)-\phi(r)} \, \delta^\mu_0 \,,\\
&&\nonumber\\
m^\mu\, & = & \,\frac{1}{\sqrt{2} \, r} \left( \delta^\mu_2
+\frac{i}{\sin{\theta}} \, \delta^\mu_3 \right) \,,\\
&&\nonumber\\
\bar{m}^\mu\, & = & \,\frac{1}{\sqrt{2} \, r} \left( \delta^\mu_2
-\frac{i}{\sin{\theta}}\delta^\mu_3 \right) \,.
\end{eqnarray}
Now we extend the set of coordinates
$x^\mu\,=\,(u,r,\theta,\phi)$ by promoting  the real
radius to the role of a complex variable. The null tetrad
then becomes\footnote{A certain degree of arbitrariness is 
present in the complexification of the functions  
$\lambda$ and $\phi$. Obviously, we must recover the metric  
(\ref{metrictensorcontro})  as soon as $r\,=\,\bar{r}$.}
\begin{eqnarray} 
l^\mu\, & = & \,\delta^\mu_1  \,,\label{tetradvectors1}\\
&&\nonumber\\
n^\mu\, & = & \,-\frac{1}{2} \, \mbox{e}^{-2\lambda (r,\bar{r})} 
\, \delta^\mu_1 +
\mbox{e}^{-\lambda(r,\bar{r})-\phi(r,\bar{r})} \, \delta^\mu_0 
\,,  \label{tetradvectors2}\\
&&\nonumber\\
m^\mu\, & = &  \,\frac{1}{\sqrt{2} \, \bar{r}} \, 
\left( \delta^\mu_2
+\frac{i}{\sin{\theta}}\delta^\mu_3 \right) \,, 
\label{tetradvectors3}\\
&&\nonumber\\
\bar{m}^\mu\, & = & \,\frac{1}{\sqrt{2} \, r} \, 
\left( \delta^\mu_2
-\frac{i}{\sin{\theta}}\delta^\mu_3 \right) \,. 
\label{tetradvectors4}
\end{eqnarray}
A new metric is obtained by performing the complex coordinate
transformation
\begin{equation}\label{transfo}
x^\mu \longrightarrow \tilde{x}^\mu=x^\mu + i \, 
y^\mu(x^\sigma)\,,
\end{equation}
where $y^\mu \left( x^\sigma \right)$ are analytic functions of 
the real coordinates $x^\sigma$,  and simultaneously letting the 
null tetrad  $Z^\mu_a\,\equiv\,(l^\mu,n^\mu,m^\mu,\bar{m}^\mu)$ 
with $a\,=\,1,2,3,4$, undergo the transformation
\begin{equation}
Z^\mu_a \longrightarrow 
\tilde{Z}^\mu_a(\tilde{x}^\sigma, 
\bar{\tilde{x}}^\sigma)\, 
=\,Z^\rho_a \, \frac{\partial\tilde{x}
^\mu}{\partial x^\rho} \,.
\end{equation} 
Obviously, one has to 
recover the old tetrad and metric as soon as
$\tilde{x}^\sigma\,=\,\bar{\tilde{x}}^\sigma$. In summary, the
effect of the ``tilde transformation'' (\ref{transfo}) is 
to generate a new metric whose components are real functions of
complex variables, 
\begin{equation}
g_{\mu\nu} \longrightarrow 
\tilde{g}_{\mu\nu}\,: 
\,\; \tilde{\mathbf{x}}\times\tilde{\mathbf{x}}\mapsto
\mathbb{R}
\end{equation} 
with
\begin{equation}
\tilde{Z}^\mu_a(\tilde{x}^\sigma, 
\bar{\tilde{x}}^\sigma)|_{\mathbf{x}=\tilde{\mathbf{x}}}
=Z^\mu_a(x^\sigma) \,.
\end{equation} 
For our purposes, we can make the choice
\begin{equation}\label{transfo_1}
\tilde{x}^\mu\,= 
\,x^\mu + ia 
\left( \delta^\mu_1-\delta^\mu_0 \right) \cos\theta 
\longrightarrow
\left\{\begin{array}{ll}
\tilde{u}\,=\,u+ia\cos\theta 
\,,\\\\
\tilde{r}\,=\,r-ia\cos\theta \,,\\\\
\tilde{\theta}\,=\,\theta \,,\\\\
\tilde{\phi}\,=\,\phi \,,\\\\
\end{array}\right.
\end{equation}
where $a$ is a constant and, with the choice
$\tilde{r}\,= \,\bar{\tilde{r}}$,  the  null 
vectors (\ref{tetradvectors1})--(\ref{tetradvectors4}) reduce to 
\begin{eqnarray} 
\tilde{l}^\mu\, & = & \,\delta^\mu_1 \,,  
\label{tetradvectors_2a}\\
&&\nonumber\\
\tilde{n}^\mu\, & = & \,-\frac{1}{2} \, 
\mbox{e}^{-2\lambda(\tilde{r},\theta)} \, \delta^\mu_1+
\mbox{e}^{-\lambda(\tilde{r},\theta) 
-\phi(\tilde{r},\theta)} \, \delta^\mu_0 
\,,\label{tetradvectors_2b}\\
&&\nonumber\\
\tilde{m}^\mu\, & = & \,\frac{1}{\sqrt{2}(\tilde{r} 
-ia \cos\theta)}\biggl[ ia(\delta^\mu_0-\delta^\mu_1) 
\sin\theta+\delta^\mu_2
+\frac{i}{\sin{\theta}}\delta^\mu_3\biggr] \,, 
\label{tetradvectors_2c}\\
&&\nonumber\\
\bar{\tilde{m}}^\mu\, & = & \,\frac{1}{\sqrt{2}(\tilde{r} 
+ia\cos\theta)}\biggl[ -ia(\delta^\mu_0 
-\delta^\mu_1)\sin\theta +\delta^\mu_2
-\frac{i}{\sin{\theta}}\delta^\mu_3\biggr] \,. 
\label{tetradvectors_2d}
\end{eqnarray}
A new metric is recovered  from the transformed null tetrad via 
Eq.~(\ref{eq:gmet}). With the null  
vectors (\ref{tetradvectors_2a})--(\ref{tetradvectors_2d}) 
and the transformation~(\ref{transfo_1}), the new 
metric in coordinates $\tilde{x}^\mu\, 
=\,(\tilde{u},\tilde{r},\theta,\phi)$ is
\begin{equation}
\tilde{g}^{\mu\nu}  = \left(
\begin{array}{cccc} \label{eqn:newmetric}
\frac{a^2 \sin^2{\theta}}{\Sigma^2} &
-\mbox{e}^{-\lambda(\tilde{r},\theta)-\phi(\tilde{r},\theta)} -
\frac{a^2\sin^2{\theta}}{\Sigma^2} & 0 & \frac{a}{\Sigma^2} \\
&&&\\
. &  \mbox{e}^{-2\lambda(\tilde{r},\theta)} +
\frac{a^2\sin^2{\theta}}{\Sigma^2}
& 0 & - \frac{a}{\Sigma^2} \\
&&&\\
. & . & \frac{1}{\Sigma^2} & 0 \\
&&&\\
. & . & . & \frac{1}{\Sigma^2\sin^2{\theta}} \\
\end{array}      \right)
\end{equation}
where $\Sigma = \sqrt{\tilde{r}^2 + a^2\cos^2{\theta}}$. The 
covariant metric $ \tilde{g}_{\mu\nu} $ is
\begin{eqnarray} 
 && \left(
\begin{array}{cccc}
-\mbox{e}^{2\phi(\tilde{r},\theta)} & 
-\mbox{e}^{\lambda(\tilde{r},\theta) -
\phi(\tilde{r},\theta)} & 0 & -a
\mbox{e}^{\phi(\tilde{r},\theta)} \left[ 
\mbox{e}^{\lambda(\tilde{r},\theta)}-
\mbox{e}^{\phi(\tilde{r},\theta)} \right] \sin^2{\theta} \\ 
&&&\\
. & 0 & 0 &  a \mbox{e}^{\phi(\tilde{r},\theta) 
-\lambda(\tilde{r},\theta)} \sin^2{\theta} \\ 
&&&\\
. & . & \Sigma^2 & 0 \\
&&&\\
. & . & . & \left[ \Sigma^2 + a^2 \sin^2{\theta} 
\mbox{e}^{\phi(\tilde{r},\theta)}(2 \mbox{e}^{\lambda(\tilde{r},
\theta)}- \mbox{e}^{\phi(\tilde{r},\theta)}) \right] 
\sin^2{\theta} \\
\end{array}      \right) \nonumber\\
&&  \label{eqn:coform}
\end{eqnarray}
The dots in the matrix  denote symmetric entries satisfying the 
metric symmetry $ g^{\mu\nu} = g^{\nu\mu}$. The 
form of this metric gives the general result of the Newman-Janis 
algorithm starting from  any 
spherical  seed metric.

The metric~(\ref{eqn:coform}) can be  simplified by 
a further gauge  transformation so that the only off-diagonal
component is $g_{\phi t}$.  This procedure makes it easier to 
compare with the standard 
Boyer-Lindquist form of the 
Kerr 
metric \cite{thorne} and to interpret physical properties such as 
frame dragging. The coordinates $\tilde{u}$ and $\varphi$ can be
redefined in such a way that the metric in the new coordinates  
has the properties described above. Explicitly, using 
\begin{equation}
d\tilde{u}\,=
\,dt+g(\tilde{r})d\tilde{r}
\end{equation}
and 
\begin{equation}
d\phi\,=\,d\phi+h(\tilde{r})d\tilde{r} \,,
\end{equation}
where
\begin{eqnarray}
g(\tilde{r})  & = & -\frac{ \mbox{e}^{\lambda(\tilde{r},\theta)}
(\Sigma^2 + a^2\sin^2{\theta} 
\mbox{e}^{\lambda(\tilde{r},\theta) +\phi(\tilde{r},\theta)})}
{ \mbox{e}^{\phi(\tilde{r},\theta)}(\Sigma^2 
+a^2 \sin^2{\theta} \mbox{e}^{2\lambda(\tilde{r},\theta)})} \,,\\
&&\nonumber\\
h(\tilde{r}) &= & -\frac{a \, 
\mbox{e}^{2\lambda(\tilde{r},\theta)}}{ \Sigma^2+a^2 
\sin^2{\theta} \, \mbox{e}^{2\lambda(\tilde{r},\theta)}} \,,
\end{eqnarray}
after algebraic manipulations the 
covariant metric~(\ref{eqn:coform}) becomes, 
in coordinates $(t,\tilde{r},\theta,\phi)$,
\begin{eqnarray} 
& & \left(
\begin{array}{cccc}
\mbox{e}^{2\phi} & 0 & 0 & a \, 
\mbox{e}^{\phi}[ \mbox{e}^{\lambda}-
\mbox{e}^{\phi}] \sin^2{\theta} \\
&&&\\
. & -\frac{\Sigma^2}{ (\Sigma^2 
\mbox{e}^{-2\lambda} +
a^2\sin^2{\theta})}
 & 0 & 0 \\
&&&\\
. & . & -\Sigma^2 & 0 \\
&&&\\
. & . & . & - \left[  \Sigma^2 +
a^2\sin^2{\theta} \, 
\mbox{e}^{\phi}(2 
\mbox{e}^{\lambda}-
\mbox{e}^{\phi}) \right] \sin^2{\theta} \\
\end{array} \right) \,, \nonumber\\
&& \label{eq:blform}
\end{eqnarray}
where $ \phi= \phi \left( \tilde{r}, \theta \right) $ and  $  
\lambda=\lambda  \left( \tilde{r}, \theta \right) $. This metric 
represents the complete family of metrics that may be 
obtained by performing the Newman-Janis 
algorithm
 on any static 
spherically symmetric seed metric, written
in Boyer-Lindquist 
coordinates.  These
transformations require that $\Sigma^2+a^2\sin^2\theta \, 
\mbox{e}^{2\lambda(\tilde{r},\theta)}\neq 0$,  where  
$ \mbox{e}^{2\lambda(\tilde{r},\theta)}> 0$.
We  now  show that this approach can be used to 
derive axially symmetric solutions also in 
$f(R)$-gravity.

Begin with the spherically symmetric  
solution~(\ref{textbf4dot89}), that we rewrite  as
\begin{equation}\label{sol_noe_2}
ds^2=- \left( \alpha+\beta r \right) dt^2 + 
\frac{\beta 
r}{ 2\left(  \alpha+\beta r \right)} \, dr^2 + r^2d\Omega_2^2 \,,
\end{equation} 
where $\alpha$ is a combination of  $\Sigma_0$,  $k$, and 
$\beta=k_1$ obtained with the Noether approach. The metric 
tensor in Eddington-Finkelstein  coordinates $ \left( 
u,r,\theta,\phi \right)$ of the form 
(\ref{metrictensorcontro}) is
\begin{equation}\label{metrictensorcontro_noe}
g^{\mu\nu} = \left(
\begin{array}{cccc}
0 & - \sqrt{\frac{2}{\beta r}} & 0 & 0     \\
&&&\\
. & 2+\frac{2\alpha}{\beta r}& 0 & 0 \\
&&&\\
. & . & \frac{1}{r^2} & 0 \\
&&&\\
. & . & . & \frac{1}{r^2\sin^2{\theta} }
\end{array}    \right).
\end{equation}
The complex null tetrad 
(\ref{tetradvectors1})--(\ref{tetradvectors4}) is now
\begin{eqnarray}
l^\mu\, & = & \,\delta^\mu_1 \,,  \label{tetradvectors_noe1}\\
&&\nonumber\\
n^\mu\, & = & \,-\biggl[1+\frac{\alpha}{\beta} 
\biggl(\frac{1}{\bar{r}}
+\frac{1}{r}\biggr)\biggr]\delta^\mu_1+
\sqrt{\frac{2}{\beta}} \, \frac{1}{ \left(\bar{r}r \right)^{1/4} 
} \,  \delta^\mu_0 \,, \label{tetradvectors_noe2}\\
&&\nonumber\\
m^\mu\, & = & \,\frac{1}{\sqrt{2}\bar{r}} \left( \delta^\mu_2
+\frac{i}{\sin{\theta}}\delta^\mu_3 \right) 
\,.  \label{tetradvectors_noe3}
\end{eqnarray}
By computing the complex coordinate  
transformation~(\ref{transfo_1}),  the null tetrad becomes
\begin{eqnarray}
\tilde{l}^\mu\, & = & \,\delta^\mu_1 \,,  
\label{tetradvectors_noe_1a}\\
&&\nonumber\\
\tilde{n}^\mu\, & = & \,-
\biggl[1+\frac{\alpha}{\beta} \, \frac{\text{ {\mbox Re}}\{ 
\tilde{r}\}}{\Sigma^2}\biggr]\delta^\mu_1+
\sqrt{\frac{2}{\beta \Sigma}} \, \delta^\mu_0 
\,,
\label{tetradvectors_noe_1b}\\
&&\nonumber\\
\tilde{m}^\mu\, & = & \,\frac{1}{\sqrt{2} \left( \tilde{r} 
+ia\cos\theta \right)}\biggl[ ia \left( 
\delta^\mu_0-\delta^\mu_1 \right) 
\sin\theta+\delta^\mu_2
+\frac{i}{\sin{\theta}}\delta^\mu_3\biggr] \,.
\label{tetradvectors_noe_1c}
\end{eqnarray}
By performing the same procedure as in GR, one  derives  
an axially symmetric metric of the 
form~(\ref{eq:blform}) but 
starting from  the spherically symmetric
covariant metric~(\ref{sol_noe_2}), 
\begin{eqnarray}
& & g_{\mu\nu}=  \left(
\begin{array}{cccc}
-\frac{r(\alpha+\beta r)+a^2\beta\cos^2\theta}{\Sigma} & 0 & 0 &
\frac{ \sigma_1}{2\Sigma} 
\\
&&&\\
. & \frac{\beta\Sigma^2}{2\alpha r+\beta(a^2+r^2+\Sigma^2)}
 & 0 & 0 \\
&&&\\
. & . & \Sigma^2 & 0 \\
&&&\\
. & . & . & \biggl( \Sigma^2 -\frac{ \sigma_2}{\Sigma}\biggr)
\sin^2\theta \\
\end{array}\,. \right) \,.\nonumber\\
&& \label{rotating}
\end{eqnarray}
where 
\begin{eqnarray}
\sigma_1 &= & a \left( 2\alpha r + 2\beta\Sigma^2 - 
\sqrt{2\beta}\Sigma^{3/2} \right) 
\sin^2\theta \,,\\
&&\nonumber\\
\sigma_2  &=&  a^2 \left( \alpha 
r+\beta\Sigma^2-\sqrt{2\beta}\Sigma^{3/2} \right) \sin^2\theta 
\,.
\end{eqnarray}
By setting  $a=0$, the metric~(\ref{sol_noe_2}) is immediately 
recovered.

The method illustrated by this example is general and can be 
extended to any spherically symmetric solution of  
$f(R)$-gravity \cite{cqgaxially}.
The key point of the method is to find out a suitable  complex transformation which, from a physical viewpoint, corresponds to the fact that we are reducing the number of independent Killing vectors. From a mathematical viewpoint, it is  useful since allows to overcome the problem of a  direct search for  axially symmetric solutions that, in $f(R)$-gravity, could be extremely cumbersome due to the fourth-order  field equations.  However,  other generating techniques  exist and all of them should be explored in order to completely extend solutions of GR to $f(R)$-gravity. They can be  more general and solid than the Newman-Janis approach. A good source for references and basic features of generating techniques is reference \cite{Stephanietal03}. In particular, the paper by Talbot  \cite{talbot},  considering the Newman-Penrose approach to twisting degenerate metrics, provides some theoretical justification for the scope and limitations of adopting the ``complex trick''.  As reported in Chap. 21 of \cite{Stephanietal03},  several techniques can be pursued to achieve axially symmetric solutions which can be particularly useful to deal with non-empty space-times ( in particular when perfect fluids are the sources of the field equations) and to deal, in general,  with problems related to  Einstein-Maxwell field equations.
We have to stress that the utility of  generating techniques is not simply to obtain a new metric, but a metric of a new space-time with specific properties as the transformation properties of the energy-momentum tensor and Killing vectors. In its original application, the Newman--Janis procedure transforms an Einstein-Maxwell solution (Reissner-Nordstrom) into another Einstein-Maxwell solution (Kerr-Newman). As a particular case (setting the charge to zero) it is possible to achieve  the transformation between  two vacuum solutions (Schwarzschild and Kerr).
Also in case of $f(R)$-gravity,  new features emerge by adopting such a technique.  In particular, it is worth studying how certain features  of spherically simmetric metrics, derived in $f(R)$-gravity, result transformed in the new axially symmetric solutions. For example,  considering the $f(R)$  spherically symmetric solution studied here, the Ricci scalar evolves as  $ r^{-2}$ and then the asymptotic flatness is recovered.
Let us consider now the axially symmetric metric achieved by the Newman-Janis method. The parameter $a\neq 0$ indicates that the spherical symmetry ($a=0$) is broken. Such a parameter can be immediately related to the presence of an axis of symmetry and then to the fact that a Killing vector, related to the angle  $\theta$, has been lost.
To conclude, we can say that once the vacuum case is discussed, more general spherical metrics can be transformed in new axially symmetric metrics  adopting more general techniques \cite{Stephanietal03}.

\section{The Post- Newtonian limit}

\subsection{The weak field and small velocities approximations}

At shorter (Galactic and Solar System)  spatial scales, ETGs 
exhibit gravitational potentials with non-Newtonian 
corrections  \cite{Will93, 
DamourFarese92, Will01,review}. This feature was discovered 
long ago 
\cite{Stelle78}, 
and recent interest arises from the possibility of explaining 
the flatness of the rotation curves of 
spiral galaxies  without 
huge amounts of dark matter. In particular, 
the rotation curves 
of a wide sample of low surface brightness spiral galaxies can be  
fitted successfully by the corrected potentials \cite{prl, 
CapCardTro07}, 
and this possibility may be extended to  other 
types of galaxies \cite{Frigerio}.

One could attempt to investigate other issues such as, for 
example,  the Pioneer anomaly 
\cite{Anderson98,  
Anderson02} with  the 
same approach \cite{bertolami}. A systematic analysis of ETGs  
at scales much smaller than the Hubble radius is then necessary. 
In this section we discuss the weak-field limit of $f(R)$-gravity 
 without specifying the form of the theory and 
highlighting the differences and similarities with the
post-Newtonian and post-Minkowskian limits of GR. The literature
contains conflicting claims
\cite{Accioly:1999a, Soussa:2003re, 
dick, Easson:2004fq,  
Navarro:2005da, navarro, Cembranos06, Shao:2005wt, 
 Olmo:2005zr,  
ppnantro,  BarrowClifton2, BarrowClifton3, 
BarrowClifton1, ppn-noi-bis, Multamaki:2006zb, 
Sotiriou06, 
Zhang:2007ne,
Hu, Baghram:2007df, Capozziello:2007id, 
Capozziello:2007eu, Multamaki:2007jk, Iorio:2007ee, 
noi-prd, Ruggiero:2006qv}, and 
clarity is needed  in order to compare theory and experiment .
Based on the scalar-tensor representation of $f(R)$-gravity 
with\footnote{Although some caution about the 
equivalence  with scalar-tensor theory in the Newtonian and the 
GR limits  is necessary \cite{Faraoni06PRD, 
Kainulainen:2007bt}, the equivalence holds in the post-Newtonian 
limit \cite{Faraoni:2007yn}.}  $\omega=0$,   Chiba 
\cite{Chiba03PLB} originally  suggested that  all $f(R)  $ 
theories are  ruled out because of the experimental limit $ 
\left| \omega \right| >40,000 $ \cite{BertottiIessTortora}. 
While this constraint can be circumvented by giving the scalar 
degree of freedom 
a large mass and, therefore, making it short-ranged, it seemed 
that its range must be at least comparable with the Hubble radius 
in order to affect the dynamics of the Universe. This conclusion 
is incorrect because of the so called {\it chameleon 
mechanism} and the 
weak-field limit is subtler than it appears \cite{Capozziello:2007eu}. 
Solar System experiments constrain the 
PPN parameter  $\gamma$, and then the Brans-Dicke parameter 
$\omega$, only when the range of the scalar degree of freedom 
is comparable to, or larger than the spatial scale of the 
experiment (for the Cassini  experiment  
providing the lower bound 
on $\omega$ \cite{BertottiIessTortora}, this is the 
size of the Solar System) \cite{Wagoner70}.  If the mass of this 
scalar is large, the parameter  $\gamma$ is close 
to unity. However, the scalar does not have a fixed range 
but, rather, its mass depends on the energy density of its  
environment, so that this field becomes short-ranged  and is 
undetectable at small (Solar System) scales, while its range is 
cosmological at cosmological densities $\rho$. This  
chameleon mechanism is widely used in 
quintessence models 
of dark energy  \cite{Khoury:2003aq, 
Khoury:2003rn}.

Again, a  direct approach independent of the 
equivalence of metric $f(R)$ and scalar-tensor gravity is 
more convincing, and was first formulated for the prototype model 
$f(R)=R-\mu^4/R$ (which, at the time,  was already ruled out by 
the Dolgov-Kawasaki  instability \cite{DolgovKawasaki03, 
mattmodgrav}) in  \cite{ErickcekSmithKamionkowski06}. The 
weak-field limit 
for  a general function $f(R)$ was presented in  
\cite{Chiba:2006jp, Olmo:2006eh}.

Weak-field experiments  such as light bending, the perihelion 
shift of planets, and  frame-dragging experiments are valuable 
tests of ETGs. There are  sufficient  theoretical predictions to 
state that certain higher  order theories of  gravity can be 
compatible with Newtonian and  post-Newtonian experiments  
\cite{ppnantro, Sotiriou06, allrugg, 
NojiriOdintsov07PLB, Chiba03PLB, BarrowClifton2,  Faraoni06PRD, 
ErickcekSmithKamionkowski06, Chiba:2006jp,clifton2008,lubini,berry,quadrupolo,naef}, 
as can be shown also by  
using  the scalar-tensor representation of $f(R)$-gravity.

In the following we outline a  formalism addressing  the  
weak-field and  small velocity limit of fourth order 
gravity allowing a  Jordan frame systematic 
discussion  
of these limits  and of spherically  symmetric solutions    
\cite{noimpla}. This discussion is valid also for   general 
higher order  theories  containing the invariants 
$R_{\mu\nu}R^{\mu\nu}$ or 
$R_{\alpha\beta\mu\nu}R^{\alpha\beta\mu\nu}$ \cite{dirk}.  The 
non-Newtonian corrections in the gravitational potentials 
could potentially  explain known astrophysical phenomenology.

A preliminary step consists of  concentrating on the 
vacuum case and then  building  a Newtonian and 
post-Newtonian formalism for $f(R)$ theories in 
the presence of matter. It is possible to  estimate the 
post-Newtonian parameter $\gamma$ by  considering
second order solutions for the metric components {\em in vacuo}. 
For  
completeness, we treat the problem also by imposing the harmonic 
gauge on the field equations. 

\subsection{General remarks on Newtonian and post-Newtonian approximations}

Certain general features must be  taken into  account when 
performing the Newtonian and post-Newtonian limits of 
a relativistic theory of gravity. For a virialized system of 
particles of total mass $\bar{M}$ interacting gravitationally, 
the kinetic energy $\bar{M} (\bar{v})^2 /2 $ is approximately of 
the same order of magnitude as the potential energy
$ U=G\bar{M}^2/\bar{r}$, where   $\bar{r}$ and 
$\bar{v}$ are typical average values of the separations  
and velocities of these particles. As a consequence, it is
\begin{equation}
\bar{v^2}\sim \frac{G\bar{M}}{\bar{r}} 
\end{equation}
(for instance, in Newtonian mechanics, a test particle in a 
circular orbit of radius $r$ about a spherically distributed  
mass $M$ has velocity $v$ given  by $v^2=GM/r$). The  
post-Newtonian approximation can be described as a method for   
obtaining the motion of the system beyond  first
({\em i.e.}, Newtonian) order  with respect to 
the quantities $G\bar{M}/\bar{r}$ and $(\bar{v})^2$, which are 
assumed to be small with respect to the square of the speed of 
light  $c^2$ (this approximation is an  expansion in inverse 
powers of $c$). 

Typical values of the Newtonian gravitational  potential $U$ 
in the Solar System  are  nowhere larger than $10^{-5}$  
(the quantity $U/c^2$ is dimensionless). Planetary 
velocities satisfy the condition $ (\bar{v})^2\lesssim U$,
while\footnote{Here the velocity $v$ is expressed in
units of $c$.} the matter pressure $ P $ 
inside the Sun and the planets is much smaller than the
energy density $\rho U$ of matter,\footnote{Typical 
values of $ P / \rho$ are $ 10^{-5}$ in the Sun and  $ 10^{-10}$ 
in the Earth \cite{Will93, Will01}.} $ P/\rho \lesssim U $. 
Furthermore, 
one  must consider that other forms of energy in the Solar 
System (stresses, radiation, thermal energy, {\em etc.})  have 
small magnitudes and their specific energy density $\Pi$ (the 
ratio of the energy density to the rest mass density) is related 
to $U$ by $\Pi\lesssim U$ ($\Pi$ is approximately $ 10^{-5}$ in 
the Sun and $ 10^{-9}$ in the Earth
\cite{Will93, Will01}). One can consider that these quantities, as 
functions  of velocity, give only second order 
contributions, 
\begin{equation}
U\sim v^2 \sim \frac{P}{\rho} \sim \Pi \sim \mathcal{O}(2)\,,
\end{equation}
therefore the velocity $v$ contributes to order $\mathcal{O}(1)$,  $U^2$ to  
order $\mathcal{O}(4)$, $Uv$ to order $\mathcal{O}(3)$, $U\Pi$ is of
order $\mathcal{O}(4)$, {\em etc.} In this approximation, one has
\begin{equation}
\frac{\partial}{\partial x^0}\sim\textbf{v}\cdot\nabla\,,
\end{equation}
and
\begin{equation}
\frac{|\partial/\partial x^0|}{|\nabla|}\sim \mathcal{O}(1)\,.
\end{equation}
Massive test particles move along geodesics given by the equation
\begin{equation}
\frac{d^2x^\mu}{ds^2}+ 
\Gamma^\mu_{\sigma\tau}\frac{dx^\sigma}{ds} 
\, \frac{dx^\tau}{ds}=0\,, 
\end{equation}
or 
\begin{equation}
\frac{d^2x^i}{dx^{0\,2}}= 
-\Gamma^i_{00}-2\Gamma^i_{0m}\frac{dx^m}{dx^0}-
\Gamma^i_{mn} \, \frac{dx^m}{dx^0}\frac{dx^n}{dx^0} 
+\biggl( \Gamma^0_{00}+
2\Gamma^0_{0m}\frac{dx^m}{dx^0} 
+2\Gamma^0_{mn}\frac{dx^m}{dx^0}\frac{dx^n}{dx^0}\biggr) 
\frac{dx^i}{dx^0}\,.  
\end{equation}
In the small velocity approximation and retaining only  
first order terms in the deviations of  $g_{\mu\nu}$ 
from the Minkowski metric $\eta_{\mu\nu}$,  the particle 
equations of  motion reduce to the Newtonian  result
\begin{equation}
\frac{d^2x^i}{d (x^0)^ 2 }\simeq-\Gamma^i_{00}\simeq 
- \frac{1}{2}\frac{\partial g_{00}}{\partial x^i}\,.
\end{equation}
The quantity $ \left( 1 + g_{00} \right) $ is of order 
$G\bar{M}/\bar{r}$, hence  the Newtonian approximation gives
$ \displaystyle{\frac{d^2x^i}{d(x^0)^2 } }$ to order
$G\bar{M}/\bar{r}^2$, that is, to order $ (\bar{v})^2/r$. As a
consequence, the post-Newtonian approximation requires one  to 
compute $\displaystyle\frac{d^2x^i}{d(x^0)^2 }$ to  order
$(\bar{v})^4/\bar{r}$. According  to the Equivalence 
Principle
and 
the local flatness of the space-time manifold, it is possible 
to find  a coordinate system in which the  metric
tensor is nearly equal to $\eta_{\mu\nu}$, with the
correction expanded in powers of $G\bar{M}/\bar{r}\sim 
( \bar{v})^2$,
\begin{eqnarray}
g_{00}(x^0,\textbf{x}) & = &  -1 + g^{(2)}_{00}(x^0,\textbf{x})
+g^{(4)}_{00}(x^0,\textbf{x})+\mathcal{O}(6) 
\,, \label{chapter5approx1a}\\
&&\nonumber\\
g_{0i}(x^0,\textbf{x}) & = &   g^{(3)}_{0i}(x^0,\textbf{x}) 
+\mathcal{O}(5) \,, \label{chapter5approx1b}\\
&&\nonumber\\
g_{ij}(x^0, \textbf{x}) & = &  \delta_{ij}+g^{(2)}_{ij}(x^0,
\textbf{x})+\mathcal{O}(4) \,,\label{chapter5approx1c}
\end{eqnarray}
and with inverse metric 
\begin{eqnarray}
g^{00}(x^0,\textbf{x}) & = &  - 1 +g^{(2)00}(x^0, 
\textbf{x})+g^{(4)00}(x^0, \textbf{x})+\mathcal{O}(6) \,,
\label{chapter5approx2a}\\
&&\nonumber\\
g^{0i}(x^0,\textbf{x}) & = &  g^{(3)0i}(x^0,\textbf{x}) 
+\mathcal{O}(5) \,, \label{chapter5approx2b}\\
&&\nonumber\\
g^{ij}(x^0,\textbf{x}) & = &  \delta^{ij} 
+g^{(2)ij}(x^0,{\textbf{x}})+\mathcal{O}(4)\,.
\label{chapter5approx2c}
\end{eqnarray}
When computing the connection  coefficients 
$\Gamma^\mu_{\alpha\beta}$, one must take into  account the 
fact that the space and time scales in the gravitational system  
are set by $\bar{r}$ and $\bar{r}/ \bar{v}$, respectively, 
hence spatial and time derivatives are  of order
\begin{equation}
\frac{\partial}{\partial x^i}\sim\frac{1}{\bar{r}}\,, \ \ \ \ \ \
\ \frac{\partial}{\partial x^0}\sim\frac{ \bar{v} }{\bar{r}} \,.
\end{equation}
Using the 
approximations~(\ref{chapter5approx1a})--(\ref{chapter5approx2c}),
we have
\begin{eqnarray}
{\Gamma^{(3)}}^{0}_{00} & = & \frac{1}{2}g^{(2),0}_{00}\, ,\\
&&\nonumber\\
{\Gamma^{(2)}}^{i}_{00} & = & \frac{1}{2}g^{(2),i}_{00} 
\,,\\
&&\nonumber\\
{\Gamma^{(2)}}^{i}_{jk} & = &  
\frac{1}{2}\biggl({g^{(2)}}^{,i}_{jk}
-{g^{(2)}}^{i}_{j,k}-{g^{(2)}}^{i}_{k,j}\biggr)\,,\\
&&\nonumber\\
{\Gamma^{(3)}}^{0}_{ij} & = & \frac{1}{2}\biggl
({g^{(3)}}^{0}_{i,j}+{g^{(3)}}^{0}_{j,i} 
-{g^{(3)}}^{,0}_{ij}\biggr) \,,\\
&&\nonumber\\
{\Gamma^{(3)}}^{i}_{0j} & = 
& \frac{1}{2}\biggl({g^{(3)}}^{,i}_{0j} 
-{g^{(3)}}^{i}_{0,j}-{g^{(2)}}^{i}_{j,0}\biggr)\,,\\
&&\nonumber\\
{\Gamma^{(4)}}^{0}_{0i} & = & \frac{1}{2}\biggl
({g^{(4)}}^{0}_{0,i}+g^{(2)00}g^{(2)}_{00,i}\biggr) \,,\\
&&\nonumber\\
{\Gamma^{(4)}}^{i}_{00} & = 
& \frac{1}{2}\biggl({g^{(4)}}^{,i}_{00} 
+g^{(2)im}g^{(2)}_{00,m}-2{g^{(3)}}^{i}_{0,0}\biggr)\,,\\
&&\nonumber\\
{\Gamma^{(2)}}^{0}_{0i} & = & \frac{1}{2}{g^{(2)}}^{0}_{0,i} \,.
\end{eqnarray}
The only non-vanishing components of the Ricci tensor are
\begin{eqnarray}
R^{(2)}_{00} & = & \frac{1}{2} \, \nabla^2 g^{(2)}_{00} 
\,,\\
&&\nonumber\\
R^{(4)}_{00} & = & \frac{1}{2} \, \nabla^2
g^{(4)}_{00}-\frac{1}{2} \, {g^{(2)}}^{mn}_{,m}g^{(2)}_{00,n} 
-\frac{1}{2} \, {g^{(2)}}^{mn}g^{(2)}_{00,mn} \nonumber\\
&&\nonumber\\
& + & \frac{1}{2} \, {g^{(2)}}^{m}_{m,00} 
-\frac{1}{4} \, {g^{(2)}}^{0,m}_{0}g^{(2)}_{00,m} 
 -  \frac{1}{4} \, {g^{(2)}}^{m,n}_{m}g^{(2)}_{00,n} 
-{g^{(3)}}^{m}_{0,m0}  \,,\\
&&\nonumber\\
R^{(3)}_{0i} & = & \frac{1}{2} \, \nabla^2
g^{(3)}_{0i}-\frac{1}{2} \, {g^{(2)}}^{m}_{i,m0} 
-\frac{1}{2} \, {g^{(3)}}^{m}_{0,mi} 
+\frac{1}{2} \, {g^{(2)}}^{m}_{m,0i} \,,\\
&&\nonumber\\
R^{(2)}_{ij} & = & \frac{1}{2} \, \nabla^2
g^{(2)}_{ij}-\frac{1}{2} \, {g^{(2)}}^{m}_{i,mj} 
-\frac{1}{2} \, {g^{(2)}}^{m}_{j,mi} 
-\frac{1}{2} \, {g^{(2)}}^{0}_{0,ij} \nonumber\\
&&\nonumber\\
& + & \frac{1}{2} \, {g^{(2)}}^{m}_{m,ij} \,.
\end{eqnarray}
In the harmonic gauge  
$g^{\rho\sigma}\Gamma^\mu_{\rho\sigma}=0$  
these expressions become 
\begin{eqnarray}
R^{(2)}_{00} & = & \frac{1}{2} \, \nabla^2 g^{(2)}_{00} \,, 
\label{chapter5PPNobja}\\
&&\nonumber\\
R^{(4)}_{00} &= & \frac{1}{2} \, \nabla^2
g^{(4)}_{00}-\frac{1}{2} \, {g^{(2)}}^{mn}g^{(2)}_{00,mn} 
-\frac{1}{2} \, {g^{(2)}}^{0}_{0,00}-\frac{1}{2} \, 
\left| \vec{\nabla} \vec{\nabla}_\eta
g^{(2)}_{00} \right|^2 \,,\label{chapter5PPNobjb}\\
&&\nonumber\\
R^{(3)}_{0i} & = & \frac{1}{2} \, \nabla^2
g^{(3)}_{0i} \,,\label{chapter5PPNobjc}\\
&&\nonumber\\
R^{(2)}_{ij} & = & \frac{1}{2} \, \nabla^2
g^{(2)}_{ij} \,,  \label{chapter5PPNobjd}
\end{eqnarray}
where $\nabla^2$ and $\vec{\nabla} $ denote the Laplacian and the 
gradient in flat space, respectively. The Ricci scalar in this 
gauge is
\begin{eqnarray}
R^{(2)} &= & {R^{(2)}}^{0}_{0}- 
{R^{(2)}}^{m}_{m}= \frac{1}{2} \, \nabla^2 
{g^{(2)}}^{0}_{0}-\frac{1}{2} \, \nabla^2 
{g^{(2)}}^{m}_{m} \,, \\
&&\nonumber\\
R^{(4)} & = & {R^{(4)}}^{0}_{0}+{g^{(2)}}^{00}
R^{(2)}_{00} + {g^{(2)}}^{mn}R^{(2)}_{mn} \nonumber\\
&&\nonumber\\
 &= &\frac{1}{2} \, \nabla^2  
{g^{(4)}}^{0}_{0}-\frac{1}{2} \, {g^{(2)}}^{0,0}_{0,0} 
-\frac{1}{2} \, {g^{(2)}}^{mn}\biggl( {g^{(2)}}^{0}_{0,mn}-
\nabla^2 g^{(2)}_{mn}\biggr) -\frac{1}{2} \, | \vec{\nabla}
{g^{(2)}}^{0}_{0}|^2 \nonumber\\
&&\nonumber\\
& + &  \frac{1}{2}\, {g^{(2)}}^{00}\nabla^2
g^{(2)}_{00} \,.
\end{eqnarray}
The inverse of the metric tensor is defined by 
$  g^{\alpha\rho}g_{\rho\beta}=\delta^\alpha_\beta $. 
The relations between terms of order  higher than first are 
\begin{eqnarray}
g^{(2)00}(x^0,\textbf{x}) & = & -g^{(2)}_{00}(x^0,\textbf{x}) 
\,, \\
&&\nonumber\\
g^{(4)00}(x^0,\textbf{x}) & = & {g^{(2)}_{00}(x^0,\textbf{x})}^2 
-g^{(4)}_{00} (x^0,\textbf{x}) 
\, , \\
&&\nonumber\\
g^{(3)0i} (x^0,\textbf{x}) & = & g^{(3)}_{0i} (x^0,\textbf{x}) 
\,, \\
&&\nonumber\\ 
g^{(2)ij}(x^0,\textbf{x}) & = & -g^{(2)}_{ij}(x^0,\textbf{x}) \,.
\end{eqnarray}
Finally, the Lagrangian of a particle in the gravitational field 
is proportional to the invariant distance $ds$, 
\begin{eqnarray}
L &= & 
\biggl(g_{\rho\sigma}\frac{dx^\rho}{dx^0} 
\frac{dx^\sigma}{dx^0}\biggr)^{1/2} = \biggl(g_{00} 
+2g_{0m}v^m+g_{mn}v^mv^n\biggr)^{1/2} \nonumber\\
&&\nonumber\\
& = & \biggl(1+g^{(2)}_{00} 
+g^{(4)}_{00}+2g^{(3)}_{0m}v^m 
-\textbf{v}^2+g^{(2)}_{mn}v^mv^n\biggr)^{1/2}\,. 
\end{eqnarray}
To second order, this expression reduces to the Newtonian
test particle Lagrangian  
$L_{\text{Newt}}=\biggl(1+g^{(2)}_{00}-\textbf{v}^2\biggr)^{1/2}$,
where $ \displaystyle{ v ^2 
=\frac{dx^m}{dx^0}\frac{dx_m}{dx^0}}$. 
Post-Newtonian physics involves terms of 
order higher than fourth in the Lagrangian.

Since the odd-order perturbation terms 
$\mathcal{O}(1)$ or $\mathcal{O}(3)$ contain  odd powers of 
the velocity $\textbf{v}$ or of time derivatives, they are 
related to the dissipation or absorption of energy  by the 
system. Mass-energy conservation prevents losses of  energy and 
mass and, as a consequence, in the Newtonian limit it prevents  
terms of order $\mathcal{O}(1)$ and $\mathcal{O}(3)$ to appear  in the  Lagrangian. When  
contributions of order  higher than $\mathcal{O}(4)$ are included, different 
theories produce different predictions. For example, due to 
the  conservation of post-Newtonian energy, GR forbids  terms of 
order $\mathcal{O}(5)$, while terms of order $\mathcal{O}(7)$ can  appear and are related 
to the energy lost due to gravitational radiation.

\subsection{Corrections to the Newtonian potential}

Let us apply the formalism of the previous section to the  
weak-field and small velocity regime  of metric  $f(R)$-gravity.
Assuming spherical symmetry in vacuum and specifying with "$t$" the time index\footnote{However, as mentioned in previous sections, Solar System tests are analyzed using isotropic coordinates \cite{gravitation}. The transformations from isotropic to Schwarzschild coordinates are non-trivial, as shown in Sec. V D. In the following discussion, we will take into account  this issue and consider consistent  orders of approximation for  the metric.}, we have 
\begin{eqnarray}
g_{tt}(t, r) & = & A(t,r) \simeq  - 1 + 
g^{(2)}_{tt}(t,r)+g^{(4)}_{tt}(t,r) \,,
\label{chapter5definexpansa} \\
&&\nonumber\\
g_{rr}(t,r) & = & B(t,r)\simeq 1 + g^{(2)}_{rr}(t,r) \,,\\
&&\nonumber\\
g_{\theta\theta}(t,r) & = &  r^2 \,, 
\label{chapter5definexpansb}\\
&&\nonumber\\
g_{\phi\phi}(t,r) & = & r^2\sin^2\theta 
\,,\label{chapter5definexpansc}
\end{eqnarray}
while the inverse metric components are 
\begin{eqnarray} 
g^{tt} & = & A(t,r)^{-1} \simeq  - 1-g^{(2)}_{tt} 
+ {g^{(2)}_{tt}}^2-g^{(4)}_{tt} \,,\\
&&\nonumber\\
g^{rr} & = & B(t,r)^{-1}\simeq 1-g^{(2)}_{rr} \,,
\end{eqnarray}
the metric determinant is
\begin{equation}
g\simeq r^4\sin^2\theta\left[ -1+ \left( g^{(2)}_{rr} 
-g^{(2)}_{tt} 
\right) + \left( g^{(2)}_{tt}g^{(2)}_{rr}-g^{(4)}_{tt} 
\right) \right] \,, \end{equation}
and the Christoffel symbols are given by
\begin{eqnarray}
&& {\Gamma^{(3)}}^{t}_{tt} =  \frac{g^{(2)}_{tt,t}}{2}\,,  
\,\,\,   {\Gamma^{(2)}}^{r}_{tt}+{\Gamma^{(4)}}^{r}_{tt}=
\frac{g^{(2)}_{tt,r}}{2}+\frac{g^{(2)}_{rr}g^{(2)}_{tt,r} 
+g^{(4)}_{tt,r}}{2} \,,\\
& & \nonumber\\
&& {\Gamma^{(3)}}^{r}_{tr} =  -\frac{g^{(2)}_{rr,t}}{2}\,,  
\,\,\,    {\Gamma^{(2)}}^{t}_{tr}+{\Gamma^{(4)}}^{t}_{tr}
=\frac{g^{(2)}_{tt,r}}{2} 
+\frac{g^{(4)}_{tt,r}-g^{(2)}_{tt}g^{(2)}_{tt,r}}{2} \,,\\
& & \nonumber\\
&& {\Gamma^{(3)}}^{t}_{rr}  =  -\frac{g^{(2)}_{rr,t}}{2}\,,  
\,\,\,   {\Gamma^{(2)}}^{r}_{rr}+{\Gamma^{(4)}}^{r}_{rr}
=-\frac{g^{(2)}_{rr,r}}{2}- 
\frac{g^{(2)}_{rr}g^{(2)}_{rr,r}}{2}\,, \\
& & \nonumber\\
&& \Gamma^{r}_{\phi\phi}  =  \sin^2\theta 
\Gamma^{r}_{\theta\theta}\,,  \,\,\,  
{\Gamma^{(0)}}^{r}_{\theta\theta}+
{\Gamma^{(2)}}^{r}_{\theta\theta}
+{\Gamma^{(4)}}^{r}_{\theta\theta}=
-r- r \, g^{(2)}_{rr}-r \, {g^{(2)}_{rr}}^2 \,.
\end{eqnarray}
The only non-vanishing components of the Ricci tensor are
\begin{eqnarray} 
R_{tt} & = & R^{(2)}_{tt}+R^{(4)}_{tt} \,,\\
&&\nonumber\\
R_{tr}& =& R^{(3)}_{tr} \,, \\
&&\nonumber\\
R_{rr}&=& R^{(2)}_{rr} \,, \\
&&\nonumber\\
R_{\theta\theta}&=& R^{(2)}_{\theta\theta} \,, \\
&&\nonumber\\
R_{\phi\phi}&=&  R^{(2)}_{\theta\theta} \sin^2\theta \,,
\end{eqnarray}
where
\begin{eqnarray}
R^{(2)}_{tt} & = & \frac{r \, 
g^{(2)}_{tt,rr}+2g^{(2)}_{tt,r}}{2r} 
\,,\\
&&\nonumber\\
R^{(4)}_{tt} & = & \biggl[ -r (g^{(2)}_{tt,r})^2
+4 \, g^{(4)}_{tt,r}+ r \, g^{(2)}_{tt,r} \, g^{(2)}_{rr,r} 
+2 \, g^{(2)}_{rr} \left( 2g^{(2)}_{tt,r}
+r \, g^{(2)}_{tt,rr}\right) \nonumber\\
&&\nonumber\\
& + &  2 r \, g^{(4)}_{tt,rr}
+2r \, g^{(2)}_{rr,tt} \biggr] \left( 4r \right)^{-1} \,,\\
&&\nonumber\\
R^{(3)}_{tr} & = & -\frac{g^{(2)}_{rr,t}}{r} \,,\\
&&\nonumber\\
R^{(2)}_{rr} & = & -\frac{r 
\, g^{(2)}_{tt,rr}+2 \, g^{(2)}_{rr,r}}{2r} 
\,,\\
&&\nonumber\\
R^{(2)}_{\theta\theta} & = & 
-\frac{2 \, g^{(2)}_{rr}+r \left( g^{(2)}_{tt,r} 
+g^{(2)}_{rr,r} \right) }{2} \,,
\end{eqnarray}
and the  post-Newtonian Ricci  scalar is
\begin{equation}
R \simeq R^{(2)}+R^{(4)}
\end{equation}
with
\begin{eqnarray}
R^{(2)} & = & \frac{2 \, g^{(2)}_{rr} 
+r \left( 2 
\, g^{(2)}_{tt,r}+2 \, g^{(2)}_{rr,r}+r \, g^{(2)}_{tt,rr} 
\right)}{r^2} \,, \\
&&\nonumber\\
R^{(4)} & = & \left\{ 4{g^{(2)}_{rr}}^2+2rg^{(2)}_{rr} 
\left( 2g^{(2)}_{tt,r} 
+4g^{(2)}_{rr,r}+rg^{(2)}_{tt,rr} \right) \right. \\
&&\nonumber\\
& + & \left. r\left[ -r{g^{(2)}_{tt,r}}^2+4g^{(4)}_{tt,r}+r
g^{(2)}_{tt,r}g^{(2)}_{rr,r} -2g^{(2)}_{tt} \left( 
2g^{(2)}_{tt,r} +rg^{(2)}_{tt,rr}\right)
 +2rg^{(4)}_{tt,rr}+2rg^{(2)}_{rr,tt} \right]\right\} \nonumber\\
&&\nonumber\\
& \cdot & \left( 2r^2 \right)^{-1}  \,.
\end{eqnarray}
We  restrict the discussion to functions $f(R)$ which are 
analytic at the value 
$R_0$ of the Ricci curvature,\footnote{At least,  the 
non-analytic part of $f(R)$ (if it is allowed to exist) must go  
to  zero faster than $R^3$ as $R\rightarrow 0$.} 
\begin{equation}\label{chapter5sertay}
f(R)=\sum_{n=0}^{+\infty} \frac{f^n(R_0)}{n!} \, \left(R-R_0 
\right)^n = f_0+f_1R+f_2R^2+f_3R^3 + \, ... \,,
\end{equation}
where in the last equality we have assumed $R_0=0$. The 
coefficient $f_1$  must be positive in order to  have a
positive gravitational coupling. The post-Newtonian 
formalism consists of  using this expansion 
in the field 
equations, which are expanded  
to  orders $\mathcal{O}(0)$, $\mathcal{O}(2)$, and  $\mathcal{O}(4)$, and then solved.

The substitution of Eq.~(\ref{chapter5sertay}) in the vacuum  
field equations and their expansion to orders $\mathcal{O}(0)$, $\mathcal{O}(2)$, and 
$\mathcal{O}(4)$   yield
\begin{eqnarray}
{\cal H}^{(0)}_{\mu\nu} =  0\,,\,\,&\,\,{\cal H}^{(0)}=0 
\,,\label{chapter5sys1a}\\
&&\nonumber\\
{\cal H}^{(2)}_{\mu\nu} =  0 \,,\,\,&\,\,{\cal H}^{(2)}=0 
\,,\label{chapter5sys1b}\\
&&\nonumber\\
{\cal H}^{(3)}_{\mu\nu}=0 \,,\,\,&\,\,{\cal H}^{(3)}=0 
\,,\label{chapter5sys1c}\\
&&\nonumber\\
{\cal H}^{(4)}_{\mu\nu}=0 
\,,\,\,&\,\,{\cal H}^{(4)}=0\,,\label{chapter5sys1d}
\end{eqnarray}
where the notations of Eqs. (\ref{highterms1}) and (\ref{highterms2}) have been adopted.
The order $\mathcal{O}(0)$  approximation gives
\begin{equation}\label{chapter5eq0}
f_0=0\,,
\end{equation}
a trivial consequence of the 
assumption (\ref{chapter5approx1a})--(\ref{chapter5approx1c}) 
that 
space-time is asymptotically Minkowskian.   If the Lagrangian is
expandable around the zero value of the Ricci scalar 
($R_0=0$), the  cosmological constant must vanish {\em in vacuo}.

If we now consider the second order approximation, the 
system (\ref{chapter5sys1a})--(\ref{chapter5sys1d}) {\em in 
vacuo} 
yields
\begin{eqnarray}
&& f_1 rR^{(2)}-2f_1g^{(2)}_{tt,r} 
+8f_2R^{(2)}_{,r}-f_1rg^{(2)}_{tt,rr}+4f_2rR^{(2)}=0 \,,
\label{chapter5eq2a}\\
&&\nonumber\\
&& f_1 r R^{(2)}-2f_1g^{(2)}_{rr,r} 
+8f_2R^{(2)}_{,r}-f_1rg^{(2)}_{tt,rr}=0 
\,,\label{chapter5eq2b}\\
&&\nonumber\\
&& 2f_1 g^{(2)}_{rr} -r \left( 
f_1rR^{(2)}-f_1g^{(2)}_{tt,r}-f_1g^{(2)}_{rr,r} 
+4f_2R^{(2)}_{,r}+4f_2rR^{(2)}_{,rr} \right)=0 
\,,\label{chapter5eq2c}\\
&&\nonumber\\
&& f_1rR^{(2)} + 6f_2 \left( 2R^{(2)}_{,r}+rR^{(2)}_{,rr} \right) 
= 0 \,,\label{chapter5eq2d}\\
&&\nonumber\\
&& 2 g^{(2)}_{rr}+r \left(  2 
g^{(2)}_{tt,r}-rR^{(2)}+2g^{(2)}_{rr,r} 
+ r g^{(2)}_{tt,rr} \right) =0 \,.\label{chapter5eq2e}
\end{eqnarray}
The trace equation~(\ref{chapter5eq2d}), in 
particular, is a differential equation for the  Ricci 
scalar which allows one to solve  the system 
(\ref{chapter5eq2a})--(\ref{chapter5eq2e}) 
to order $\mathcal{O}(2)$ as
\begin{eqnarray}
 g^{(2)}_{tt} &= & - \delta_0  + \frac{\delta_1(t) }{3\xi r}\, 
\mbox{e}^{-r\sqrt{-\xi}}  - \frac{\delta_2(t)}{6({-\xi)}^{3/2}r} 
\,\mbox{e}^{r\sqrt{-\xi}} \,,\label{chapter5sola}\\
&&\nonumber\\
 g^{(2)}_{rr} & = & - \frac{\delta_1(t) 
\left( r\sqrt{-\xi}+1 \right) }{3\xi
r} \, \mbox{e}^{-r\sqrt{-\xi}} 
+ \frac{\delta_2(t) \left( \xi    r+\sqrt{-\xi} \right)}{ 
6\xi^2r} \, \mbox{e}^{r\sqrt{-\xi}} \,,\label{chapter5solb}\\
&&\nonumber\\
R^{(2)} & = & \frac{\delta_1(t) }{r} \, \mbox{e}^{-r\sqrt{-\xi}} 
-\frac{\delta_2(t)\sqrt{-\xi} }{2\xi r} \, 
\mbox{e}^{r\sqrt{-\xi}} \,,\label{chapter5solc}
\end{eqnarray}
where 
\begin{equation} 
 \xi= \frac{ f_1}{ 6f_2} \,,
\end{equation}  
and $f_1$ and $f_2$  are expansion  coefficients of $f(R)$. 
The  integration constant  $ \delta_0 $ is 
dimensionless, while the two arbitrary functions of time 
$\delta_1(t)$ and $\delta_2(t)$ have the dimensions of an 
inverse length and an inverse length squared, respectively, 
and  $\xi$ has the  dimensions on an inverse length squared. 
These functions  are completely arbitrary 
because the differential system 
(\ref{chapter5eq2a})--(\ref{chapter5eq2e}) contains 
only spatial derivatives.  The additive quantity $\delta_0$ can  
be set to zero.

The gravitational potential for a  generic analytic $f(R)$ can 
now be obtained. Eqs. (\ref{chapter5sola})--(\ref{chapter5solc})  
provide the 
second order solution in term of the metric expansion (see the 
definition 
(\ref{chapter5definexpansa})--(\ref{chapter5definexpansc}))  but, 
as said above, this term  coincides with the gravitational 
potential at the Newtonian order, $g_{tt}\,= - 1 - 
2\phi \,= -1+g_{tt}^{(2)}$. The  
gravitational potential of a fourth order theory of gravity 
analytic in $R$ is
\begin{equation}\label{chapter5gravpot}
\phi^{(FOG)}\,=\,\frac{K_1 }{3\xi r} \, 
\mbox{e}^{-r\sqrt{-\xi}} + \frac{K_2  }{6{(-\xi)}^{3/2}r} \, 
\mbox{e}^{r\sqrt{-\xi}} 
\end{equation}
with $K_1=\delta_1(t)$ and $K_2=\delta_2(t)$.

For a given $f(R)$ theory, the structure of the potential 
is 
determined by  the parameter $\xi$, which depends on the first 
and second derivatives of $f(R)$ at $R_0$. The 
potential~(\ref{chapter5gravpot}) is valid for non-vanishing 
$f_2$, since we manipulated Eqs.  
(\ref{chapter5eq2a})--(\ref{chapter5eq2e}) dividing 
by $f_2$. The Newtonian limit of GR cannot be obtained  directly 
from the solution~(\ref{chapter5gravpot}) but requires the field 
equations (\ref{chapter5eq2a})--(\ref{chapter5eq2e}) once  the 
appropriate expressions 
in terms of the constants $f_i$ are derived.

The solution (\ref{chapter5gravpot}) must be  discussed in relation to
the sign of the term under square root in  the  exponents. If 
this sign is positive (which means that $f_1$ and $f_2$ have 
opposite signature), the solutions 
(\ref{chapter5sola})--(\ref{chapter5solc}) and  
(\ref{chapter5gravpot}) can be rewritten  introducing the  scale 
parameter 
$l=|\xi|^{-1/2}$. In particular, considering $\delta_0\,=\,0$, 
the functions $\delta_i(t)$  as
constants, $k_1\,= l \delta_1(t)/3 $ and
$k_2(t)=\,l^2  \, \delta_2(t)/6 $ and  introducing a  radial
coordinate $\tilde{r}$ in units of $l$, we have
\begin{eqnarray}
g^{(2)}_{tt} & = & - \delta_0 - \frac{\delta_1(t)l}{3} \, 
\frac{\mbox{e}^{-r/l}}{r/l} - \frac{\delta_2(t)l^2}{6} 
\, \frac{\mbox{e}^{r/l}}{r/l}= 
\frac{k_1}{\tilde{r}} \, \mbox{e}^{-\tilde{r}} 
+\frac{k_2}{\tilde{r}} \, \mbox{e}^{\tilde{r}}
\,,\label{chapter5sol1a}\\
&&\nonumber\\
g^{(2)}_{rr} & = & \frac{\delta_1(t)l}{3} 
\,  \frac{(r/l+1) }{r/l} \, \mbox{e}^{-r/l}  
- \frac{\delta_2(t)l^2}{6}\frac{(r/l-1) }{r/l} \, 
\mbox{e}^{r/l} \nonumber\\
&&\nonumber\\
& = & -
k_1 \, \frac{(\tilde{r}+1) \, \mbox{e}^{-\tilde{r}}}{\tilde{r}}
+k_2 \, \frac{(\tilde{r}-1) \, \mbox{e}^{\tilde{r}}}{\tilde{r}} 
\,,\label{chapter5sol1b} \\
&&\nonumber\\
R^{(2)} & = & \frac{\delta_1(t)}{l} \, \frac{ 
\mbox{e}^{-r/l}}{r/l}
+\frac{\delta_2(t)}{2} \, \frac{ 
\mbox{e}^{r/l}}{r/l} =\frac{3}{l^2}\biggl( k_1\frac{ 
\mbox{e}^{-\tilde{r}}}
{\tilde{r}} + k_2 \, 
\frac{\mbox{e}^{\tilde{r}}}{\tilde{r}}\biggr) 
\,. \label{chapter5sol1c}
\end{eqnarray}
The  gravitational potential can then be rewritten as 
\begin{equation}\label{chapter5gravpot*}
\phi^{(FOG)}\,=\,\frac{k_1 }{ \tilde{r}} \, 
\mbox{e}^{-\tilde{r}} + \frac{k_2}{\tilde{r}} \, 
\mbox{e}^{\tilde{r}} \,,
\end{equation}
which is analogous to the result or Ref.~\cite{Stelle78} derived 
for the theory $R+\alpha R^2+\beta R_{\mu\nu}R^{\mu\nu}$ 
and consistent\footnote{In a spatially homogeneous and isotropic 
space-time manifold, the higher order curvature invariants  
$R_{\mu\nu}R^{\mu\nu}$ and 
$R_{\alpha\beta\mu\nu}R^{\alpha\beta\mu\nu}$ can be  written in 
terms of $R^2$.}
with  Refs.~\cite{Qua91, Schmidt07} discussing higher order
Lagrangians such as $ \displaystyle{ f(R,\Box 
R)\,=\,R+\sum_{k=0}^p\,a_kR\,\Box^kR } $. In
this last  case, it was demonstrated that the number of Yukawa
corrections  to the gravitational 
potential is  related to
the order of the theory \cite{Qua91}.  
However,  it is straightforward to show 
\cite{dirk} that the usual form Newton plus 
Yukawa is recovered 
in Eq.~(\ref{chapter5gravpot*}) using a coordinate change, and 
Eq.~(\ref{chapter5gravpot*}) assumes the form
\begin{eqnarray}\label{chapter5gravpotyuk}
\phi^{(FOG)}\,=\,-\left(\frac{GM}{f_1r} 
+\frac{\delta_1(t) }{6\xi
r} \, \mbox{e}^{-r\sqrt{-\xi}} \right)\,,
\end{eqnarray} where  $\delta_1(t)$ is again an arbitrary 
function of time and the parameters depend on the 
Taylor coefficients. An effective Newton 
constant  
$G_{eff}=G/f_1$ and a range  $l=|\xi|^{-1/2}$ emerge, and depend  
on the form of the function $f(R)$.

The inspection of  Eqs. 
(\ref{chapter5sola})--(\ref{chapter5solc}) 
and (\ref{chapter5sol1a})--(\ref{chapter5sol1c}) reveals that  
the  Newtonian limit of an analytic $f(R)$ theory depends only on  
the first and second  coefficients of  the Taylor  
expansion of $f(R)$. The gravitational potential 
is always characterized by two Yukawa 
corrections determined  
only  by the first two terms of the Taylor expansion.

The diverging  contribution, arising from the exponentially 
growing
mode, has to be analyzed carefully  and, in particular, the
physical relevance of this term must be evaluated in relation to
the length scale $(-\xi)^{-1/2}$. For   $r \gg (-\xi)^{-1/2}$,    
the  weak-field approximation turns  out to be unphysical 
and (\ref{chapter5sola})--(\ref{chapter5solc}) no longer holds. 
One can obtain a modified 
gravitational potential which can work as a standard Newtonian 
one in the appropriate limit and provides interesting behavior 
at larger scales, even in the presence
of the growing mode, once the  constants in 
eq.~(\ref{chapter5gravpot})
have been suitably adjusted. Once the growing exponential term is 
discarded, this potential reproduces the Yukawa-like  
potential  
phenomenologically introduced in order to explain the flat 
rotation curves  of spiral galaxies without 
dark  matter 
\cite{Sanders90}.

Yukawa-like corrections  to the 
gravitational potential
have been suggested in several contexts, for example,  in a model 
describing the gravitational interaction between dark 
matter  and 
baryons. In this model  the interaction suppressed at  small 
(subgalactic) scales is described by a 
Yukawa contribution 
to  the standard Newtonian potential. This behaviour is suggested 
by
observations of the inner rotation curves
of  low-mass
galaxies and provides a natural scenario in which to interpret the
cuspy profile of dark matter halos  arising in 
$N$-body 
simulations \cite{piazza}.

The result outlined here  is consistent with other  calculations. 
Since an exponential potential is expanded in a power-law 
series, it is not surprising to find a power-law correction to 
the Newtonian potential \cite{prl, CapCardTro07} when a less rigorous 
approach is considered in order to calculate the weak-field limit 
of a generic $f(R)$ theory, and perturbative  calculations will 
provide effective potentials which can be  recovered by means of 
an appropriate approximation from the  general 
case~(\ref{chapter5gravpot*}).

Let us consider now a negative sign of
$\xi$, when the two Yukawa corrections in 
(\ref{chapter5sol1a})--(\ref{chapter5sol1c}) are 
complex. Using the form of $g_{tt}$, the  gravitational 
potential~(\ref{chapter5gravpot*}) is
\begin{equation}\label{chapter5gravpottrig}
\phi^{(FOG)}\, = \,\frac{k_1 }{ \tilde{r}} \, 
\exp \left(- i \, \tilde{r} \right) +\frac{k_2  }{\tilde{r}} 
\, \exp \left( i  \, \tilde{r} \right) \,,
\end{equation}
which can be recast as
\begin{equation}\label{chapter5gravpottrig1}
\phi^{(FOG)}\,=\,\frac{1}{\tilde{r}} 
\biggl[(k_1+k_2)\cos{{\tilde{r}}} 
+ i \left( k_2-k_1 \right) \sin{\tilde{r}}\biggr]\,.
\end{equation}
This gravitational potential, which could {\em a priori} be 
discarded as physically irrelevant, satisfies the 
Helmholtz equation
$ {\displaystyle \nabla^2\phi+\vec{k}^2\phi=4\pi G\rho } $, where  
$\rho$ is a real
function acting both as matter and antimatter density. As
discussed in \cite{Bartlett94, Bartlett01}, 
 {\em Re}$\left\{ \phi^{(FOG)} \right\}$ can be
seen as a classically modified Newtonian potential corrected
by  a Yukawa factor while {\em Im}$ 
\left\{ \phi^{(FOG)} 
\right\} $ could have implications for quantum mechanics. This 
term can provide an astrophysical  origin for the puzzling decay
$K_L\rightarrow \pi^+\pi^-$, whose phase is related to an 
imaginary
potential in the kaon mass matrix. Of course, these
considerations are purely speculative but it could be 
interesting to pursue them.

Let us  consider now third  order contributions in the 
system~(\ref{chapter5sys1a})--(\ref{chapter5sys1d}); at this 
order 
the off-diagonal equation
\begin{equation}\label{chapter5off-d}
f_1 \, g^{(2)}_{rr,t}+2f_2r \, R^{(2)}_{,tr}=0 
\end{equation}
relating the time derivatives of $R$ and  $g^{(2)}_{rr}$ must be 
taken into account. If the Ricci scalar depends on time, also  
the metric components and the gravitational potential do. This 
result agrees with the analysis of~\cite{noimpla} in terms of 
the dynamical evolution of $R$ and demonstrating that a 
time-independent Ricci  scalar implies static spherically 
symmetric solutions, which is confirmed (and explained) by 
eq.~(\ref{chapter5off-d}).  In conjunction with Eqs. 
(\ref{chapter5sol1a})--(\ref{chapter5sol1c}),  
eq.~(\ref{chapter5off-d})  suggests that 
if one considers the problem 
to lower (second) order,  the background metric can have static 
solutions according to the Jebsen-Birkhoff 
theorem, but this is no 
longer  true when higher orders are considered.  The validity of 
the Jebsen-Birkhoff  theorem in 
higher order theories of gravity  
depends 
on the approximation order considered. 
This theorem holds in metric $f(R)$-gravity only when the 
Ricci scalar is time-independent, and to second order 
in a $v/c$ expansion of the metric coefficients. According to 
Eqs. (\ref{chapter5sola})--(\ref{chapter5solc}) and  
(\ref{chapter5gravpot}), it is only in 
the limit of small velocities and weak fields that  the 
gravitational potential is effectively time-independent.
But, contrary to GR, in metric $f(R)$-gravity a spherically 
symmetric background can have time-dependent evolution.

The next step is  the order $\mathcal{O}(4)$ analysis of the 
system~(\ref{chapter5sys1a})--(\ref{chapter5sys1d}) providing the 
solutions in terms of 
$g_{tt}^{(4)}$,  the order necessary to compute the 
post-Newtonian  parameters. Unfortunately, 
at this order the 
system is much more complicated  and a general solution is not 
possible. One sees from  
Eqs. (\ref{chapter5sys1a})--(\ref{chapter5sys1d}) that  the  
general 
solution is  characterized only by the first three orders  of the 
$f(R)$ expansion, in agreement with the $f(R)$ reconstruction 
using the  post-Newtonian parameters  in the 
scalar-tensor 
representation  \cite{ppnantro, ppn-noi-bis}. Although a  
complete description is difficult, an  estimate  of the 
post-Newtonian parameter $\gamma$ can be 
obtained from  
the order $\mathcal{O}(2)$ evaluation of the metric coefficients {\em in 
vacuo}.
Since (\ref{chapter5sola})--(\ref{chapter5solc}) suggest a 
non-Newtonian gravitational 
potential  as a general solution of  analytic $f(R)$-gravity,  
there is no reason to ask for a post-Newtonian 
description
of these theories. In fact, as said earlier, the
post-Newtonian analysis presupposes to 
evaluate deviations from
the Newtonian potential at a higher than second order
approximation in $v/c$. Thus, if the
gravitational potential deduced from a given $f(R)$ theory  is a  
general function of the radial coordinate 
displaying a Newtonian behaviour only in a certain regime (or in a
given range of the radial coordinate), it would be meaningless to
develop a general post-Newtonian formalism as in GR
\cite{Will93, Will01, Nordtvedt68}. Of course, by a proper 
expansion of the
gravitational potential for small values of the radial coordinate,
and only in this limit, one can develop an analog of the
post-Newtonian limit for these theories.

In order to estimate  the post-Newtonian
parameter $\gamma$, one proceeds by expanding
$g_{tt}$ and $g_{rr}$, obtained to second order in
(\ref{chapter5sol1a})--(\ref{chapter5sol1c}), with respect to the 
dimensionless  coordinate
$\tilde{r}$, obtaining
\begin{eqnarray} 
g^{(2)}_{tt} & = & k_2-k_1 + \frac{k_1+k_2}{\tilde{r}} 
+\frac{k_1+k_2}{2} \, \tilde{r}+ \mathcal{O}(2)\,,\\
&& \label{chapter5taylorppn}\\
&&\nonumber\\
g^{(2)}_{rr} & = & -\frac{k_1+k_2}{\tilde{r}} 
+\frac{k_1+k_2}{2}\tilde{r}+ \mathcal{O}(2) \,, 
\end{eqnarray}
where $k_1+k_2=GM$ and $k_1=k_2$ in the standard case.
When $\tilde{r}\rightarrow 0$ ({\em i.e.}, when 
$r\ll \sqrt{-\xi}$) the linear and successive order terms
are small and  the first (Newtonian) term dominates.
Since the post-Newtonian parameter  $\gamma$ 
is related to
the coefficients of the $1/r$ terms  in 
$g_{tt}$ and $g_{rr}$, one can estimate  this quantity 
by comparing the coefficients of the
Newtonian terms relative to both expressions in
(\ref{chapter5taylorppn}). Since $\gamma\,=\,1$ in GR, the difference
between these two coefficients  gives the effective deviation from
the GR value.

A generic fourth order gravity theory provides a post-Newtonian 
parameter $\gamma$  consistent
with the GR prescription  if $k_1\,=\,k_2$.
Conversely, deviations from this  behavior can be accommodated by
tuning the relation between the two integration constants $k_1$
and $k_2$. This is equivalent to adjusting the form of the $f(R)$
theory to obtain the correct GR limit first, and 
then the Newtonian potential. This result agrees with  recovering 
the GR behavior from  generic $f(R)$ theories in the 
post-Newtonian limit \cite{Zhang:2007ne,sotiriou}. 
This is particularly 
true when the $f(R)$ Lagrangian behaves, in the weak-field and 
small velocity regime, as the Hilbert-Einstein
Lagrangian. If deviations from this regime are observed, an
$f(R)$ Lagrangian which is a third order polynomial in the
Ricci scalar can be more appropriate \cite{ppn-noi-bis}.

The degeneracy in the integration constants can be partially 
broken once a complete post-Newtonian 
parameterization is 
developed in the presence of matter. Then, the integration 
constants are constrained by the 
Boltzmann-Vlasov equation
describing conservation of matter at small scales 
\cite{BinneyTremaine87}.

So far, no specific gauge choice has been made, however 
particular  
gauges can be considered  to simplify the calculations.   A 
natural choice consists of the 
conditions~(\ref{chapter5PPNobja})--(\ref{chapter5PPNobjd}), 
which  coincide with  the standard  
post-Newtonian gauge
\begin{eqnarray}
&& h_{jk}{}^{,k} - \frac{1}{2} h_{,j} = \mathcal{O}(4) \,, \\
&& \nonumber\\
&& h_{0k}{}^{,k} - \frac{1}{2} h^{k}{}_{k,0} = \mathcal{O}(5)\,, 
\end{eqnarray}
where $h_{\mu\nu} \equiv   g_{\mu\nu} - \eta_{\mu\nu}$. In 
this gauge the Ricci  tensor becomes\footnote{We denote harmonic 
gauge quantities  with the subscript 
$hg$.}
\begin{eqnarray} 
R_{tt|_{hg}} & = &  R^{(2)}_{tt|_{hg}}+R^{(4)}_{tt|_{hg}} 
\,,\\
&&\nonumber\\
R_{rr|_{hg}}& = &  R^{(2)}_{rr|_{hg}} \,,
\end{eqnarray}
where
\begin{eqnarray}
R^{(2)}_{tt|_{hg}} & = & 
\frac{r \, g^{(2)}_{tt,rr} + 2g^{(2)}_{tt,r}}{2r} \,,  \\
&&\nonumber\\
R^{(4)}_{tt|_{hg}} & = & \frac{ r \, g^{(4)}_{tt,rr}+ 
2 \, g^{(4)}_{tt,r} 
+r \left( g^{(2)}_{rr} g^{(2)}_{tt,rr} - g^{(2)}_{tt,tt}-{g^{(2)}
_{tt,rr}}^2 \right) }{2r} \,, \\
&&\nonumber\\
R^{(2)}_{rr|_{hg}} & = 
& \frac{r \, g^{(2)}_{rr,rr} + 2 \, g^{(2)}_{rr,r}}{2r} 
\,, \\
&&\nonumber\\
R^{(2)}_{\theta\theta|_{hg}} & = & R^{(2)}_{\phi\phi|_{hg}}=0 \,,
\end{eqnarray}
while the Ricci scalar to order$ \mathcal{O}(2)$ and $\mathcal{O}(4)$  is
\begin{eqnarray}
R^{(2)}_{|hg} & = & \frac{rg^{(2)}_{tt,rr}+2g^{(2)}_{tt,r} 
-rg^{(2)}_{rr,rr}-2g^{(2)}_{rr,r}}{2r} \,, \\
&&\nonumber\\
R^{(4)}_{|hg} & = & \biggl[ r \, g^{(4)}_{tt,rr} + 
2 \, g^{(4)}_{tt,r}  +r 
\left( g^{(2)}_{rr}g^{(2)}_{tt,rr}-g^{(2)}_{tt,tt}- 
{g^{(2)}_{tt,rr}}^2 \right) 
-g^{(2)}_{tt} \left( r \, g^{(2)}_{tt,rr} + 2 \, g^{(2)}_{tt,r} 
\right)  \nonumber\\
&&\nonumber\\
& - & g^{(2)}_{rr} \left( r \, g^{(2)}_{rr,rr} + 2 
\, g^{(2)}_{rr,r} \right)  \biggr] (2r)^{-1}  \,.
\end{eqnarray}
The gauge choice does not affect the connection coefficients. 
The  solution of the  system  
(\ref{chapter5sys1a})--(\ref{chapter5sys1d}) in this  gauge is
\begin{eqnarray}
g_{tt_{|_{hg}}} \left( t,r \right) & = &  - 1 + \frac{k_1}{r} 
+\frac{k_2}{r^2} + \frac{k_3 \log r}{r} \,, 
\label{chapter5armonica}\\
&&\nonumber\\
g_{rr_{|_{hg}}} \left( t,r \right) & = &  1+\frac{k_4}{r} \,,
\label{chapter5armonicb} 
\end{eqnarray}
where the constants $k_1$ and $k_4$ pertain  to the order $\mathcal{O}(2)$,  
while $k_2$ and $k_3$ pertain  to the 
order $\mathcal{O}(4)$. The Ricci scalar vanishes to orders $\mathcal{O}(2)$ and  
$\mathcal{O}(4)$.

Using Eqs.~(\ref{chapter5armonica}) and~(\ref{chapter5armonicb}), 
it 
is easy to check that the GR 
prescriptions are immediately recovered for
$k_1\,=\,k_4$ and $k_2\,=\,k_3\,=0$.  The $g_{rr}$ component
contains only the second order term, as required by a GR-like
behavior, while the $g_{tt}$ component exhibits also the fourth 
order corrections which determine the second post-Newtonian 
parameter
$\beta$ \cite{Will93, Will01}. A full
post-Newtonian formalism requires the 
consideration of  matter in
the system ({\ref{chapter5sys1a})--(\ref{chapter5sys1d}): the 
presence of matter links the 
second and fourth order contributions in the metric coefficients
\cite{Will93, Will01}. A relevant application of these considerations is the following.

\subsection{An example: Neutrino oscillation phase dynamically induced by non-Newtonian corrections }
Neutrinos are elementary particles that travel at the speed of light (or close to it if massive), are electrically neutral and are capable of passing through ordinary matter with minimal interaction. Due to these properties, they can be investigated, in principle, 
  at all length scales, ranging from nuclei \cite{Reines}, to
molecular structures \cite{Collar}, up to galaxies \cite{Weinberg72,CapLam} and to the whole Universe \cite{Blasone}. 
They results from  radioactive decays or nuclear reactions such as those that take place in the Sun, stars or
 nuclear reactors. In particular, they are generated   when cosmic rays hit atoms.
Current evidences of dark matter and dark energy can be related to the issue that
neutrinos have masses and that  mass eigenstates mix and/or superimpose \cite{Barenboim, Barenboim3, nieuwen}.
The observation of such a mixing is related to suitable  constraints.  Such constraints should work  on
 observables sensitive to  the effective neutrino
mass as the mass in Tritium-beta decay, the sum of neutrinos masses in cosmology and  the
effective Majorana neutrino mass in neutrinoless double-beta decay  \cite{Fogli}. 

A key role  is played by the neutrino oscillations
that allow the transition among the three  types or {\it "flavor"} eigenstates, that is the electron, muon, and tauon neutrinos.
It is well known that such a problem is still open and the research of new effects,
in which the oscillations could manifest is one of the main goal of modern
physics. For this reason, the quantum mechanical phase of neutrinos, propagating
in gravitational field, has been discussed by several authors, also in view of the
astrophysical consequences. More controversial is the debate concerning the redshift
of flavor oscillation clocks, in the framework of the weak gravitational field
of a star \cite{Ahluwalia}. It has also been suggested that the gravitational oscillation phase
might have a significant effect in supernova explosions due to the extremely large
fluxes of neutrinos produced with different energies, corresponding to the flavor
states. This result has been confirmed in \cite{Grossman}, and it has been also derived under
the assumption that the radial momentum of neutrinos is constant along the trajectory
of the neutrino itself \cite{Konno}. Besides, neutrino oscillations, in particular the
gravitational part of the oscillation phase, could straightforwardly come into the
debate to select what is the correct theory of gravity
\cite{Will93,gaetano}. 
Further gravitational interaction lengths, emerging from ETGs, could be related to neutrino oscillation phase. On the other hand, the experimental identification of such a gravitational phase could be a formidable probe both for confirming or ruling out such theories at a fundamental level.

Let us start our discussion considering  how the gravitational field contributes to the neutrino oscillations. The approach has been firstly developed in \cite{Ahluwalia} and we will outline the main results reported there.
If $R_A$ is the size of a physical region where neutrinos are generated, a neutrino energy eigenstate $E_{\nu}$ can be denoted by $|\nu_l,R_A\rangle$ (where l = $e,\mu,\tau$ represents the weak flavor eigenstates). The three neutrino mass eigenstates can be represented by $|\nu_i \rangle$ with $i=1,2,3$ corresponding to the masses $m_1, m_2, m_3$. The mixing between mass and flavor
eigenstates is achieved by the unitary transformation
\begin{equation}
|\nu_l',R_A\rangle=\sum_{i=1,2,3}U_{li}|\nu_i\rangle\label{15},
\end{equation}
where
\begin{equation}
U\left(\theta,\beta,\psi\right)=\left(\begin{array}{ccc}
c_{\theta}c_{\beta} & s_{\theta}c_{\beta} & s_{\beta}\\
-c_{\theta}s_{\beta}s_{\psi}-s_{\theta}c_{\psi} & c_{\theta}c_{\psi}-s_{\theta}s_{\beta}s_{\psi} & c_{\beta}s_{\psi}\\
-c_{\theta}s_{\beta}c_{\psi}-s_{\theta}s_{\psi} & -s_{\theta}s_{\beta}c_{\psi}-c_{\theta}s_{\psi} & c_{\beta}c_{\psi}\end{array}\right)\label{16}
\end{equation}
is a $3\times3$ unitary matrix parametrized by the three mixing angles $\eta =\theta, \beta,\psi$ with
$c_\eta=\cos\eta$ and $s_\eta=\sin\eta$. At time $t = t_B > t_A$, the weak flavor eigenstates
can be detected in a region $R_B$ and, in general, the evolution is given by
\begin{equation}
|\nu_{l},R_{B}\rangle=exp\left(-\frac{i}{\hbar}\int_{t_{A}}^{t_{B}}\mathcal{H}dt+\frac{i}{\hbar}\int_{r_{A}}^{r_{B}}\overrightarrow{P}\cdot d\overrightarrow{x}\right)|\nu_l,R_A\rangle\label{17},
\end{equation}
where $\mathcal{H}$ is the Hamiltonian operator associated to the system  representing the time translation
operator and $\overrightarrow{P}$ is the momentum operator  representing the spatial translation operator.
The phase change in Eq.(\ref{17}) is  the argument of the exponential function. It can be
recast in the form
\begin{equation}
\phi_{\nu}=\frac{1}{\hbar}\int_{r_{A}}^{r_{B}}\left[E\frac{dt}{dr}-p_{r}\right]dr\label{18}\,.
\end{equation}
 The covariant formulation is
\begin{equation}
\phi_{\nu}=\frac{1}{\hbar}\int_{r_A}^{r_B}mds=\frac{1}{\hbar}\int_{r_A}^{r_B}p_{\mu}dx^{\mu},\label{19}
\end{equation}
where $\displaystyle{p_\mu=mg_{\mu\nu}\frac{dx^\nu}{ds}}$ is the $4$-momentum of the particle. The effect of
 gravitational field is given by $g_{\mu\nu}$ and, in general, the neutrino oscillation
probability from a state $|\nu_l,R_A\rangle$  to another state $|\nu_l,R_B\rangle$	 is given by
\begin{eqnarray*}
\mathcal{P}\left[\left|\nu_{l},R_{A}\right\rangle \rightarrow\left|\nu_{l'},R_{B}\right\rangle \right] & = & \delta_{ll'}-4U_{l'1}U_{l1}U_{l'2}U_{l2}\sin^{2}\left[\phi_{0}^{21}+\right.\\ 
 &  &+\left.\phi_{G}^{21}\right] -4U_{l'1}U_{l1}U_{l'3}U_{l3}\cdot\\
 &  &\cdot \sin^{2}\left[\phi_{0}^{31}+\phi_{G}^{31}\right] -4U_{l'1}U_{l1}U_{l'3}\cdot \\
 &  &\cdot U_{l3}\sin^{2}\left[\phi_{0}^{31}+\phi_{G}^{31}\right]\label{20},\end{eqnarray*}
 
 where $\phi_{0}^{ij}$ are the usual kinematic phase while $\phi_{G}^{ij}$ are the gravitational contributions. It can be shown that, in a flat space-time, the $\phi_{G}^{ij}$ contributions are zero. In fact, a particle passing nearby a point mass feels a Schwarzschild geometry so  the trajectories is
\begin{equation}
dx\simeq\left[1-\frac{2G_{N}M}{c^{2}r}\right]cdt\label{21}\,.\end{equation}
If the effects of gravitational field are vanishing, Eq. (\ref{21})  becomes $dx\simeq cdt$. Considering two generic neutrino mass eingestates in a Schwarzshild geometry, the standard phase of neutrino oscillation is

\begin{equation}
\phi_0=\frac{\Delta m^2 c^3}{4E \hbar}\left(r_B-r_A\right),       \label{23}
\end{equation}
while the total gravitational phase shift is

\begin{equation}
\phi_{grav}=\frac{G_{N}\bigtriangleup m^{2}M c}{4\hbar E}\log\frac{r_{B}}{r_{A}}\,,\label{22}\end{equation}
as shown in \cite{Ahluwalia}, where $\bigtriangleup m^{2}$ is the mass squared difference, $\bigtriangleup m^{2}=\left|m_{2}^{2}-m_{1}^{2}\right|$, $E$ the neutrino energy, $r_A$ and $r_B$ the point where neutrinos are created and detected, respectively. Nevertheless, assuming that the neutrino energy is constant along the trajectory,  the term (\ref{22}) could be cancelled out at  typical astrophysical scales \cite{Bhattacharya}. 

With this considerations in mind, let us take into account how possible corrections to the Newtonian potential could affect this result. 
We will  follow the discussion developed above for the post-Newtonian limit of $f(R)$-gravity. We do not want to  impose a particular forms for  $f(R)$-model but
only consider analytic Taylor expansion where the cosmological term and terms higher than second are discarded. The Lagrangian is then
\begin{equation}
\label{fRf}
f(R) \sim a_1 R + a_2 R^2 + ... 
\end{equation}
where the parameters $a_{1,2}$ specifies the particular models. Adopting the argument of previous subsection, Eq. (\ref{chapter5gravpotyuk}) gives
\begin{equation}
\label{gravpot1} \Phi(r) = -\frac{3 G_{N} M}{4 a_1
r}\left(1+\frac{1}{3}e^{-\frac{r}{L}}\right)=\Phi(r)_{Newton}+\Phi(r)_{Yukawa}\,,
\end{equation}
where
\begin{equation}\label{lengths model}
L \equiv L(a_{1},a_{2}) =  \left( -\frac{6 a_2}{a_1}
\right)^{1/2}\,.
\end{equation}
$L$ is an  {\it interaction gravitational length}  due to the correction to the Newtonian potential. However, as soon as $a_1= 3/4,$  $a_2 =0$ and $\Phi(r)_{Yukawa}\rightarrow 0$,  the standard Newtonian limit of GR is fully recovered (for a discussion on this point, see \cite{mnras2}).

Now we calculate the gravitational phase shift of neutrino oscillation in $f(R)$-gravity using the potential (\ref{gravpot1}) in the Eq. (\ref{18}). We obtain the general expression

\begin{equation}
\phi_{grav}=\frac{\Delta m^2 M c}{4 \hbar E}\left( \frac{3 G_{N}}{4 a_{1}}\right)\int_{r_A}^{r_B}\left(\frac{1}{r}+\frac{1}{3r} e^{-\frac{r}{L}}\right),
\end{equation}

from which we have the following result
\begin{equation}
\phi_{grav}=\frac{\Delta m^2 M c}{4 \hbar E}\left( \frac{3 G_{N}}{4 a_{1}}\right)\left[ \log \frac{r_B}{r_A}+ \sum_{n=1}^{\infty}(-1)^{n}\left(\frac{\left(\frac{r_B-r_A}{L}\right)^n}{n \cdot n!}\right)\right]
=\phi_{Newton}+\phi_{Yukawa}\, ,
\end{equation}

where $\phi_{grav}$ is the total gravitational phase shift in Eq. (\ref{22}) and 
\begin{equation}
\label{corr}
\phi_{Yukawa}=\frac{\Delta m^2 M c}{4 \hbar E}\left( \frac{3 G_{N}}{4 a_{1}}\right)\left[ \sum_{n=1}^{\infty}(-1)^{n}\left(\frac{\left(\frac{r_B-r_A}{L}\right)^n}{n \cdot n!}\right)\right]\,  .
\end{equation}
The Yukawa term disappears in standard Einstein gravity, that is for $f(R)=R$.
Note that the series in above equation is absolutely convergent. If we consider Solar neutrinos,  we can  use the following values: $M\sim M_\odot \sim 1.9891\times 10^{30}$Kg, $r_{A}\sim r_{\oplus}\sim6.3 \times 10^{3}$Km, and $r_{B}\sim r_A+D$, where $D\sim 1.5\times 10^8$Km is the Sun-Earth distance. 
In order to estimate the  phases differences (\ref{23}), (\ref{22}) and (\ref{corr}),  we introduce the ratio $Q_{grav}$ defined as
\begin{equation}
Q_{grav}=\frac{\phi_{Newton}}{\phi_0}\sim \frac{G_N M \log \frac{r_B}{r_A}}{c^2 (r_B-r_A)}\sim 10^{-7},   
\end{equation}
and the ratio $Q_{Yukawa}$ defined as
\begin{equation}
Q_{Yukawa}=\frac{\phi_{Yukawa}}{\phi_0}\sim \frac{G_N M}{c^2 (r_B-r_A)}\sum_{n=1}^{\infty}(-1)^{n}\left(\frac{\left(\frac{r_B-r_A}{L}\right)^n}{n \cdot n!}\right), \label{fi2}
\end{equation}
where we have assumed that $3/4a_1\sim1$. Note that both $Q_{Newton}$ and  $Q_{Yukawa}$ do not depend on the squared-mass difference $\Delta m^2$ and on the neutrino energy $E$. The ratio $Q_{Yukawa}$ can be calculated for different values of the {\it interac-\\tion lenght} $L$.  
For example from Eq. (\ref{fi2}), after summing the series, we obtain the results:
\begin{equation}
L\sim 1.5\cdot10^7 Km \Longrightarrow Q_{Yukawa}\sim -2.9\cdot10^{-8},
\end{equation}
\begin{equation}
L\sim 1.5\cdot10^8 Km \Longrightarrow Q_{Yukawa}\sim -8\cdot10^{-9}.
\end{equation}
In this way the values of $Q_{Yukawa}$  can be seen as corrections to the standard gravitational phase shift of neutrino oscillations depending on the particular choice of $L$ and so, through Eq.(\ref{lengths model}), directly on the particular $f(R)$-model considered.

We remark that the calculated correction to gravitational phase shift in Eq. (\ref{corr}) depends on the  {\it interaction lenght} $L$ defined in Eq. (\ref{lengths model}).  This is directly related to the $f(R)$-gravity model through the coefficients $a_1$ and $a_2$ in Eq. (\ref{fRf}). This fact could be used as an experimental test to probe a  given gravity theory through the neutrino oscillation induced by means of the gravitational field itself. On the other hand,   interpreting $L$ as the characteristic wavelenght of the neutrino interaction with the gravitational field, the gravitational phase correction could be used as a method to constrain the mass of electronic neutrinos travelling from the Sun to the Earth surface, or, eventually,  also from other neutrinos sources as Supernovae or neutron stars.

\section{The Post- Minkowskian limit}

\subsection{The weak field limit in Minkowski space-time}

We have developed a general analytic
procedure to deduce the Newtonian and post-Newtonian limits of
$f(R)$-gravity outside matter sources. Now we discuss a 
different limit of these theories, obtained  when the small 
velocity assumption is relaxed and only the weak-field 
approximation is retained. Again, we assume spherical symmetry 
of the metric, considering gravitational potentials $A$ 
and  $B$ of the form
\begin{eqnarray}
A \left(t,r \right) & = &  - 1 + a \left( t,r \right) \,, \\
&&\nonumber\\
B\left( t,r \right) & = &  1+b \left( t,r \right) \,,
\end{eqnarray}
with $ |a(t,r)|, |b(t,r)| \ll 1$. Let us perturb the field 
equations
considering again  the Taylor expansion~(\ref{chapter5sertay}).  
{\em In  vacuo}  and to  first order in $a$ and $b$, one obtains
\begin{eqnarray}
&& f_0=0 \,, \label{chapter5eq4a}\\
&&\nonumber\\
&& f_1 \left( R^{(1)}_{\mu\nu}-\frac{1}{2} 
\, g^{(0)}_{\mu\nu}R^{(1)}  \right) +\mathcal{H}^{(1)}_{\mu\nu}=0 
\,,\label{chapter5eq4b}
\end{eqnarray}
where
\begin{eqnarray}
\mathcal{H}^{(1)}_{\mu\nu} & = & -f_2\biggl[ R^{(1)}_{,\mu\nu}
-{\Gamma^{(0)}}^{\rho}_{\mu\nu}R^{(1)}_{,\rho}-g^{(0)}_{\mu\nu} 
\biggl( {g^{(0)\rho\sigma}}_{,\rho}R^{(1)}_{,\sigma} 
+g^{(0)\rho\sigma}R^{(1)}_{,\rho\sigma}  \nonumber\\
&&\nonumber\\
&+&  g^{(0)\rho\sigma} 
\ln \sqrt{-g}^{(0)}_{,\rho}R^{(1)}_{,\sigma}\biggr) 
\biggr] \,.
\end{eqnarray}
In this approximation the Ricci scalar vanishes and  the 
derivatives are evaluated at $R=0$. Let us consider now the large 
$r$ limit  far from the sources of 
the gravitational field:  Eqs.~(\ref{chapter5eq4a}) 
and~(\ref{chapter5eq4b}) become
\begin{eqnarray}
&& \frac{\partial^2a(t,r)}{\partial r^2} 
-\frac{\partial^2b(t,r)}{\partial t^2}=0 \,,\label{chapter5eq5a}\\
&&\nonumber\\
&& f_1\biggl[a(t,r)-b(t,r)\biggr] 
-8f_2\biggl[\frac{\partial^2b(t,r)}{\partial
r^2}+\frac{\partial^2a(t,r)}{\partial
t^2}-2\frac{\partial^2b(t,r)}{\partial
t^2}\biggr]=\Psi(t) \,,\nonumber\\
&& \label{chapter5eq5b}
\end{eqnarray}
where $\Psi(t)$ is a generic time-dependent function.
Eqs.~(\ref{chapter5eq5a}) and (\ref{chapter5eq5b}) are 
coupled wave equations for  $a(t,r)$ and 
$b(t,r)$, therefore we search for a  wave-like solution 
\begin{eqnarray}
a\left( t,r \right) & = & \int\frac{d\omega \, dk}{2\pi} \, \, 
\tilde{a}(\omega,k) 
\, \mbox{e}^{i \, \left(kr \omega t \right)} \,,\\
&&\nonumber\\
b \left( t,r \right) & = & \int\frac{d\omega \, dk}{2\pi} \, \, 
\tilde{b}(\omega,k) 
\, \mbox{e}^{i \, \left(kr \omega t \right)} \,, 
\end{eqnarray}
where $k \equiv \left| \vec{k} \right|$, and we substitute these 
into Eqs.~(\ref{chapter5eq5a}) and 
(\ref{chapter5eq5b}), setting 
$\Psi(t)=0$. Eqs.~(\ref{chapter5eq5a}) and 
(\ref{chapter5eq5b}) are satisfied if
\begin{eqnarray}
\tilde{a}(\omega,k) & = & \tilde{b}(\omega,k)\,,  \;\;\;\;\;\; 
\omega=\pm  k \,,\\
&&\nonumber\\
\tilde{a}(\omega,k) & = &  \biggl( 1-\frac{3\xi}{4k^2}\biggr)  
\tilde{b}(\omega,k)\,,   \;\;\;\;\;\;  \omega 
=\pm\sqrt{k^2-\frac{3\xi}{4}} 
\end{eqnarray}
where, as before, $ \xi=f_1/ 6f_2 $.  In
particular, for $f_1=0$ or $f_2=0$, one  obtains solutions with 
dispersion relation  $\omega=\pm k$. 
For $f_i \neq
0$ ($i=1,2$), the dispersion relation  suggests that 
massive modes are present. In particular, for $\xi<0$,  the mass 
of the scalar graviton is
${\displaystyle m_{grav}\,=\,-3\xi/ 4}$ and, accordingly, 
it is obtained for a modified real gravitational potential. A  
non-Newtonian gravitational potential describes a massive 
degree of freedom in the particle spectrum of the gravity 
sector with interesting perspectives for the detection and 
production of gravitational waves \cite{CCD}. The presence of 
massive modes in higher order gravity is 
well known  \cite{Stelle78}.

If $\xi>0$,  the solution
\begin{eqnarray}
a\left( \tilde{t},\tilde{r} \right) & = & \left( 
a_0+a_1\tilde{r} \right) 
\, \mbox{e}^{\pm\frac{\sqrt{3}}{2} \, \tilde{t}} \,, \\
&&\nonumber\\
b\left( \tilde{t},\tilde{r} \right)  
& = & \left( b_0 + b_1\tilde{t} \right) \cos\biggl( 
\frac{\sqrt{3}}{2} \,   \tilde{r}\biggr) + \left( 
b'_0+b'_1\tilde{t} \right)  \sin\biggl( 
\frac{\sqrt{3}}{2}\tilde{r}\biggr) +b''_0+
b''_1\tilde{t} \,,\nonumber\\
&&
\end{eqnarray}
with $a_0$, $a_1$, $b_0$, $b_1$, $b'_0$, $b'_1$, $b''_0$, $b''_1$
constants is admitted. The variables $\tilde{r}$ and $\tilde{t}$
are expressed in units of $\xi^{-1/2}$. In the post-Minkowskian
approximation, as expected, the 
gravitational field propagates 
via  wave-like solutions. The gravitational wave  content  of 
fourth order gravity originates new phenomenology  (massive 
modes) to be  taken  into account by the gravitational wave 
community. These massive degrees of freedom could also constitute 
a potential candidate for cold dark matter 
\cite{massivegrav}.

\subsection{The $f(R)$-gravity energy-momentum pseudo-tensor and gravitational radiation }

As we have seen, higher order theories of gravity introduce 
extra degrees of freedom  which can be described by
writing the field equations as effective Einstein 
equations and introducing  an additional curvature ``effective 
source'' in their right  hand side. This quantity behaves as 
an effective energy-momentum tensor contributing to the 
energy loss of a system due to the emission of gravitational 
radiation. The procedure  to calculate the 
stress-energy  pseudo-tensor
of gravitational waves in GR can 
be extended to more  general theories and this quantity can be 
obtained by varying  the gravitational Lagrangian. In GR this 
quantity is known as the Landau-Lifshitz 
pseudo-tensor \cite{LandauLifschitz}.

Let us consider $f(R)$-gravity, for which 
\begin{eqnarray}
\delta \int d^4 x \, \sqrt{-g} \, f(R) & = &  \delta \int
d^4x \mathcal{L}\left( g_{\mu\nu}, g_{\mu\nu,\rho} , 
g_{\mu\nu,\rho\sigma} \right) \nonumber\\
&&\nonumber\\
&  \approx & \int
d^4x \biggl[ \frac{ \partial \mathcal{L}}{\partial 
g_{\rho\sigma}}   - \partial_{\lambda} \left(  \frac{\partial 
\mathcal{L}}{\partial g_{\rho\sigma,\lambda}} \right) 
+\partial^2_{\lambda\xi}  \left( \frac{\partial 
\mathcal{L}}{\partial
g_{\rho\sigma,\lambda\xi}} \right)  \biggr] \delta
g_{\rho\sigma}  \nonumber\\ 
&&\nonumber\\
& \equiv & \int d^4x \, \sqrt{-g} \, {\cal H}^{\rho\sigma}\delta 
g_{\rho\sigma} = 0 \,.
\end{eqnarray}
The Euler-Lagrange equations 
\begin{eqnarray}
\frac{\partial \mathcal{L} }{\partial  g_{\rho\sigma}} 
-\partial_{\lambda} \left( \frac{\partial \mathcal{L}}{\partial
g_{\rho\sigma,\lambda}} \right) + 
\partial^2_{\lambda\xi} \left( 
\frac{\partial \mathcal{L}}{\partial
g_{\rho\sigma,\lambda\xi}} \right) = 0 
\end{eqnarray}
coincide with the vacuum field equations. Even in the case of 
more general theories,  it is possible to
define the energy-momentum pseudo-tensor 
\begin{eqnarray}
t^{ \lambda }_{\alpha} = \frac{1}{\sqrt{-g}}  
\left\{ \biggl[  \frac{\partial \mathcal{L} }{\partial
g_{\rho\sigma,\lambda} } - \partial_{\xi} 
\left( \frac{\partial \mathcal{L} }{\partial 
g_{\rho\sigma,\lambda\xi} } \right) \biggr]  
g_{\rho\sigma,\alpha} 
+ \frac{\partial \mathcal{L} }{ \partial  
g_{\rho\sigma,\lambda\xi} } \, 
g_{\rho\sigma,\xi\alpha } 
- \delta^{\lambda}_{\alpha} \, \mathcal{L} \right\} 
\,.\nonumber\\
&& 
\end{eqnarray}
This quantity, together with the matter energy-momentum tensor  
$T_{\mu\nu}^{(m)} $, satisfies a conservation law as required by 
the contracted Bianchi identities in conjunction with the 
effective Einstein  equations. In fact, in the presence of 
matter one has
${\displaystyle 
{\cal H}_{\mu\nu}\, = \,\displaystyle \frac{\kappa}{2} \, 
T_{\mu\nu}^{(m)} }$  
and 
\begin{eqnarray}
&& \left( \sqrt{-g} \, t^\lambda_{\alpha} \right)_{,\lambda}  
= -\sqrt{-g} \, H^{\rho\sigma} \, g_{\rho\sigma,\alpha} 
= -\frac{\kappa}{2} \, \sqrt{-g} \, T^{\rho\sigma}_{(m)}
g_{\rho\sigma,\alpha} 
=-\kappa \left( \sqrt{-g} \, 
{T_{(m)} }^\lambda_{\alpha} \right)_{,\lambda} \,;\nonumber\\
&&
\end{eqnarray}
as a consequence,
\begin{eqnarray}
\left[ \sqrt{-g} \left( t^{\lambda}_{\alpha}  
+ \kappa \, {T^{(m)} }^\lambda_\alpha \right) \right]_{,\lambda} 
=0 
\,,
\end{eqnarray}
which is  the conservation law given by the contracted Bianchi 
identities.  We can now write  the expression of the 
energy-momentum pseudo-tensor $t^\lambda_\alpha $  in terms of  
$f(R)$  and its derivatives
\begin{eqnarray} 
t^{\lambda}_{\alpha}& = & f'\biggl\{\biggl[ 
\frac{\partial R}{\partial g_{\rho\sigma,\lambda}} 
-\frac{1}{\sqrt{-g}} \,  \partial_\xi
\biggl(\sqrt{-g} \, \frac{\partial R}{\partial
g_{\rho\sigma,\lambda\xi}} 
\biggr) \biggl] g_{\rho\sigma,\alpha}
+\frac{\partial R }{\partial g_{\rho\sigma,\lambda\xi}} \,
g_{\rho\sigma,\xi\alpha}\biggr\} \nonumber\\
&&\nonumber\\
& - &  f''R_{,\xi} \, \frac{\partial R }{\partial
g_{\rho\sigma,\lambda\xi}} \,  
g_{\rho\sigma,\alpha} - \delta^{\lambda}_{\alpha} \, f \,.
\label{tens-f(R)}
\end{eqnarray}
$t^{\lambda}_{\alpha}$ is a non-covariant quantity in GR while 
its generalization to fourth order gravity turns out to be 
covariant. This  expression  reduces to the Landau-Lifshitz 
pseudo-tensor of GR in the 
limit $f(R) \rightarrow  R$,  in which
\begin{eqnarray}
{t^\lambda_\alpha}_{|_{\text{GR}}} 
=\frac{1}{\sqrt{-g}}\biggl(\frac{ 
\partial\mathcal{L}_{\text{GR}}}{ 
\partial g_{\rho\sigma,\lambda}} \, g_{\rho\sigma
,\alpha}-\delta^{\lambda}_{\alpha} \mathcal{L}_{\text{GR}}\biggr)
\end{eqnarray}
and where the GR Lagrangian has been considered in its effective 
form containing the symmetric part of the Ricci tensor which 
 leads to the equations of motion 
\begin{equation}
\mathcal{L}_{\text{GR}} = \sqrt{-g} \, g^{\mu\nu}\left( 
\Gamma^\rho_{\mu\sigma}\Gamma^\sigma_{\rho\nu} 
-\Gamma^\sigma_{\mu\nu}\Gamma^\rho_{\sigma\rho} \right) \,.
\end{equation}
The definitions of energy-momentum 
pseudo-tensor
 in GR and in 
$f(R)$-gravity are  different. The difference is due to the 
fact that   terms of order higher than second 
are present in  $f(R)$-gravity and they  cannot be discarded as 
boundary terms following  integration by parts, as 
is done  in GR. The effective Lagrangian of GR turns out to be 
the symmetric part of the Ricci scalar since the second order 
terms appearing in the expression of $R$ can be 
removed  integrating by parts. An  analytic $f(R)$ 
Lagrangian can be rewritten,  to  linear order,  as $f\sim 
f'_0R+ \mathcal{F}(R)$, where the  function $ \mathcal{F}$ 
is such that  
${\displaystyle   \mathcal{F} (R) \approx R^2 } $ as 
$R\rightarrow 0$. As a  consequence, one can rewrite 
$t^\lambda_\alpha$ as 
\begin{eqnarray}
t^\lambda_\alpha & = & f'_0{t^\lambda_\alpha}_{|_{\text{GR}}} 
+\mathcal{F}'\biggl\{\biggl[ \frac{\partial R}{\partial g_{\rho
\sigma,\lambda}} - \frac{1}{\sqrt{-g}} \,  
\partial_\xi\biggl(\sqrt{-g}\frac{\partial R}{
\partial  g_{\rho\sigma,\lambda\xi}}\biggr) 
\biggl]g_{\rho\sigma,\alpha} \nonumber\\
&&\nonumber\\ 
& + &  \frac{\partial R }{\partial
g_{\rho\sigma,\lambda\xi}}g_{\rho\sigma, 
\xi\alpha}\biggr\}- \mathcal{F}''R_{,\xi}\frac{\partial
R }{\partial g_{\rho\sigma,\lambda\xi}} 
g_{\rho\sigma,\alpha} - \delta^\lambda_\alpha \mathcal{F} \,.
\label{tensorf(R).1}
\end{eqnarray}
The general expression of the Ricci scalar obtained by 
splitting its linear ($R^*$) and quadratic ($\bar{R}$) parts  
in the metric perturbations  is
\begin{eqnarray}\label{defRicciscalar}
R = g^{\mu\nu}(\Gamma^\rho_{\mu\nu,\rho} 
-\Gamma^\rho_{\mu\rho,\nu})+g^{\mu\nu}( 
\Gamma^{\rho}_{\sigma\rho}\Gamma^
{\sigma}_{\mu\nu}-\Gamma^{\sigma}_{\rho\mu}\Gamma^{\rho}
_{\nu\sigma})= R^*+\bar{R} 
\end{eqnarray}
(note that $ \mathcal{L}_{\text{GR}} = -\sqrt{-g} \, \bar{R}$). 
In the GR case  ${t^\lambda_\alpha}_{|_{\text{GR}}}$, the 
first non-vanishing term of the Landau-Lifshitz 
pseudo-tensor is of order $ 
h^2 $ \cite{LandauLifschitz, Isaacson68}. A similar 
result  can be obtained in $f(R)$-gravity: using 
Eq.~(\ref{tensorf(R).1}) one 
obtains that, to lowest order, 
\begin{eqnarray} 
t^\lambda_\alpha & \sim & {t^\lambda_\alpha}_{|h^2}  
= f'_0{t^\lambda_\alpha}_{|_{\text{GR}}}
+f''_0R^*\biggl[\biggl(- \partial_\xi \frac{\partial 
R^*}{\partial g_{\rho\sigma,\lambda\xi}} 
\biggr)g_{\rho\sigma,\alpha} + \frac{\partial
R^*}{\partial g_{\rho\sigma, \lambda\xi}} \, 
g_{ \rho\sigma,\xi\alpha}\biggr] \nonumber\\
&&\nonumber\\
& - &  f''_0 
\, R^*_{,\xi} \, 
\frac{\partial R^*}{\partial g_{\rho\sigma,\lambda\xi}} \, 
g_{\rho\sigma,\alpha}-\frac{f''_0}{2} \, 
\delta^{\lambda}_{\alpha} \, {R^*}^2 \nonumber\\
&&\nonumber\\
&=& f'_0{t^\lambda_\alpha}_{|_{\text{GR}}} 
+ f''_0\biggl[R^*\biggl(\frac{\partial R^*}{
\partial g_{\rho\sigma,\lambda\xi}} g_{\rho\sigma,  \xi\alpha}
-\frac{1}{2}R^*\delta^\lambda_\alpha\biggr) \nonumber\\
&&\nonumber\\
& - & \partial_\xi\biggl(R^*\frac{\partial R^*}{\partial
g_{\rho\sigma,\lambda\xi}} \biggr)g_{\rho\sigma,\alpha}\biggr]\,.
\label{tensorf(R).2}
\end{eqnarray}
Using the perturbed metric we have $ R^*\sim R^{(1)}$, 
where $R^{(1)}$ is defined by
\begin{eqnarray} 
R^{(1)}_{\mu\nu}  & = &  h^\sigma_{(\mu,\nu)\sigma}
-\frac{1}{2} \, \Box h_{\mu\nu} -\frac{1}{2} \, h_{,\mu\nu} \,,
\label{approx1a}\\
&&\nonumber\\
R^{(1)} & = & {h_{\sigma\tau}}^{,\sigma\tau}-\Box h  
\label{approx1b}
\end{eqnarray}
with $ h \equiv {h^\sigma}_\sigma$. In terms of $h$ and 
$\eta_{\mu\nu}$, one obtains 
\begin{eqnarray}
&& \frac{\partial R^*}{\partial g_{\rho\sigma,\lambda\xi}} 
\sim \frac{\partial R^{(1)}}{\partial
h_{\rho\sigma,\lambda\xi}}= 
\eta^{\rho\lambda}\eta^{\sigma\xi}- 
\eta^{\lambda\xi}\eta^{\rho\sigma} \,, \\
&&\nonumber\\
&& \frac{\partial R^*}{\partial g_{\rho\sigma,\lambda\xi}} 
\, g_{\rho\sigma,\xi\alpha}\sim
h^{\lambda\xi}_{\,\,\,\,\,\,, 
\xi\alpha}-h^{,\lambda}_{\,\,\,\,\,\alpha} \,.
\end{eqnarray}
The first significant term in Eq.~(\ref{tensorf(R).2}) 
is of second order in the perturbations. We can now  write  the 
explicit expression of the  pseudo-tensor  in 
terms of the perturbation $h $, 
\begin{eqnarray}
t^{\lambda}_{\alpha}  & \sim & f'_0  
{t^{\lambda}_{\alpha}}_{|_{\text{GR}}} 
+f''_0 \left\{ 
\left( h^{\rho\sigma}_{\,\,\,\,\,\,\,,\rho\sigma}-\Box
h \right) \left[ h^{\lambda\xi}_{\, \,\,\,\,\,\,,\xi\alpha} 
-h^{,\lambda}_{\,\,\,\,\,\,\,\alpha} 
-\frac{1}{2} \, 
\delta^{\lambda}_{\alpha} \left( h^{\rho\sigma}_{\,\,\,\,\,\,\,,
\rho\sigma}-\Box h \right) \right] \right.\nonumber\\
\nonumber\\
& - & \left.  h^{\rho\sigma}_{\, 
\,\,\,\,\,\,,\rho\sigma\xi} h^{\lambda\xi}_{\,\, 
\,\,\,\,\,,\alpha}+ 
h^{\rho\sigma\,\,\,\,\,\,\,\,\, 
\,\lambda}_{\,\,\,\,\,\,\,,\rho\sigma} 
h_{,\alpha} + h^{\lambda\xi}_{\,\,\,\,\,\,\,,\alpha} \Box
h_{,\xi}-\Box h^{,\lambda}h_{,\alpha}\right\} \,.
\end{eqnarray}
This expression can be put in compact form using  the metric 
perturbation $\tilde{h}_{\mu\nu}$ as
\begin{eqnarray}
{t^\lambda_\alpha}_{|_f}= 
\frac{1}{2}\biggl[ \frac{1}{2} \, \tilde{h}^{,\lambda}_{\,\,\,\, 
\alpha} \Box \tilde{h}-\frac{1}{2} \, \tilde{h}_{,\alpha}
\Box\tilde{h}^{,\lambda} - \tilde{h}^{\lambda}_{\,\,\, 
\,\,\sigma,\alpha} \Box\tilde{h}^{,\sigma} 
-\frac{1}{4} \left( \Box\tilde{h} \right)^2 
\delta^{\lambda}_{\alpha} \biggr]\,.
\end{eqnarray}
The energy-momentum pseudo-tensor of the  gravitational field 
describing the energy transport  during propagation has a natural 
generalization to $f(R)$-gravity. Here we have adopted the
Landau-Lifshitz construct, but  many other 
pseudo-tensors
  can be used  \cite{multamaki,stein}. The general 
definition of $ {t^\lambda_\alpha}$ obtained above consists of 
the sum of a GR contribution plus a term characteristic of 
$f(R)$-gravity,
\begin{eqnarray}
t^{\lambda}_{\alpha} = f'_0 \,\, 
{t^\lambda_\alpha}_{|_{\text{GR}}} 
+f''_0 \,\, {t^\lambda_\alpha}_{|_f}\,.
\end{eqnarray}
In the limit  $f(R) \rightarrow R$ one obtains
$ t^\lambda_\alpha = {t^\lambda_\alpha}_{|_{\text{GR}}}$. Massive 
gravitational modes are contained in $ 
{{t^{\lambda}_\alpha}}_{|_f} $
since $\Box\tilde{h}$ can be considered as an effective scalar 
field degree of freedom evolving in a potential and  $ 
t^\lambda_{\alpha} $ describes the transport of energy and 
momentum.

\subsection{Ghosts, massless and massive gravitational modes}
Detecting new gravitational modes  could  be a crucial 
experiment able to discriminate among theories since these modes  
would constitute evidence  that GR must be 
enlarged or modified  
\cite{BellucciCapozzielloDeLaurentisFaraoni09, elizalde}. In 
general, field equations containing  higher order terms describe, 
in addition to the massless spin two field (the standard 
graviton of GR), also spin zero  and spin two  massive modes, 
the latter possibly being ghosts. This result 
is general and can 
be  obtained by means of a straightforward generalization of 
the above discussion for $f(R)$-gravity.

Let us generalize the Hilbert-Einstein action by adding
curvature invariants different from the Ricci scalar, 
\be \label{GWHIGHERaction}
S= \int d^4x \, \sqrt{-g} \, f \left( R, P, Q \right) \,,
\ee 
where 
\begin{eqnarray}
 P & \equiv & R_{\mu\nu}R^{\mu\nu} \,, \\
&&\nonumber\\
Q &\equiv & R_{\mu\nu\rho\sigma}R^{\mu\nu\rho\sigma} \,. 
\end{eqnarray} 
By varying the action~(\ref{GWHIGHERaction}) with respect to 
$g^{\mu\nu}$, one obtains the  field equations 
\cite{Carroll:2004de} 
\begin{eqnarray}
F \, G_{\mu\nu} & = & \frac{1}{2} \,  g_{\mu\nu}\left( f-
R \,F \right) - \left( g_{\mu\nu} \Box - 
\nabla_\mu \nabla_\nu \right) F \nonumber\\
&&\nonumber\\
& - &2\left( f_P R^{\alpha}_\mu
R_{\alpha\nu}+f_Q \, R_{\alpha \beta 
\gamma\mu}R^{\alpha\beta\gamma}_{~~~\nu} \right)\nonumber\\
&&\nonumber\\
&- & g_{\mu\nu} \nabla_{\alpha}\nabla_{\beta} \left( f_P 
R^{\alpha\beta} \right)- \Box \left( f_P
R_{\mu\nu} \right) \nonumber\\ 
&&\nonumber\\
& + & 2\nabla_{\alpha} \nabla_{\beta} \left( 
f_P~R^{\alpha}_{~(\mu}\delta^{\beta}_{~\nu)}+2
f_Q \, R^{\alpha~~~~\beta}_{~(\mu\nu)}\right) \,,
\label{GWHIGHERfieldeqs}
\end{eqnarray}
where 
\be 
F \equiv \frac{\partial  f}{\partial  R} \,, 
~~~~~f_P \equiv \frac{\partial f}{\partial P} \,, 
~~~~~f_Q \equiv\frac{\partial f}{\partial Q} \,.
\ee 
The trace of Eq.~(\ref{GWHIGHERfieldeqs}) yields
\begin{eqnarray}
&& \Box \left( F+\frac{f_P}{3} R \right) \nonumber\\
&&\nonumber\\
&& = \frac{1}{3}\left\{ 2 f-RF-2 
\nabla_{\alpha}\nabla_{\beta} \left[ (f_P+2f_Q)R^{\alpha\beta} 
\right] -2 \left( f_P P + f_Q Q \right) \right\} 
\,.  \label{GWHIGHERtrace}
\end{eqnarray}
Expanding the third term on the right hand side 
of~(\ref{GWHIGHERtrace}) 
and using the contracted Bianchi identities, one obtains 
\begin{eqnarray}
&& \Box  \left(F+\frac{2}{3}(f_P+f_Q) R\right) 
 \nonumber\\ 
&&\nonumber\\
& = & \frac{1}{3} \,  \Bigg[ 2 f-RF-2R^{\mu\nu}\nabla_{\mu} 
\nabla_{\nu} 
\left( f_P+ 2f_Q \right) 
-R \Box(f_P+2f_Q) \nonumber\\ 
&&\nonumber\\
&- & 2  \left( f_P P+f_Q Q \right) \Bigg]  \,.
\label{GWHIGHERtrace1}
\end{eqnarray}
By defining 
\be
 \Phi  \equiv  F+ \frac{2}{3} \left( f_P+f_Q \right) R 
\label{GWHIGHERphidef} 
\ee
and 
\be
\frac{dV}{d\Phi}  \equiv  
\frac{1}{3} \left[ 2 f-RF-2R^{\mu\nu}\nabla_{\mu} 
\nabla_{\nu} \left( f_P+ 2f_Q \right) 
-R \, \Box(f_P+2f_Q) 
-  2  \left( f_P P+f_Q Q \right) \right] \,,
\ee
the Klein-Gordon equation 
\be 
\Box \Phi  - \frac{dV}{d\Phi}=0 
\ee 
is obtained.  In order to find the modes of the gravity 
waves of this theory, we linearize  around the Minkowski 
background, 
\begin{eqnarray}
g_{\mu\nu} & = & \eta_{\mu\nu}+h_{\mu\nu} \,, \\
&&\nonumber\\
\Phi & = & \Phi_0+\delta \Phi \,;
\end{eqnarray}
then Eq.~(\ref{GWHIGHERphidef}) yields
\be 
\delta \Phi = \delta F+\frac{2}{3} \left( \delta f_P+\delta f_Q 
\right) R_0 + \frac{2}{3} \left( f_{P0}+f_{Q0} \right) \delta R 
\,, \label{GWHIGHERpertphi1}
\ee 
where $R_0 \equiv R (\eta_{\mu\nu})=0$ and, similarly,  
$ {\displaystyle f_{P0}=\frac{\partial   f}{\partial P} 
\left.\right|_{\eta_{\mu\nu}} } $, which 
is either constant or  zero  (a zero subscript  denoting  
quantities  evaluated with the Minkowski  metric). $\delta R$   
denotes the first order perturbation of the Ricci scalar which, 
together with the perturbed parts of the Riemann and Ricci 
tensors, is given by 
\begin{eqnarray}
\delta R_{\mu\nu\rho\sigma}&=&\frac{1}{2}\left(\partial_\rho 
\partial_\nu h_{\mu \sigma}+\partial_\sigma \partial_\mu h_{\nu 
\rho}-\partial_\sigma  \partial_\nu h_{\mu
\rho}-\partial_\rho \partial_\mu h_{\nu \sigma} \right) \,,\\
&&\nonumber\\
\delta R_{\mu\nu} &=& \frac{1}{2}\left(\partial_\sigma 
\partial_\nu
h^\sigma_{~\mu} +\partial_\sigma \partial_\mu 
h^\sigma_{~\nu}-\partial_\mu \partial_\nu
h-\Box h_{\mu \nu} \right) \,,\\
&&\nonumber\\
\delta R &=&  \partial_\mu \partial_\nu h^{\mu \nu}-\Box 
h \,,
\end{eqnarray}
where
$ h \equiv  \eta^{\mu \nu} h_{\mu \nu}$. The first term of 
Eq.~(\ref{GWHIGHERpertphi1}) is 
\be 
\delta F=\frac{\partial F}{\partial  
R}\left.\right|_0 \, \delta R +\frac{\partial F}{\partial  
P}\left.\right|_0 \, \delta P +\frac{\partial   F}{\partial  
Q}\left.\right|_0 \, \delta Q \,,
\ee
however since $\delta P$ and $\delta Q$ are second order, it is
$\delta F \simeq F_{,R0}~ \delta R $ and 
\be 
\delta \Phi = \left[ F_{,R0} +\frac{2}{3} 
\left( f_{P0}+f_{Q0} \right) \right] 
\delta R \,. \label{GWHIGHERpertphi2}
\ee 
Eq.~(\ref{GWHIGHERtrace1}) then yields the Klein-Gordon equation 
for the scalar perturbation $\delta \Phi$
\begin{eqnarray}
\Box \delta \Phi & = & \frac{1}{3} \, \frac{F_0}{F_{,R0} 
+\frac{2}{3} \left( f_{P0}+f_{Q0} \right)} \, \delta 
\Phi \nonumber\\ 
&&\nonumber\\
& - & \frac{2}{3} \, \delta
{R}^{\alpha\beta} 
\partial_{\alpha} 
\partial_{\beta} \left( f_{P0}+2f_{Q0} \right) -\frac{1}{3} 
\, \delta {R} \, \Box(f_{P0}+2f_{Q0})\nonumber \\ 
&&\nonumber\\
&=& m_s^2 \, \delta \Phi \;.  \label{GWHIGHERkgordon1}  
\end{eqnarray}
The  second  line of Eq.~(\ref{GWHIGHERkgordon1}) vanishes 
because   $ f_{P0}$ and  $f_{Q0}$ are constant and   
the scalar mass is defined as 
\be
 m_s^2\equiv  \, \frac{F_0}{ 3F_{,R0} 
+2 \left( f_{P0}+f_{Q0} \right)} \,.
\ee
Perturbing the field equations (\ref{GWHIGHERfieldeqs}), one 
obtains
\begin{eqnarray}
&& F_0 \left( \delta{R}_{\mu\nu}-\frac{1}{2}\eta_{\mu\nu}
\delta{R} \right)  \nonumber\\ 
&&\nonumber\\
& = & - \left( \eta_{\mu\nu}\Box - \partial_\mu 
\partial_\nu \right) \left[ \delta
\Phi-\frac{2}{3} \left( f_{P0}+f_{Q0} \right) \delta{R} \right] 
\nonumber\\ 
&&\nonumber\\
&- & \eta_{\mu\nu} \partial_{\alpha} \partial_{\beta} \left( 
f_{P0} \delta{R}^{\alpha\beta} \right)-\Box \left( f_{P0}
\delta{R}_{\mu\nu} \right) \nonumber\\ 
&&\nonumber\\
& + & 2\, \partial_{\alpha} \partial_{\beta} 
\left( f_{P0}~\delta{R}^{\alpha}_{~(\mu}\delta^{\beta}_{~\nu ) } 
+2  f_{Q0}~\delta{R}^{\alpha~~~~\beta}_{~(\mu\nu )} \right) 
\,.\label{GWHIGHER15}
\end{eqnarray}
It is convenient to work in Fourier space so that, for example, 
$\partial_\gamma  h_{\mu\nu}\rightarrow i k_\gamma h_{\mu\nu}$ 
and $\Box h_{\mu\nu} \rightarrow - k^2 h_{\mu\nu}$, where now 
$ k^2 \equiv k^{\mu} k_{\mu}$. Then,  
Eq.~(\ref{GWHIGHER15}) becomes 
\begin{eqnarray}
&& F_0 \left( \delta{R}_{\mu\nu}-\frac{1}{2}\eta_{\mu\nu}
\delta{R} \right) \nonumber\\ 
&&\nonumber\\
&=& \left( \eta_{\mu\nu}k^2 -k_\mu k_\nu \right) \left[ \delta
\Phi-\frac{2}{3} \left( f_{P0}+f_{Q0} \right)\delta{R} 
\right] \nonumber\\
&&\nonumber\\
&+ & \eta_{\mu\nu} k_a k_b (f_{P0} \delta{R}^{ab})+k^2(f_{P0}
\delta{R}_{\mu\nu})\nonumber\\ 
&&\nonumber\\
&- & 2 k_{\alpha} 
k_{\beta} \left( 
f_{P0}~\delta{R}^{\alpha}_{~(\mu}\delta^{\beta}_{~\nu)} \right) 
-4 k_{\alpha} k_{\beta} \left( 
f_{Q0}~\delta{R}^{\alpha~~~~\beta}_{~(\mu\nu)} \right) 
\,.\nonumber\\ 
&& \label{GWHIGHERfields2}
\end{eqnarray}
We rewrite the metric perturbation as 
\be 
h_{\mu\nu}=\bar{h}_{\mu\nu}-\frac{\bar{h}}{2}~
\eta_{\mu\nu}+\eta_{\mu\nu} h_f \label{GWHIGHERgauge}
\ee 
and use the gauge freedom to demand that the usual 
conditions $ \partial_\mu
\bar{h}^{\mu\nu} = 0 $ and $\bar{h}=0$ hold. The first 
condition implies that $k_\mu \bar{h}^{\mu\nu} =0$, while the
second one gives 
\begin{eqnarray}
 h_{\mu\nu}& = &\bar{h}_{\mu\nu} +\eta_{\mu\nu} h_f \,,\\
&&\nonumber\\
h & = & 4 h_f \,. 
\end{eqnarray}
With these conditions in mind, we have 
\begin{eqnarray}
&&  \delta
R_{\mu\nu}=\frac{1}{2}\left( 2k_\mu k_\nu h_f+k^2 \eta 
_{\mu\nu} h_f + k^2 \bar{h}_{\mu\nu}\right) \,,\\
&&\nonumber\\
&&\delta R = 3 k^2 h_f \,,\\
&&\nonumber\\
&& k_{\alpha} k_{\beta} \, \delta
R^{\alpha~~~~~\beta}_{~~(\mu\nu)~} =  -\frac{1}{2} \left[ 
\left(k^4 \eta_{\mu\nu}-k^2 k_\mu k_\nu \right) h_f + k^4  
\bar{h}_{\mu\nu}\right] \,, \\
&&\nonumber\\
&& k_{\alpha}  k_{\beta}  
\, \delta{R}^{\alpha}_{~(\mu}\delta^{\beta}_{~\nu)} 
=   \frac{3}{2} \, k^2 k_\mu k_\nu h_f \,. 
\label{GWHIGHERresults1}
\end{eqnarray}
Using Eqs.~(\ref{GWHIGHERgauge})--(\ref{GWHIGHERresults1}) in  
Eq.~(\ref{GWHIGHERfields2}),  a little  algebra yields 
\begin{eqnarray}
&&  \frac{1}{2}\left(k^2-k^4 \, \frac{f_{P0}+4f_{Q0}}{F_0}\right)
\bar{h}_{\mu\nu} \nonumber\\
&&\nonumber\\
&& =(\eta_{\mu\nu}k^2 -k_\mu k_\nu)\frac{\delta \Phi}{F_0}
+(\eta_{\mu\nu}k^2 -k_\mu k_\nu)h_f \,.
\end{eqnarray}
Defining now $ h_f \equiv - \delta \Phi / F_0$, we find the 
perturbation equation  
\be
k^2 \left( 1 +\frac{k^2 }{m^2_{spin~2}}\right)
\bar{h}_{\mu\nu}=0 \,, \label{GWHIGHERsolution} 
\ee 
where
\be
 m^2_{spin~2}\equiv -\frac{F_0}{f_{P0}+4f_{Q0}} \,,
\ee
while Eq.~(\ref{GWHIGHERkgordon1}) gives 
\be 
\Box h_f=m_s^2 \, h_f \, . \label{GWHIGHERkgordon3} 
\ee 
It is easy to see from Eq.~(\ref{GWHIGHERsolution})  that we have 
a  modified dispersion relation
corresponding to a massless  spin two field ($k^2=0$) and a 
massive spin two ghost mode with 
\be
k^2 = \frac{ 2 F_0}{ \, f_{P0} + 4 f_{Q0}} \equiv 
-m^2_{spin~2}
\ee
with mass $m^2_{spin~2}$. In fact, the propagator of 
$\bar{h}_{\mu\nu}$ can be rewritten as 
\be 
G(k) \propto \frac{1}{k^2}- \, \frac{1}{k^2+m^2_{spin2}} \,.
\ee 
The negative  sign of the second term  indicates its 
ghost nature, which agrees with the results  found in the 
literature for this class of theories 
\cite{Nunez:2004ts, Chiba:2005nz, Stelle78,odi20052,cognola,bazeia1}. As a  check, we 
can see that for the Gauss-Bonnet Lagrangian density   
$\mathcal{G} = Q-4P+R^2$, we have $f_{P0}=-4$ and 
$f_{Q0}=1$, then Eq.~(\ref{GWHIGHERsolution}) simplifies to $k^2
\bar{h}_{\mu\nu}=0$ and in this case we have no 
ghosts,  as
expected.

The solution of  Eqs.~(\ref{GWHIGHERsolution}) 
and~(\ref{GWHIGHERkgordon3}) can 
be expanded  in plane waves as
\begin{eqnarray}
 \bar{h}_{\mu\nu}&=&A_{\mu\nu}
(\overrightarrow{p}) \,  \exp \left( ik^\alpha x_\alpha 
\right)+ \mbox{c.c.} 
\,, \label{GWHIGHERpw1}\\
&&\nonumber\\
h_f &=& a (\overrightarrow{p}) \, \exp \left( iq^\alpha
x_\alpha \right)+ \mbox{c.c.} \,, \label{GWHIGHERpw2} 
\end{eqnarray}
where
\begin{eqnarray}
k^{\alpha} & \equiv & \left( \omega_{m_{spin~2}}, 
\overrightarrow{p} \right) \,, \;\;\;\;\;\;\;\; 
\omega_{m_{spin~2}} = \sqrt{m_{spin~2}^{2}+p^{2}} \,,
\label{GWHIGHEReq:keq1}\\
&& \nonumber\\
q^{\alpha} & \equiv & \left( \omega_{m_s}, \overrightarrow{p} 
\right)  \,, \;\;\;\;\;\;\;\;
\omega_{m_s} = \sqrt{m_s^{2} + p^{2}} \,,
\label{GWHIGHEReq:keq2}
\end{eqnarray} 
and where $ m_{spin~2}$ is zero (respectively, non-zero) in the 
case of massless (respectively, massive) spin two modes and the 
polarization tensor $A_{\mu\nu} (\overrightarrow{p})$ is given 
by Eqs.~(21)--(23) of Ref.~\cite{vanDam:1970vg}.  In  
Eqs.~(\ref{GWHIGHERsolution}) and~(\ref{GWHIGHERpw1}),  
the equation and the solution for the standard waves
of GR \cite{thorne} have been obtained while 
 Eqs.~(\ref{GWHIGHERkgordon3}) and~(\ref{GWHIGHERpw2}) are 
the equation and the solution for the massive mode, respectively  
(see also Ref.~\cite{CapozzielloCordaDeLaurentis08}).

The fact that the dispersion law for 
the modes of the massive
field $h_{f}$ is not linear has to be emphasized. The velocity of
every "ordinary'' ({\em i.e.},  arising  from GR) mode 
$\bar{h}_{\mu\nu}$ is the light speed $c$, but the dispersion
law~(\ref{GWHIGHEReq:keq2}) for the 
modes of  $h_{f}$ is that of a massive field which can be 
discussed like a wave packet 
\cite{CapozzielloCordaDeLaurentis08}.  The group 
velocity of a wave packet of $h_{f}$ centered in 
$\overrightarrow{p}$ is
\begin{equation}
\overrightarrow{v_{g}}=\frac{\overrightarrow{p}}{\omega} \,, 
\label{GWHIGHEReq:velocita'digruppo} 
\end{equation}
which is exactly the velocity of a massive particle with mass 
$m$ and momentum $\overrightarrow{p}$.
From Eqs.~(\ref{GWHIGHEReq:keq2})  
and~(\ref{GWHIGHEReq:velocita'digruppo}), it is easy to obtain
\begin{equation}
v_{g}=\frac{\sqrt{\omega^{2}-m^{2}}}{\omega} \,.
\label{GWHIGHEReq:velocita'digruppo2}
\end{equation}
In order for the wave packet to have constant speed, it must 
be \cite{CapozzielloCordaDeLaurentis08}
\begin{equation}
m=\sqrt{(1-v_{g}^{2})} \,  \omega \,.
\label{GWHIGHEReq:relazionemassa-frequenza}
\end{equation}
Before proceeding, we discuss the phenomenological constraints on 
the mass of the gravitational wave field. For  
frequencies  in the range relevant for  space-based and 
terrestrial  gravitational antennas, {\em i.e.},  
$10^{-4} $~Hz$ \leq f \leq10 $~kHz  \cite{acernese, willke, sigg, 
abbott, ando,  tatsumi, lisa2}, a strong constraint  
is available. For a massive gravitational wave it is  
\begin{equation}
 \omega=\sqrt{m^{2}+p^{2}} \,,
\label{GWHIGHEReq:frequenza-massa}
\end{equation}
and then 
\begin{equation}
0 \,\, \mbox{eV} \leq m\leq10^{-11} \,\, \mbox{eV} \,.
\label{GWHIGHEReq:rangedimassa}
\end{equation}
A stronger bound comes from  cosmology and Solar System tests, 
which provide
\begin{equation}
0 \,\, \mbox{eV} \leq m\leq10^{-33}\,\, \mbox{eV} \,.
\label{GWHIGHEReq:rangedimassa2}
\end{equation}
The effects of  these light scalars can be discussed as those of 
a coherent gravitational wave.

\subsubsection{New polarization states of gravitational radiation}

Looking at  Eq.~(\ref{GWHIGHERkgordon1}) we see that  we can have 
a $k^2=0$  mode corresponding to a massless spin two field with 
two independent polarizations plus a scalar mode while, if  
$k^2\neq 0$, we have a massive spin two ghost mode 
(``poltergeist'') and there are five 
independent  polarization tensors plus a scalar mode. First, let 
us consider the case in which the  spin two field is massless.

Taking $\overrightarrow{p}$ in the $z$-direction, a gauge in 
which only $A_{11}$, $A_{22}$, and $A_{12}=A_{21}$ are different 
from zero can be chosen. The condition $\bar{h}=0$ gives 
$A_{11}=-A_{22}$. In this frame, we can take the 
polarization bases\footnote{These polarizations are defined in 
the physical three-space.  The  polarization vectors are 
orthogonal to each  another  and are normalized according to 
$e_{\mu\sigma}e^{\sigma\nu} =2\delta_{\mu}^{\nu}$. The other 
modes are not traceless, in contrast to the ordinary "plus'' and 
"cross'' polarization modes of GR.}
\begin {equation}
e_{\mu\nu}^{(+)}=\frac{1}{\sqrt{2}}\left(
\begin{array}{ccc}
1 & 0 & 0 \\
0 & -1 & 0 \\
0 & 0 & 0
\end{array}
\right) \,,  \qquad e_{\mu\nu}^{(\times)} 
=\frac{1}{\sqrt{2}}\left(
\begin{array}{ccc}
0 & 1 & 0 \\
1 & 0 & 0 \\
0 & 0 & 0
\end{array}
\right) \,,
\end{equation}

\begin {equation}
e_{\mu\nu}^{(s)}=\frac{1}{\sqrt{2}}\left(
\begin{array}{ccc}
0 & 0 & 0 \\
0 & 0 & 0 \\
0 & 0 & 1
\end{array}\right) \,.
\end{equation}
Substituting these expressions into Eq.~(\ref{GWHIGHERgauge}), 
it follows that
\begin{eqnarray}
 h_{\mu\nu}(t,z)&=&A^{+}(t-z) \, e_{\mu\nu}^{(+)}
+A^{\times}(t-z) \, e_{\mu\nu}^{(\times)}\nonumber\\
&&\nonumber\\
&+&h_{s}(t-v_{g}z) \, e_{\mu\nu}^{s} \,.
\label{GWHIGHEReq:perturbazionetotale}
\end{eqnarray}
The terms $ A^{+}(t-z) \, e_{\mu\nu}^{(+)} $ and 
$ A^{\times}(t-z)  \, e_{\mu\nu}^{(\times)}$
describe the two standard polarizations of gravitational waves
which arise in GR, while the term $ h_{s}(t-v_{g}z) \, 
\eta_{\mu\nu}$ is the massive field arising 
from the generic $f(R)$ theory.

When the spin two field is massive, the bases of 
the  six polarizations are defined by 
\begin {equation}
e_{\mu\nu}^{(+)}=\frac{1}{\sqrt{2}}\left(
\begin{array}{ccc}
1 & 0 & 0 \\
0 & -1 & 0 \\
0 & 0 & 0
\end{array}
\right) \,, \qquad e_{\mu\nu}^{(\times)} 
=\frac{1}{\sqrt{2}}\left(
\begin{array}{ccc}
0 & 1 & 0 \\
1 & 0 & 0 \\
0 & 0 & 0
\end{array}
\right) \,,
\end{equation}

\begin {equation}
e_{\mu\nu}^{(B)}=\frac{1}{\sqrt{2}}\left(
\begin{array}{ccc}
0 & 0 & 1 \\
0 & 0 & 0 \\
1 & 0 & 0
\end{array}
\right)\,, \qquad 
e_{\mu\nu}^{(C)}=\frac{1}{\sqrt{2}}\left(
\begin{array}{ccc}
0 & 0 & 0 \\
 0 & 0 & 1 \\
0 & 1 & 0
\end{array}
\right) \,, 
\end{equation}

\begin {equation}
e_{\mu\nu}^{(D)}=\frac{\sqrt{2}}{3}\left(
\begin{array}{ccc}
\frac{1}{2} & 0 & 0 \\
0 & \frac{1}{2} & 0 \\
0 & 0 & -1
\end{array}
\right) \,,\qquad 
e_{\mu\nu}^{(s)}=\frac{1}{\sqrt{2}}\left(
\begin{array}{ccc}
0 & 0 & 0 \\
0 & 0 & 0 \\
0 & 0 & 1
\end{array}
\right)\,, \label{GWHIGHERtensorpol}
\end{equation}
and the amplitude can be written in terms of the  six 
polarization states as
\begin{eqnarray}
h_{\mu\nu}(t,z) & = & A^{+}(t-v_{g_{s2}} z) \, 
e_{\mu\nu}^{(+)}+A^{\times}(t-v_{g_{s2}} z) \, 
e_{\mu\nu}^{(\times)} \nonumber\\
&&\nonumber\\
&+ & B^{B}(t-v_{g_{s2}} z) \, e_{\mu\nu}^{(B)} 
+C^{C}(t-v_{g_{s2}} z) \, e_{\mu\nu}^{(C)}\nonumber\\
&&\nonumber\\
& + & D^{D}(t-v_{g_{s2}}
z) \, e_{\mu\nu}^{(D)}+h_{s}(t-v_{g}z) \, e_{\mu\nu}^{s} \,,
\end{eqnarray}
where
\be
v_{g_{s2}}=\frac{\sqrt{\omega^{2}-m_{s2}^{2}}}{\omega} 
\label{GWHIGHERspin2group}
\ee
is the group velocity of the massive spin two field.  The first 
two polarizations are the same as in the massless case,
inducing tidal deformations of the $\left( x, y \right)$ plane.  
Fig.~\ref{GWHIGHERfig1} illustrates how each 
gravitational wave  polarization  affects test masses arranged on 
a circle before the wave impinges on them.
\begin{figure}
\begin{center}
\includegraphics[scale=0.4]{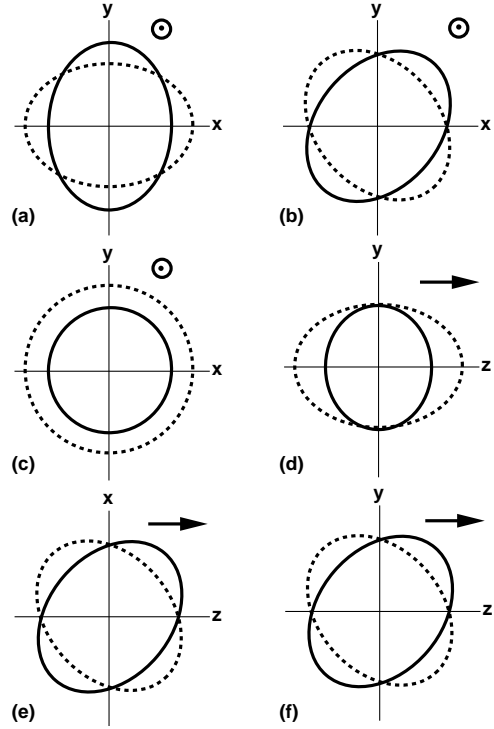}
\caption{\label{GWHIGHERfig1}The six polarization modes of  
gravitational waves. We illustrate the displacement induced at  
phases spaced  by  $\pi$ radians by each mode  on a circle of 
test particles  at rest before the wave impinges upon them. The 
wave propagates out of the plane of the page in~(a), 
(b), and~(c)  and  into this plane in~(d), (e), and~(f).  
While~(a)  and~(b) describe the ``plus'' and  ``cross'' modes, 
respectively, (c) corresponds to  the scalar mode, and~(d), 
(e), and~(f) to the~D, B, and~C modes.}
\end{center}
\end{figure}

From a purely quantum-mechanical point of view, the presence of 
the ghost mode may seem as a pathology of the 
theory. There are
several reasons why this mode is  problematic in  the particle 
interpretation of the metric perturbations. The ghost 
mode can be 
viewed as either a particle state with  positive energy and 
negative probability density, or as a positive  probability 
density state with negative energy. In 
the first case, allowing the presence of such a particle will 
induce violations of unitarity, while the negative energy 
scenario leads to a theory without ground state and 
the system  becomes unstable. Vacuum can decay into pairs 
of ordinary and ghost gravitons leading to a 
catastrophic 
instability.

A way out of these problems consists of imposing a very weak 
coupling of the ghost  with the other 
particles in the 
theory, such that the decay rate of the vacuum  becomes  
comparable to the inverse of the Hubble time. The present vacuum 
state will then appear to be sufficiently stable. This is not a 
viable option in our theory because the ghost state appears in 
the gravitational sector, which is bound to couple to all forms 
of  matter present and it seems physically and mathematically 
unlikely for the ghost graviton to couple 
differently than the 
ordinary massless graviton does.

Another possibility consists of  assuming that this picture does 
not hold up to arbitrarily high energies and that at some cutoff 
scale $M_{cutoff}$ the theory gets modified appropriately to 
ensure a ghost-free behavior and a stable ground state. This can 
happen, for example,  if we assume that Lorentz-invariance is 
violated at $M_{cutoff}$, thereby restricting any potentially 
harmful decay \cite{Emparan:2005gg}. However, there is no 
guarantee that modified gravities  like  the one investigated 
here are valid to arbitrarily high  energies. Such models are 
plagued at the quantum level by the same problems of ordinary GR, 
{\em i.e.},  they are not renormalizable. It is, therefore, not 
necessary for them to be considered as genuine candidates for a 
quantum gravity theory and the corresponding ghost particle 
interpretation becomes ambiguous. At the classical level, the 
perturbation $h_{\mu \nu}$ should be viewed as nothing more than 
a tensor representing the stretching of space-time away from 
flatness. A ghost mode  then makes sense as 
just another way of 
propagating this perturbation of the space-time geometry, one 
which, in the propagator, carries  a sign opposite to that of 
an ordinary massive graviton. Viewed in this way, the presence of 
the massive ghost graviton 
will induce on an interferometer the same effects as an ordinary
massive graviton transmitting the perturbation, but with the
opposite sign of the displacement. Tidal stretching 
of the polarization plane by a polarized wave  will  turn 
into shrinking  and {\em vice-versa}. Eventually, the signal 
will be  a superposition of the displacements coming from the 
ordinary massless spin two graviton and the 
massive ghost. Since these two 
modes  induce competing effects, their superposition will lead to 
a less  pronounced signal than the one expected were the ghost 
mode absent, setting  less stringent  
constraints on the 
theory. However, the presence of the new modes will also affect 
the total energy density carried by the gravitational waves and 
this may  also appear as a candidate signal in stochastic 
gravitational wave 
backgrounds.

\subsubsection{The detector response}

Let us consider now the possible response of a  detector 
in the presence of   gravitational waves coming from a definite 
direction. The detector output depends on the gravitational wave  
amplitude, which is determined by specific
theoretical models. However, one can study the detector response 
to each gravitational wave polarization without specifying {\em a 
priori} the theoretical model. Following Refs.~\cite{abio, abio2,  
bonasia, babusci, vicere, leaci, Maggiore}, the angular pattern 
function  of a detector of 
gravitational waves is given by
\begin{eqnarray}
&& F_A (\hat{ \Omega}) = \mathbf{D} : \mathbf{e}_A
(\hat{\Omega})\,,  \label{GWHIGHEReq2} \\
&&\nonumber\\
&& \mathbf{D} =  \frac{1}{2}\left( \hat{\mathbf{u}} \otimes \hat
{\mathbf{u}}- \hat{\mathbf{v}}
\otimes \hat{\mathbf{v}}\right) \,, \label{GWHIGHEReq2bis}
\end{eqnarray}
where $A=+,\times,B,C,D,s$ and~:  denotes
a contraction between tensors.  $\mathbf{D}$ is the
{\it detector tensor} representing  the response of a
laser-interferometric detector. It maps the  metric
perturbation in a  signal on the detector. The vectors
$ \hat{\mathbf{u}}$ and $\hat{\mathbf{v}}$ are unitary  and 
orthogonal to each other, they are directed to each detector 
arm,  and they form an orthonormal coordinate basis together with 
the unit vector $ \hat{\mathbf{w}}$ (see  
Fig.~\ref{GWHIGHERfig2}).
$\hat{\Omega}$ is the unit vector directed along the 
direction of  propagation of the gravitational wave.
Eq.~(\ref{GWHIGHEReq2}) holds only when the arm length of the
detector is  much  smaller than the gravitational wave   
wavelength, a condition satisfied by ground-based laser 
interferometers but  not by   space interferometers such as 
{\em LISA}.
\begin{figure}[h]
\begin{center}
\includegraphics[width=6.5cm]{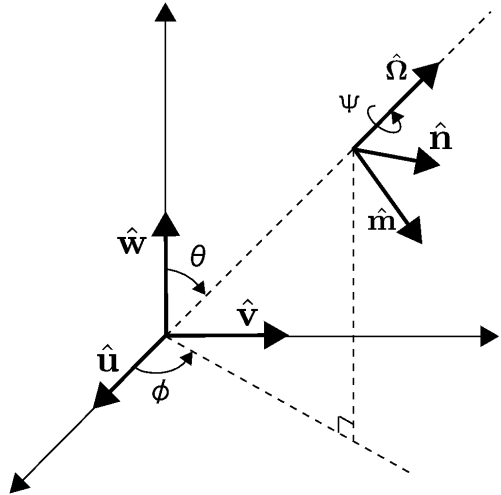}
\caption{The coordinate systems used to calculate the 
polarization  tensors and a view of the coordinate 
transformation.}
\label{GWHIGHERfig2}
\end{center}
\end{figure}
A standard orthonormal coordinate system for the detector is
\begin{eqnarray}
\hat{\mathbf{u}} & = &  \left( 1,0,0 \right) \,,\\
&&\nonumber\\
\hat{\mathbf{v}} & = &  \left( 0,1,0 \right) \,,\\
&&\nonumber\\
\hat{\mathbf{w}}  & = & \left( 0,0,1 \right) \,,
\end{eqnarray}
and the coordinate system for the gravitational 
wave, rotated by  $(\theta, \phi)$,  is given by
\begin{eqnarray}
\hat{\mathbf{u}}^{\prime} & = & \left( \cos \theta \cos \phi , 
\cos \theta \sin \phi , -\sin \theta \right) \,,\\
&&\nonumber\\
\hat{\mathbf{v}}^{\prime} & = &  \left( - \sin \phi , \cos \phi , 
0 \right) \,,\\
&&\nonumber\\
\hat{\mathbf{w}}^{\prime} & = &  \left( \sin \theta \cos \phi , 
\sin \theta \sin \phi , \cos \theta \right) \,.
\end{eqnarray}
A rotation by the angle $\psi$ around the direction of 
propagation of the gravitational wave  gives the 
most general choice of coordinates, that is
\begin{eqnarray}
\hat{\mathbf{m}} & = &  \hat{\mathbf{u}}^{ \prime} \cos \psi + 
\hat{\mathbf{v}}^{\prime} \sin \psi \,,\\
&&\nonumber\\
\hat{\mathbf{n}} &  =  & - \hat{\mathbf{v}}^{ \prime} \sin \psi + 
\hat{\mathbf{u}} ^{\prime} \cos \psi \,,\\
&&\nonumber\\
\hat{\Omega} & = &  \hat{\mathbf{w}}^{ \prime} \,.
\end{eqnarray}
The coordinates $ 
\left( \hat{\mathbf{u}}, \hat{\mathbf{v}}, \hat{\mathbf{w}} 
\right) $ are related to 
$ \left( \hat{\mathbf{m}}, 
\hat{\mathbf{n}}, \hat{\Omega} \right) $ by the
rotation angles $ \left( \phi,\,\theta,\,\psi \right)$, as 
shown in Fig.~\ref{GWHIGHERfig2}. Using the vectors 
$\hat{\mathbf{m}}$,
$\hat{\mathbf{n}}$, and $\hat{\Omega}$, the polarization
tensors are
\begin{eqnarray}
\mathbf{e}_{+} &=& \frac{1}{\sqrt{2}}\left(\hat{\mathbf{m}} 
\otimes \hat{\mathbf{m}} -\hat{\mathbf{n}}  \otimes 
\hat{\mathbf{n}}\right) \,,  \\
&&\nonumber\\
\mathbf{e}_{\times} &=& \frac{1}{\sqrt{2}}\left(  
\hat{\mathbf{m}} \otimes \hat{\mathbf{n}}  +\hat{\mathbf{n}} 
\otimes \hat{\mathbf{m}}\right) \,,  \\
&&\nonumber\\
\mathbf{e}_{B} &=& \frac{1}{\sqrt{2}}\left(\hat{\mathbf{m}}  
\otimes \hat{\Omega}  +\hat{\Omega} \otimes 
\hat{\mathbf{m}}\right) \,,  \\
&&\nonumber\\
\mathbf{e}_{C} &=&\frac{1}{\sqrt{2}} \left(\hat{\mathbf{n}}  
\otimes \hat{\Omega} +\hat{\Omega}  \otimes 
\hat{\mathbf{n}}\right) \,. \\
&&\nonumber\\
\mathbf{e}_{D} &=&\frac{\sqrt{3}}{2}\left(  
\hat{\mathbf{\frac{m}{2}}} \otimes \hat{\mathbf{\frac{m}{2}}}  + 
\hat{\mathbf{\frac{n}{2}}} \otimes \hat{\mathbf{\frac{n}{2}}}+  
\hat{\Omega} \otimes \hat{\Omega} \right) \,,\\
&&\nonumber\\
\mathbf{e}_{s} &=& \frac{1}{\sqrt{2}}\left( \hat{\Omega}  
\otimes \hat{\Omega}\right) \,.  
\end{eqnarray}
Taking into account Eqs.~(\ref{GWHIGHEReq2}) 
and~(\ref{GWHIGHEReq2bis}), the angular 
patterns for each polarization 
are
\begin{eqnarray}
F_{+}(\theta, \phi, \psi) &=&  \frac{1}{\sqrt{2}}  (1+ \cos ^2 
\theta ) \cos 2\phi \cos 2 \psi \nonumber\\
&&\nonumber\\
& - & \cos \theta \sin 2\phi \sin 2 \psi 
\,,\label{GWHIGHEReq5a}\\
&&\nonumber\\
F_{\times}(\theta,  \phi, \psi) &=& - \frac{1}{\sqrt{2}} (1+ \cos 
^2 \theta ) \cos 2\phi \sin 2 \psi \nonumber \\
&&\nonumber\\
& - & \cos \theta \sin 2\phi \cos 2 \psi \,,\label{GWHIGHEReq5b} 
\\
&&\nonumber\\
F_{B}(\theta,  \phi, \psi) &=& \sin \theta \left( \cos \theta 
\cos 2  \phi \cos \psi -\sin 2\phi \sin \psi \right) \,,  
\label{GWHIGHEReq5c}\\
&&\nonumber\\
F_{C}(\theta, \phi, \psi) &=& \sin \theta  \left(\cos \theta \cos 
2  \phi \sin \psi +\sin 2\phi \cos \psi \right) 
\,,\label{GWHIGHEReq5d} \\
&&\nonumber\\
F_{D}(\theta, \phi) &=& \frac{ \sqrt{3}}{32} \cos 2 \phi 
\left[ 6 \sin ^2\theta + \left( \cos 2 \theta +3 \right) \cos 2 
\psi \right] \,,
\label{GWHIGHEReq5e}\\
&&\nonumber\\
F_{s}(\theta,  \phi) &=& \frac{1}{\sqrt{2}} \sin^2 \theta \cos 
2\phi \,. \label{GWHIGHEReq5f}
\end{eqnarray}
The angular pattern functions 
for each  polarization are plotted  in Fig.~\ref{GWHIGHERfig:3}. 
Even if  we have  considered a  different model, these results 
are consistent,  for example,   with those of Refs.~\cite{abio, 
abio2, nishi,tobar,alggen1,alggen2}.

Another area of research  which we do not discuss here consists 
of the study of the stochastic background of gravitational 
waves  
which may contain  the possible  signature of extra 
gravitational wave modes and be relevant for the detectability of 
these contributions to gravitational radiation.

\begin{figure}[t]
\centering
{\includegraphics[width=0.4\textwidth]{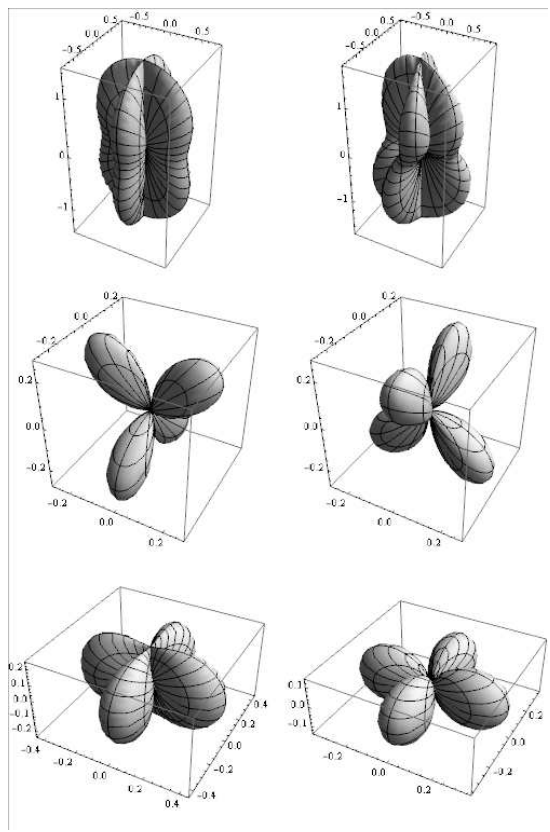}}
\caption{Angular pattern functions of an interferometric detector 
for the various polarizations. From  left to right and from top 
to bottom, one sees constant level surfaces 
corresponding to the ``plus'',  ``cross'',  B, C, D, and scalar 
modes.} 
\label{GWHIGHERfig:3}
\end{figure}

The above analysis covers extended gravity models with a generic
class of higher order Lagrangian densities and Lagrangian terms 
of the
form $ f \left( R,P,Q \right)$.  We have linearized the field 
equations of these theories around a Minkowski  background and 
found that, in addition to a massless spin two field,  the theory 
contains also spin zero  and  two massive modes with the 
latter being, in general,  ghosts. If the 
interferometer is 
directionally 
sensitive and we also know the orientation of the source (and, of 
course, if the source is coherent) the discussion is 
straightforward. In this case, the massive mode coming from the 
simplest extension of GR, {\em i.e.}, $f(R)$ gravity, would 
induce longitudinal  displacements along the direction of 
propagation of the 
wave, which should be detectable, and only the  scalar mode would 
be the detectable truly new signal 
\cite{CapozzielloCordaDeLaurentis08}. But, even 
in  this case, there could be a second scalar mode inducing a 
similar effect and representing a  massive
ghost, although with a negative sign.

For the situation considered here,  massive modes are certainly of
interest for the {\em LISA}  space interferometer. It 
is in principle possible that massive  gravitational wave  modes
could be produced in more significant quantities in cosmological
or early astrophysical processes in alternative theories of
gravity, a possibility which is still largely unexplored. This 
situation should be kept in mind when looking for a signature 
capable of distinguishing these theories from GR, and it seems to 
deserve further investigation.

\subsection{Concluding remarks}

The weak-field limit of ETGs shows new aspects of gravitation 
which are not present in GR. The Newtonian and post-Newtonian 
limits give weak-field potentials which are not of the standard 
Newtonian form. The corrections, in general, are Yukawa-like 
terms which could 
explain in a very natural fashion several astrophysical and 
cosmological observations \cite{prl,mnras2,noipla}.

The post-Minkowskian limit of ETGs exhibits new gravitational 
field modes which can easily be interpreted as massive gravitons.

The study of generation, propagation, and detection of 
GWs in the weak-field limit of a given 
relativistic theory of gravity is an important part of 
astrophysics.   Primordial gravitational waves  generated during 
the early  epochs of the Universe (especially during inflation) 
would  allow, when detected, to rule out or constrain certain 
theories and investigate others. The detection of GWs of  
astrophysical or cosmological origin can hardly be 
over emphasized because it would open a new branch of 
astronomy providing information which is not accessible with 
visible, infrared, optical, X-ray, or $\gamma$-ray astronomy. 
In fact,  GWs can be generated  in regions deep 
inside supernovae, near black hole horizons, or 
very early in 
the history of the Universe when the latter is completely opaque 
to  photons.  The study of relativistic astrophysics 
related to GWs in ETGs is a  broad and  complex 
subject  for which could open new interesting scenarios.

\section{ Conclusions and Perspectives }

In this Report, we have outlined the theoretical foundations of Extended Theories of Gravity, a new approach aimed to address and solve shortcomings and inconsistencies of General Relativity. These problems essentially come out at infra-red and ultra-violet scales, that is at cosmological and astrophysical scales from one side and at quantum scales from the other side. The issue is that to overcome the lack of explanation (at fundamental level) of cosmic dark sector and the lack of a self-consistent theory of quantum gravity, people are driving new patterns with two objectives: $i)$ maintaining the good results of GR by extending its approach; $ii)$ avoiding the introduction of new ingredients in the cosmic pie since no final evidence exists, up to now, for them. This could appear nothing else but a trial and error approach if a robust theoretical structure is missing.

We have shown, that ETGs can be traced back within the fold of gauge theories like GR. In fact gauge invariants, local gauge transformations, and symmetries plays the same fundamental role as in GR. As final result, GR is nothing else but a particular case of a more extended class of theories were higher-order curvature invariants and non-minimal couplings have to be considered.

Besides, space-time deformations play an interesting role in this line of thinking since it is clear that they are related not only with GR but with large families of theories. However the well founded results of GR must be recovered in any case. This means that the Equivalence Principle, the geodesic and causal structures, the post-Newtonian dynamics and the post-Minkowskian limit have to be constistently addressed by any ETGs. In our opinion, a part the urgency of a full quantization of gravity, this issues are the test-bed of any theory of gravity. In any case such features could result very different with respect to the Einstein theory but they have to be recovered as soon as ETGs $\longrightarrow$ GR. The first need of "corrections" to GR emerges when Quantum Field Theory is formulated on curved space-time. Normalization and regularization processes lead to non-minimal couplings and higher-order corrections in curvature invariants. These new terms do not give rise to finite renormalizable series but ensure regular perturbative behaviours at least at one-loop level. Starting from these considerations, it straightforward to observe that any unification theory predicts effective actions where higher-order curvature terms or non-minimal coupling are present. On the other hand, at larger scales, the Mach Principle asks for non-minimal couplings and cosmological issues as inflation and dark energy can be addressed in the realm of these enlarged schemes. In this sense, ETGs have become a paradigm in modern theoretical physics. In the track of GR, variational principles and field equations can be obtained for any ETGs. However the field equations can result of order higher than second and very difficult to handle. Instead of GR, metric structure and geodesic structure cannot coincide and  then dynamics is described by the metric field $g$ and the connection field $\Gamma$. In this case, we deal with the Palatini or metric-affine formalism. The standard metric formalism is recovered by observing that the Palatini approaches endowed with a bi-metric structure  is strictly connected to the conformal structure.   However we have to point out that, in theories like $f(R, R_{\mu\nu}R^{\mu\nu})$ as those discussed in \cite{barraolmo,limota,olmoalepuz}, the bi-metric structure is much more general than the conformal structure. Specifically, it  can be discussed in the most general frame of deformations considered in Sec IV.

At this point, it is worth stressing that the whole geometric budget of a generic relativistic theory of gravity could include also torsion beside curvature. It is well known that such an ingredient can represent the geometric counterpart of spin as mass-energy is the source of curvature. We discussed in detail the role of torsion in $f(R)$-gravity showing the straightforward relation with the metric-affine formalism and scalar tensor theories. In a genuine gauge-formalism, torsion, curvature, and matter can be represented in the ${\cal J}$-bundle framework. This picture is particularly relevant if we want to put in evidence symmetries and conserved quantities. We have taken into account the case of  $f(R)$-gravity with different matter fields acting as source in the field equations.

Also the Hamiltonian representation can give an interesting insight  for ETGs. For example, the cosmological constant can be recovered as an eigenvalue of the $f(R)$-Hamiltonian constructed  by the ADM formalism. It is straightforward to achieve one-loop energy regularization and renormalization by this approach. 

One the most important issues related to ETGs is the Initial Value Problem. In order to construct a self-consistent theory of physics, as GR or Electromagnetism, this problem needs to be well-formulated and well-posed. We have discussed the Cauchy problem for scalar tensor and $f(R)$-gravity adopting the ADM formulation and the Gaussian normal coordinates approach. It results that well-formulation and well-position are always possible in metric formalism. In the metric-affine approach, the well-formulation is recovered while the well-position strictly depends on the matter-field acting as source. The Initial Value Poblem is still an open question that needs to be fully addressed for the self-consistency of ETGs.  However, as recently reported in \cite{olmoalepuz2}, the ADM methods results extremely suitable to investigate  such a problem in the Palatini formalism indicating that not only the well-formulation but also the  well-position is at hand. 

One of the crucial achievement of a give theory of physics is the possibility to get exact solutions. In this case, one has the full control of initial conditions and dynamics. Among the applications of ETGs, we have discussed the spherical and axial symmetry with the aim to get exact solutions. It is interesting to see that while spherical solutions are essentially Schwarzschild-like in GR (with constant or null curvature), in $f(R)$-gravity we have several possibilities essentially related to the fact that also curvature dependent on radial coordinate has to be considered. We have developed various techniques to get exact spherical solutions and have shown how it is possible to generate axially symmetric solutions, starting from spherical ones, by the Newman-Janis algorithm.

The post-Newtonian limit of ETGs gives rise to several results that cannot be found in standard GR. The main achievement is the fact that Newtonian potential has to be corrected thanks to the higher-order and non-minimal coupling terms in the field equations. The corrections are essentially Yukawa-like and the physical implication is that new characteristic lengths emerge beside the standard Schwarzschild radius related to the gravitational mass. This feature can have dramatic consequences in astrophysics and cosmology as discussed in \cite{CapCardTro07,mauro,mnras2}. An implication could emerge also for quantum effects like neutrino oscillations when considered in presence of gravitational fields \cite{vernieri}.

 Also the post-Minkowskian limit of ETGs deserves an accurate consideration, in particular with respect to the problem of gravitational radiation. In fact, new polarizations emerge as soon as the gravitational action is not the Hilbert-Einstein one: in general, massive, massless, and ghost modes have to be taken into account, while only massless modes and two polarizations are present in GR. This result implies a revision of the physics of GWs. In fact, detectors should be designed and calibrated also in view of these possibilities. On the other hand, the detection of gravitational massive modes (or modes different from those predicted by GR) could be the final "experimentum crucis" for ETGs.

 In conclusion, we can say that such theories seem a reliable approach to address the gravity problem at various scales also if no final result is available up to day. In this Report we have investigated the theoretical foundations of ETGs in order to show that they are not only phenomenological toy models but satisfy the issues of modern gauge theories. The validity of this approach could be definitely confirmed or ruled out by forthcoming experiments ({\it e.g.} at LHC see \cite{lhcmasses}) or by fine and comprehensive astrophysical observations. 
 
 \section*{Acknowledgments}
We are indebted to many collaborators and colleagues for 
discussions, results and remarks reported in this work. 
 In particular we thank  G. Basini, T. Bernal, C. Bogdanos, M. Bouhmadi L\'opez, S. Carloni, V.F. Cardone, R. Cianci, F. Darabi, A. De Felice, M. Demianski, R. de Ritis, P.K.S. Dunsby, E. Elizalde, L. Fabbri, V. Faraoni, L. Fatibene, M. Francaviglia, R. Garattini, X. Hernandez, J. C. Hidalgo, G. Lambiase, S. Mendoza, S. Mercadante, D.F. Mota, S. Nesseris, S. Nojiri, S.D. Odintsov, G. J. Olmo, L. Perivolaropoulos, E. Piedipalumbo, C. Rubano, D. Saez-Gomez, V. Salzano, P. Scudellaro, A. Stabile, C. Stornaiolo, A. Troisi, S. Tsujikawa, D. Vernieri, S. Vignolo.  
 We wish to thank Prof. R. Petronzio for his kind invitation to write this Report. We are supported by {\it Istituto Nazionale di Fisica Nucleare} (INFN) and by the {\it Università di Napoli  "Federico~II''}.

\section*{Appendix A. Notations}

\begin{tabular}{ll}

Mathematical symbols &\\
\hline & \\
$\partial _{\mu }=\frac{\partial }{\partial x^{\mu }}$:  & Partial derivative
with respect to $\left\{ x_{\mu }\right\} $\\

$\left\{ e_{\mu }\right\} $ : & Set with elements $e_{\mu }$\\

$\nabla _{\mu }=\partial _{\mu }+\Gamma _{\mu }$ : & Gauge covariant derivative operator\\

$\Gamma _{\mu }$ : & Gauge potential 1-form \\

$d$ : & Exterior derivative operator\\

$\left\langle V|e\right\rangle $ :  & Inner multiplication between vector $e$
and 1-form $V$\\

$\left[ A\text{, }B\right] $ : & Commutator of operators $A$ and $B$\\

$\left\{ A\text{, }B\right\} $ : & Anti-commutator of operators $A$ and $B$\\

$\wedge $ : & Exterior multiplication operator\\

$\rtimes $ : & Semi-direct product\\

${\cal M}$ :  & differential manifold\\

${\cal U}_4$ : & $4D$-manifold with torsion\\

${\cal V}_4$ : & $4D$ pseudo-Riemaniann manifold without torsion\\

${\cal E}$ : & co-frame bundle of ${\cal M}$\\

${\cal J}$ : & jet-bundle\\

$\times $ : & Direct product\\

$\times _{\cal M}$ : & Fibered product over manifold ${\cal M}$\\

$\interior$: & Inner product\\

$\oplus $ : & Direct sum\\

$\otimes $: & Tensor product\\

$A\cup B$ :  & Union of $A$ and $B$\\

$A\cap B$ : & Intersection of $A$ and $B$\\

$\Theta $ ($\overline{\Theta }$) : & Right (left) invariant Maurer-Cartan
1-form\\

$\circ $ : & Group (element) composition operator\\

$\eta_{\alpha \beta }=diag(-1$, $1$, $1$, $1)$ : & Lorentz group metric (flat metric)\\

$GL\left( 4\text{, }
\mathbb{R}\right) $ : & Group of real $4\times 4$ invertible matrices\\

$SO(3$, $1)$ : & Lorentz group\\

$P(3$, $1)$ : & Poincar\'{e} group\\

$\mathfrak{g}$ : & Lie algebra of group\ $G$\\

$g\in G$ :& Element $g$ of $G$\\

$\left\{ \mathcal{U}\right\} \subset {\cal M}$ : & Set $\mathcal{U}$ is a subset of ${\cal M} $\\

$\mathbf{G}$ : &Algebra generator of group $G$\\

$\rho \left( \mathbf{G}\right) $ : & Representation of $G$-algebra\\

$^{\ast }A$ : & Dual of $A$ with respect to (coordinate) basis indices\\

$\epsilon _{a_{1}...a_{n}}$ or $\varepsilon _{a_{1}...a_{n}}$ :  & Levi-Civita
totally skew tensor density\\

$\eta _{a_{1}...a_{n}}$ : & Eta basis volume $n$-form density\\

$\sigma ^{\ast }$ : & Pullback by local section $\sigma $\\

$L_{h\ast }$ : & Differential (pushforward) map induced by $L_{h}$\\

$T_{\left( a_{1}...a_{n}\right) }$ : & Symmetrization of indices \\

$T_{\left[ a_{1}...a_{n}\right] }$ :  & Antisymmetrization of indices\\

$T({\cal M})$ :  & Tangent space to manifold ${\cal M}$\\

$f:A\rightarrow B$ : & Map $f$ taking elements $\left\{ a\right\} \in A$ to $
\left\{ b\right\} \in B$\\

\end{tabular}
\vskip+1cm 


\begin{tabular}{ll}

Convention for the gravitational field&\\
\hline & \\
Standard gravitational coupling & $\displaystyle{\kappa=\frac{8\pi G_N}{c^4}}$\\  
Index ranges & $\alpha,\beta\,=\,0,1,2,3$; $i,j\,=\,1,2,3$\\
Coordinates &$x^\alpha=(x^0,x^1,x^2,x^3)=(ct,x^1,x^2,x^3)$\\
Vectors & \b{v} = $(v^1,v^2,v^3)$; $\nabla=(\partial/\partial x^1,\partial/\partial x^2,\partial/\partial x^3) $ \\
Symmetrization & $T_{(\alpha | \beta \dots \gamma |
\delta)}=\frac{1}{2}(T_{\alpha \beta \dots
\gamma\delta}+T_{\delta \beta\dots \gamma\alpha})$ \\
Kronecker &  $\delta^\alpha_\beta$ = 1 if $\alpha = \beta$, $0$ else\\
Connection & $\Gamma^\alpha_{\mu \nu}=\frac{1}{2}g^{\alpha\sigma}\left( g_{\mu\sigma,\nu}+g_{\nu\sigma,\mu}-g_{\mu\nu,\sigma}\right)$\\
Riemann tensor & $R^{\alpha }{}_{\beta \mu \nu } =\Gamma _{\beta
\nu ,\mu }^{\alpha }-\Gamma _{\beta \mu ,\nu }^{\alpha}+\Gamma
_{\beta \nu }^{\sigma }\Gamma _{\sigma \mu}^{\alpha
}-\Gamma_{\beta \mu }^{\sigma }
\Gamma_{\sigma \nu }^{\alpha }$ \\
Ricci tensor & $R_{\mu \nu } =R^{\sigma }{}_{\mu \sigma \nu  }$\\

\end{tabular}

\section*{Appendix B. Zeta function regularization}

\label{app}In this appendix, we report details on computation leading to
expression $\left(  \ref{zeta}\right)  $. We begin with the following integral%
\begin{equation}
\rho\left(  \varepsilon\right)  =\left\{
\begin{array}
[c]{c}%
I_{+}=\mu^{2\varepsilon}\int_{0}^{+\infty}d\omega\frac{\omega^{2}}{\left(
\omega^{2}+m^{2}\left(  r\right)  \right)  ^{\varepsilon-\frac{1}{2}}}\\
\\
I_{-}=\mu^{2\varepsilon}\int_{0}^{+\infty}d\omega\frac{\omega^{2}}{\left(
\omega^{2}-m^{2}\left(  r\right)  \right)  ^{\varepsilon-\frac{1}{2}}}%
\end{array}
\right.  , \label{rho}%
\end{equation}
with $m^{2}\left(  r\right)  >0$.

\subsection*{$I_{+}$ computation}

\label{app1}If we define $t=\omega/\sqrt{m^{2}\left(  r\right)  }$, the
integral $I_{+}$ in Eq.$\left(  \ref{rho}\right)  $ becomes%
\[
\rho\left(  \varepsilon\right)  =\mu^{2\varepsilon}m^{4-2\varepsilon}\left(
r\right)  \int_{0}^{+\infty}dt\frac{t^{2}}{\left(  t^{2}+1\right)
^{\varepsilon-\frac{1}{2}}}%
\]%
\[
=\frac{1}{2}\mu^{2\varepsilon}m^{4-2\varepsilon}\left(  r\right)  B\left(
\frac{3}{2},\varepsilon-2\right)
\]%
\[
\frac{1}{2}\mu^{2\varepsilon}m^{4-2\varepsilon}\left(  r\right)  \frac
{\Gamma\left(  \frac{3}{2}\right)  \Gamma\left(  \varepsilon-2\right)
}{\Gamma\left(  \varepsilon-\frac{1}{2}\right)  }%
\]%
\begin{equation}
=\frac{\sqrt{\pi}}{4}m^{4}\left(  r\right)  \left(  \frac{\mu^{2}}%
{m^{2}\left(  r\right)  }\right)  ^{\varepsilon}\frac{\Gamma\left(
\varepsilon-2\right)  }{\Gamma\left(  \varepsilon-\frac{1}{2}\right)  },
\end{equation}
where we have used the following identities involving the beta function%
\begin{equation}
B\left(  x,y\right)  =2\int_{0}^{+\infty}dt\frac{t^{2x-1}}{\left(
t^{2}+1\right)  ^{x+y}}\qquad\operatorname{Re}x>0,\operatorname{Re}y>0
\end{equation}
related to the gamma function by means of%
\begin{equation}
B\left(  x,y\right)  =\frac{\Gamma\left(  x\right)  \Gamma\left(  y\right)
}{\Gamma\left(  x+y\right)  }.
\end{equation}
Taking into account the following relations for the $\Gamma$-function%
\begin{equation}%
\begin{array}
[c]{c}%
\Gamma\left(  \varepsilon-2\right)  =\frac{\Gamma\left(  1+\varepsilon\right)
}{\varepsilon\left(  \varepsilon-1\right)  \left(  \varepsilon-2\right)  }\\
\\
\Gamma\left(  \varepsilon-\frac{1}{2}\right)  =\frac{\Gamma\left(
\varepsilon+\frac{1}{2}\right)  }{\varepsilon-\frac{1}{2}}%
\end{array}
, \label{gamma}%
\end{equation}
and the expansion for small $\varepsilon$%
\begin{equation}%
\begin{array}
[c]{cc}%
\Gamma\left(  1+\varepsilon\right)  = & 1-\gamma\varepsilon+O\left(
\varepsilon^{2}\right) \\
& \\
\Gamma\left(  \varepsilon+\frac{1}{2}\right)  = & \Gamma\left(  \frac{1}%
{2}\right)  -\varepsilon\Gamma\left(  \frac{1}{2}\right)  \left(  \gamma
+2\ln2\right)  +O\left(  \varepsilon^{2}\right) \\
& \\
x^{\varepsilon}= & 1+\varepsilon\ln x+O\left(  \varepsilon^{2}\right)
\end{array}
,\qquad
\end{equation}
where $\gamma$ is the Euler's constant, we find%
\begin{equation}
\rho\left(  \varepsilon\right)  =-\frac{m^{4}\left(  r\right)  }{16}\left[
\frac{1}{\varepsilon}+\ln\left(  \frac{\mu^{2}}{m^{2}\left(  r\right)
}\right)  +2\ln2-\frac{1}{2}\right]  .
\end{equation}

\subsection*{$I_{-}$ computation}

\label{app2}If we define $t=\omega/\sqrt{m^{2}\left(  r\right)  }$, the
integral $I_{-}$ in Eq.$\left(  \ref{rho}\right)  $ becomes%
\[
\rho\left(  \varepsilon\right)  =\mu^{2\varepsilon}m^{4-2\varepsilon}\left(
r\right)  \int_{0}^{+\infty}dt\frac{t^{2}}{\left(  t^{2}-1\right)
^{\varepsilon-\frac{1}{2}}}%
\]%
\[
=\frac{1}{2}\mu^{2\varepsilon}m^{4-2\varepsilon}\left(  r\right)  B\left(
\varepsilon-2,\frac{3}{2}-\varepsilon\right)
\]%
\[
\frac{1}{2}\mu^{2\varepsilon}m^{4-2\varepsilon}\left(  r\right)  \frac
{\Gamma\left(  \frac{3}{2}-\varepsilon\right)  \Gamma\left(  \varepsilon
-2\right)  }{\Gamma\left(  -\frac{1}{2}\right)  }%
\]%
\begin{equation}
=-\frac{1}{4\sqrt{\pi}}m^{4}\left(  r\right)  \left(  \frac{\mu^{2}}%
{m^{2}\left(  r\right)  }\right)  ^{\varepsilon}\Gamma\left(  \frac{3}%
{2}-\varepsilon\right)  \Gamma\left(  \varepsilon-2\right)  ,
\end{equation}
where we have used the following identity involving the beta function%
\begin{equation}%
\begin{array}
[c]{c}%
\frac{1}{p}B\left(  1-\nu-\frac{\mu}{p},\nu\right)  =\int_{1}^{+\infty
}dtt^{\mu-1}\left(  t^{p}-1\right)  ^{\nu-1}\\
\\
p>0,\operatorname{Re}\nu>0,\operatorname{Re}\mu<p-p\operatorname{Re}\nu
\end{array}
\end{equation}
and the reflection formula%
\begin{equation}
\Gamma\left(  z\right)  \Gamma\left(  1-z\right)  =-z\Gamma\left(  -z\right)
\Gamma\left(  z\right)
\end{equation}
From the first of Eqs.$\left(  \ref{gamma}\right)  $ and from the expansion
for small $\varepsilon$%
\[
\Gamma\left(  \frac{3}{2}-\varepsilon\right)  =\Gamma\left(  \frac{3}%
{2}\right)  \left(  1-\varepsilon\left(  -\gamma-2\ln2+2\right)  \right)
+O\left(  \varepsilon^{2}\right)
\]%
\begin{equation}
x^{\varepsilon}=1+\varepsilon\ln x+O\left(  \varepsilon^{2}\right)  ,
\end{equation}
we find%
\begin{equation}
\rho\left(  \varepsilon\right)  =-\frac{m^{4}\left(  r\right)  }{16}\left[
\frac{1}{\varepsilon}+\ln\left(  \frac{\mu^{2}}{m^{2}\left(  r\right)
}\right)  +2\ln2-\frac{1}{2}\right]  .
\end{equation}

\section*{Appendix C. The field equations and the Noether vector for 
spherically symmetric $f(R)$-gravity}

The field equations of metric $f(R)$-gravity with spherical 
symmetry are
\begin{eqnarray}
H_{00} \, & = & \,  
2A^2B^2Mf+\left\{B M A'^2 - A\left[ 2BA'M' + 
M\left( 2BA''-A'B' \right) \right] \right\} f_{R} \nonumber\\
\nonumber\\
& + & \left( -2A^2 MB'R' + 4A^2BM'R'+ 4 A^2BMR'' \right) f_{RR} 
\nonumber\\
&& \nonumber\\
& + & \, 4 A^2 BMR'^2 f_{RRR}=0 \,,
\end{eqnarray}
\begin{eqnarray}
H_{rr} \, &  = & \, 2\,A^2 B^2M^2 f + \left( BM^2A'^2 + AM^2A'B' 
+ 2A^2MB'M' + 2A^2BM'^2 \right. \nonumber\\
\nonumber\\
&- & \left. 2ABM^2A''-4A^2BMM'' \right) f_{R}+ \left( 2ABM^2A'R' 
\right. \nonumber\\
&& \nonumber\\
& + & \left. 4A^2BMM'R' \right) f_{RR}=0\,,
\end{eqnarray}

\begin{eqnarray}
H_{\theta\theta} \, & = &  \, 2AB^2Mf +  \left( 
4AB^2 - BA'M'+AB'M'-2ABM'' 
\right) f_{R} \nonumber\\
&& \nonumber\\
& + &  \left( 2BMA'R'-2AMB'R'+2ABM'R'+4ABMR'' \right)
f_{RR} \nonumber\\
&& \nonumber\\
& + & 4ABMR'^2f_{RRR}=0\,,\nonumber\\
&&\nonumber\\
H_{\varphi\varphi} & = & \, \sin^2 \theta \, 
H_{\theta\theta}=0\,.
\end{eqnarray}
The trace of the field equations is
\begin{eqnarray}
H & = &  \, g^{\mu\nu} H_{\mu\nu}=4AB^2Mf-2AB^2MRf_{R}+ 
3\left( BMA'R'-AMB'R' \right. \nonumber\\
&& \nonumber\\
& + & \left. 2ABM'R' +2ABMR'' \right) f_{RR}+ 6 ABMR'^2f_{RRR}=0 
\,.
\end{eqnarray}
The system~(\ref{NSsys}) is derived from  the condition for the 
existence  of a Noether symmetry ${\cal L}_{\mathbf{X}} L=0$.  
Considering the  configuration space 
$\underline{q}= \left( A\,,M\,,R \right)$ and defining the 
Noether vector
 $\underline{\alpha}=(\alpha_1\,,\alpha_2\,,\alpha_3)$,
the system (\ref{NSsys}) assumes the explicit form
\begin{eqnarray}
&& \xi \biggl( \frac{\partial\alpha_2}{\partial A}f_{R}
+M \, \frac{\partial\alpha_3}{\partial A}f_{RR}\biggr)=0 \,,\\
&&\nonumber\\
&& \frac{A}{M}\biggl[ \left( 2 +MR \right) \alpha_3 
f_{RR}-\frac{2\alpha_2}{M} f_{R}\biggr] f_{R}\nonumber\\
&& \nonumber\\
&& +\xi\biggl[\biggl( \frac{\alpha_1}{M} 
+ 2\, \frac{\partial\alpha_1}{\partial M} 
+ \frac{2A}{M} \, \frac{\partial\alpha_2}{\partial M} 
\biggr) f_{R} + A\biggl( \frac{\alpha_3}{M} + 4\, \frac{\partial 
\alpha_3 }{\partial M} \biggr)f_{RR}\biggr] = 0 \,,\\
&&\nonumber\\
&& \xi\biggl(M\frac{\partial \alpha_1}{\partial R}
+ 2 A \frac{\partial \alpha_2}{\partial R} \biggr) f_{RR}=0 
\,,\\
&&\nonumber\\
&& \alpha_2 \left( f-Rf_{R}\right) f_{R} 
-\xi \left[ \biggl( \alpha_3+M \, \frac{\partial\alpha_3}
{\partial M} + 2A \, \frac{\partial\alpha_3}{\partial
A}\biggr) f_{RR} \right. \nonumber\\
&&\nonumber\\
&& \left. + \biggl( \frac{\partial\alpha_2}{\partial M} + 
\frac{\partial\alpha_1}{\partial A} 
+\frac{A}{M}\frac{\partial \alpha_2}{\partial A} 
\biggr) f_{R} \right]=0 \,,\\
&&\nonumber\\
&& \left[ M \left( 2+MR \right) \alpha_3 f_{RR}-2\alpha_2 
f_{R} \right] f_{RR} + \xi \left[ f_{R} \, 
\frac{\partial \alpha_2}{\partial R} \right.\nonumber\\
&& \nonumber\\
&&\left. + \biggl( 2\alpha_2 + M \, 
\frac{\partial\alpha_1 }{ \partial A} + 2 A\, 
\frac{\partial\alpha_2}{\partial A} 
+ M \, \frac{\partial \alpha_3}{\partial R} \biggr) \alpha_3 
f_{RR} + M f_{RRR} \right] =0 \,,
\end{eqnarray}
\begin{eqnarray}
&& 2A \left[ \left( 2+MR \right) \alpha_3 
f_{RR}-\left( f-Rf_{R} \right) \alpha_2 \right] 
f_{RR} + \xi\left[  \biggl( \frac{\partial\alpha_1}{\partial
R}+\frac{A}{M} \, \frac{\partial\alpha_2}{\partial
R} \biggr) f_{R}  \right. \nonumber\\
&& \nonumber\\
&& \left. +\biggl( 2\alpha_1 +2A \, 
\frac{\partial\alpha_3}{\partial R} + M\, 
\frac{\partial \alpha_1}{\partial M} + 2A 
\, \frac{\partial \alpha_2}{\partial M} \biggr) f_{RR} 
+ 2A \, \alpha_3 f_{RRR} \right]=0  \,,
\end{eqnarray}
with the condition $\xi= \left( 2+MR \right) f_{R}-M f\neq 0$ to 
guarantee the  non-vanishing of  the
Hessian of the Lagrangian~(\ref{NSlag2}).


\end{document}